\documentclass[twocolumn]{book}

\usepackage[utf8]{inputenc}
\usepackage[T1]{fontenc}

\usepackage{makeidx}
\usepackage{import}
\usepackage{bigints}
\usepackage{algorithm}
\usepackage{algpseudocode}
\usepackage{mathtools}
\usepackage[columnsep=1cm, top=2cm, bottom=2cm, left=1cm, right=1cm]{geometry}
\usepackage{amsmath,amsthm,amssymb}
\usepackage{graphicx}
\usepackage{float}
\usepackage{amsmath, amssymb, amscd}
\usepackage{alltt}
\usepackage{textcomp}
\usepackage{gensymb}
\usepackage{multicol}
\usepackage{tabularx}
\usepackage{subcaption}
\usepackage[nice]{units}
\usepackage[mathscr]{euscript}
\usepackage{fouriernc}
\usepackage{bm}
\DeclareMathAlphabet{\mathcal}{OMS}{cmsy}{m}{n}   % or txsy
\SetMathAlphabet{\mathcal}{bold}{OMS}{cmsy}{b}{n} % or txsy
\DeclareMathAlphabet{\pazocal}{OMS}{zplm}{m}{n}

\newcommand{\R}{\mathbb{R}}
\newcommand{\bigo}{\mathcal{O}}

\newcommand{\Lu}{\mathcal{L}}
\newcommand{\T}{^\intercal}

\newcommand{\tr}{\mathrm{tr}}

\newcommand{\testspace}{S_0^h \subset \mathcal{H}_{E_0}^1(\Omega)}
\newcommand{\trialspace}{S_E^h \subset \mathcal{H}_E^1(\Omega)}
\newcommand{\ltwospace}{M^h \subset L^2(\Omega)}

\DeclareMathOperator*{\argmin}{argmin}

\newcommand*{\argminl}{\argmin\limits}

\newfloat{myalgo}{tbhp}{mya}

{\begin{myalgo}[#1]
\centering
\begin{minipage}{#2}
\begin{algorithm}[H]}%
{\end{algorithm}
\end{minipage}
\end{myalgo}}

\newcommand{\sups}[1]{\ensuremath{^{\textrm{#1}}}}

\makeatletter
\newsavebox{\mybox}\newsavebox{\mysim}
\newcommand{\distras}[1]
{
  \savebox{\mybox}{\hbox{\kern3pt$\scriptstyle#1$\kern3pt}}%
  \savebox{\mysim}{\hbox{$\sim$}}%
  \mathbin{\overset{#1}{\kern\z@\resizebox{\wd\mybox}{\ht\mysim}{$\sim$}}}%
}
\makeatother

\makeatletter
\renewcommand*\env@matrix[1][c]{\hskip -\arraycolsep
  \let\@ifnextchar\new@ifnextchar
  \array{*\c@MaxMatrixCols #1}}
\makeatother

\usepackage[final]{listings}

\usepackage{color}
\definecolor{bg}{rgb}{0.96,0.96,0.85}
\definecolor{deepblue}{rgb}{0,0,0.5}
\definecolor{deepred}{rgb}{0.6,0,0}
\definecolor{deepgreen}{rgb}{0,0.5,0}

\usepackage{xcolor}

\definecolor{gray}{gray}{0.5}
\colorlet{commentcolour}{green!50!black}

\colorlet{stringcolour}{red!60!black}
\colorlet{keywordcolour}{magenta!90!black}
\colorlet{exceptioncolour}{yellow!50!red}
\colorlet{commandcolour}{blue!60!black}
\colorlet{numpycolour}{blue!60!green}
\colorlet{literatecolour}{magenta!90!black}
\colorlet{promptcolour}{green!50!black}
\colorlet{specmethodcolour}{violet}
\colorlet{indendifiercolour}{green!70!white}

\newcommand{\literatecolour}{\textcolor{literatecolour}}

\newcommand\Small{\fontsize{1.00}{5.0}\selectfont}

\newcommand\pythonstyle{\lstset{
language=python,
showtabs=true,
tab=,
tabsize=2,
basicstyle=\ttfamily\Small,%\setstretch{.5},
stringstyle=\color{stringcolour},
showstringspaces=false,
alsoletter={1234567890},
otherkeywords={\ , \}, \{, \%, \&, \|},
keywordstyle=\color{keywordcolour}\bfseries,
emph={and,break,class,continue,def,yield,del,elif ,else,%
except,exec,finally,for,from,global,if,import,in,%
lambda,not,or,pass,print,raise,return,try,while,assert},
emphstyle=\color{blue}\bfseries,
emph={[2]True, False, None},
emphstyle=[2]\color{keywordcolour},
emph={[3]object,type,isinstance,copy,deepcopy,zip,enumerate,reversed,list,len,dict,tuple,xrange,append,execfile,real,imag,reduce,str,repr},
emphstyle=[3]\color{commandcolour},
emph={Exception,NameError,IndexError,SyntaxError,TypeError,ValueError,OverflowError,ZeroDivisionError},
emphstyle=\color{exceptioncolour}\bfseries,
morestring=[s]{"""}{"""},
morestring=[s]{'''}{'''},
commentstyle=\color{commentcolour}\slshape,
emph={[4]ode, fsolve, sqrt, exp, sin, cos, arccos, pi,  array, norm, solve, dot, arange, , isscalar, max, sum, flatten, shape, reshape, find, any, all, abs, linspace, legend, quad, polyval,polyfit, hstack, concatenate,vstack,column_stack,empty,zeros,ones,rand,vander,grid,pcolor,eig,eigs,eigvals,svd,qr,tan,det,logspace,roll,min,mean,cumsum,cumprod,diff,vectorize,lstsq,cla,eye,xlabel,ylabel,squeeze,plot,median,std,hist},
emphstyle=[4]\color{numpycolour},
emph={[5]__init__,__add__,__mul__,__div__,__sub__,__call__,__getitem__,__setitem__,__eq__,__ne__,__nonzero__,__rmul__,__radd__,__repr__,__str__,__get__,__truediv__,__pow__,__name__,__future__,__all__},
emphstyle=[5]\color{specmethodcolour},
emph={[6]assert,range,yield},
emphstyle=[6]\color{keywordcolour}\bfseries,
literate=*%
{:}{{\literatecolour:}}{1}%
{=}{{\literatecolour=}}{1}%
{-}{{\literatecolour-}}{1}%
{+}{{\literatecolour+}}{1}%
{*}{{\literatecolour*}}{1}%
{/}{{\literatecolour/}}{1}%
{!}{{\literatecolour!}}{1}%
{[}{{\literatecolour[}}{1}%
{]}{{\literatecolour]}}{1}%
{<}{{\literatecolour<}}{1}%
{>}{{\literatecolour>}}{1}%
{>>>}{{\textcolor{promptcolour}{>>>}}}{1}%
,%
breaklines=true,
breakatwhitespace= true,
aboveskip=1ex,
frame=trbl,
rulecolor=\color{black!40},
backgroundcolor=\color{yellow!10}
}}

\lstnewenvironment{python}[1][]
{
  \pythonstyle
  \lstset{#1}
}
{}

\newcommand{\pythonexternal}[2][]
{{
  \pythonstyle
  \lstinputlisting[#1]{#2}
}}

\newcommand\pythoninline[1]
{{
  \pythonstyle
  \lstinline!#1!
}}

\usepackage{etoolbox}
\let\bbordermatrix\bordermatrix
\patchcmd{\bbordermatrix}{8.75}{4.75}{}{}
\patchcmd{\bbordermatrix}{\left(}{\left[}{}{}
\patchcmd{\bbordermatrix}{\right)}{\right]}{}{}

\usepackage[backend=bibtex, style=authoryear, sorting=nyt, natbib]{biblatex}

\usepackage{tocloft}
\cftsetindents{section}{1em}{3em}
\cftsetindents{subsection}{3em}{3em}

\makeindex

\begin{document}

\frontmatter

\begin{titlepage}
  
  \begin{center}

  MODELING THE CRYOSPHERE WITH FEniCS
  
  \vspace{8mm} 
  
  By
  
  \vspace{8mm}
  
  EVAN MICHAEL CUMMINGS

  Associates of Applied Art in Audio Production, The Art Institute of Seattle, WA, 2008

  Bachelor of Art in Mathematics, University of Montana, Missoula, 2013

  Bachelor of Science in Computer Science, University of Montana, Missoula, 2013

  \vspace{8mm}

  Thesis Paper

  \vspace{8mm}

  presented in partial fulfillment of the requirements for the degree of

  \vspace{8mm}

  Master of Science in Computer Science

  \vspace{8mm}

  The University of Montana

  Missoula, Montana

  \vspace{8mm}

  August 2016

  \vspace{8mm}

  Approved by:

  \vspace{8mm}

  Scott Whittenburg, Dean of The Graduate School

  Graduate School

  \vspace{8mm}

  Douglas Raiford, Chair

  Department of Computer Science

  \vspace{8mm}

  Travis Wheeler

  Department of Computer Science

  \vspace{8mm}

  Johnathan Bardsley

  Department of Mathematics
  
  \end{center}

  \newpage

  \noindent Cummings, Evan, Master of Science, August 2016 \hspace{88mm} Computer Science

  \vspace{8mm}

  \noindent Modeling the Cryosphere with FEniCS

  \vspace{8mm}

  \noindent Chairperson: Douglas Raiford

  \vspace{8mm}

  This manuscript is a collection of problems and solutions related to modeling the cryosphere using the finite element software FEniCS.  Included is an introduction to the finite element method; solutions to a variety of problems in one, two, and three dimensions; an overview of popular stabilization techniques for numerically-unstable problems; and an introduction to the governing equations of ice-sheet dynamics with associated FEniCS implementations.  The software developed for this project, Cryospheric Problem Solver (CSLVR), is fully open-source and has been designed with the goal of simplifying many common tasks associated with modeling the cryosphere.  CSLVR possesses the ability to download popular geological and geographical data, easily convert between geographical projections, develop sophisticated two- or three-dimensional finite-element meshes, convert data between many popular formats, and produce production-quality images of data.  Scripts are presented which model the flow of ice using geometry defined by mathematical functions and observed Antarctic and Greenland ice-sheets data.  A new way of solving the internal energy distribution of ice to match observed intra-ice water contents within temperate regions is thoroughly explained.

\vspace{20mm}

\begin{figure}[H]
  \centering
    \includegraphics[width=0.6\linewidth,draft=false]{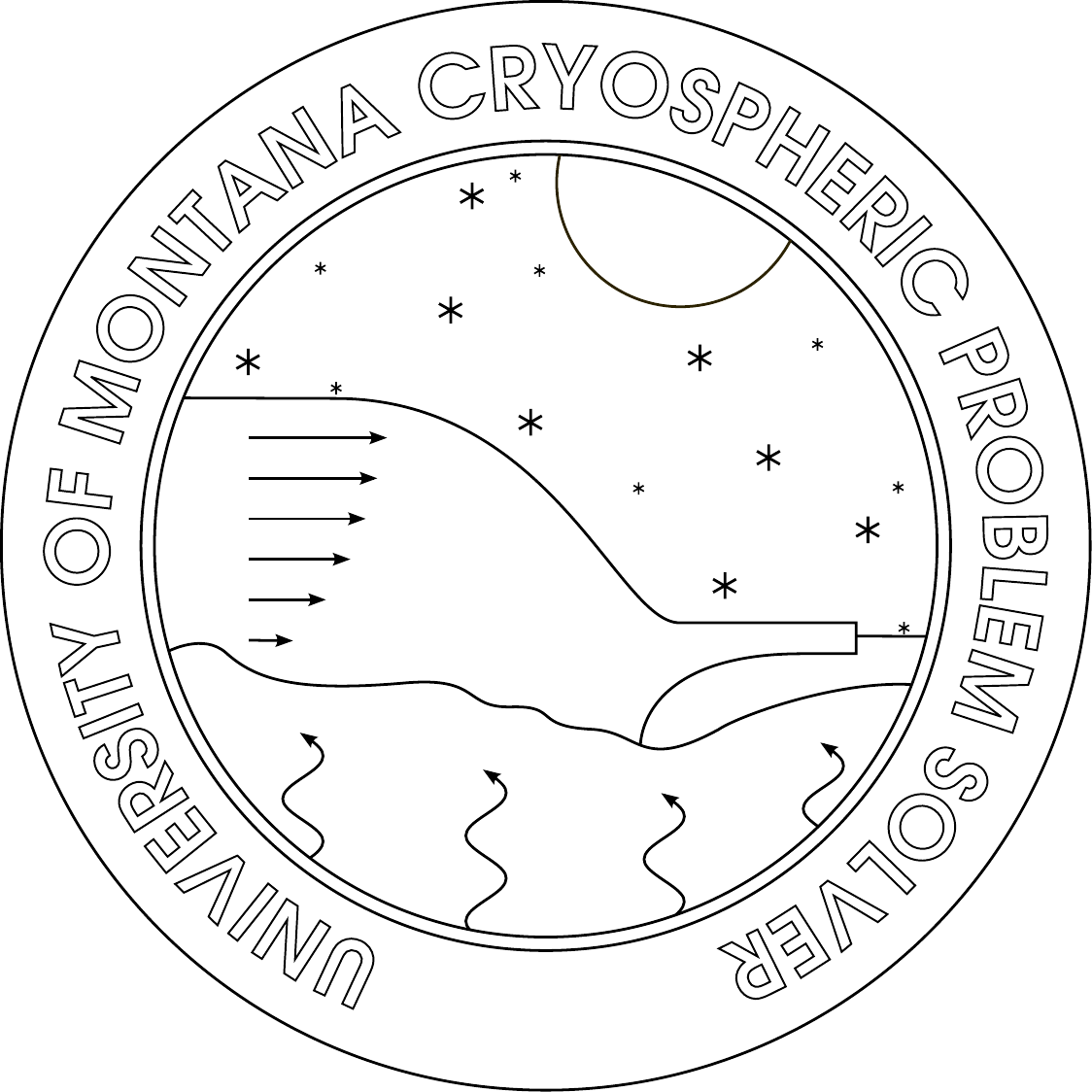}
\end{figure}

\newpage

\begin{center}

\

\vspace{8cm}

\copyright\ COPYRIGHT

\vspace{5mm}

by

\vspace{5mm}

Evan Michael Cummings

\vspace{5mm}

2016

\vspace{5mm}

All Rights Reserved

\end{center}

\newpage

\begin{table*}[t]
\centering
\begin{tabular}{p{50mm}}
  \begin{flushleft}
    To my parents, without whom this project may never have been completed.
  \end{flushleft}
\end{tabular}
\end{table*}

\end{titlepage}

\setcounter{tocdepth}{-1}
\chapter*{Acknowledgments}
\setcounter{tocdepth}{3}

The software newly presented in this manuscript, CSLVR, is largely built from the finite-element software FEniCS \citep{logg}.  CSLVR solves PDE-constrained-optimization problems through the use of Dolfin-Adjoint \citep{farrell}, which in turn utilizes the IPOPT framework \citep{wachter} compiled with the Harwell Subroutine Library, a collection of Fortran codes for large scale scientific computation (\texttt{http://www.hsl.rl.ac.uk/}).  All CSLVR source code is written in the Python programming language (Python Software Foundation, \texttt{http://www.python.org}).  Countless numerical calculations were accomplished throughout the code-generation process using NumPy and SciPy \citep{numpy} via the interactive terminal IPython \citep{ipython}.  The meshes used for the Greenland and Antarctic simulations were created with GMSH \citep{gmsh}.  All of the code-generated figures were created with Matplotlib \citep{matplotlib}.  All hand-drawn-vector images were created with the open-source-vector-graphics-software Inkscape.  Paraview \citep{paraview} was used extensively to investigate simulation results and generate figures of three-dimensional data.  Finally, this manuscript was compiled by the invaluable document-creation software \LaTeX.

Special thanks are given to the following people: my thesis committee for taking the time to provide insight and criticism; Robyn Berg for her thoughtful support throughout my undergraduate and graduate career at the University of Montana; all the faculty in the Departments of Computer Science and Mathematics; The University of Montana Group for Quantitative Study of Snow and Ice for providing many thought-provoking discussions and intuition pertaining to the nature of ice-sheets and glaciers; Douglas Brinkerhoff for creating the software VarGlaS \citep{brinkerhoff} from which CSLVR evolved; and finally Jesse Johnson for his encouragement, support, and guidance.

\tableofcontents

\addcontentsline{toc}{chapter}{List of Figures}
\listoffigures
\addcontentsline{toc}{chapter}{List of Tables}
\listoftables
\addcontentsline{toc}{chapter}{List of Algorithms}
\listofalgorithms

\mainmatter

\part{Familiarization with FEniCS}

%===============================================================================
%===============================================================================

\chapter{Basics of finite elements}

Many differential equations of interest cannot be solved exactly; however, they may be solved approximately if given some simplifying assumptions.  For example, perturbation methods approximate an \emph{inner} and \emph{outer} solution to a problem with different characteristic length or time scales, Taylor-series methods determine a locally convergent approximation, Fourier-series methods determine a globally-convergent approximation, and finite-difference methods provide an approximation over a uniformly discretized domain.  The finite element method is a technique which non-uniformly discretizes the domain of a variational or weighted-residual problem into \emph{finite elements}, which may then be assembled into a single matrix equation and solved for an approximate solution.

%===============================================================================

\section{Motivation: weighted integral approximate solutions} \label{ssn_intro_motivation}

Following the explanation in \citet{reddy}, the approximation of a differential equation with unknown variable $u$ which we seek is given by the linear expansion
\begin{align}
  \label{intro_motivation_approximation}
  u(x) \approx \sum_{i=0}^N u_i \psi_i,
\end{align}
where $u_i$ are coefficients to the solution, $N$ is the number of parameters in the approximation, and $\psi$ is a set of linearly independent functions which satisfy the boundary conditions of the equation.  For example, consider the second-order differential equation
\begin{align}
  \label{intro_motivation_ode}
-\frac{d}{dx} \left[ \frac{du}{dx} \right] + u = 0, \hspace{5mm} 0 < x < 1,
\end{align}
\begin{align}
  \label{intro_motivation_bcs}
  u(0) = 1, \hspace{10mm} \left( \frac{du}{dn} \right) \Bigg|_{x=1} = 0,
\end{align}
where $n$ is the outward-pointing normal to the domain.  In the 1D case here, $n(0) = -1$ and $n(1) = 1$.  The $N=2$ parameter approximation with
\begin{align*}
  \psi_0 = 1, \hspace{5mm} \psi_1 = x^2 - 2x, \hspace{2.5mm} \text{and} \hspace{2.5mm} \psi_2 = x^3 - 3x,
\end{align*}
in (\ref{intro_motivation_approximation}) gives the approximate solution
$$u(x) \approx U_N = u_0 + u_1 (x^2 - 2x) + u_2 (x^3 - 3x).$$
This approximation satisfies the \emph{Neumann} or \emph{natural} boundary condition
\index{Boundary conditions!Essential} \index{Boundary conditions!Natural}
at $x=1$, and in order to satisfy the \emph{Dirichlet} or \emph{essential} boundary condition at $x=0$, we make $u_0 = 1$, producing
\begin{align}
  \label{intro_motivation_expansion}
  u(x) \approx U_N = 1 + u_1 (x^2 - 2x) + u_2 (x^3 - 3x).
\end{align}
Substituting this approximation into differential equation (\ref{intro_motivation_ode}) results in
\begin{align*}
  - 2u_1(x - 1) - 3u_2(x^2 - 1) + 1 + u_1(x^2 - 2x) + u_2 (x^3 - 3x) &= 0 \\
  (2u_1 + 3u_2 + 1) - (2u_1 + 2u_1 + 3u_2)x - (3u_2 - u_1)x^2 + u_2x^3 &= 0,
\end{align*}
implying that
\begin{align*}
  2u_1 + 3u_2 + 1 &= 0 \\
  4u_1 + 3u_2 &= 0 \\
  3u_2 - u_1 &= 0 \\
  u_2 &= 0.
\end{align*}
This system of equations has only the trivial solution $u=0$ and is hence inconsistent with differential equation (\ref{intro_motivation_ode}, \ref{intro_motivation_bcs}).  However, if the problem is evaluated as a \emph{weighted integral} it can be guaranteed that the number of parameters equal the number of linearly independent equations.  This weighted integral relation is
$$\int_0^1 w R dx = 0,$$
where $R$ is the approximation residual of Equation (\ref{intro_motivation_ode}),
$$R = - \frac{d^2 U_N}{dx^2} + U_N,$$
and $w$ are a set of $N$ linearly independent \emph{weight functions}.  For this example we use
$$w_1 = x, \hspace{2.5mm} \text{and} \hspace{2.5mm} w_2 = x^2,$$
and two integral relations to evaluate,
\footnotesize
\begin{align*}
  0 &= \int_0^1 w_1 R dx = \int_0^1 x R dx \\
    &= \left[ \frac{1}{2}(2u_1 + 3u_2 + 1)x^2 - \frac{1}{3}(4u_1 + 3u_2)x^3 - \frac{1}{4}(3u_2 - u_1)x^4 + \frac{1}{5}u_2x^5 \right]_0^1 \\
    &= \frac{1}{2}(2u_1 + 3u_2 + 1) - \frac{1}{3}(4u_1 + 3u_2)x^3 - \frac{1}{4}(3u_2 - u_1) + \frac{1}{5}u_2 \\
    &= \frac{1}{2} + \left( 1 - \frac{4}{3} + \frac{1}{4} \right) u_1 + \left( \frac{3}{2} - 1 - \frac{3}{4} + \frac{1}{5} \right) u_2 \\
    &= \frac{1}{2} - \frac{1}{12} u_1 - \frac{1}{20} u_2, \\
  0 &= \int_0^1 w_2 R dx = \int_0^1 x^2 R dx \\
    &= \left[ \frac{1}{3}(2u_1 + 3u_2 + 1)x^3 - \frac{1}{4}(4u_1 + 3u_2)x^4 - \frac{1}{5}(3u_2 - u_1)x^5 + \frac{1}{6}u_2x^6 \right]_0^1 \\
    &= \frac{1}{3}(2u_1 + 3u_2 + 1) - \frac{1}{4}(4u_1 + 3u_2) - \frac{1}{5}(3u_2 - u_1) + \frac{1}{6}u_2 \\
    &= \frac{1}{3} + \left( \frac{2}{3} - 1 + \frac{1}{5} \right) u_1 + \left( 1 - \frac{3}{4} - \frac{3}{5} + \frac{1}{6} \right) u_2 \\
    &= \frac{1}{3} - \frac{2}{15} u_1 - \frac{11}{60} u_2,
\end{align*}
\normalsize
giving a system of equations for the coefficients $u_1$ and $u_2$,
\begin{align*}
  \begin{bmatrix}
    - \frac{1}{12} & - \frac{1}{20} \\
    - \frac{2}{15} & - \frac{11}{60} \\
  \end{bmatrix} \cdot
  \begin{bmatrix}
    u_1 \\
    u_2
  \end{bmatrix} = 
  \begin{bmatrix}
    -\frac{1}{2} \\
    -\frac{1}{3}
  \end{bmatrix}.
\end{align*}
Solving this system produces $u_1 = \frac{270}{31}$ and $u_2 = \frac{-140}{31}$, and thus approximation (\ref{intro_motivation_expansion}) is given by
\begin{align}
  \label{intro_motivation_approximation_final}
  u_N(x) = 1 + \frac{270}{31} (x^2 - 2x) - \frac{140}{31} (x^3 - 3x).
\end{align}

\subsection{Exact solution}

Differential equation (\ref{intro_motivation_ode}) is easily solved exactly:
$$\frac{d^2 u}{dx^2} - u = 0 \hspace{2.5mm} \implies \hspace{2.5mm} u = c_1 \cosh(x) + c_2 \sinh(x)$$ 
$$u(0) = c_1 = 1, \hspace{5mm} u'(1) = \sinh(1) + c_2 \cosh(1) = 0$$
$$\implies c_2 = -\frac{\sinh(1)}{\cosh(1)} = -\tanh(1),$$
and 
\begin{align}
  \label{intro_motivation_exact}
  u(x) = \cosh(x) - \tanh(1)\sinh(x).
\end{align}

\begin{figure}
  \centering
    \includegraphics[width=\linewidth]{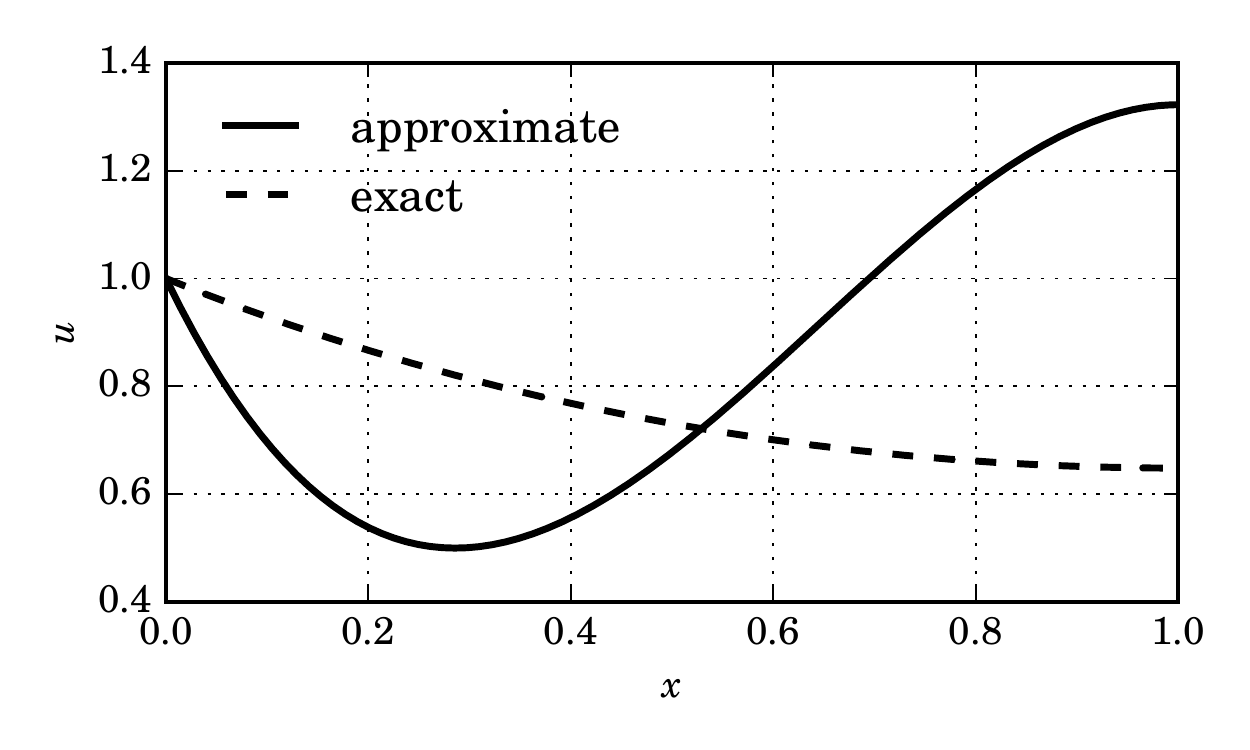}
  \caption[FEM intro scratch example]{Weighted integral approximation (solid) and exact solution (dashed).  Note that the approximation will be improved by increasing $N$.}
  \label{scratch_example_image}
\end{figure}

Weighted integral approximation (\ref{intro_motivation_approximation_final}) and exact solution (\ref{intro_motivation_exact}) are shown in Figure \ref{scratch_example_image}.

\pythonexternal[label=weight_int_plot, caption={Scipy source code used to generate Figure \ref{scratch_example_image}}]{scripts/fenics_intro/weight_int.py}

%===============================================================================

\section{The finite element method}

The finite element method combines variational calculus, Galerkin approximation methods, and numerical analysis to solve initial and boundary value differential equations \citep{reddy}.  The steps involved in the finite element approximation of a typical problem are
\begin{enumerate}
  \item Discretization of the domain into finite elements.
  \item Derivation of element equations over each element in the mesh.
  \item Assembly of local element equations into a global system of equations.
  \item Imposition of boundary conditions.
  \item Numerical solution of assembled equations.
  \item Post processing of results.
\end{enumerate}

In the following sections, we examine each of these steps for the 1D-boundary-value problem \citep{davis} over the domain $\Omega \in (0,\ell)$ with essential boundary $\Gamma_D$ at $x=0$ and natural boundary $\Gamma_N$ at $x=\ell$
\begin{align}
  \label{intro_ode}
  -\frac{d}{dx} \left[ k(x) \frac{du}{dx} \right] &= f(x) && \text{ in } \Omega \\
  \label{intro_ode_nbc}
  \left( k(x) \frac{du}{dn} \right) &= g_N && \text{ on } \Gamma_N \\
  \label{intro_ode_ebc}
  u &= g_D && \text{ on } \Gamma_D,
\end{align}
and develop a finite-element model from scratch.

\subsection{Variational form} \label{ssn_variational_form}

The variational problem \index{Variational form} corresponding to (\ref{intro_ode} -- \ref{intro_ode_ebc}) is formed by multiplying Equation (\ref{intro_ode}) by the weight function $w(x)$ and integrating over the $x$-coordinate domain $\Omega \in (0,\ell)$,
$$-\int_{\Omega} \frac{d}{dx} \left[ k(x) \frac{du}{dx} \right] w(x) d\Omega = \int_{\Omega} f(x) w(x) d\Omega,$$
with no restrictions on $w(x)$ made thus far.  Integrating the left-hand side by parts,
\begin{align*}
  \int_0^{\ell} k \frac{du}{dx} \frac{dw}{dx} dx - \left[ wk\frac{du}{dn}\right]_0^{\ell} &= \int_0^{\ell} f w dx.
\end{align*}
This formulation is called the \emph{weak form} \index{Weak form} of the differential equation due to the ``weakened'' conditions on the approximation of $u(x)$.

In the language of distributional solutions in mathematical analysis, the \index{Trial functions} \emph{trial} or \emph{solution function} $u$ is a member of the \emph{trial} or \emph{solution space} that satisfies the essential boundary condition $g_D$ on Dirichlet boundary $\Gamma_D$,
\begin{align}
  \label{trial_space}
  \mathcal{H}_E^1(\Omega) &= \left\{ u \in \mathcal{H}^1(\Omega)\ |\ u = g_D\ \text{on}\ \Gamma_D \right\},
\end{align}
while the \index{Test functions} \emph{test function} $w$ is member of the \emph{test space}
\begin{align}
  \label{test_space}
  \mathcal{H}_{E_0}^1(\Omega) &= \left\{ u \in \mathcal{H}^1(\Omega)\ |\ u = 0\ \text{on}\ \Gamma_D \right\}.
\end{align}
These spaces are both defined over the space of square-integrable functions whose first derivatives are also square integrable; the $\Omega \subset \R$ \index{Sobolev space} \emph{Sobolev space} \citep{elman}
\begin{align}
  \label{sobolev_space}
  \mathcal{H}^1(\Omega) &= \left\{ u\ :\ \Omega \rightarrow \R\ \Bigg|\ u, \frac{du}{dx} \in L^2(\Omega) \right\},\\
  \label{l2_space}
  L^2(\Omega) &= \left\{ u\ :\ \Omega \rightarrow \R\ \Bigg|\ \int_{\Omega} u^2\ d\Omega < \infty \right\},
\end{align}
where the space of functions in \index{L@$L^2$ space} $L^2(\Omega)$ is defined with the measure
\begin{align}
  \label{l2_norm}
  \Vert u \Vert_2 = \left( \int_{\Omega} u^2\ d\Omega \right)^{\nicefrac{1}{2}},
\end{align}
and the \index{Inner product} $L^2$ \emph{inner product} $(f,g) = \int_{\Omega} f g d\Omega$.

The variational problem consists of finding $u \in \mathscr{H}_E^1(\Omega)$ such that
\begin{align}
  \label{intro_variational_problem}
  a(u,w) &= l(w) && \forall w \in \mathscr{H}_E^1(\Omega),
\end{align}
with \index{Bilinear form} \emph{bilinear} term $a(u,w)$ and \emph{linear} term $l(w)$ \citep{reddy}
\begin{align*}
  a(u,w) &= \int_0^{\ell} k \frac{du}{dx} \frac{dw}{dx} dx - \left[ wk\frac{du}{dn}\right]_0^{\ell} \\
  l(w)   &= \int_0^{\ell} f w dx.
\end{align*}

The next section demonstrates how to solve the finite-dimensional analog of (\ref{intro_variational_problem}) for $U \in S_E^h \subset \mathscr{H}_E^1(\Omega)$ such that
\begin{align}
  \label{intro_discrete_variational_problem}
  \int_0^{\ell} k \frac{d U}{dx} \frac{d \psi}{dx} dx - \left[ \psi k \frac{d U}{dn}\right]_0^{\ell} &= \int_0^{\ell} f \psi dx
\end{align}
for all $\psi \in S_0^h \subset \mathscr{H}_{E_0}^1(\Omega)$.

\subsection{Galerkin element equations} \label{ssn_intro_galerkin_equations}

Similarly to \S \ref{ssn_intro_motivation}, the \index{Galerkin method} \emph{Galerkin approximation method} seeks to derive an $n$-node approximation over a single element $e$ of the form
\begin{align}
  \label{intro_approximation}
  u(x) \approx U^e(x) = \sum_{j=1}^n \psi_j^e(x) u_j^e,
\end{align}
where $u_j^e$ is the unknown value at node $j$ of element $e$ and $\psi^e$ is a set of $n$ linearly independent approximation functions, otherwise know as \index{Shape functions} \index{Finite-element interpolation} \emph{interpolation} , \emph{basis}, or \emph{shape} functions, for each of the $n$ nodes of element $e$.  The approximation functions must be continuous over the element and be differentiable to the same order as the equation.

\begin{figure}
  \centering
    \def\svgwidth{\linewidth}
    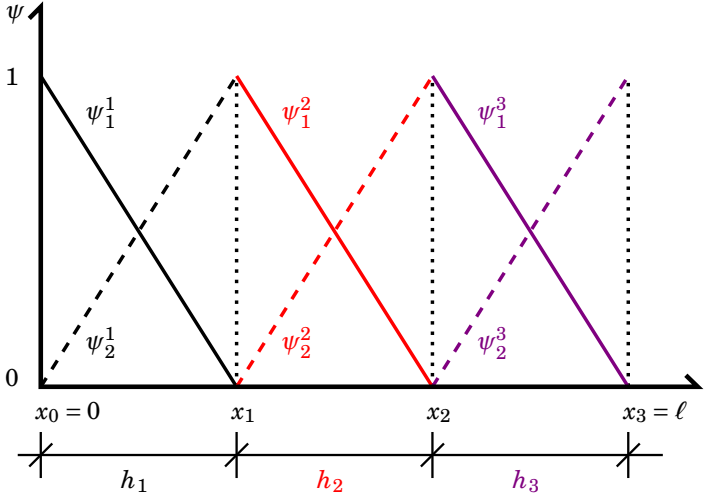
  \caption[linear Lagrange shape functions]{Linear Lagrange interpolation functions $\psi^e$, where the superscript is the element number and subscript the element function number.  The even element functions are solid and the odd functions dashed, color-coded by element number.}
    \label{lagrange_ftns_image}
\end{figure}

For the simplest example, the linear interpolation functions with $C^0$ continuity, known as \index{Lagrange interpolation functions|seealso{Shape functions}} \emph{Lagrange} interpolation functions defined only over the element interval $x \in [x_i, x_{i+1}]$,
\begin{align}
  \label{linear_lagrange_functions}
  \psi_1^e(x) = 1 - \frac{x^e}{h_e} \hspace{10mm} \psi_2^e(x) = \frac{x^e}{h_e},
\end{align}
where $x^e = x - x_i$ is the $x$-coordinate local to element $e$ with first node $i$ and last node $i+1$, and $h_e = x_{i+1}^e - x_i^e$ is the width of element $e$ (Figure \ref{lagrange_ftns_image}).  Note that these functions are once differentiable as required by the weak form of our example equation, and satisfies the required \index{Interpolation properties} \emph{interpolation properties}
\begin{align}
  \label{interpolation_properties}
  \psi_i^e(x_j^e) = \delta_{ij} \hspace{10mm} \sum_{j=1}^n \psi_j^e(x^e) = 1,
\end{align}
where $\delta_{ij}$ is the \index{Kronecker delta} \emph{Kronecker delta},
\begin{align*}
  \delta_{ij} = \begin{cases}
                  0 & \text{ if } i \neq j,\\
                  1 & \text{ if } i = j
                \end{cases}.
\end{align*}
The second property in (\ref{interpolation_properties}) implies that the set of functions $\psi$ form a \emph{partition of unity}; this explains how the unknown coefficients $u_j^e$ of approximation (\ref{intro_approximation}) are equal to the value of approximation $u$ at node $j$ of element $e$.

Inserting approximation (\ref{intro_approximation}) into weak form (\ref{intro_discrete_variational_problem}) integrated over a single element with (not necessarily linear Lagrange) weight functions $w = \psi_i, i = 1,\ldots,n$ and add terms for the flux variables interior to the nodes,
\begin{align}
  \label{intro_element_form}
  \int_x k \left( \sum_{j=1}^n u_j^e \frac{d\psi_j^e}{dx} \right) \frac{d\psi_i^e}{dx} dx = \int_x f \psi_i^e dx + \sum_{j=1}^n \psi_i^e(x_j^e) Q_j^e,
\end{align}
where $Q_j^e$ is the outward flux from node $j$ of element $e$,
\begin{align*}
  Q_j^e = k \frac{du_j^e}{dn}.
\end{align*}

Using the second interpolation property in (\ref{interpolation_properties}) the last term in (\ref{intro_element_form}) is evaluated,
\begin{align*}
  \sum_{j=1}^n \psi_i^e(x_j^e) Q_j^e = \sum_{j=1}^n \delta_{ij} Q_j^e = Q_i^e.
\end{align*}
Next, using the fact that the $u_j^e$ are constant the left-hand-side of (\ref{intro_element_form}) is re-written
\begin{align*}
  \int_x k \left( \sum_{j=1}^n u_j^e \frac{d\psi_j^e}{dx} \right) \frac{d\psi_i^e}{dx} dx =  \sum_{j=1}^n u_j^e \int_x k \frac{d\psi_j^e}{dx} \frac{d\psi_i^e}{dx} dx, 
\end{align*}
Therefore, system (\ref{intro_element_form}) is re-written as
\begin{align*}
  \sum_{j=1}^n K_{ij}^e u_j^e = f_i^e + Q_i^e, \hspace{5mm} i = 1,2,\ldots,n,
\end{align*}
with bilinear and linear terms
\begin{align*}
  K_{ij}^e = a\left( \psi_i^e, \psi_j^e \right) = \int_x k \frac{d\psi_j^e}{dx} \frac{d\psi_i^e}{dx} dx, \hspace{8mm} f_i^e = l\left( \psi_i^e \right) = \int_x f \psi_i^e dx.
\end{align*}
This is sum is also expressed as the matrix equation
\begin{align}
  \label{intro_local_system}
  K^e \mathbf{u}^e = \mathbf{f}^e + \mathbf{q}^e.
\end{align}

Approximations of this kind are referred to as \emph{Galerkin approximations}.

\subsection{Local element Galerkin system} \label{ssn_local_galerkin_assembly}

Using linear Lagrange interpolation functions (\ref{linear_lagrange_functions}) in weak form (\ref{intro_local_system}) integrated over a single element $e$ of width $h_e$,
\begin{align*}
  K_{ij}^e = \int_0^{h_e} k_e \frac{d\psi_j^e}{dx} \frac{d\psi_i^e}{dx} dx, \hspace{5mm} f_i^e = \int_0^{h_e} f_e \psi_i^e dx, \hspace{5mm} i,j \in \{1,2\}.
\end{align*}
Evaluating the \index{Stiffness matrix} \emph{stiffness matrix} for the element first,

\begin{table}[H]
\centering
\begin{tabularx}{\linewidth}{XX}
{\begin{align*}
  K_{11}^e &= \int_0^{h_e} k_e \frac{d\psi_1^e}{dx} \frac{d\psi_1^e}{dx} dx \\
           &= \int_0^{h_e} k_e \left( -\frac{1}{h_e} \right) \left( -\frac{1}{h_e} \right) dx \\
           &= k_e \left(\frac{x}{h_e^2} \right) \Bigg|_0^{h_e} \\
           &= \frac{k_e}{h_e},
\end{align*}}
&
{\begin{align*}
  K_{12}^e &= \int_0^{h_e} k_e \frac{d\psi_2^e}{dx} \frac{d\psi_1^e}{dx} dx \\
           &= \int_0^{h_e} k_e \left( -\frac{1}{h_e} \right) \left( \frac{1}{h_e} \right) dx \\
           &= -k_e \left(\frac{x}{h_e^2} \right) \Bigg|_0^{h_e} \\
           &= -\frac{k_e}{h_e},
\end{align*}} \\ 
{\begin{align*}
  K_{21}^e &= \int_0^{h_e} k_e \frac{d\psi_1^e}{dx} \frac{d\psi_2^e}{dx} dx \\
           &= \int_0^{h_e} k_e \left( \frac{1}{h_e} \right) \left( -\frac{1}{h_e} \right) dx \\
           &= -k_e \left(\frac{x}{h_e^2} \right) \Bigg|_0^{h_e} \\
           &= -\frac{k_e}{h_e},
\end{align*}}
&
{\begin{align*}
  K_{22}^e &= \int_0^{h_e} k_e \frac{d\psi_2^e}{dx} \frac{d\psi_2^e}{dx} dx \\
           &= \int_0^{h_e} k_e \left( \frac{1}{h_e} \right) \left( \frac{1}{h_e} \right) dx \\
           &= k_e \left(\frac{x}{h_e^2} \right) \Bigg|_0^{h_e} \\
           &= \frac{k_e}{h_e},
\end{align*}}
\end{tabularx}
\end{table}
and the source term $f$,
\begin{table}[H]
\centering
\begin{tabularx}{\linewidth}{XX}
{\begin{align*}
  f_1^e &= \int_0^{h_e} f_e \psi_1^e dx \\
        &= \int_0^{h_e} f_e \left( 1 - \frac{x}{h_e} \right) dx \\
        &= f_e \left( x - \frac{x^2}{2h_e} \right) \Bigg|_0^{h_e} \\
        &= f_e \left( h_e - \frac{h_e^2}{2h_e} \right) \\
        &= f_e \left( h_e - \frac{h_e}{2} \right) \\
        &= \frac{1}{2} f_e h_e,
\end{align*}}
&
{\begin{align*}
  f_2^e &= \int_0^{h_e} f_e \psi_2^e dx \\
        &= \int_0^{h_e} f_e \left( \frac{x}{h_e} \right) dx \\
        &= f_e \left( \frac{x^2}{2h_e} \right) \Bigg|_0^{h_e} \\
        &= f_e \left( \frac{h_e^2}{2h_e} \right) \\
        &= f_e \left( \frac{h_e}{2} \right) \\
        &= \frac{1}{2} f_e h_e.
\end{align*}}
\end{tabularx}
\end{table}

Finally, the \index{Local element matrix} \emph{local} element matrix \emph{Galerkin system} corresponding to (\ref{intro_local_system}) with linear-Lagrange elements is
\begin{align*}
 \frac{k_e}{h_e}
 \begin{bmatrix}[r]
   1 & -1 \\
   -1 & 1 
 \end{bmatrix} \cdot 
 \begin{bmatrix}
   u_1^e \\ u_2^e
 \end{bmatrix} &= 
 \frac{f_e h_e}{2}
 \begin{bmatrix}
   1 \\ 1
 \end{bmatrix} + 
 \begin{bmatrix}
   Q_1^e \\ Q_2^e
 \end{bmatrix}.
\end{align*}

\subsection{Globally assembled Galerkin system} \label{ssn_global_galerkin_assembly}

In order to connect the set of elements together, extra constraints are imposed on the values interior to the domain.  These are
\begin{enumerate}
  \item The \emph{primary variables} are continuous between nodes such that the last nodal value of an element is equal to its adjacent element's first nodal value,
  \begin{align}
    \label{primary_variable_continuity}
    u_n^e = u_1^{e+1}.
  \end{align}
  \item The \emph{secondary variables} are balanced between nodes such that outward flux from a connected element is equal to the negative outward flux of its neighboring node,
  \begin{align}
    \label{secondary_variable_continuity}
    Q_n^e + Q_1^{e+1} = 0.
  \end{align}
  If a point source is applied or it is desired to make $Q$ an unknown to be determined,
  $$Q_n^e + Q_1^{e+1} = Q_0.$$
\end{enumerate}

First, for global node $N$,
$$U_N = u_n^{N} = u_1^{N+1}, \hspace{5mm} f_N = f_n^{N} + f_1^{N+1}, \hspace{2.5mm} \text{and} \hspace{2.5mm} Q_N = Q_n^{N} + Q_1^{N+1},$$
and add the last equation from element $e$ to the first equation of element $e+1$,
\begin{align*}
  \sum_{j=1}^n K_{nj}^e u_j^e + \sum_{j=1}^n K_{1j}^{e+1} u_j^{e+1} &= \left( f_n^e + Q_n^e \right) + \left( f_1^{e+1} + Q_1^{e+1} \right) \\
  \sum_{j=1}^n \left( K_{nj}^e u_j^e + K_{1j}^{e+1} u_j^{e+1} \right) &= f_n^e + f_1^{e+1} + Q_n^e + Q_1^{e+1} \\
  \sum_{j=1}^n U_j \left( K_{nj}^e + K_{1j}^{e+1} \right) &= f_e + Q_e,
\end{align*}
which can be transformed into the \index{Global element matrix} \emph{global} matrix equation; the Galerkin system
\begin{align}
  \label{intro_galerkin_system}
  K\mathbf{u} = \mathbf{f} + \mathbf{q},
\end{align}
where
\footnotesize
\begin{align*}
  K &=
  \begin{bmatrix}
    K_{11}^1 & K_{12}^1            &                                                   &                         &            \\
    K_{21}^1 & K_{22}^1 + K_{11}^2 & K_{12}^2            &                             &                         &            \\
             & K_{21}^2            & K_{22}^2 + K_{11}^3 & K_{12}^3                    &                         &            \\
             &                     & \hspace{10mm}\ddots &                             &                         &            \\ 
             &                     & K_{21}^{E-2}        & K_{22}^{E-2} + K_{11}^{E-1} & K_{12}^{E-1}            &            \\
             &                     &                     & K_{21}^{E-1}                & K_{22}^{E-1} + K_{11}^E & K_{12}^E   \\
             &                     &                     &                             & K_{21}^{E}              & K_{22}^{E} 
  \end{bmatrix} \\
  \mathbf{u} &= \begin{bmatrix} U_1 & U_2 & \cdots & U_E \end{bmatrix}\T \\
  \mathbf{f} &= \begin{bmatrix} f_1 & f_2 & \cdots & f_E \end{bmatrix}\T \\
  \mathbf{q} &= \begin{bmatrix} Q_1 & Q_2 & \cdots & Q_E \end{bmatrix}\T.
\end{align*}
\normalsize

Applying Lagrange element equations (\ref{linear_lagrange_functions}), subdividing the domain $x \in [0,\ell]$ into three equal width parts, and making coefficients $k$ and source term $f$ constant throughout the domain, system of equations (\ref{intro_galerkin_system}) is 
\begin{align} 
  \label{intro_expanded_global_system}
  \frac{k}{h_e}
  \begin{bmatrix}[r]
     1 & -1 &  0 &  0 \\
    -1 &  2 & -1 &  0 \\
     0 & -1 &  2 & -1 \\
     0 &  0 & -1 &  1 \\
  \end{bmatrix} \cdot
  \begin{bmatrix} U_1 \\ U_2 \\ U_3 \\ U_4 \end{bmatrix} &=
  \frac{f h_e}{2}
  \begin{bmatrix} 1   \\ 2   \\ 2   \\ 1   \end{bmatrix} +
  \begin{bmatrix} Q_1 \\ Q_2 \\ Q_3 \\ Q_4 \end{bmatrix},
\end{align}
a system of four equations and eight unknowns.  In the next section, this under-determined system is made solvable by applying boundary conditions and continuity requirements on the internal element flux terms $Q_e$.

\subsection{Imposition of boundary conditions}
  
Recall Equation (\ref{intro_ode}) is defined with use the \index{Boundary conditions!Essential} essential boundary condition (\ref{intro_ode_nbc}) and \index{Boundary conditions!Natural} natural boundary condition (\ref{intro_ode_ebc}).  In terms of approximation (\ref{galerkin_approximation}), these are respectively
$$u(0) = U_1 = g_D, \hspace{10mm} \left( k \frac{du}{dx} \right) \Bigg|_{x=\ell} = Q_4 = g_N.$$
Applying continuity requirement for interior nodes (\ref{secondary_variable_continuity}),
$$Q_e = Q_n^e + Q_1^{e+1} = 0, \hspace{10mm} e = 2,3,$$
to global matrix system (\ref{intro_expanded_global_system}) results in
\begin{align}
  \label{intro_final_global_system}
  \frac{k}{h_e}
  \begin{bmatrix}[r]
     1 & -1 &  0 &  0 \\
    -1 &  2 & -1 &  0 \\
     0 & -1 &  2 & -1 \\
     0 &  0 & -1 &  1 \\
  \end{bmatrix} \cdot
  \begin{bmatrix} g_D \\ U_2 \\ U_3 \\ U_4 \end{bmatrix} &=
  \frac{f h_e}{2}
  \begin{bmatrix} 1   \\ 2   \\ 2   \\ 1   \end{bmatrix} +
  \begin{bmatrix} Q_1 \\ 0   \\ 0   \\ g_N   \end{bmatrix},
\end{align}
a system of four equations and four unknowns $U_2$, $U_3$, $U_4$, and $Q_1$.

\subsection{Solving procedure} \label{ssn_galerkin_solve}

Before solving global system (\ref{intro_final_global_system}), values must be chosen for the known variables and length of the domain.  For simplicity, we use
$$g_D = 0, \hspace{5mm} g_N = 0, \hspace{5mm} k = 1, \hspace{5mm} f=1, \hspace{5mm} h_e = \nicefrac{1}{3}.$$
With this, system (\ref{intro_final_global_system}) simplifies to
\begin{align}
  \begin{bmatrix}[r]
     3 & -3 &  0 &  0 \\
    -3 &  6 & -3 &  0 \\
     0 & -3 &  6 & -3 \\
     0 &  0 & -3 &  3 \\
  \end{bmatrix} \cdot
  \begin{bmatrix} 0   \\ U_2 \\ U_3 \\ U_4 \end{bmatrix} &=
  \frac{1}{6}
  \begin{bmatrix} 1   \\ 2   \\ 2   \\ 1   \end{bmatrix} +
  \begin{bmatrix} Q_1 \\ 0   \\ 0   \\ 0   \end{bmatrix} \notag \\
  \label{intro_full_matrix}
  \begin{bmatrix}[r]
     3 & -3 &  0 &  0 \\
    -3 &  6 & -3 &  0 \\
     0 & -3 &  6 & -3 \\
     0 &  0 & -3 &  3 \\
  \end{bmatrix} \cdot
  \begin{bmatrix} 0   \\ U_2 \\ U_3 \\ U_4 \end{bmatrix} &=
  \begin{bmatrix} Q_1 + \nicefrac{1}{6} \\ \nicefrac{1}{3} \\ \nicefrac{1}{3} \\ \nicefrac{1}{6} \end{bmatrix}.
\end{align}

This equation is easily reduced to include only the unknown primary \emph{degrees of freedom} $U_e$,
\begin{align*}
  \begin{bmatrix}[r]
     6 & -3 &  0 \\
    -3 &  6 & -3 \\
     0 & -3 &  3 \\
  \end{bmatrix} \cdot
  \begin{bmatrix} U_2 \\ U_3 \\ U_4 \end{bmatrix} &=
  \begin{bmatrix} \nicefrac{1}{3} \\ \nicefrac{1}{3} \\ \nicefrac{1}{6} \end{bmatrix}.
\end{align*}
Because this matrix is square and non-singular, $K = LU$ where $L$ and $U$ are lower- and upper-triangular matrices.  Thus the system of equations can be solved by forward and backward substitutions \citep{watkins}
\begin{align*}
  L\mathbf{y} &= \mathbf{q} \hspace{5mm} \leftarrow \hspace{5mm} \text{forward substitution}, \\
  U\mathbf{u} &= \mathbf{y} \hspace{5mm} \leftarrow \hspace{5mm} \text{backward substitution.}
\end{align*}
For stiffness matrix $K$,
\begin{multicols}{2}
\begin{align*}
  L &= 
  \begin{bmatrix}[r]
     1               &  0               &  0 \\
    -\nicefrac{1}{2} &  1               &  0 \\
     0               & -\nicefrac{2}{3} &  1
  \end{bmatrix}
\end{align*}

\begin{align*}
  U &=
  \begin{bmatrix}[r]
     6 & -3               &  0 \\
     0 &  \nicefrac{9}{2} & -3 \\
     0 &  0               &  1
  \end{bmatrix},
\end{align*}
\end{multicols}
and thus
\begin{align*}
  L\mathbf{y} &= \mathbf{q} \\
  \begin{bmatrix}[r]
     1           &  0           &  0 \\
    -\frac{1}{2} &  1           &  0 \\
     0           & -\frac{2}{3} &  1 \\
  \end{bmatrix} \cdot
  \begin{bmatrix} y_1 \\ y_2 \\ y_3 \end{bmatrix} &=
  \begin{bmatrix} \nicefrac{1}{3} \\ \nicefrac{1}{3} \\ \nicefrac{1}{6} \end{bmatrix}
\end{align*}
provides $\mathbf{y} = \begin{bmatrix} \nicefrac{11}{54} & \nicefrac{8}{27} & \nicefrac{2}{3} \end{bmatrix}\T$, which can then be used in backward substitution
\begin{align*}
  U\mathbf{u} &= \mathbf{y} \\
  \begin{bmatrix}[r]
     6 & -3               &  0 \\
     0 &  \nicefrac{9}{2} & -3 \\
     0 &  0               &  1 \\
  \end{bmatrix} \cdot
  \begin{bmatrix} U_2 \\ U_3 \\ U_4 \end{bmatrix} &=
  \begin{bmatrix} \nicefrac{11}{54} \\ \nicefrac{8}{27} \\ \nicefrac{2}{3} \end{bmatrix},
\end{align*}
producing $\mathbf{u} = \begin{bmatrix} \nicefrac{5}{18} & \nicefrac{4}{9} & \nicefrac{1}{2} \end{bmatrix}\T$.  Finally, $Q_1$ is solved from the first equation of full system (\ref{intro_full_matrix}),
\begin{align*}
  -3U_2 &= Q_1 + \frac{1}{6} \\
  -\frac{15}{18} - \frac{1}{6} &= Q_1 \\
  \implies Q_1 &= - 1.
\end{align*}

Note that this term is not required to be computed, as the nodal values have been fully discovered.  The final three-element solution to this problem is
$$\mathbf{u} = \begin{bmatrix} 0 & \nicefrac{5}{18} & \nicefrac{4}{9} & \nicefrac{1}{2} \end{bmatrix}\T.$$
The flux of quantity $u$ at the left endpoint $x = 0$ is easily calculated:
$$Q_1 = \left(k \frac{du}{dn} \right) \Bigg|_{x=0} = - \left(\frac{du}{dx} \right) \Bigg|_{x=0} = -1$$ 
$$\implies \left(\frac{du}{dx} \right) \Bigg|_{x=0} = 1.$$

\subsection{Exact solution}

The differential equation
$$- \frac{d^2u}{dx^2} = 1, \hspace{5mm} 0 < x < 1,$$
$$u(0) = 0, \hspace{10mm} \frac{du}{dn} \Bigg|_{x=1} = 0$$
is easily solved for the exact solution
$$u_e(x) = -\frac{x^2}{2} + x.$$

\begin{figure}
  \centering
    \includegraphics[width=\linewidth]{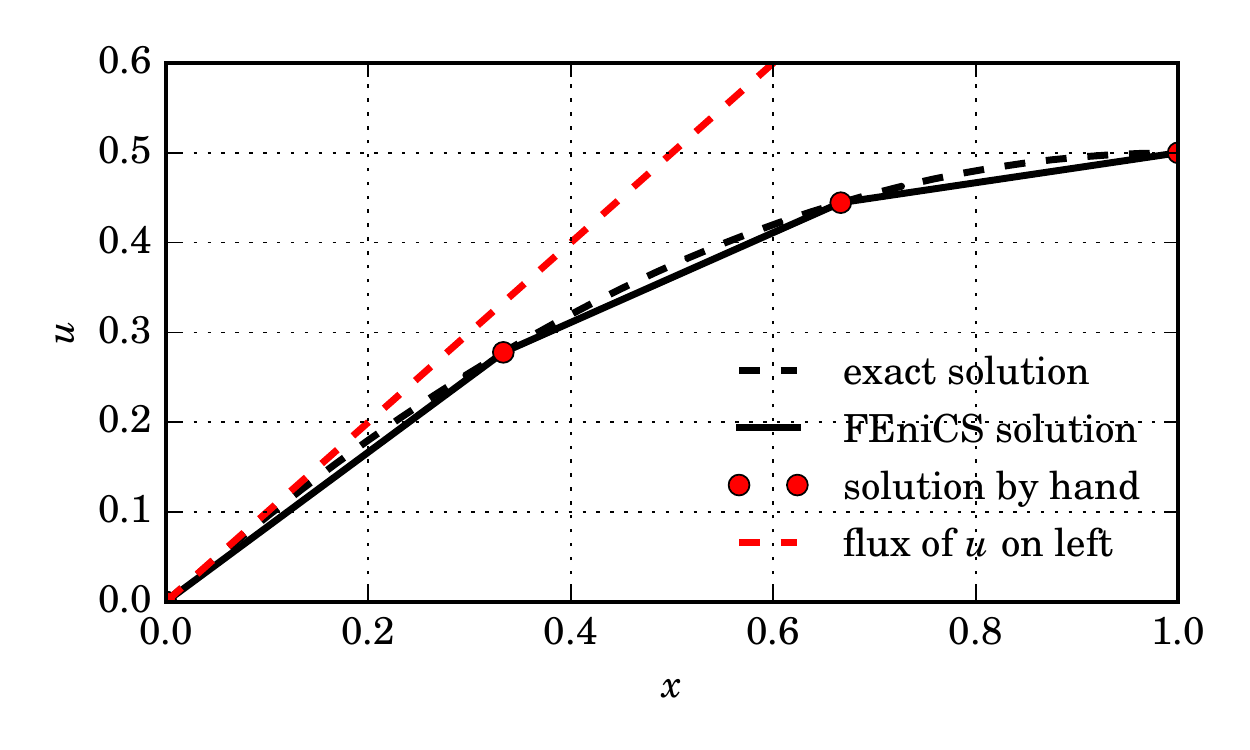}
  \caption[Introductory FEM example solution]{Finite element solution computed with FEniCS (solid black), solved exactly (dashed black), and manual finite element (red dots).  The slope $Q_1 = 1$ is shown in dashed red.}
  \label{scratch_ex_image}
\end{figure}

The results obtained by hand are compared to the exact solution in Figure \ref{scratch_ex_image}.

%===============================================================================

\section{FEniCS framework}

The FEniCS ([F]inite [E]lement [ni] [C]omputational [S]oftware) \index{FEniCS} package for python and C++ is a set of packages for easily formulating finite element solutions for differential equations \citep{logg}.  This software includes tools for automatically creating a variety of finite element function spaces, differentiating variational functionals, creating finite element meshes, and much more.  It also includes several linear algebra packages for solving the element equations, including PETSc, uBLAS, Epetra, and MTL4.

For example, the finite element code for introductory problem (\ref{intro_ode} -- \ref{intro_ode_ebc}) is presented in Code Listing \ref{scratch_example_code}.  Notice that only the variational form of the problem is required to find the approximate solution.

\pythonexternal[label=scratch_example_code, caption={FEniCS source code for the introductory problem.}]{scripts/fenics_intro/scratch_example.py}

%===============================================================================
%===============================================================================

\chapter{Problems in one dimension}

To develop an understanding of finite elements further, we investigate several common one-dimensional problems.

%===============================================================================

\section{Second-order linear equation}

\index{Linear differential equations!1D}
  For the first example, we present the inner and outer singular-perturbation solution \index{Singular perturbation} \citep{logan} to the second-order linear equation with non-constant coefficients, 
  \begin{align}
    \label{1d_example_1_ode}
    \epsilon \ddot{u} - (2t + 1)\dot{u} + 2u = 0, \hspace{5mm} 0 < t < 1, \hspace{5mm} 0 < \epsilon \ll 1,
  \end{align}
  \begin{align}
    \label{1d_example_1_bcs}
    u(0) = 1, \hspace{5mm} u(1) = 0,
  \end{align}
  and compare it to the solution obtained using the finite element method.

  \subsection{Singular perturbation solution}

  The solution to the unperturbed problem ($\epsilon = 0$) is found with the left boundary condition $u(0) = 1$:
  \begin{align*}
    - (2t + 1)\dot{u} + 2u = 0, \\
    \implies u(t) = c_1(2t + 1),
  \end{align*}
  applying the boundary condition $u(0) = 1$ results in the outer solution
  \begin{align*}
    u_o(t) = 2t + 1.
  \end{align*}
  In order to determine the width $\delta(\epsilon)$ of the \index{Boundary layer} boundary layer we re-scale near $t=1$ via
  \begin{align*}
    \xi = \frac{1 - t}{\delta(\epsilon)}, \hspace{10mm} U(\xi) = u(t).
  \end{align*}
  In scaled variables the differential equation becomes
  \begin{align*}
    \left( \frac{\epsilon}{\delta(\epsilon)^2} \right) \ddot{U} + \left( \frac{2 - 2\xi\delta(\epsilon) + 1}{\delta(\epsilon)}\right)\dot{U} + 2U &= 0 
  \end{align*}
  For this problem the second derivative term may be retained by making $\delta(\epsilon) = \bigo\left(\epsilon\right)$ resulting in the scaled differential equation
  \begin{align*}
    \left( \frac{\epsilon}{\epsilon^2} \right) \ddot{U} + \left( \frac{2 - 2\xi\epsilon + 1}{\epsilon}\right)\dot{U} + 2U &= 0 \\ 
    \ddot{U} + 3\dot{U} - 2\xi\epsilon \dot{U} + 2\epsilon U &= 0.
  \end{align*}
  The inner approximation to first-order satisfies
  \begin{align*}
    \ddot{U} + 3\dot{U} = 0,
  \end{align*}
  with general solution
  \begin{align*}
    U(\xi) = C_1 + C_2 e^{-3\xi},
  \end{align*}
  and also in terms of $u$ and $t$,
  \begin{align*}
    u(t) = C_1 + C_2 \exp\left( -3\left( \frac{1 - t}{\epsilon} \right) \right).
  \end{align*}
  Applying the boundary condition $u(1) = 0$ in the boundary layer gives $C_1 = -C_2$, and the inner approximation is
  \begin{align*}
    u_i(t) = C_2 \left( \exp\left( \frac{3t - 3}{\epsilon} \right) - 1 \right).
  \end{align*}

  To find $C_2$, an overlap domain of order $\sqrt{\epsilon}$ and an appropriate intermediate scaled variable
  $$\eta = \frac{1 - t}{\sqrt{\epsilon}}.$$
  are introduced.  Thus $t = 1 - \eta\sqrt{\epsilon}$ and the matching conditions becomes (with $\eta$ fixed)
  \begin{align*}
    \lim_{\epsilon \rightarrow 0^+} u_o\left(1 - \eta\sqrt{\epsilon}\right) = \lim_{\epsilon \rightarrow 0^+} u_i\left(1 - \eta\sqrt{\epsilon}\right),
  \end{align*}
  or
  \begin{align*}
    0 = &+ \lim_{\epsilon \rightarrow 0^+} \exp\left( 2(1 - \eta\sqrt{\epsilon}) + 1 \right) \\
        &- \lim_{\epsilon \rightarrow 0^+} C_2\left(\exp\left(\frac{3(1 - \eta\sqrt{\epsilon}) - 3}{\epsilon} \right) - 1 \right) \\
    0 = &+ \lim_{\epsilon \rightarrow 0^+} \exp\left( 3 - 2\eta\sqrt{\epsilon} \right) \\
        &- \lim_{\epsilon \rightarrow 0^+} C_2\left(\exp\left(\frac{- 3\eta\sqrt{\epsilon}}{\epsilon} \right) - 1 \right) \\
    0 = &\ 3 + C_2 \implies C_2 = -3.
  \end{align*}
  A uniform approximation $y_u(t)$ is found by adding the inner and outer approximations and subtracting the common limit in the overlap domain, which is $3$ in this case.  Consequently,
  \begin{align*}
    u_u(t) &= 2t + 1 - 3 \left( \exp\left( \frac{3t - 3}{\epsilon} \right) - 1 \right) - 3 \\
    &= 2t - 3 \exp\left( \frac{3t - 3}{\epsilon} \right) + 1.
  \end{align*}

  \subsection{Finite element solution} \label{ssn_intro_sing_perp_one}
  
  We arrive at the weak form by taking the inner product of Equation (\ref{1d_example_1_ode}) with the test function $\phi$, integrating over the domain of the problem $\Omega$ and integrating the second derivative term by parts,
  \begin{align*}
    0 = &\int_{\Omega} \left[ \epsilon \ddot{u} - (2t + 1)\dot{u} + 2u \right] \phi d\Omega \\
    0 = &\epsilon \int_{\Omega} \frac{d^2 u}{dt^2} \phi d\Omega - \int_{\Omega} (2t + 1) \frac{du}{dt} \phi d\Omega + 2\int_{\Omega} u \phi d\Omega \\
    0 = &+\epsilon \int_{\Gamma} \left( \frac{du}{dt} \phi \right) n d\Gamma - \epsilon \int_{\Omega} \frac{du}{dt} \frac{d\phi}{dt} d\Omega \\
    &- \int_{\Omega} (2t + 1) \frac{du}{dt} \phi d\Omega + 2\int_{\Omega} u \phi d\Omega,
  \end{align*}
  where $n$ is the outward-pointing normal to the boundary $\Gamma$.  Because the boundary conditions are both Dirichlet, the integral over the boundary are all zero (see test space \ref{test_space}).  Therefore, the variational problem reads: find $u \in \trialspace$ such that 
  \begin{align*}
    0 = - \epsilon \int_{\Omega} \frac{du}{dt} \frac{d\phi}{dt} d\Omega - \int_{\Omega} (2t + 1) \frac{du}{dt} \phi d\Omega + 2\int_{\Omega} u \phi d\Omega.
  \end{align*}
  for all $\phi \in \testspace$.
    
\begin{figure}
    \centering
      \includegraphics[width=\linewidth]{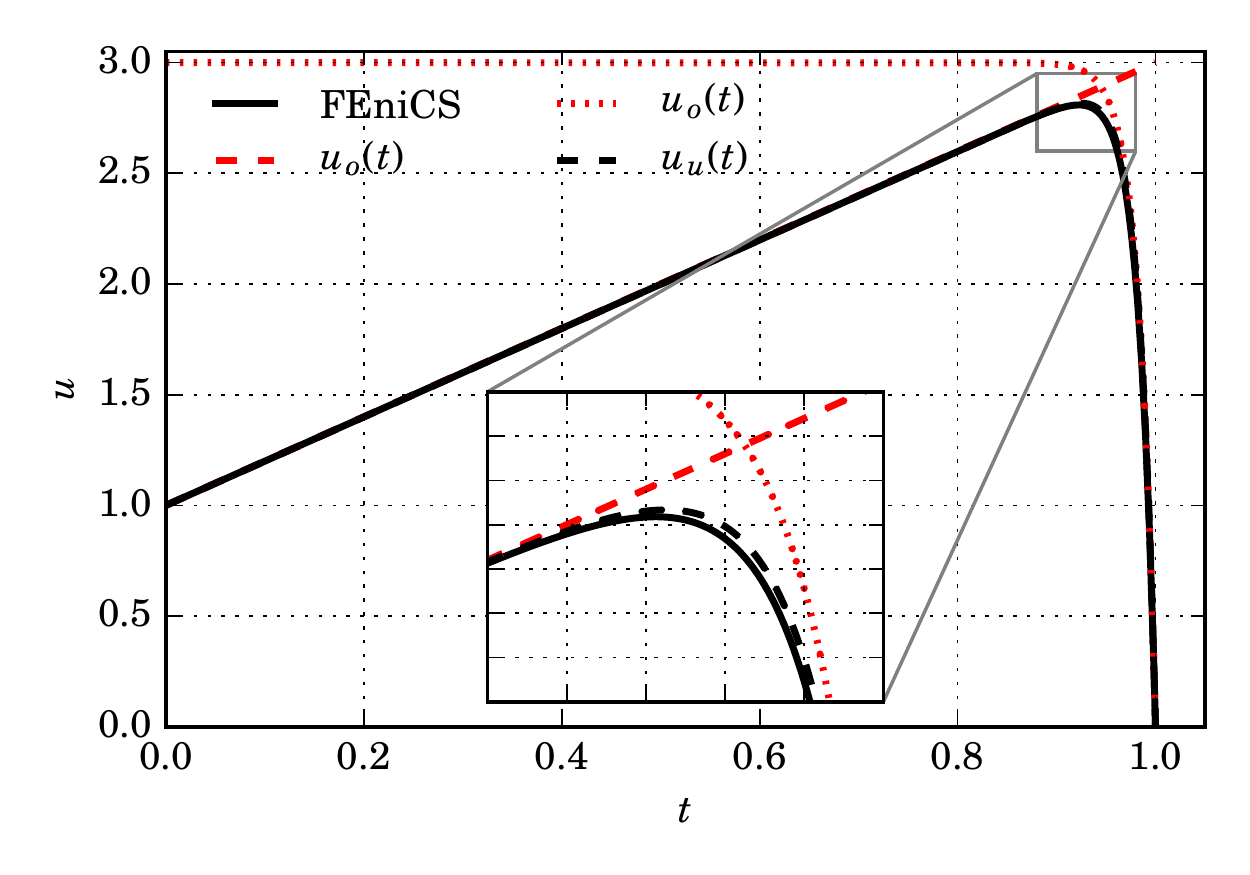}
    \caption[Singular-perturbation solution]{Finite element (solid black) and singular perturbation (dashed black) solutions.}
    \label{1d_bvp_1_image}
  \end{figure}

  A weak solution to this weak problem using linear Lagrange interpolation functions (\ref{linear_lagrange_functions}) is shown in Figure \ref{1d_bvp_1_image}, and was generated from Code Listing \ref{1d_bvp_1_code}.

\pythonexternal[label=1d_bvp_1_code, caption={FEniCS solution to BVP (\ref{1d_example_1_ode}, \ref{1d_example_1_bcs}).}]{scripts/fenics_intro/1D_BVP_1.py}
  
%===============================================================================

  \section{Neumann-Dirichlet problem}

\index{Linear differential equations!1D}
  It may also be of interest to solve a problem possessing a Neumann boundary condition.  For example, boundary conditions (\ref{1d_example_1_bcs}) for differential equation (\ref{1d_example_1_ode}) may be altered to 
  \begin{align}
    \label{1d_example_2_bcs}
    u(0) = 0, \hspace{10mm} \dot{u}(1) = u_r = -10.
  \end{align}
  In this case the weak form is derived similarly to \S \ref{ssn_intro_sing_perp_one}, and consists of finding $u \in \trialspace$ such that
  \begin{align*}
    0 = &+\epsilon \int_{\Gamma} \left( \frac{du}{dt} \phi \right) n d\Gamma - \epsilon \int_{\Omega} \frac{du}{dt} \frac{d\phi}{dt} d\Omega \\
    &- \int_{\Omega} (2t + 1) \frac{du}{dt} \phi d\Omega + 2\int_{\Omega} u \phi d\Omega \\
    0 = &+\epsilon \int_{\Gamma_r} u_r \phi d\Gamma_r - \epsilon \int_{\Omega} \frac{du}{dt} \frac{d\phi}{dt} d\Omega \\
    &- \int_{\Omega} (2t + 1) \frac{du}{dt} \phi d\Omega + 2\int_{\Omega} u \phi d\Omega,
  \end{align*}
  for all $\phi \in \testspace$, where the fact that $n = 1$ on the right boundary $\Gamma_r$.
    
\begin{figure}
    \centering
      \includegraphics[width=\linewidth]{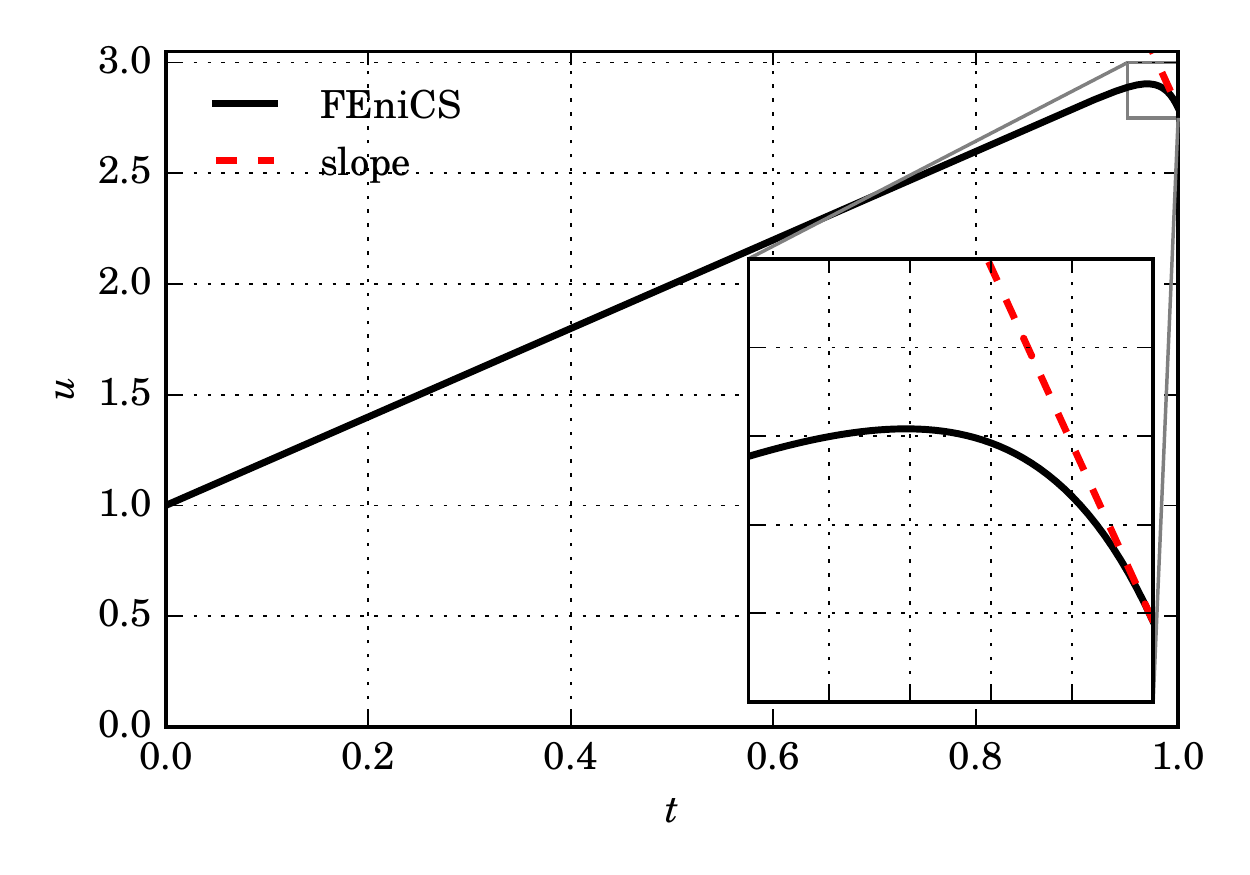}
    \caption[BVP example one solution]{Finite element solution (solid black) and slope (dashed red).}
    \label{1d_bvp_2_image}
  \end{figure}

    The weak solution using linear Lagrange shape functions (\ref{linear_lagrange_functions}) is shown in Figure \ref{1d_bvp_2_image}, and was generated from Code Listing \ref{1d_bvp_2_code}.

\pythonexternal[label=1d_bvp_2_code, caption={FEniCS solution to BVP (\ref{1d_example_1_ode}, \ref{1d_example_2_bcs}).}]{scripts/fenics_intro/1D_BVP_2.py}

%===============================================================================

  \section{Integration} \label{ssn_integration}
  
  \index{Numerical integration}
  An interesting problem easily solved with the finite element method is the integration of a function $u$ over domain $\Omega = [a,b]$,
  \begin{align}
    \label{numerical_integration_example}
    v(x) &= \int_a^x u(s) ds = U(x) - U(a),
  \end{align}
  where $U$ is an anti-derivative of $u$ such that
  \begin{align*}
    \frac{dU}{dx} = u(x).
  \end{align*}
  Because the integral is from $a$ to $x$, $U(a) = 0$; hence $v(x) = U(x)$ and the equivalent problem to (\ref{numerical_integration_example}) is the first-order boundary-value problem
  \begin{align}
    \label{1d_example_4}
    \frac{dv}{dx} = u(x), \hspace{5mm} v(a) = 0.
  \end{align}

  The corresponding variational problem reads: find $v \in \trialspace$ such that
  \begin{align*}
    \int_{\Omega} \frac{dv}{dx} \phi d\Omega &= \int_{\Omega} u \phi d\Omega,
  \end{align*}
  for all $\phi \in \testspace$.
  
\begin{figure}
    \centering
      \includegraphics[width=\linewidth]{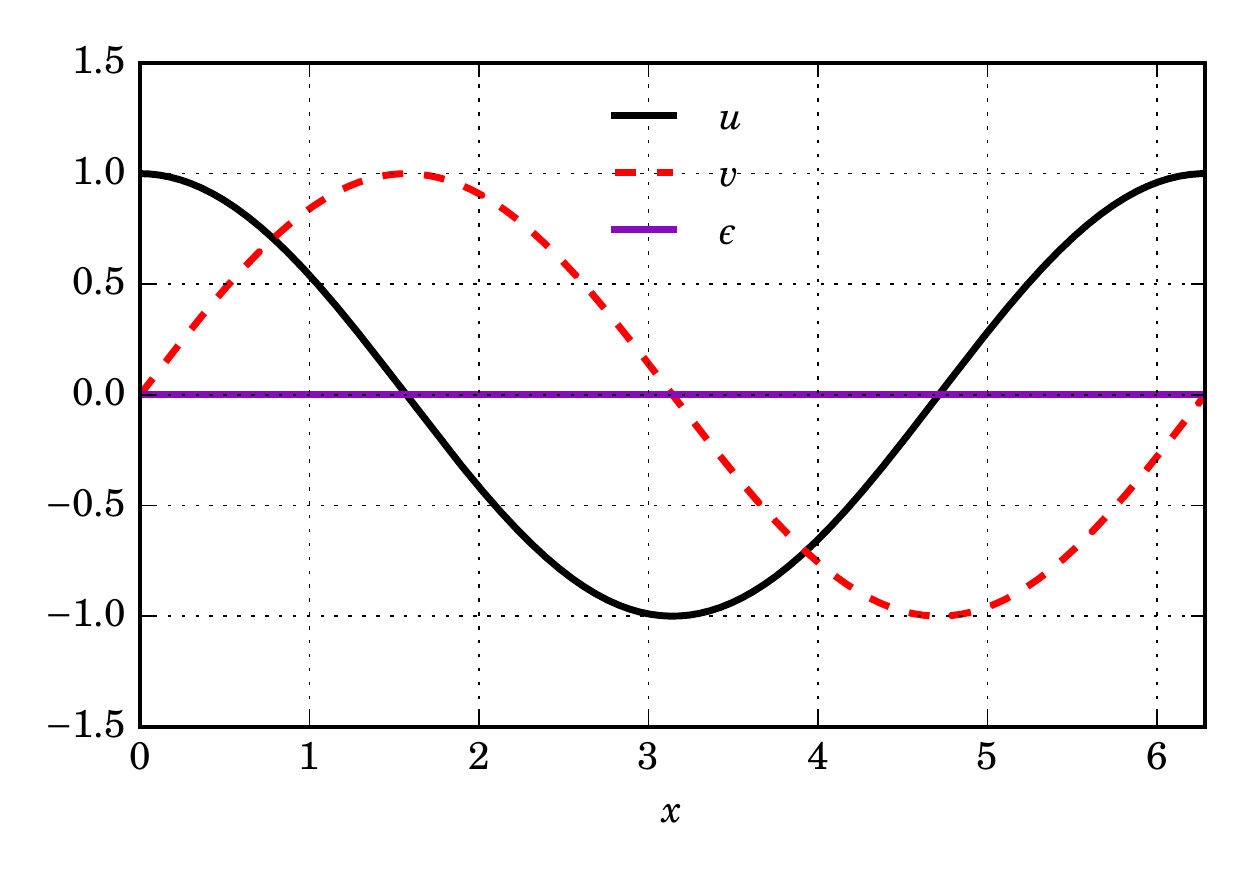}
    \caption[BVP example two solution]{The function $u(x) = \cos(x)$ for $a=0$, $b=2\pi$ (solid black); finite element solution $v(x) \approx \sin(x)$ (dashed red); and error $\epsilon(x) = v(x) - \sin(x)$ (purple solid).}
    \label{1d_bvp_4_image}
  \end{figure}

  The linear-Lagrange-element-basis-weak solution to this problem with $u(x) = \cos(x)$ over the domain $\Omega = [0,2\pi]$ is shown in Figure \ref{1d_bvp_4_image}, and was generated from Code Listing \ref{1d_bvp_4_code}. 

\pythonexternal[label=1d_bvp_4_code, caption={FEniCS solution to BVP (\ref{1d_example_4})}]{scripts/fenics_intro/1D_BVP_4.py}
  
%===============================================================================
  
  \section{Directional derivative} \label{ssn_intro_directional_deriavative}
 
  \index{Directional derivative}
\index{Linear differential equations!1D}
  It is often important to compute the derivative of one function with respect to another, say
  \begin{align}
    \label{1d_example_5}
    w(x) = \frac{du}{dv} = \frac{du}{dx} \frac{dx}{dv} = \frac{du}{dx} \left( \frac{dv}{dx} \right)^{-1},
  \end{align}
  for continuous functions $u(x)$ and $v(x)$ defined over the interval $\Omega = [a,b]$.  The variational form for this problem with trial function $w \in \ltwospace$ (see $L^2$ space (\ref{l2_space})), test function $\phi \in \testspace$, is simply the inner product
   \begin{align*}
     \int_{\Omega} w \phi d\Omega &= \int_{\Omega} \frac{du}{dx} \left( \frac{dv}{dx} \right)^{-1} \phi d\Omega, \hspace{5mm} \forall \phi in \testspace,
   \end{align*}
   where no restrictions are made on the boundary; this is possible because both $u$ and $v$ are known \emph{a priori} and hence their derivatives may be computed directly and estimated throughout the entire domain.
   
   Solving problems of this type are referred to in the literature as \index{Projection} \emph{projections}, due to the fact that they simply project a known solution onto a finite element basis.
   
\begin{figure}
    \centering
      \includegraphics[width=\linewidth]{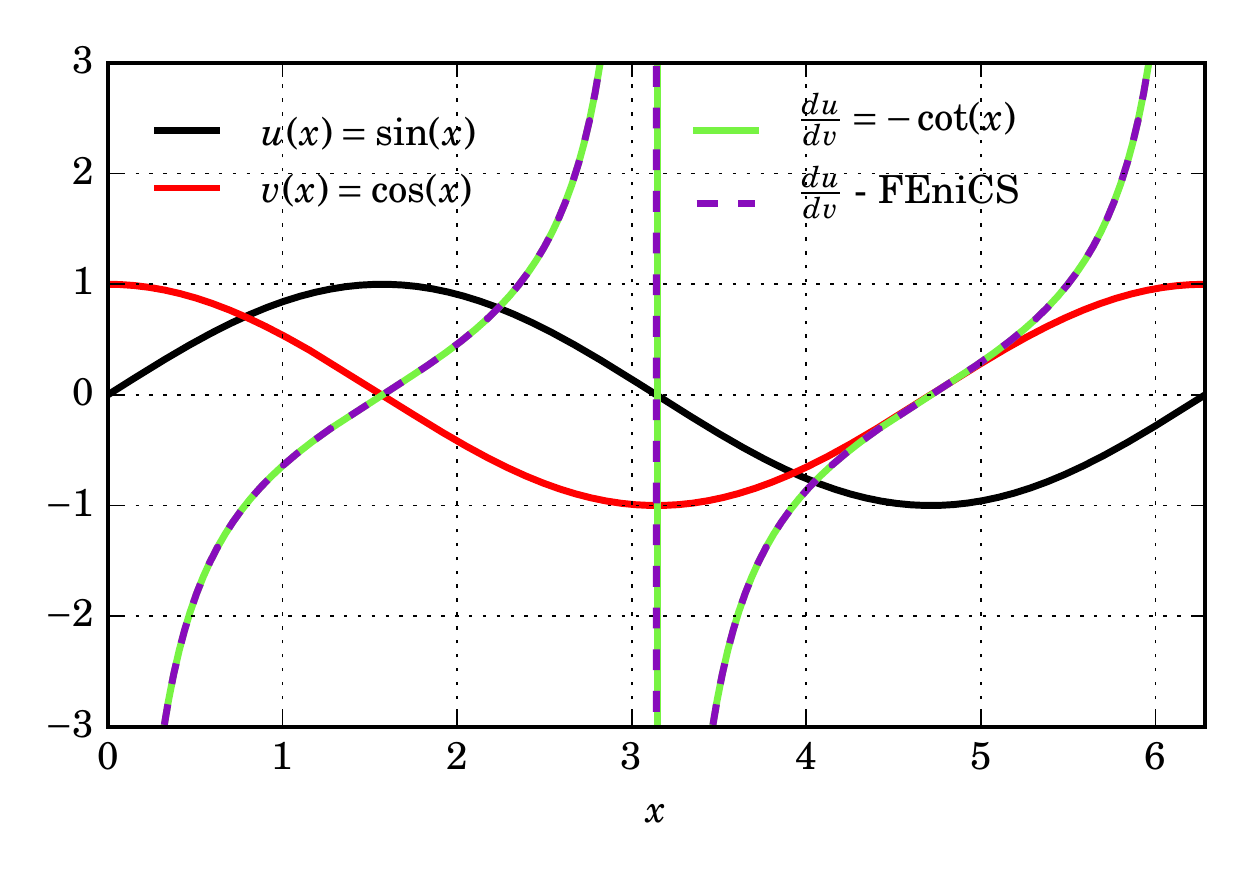}
    \caption[Functional derivative example one solution]{Both analytic (green) and numerical (purple) functional derivative $\frac{du}{dv} = -\cot(x)$ for $u(x) = \sin(x)$ (black), and $v(x) = \cos(x)$ (red).}
    \label{1d_dir_dir_1_image}
  \end{figure}

\begin{figure}
    \centering
      \includegraphics[width=\linewidth]{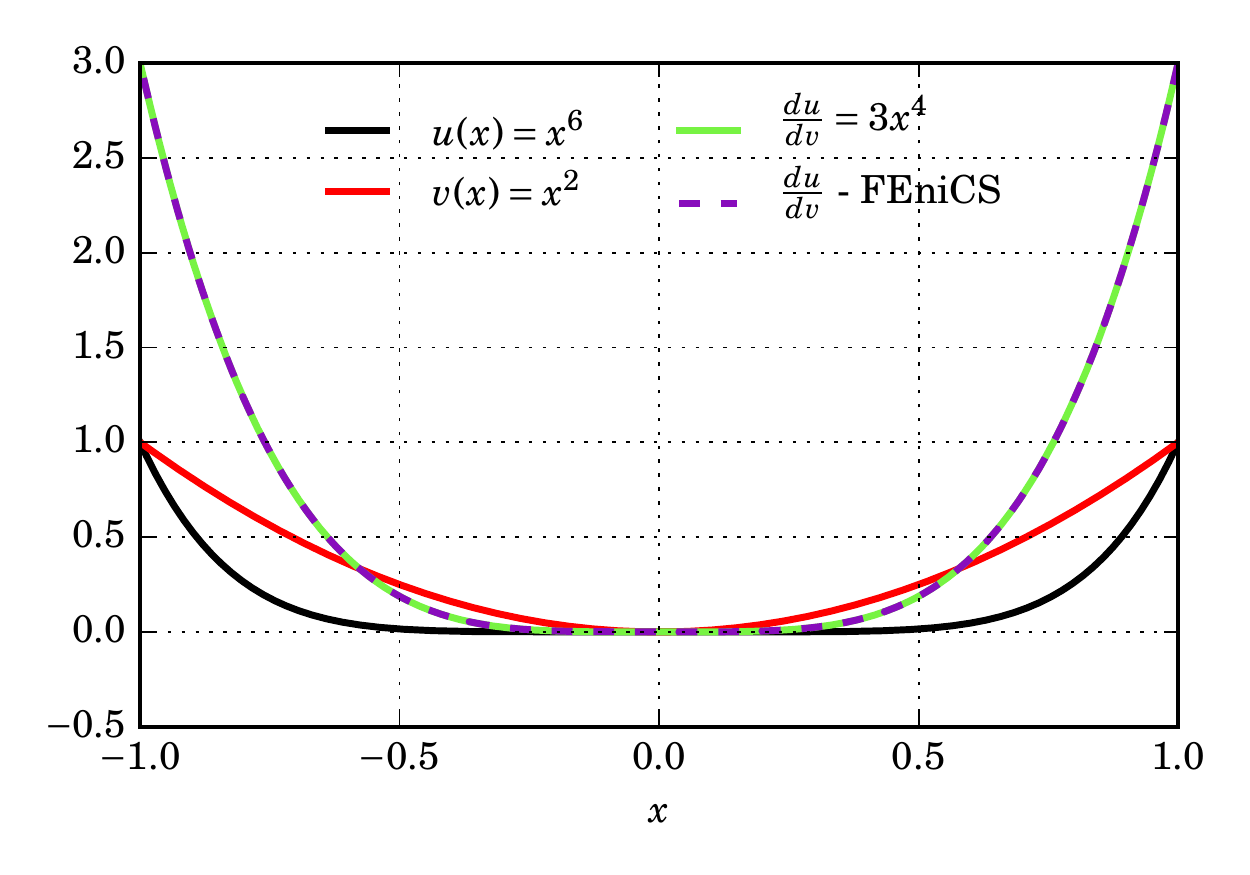}
    \caption[Functional derivative example two solution]{Both analytic (green) and numerical (purple) functional derivative $\frac{du}{dv} = 3x^4$ for $u(x) = x^6$ (black), and $v(x) = x^2$ (red).}
    \label{1d_dir_dir_2_image}
  \end{figure}

   An example solution with $u(x) = \sin(x)$, $v(x) = \cos(x)$ is generated using linear-Lagrange elements from Code Listing \ref{1d_dir_dir_1_code} and depicted in Figure \ref{1d_dir_dir_1_image},  and another with $u(x) = 3x^4$, $v(x) = x^6$ generated from Code Listing \ref{1d_dir_dir_2_code} and depicted in Figure \ref{1d_dir_dir_2_image}.

\pythonexternal[label=1d_dir_dir_1_code, caption={FEniCS solution to BVP (\ref{1d_example_5}) with $u(x) = \sin(x)$, $v(x) = \cos(x)$.}]{scripts/fenics_intro/1D_dir_dir_1.py}
  
\pythonexternal[label=1d_dir_dir_2_code, caption={FEniCS solution to BVP (\ref{1d_example_5}) with $u(x) = 3x^4$, $v(x) = x^6$.}]{scripts/fenics_intro/1D_dir_dir_2.py}
  
%===============================================================================

\section{Eigenvalue problem}

\index{Transient problem}
\index{Eigenvalue problem}
The 1D initial-boundary-value problem for the heat equation with heat conductivity $k$ is
\begin{align}
  \label{intro_eigen_pde}
  \frac{\partial u}{\partial t} &= \frac{\partial}{\partial x} \left[ k \frac{\partial u}{\partial x} \right], &&0 < x < \ell,\ t > 0, \\
  \label{intro_eigen_bcs}
  u(0,t) &= u(\ell,t) = 0, &&t > 0, \\
  \label{intro_eigen_ic}
  u(x,0) &= f(x),       &&0 < x < \ell.
\end{align}
We examine this problem in the context of exact and approximate Eigenvalues and Eigenvectors in the following sections.

\subsection{Fourier series approximation}

\index{Fourier series method}
The solution of system (\ref{intro_eigen_pde} -- \ref{intro_eigen_ic}) can be approximated by the method of separation of variables, developed by Joseph Fourier in 1822 \citep{davis}.  Using this method, we first assume a solution of the form $u(x,t) = X(x)T(t)$, so that equation (\ref{intro_eigen_pde}) reads
\begin{align*}
  X(x)\frac{\partial T(t)}{\partial t}  &= \frac{\partial}{\partial x} \left[ k \frac{\partial X(x)}{\partial x} \right] T(t) \\ 
  \frac{1}{k T(t)}\frac{\partial T(t)}{\partial t}  &= \frac{1}{X(x)} \frac{\partial}{\partial x} \left[ \frac{\partial X(x)}{\partial x} \right], 
\end{align*}
For two functions with two different independent variables to be equal, they must both equal to the same constant $-\lambda$, $\lambda > 0$,
\begin{align*}
  \frac{1}{k T(t)}\frac{\partial T(t)}{\partial t} = \frac{1}{X(x)} \frac{\partial}{\partial x} \left[ \frac{\partial X(x)}{\partial x} \right] = - \lambda, 
\end{align*}
so
\begin{align}
  \label{intro_eigen_separated}
  \frac{\partial T(t)}{\partial t} + \lambda k T(t) = 0, \hspace{10mm} \frac{\partial^2 X(x)}{\partial x^2} + \lambda X(x) = 0.
\end{align}
The solution to the first-order $T$ equation is 
\begin{align}
  \label{intro_eigen_t}
  T(t) = Ae^{-\lambda k t},
\end{align}
while the solution to the second-order $X$ equation is
\begin{align*}
  X(x) = B_1 \sin\left( \sqrt{\lambda} x \right) + B_2 \cos\left( \sqrt{\lambda} x \right),
\end{align*}
for some coefficients $A$, $B_1$, and $B_2$ where $B_1$ and $B_2$ cannot both be zero since then only the trivial solution exists.  Applying boundary conditions (\ref{intro_eigen_bcs}), $X(0) = X(\ell) = 0$,
\begin{align*}
  X(0) &= B_2 = 0, \\
  X(\ell) &= B_1 \sin\left( \sqrt{\lambda} \ell \right) = 0 &&\\
  \implies \sqrt{\lambda} \ell &= n\pi, &&n = 1,2,3,\dots \\
  \implies \lambda &= \left( \frac{n \pi}{\ell} \right)^2, &&n = 1,2,3,\dots
\end{align*}
Therefore, the Eigenfunctions and Eigenvalues for the steady-state problem are
\begin{align}
  \label{intro_eigen_x}
  X_n(x) = \sin\left( \sqrt{\lambda_n} x \right) \hspace{5mm}
  \lambda_n = \left( \frac{n \pi}{\ell} \right)^2, &&n = 1,2,3,\dots
\end{align}
By the Superposition Principle, any linear combination of solutions is again a solution, and solution to the transient problem is
\begin{align}
  u(x,t) &= \sum_{n=1}^{\infty} c_n X_n(x) T_n(t) \notag \\
  \label{intro_eigen_fourier_soln}
         &= \sum_{n=1}^{\infty} c_n \sin\left( \sqrt{\lambda_n} x \right) e^{-\lambda_n k t},
\end{align}
where $c_n = A B_1$ and (\ref{intro_eigen_t}) has been used with $\lambda_n = \left( \frac{n \pi}{\ell} \right)^2$.  The coefficient $c_n$ may be discovered by inspecting initial condition (\ref{intro_eigen_ic})
\begin{align*}
  u(x,0) &= \sum_{n=1}^{\infty} c_n \sin\left( \sqrt{\lambda_n} x \right) = f(x).
\end{align*}
Multiplying both sides of this function by $\sin(m\pi x/\ell)$ for arbitrary $m$ and utilizing the fact that
\begin{align*}
  \int_{0}^{\ell} \sin\left(\frac{n\pi x}{\ell}\right) \sin\left(\frac{m\pi x}{\ell}\right) dx =
  \begin{cases}
    0,      & n \neq m, \\
    \ell/2, & n = m,
  \end{cases}
\end{align*}
results in
\begin{align*}
  \sum_{n=1}^{\infty} c_n \sin\left( \frac{n\pi x}{\ell} \right) \sin\left(\frac{m\pi x}{\ell} \right) &= f(x) \sin\left( \frac{m\pi x}{\ell} \right) \\
  \int_{0}^{\ell} \sum_{n=1}^{\infty} c_n \sin\left( \frac{n\pi x}{\ell} \right) \sin\left(\frac{m\pi x}{\ell} \right) dx &= \int_{0}^{\ell} f(x) \sin\left( \frac{m\pi x}{\ell} \right) dx \\
  \sum_{n=1}^{\infty} c_n \int_{0}^{\ell} \sin\left( \frac{n\pi x}{\ell} \right) \sin\left(\frac{m\pi x}{\ell} \right) dx &= \int_{0}^{\ell} f(x) \sin\left( \frac{m\pi x}{\ell} \right) dx \\
  c_m \int_{0}^{\ell} \sin\left( \frac{m\pi x}{\ell} \right) \sin\left(\frac{m\pi x}{\ell} \right) dx &= \int_{0}^{\ell} f(x) \sin\left( \frac{m\pi x}{\ell} \right) dx \\
  c_m \left( \frac{\ell}{2} \right) &= \int_{0}^{\ell} f(x) \sin\left( \frac{m\pi x}{\ell} \right) dx.
\end{align*}
Finally, replacing the dummy variable $m$ with $n$,
\begin{align}
  \label{intro_eigen_fourier_coef}
  c_n = \frac{2}{\ell} \int_{0}^{\ell} f(x) \sin\left( \frac{n\pi x}{\ell} \right) dx.
\end{align}

Therefore, the Fourier series approximation to (\ref{intro_eigen_pde} -- \ref{intro_eigen_ic}) is (\ref{intro_eigen_fourier_soln}) with coefficients given by (\ref{intro_eigen_fourier_coef}).

\subsection{Finite element approximation}

Investigating the Eigenvalue problem of separated equations (\ref{intro_eigen_separated}) for $X$,
\begin{align*}
  \frac{\partial^2 X(x)}{\partial x^2} + \lambda X(x) = 0, \hspace{5mm} X(0) = X(\ell) = 0,
\end{align*}
suggests that a weak form can be developed by making $X \in \trialspace$ and taking the inner product of this equation with the test function $\phi \in \testspace$
\begin{align}
  \int_0^{\ell} \left[ \frac{\partial^2 X}{\partial x^2} + \lambda X \right]\phi dx &= 0 \notag \\
  \int_0^{\ell} \left[ \lambda X \phi - \frac{\partial X}{\partial x} \frac{\partial \phi}{\partial x} \right] dx - \frac{\partial X}{\partial x} \phi(0) + \frac{\partial X}{\partial x} \phi(\ell) &= 0 \notag \\
  \label{intro_eigen_weak_form}
  \int_0^{\ell} \left[ \lambda X \phi - \frac{\partial X}{\partial x} \frac{\partial \phi}{\partial x} \right] dx &= 0,
\end{align}
where the fact that the boundary conditions are all essential has been used.

Substituting the Galerkin approximation
\begin{align*}
  X_e = \sum_{j=1}^n u_j^e \psi_j^e(x),
\end{align*}
where $u_j^e$ is the $j$th nodal value of $X$ at element $e$ and $\psi_j^e(x)$ is the element's associated interpolation function, into Equation (\ref{intro_eigen_weak_form}) results in the Galerkin system
\begin{align*}
  \left[ \lambda M_{ij}^e - K_{ij}^e \right] \cdot \mathbf{u}^e = \mathbf{0},
\end{align*}
with stiffness tensor $K$ and \index{Mass matrix} \emph{mass tensor} $M$
\begin{align*}
  K_{ij}^e = \int_0^{\ell} \frac{\partial \psi_i^e}{\partial x} \frac{\partial \psi_j^e}{\partial x} dx, \hspace{10mm} M_{ij}^e = \int_0^{\ell} \psi_i^e \psi_j^e dx.
\end{align*}
Expanding the element equation tensors as in \S \ref{ssn_local_galerkin_assembly} results in
\begin{align*}
  \left( \frac{\lambda h_e}{6} \begin{bmatrix}[r]
                                 2 & 1 \\
                                 1 & 2
                               \end{bmatrix}
         - \frac{1}{h_e} \begin{bmatrix}[r]
                            1 & -1 \\
                           -1 & 1
                         \end{bmatrix}
  \right) \cdot
  \begin{bmatrix}
    u_1^e \\
    u_2^e\
  \end{bmatrix} &=
  \begin{bmatrix}
    0 \\
    0
  \end{bmatrix}.
\end{align*}

For a concrete example, we assemble this local system over an equally-space $n=3$-element function space, implying that $h_e = 1/3$ and (see \S \ref{ssn_global_galerkin_assembly})
\begin{align*}
  \left( \frac{\lambda}{18} \begin{bmatrix}[r]
                              2 & 1 & 0 & 0 \\
                              1 & 4 & 1 & 0 \\
                              0 & 1 & 4 & 1 \\
                              0 & 0 & 1 & 2 \\
                            \end{bmatrix}
         - 3 \begin{bmatrix}[r]
                1 & -1 &  0 &  0 \\
               -1 &  2 & -1 &  0 \\
                0 & -1 &  2 & -1 \\
                0 &  0 & -1 &  1
             \end{bmatrix}
  \right) \cdot
  \begin{bmatrix}
    X_1 \\
    X_2 \\
    X_3 \\
    X_4 \\
  \end{bmatrix} &=
  \begin{bmatrix}
    0 \\
    0 \\
    0 \\
    0
  \end{bmatrix}.
\end{align*}
The boundary conditions $X(0) = X(\ell) = 0$ imply that $X_1 = X_4 = 0$ and thus this system reduces to a system of two linear equations,
\begin{align*}
  \left( \lambda \frac{4}{18} - 6 \right) X_2 + \left( \lambda\frac{1}{18} + 3 \right) X_3 &= 0 \\
  \left( \lambda \frac{1}{18} + 3 \right) X_2 + \left( \lambda\frac{4}{18} - 6 \right) X_3 &= 0,
\end{align*}
or
\begin{align*}
  \begin{bmatrix}
    \left( \lambda \frac{2}{9} - 6 \right) & \left( \lambda \frac{1}{18} + 3 \right) \\
    \left( \lambda \frac{1}{18} + 3 \right) & \left( \lambda \frac{2}{9} - 6 \right) \\
  \end{bmatrix} \cdot
  \begin{bmatrix}
    X_2 \\
    X_3 
  \end{bmatrix} &=
  \begin{bmatrix}
    0 \\
    0
  \end{bmatrix}.
\end{align*}
The characteristic polynomial is found by setting the determinant of the coefficient matrix equal to zero,
\begin{align*}
  \begin{vmatrix}
    \left( \lambda \frac{2}{9} - 6 \right) & \left( \lambda \frac{1}{18} + 3 \right) \\
    \left( \lambda \frac{1}{18} + 3 \right) & \left( \lambda \frac{2}{9} - 6 \right) \\
  \end{vmatrix} &= 0 \\
  \left( \lambda \frac{2}{9} - 6 \right) \left( \lambda \frac{2}{9} - 6 \right) - \left( \lambda \frac{1}{18} + 3 \right) \left( \lambda \frac{1}{18} + 3 \right) &= 0 \\
  \left( \lambda^2 \frac{4}{81} - \lambda \frac{24}{9} + 36 \right) - \left( \lambda^2 \frac{1}{324} + \lambda \frac{6}{18} + 9 \right) &= 0 \\
  \left( \frac{4}{81} - \frac{1}{324} \right) \lambda^2 - \left( \frac{24}{9} + \frac{6}{18} \right) \lambda + 27 &= 0 \\
  \left( \frac{5}{108} \right) \lambda^2 - 3 \lambda + 27 &= 0,
\end{align*}
with roots
\begin{align*}
  \lambda &= \frac{3 \pm \sqrt{9 - 108\left( \frac{5}{108} \right)}}{2\left( \frac{5}{108} \right)} = \frac{162}{5} \pm \frac{108}{5},
\end{align*}
providing the two Eigenvalue approximations $\widetilde{\lambda}_1 = 54/5 = 10.8$ and $\widetilde{\lambda}_2 = 54$.  The Eigenvectors may then be derived from the linear systems
\begin{align*}
  \begin{cases}
  \begin{bmatrix}
    \left( \widetilde{\lambda}_1 \frac{2}{9} - 6 \right) & \left( \widetilde{\lambda}_1 \frac{1}{18} + 3 \right) \\
    \left( \widetilde{\lambda}_1 \frac{1}{18} + 3 \right) & \left( \widetilde{\lambda}_1 \frac{2}{9} - 6 \right) \\
  \end{bmatrix} \cdot
  \begin{bmatrix}
    X_2^1 \\
    X_3^1 
  \end{bmatrix} &=
  \begin{bmatrix}
    0 \\
    0
  \end{bmatrix} \\
  \begin{bmatrix}
    \left( \widetilde{\lambda}_2 \frac{2}{9} - 6 \right) & \left( \widetilde{\lambda}_2 \frac{1}{18} + 3 \right) \\
    \left( \widetilde{\lambda}_2 \frac{1}{18} + 3 \right) & \left( \widetilde{\lambda}_2 \frac{2}{9} - 6 \right) \\
  \end{bmatrix} \cdot
  \begin{bmatrix}
    X_2^2 \\
    X_3^2 
  \end{bmatrix} &=
  \begin{bmatrix}
    0 \\
    0
  \end{bmatrix}
  \end{cases}&\\
  \begin{cases}
  \begin{bmatrix}
    -\frac{18}{5} &  \frac{18}{5} \\
     \frac{18}{5} & -\frac{18}{5} \\
  \end{bmatrix} \cdot
  \begin{bmatrix}
    X_2^1 \\
    X_3^1 
  \end{bmatrix} &=
  \begin{bmatrix}
    0 \\
    0
  \end{bmatrix} \\
  \begin{bmatrix}
    6 & 6 \\
    6 & 6 \\
  \end{bmatrix} \cdot
  \begin{bmatrix}
    X_2^2 \\
    X_3^2 
  \end{bmatrix} &=
  \begin{bmatrix}
    0 \\
    0
  \end{bmatrix}
  \end{cases}&
\end{align*}
providing $X^1 = \begin{bmatrix} 1 & 1 \end{bmatrix}\T$ and $X^2 = \begin{bmatrix} 1 & -1 \end{bmatrix}\T$.

Similar to the derivation of Fourier series solution (\ref{intro_eigen_fourier_soln}), these approximate Eigenvectors and Eigenvalues are used to create the approximate three-element transient solution
\begin{align}
  u(x,t) \approx \widetilde{u}(x,t) &= \sum_{e=1}^n X^e(x) T^e(t, \widetilde{\lambda}_e) \notag \\
  \label{intro_eigen_fem_soln}
  &= \sum_{e=1}^3 \left[ \sum_{j=1}^2 X^e_j \psi_j^e(x) \right] c_e e^{-\widetilde{\lambda}_e k t},
\end{align}
where $c_e = c_n$ are the same coefficients of orthogonality as (\ref{intro_eigen_fourier_coef}) from the Fourier series approximation at element $e = n$, $n=1,2,3$, and $X^e$ is the Eigenvector associated with Eigenvalue $\widetilde{\lambda}_e$.

\begin{figure*}
  \centering
    \includegraphics[width=0.7\linewidth]{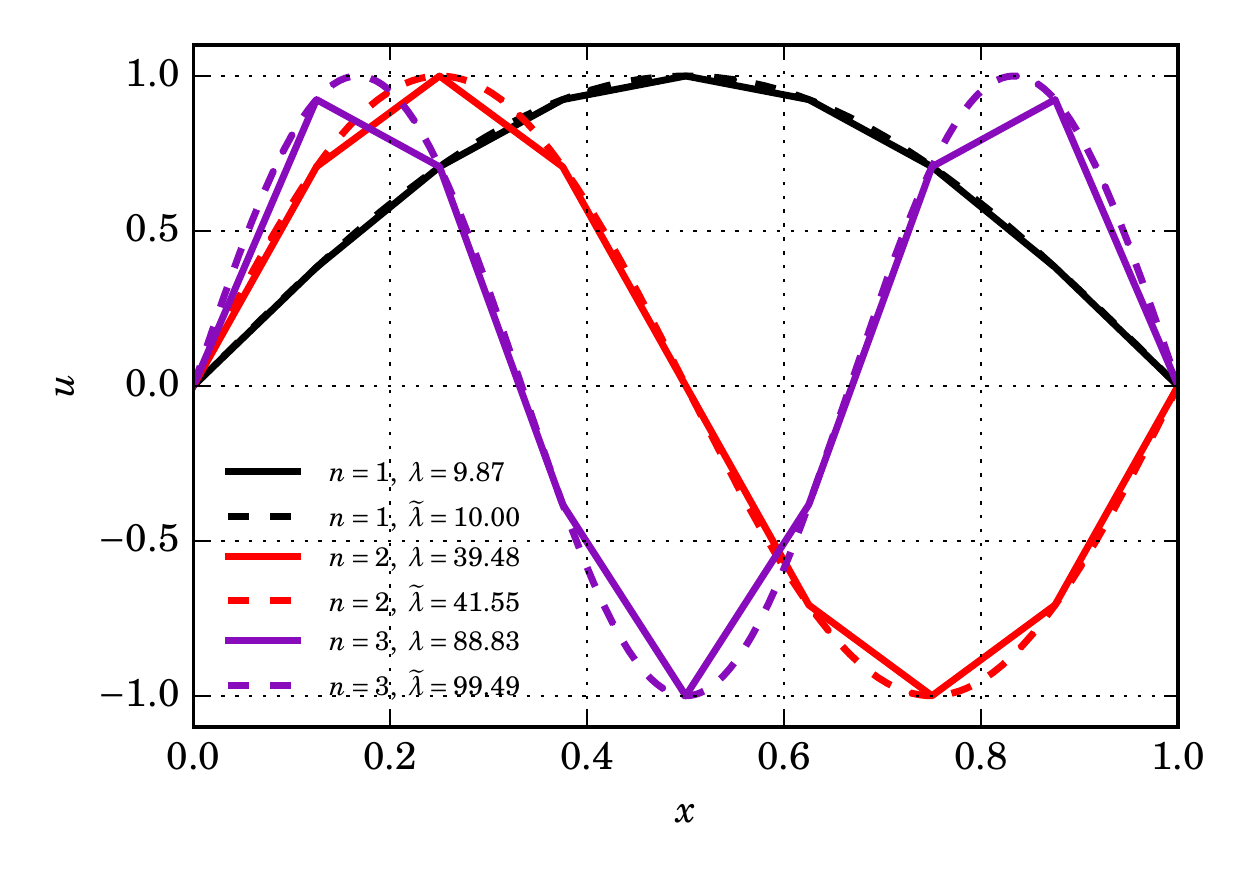}
  \caption[Eigenvector example solution]{Fourier series Eigenfunctions (dashed) and 8 linear element finite element Eigenvectors (solid) for $n=1,2,3$.  The associated exact Eigenvalue $\lambda$ and FEM-approximated Eigenvalues $\tilde{\lambda}$ are listed in the legend.}
  \label{1d_example_eigenvectors}
\end{figure*}

\begin{figure*}
  \centering
    \includegraphics[width=0.7\linewidth]{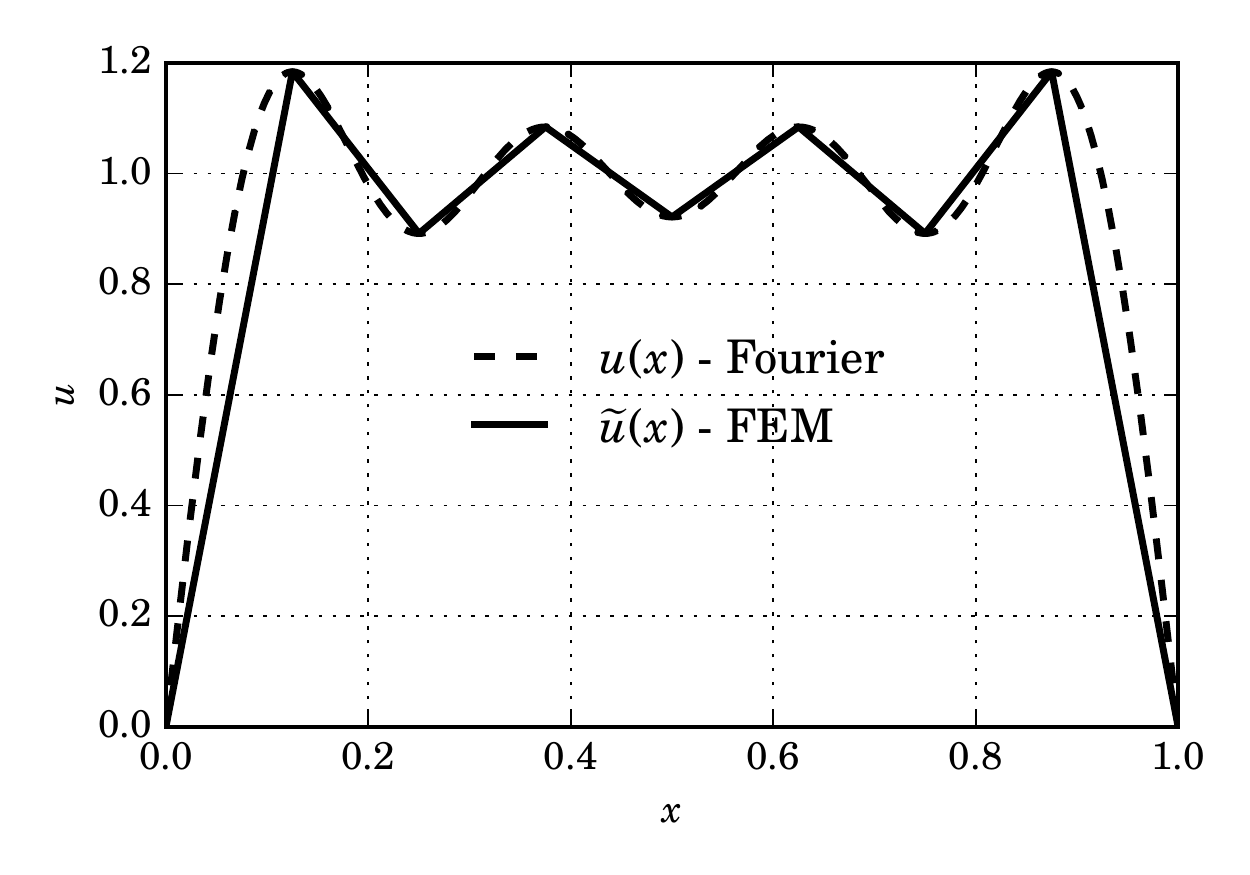}
  \caption[Comparison of FEM with Fourier series solution]{Eight-term Fourier series approximation (dashed) and eight-element finite element approximation (solid) to transient problem (\ref{intro_eigen_pde} -- \ref{intro_eigen_ic}) with $\ell = 1$ with $f(x)=1$ at $t=0$.}
  \label{1d_example_eigensolution}
\end{figure*}

Results derived using initial condition $f(x) = 1$ in (\ref{intro_eigen_ic}) for an eight-term Fourier series approximation (\ref{intro_eigen_fourier_soln}, \ref{intro_eigen_fourier_coef}) and an 8-element finite element approximation analogous to (\ref{intro_eigen_fem_soln} -- \ref{intro_eigen_fourier_coef}) are generated with Code Listing \ref{eigenvalue_code}.  The resulting Eigenfunctions are plotted in Figure \ref{1d_example_eigenvectors} and resulting approximations in Figure \ref{1d_example_eigensolution}.

\pythonexternal[label=eigenvalue_code, caption={FEniCS code for solution of the Eigenvalue problem.}]{scripts/fenics_intro/eigenvalues.py}

%===============================================================================
%===============================================================================

\chapter{Problems in two dimensions}

For two-dimensional equations, the domain is $\Omega \in \R^2$ and Sobolev space (\ref{sobolev_space}) becomes
\begin{align}
  \label{2d_sobolev_space}
  \mathcal{H}^1(\Omega) &= \left\{ u\ :\ \Omega \rightarrow \R\ \Bigg|\ u, \frac{\partial u}{\partial x}, \frac{\partial u}{\partial y} \in L^2(\Omega) \right\}.
\end{align}
The two-dimensional domain is discretized into either triangular or quadrilateral elements, with new interpolation functions that satisfy a two-dimensional analog of interpolation properties (\ref{interpolation_properties}).  For an excellent explanation of the resulting 2D Galerkin system analogous to (\ref{intro_galerkin_system}), see \citet{elman}.

In this chapter, we solve the two-dimensional variant of heat equation (\ref{intro_ode}) and the Stokes equations.

%===============================================================================

\section{Poisson equation}

\index{Linear differential equations!2D}
\index{Poisson equation}
The Poisson equation to be solved over the domain $\Omega = [0,1] \times [0,1]$ is
\begin{align}
  \label{2d_poisson_start}
  -\nabla^2 u &= f &&\text{ in } \Omega \\
  f &= 10 \sin\left(\frac{2\pi x}{L}\right) \sin\left(\frac{2\pi y}{L}\right) &&\text{ in } \Omega \\
  \nabla u \cdot \mathbf{n} &= g_N = \sin(x) &&\text{ on } \Gamma_N, \Gamma_S \\
  \label{2d_poisson_end}
  u &= g_D = 1 &&\text{ on } \Gamma_E, \Gamma_W,
\end{align}
where $\Gamma_N$, $\Gamma_S$, $\Gamma_E$, and $\Gamma_W$ are the North, South, East and West boundaries, $L=1$ is the length of the square side, and $\mathbf{n}$ is the outward normal to the boundary $\Gamma$.

\begin{figure}
  \centering
    \includegraphics[width=\linewidth]{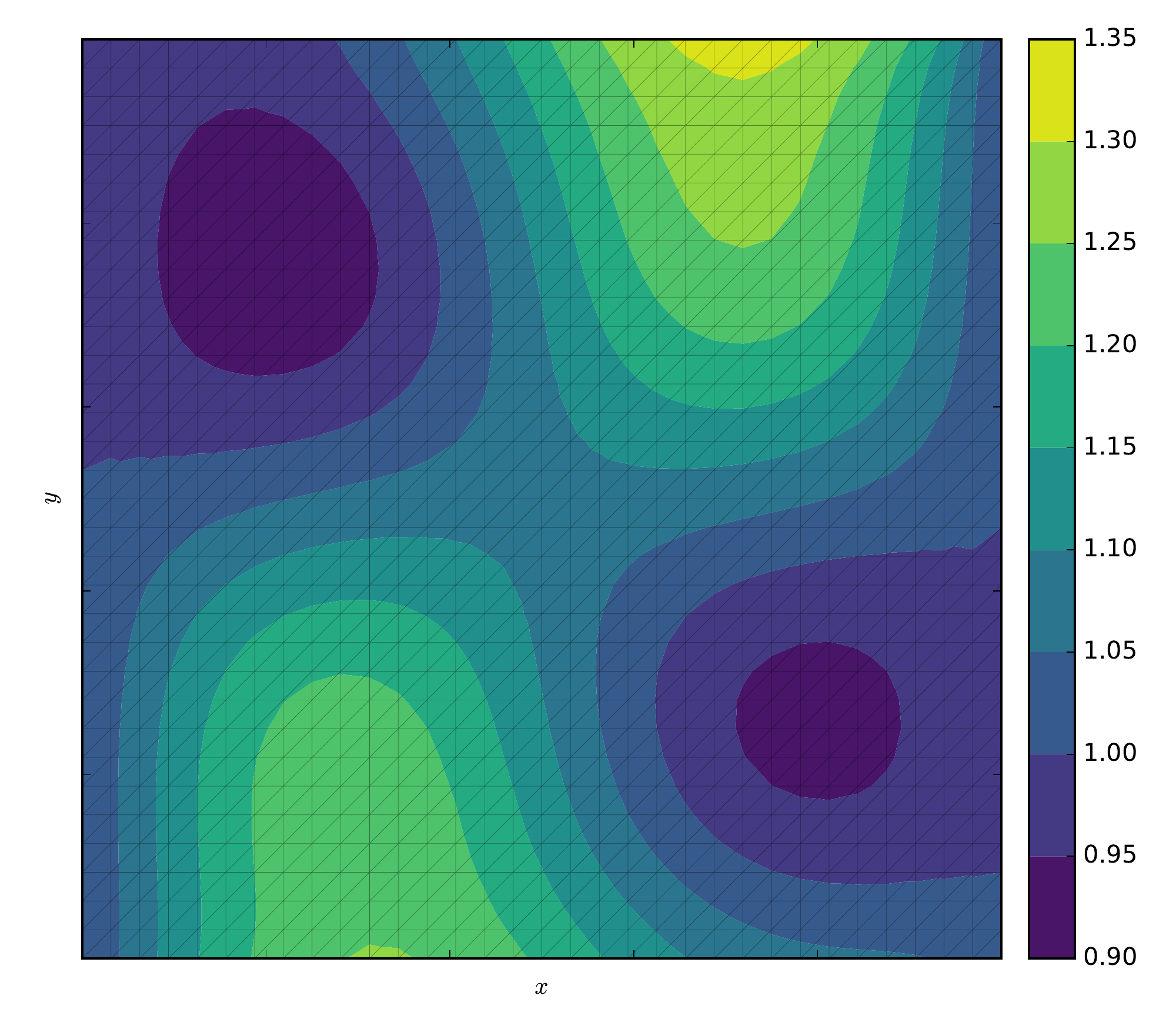}
  \caption[Two-dimension Poisson example]{Poisson solution for a uniform $32 \times 32$ node mesh.}
  \label{2d_poisson_image}
\end{figure}

The associated Galerkin weak form with test function $\phi \in \testspace$ defined by (\ref{test_space}) is
\begin{align*}
  -\int_{\Omega} \nabla^2 u \phi d\Omega &= \int_{\Omega} f \phi d\Omega \\
  \int_{\Omega} \nabla u \cdot \nabla \phi d\Omega - \int_{\Gamma} \phi \nabla u \cdot \mathbf{n} d\Gamma &= \int_{\Omega} f \phi d\Omega \\
  \int_{\Omega} \nabla u \cdot \nabla \phi d\Omega - \int_{\Gamma_N} \phi g_N d\Gamma_N - \int_{\Gamma_S} \phi g_N d\Gamma_S &= \int_{\Omega} f \phi d\Omega,
\end{align*}
and so the variational problem consists of finding $u \in \trialspace$ (see trial space \ref{trial_space}) such that
\begin{align*}
  a(u,\phi) &= L(\phi) && \forall \phi \in S_0^h \subset \mathcal{H}_{E_0}^1(\Omega),
\end{align*}
subject to Dirichlet condition (\ref{2d_poisson_end}), where
\begin{align*}
  L(\phi) &= \int_{\Gamma_N} \phi g_N d\Gamma_N + \int_{\Gamma_S} \phi g_N d\Gamma_S + \int_{\Omega} f \phi d\Omega, \\
  a(u,\phi) &= \int_{\Omega} \nabla u \cdot \nabla \phi d\Omega.
\end{align*}
This form is all that is required to derive an approximate solution with FEniCS, as demonstrated by Code Listing \ref{2d_poisson_code} and Figure \ref{2d_poisson_image}.

\pythonexternal[label=2d_poisson_code, caption={FEniCS solution to two-dimensionial Poisson problem (\ref{2d_poisson_start} -- \ref{2d_poisson_end}).}]{scripts/fenics_intro/2D_poisson.py}

%===============================================================================

\section{Stokes equations with no-slip boundary conditions} \label{ssn_intro_stokes_2d}

\index{Linear differential equations!2D}
\index{Stokes equations!No-slip}
The Stokes equations for incompressible fluid over the domain $\Omega = [0,1] \times [0,1]$ are
\begin{align}
  \label{intro_stokes_momentum}
  -\nabla \cdot \sigma &= \mathbf{f} &&\text{ in } \Omega &&\leftarrow \text{ conservation of momentum} \\
  \label{intro_stokes_mass}
  \nabla \cdot \mathbf{u} &= 0 &&\text{ in } \Omega &&\leftarrow \text{ conservation of mass},
\end{align}
where $\sigma$ is the Cauchy-stress tensor defined as $\sigma = 2\eta \dot{\epsilon} - pI$; viscosity $\eta$, strain-rate tensor
\begin{align}
  \label{intro_strain_rate_tensor}
  \dot{\epsilon} &= \frac{1}{2}\left[\nabla \mathbf{u} + (\nabla \mathbf{u})\T \right] \notag \\
  &= \begin{bmatrix}
       \frac{\partial u}{\partial x} & \frac{1}{2} \left( \frac{\partial u}{\partial y} + \frac{\partial v}{\partial x} \right) \\
       \frac{1}{2} \left( \frac{\partial v}{\partial x} + \frac{\partial u}{\partial y} \right) & \frac{\partial v}{\partial y}
     \end{bmatrix};
\end{align}
velocity $\mathbf{u}$ with components $u$, $v$ in the $x$ and $y$ directions, and pressure $p$; and vector of internal forces $\mathbf{f}$.  For our example, we take $\mathbf{f}=\mathbf{0}$ and boundary conditions
\begin{align}
  \label{intro_stokes_gD_N_S_D}
  \mathbf{u} &= \bm{g_D} = \mathbf{0} &&\text{ on } \Gamma_N, \Gamma_S, \Gamma_D \\
  \label{intro_stokes_gD_E}
  \mathbf{u} &= \bm{g_D} = [-\sin(\pi y)\ 0]\T &&\text{ on } \Gamma_E \\
  \label{intro_stokes_gN}
  \sigma \cdot \mathbf{n} &= \bm{g_N} = [g_{N_x}\ g_{N_y}]\T = \mathbf{0} &&\text{ on } \Gamma_W,
\end{align}
where $\Gamma_E$, $\Gamma_W$, $\Gamma_N$, and $\Gamma_S$ are the East, West, North, and South boundaries, $\Gamma_D$ is the dolphin boundary (Figure \ref{intro_stokes_2d}) and $\mathbf{n}$ is the outward-pointing normal vector to these faces. For obvious reasons, velocity boundary condition (\ref{intro_stokes_gD_N_S_D}) is referred to as a \emph{no-slip} boundary.

It may be of interest to see how the conservation of momentum equations look in their expanded form,
\begin{align*}
  -\nabla \cdot \sigma &= \mathbf{f} \\
  \begin{bmatrix}
    \frac{\partial \sigma_{xx}}{\partial x} + \frac{\partial \sigma_{xy}}{\partial y} \\ 
    \frac{\partial \sigma_{yx}}{\partial x} + \frac{\partial \sigma_{yy}}{\partial y} \\ 
  \end{bmatrix} &=
  \begin{bmatrix}
    0 \\ 0
  \end{bmatrix},
\end{align*}
providing two equations,
\begin{align*}
    \frac{\partial}{\partial x} \left[2\eta \frac{\partial u}{\partial x} \right]  - \frac{\partial p}{\partial x} + \frac{\partial}{\partial y} \left[\eta \left( \frac{\partial u}{\partial x} + \frac{\partial v}{\partial y} \right) \right] &= 0 \\ 
    \frac{\partial}{\partial x} \left[\eta \left( \frac{\partial v}{\partial y} + \frac{\partial u}{\partial x} \right) \right]  + \frac{\partial}{\partial y} \left[2\eta \frac{\partial v}{\partial y} \right] - \frac{\partial p}{\partial y} &= 0, 
\end{align*}
which along with conservation of mass equation (\ref{intro_stokes_mass}), also referred to as the \emph{incompressibility constraint}
\begin{align*}
  \nabla \cdot \mathbf{u} &= \frac{\partial u}{\partial x} + \frac{\partial v}{\partial y} = 0, 
\end{align*}
makes three equations with three unknowns $u$, $v$, and $p$.

Neumann condition (\ref{intro_stokes_gN}) may be expanded into
\begin{align}
  \label{intro_stokes_outflow}
  2 \eta \dot{\epsilon}_{xx} n_x - p n_x + 2 \eta \dot{\epsilon}_{xy} n_y &= g_{N_x} \\
  2 \eta \dot{\epsilon}_{yx} n_x + 2 \eta \dot{\epsilon}_{yy} n_y - p n_y &= g_{N_y}.
\end{align}
The pressure boundary condition on the outflow boundary $\Gamma_W$ may be discovered by integrating boundary condition (\ref{intro_stokes_outflow}) along $\Gamma_W$ with $\mathbf{n} = [-1\ 0]\T$, assuming a constant viscosity $\eta$, and strain-rate tensor definition (\ref{intro_strain_rate_tensor}),
\begin{align*}
  - 2 \eta \int_{\Gamma_W} \frac{\partial u}{\partial x}\ d\Gamma_W + \int_{\Gamma_W} p\ d\Gamma_W &= \int_{\Gamma_W} g_{N_x}\ d\Gamma_W.
\end{align*}
Next, constraint (\ref{intro_stokes_mass}) implies that $\partial_x u = - \partial_y v$ and thus
\begin{align*}
  2 \eta \int_0^1 \frac{\partial v}{\partial y}\ dy + \int_0^1 p\ dy &= \int_{\Gamma_W} g_{N_x}\ d\Gamma_W \\
  2 \eta v(0,1) - 2 \eta v(0,0) + \int_0^1 p\ dy &= \int_{\Gamma_W} g_{N_x}\ d\Gamma_W \\
  \int_0^1 p\ dy &= \int_{\Gamma_W} 0\ d\Gamma_W,
\end{align*}
which implies that $p = 0$ over the entire outflow boundary $\Gamma_W$ (for further illustration, see \citet{elman}).

The weak form for problem (\ref{intro_stokes_momentum}, \ref{intro_stokes_mass}, \ref{intro_stokes_gD_N_S_D} -- \ref{intro_stokes_gN}) is formed by taking the inner product of both sides of the conservation of momentum equation with the vector test function $\bm{\Phi} = [\phi\ \psi ]\T \in \mathbf{S_0^h} \subset \left( \mathcal{H}_{E_0}^1(\Omega) \right)^2$ (see test space (\ref{test_space})) integrating over the domain $\Omega$,
\begin{align*}
  -\int_{\Omega} \nabla \cdot \sigma \cdot \bm{\Phi}\ d\Omega &= \int_{\Omega} \mathbf{f} \cdot \bm{\Phi}\ d\Omega,
\end{align*}
then integrate by parts to get 
\begin{align*}
  \int_{\Omega} \sigma : \nabla \bm{\Phi}\ d\Omega - \int_{\Gamma} \sigma \cdot \mathbf{n} \cdot \bm{\Phi}\ d\Gamma &= \int_{\Omega} \mathbf{f} \cdot \bm{\Phi}\ d\Omega \\
  \int_{\Omega} \sigma : \nabla \bm{\Phi}\ d\Omega &= \int_{\Omega} \mathbf{f} \cdot \bm{\Phi}\ d\Omega,
\end{align*}
where the facts that $\sigma \cdot \mathbf{n} = \mathbf{0}$ on the West boundary and Dirichlet conditions exist on the North, South, East, and dolphin boundaries has been used.  Next, taking the inner product of incompressibility (conservation of mass) equation (\ref{intro_stokes_mass}) with the test function $\xi \in M^h \subset L^2(\Omega)$ (see $L^2$ space (\ref{l2_space})) integrating over $\Omega$,
\begin{align*}
  \int_{\Omega} \left( \nabla \cdot \mathbf{u} \right) \xi\ d\Omega &= 0.
\end{align*}

Finally, using the fact that the right-hand side of incompressibility equation (\ref{intro_stokes_mass}) is zero, the \index{Mixed methods} \emph{mixed formulation} (see for example \citet{johnson}) consists of finding \emph{mixed approximation} $\mathbf{u} \in \mathbf{S_E^h} \subset \left( \mathcal{H}_E^1(\Omega) \right)^2$ and $p \in M^h \subset L^2(\Omega)$ such that
\begin{align}
  \label{intro_stokes_mixed_problem}
  a(\mathbf{u},p,\bm{\Phi},\xi) &= L(\bm{\Phi}) && \forall \bm{\Phi} \in \mathbf{S_0^h} \subset \left( \mathcal{H}_{E_0}^1(\Omega) \right)^2,\ \xi \in M^h \subset L^2(\Omega),
\end{align}
subject to Dirichlet conditions (\ref{intro_stokes_gD_N_S_D} -- \ref{intro_stokes_gN}) and
\begin{align*}
  a(\mathbf{u},p,\bm{\Phi},\xi) &= \int_{\Omega} \sigma : \nabla \bm{\Phi}\ d\Omega + \int_{\Omega} \left( \nabla \cdot \mathbf{u} \right) \xi\ d\Omega, \\
  L(\bm{\Phi}) &= \int_{\Omega} \mathbf{f} \cdot \bm{\Phi}\ d\Omega.
\end{align*}

\subsection{Stability}

\index{Stokes equations!Stability}
In order to derive a unique solution for pressure $p$, the trial and test spaces must be chosen in such a way that the \emph{inf-sup condition}
\begin{align}
  \label{inf_sup_condition}
  \min_{\xi \neq \text{constant}} \left\{ \max_{\bm{\Phi} \neq \mathbf{0}} \left\{ \frac{\left| \left( \xi, \nabla \cdot \bm{\Phi} \right) \right|}{\Vert \bm{\Phi} \Vert_{1,\Omega} \Vert \xi \Vert_{0,\Omega} } \right\} \right\} \geq \gamma
\end{align}
is satisfied for any conceivable grid and some constant $\gamma$ \citep{elman}.  The notation $(f,g) = \int_{\Omega} f g d\Omega$ is the inner product, $\Vert \mathbf{f} \Vert_{1,\Omega} = \left( \int_{\Omega} \left[ \mathbf{f} \cdot \mathbf{f} + \nabla \mathbf{f} : \nabla \mathbf{f} \right] d\Omega \right)^{\nicefrac{1}{2}}$ is the $\mathbf{S_E^h}$-norm, and $\Vert f \Vert_{0,\Omega} = \Vert f - \frac{1}{|\Omega|} \int_{\Omega} f d\Omega \Vert$ is a so-called \emph{quotient space norm} \citep{elman}.

One way of satisfying inf-sup condition (\ref{inf_sup_condition}) is through the use of Taylor-Hood finite element space \citep{taylor}, which utilize a quadratic function space for the velocity vector components and the linear Lagrange function space for the pressure.

\begin{figure*}
  \centering
  \begin{minipage}[b]{0.60\linewidth}
    \includegraphics[width=\linewidth]{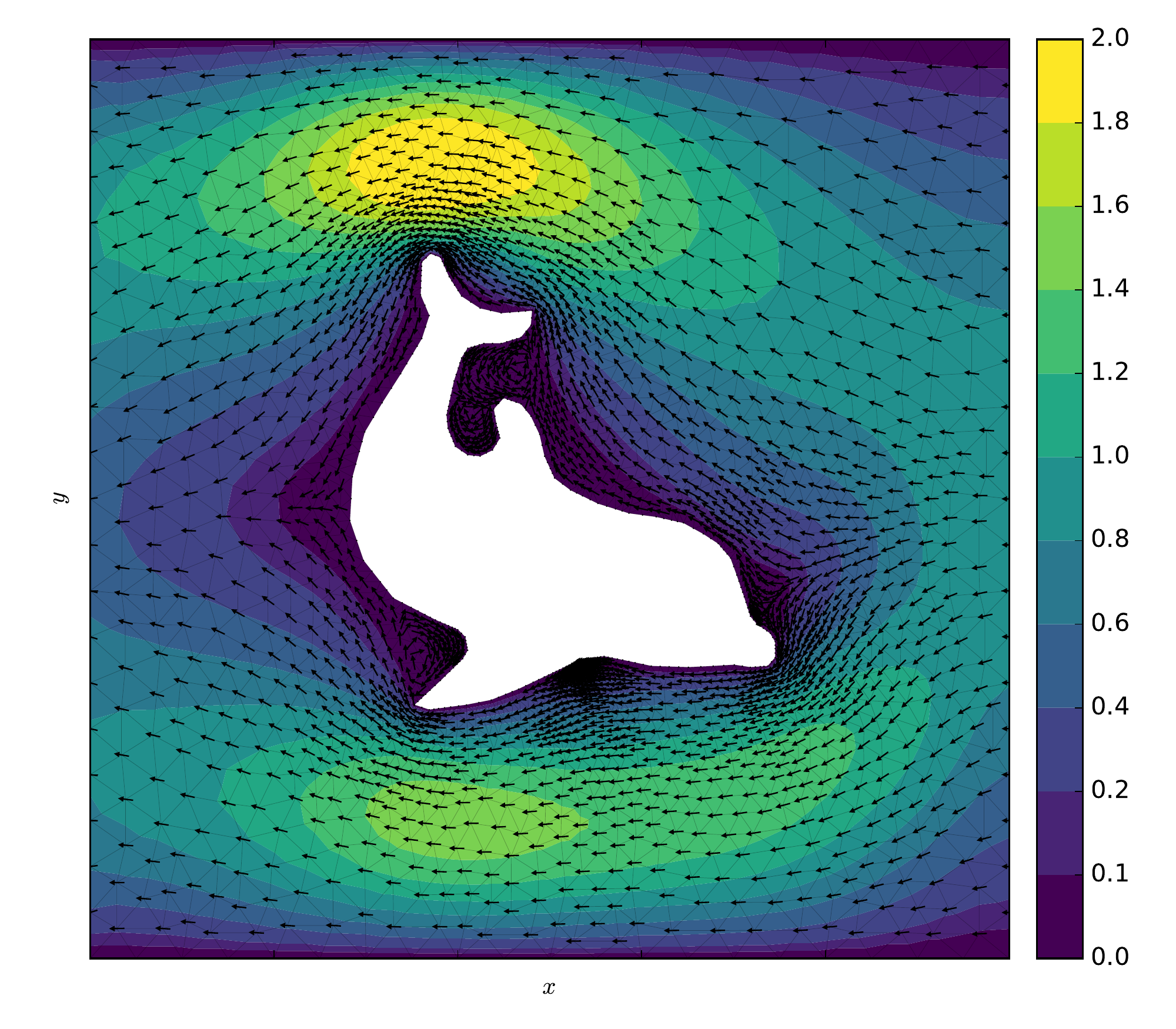}
  \end{minipage}
  \quad
  \begin{minipage}[b]{0.60\linewidth}
    \includegraphics[width=\linewidth]{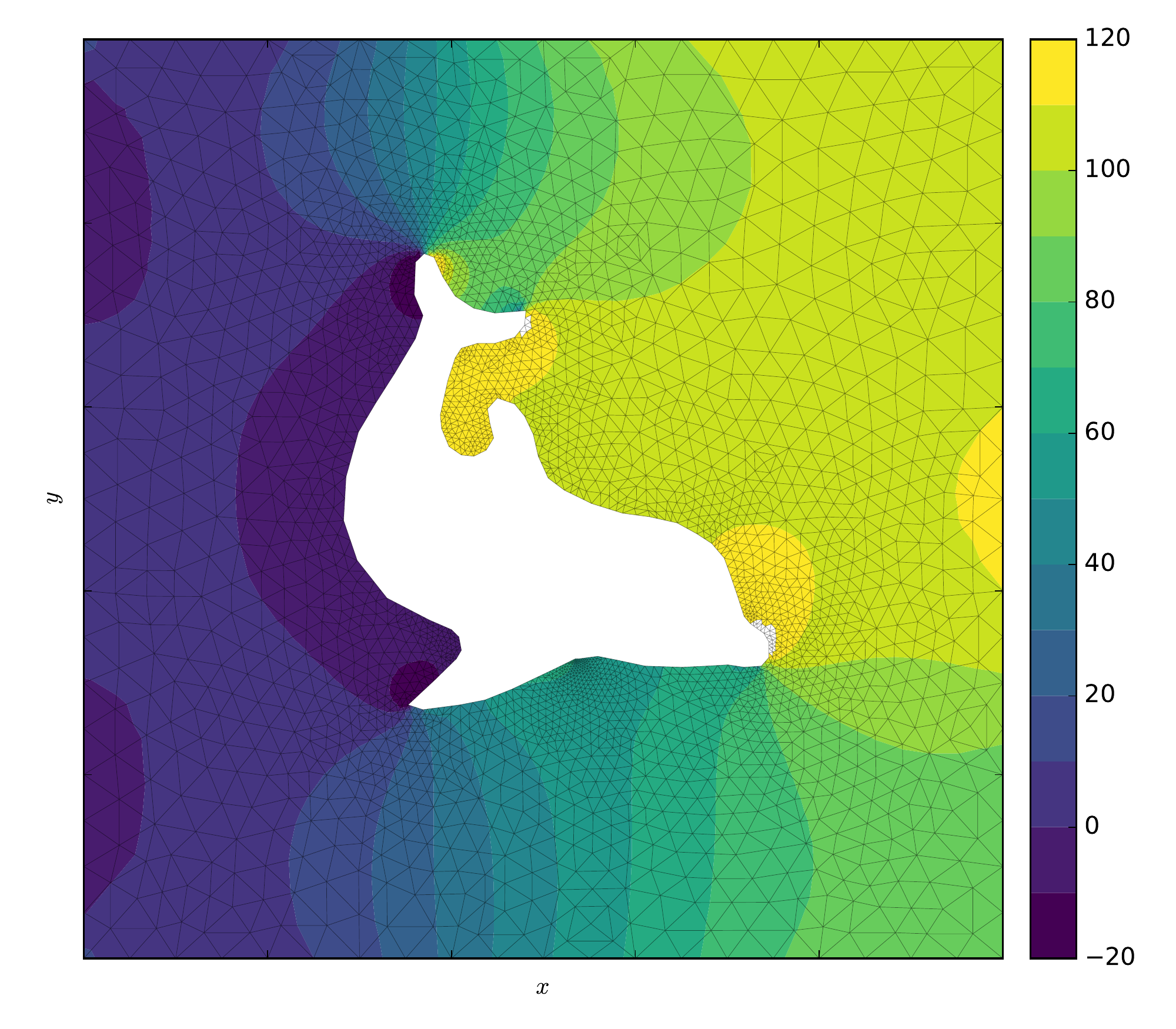}
  \end{minipage}
  \caption[Two-dimensional-no-slip Stokes example]{Velocity field $\mathbf{u}$ (top) and pressure $p$ (bottom) for the no-slip formulation given in \S \ref{ssn_intro_stokes_2d} using the Taylor-hood element (referred to as the P2 -- P1 approximation).}
  \label{intro_stokes_2d}
\end{figure*}

The velocity and pressure solutions to (\ref{intro_stokes_mixed_problem}) using Taylor-Hood elements are depicted in Figure \ref{intro_stokes_2d} and generated by Code Listing \ref{2d_stokes_no_slip_code}.

\section{Stokes equations with slip-friction boundary conditions} \label{ssn_intro_stokes_2d_slip}

\index{Linear differential equations!2D}
\index{Stokes equations!Slip-friction}
A \emph{slip-friction} boundary condition for Stokes equations (\ref{intro_stokes_momentum}, \ref{intro_stokes_mass}) using an identical domain $\Omega$ as in \S \ref{ssn_intro_stokes_2d} may be generated by replacing no-slip boundary condition (\ref{intro_stokes_gD_N_S_D}) with the pair of boundary conditions
\begin{align}
  \label{intro_stokes_gD_N_S_D_slip}
  \mathbf{u} \cdot \mathbf{n} &= g_D = 0 &&\text{ on } \Gamma_N, \Gamma_S, \Gamma_D \\
  \label{intro_stokes_gN_N_S_D_fric}
  \left( \sigma \cdot \mathbf{n} \right)_{\Vert} &= \bm{g_N} = -\beta \mathbf{u} &&\text{ on } \Gamma_N, \Gamma_S, \Gamma_D,
\end{align}
where $(\mathbf{v})_{\Vert} = \mathbf{v} - \left( \mathbf{v} \cdot \mathbf{n} \right) \mathbf{n}$ denotes the tangential component of a vector $\mathbf{v}$ and $\beta \geq 0$ is a friction coefficient.  Boundary conditions (\ref{intro_stokes_gD_N_S_D_slip}, \ref{intro_stokes_gN_N_S_D_fric}) are equivalent to no-slip boundary condition (\ref{intro_stokes_gD_N_S_D}) as $\beta$ approaches infinity.  Note also that impenetrability condition (\ref{intro_stokes_gD_N_S_D_slip}) specifies one component of velocity and is an essential boundary condition, while friction condition (\ref{intro_stokes_gN_N_S_D_fric}) specifies the other component (in three dimension it would specify the other two components) and is a natural boundary condition.  For comparison purposes, we use the same inflow boundary condition (\ref{intro_stokes_gD_E}) and outflow boundary condition (\ref{intro_stokes_gN}).

The weak form for problem (\ref{intro_stokes_momentum}, \ref{intro_stokes_mass}, \ref{intro_stokes_gD_N_S_D_slip}, \ref{intro_stokes_gN_N_S_D_fric}, \ref{intro_stokes_gD_E} \ref{intro_stokes_gN}) is formed by taking the inner product of both sides of the conservation of momentum equation with the vector test function $\bm{\Phi} = [\phi\ \psi ]\T \in \mathbf{S_0^h} \subset \left( \mathcal{H}_{E_0}^1(\Omega) \right)^2$ (see test space (\ref{test_space})) integrating over the domain $\Omega$,
\begin{align*}
  -\int_{\Omega} \nabla \cdot \sigma \cdot \bm{\Phi}\ d\Omega &= \int_{\Omega} \mathbf{f} \cdot \bm{\Phi}\ d\Omega,
\end{align*}
then integrate by parts to get and add the incompressibility constraint as performed in \S \ref{ssn_intro_stokes_2d},
\begin{align}
  \label{intro_stokes_slip_first_var_form}
  \int_{\Omega} \sigma : \nabla \bm{\Phi}\ d\Omega - \int_{\Gamma} \sigma \cdot \mathbf{n} \cdot \bm{\Phi}\ d\Gamma + \int_{\Omega} \left( \nabla \cdot \mathbf{u} \right) \xi\ d\Omega &= \int_{\Omega} \mathbf{f} \cdot \bm{\Phi}\ d\Omega.
\end{align}
Expanding tangential stress condition (\ref{intro_stokes_gN_N_S_D_fric}), we have
\begin{align*}
  \sigma \cdot \mathbf{n} &= \left( \mathbf{n} \cdot \sigma \cdot \mathbf{n} \right) \mathbf{n} - \beta \mathbf{u},
\end{align*}
producing
\begin{align}
  \label{intro_stokes_slip_intermediate_var_form}
  \mathcal{B}_{\Omega} + \mathcal{B}_{\Gamma_G} + \mathcal{B}_{\Gamma_E} = \mathcal{F}
\end{align}
with individual terms
\begin{align*}
  \mathcal{B}_{\Omega} &= + \int_{\Omega} \sigma(\mathbf{u},p) : \nabla \bm{\Phi}\ d\Omega + \int_{\Omega} \left( \nabla \cdot \mathbf{u} \right) \xi\ d\Omega \\
  \mathcal{B}_{\Gamma_G} &= - \int_{\Gamma_G} \left( \mathbf{n} \cdot \sigma(\mathbf{u},p) \cdot \mathbf{n} \right) \mathbf{n} \cdot \bm{\Phi}\ d\Gamma_G + \int_{\Gamma_G} \beta \mathbf{u} \cdot \bm{\Phi}\ d\Gamma_G \\
  \mathcal{B}_{\Gamma_E} &= - \int_{\Gamma_E} \sigma(\mathbf{u},p) \cdot \mathbf{n} \cdot \bm{\Phi}\ d\Gamma_E \\
  \mathcal{F} &= + \int_{\Omega} \mathbf{f} \cdot \bm{\Phi}\ d\Omega, 
\end{align*}
where $\Gamma_G = \Gamma_N \cup \Gamma_S \cup \Gamma_D$ is the entire slip-friction boundary, and the fact that $\sigma \cdot \mathbf{n} = \mathbf{0}$ on the West boundary $\Gamma_W$ has been used.

A method devised by \index{Nitsche method} \cite{nitsche} and further explained by \cite{freund} imposes Dirichlet conditions (\ref{intro_stokes_gD_E}, \ref{intro_stokes_gD_N_S_D_slip}) in a weak form by adjoining symmetric terms to (\ref{intro_stokes_slip_intermediate_var_form}),
\begin{align}
  \label{intro_stokes_slip_var_form}
  \mathcal{B}_{\Omega} + \mathcal{B}_{\Gamma_G} + \mathcal{B}_{\Gamma_G}^W + \mathcal{B}_{\Gamma_E} + \mathcal{B}_{\Gamma_E}^W = \mathcal{F} + \mathcal{F}^W,
\end{align}
where
\begin{align*}
  \mathcal{B}_{\Gamma_G}^W = &- \int_{\Gamma_G} \left( \mathbf{n} \cdot \sigma(\bm{\Phi},\xi) \cdot \mathbf{n} \right) \mathbf{n} \cdot \mathbf{u} \ d\Gamma_G + \gamma \int_{\Gamma_G} \frac{1}{h} \left( \mathbf{u} \cdot \mathbf{n} \right) \left( \bm{\Phi} \cdot \mathbf{n} \right)\ d\Gamma_G \\
  \mathcal{B}_{\Gamma_E}^W = &- \int_{\Gamma_E} \sigma(\bm{\Phi},\xi) \cdot \mathbf{n} \cdot \mathbf{u}\ d\Gamma_E + \gamma \int_{\Gamma_E} \frac{1}{h} \left( \bm{\Phi} \cdot \mathbf{u} \right)\ d\Gamma_E \\
  \mathcal{F}^W = &- \int_{\Gamma_G} \left( \mathbf{n} \cdot \sigma(\bm{\Phi},\xi) \cdot \mathbf{n} \right) g_D\ d\Gamma_G + \gamma \int_{\Gamma_G} \frac{1}{h} g_D \bm{\Phi} \cdot \mathbf{n}\ d\Gamma_G \\
  &- \int_{\Gamma_E} \sigma(\bm{\Phi},\xi) \cdot \mathbf{n} \cdot \bm{g_D}\ d\Gamma_E + \gamma \int_{\Gamma_E} \frac{1}{h} \left( \bm{\Phi} \cdot \bm{g_D} \right)\ d\Gamma_E.
\end{align*}
with element diameter $h$ and application-specific parameter $\gamma > 0$ normally derived by experimentation.  Variational form (\ref{intro_stokes_slip_var_form}) is justified using the properties of the self-adjoint linear differential operator $\sigma$:
\begin{align*}
  \int_{\Gamma_G} \left( \mathbf{n} \cdot \sigma(\mathbf{u},p) \cdot \mathbf{n} \right) \mathbf{n} \cdot \bm{\Phi}\ d\Gamma_G &= \int_{\Gamma_G} \left( \mathbf{n} \cdot \sigma(\bm{\Phi},\xi) \cdot \mathbf{n} \right) \mathbf{n} \cdot \mathbf{u} \ d\Gamma_G \\
  \int_{\Gamma_E} \sigma(\mathbf{u},p) \cdot \mathbf{n} \cdot \bm{\Phi}\ d\Gamma_E &= \int_{\Gamma_E} \sigma(\bm{\Phi},\xi) \cdot \mathbf{n} \cdot \mathbf{u}\ d\Gamma_E,
\end{align*}
and using boundary conditions (\ref{intro_stokes_gD_N_S_D_slip},\ref{intro_stokes_gN_N_S_D_fric}), 
\begin{align*}
  \int_{\Gamma_G} \left( \mathbf{n} \cdot \sigma(\bm{\Phi},\xi) \cdot \mathbf{n} \right) \mathbf{n} \cdot \mathbf{u} \ d\Gamma_G &= \int_{\Gamma_G} \left( \mathbf{n} \cdot \sigma(\bm{\Phi},\xi) \cdot \mathbf{n} \right) g_D\ d\Gamma_G \\
  \int_{\Gamma_E} \sigma(\mathbf{u},p) \cdot \mathbf{n} \cdot \bm{\Phi}\ d\Gamma_E &= \int_{\Gamma_E} \sigma(\bm{\Phi},\xi) \cdot \mathbf{n} \cdot \bm{g_D}\ d\Gamma_E.
\end{align*}
The extra terms
\begin{align*}
  \gamma \int_{\Gamma_G} \frac{1}{h} \left( \mathbf{u} \cdot \mathbf{n} \right) \left( \bm{\Phi} \cdot \mathbf{n} \right)\ d\Gamma_G &= \gamma \int_{\Gamma_G} \frac{1}{h} g_D \bm{\Phi} \cdot \mathbf{n}\ d\Gamma_G \\
  \gamma \int_{\Gamma_E} \frac{1}{h} \left( \bm{\Phi} \cdot \mathbf{u} \right)\ d\Gamma_E &= \gamma \int_{\Gamma_E} \frac{1}{h} \left( \bm{\Phi} \cdot \bm{g_D} \right)\ d\Gamma_E
\end{align*}
have been added to enable the simulator to enforce boundary conditions (\ref{intro_stokes_gD_N_S_D_slip},\ref{intro_stokes_gN_N_S_D_fric}) to the desired level of accuracy.  

Finally, the mixed formulation \index{Mixed methods} consistent with problem (\ref{intro_stokes_momentum}, \ref{intro_stokes_mass}, \ref{intro_stokes_gD_N_S_D_slip}, \ref{intro_stokes_gN_N_S_D_fric}, \ref{intro_stokes_gD_E} \ref{intro_stokes_gN})  reads: find mixed approximation $\mathbf{u} \in \mathbf{S_E^h} \subset \left( \mathcal{H}_E^1(\Omega) \right)^2$ and $p \in M^h \subset L^2(\Omega)$ subject to (\ref{intro_stokes_slip_var_form}) for all $\bm{\Phi} \in \mathbf{S_0^h} \subset \left( \mathcal{H}_{E_0}^1(\Omega) \right)^2$ and $\xi \in M^h \subset L^2(\Omega)$.

\begin{figure*}
  \centering
  \begin{minipage}[b]{0.60\linewidth}
    \includegraphics[width=\linewidth]{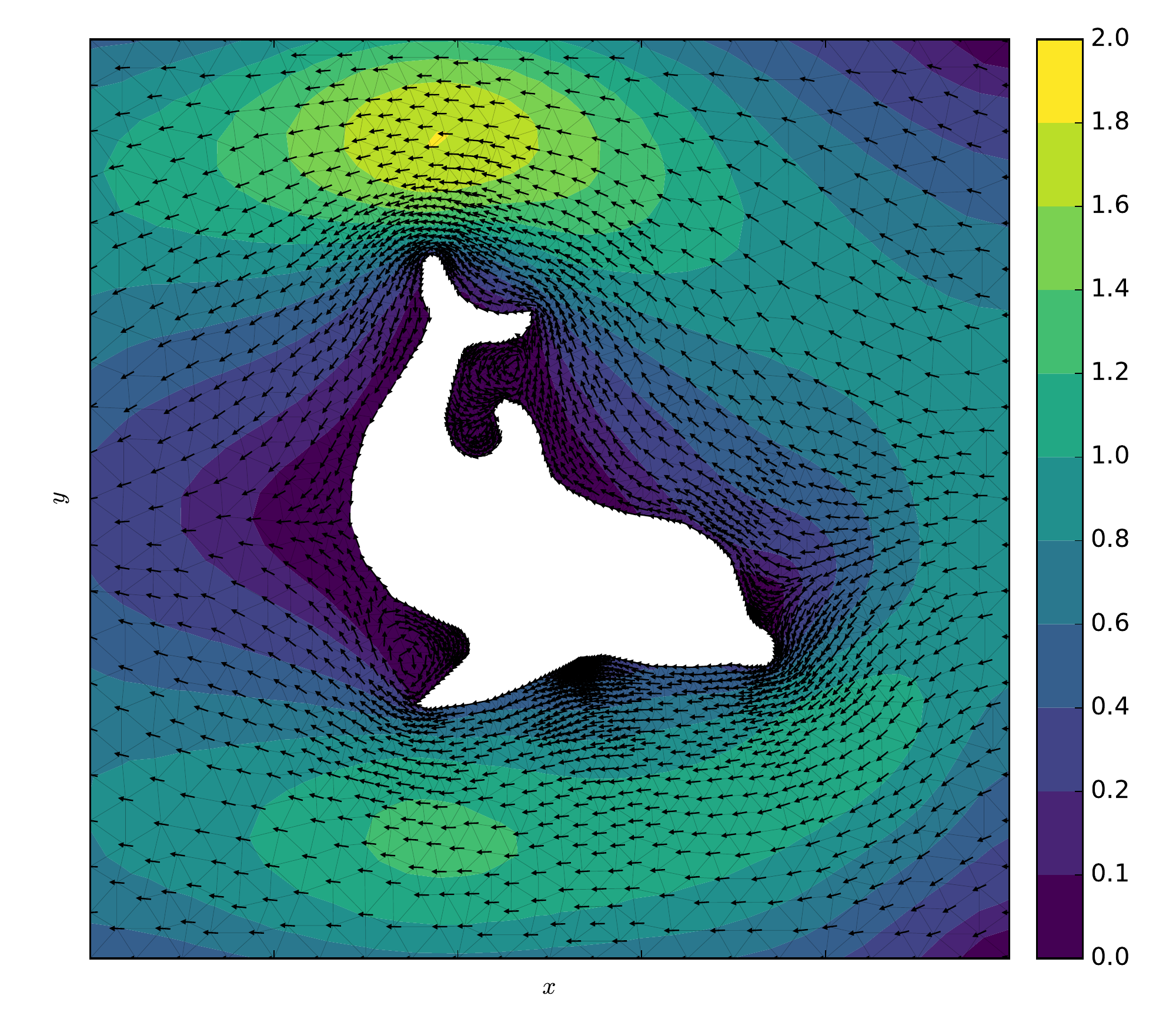}
  \end{minipage}
  \quad
  \begin{minipage}[b]{0.60\linewidth}
    \includegraphics[width=\linewidth]{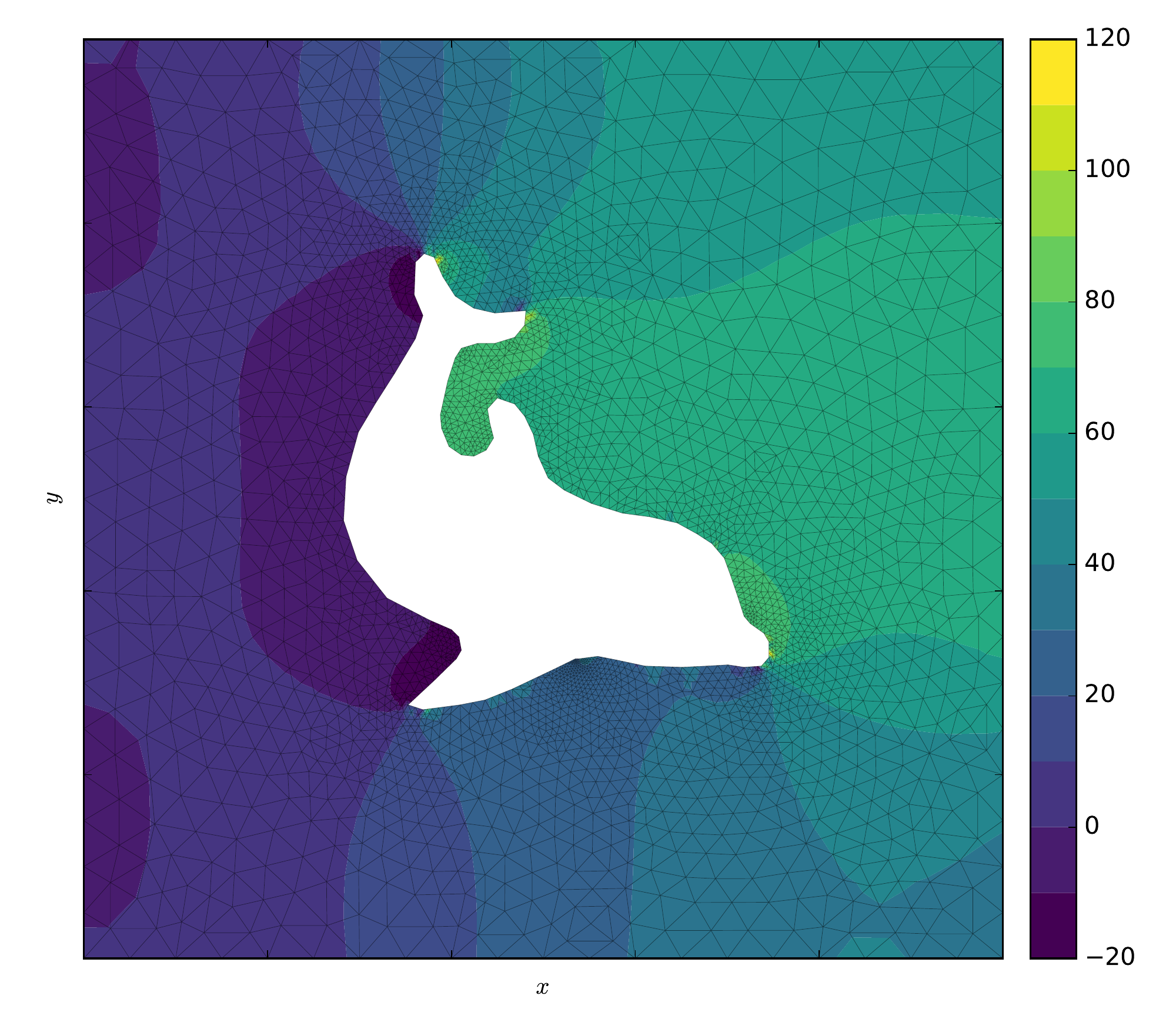}
  \end{minipage}
  \caption[Two-dimensional-slip-friction Stokes example]{Velocity field $\mathbf{u}$ (top) and pressure $p$ (bottom) for the slip-friction formulation given in \S \ref{ssn_intro_stokes_2d_slip} using the same Taylor-Hood element as model \S \ref{ssn_intro_stokes_2d}, with results depicted in Figure \ref{intro_stokes_2d}.}
  \label{intro_stokes_2d_nitsche}
\end{figure*}

Identically to \S \ref{ssn_intro_stokes_2d_slip}, we use the Taylor-Hood element to satisfy inf-sup condition (\ref{inf_sup_condition}).  The friction along the dolphin, North, and South boundaries was taken to be $\beta = 10$, and the Nitsche parameter $\gamma = 100$ was derived by experimentation.  The velocity $\mathbf{u}$ and pressure $p$ solutions to this problem are depicted in Figure \ref{intro_stokes_2d_nitsche} and were generated from Code Listing \ref{2d_stokes_slip_friction_code}.

\pythonexternal[label=2d_stokes_no_slip_code, caption={FEniCS code used to solve 2D-Stokes-no-slip problem (\ref{intro_stokes_mixed_problem}).}]{scripts/fenics_intro/2D_stokes.py}

\pythonexternal[label=2d_stokes_slip_friction_code, caption={FEniCS code used to approximate the solution to the-2D-Stokes-slip-friction problem of \S \ref{ssn_intro_stokes_2d_slip}.}]{scripts/fenics_intro/2D_stokes_nitsche.py}

%===============================================================================
%===============================================================================

\chapter{Problems in three dimensions}

For three dimensional equations, the domain of the system $\Omega \in \R^3$ and Sobolev space (\ref{sobolev_space}) becomes
\begin{align}
  \label{3d_sobolev_space}
  \mathcal{H}^1(\Omega) &= \left\{ u\ :\ \Omega \rightarrow \R\ \Bigg|\ u, \frac{\partial u}{\partial x}, \frac{\partial u}{\partial y}, \frac{\partial u}{\partial z} \in L^2(\Omega) \right\},
\end{align}
and the domain is discretized into either tetrahedral or brick elements, with new interpolation functions that satisfy 3D interpolation properties analogous to (\ref{interpolation_properties}).  For an excellent explanation of the resulting 3D Galerkin system analogous to (\ref{intro_galerkin_system}), see \citet{elman}.

In this chapter, we solve the three-dimensional variant of heat equation (\ref{intro_ode}) and the Stokes equations.

%===============================================================================

\section{Poisson equation}

\index{Linear differential equations!3D}
\index{Poisson equation}
The Poisson equation to be solved over the domain $\Omega = [-\pi,\pi] \times [-\pi,\pi] \times [-\pi,\pi]$ is
\begin{align}
  \label{3d_poisson_start}
  -\nabla^2 u &= f &&\text{ in } \Omega \\
  f &= 10 \exp\left( -\frac{x^2}{2} - \frac{y^2}{2} - \frac{z^2}{2}\right) &&\text{ in } \Omega \\
  \nabla u \cdot \mathbf{n} &= g_{N_y} = \sin(x) &&\text{ on } \Gamma_N, \Gamma_S \\
  \nabla u \cdot \mathbf{n} &= g_{N_x} = 0 &&\text{ on } \Gamma_E, \Gamma_W \\
  \label{3d_poisson_end}
  u &= g_D = 0 &&\text{ on } \Gamma_T, \Gamma_B,
\end{align}
where $\Gamma_N$, $\Gamma_S$, $\Gamma_E$, and $\Gamma_W$ are the North, South, East and West boundaries, $\Gamma_T$ and $\Gamma_B$ are the top and bottom boundaries, and $\mathbf{n}$ is the outward unit normal to the boundary $\Gamma$.

\begin{figure}
  \centering
    \includegraphics[width=\linewidth]{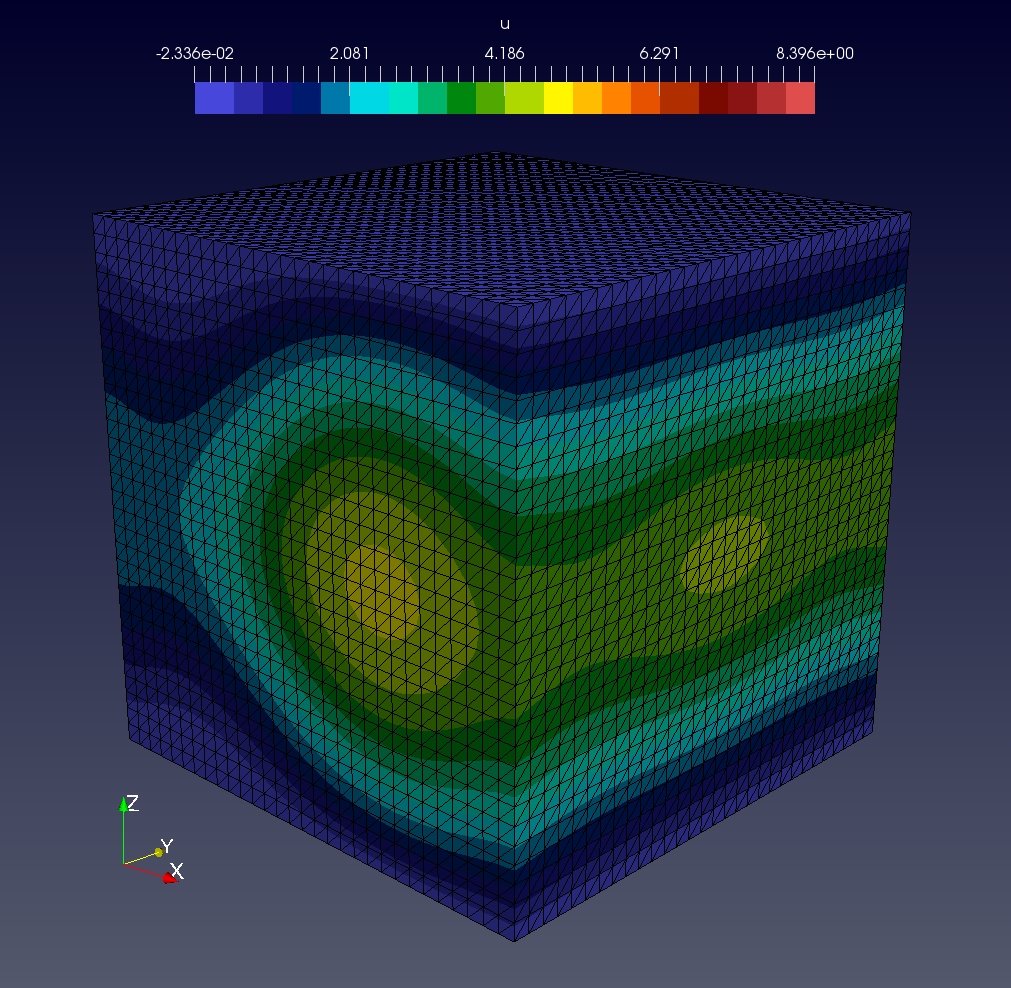}
  \caption[Three-dimensional Poisson solution]{Poisson equation solution to (\ref{3d_poisson_start} -- \ref{3d_poisson_end}) defined over a $30 \times 30 \times 30$ element mesh.}
  \label{3d_poisson_image_1}
\end{figure}

\begin{figure}
  \centering
    \includegraphics[width=\linewidth]{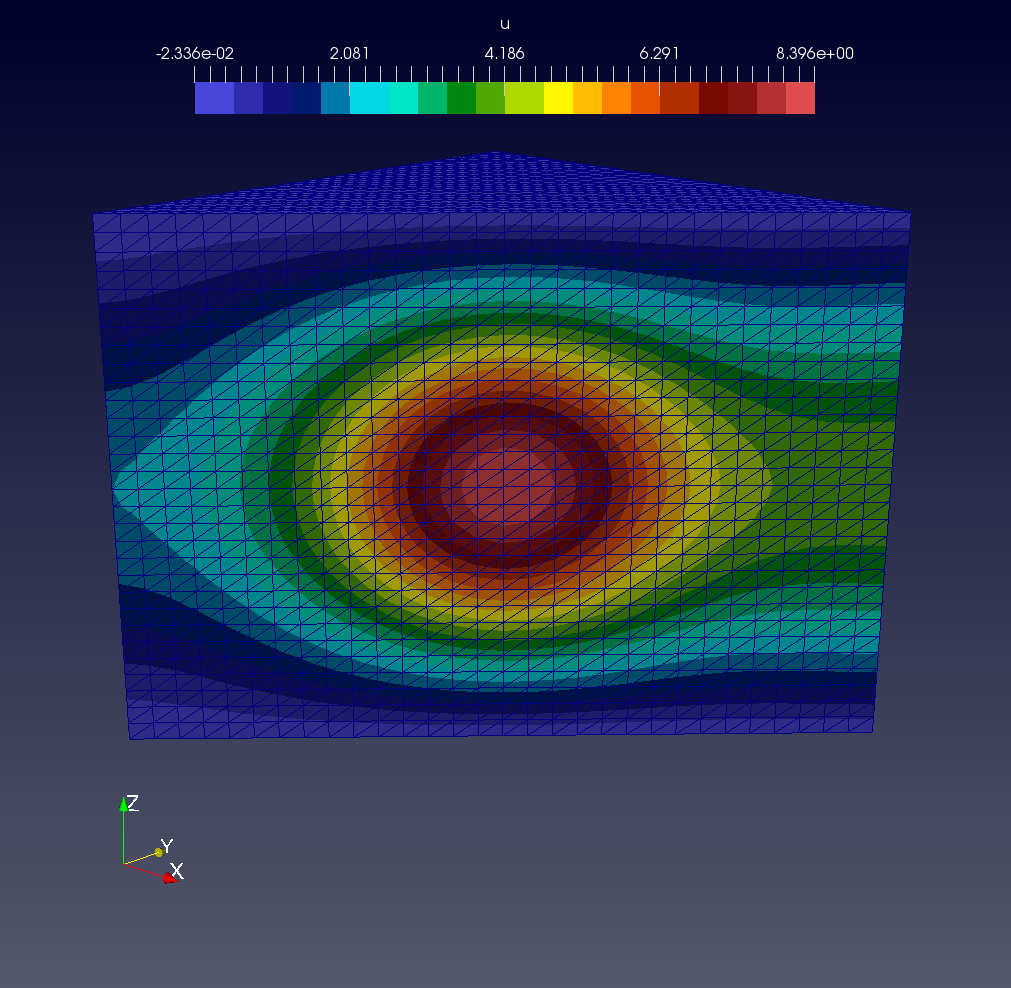}
  \caption[Inside the three-dimensional Poisson solution]{Inside the Poisson solution represented by Figure \ref{3d_poisson_image_1}.}
  \label{3d_poisson_image_2}
\end{figure}

The associated Galerkin weak form with test function $\phi \in \testspace$ defined by (\ref{test_space}) is
\begin{align*}
  -\int_{\Omega} \nabla^2 u \phi d\Omega &= \int_{\Omega} f \phi d\Omega \\
  \int_{\Omega} \nabla u \cdot \nabla \phi d\Omega - \int_{\Gamma} \phi \nabla u \cdot \mathbf{n} d\Gamma &= \int_{\Omega} f \phi d\Omega \\
  \int_{\Omega} \nabla u \cdot \nabla \phi d\Omega - \int_{\Gamma_N} \phi g_{N_y} d\Gamma_N - \int_{\Gamma_S} \phi g_{N_y} d\Gamma_S &= \int_{\Omega} f \phi d\Omega,
\end{align*}
and so the discrete variational problem consists of finding $u \in \trialspace$ given by (\ref{trial_space}) such that
\begin{align*}
  a(u,\phi) &= l(\phi) && \forall \phi \in S_0^h \subset \mathcal{H}_{E_0}^1(\Omega),
\end{align*}
subject to Dirichlet condition (\ref{3d_poisson_end}), where
\begin{align*}
  l(\phi) &= \int_{\Gamma_N} \phi g_{N_y} d\Gamma_N + \int_{\Gamma_S} \phi g_{N_y} d\Gamma_S + \int_{\Omega} f \phi d\Omega, \\
  a(u,\phi) &= \int_{\Omega} \nabla u \cdot \nabla \phi d\Omega.
\end{align*}
This is as far as is needed to proceed in order to solve this simple problem with FEniCS, as demonstrated by Code Listing \ref{3d_poisson_code} and Figures \ref{3d_poisson_image_1} and \ref{3d_poisson_image_2}.

\pythonexternal[label=3d_poisson_code, caption={FEniCS code used to solve 3D-Poisson problem (\ref{3d_poisson_start} -- \ref{3d_poisson_end}).}]{scripts/fenics_intro/3D_poisson.py}
  
%===============================================================================

\section{Stokes equations} \label{ssn_intro_stokes_3d}

\index{Linear differential equations!3D}
\index{Stokes equations!No-slip}
Recall from \S \ref{ssn_intro_stokes_2d} that the Stokes equations for an incompressible fluid are
\begin{align}
  \label{intro_stokes_3d_momentum}
  -\nabla \cdot \sigma &= \mathbf{f} &&\text{ in } \Omega &&\leftarrow \text{ conservation of momentum} \\
  \label{intro_stokes_3d_mass}
  \nabla \cdot \mathbf{u} &= 0 &&\text{ in } \Omega &&\leftarrow \text{ conservation of mass},
\end{align}
where $\sigma$ is the Cauchy-stress tensor defined as $\sigma = 2\eta \dot{\epsilon} - pI$; viscosity $\eta$, strain-rate tensor
\begin{align}
  \label{intro_3d_strain_rate_tensor}
  \dot{\epsilon} &= \frac{1}{2}\left[\nabla \mathbf{u} + (\nabla \mathbf{u})\T \right] \notag \\
  &= \begin{bmatrix}
       \frac{\partial u}{\partial x} & \frac{1}{2}\left( \frac{\partial u}{\partial y} + \frac{\partial v}{\partial x} \right) & \frac{1}{2}\left( \frac{\partial u}{\partial z} + \frac{\partial w}{\partial x} \right) \\
       \frac{1}{2}\left( \frac{\partial v}{\partial x} + \frac{\partial u}{\partial y} \right) & \frac{\partial v}{\partial y} & \frac{1}{2}\left( \frac{\partial v}{\partial z} + \frac{\partial w}{\partial y} \right) \\
       \frac{1}{2}\left( \frac{\partial w}{\partial x} + \frac{\partial u}{\partial z} \right) & \frac{1}{2}\left( \frac{\partial w}{\partial y} + \frac{\partial v}{\partial z} \right) & \frac{\partial w}{\partial z}
     \end{bmatrix},
\end{align}
velocity $\mathbf{u} = [u\ v\ w]\T$ with components $u$, $v$, $w$ in the $x$, $y$, and $z$ directions, and pressure $p$; and vector of internal forces $\mathbf{f} = \rho \mathbf{g}$ composed of material density $\rho$ and gravitational acceleration vector $\mathbf{g} = [0\ 0\ \text{-}g]\T$.  For our example we use boundary conditions
\begin{align}
  \label{intro_stokes_3d_bc_n}
  \sigma \cdot \mathbf{n} &= \bm{g_N} = \mathbf{0} &&\text{ on } \Gamma_T, \Gamma_B \\
  \label{intro_stokes_3d_bc_d}
  \mathbf{u} &= \bm{g_D} = \mathbf{0} && \text{ on } \Gamma_L,
\end{align}
where $\Gamma_T$, $\Gamma_B$, and $\Gamma_L$ are the top, bottom, and lateral boundaries and $\mathbf{n}$ is the outward-pointing normal vector.

It may be of interest to see how the conservation of momentum equations look in their expanded form,
\begin{align*}
  -\nabla \cdot \sigma &= \mathbf{f} \\
  \begin{bmatrix}
    \frac{\partial \sigma_{xx}}{\partial x} + \frac{\partial \sigma_{xy}}{\partial y} + \frac{\partial \sigma_{xz}}{\partial z} \\ 
    \frac{\partial \sigma_{yx}}{\partial x} + \frac{\partial \sigma_{yy}}{\partial y} + \frac{\partial \sigma_{yz}}{\partial z} \\ 
    \frac{\partial \sigma_{zx}}{\partial x} + \frac{\partial \sigma_{zy}}{\partial y} + \frac{\partial \sigma_{zz}}{\partial z} \\ 
  \end{bmatrix} &=
  \begin{bmatrix}
    0 \\ 0 \\ \rho g
  \end{bmatrix}
\end{align*}
so that we have three equations,
\footnotesize
\begin{subequations}
  \label{3d_stokes_expanded}
  \begin{eqnarray}
  \frac{\partial}{\partial x} \left[2\eta \frac{\partial u}{\partial x} \right]  - \frac{\partial p}{\partial x} + \frac{\partial}{\partial y} \left[\eta \left( \frac{\partial u}{\partial x} + \frac{\partial v}{\partial y} \right) \right] + \frac{\partial}{\partial z} \left[\eta \left( \frac{\partial u}{\partial x} + \frac{\partial w}{\partial z} \right) \right] &= 0 \\ 
  \frac{\partial}{\partial x} \left[\eta \left( \frac{\partial v}{\partial y} + \frac{\partial u}{\partial x} \right) \right]  + \frac{\partial}{\partial y} \left[2\eta \frac{\partial v}{\partial y} \right] - \frac{\partial p}{\partial y} + \frac{\partial}{\partial z} \left[\eta \left( \frac{\partial v}{\partial y} + \frac{\partial w}{\partial z} \right) \right] &= 0 \\ 
  \frac{\partial}{\partial x} \left[\eta \left( \frac{\partial w}{\partial z} + \frac{\partial u}{\partial x} \right) \right]  + \frac{\partial}{\partial y} \left[\eta \left( \frac{\partial w}{\partial z} + \frac{\partial v}{\partial y} \right) \right] + \frac{\partial}{\partial z} \left[2\eta \frac{\partial w}{\partial z} \right] - \frac{\partial p}{\partial z} &= \rho g,
  \end{eqnarray}
\end{subequations}
\normalsize
which when combined with the conservation of mass equation,
\begin{align*}
  \nabla \cdot \mathbf{u} &= \frac{\partial u}{\partial x} + \frac{\partial v}{\partial y} + \frac{\partial w}{\partial z} = 0, 
\end{align*}
gives four equations and four unknowns $u$, $v$, $w$, and $p$.

The weak form for this problem is formed by taking the inner product of both sides of conservation of momentum equation (\ref{intro_stokes_3d_momentum}) with the vector test function $\bm{\Phi} = [\phi\ \psi\ \chi]\T \in \mathbf{S_0^h} \subset \left( \mathcal{H}_{E_0}^1(\Omega) \right)^3$ (see test space (\ref{test_space})) integrating over the domain $\Omega$,
\begin{align*}
  -\int_{\Omega} \nabla \cdot \sigma \cdot \bm{\Phi}\ d\Omega &= \int_{\Omega} \mathbf{f} \cdot \bm{\Phi}\ d\Omega,
\end{align*}
then integrate by parts to get 
\begin{align*}
  \int_{\Omega} \sigma : \nabla \bm{\Phi}\ d\Omega - \int_{\Gamma} \sigma \cdot \mathbf{n} \cdot \bm{\Phi}\ d\Gamma &= \int_{\Omega} \mathbf{f} \cdot \bm{\Phi}\ d\Omega \\
  \int_{\Omega} \sigma : \nabla \bm{\Phi}\ d\Omega &= \int_{\Omega} \mathbf{f} \cdot \bm{\Phi}\ d\Omega,
\end{align*}
where the fact that all boundaries are either homogeneous Neumann or Dirichlet has been used to eliminate boundary integrals.  Next, multiplying incompressibility (conservation of mass) equation (\ref{intro_stokes_3d_mass}) by the test function $\xi \in M^h \subset L^2(\Omega)$ (see $L^2$ space (\ref{l2_space})) also integrating over $\Omega$,
\begin{align*}
  \int_{\Omega} \left( \nabla \cdot \mathbf{u} \right) \xi\ d\Omega &= 0.
\end{align*}

Finally, using the fact that the right-hand side of incompressibility equation (\ref{intro_stokes_3d_mass}) is zero, the mixed formulation consists of finding mixed approximation $\mathbf{u} \in \mathbf{S_E^h} \subset \left( \mathcal{H}_E^1(\Omega) \right)^3$ (see trial space (\ref{trial_space})) and $p \in M^h \subset L^2(\Omega)$ such that
\begin{align}
  \label{intro_stokes_3d_mixed_problem}
  a(\mathbf{u},p,\bm{\Phi},\xi) &= L(\bm{\Phi}) && \forall \bm{\phi} \in \mathbf{S_0^h} \subset \left( \mathcal{H}_{E_0}^1(\Omega) \right)^3,\ \xi \in M^h \subset L^2(\Omega),
\end{align}
subject to Dirichlet condition (\ref{intro_stokes_3d_bc_d}) and
\begin{align*}
  a(\mathbf{u},p,\bm{\Phi},\xi) &= \int_{\Omega} \sigma : \nabla \bm{\Phi}\ d\Omega + \int_{\Omega} \left( \nabla \cdot \mathbf{u} \right) \xi\ d\Omega, \\
  L(\bm{\Phi}) &= \int_{\Omega} \mathbf{f} \cdot \bm{\Phi}\ d\Omega.
\end{align*}

\begin{figure*}
  \centering
    \includegraphics[width=0.8\linewidth]{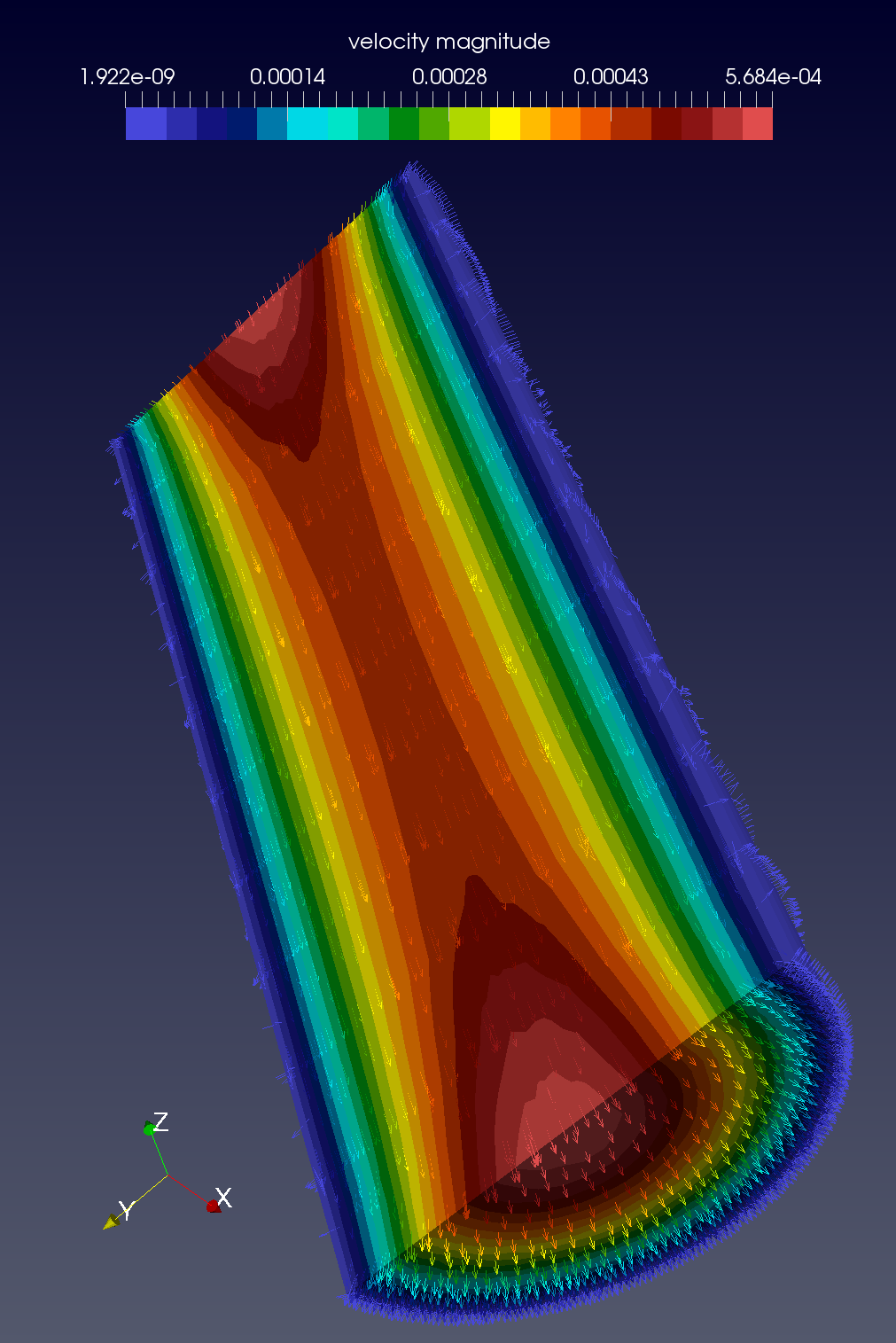}
  \caption[Three-dimensional-no-slip Stokes example]{Inside Stokes velocity solution $\mathbf{u}$ in m s\sups{-1} within a tube of diameter 1 mm and height 5 mm filled with honey at density $\rho = 1420$ kg m$^{-3}$ and viscosity $\eta = 8$ Pa s.  Gravity forces the honey in the $-z$ direction, as indicated by the overlain velocity vectors.}
  \label{intro_3d_stokes_image}
\end{figure*}

For our solution satisfying inf-sup condition (\ref{inf_sup_condition}), we enrich the finite element space with bubble functions (see Chapter \ref{ssn_subgrid_scale_effects}), thus creating MINI elements \citep{arnold}.  The solution is generated with Code Listing \ref{intro_3d_stokes_code} and depicted in Figure \ref{intro_3d_stokes_image}.

\pythonexternal[label=intro_3d_stokes_code, caption={FEniCS solution to 3D-Stokes-no-slip problem (\ref{intro_stokes_3d_mixed_problem}).}]{scripts/fenics_intro/3D_stokes.py}

%===============================================================================
%===============================================================================

\chapter{Subgrid scale effects} \label{ssn_subgrid_scale_effects}

The discretization of a domain for which a differential equation is to be solved often results in numerically unstable solutions.  These instabilities result from \emph{subgrid}-scale effects that cannot be accounted for at the resolution of the reduced domain.  Here we review several techniques for stabilizing the solution to such problems through the development of a distributional formulation including an approximation of these subgrid-scale effects.

%===============================================================================

\section{Subgrid scale models}

\begin{figure}
  \centering
    \def\svgwidth{\linewidth}
    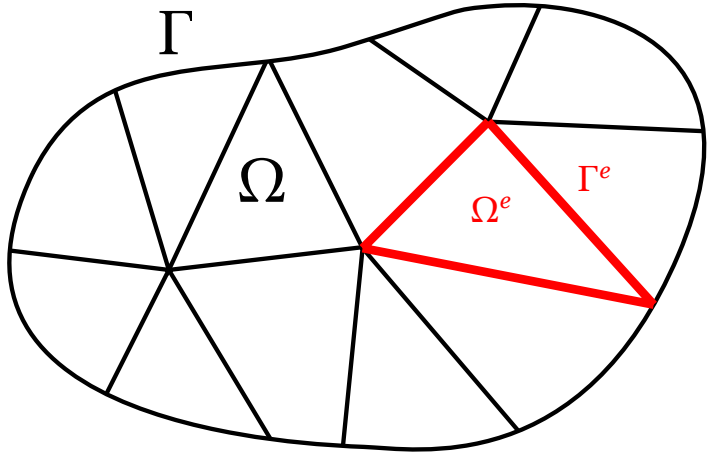
  \caption[The discrete model]{The domain of the partitioned problem.}
  \label{subgrid_domain_image}
\end{figure}

As presented by \citet{hughes}, consider the bounded domain $\Omega$ with boundary $\Gamma$ discretized into $N$ element subdomains $\Omega^e$ with boundaries $\Gamma^e$ (Figure \ref{subgrid_domain_image}).  Let
\begin{align*}
  \Omega' &= \bigcup_{e=1}^{N} \Omega^e  &&\leftarrow \text{element interiors} \\
  \Gamma' &= \bigcup_{e=1}^{N} \Gamma^e  &&\leftarrow \text{element boundaries} \\
  \Omega &= \bar{\Omega} = \mathrm{closure}(\Omega'). 
\end{align*}

\index{Subgrid scales}
The abstract problem that we wish to solve consists of finding a function $u \in L^2(\Omega)$, such that for given functions $f \in L^2(\Omega)$, $q \in L^2(\Gamma)$,
\begin{align}
  \label{bubble_diff_eq}
  \Lu u &= f &&\text{ in } \Omega \\
  \label{bubble_bc}
  u &= q     &&\text{ on } \Gamma,
\end{align}
\index{Differential operator}
where $\mathcal{L}$ is a possibly non-symmetric differential operator and the unknown $u$ is composed of overlapping resolvable scales $\bar{u}$ and unresolvable, or subgrid scales $u'$, i.e.\ $u = \bar{u} + u'$.

The variational form of this system may be stated using the definition of the \index{Adjoint operator} \index{Differential operator} adjoint $\Lu^*$ of the operator $\Lu$
\begin{align}
  \label{subgrid_variational_form}
  a(w,u) = (w, \Lu u) = (\Lu^* w, u)
\end{align}
for all sufficiently smooth $u$, $w$ such that
$$u = q, \hspace{5mm} w = 0 \hspace{5mm} \text{ on } \Gamma,$$
where
\begin{align*}
  u = \bar{u} + u'  \hspace{5mm} \text{and} \hspace{5mm}
  w = \bar{w} + w'.
\end{align*}
Making the assumption that the unresolvable scales vanish on element boundaries,
$$u' = w' = 0 \text{ on } \Gamma',$$
variational Equation (\ref{subgrid_variational_form}) is transformed into
\begin{align*}
  a(w,u) &= (w,f) \\
  a(\bar{w} + w', \bar{u} + u') &= (\bar{w} + w', f) \\
  a(\bar{w}, \bar{u}) + a(\bar{w}, u') + a(w', \bar{u}) + a(w',u')  &= (\bar{w}, f) + (w',f).
\end{align*}
providing two equations, which we collect in terms of $\bar{w}$ and $w'$,
\begin{align}
  \label{bubble_relation_1}
  \begin{cases}
    a(\bar{w}, \bar{u}) + a(\bar{w}, u') = (\bar{w}, f) \\
    (\bar{w}, \Lu \bar{u}) + (\bar{w}, \Lu u') = (\bar{w}, f) \\
    (\bar{w}, \Lu \bar{u}) + (\Lu^* \bar{w}, u') = (\bar{w}, f)
  \end{cases}
\end{align}
and
\begin{align}
  \label{bubble_relation_2}
  \begin{cases}
    a(w', \bar{u}) + a(w',u')  = (w',f) \\
    (w', \Lu \bar{u}) + (w', \Lu u')  = (w',f).
  \end{cases}
\end{align}
The Euler-Lagrange equations corresponding to second subproblem (\ref{bubble_relation_2}) are
\begin{align}
  \Lu \bar{u} + \Lu u' &= f && \notag \\
  \label{bubble_euler_lagrange}
  \implies \Lu u' &= - (\Lu\bar{u} - f) &&\text{ in } \Omega^e \\
               u' &= 0 &&\text{ on } \Gamma^e. \notag
\end{align}
Thus, the differential operator $\Lu$ applied to the unresolvable scales $u'$ is equal to the residual of the resolved scales, $f - \Lu \bar{u}$ when we assume that the unresolvable scales vanish on element boundaries.

%===============================================================================

\section{Green's function for $\Lu$}

\index{Green's functions}
The Green's function problem for a linear operator $\Lu$ in problem (\ref{bubble_diff_eq}) seeks to find $g(x,y)$ such that
$$u(y) = \left(\Lu^{-1}f\right)(y) = \int_{\Omega'} g(x,y)f(x)d\Omega_x,$$
with Green's functions satisfying
\begin{align*}
  \Lu g &= \delta &&\text{ in } \Omega^e \\
  g     &= 0 &&\text{ on } \Gamma^e,
\end{align*}
where $\delta$ is the Dirac delta distribution.  An expression for the unresolvable scales may be formed in terms of the resolvable scales from (\ref{bubble_euler_lagrange}):
\begin{align}
  \label{greens_ftn_prob}
  u'(y) = - \int_{\Omega'} g(x,y) \left(\Lu\bar{u} - f\right)(x) d\Omega_x'.
\end{align}
Substituting this expression into (\ref{bubble_relation_1}) results in
\begin{align}
  \label{bubble_relation_1_resolve}
  (\bar{w}, \Lu \bar{u}) + (\Lu^* \bar{w}, M(\Lu\bar{u} - f)) &= (\bar{w}, f),
\end{align}
where
\begin{align}
  \label{subgrid_scale_integral}
  M v(x) &= - \int_{\Omega'} g(x,y) v(x) d\Omega_x'
\end{align}
and
\begin{align*}
  (\Lu^* \bar{w}, M(\Lu\bar{u} - f)) &= - \int_{\Omega'} \int_{\Omega'} \left( \Lu^* \bar{w} \right)(y) g(x,y) \left( \Lu \bar{u} - f \right)(x) d\Omega_x d\Omega_y.
\end{align*}

Expression (\ref{bubble_relation_1_resolve}) may be stated in the bilinear form
\begin{align*}
  B(\bar{w}, \bar{u}; g) &= L(\bar{w}; g),
\end{align*}
where
\begin{align*}
  B(\bar{w}, \bar{u}; g) &= (\bar{w}, \Lu \bar{u}) + (\Lu^* \bar{w}, M \Lu\bar{u}) \\
  L(\bar{w}; g) &= (\Lu^* \bar{w}, Mf) + (\bar{w}, f).
\end{align*}

Thus all the effects of the unresolvable scales have been accounted for up to the assumption that $u'$ vanish on element boundaries.  Next is derived an approximation of Green's function $g$ and a development of a finite-dimensional analog of (\ref{bubble_relation_1_resolve}).

%===============================================================================

\section{Bubbles}

\index{Bubble functions}
The space of bubble functions consists of the set of functions that vanish on element boundaries and whose maximum values is one, the space
\begin{align}
  \label{bubble_space}
  \mathcal{B}_0^k(\Omega) &= \left\{ u \in \mathcal{H}^k(\Omega)\ |\ u = 0\ \text{on}\ \Gamma^e, \Vert u \Vert_{\infty} = 1 \right\}.
\end{align}

\begin{figure}
  \centering
    \includegraphics[width=\linewidth]{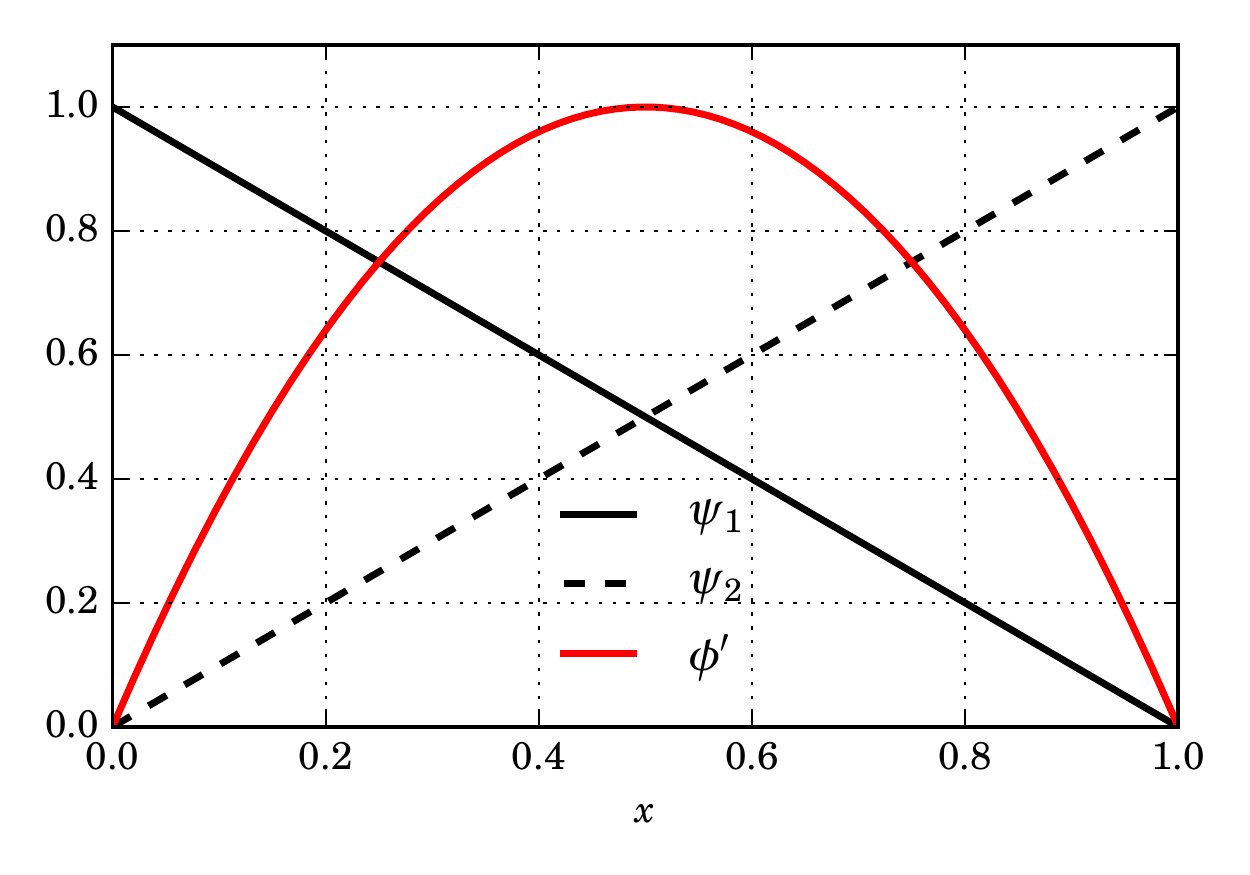}
  \caption[One-dimensional bubble function]{The lowest order one-dimensional triangular element reference bubble function given by (\ref{bubble_function}) with $h_e = 1$.}
  \label{1d_bubble_image}
\end{figure}

For a concrete example, the lowest order -- corresponding to $k = 2$ in (\ref{bubble_space}) -- one-dimensional reference bubble function is defined as
\begin{align}
  \label{bubble_function}
  \phi'_e(x) = 4 \psi_1^e(x) \psi_2^e(x),
\end{align}
with basis given by the one-dimensional linear Lagrange interpolation functions described previously in \S \ref{ssn_local_galerkin_assembly}, 
\begin{align*}
  \psi_1^e(x) = 1 - \frac{x}{h_e} \hspace{10mm} \psi_2^e(x) = \frac{x}{h_e},
\end{align*}
where $h_e$ is the width of element $e$.  This basis satisfies the required interpolation properties
\begin{align*}
  \psi_i^e(x_j) = \delta_{ij} \hspace{10mm} \sum_{j=1}^n \psi_j^e(x) = 1,
\end{align*}
where $n$ is the number of element equations.  Note that $\phi'$ has the properties that $\Vert \phi' \Vert_{\infty} = 1$ and is zero on the element boundaries (Code Listing \ref{bubble_1d_ftn_code} and Figure \ref{1d_bubble_image}).

\begin{figure}
  \centering
    \includegraphics[width=\linewidth]{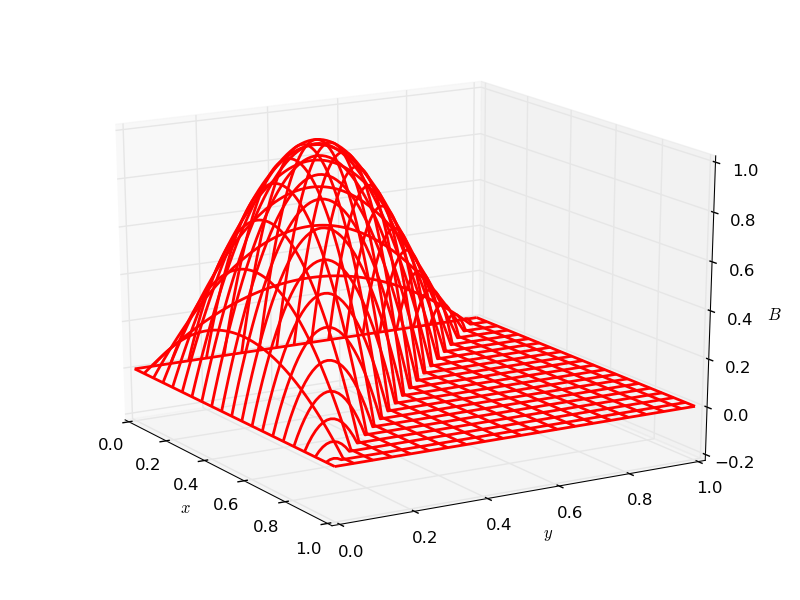}
  \caption[Two-dimensional bubble function]{The lowest order two-dimensional reference bubble function given by (\ref{2d_bubble_function}) with $h_x = h_y = 1$.}
  \label{2d_bubble_image}
\end{figure}

The lowest order -- corresponding to $k = 2$ in (\ref{bubble_space}) -- two-dimensional triangular reference element bubble function is defined as
\begin{align}
  \label{2d_bubble_function}
  \phi'_{2e}(x,y) = 27 \psi_1^{2e}(x,y) \psi_2^{2e}(x) \psi_3^{2e}(y),
\end{align}
with basis given by the quadratic Lagrange interpolation functions
\begin{align*}
  \psi_1^{2e}(x,y) = 1 - \frac{x}{h_x} - \frac{y}{h_y}, \hspace{10mm} \psi_2^{2e}(x) = \frac{x}{h_e}, \hspace{10mm} \psi_3^{2e}(y) = \frac{y}{h_e},
\end{align*}
where $h_x$ is the max height in the $x$ direction and $h_y$ is the max height of in the $y$ direction of the reference element $e$.  Again, $\phi'_{2e}(x,y)$ has the properties that $\Vert \phi'_{2e} \Vert_{\infty} = 1$ and is zero on the element boundaries (Code Listing \ref{bubble_2d_ftn_code} and Figure \ref{2d_bubble_image}).

\pythonexternal[label=bubble_1d_ftn_code, caption={Python code used to generate Figure \ref{1d_bubble_image}.}]{scripts/bubbles/bubble_ftn.py}

\pythonexternal[label=bubble_2d_ftn_code, caption={Python code used to generate Figure \ref{2d_bubble_image}.}]{scripts/bubbles/bubble_2.py}

%===============================================================================

\section{Approximation of Green's function for $\Lu$ with bubbles}

Following the work of \citet{hughes}, if $\phi'_j(x)$, $j \in 1,2,\ldots,N_b$ is a set of $N_b$ linearly independent bubble functions, a single element's unresolvable scales can be approximated -- in a process referred to as \index{Static condensation} \emph{static condensation} -- by linearly expanding $u'$ into $N_b$ nodes and $\bar{u}$ into $N_n$ nodes,
\begin{align}
  \label{bubble_u_approximation}
  u'(x) \approx u_h'(x) = \sum_{j=1}^{N_b} \phi_j'(x) u_j', \hspace{5mm} \bar{u}(x) \approx \bar{u}_h(x) = \sum_{j=1}^{N_n} \psi_j(x) \bar{u}_j,
\end{align}
where $u_j'$ is the coefficient associated with bubble function $j$ and $\bar{u}_j$ is the coefficient associated with a finite-element shape function $j$.

Fixing $w_k' = \phi_k'$ and inserting $u_h'(x)$ into (\ref{bubble_relation_2}),
\begin{align*}
  a(w_k', \bar{u}_h) + a(w_k',u_h')  &= (w_k',f) \\
  a(\phi_k', \bar{u}_h) + a\left( \phi_k', \sum_{j=1}^{N_b} \phi_j' u_j' \right)  &= (\phi_k',f) \\
  \left( \phi_k', \Lu \left( \sum_{j=1}^{N_b} \phi_j' u_j' \right) \right)  &= (\phi_k',f) - (\phi_k', \Lu \bar{u}_h) \\
  \left( \phi_k', \sum_{j=1}^{N_b} \Lu \phi_j' u_j' \right)  &= - \left( \phi_k', \Lu \bar{u}_h - f \right) \\
  \sum_{j=1}^{N_b} \left( \phi_k', \Lu \phi_j' \right) u_j'  &= - \left( \phi_k', \Lu \bar{u}_h - f \right) \\
  \sum_{j=1}^{N_b} a\left( \phi_k', \phi_j' \right) u_j'  &= - \left( \phi_k', \Lu \bar{u}_h - f \right),
\end{align*}
for $k = 1,2,\ldots,N_b$.  This implies that the $j$ nodal values of $u_h'(x)$ in (\ref{bubble_u_approximation}) are given by the linear system of equations
\begin{align*}
  u_j' &= - \sum_{k=1}^{N_b} a\left( \phi_j', \phi_k' \right)^{-1} \left( \phi_k'(x), \Lu \bar{u}_h(x) - f(x) \right),
\end{align*}
where $a\left( \phi_j', \phi_k' \right)^{-1}$ is the $jk$ component of the inverse of matrix $a\left( \phi_j', \phi_k' \right)$.  Inserting this into $u_h'(y)$ given by (\ref{bubble_u_approximation}),
\begin{align*}
  u_h'(y) &= \sum_{j=1}^{N_b} \phi_j'(y) u_j' \\
  &= \sum_{j=1}^{N_b} \phi_j'(y) \left[ - \sum_{k=1}^{N_b} a\left( \phi_j', \phi_k' \right)^{-1} \left( \phi_k'(x), \Lu \bar{u}_h(x) - f(x) \right) \right] \\
  &= -\sum_{j,k=1}^{N_b} \phi_j'(y) \left[ a\left( \phi_j', \phi_k' \right)^{-1} \left( \phi_k'(x), \Lu \bar{u}_h(x) - f(x) \right) \right] \\
  &= - \sum_{j,k=1}^{N_b} \int_{\Omega'} \phi_j'(y) \left[ a\left( \phi_j', \phi_k' \right)^{-1} \right] \phi_k'(x) \left(\Lu \bar{u}_h - f \right)(x) d\Omega_x' \\
  &= - \int_{\Omega'} \left( \sum_{j,k=1}^{N_b} \phi_j'(y) \left[ a\left( \phi_j', \phi_k' \right)^{-1} \right] \phi_k'(x) \right) \left(\Lu \bar{u}_h - f \right)(x) d\Omega_x' \\
  &= - \int_{\Omega'} \tilde{g}(x,y) \left(\Lu \bar{u}_h - f \right)(x) d\Omega_x',
\end{align*}
thus providing the Green's function approximation in (\ref{greens_ftn_prob})
\begin{align}
  \label{greens_ftn_approx}
  g(x,y) \approx \tilde{g}(x,y) = \sum_{j,k=1}^{N_b} \phi_j'(y) \left[ a\left( \phi_j', \phi_k' \right)^{-1} \right] \phi_k'(x).
\end{align}

Finally, inserting resolvable scale approximation $\bar{u}_h$ defined by (\ref{bubble_u_approximation}) and Green's function approximation (\ref{greens_ftn_approx}) into (\ref{bubble_relation_1_resolve}), we have
\begin{align}
  \label{bubble_relation_1_resolve_approx}
  (\bar{w}_h, \Lu \bar{u}_h) + (\Lu^* \bar{w}_h, \tilde{M}(\Lu\bar{u}_h - f)) &= (\bar{w}_h, f),
\end{align}
where
\begin{align*}
  M v(x) \approx \tilde{M} v(x) &= - \int_{\Omega'} \tilde{g}(x,y) v(x) d\Omega_x',
\end{align*}
and the associated approximate bilinear form
\begin{align*}
  B(\bar{w}_h, \bar{u}_h; \tilde{g}) &= L(\bar{w}_h; \tilde{g}),
\end{align*}
where
\begin{align*}
  B(\bar{w}_h, \bar{u}_h; \tilde{g}) &= (\bar{w}_h, \Lu \bar{u}_h) + (\Lu^* \bar{w}_h, \tilde{M} \Lu\bar{u}_h) \\
  L(\bar{w}_h; \tilde{g}) &= (\Lu^* \bar{w}_h, \tilde{M}f) + (\bar{w}_h, f).
\end{align*}

%===============================================================================

\section{Stabilized methods} \label{ssn_stabilized_methods}

\index{Stabilization methods!Streamline-upwind/Petrov-Galerkin}
\index{Stabilization methods!Galerkin/least-squares}
\index{Stabilization methods!Subgrid-scale-model}
As described by \citet{hughes} and \citet{codina}, stabilized methods are \emph{generalized Galerkin methods} of the form
\begin{align}
  \label{generalized_form}
  (\bar{w}_h, \Lu \bar{u}_h) + (\mathbb{L} \bar{w}_h, \tau(\Lu\bar{u}_h - f)) &= (\bar{w}_h, f),
\end{align}
where operator $\mathbb{L}$ is a differential operator typically chosen from
\begin{align}
  \label{bubble_gls_operator}
  \mathbb{L} &= + \Lu && \text{Galerkin/least-squares (GLS)} \\
  \label{bubble_supg_operator}
  \mathbb{L} &= + \Lu_{\text{adv}} && \text{SUPG} \\
  \label{bubble_ssm_operator}
  \mathbb{L} &= - \Lu^* && \text{subgrid-scale model (SSM)}
\end{align}
where $\Lu_{\text{adv}}$ is the advective part of the operator $\Lu$.

Note that when using differential operator (\ref{bubble_ssm_operator}), stabilized form (\ref{generalized_form}) implies that $\tau = - \tilde{M} \approx -M$, and therefore the \index{Intrinsic-time parameter!General form} \emph{intrinsic-time} parameter $\tau$ approximates integral operator (\ref{subgrid_scale_integral}).  Equivalently,
\begin{align*}
  \tau \cdot \delta(y-x) = \tilde{g}(x,y) \approx g(x,y),
\end{align*}
and we can generate an explicit formula for $\tau$ by integrating over a single element $\Omega^e$,
\begin{align*}
  \int_{\Omega^e} \int_{\Omega^e} \tilde{g}(x,y)\ d\Omega^e_x d\Omega^e_y &= \int_{\Omega^e} \int_{\Omega^e} \tau \cdot \delta(y-x)\ d\Omega^e_x d\Omega^e_y = \tau h,
\end{align*}
\begin{align*}
  \implies \tau &= \frac{1}{h} \int_{\Omega^e} \int_{\Omega^e} \tilde{g}(x,y)\ d\Omega^e_x d\Omega^e_y,
\end{align*}
where $h$ is the element diameter.  Therefore, the parameter $\tau$ will depend both on the operator $\Lu$ and the basis chosen for Green's function approximation $\tilde{g}$ as evident by (\ref{greens_ftn_approx}).

For example, when $\Lu$ is the advective-diffusive operator $\Lu u = -\nabla \cdot \sigma(u) = -\nabla \cdot (k \nabla u - \mathbf{a} u)$ and linear-Lagrange elements are used, the optimal expression for $\tau$ is the \emph{streamline upwind/Petrov-Galerkin} (SUPG) coefficient \citep{brooks, hughes}. 
\begin{align}
  \label{tau_supg}
  \tau_{\text{SUPG}} = \frac{h}{2|\mathbf{a}|} \left( \coth(P_{\'e}) - \frac{1}{P_{\'e}} \right), \hspace{5mm} P_{\'e}  = \frac{h |\mathbf{a}|}{2 \kappa},
\end{align}
where \index{Peclet@P\`eclet number!General form} $P_{\'e}$ is the element P\'eclet number and $\mathbf{a}$ is the material velocity vector. 

On the other hand, if $\Lu$ is the diffusion-reaction operator $\Lu u = - \nabla \cdot \left( k \nabla u \right) + su$ with absorption coefficient $s \geq 0$, $P_{\'e}=0$ and $\tau$ is given by the coefficient \citep{hughes_VIII}
\begin{align}
  \label{tau_dr}
  \tau_{\text{DR}} = \alpha \frac{h^2}{\kappa},
\end{align}
where $\alpha$ is a mesh-size-independent parameter dependent on the specific model used.  

When $\Lu$ is the advective-diffusion-reaction equation $\Lu u = - \nabla \cdot \left( k \nabla u \right) - \mathbf{a} \cdot \nabla u + s u$, \citet{codina} used the coefficient
\begin{align}
  \label{tau_adr}
  \tau_{\text{ADR}} = \frac{1}{\frac{4 \kappa}{h^2} + \frac{2 |\mathbf{a}|}{h} + s},
\end{align}
to stabilize the formulation over a range of values for $s$ and $|\mathbf{a}|$ using the space of linear Lagrange interpolation functions $\psi$.

Finally, when $\Lu$ is the Stokes operator $\Lu (\mathbf{u},p) = -\nabla \cdot \sigma(\mathbf{u},p) = -\nabla \cdot \left( 2\eta\dot{\epsilon}(\mathbf{u}) - pI \right)$, $\tau$ has been found to be the coefficient \citep{hughes_V}
\begin{align}
  \label{tau_stokes}
  \tau_{\text{S}} = \alpha \frac{h^2}{2\eta},
\end{align}
where the unknowns consist of the material velocity $\mathbf{u}$ and pressure $p$ (see \S \ref{ssn_intro_stokes_2d}, \S \ref{ssn_intro_stokes_2d_slip}, and \S \ref{ssn_intro_stokes_3d}), and $\alpha > 0$ may or may not depend on the basis used for $\mathbf{u}$.

%===============================================================================

\section{Diffusion-reaction problem}

\index{Linear differential equations!1D}
For an example, consider the steady-state advection-diffusion equation defined over the domain $\Omega \in [0,1]$
\begin{align}
  \label{bubble_example_1}
  \Lu u = - \kappa \frac{d^2u}{dx^2} + s u = f, \hspace{4mm} u(0) = 1,\ u'(1) = 0,
\end{align}
with diffusion coefficient $\kappa$, absorption coefficient $s \geq 0$, and source term $f$ are constant throughout the domain.

\subsection{Bubble-enriched solution}

The bubble-function-enriched distributional form of the equation consists of finding $\hat{u} = \bar{u} + u'$ where $\bar{u} \in S_E^h \subset \mathcal{H}_E^1(\Omega)$ (see trial space (\ref{trial_space})) and $u' \in B_0^h \subset \mathcal{B}_0^2(\Omega)$ (see bubble space (\ref{bubble_space})) such that
\begin{align*}
  (\hat{\psi}, \Lu \hat{u}) &= (\hat{\psi}, f) \\
  -\kappa \int_{\Omega} \frac{d^2\hat{u}}{dx^2} \hat{\psi} d\Omega + s\int_{\Omega} \hat{u} \hat{\psi} d\Omega &= \int_{\Omega} f \hat{\psi} d\Omega \\
  \kappa \int_{\Omega} \frac{d\hat{u}}{dx} \frac{d\hat{\psi}}{dx} d\Omega - \kappa \int_{\Gamma} \frac{d\hat{u}}{dx} \hat{\psi} d\Gamma + s\int_{\Omega} \hat{u} \hat{\psi} d\Omega &= \int_{\Omega} f \hat{\psi} d\Omega \\
  \kappa \int_{\Omega} \frac{d\hat{u}}{dx} \frac{d\hat{\psi}}{dx} d\Omega + s\int_{\Omega} \hat{u} \hat{\psi} d\Omega &= \int_{\Omega} f \hat{\psi} d\Omega.
\end{align*}
for all \emph{enriched} or \emph{augmented} test functions $\hat{\psi} = \psi + \phi$, where $\psi \in S_0^h \subset \mathcal{H}_E^1(\Omega)$ (see test space (\ref{test_space})) and $\phi \in B_0^h \subset \mathcal{B}_0^2(\Omega)$ (bubble space (\ref{bubble_space})).

\subsection{SSM-stabilized solution}

The subgrid-scale stabilized distributional form of the equation is derived by using subgrid-scale-model operator (\ref{bubble_ssm_operator}) and DR stability parameter (\ref{tau_dr}) within the general stabilized form (\ref{generalized_form}).  Thus, the stabilized problem consists of finding $\tilde{u} \in S_E^h \subset \mathcal{H}_E^1(\Omega)$ (see trial space (\ref{trial_space})) such that
\begin{align*}
  (\psi, \Lu \tilde{u}) - (\Lu^* \psi, \tau_{\text{DR}} (\Lu \tilde{u} - f)) &= (\psi, f),
\end{align*}
for all test functions $\psi \in S_0^h \subset \mathcal{H}_E^1(\Omega)$ (see test space (\ref{test_space})).  Using the fact that the diffusion-reaction operator (\ref{bubble_example_1}) is self adjoint, we have the bilinear form
\begin{align*}
  B(\tilde{u},\psi) &= L(\psi),
\end{align*}
where
\begin{align*} 
  B(u,\psi) &= \kappa \int_{\Omega} \frac{d\tilde{u}}{dx} \frac{d\psi}{dx} d\Omega + s\int_{\Omega} \tilde{u} \psi d\Omega - \frac{\alpha}{\kappa} \int_{\Omega} h^2 \left( \Lu^* \psi \right) \left( \Lu \tilde{u} \right) d\Omega \\
  L(\psi) &= \int_{\Omega} f \psi d\Omega - \frac{\alpha}{\kappa} \int_{\Omega} h^2 \Lu^* \psi f d\Omega.
\end{align*}

Note if linear Lagrange elements are used as a basis for $\tilde{u}$ and $\psi$, the diffusive terms with coefficient $\kappa$ in $\Lu \tilde{u}$ and $\Lu^* \psi$ will be zero.

\subsection{Analytic solution}

With the constants
\begin{align*}
  \kappa = \frac{1}{500}, \hspace{4mm} s = 1, \hspace{5mm} f = 0,
\end{align*}
the analytic solution to (\ref{bubble_example_1}) is 
$$u_{\text{a}}(x) = \frac{\exp\left(-10\sqrt{5}(x-2)\right) + \exp\left(10 \sqrt{5}x\right)}{1 + \exp\left(20\sqrt{5}\right)}.$$

\begin{figure}
  \centering
    \includegraphics[width=\linewidth]{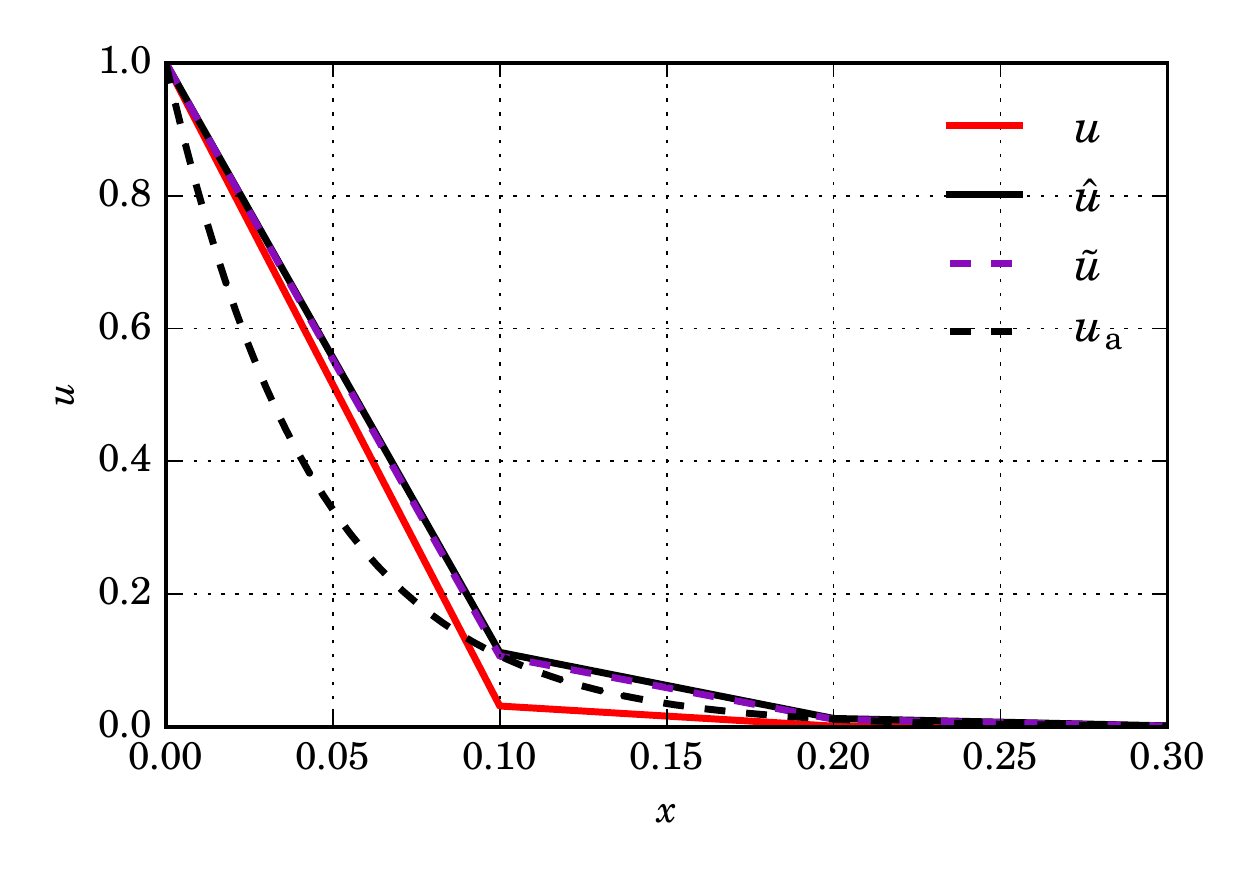}
  \caption[Diffusion-reaction stabilization example]{The analytic solution $u_{\text{a}}$ (black dashed) plotted against the unstabilized method solution $u$ (red), bubble-enriched solution $\hat{u}$ (dashed purple), and SSM-stabilized solution $\tilde{u}$ (black).  The parameter $\alpha = \frac{1}{15}$ was found by experimentation.}
  \label{dr_analytic_image}
\end{figure}

Note that this is a heavily reaction-dominated problem, resulting in high gradients in the solution, referred to as a \index{Boundary layer} \emph{boundary layer}, near $x=0$.  The analytic solution is plotted against solutions determined with the standard Galerkin method with linear Lagrange elements $\psi$, quadratic-bubble-enriched linear Lagrange elements $\hat{\psi}$, and the SSM-stabilized formulation in Figure \ref{dr_analytic_image}, generated by Code Listing \ref{dr_analytic_code}.

\pythonexternal[label=dr_analytic_code, caption={FEniCS code used solve diffusion-reaction problem (\ref{bubble_example_1}).}]{scripts/bubbles/DR_analytic.py}

%===============================================================================

\section{Advection-diffusion-reaction example}

\index{Linear differential equations!1D}
Consider the model defined over the domain $\Omega \in [0,1]$
\begin{align}
  \label{bubble_example_2}
  \Lu u = -\kappa \frac{d^2u}{dx^2} + d \frac{du}{dx} + su = f, \hspace{4mm} u(0) = 0,\ u'(1) = 0,
\end{align}
where $\kappa$ is the diffusion coefficient, $d$ is the velocity of the material, $s \geq 0$ is an absorption coefficient, and $f$ is a source term.

\subsection{GLS-stabilized solution}

The Galerkin/least-squares  stabilized distributional form of the equation is derived by using GLS operator (\ref{bubble_gls_operator}) and ADR stability parameter (\ref{tau_adr}) within the general stabilized form (\ref{generalized_form}).  Thus, the stabilized problem consists of finding $\tilde{u} \in S_E^h \subset \mathcal{H}_E^1(\Omega)$ (see trial space (\ref{trial_space})) such that
\begin{align*}
  (\psi, \Lu \tilde{u}) + (\Lu \psi, \tau_{\text{ADR}} (\Lu \tilde{u} - f)) &= (\psi, f),
\end{align*}
for all test functions $\psi \in S_0^h \subset \mathcal{H}_E^1(\Omega)$ (see test space (\ref{test_space})).  Using ADR stability parameter (\ref{tau_adr}), we have the bilinear form
\begin{align*}
  B(\tilde{u},\psi) &= L(\psi),
\end{align*}
where
\begin{align*}
  B(\tilde{u}, \psi) = &+ \kappa \int_{\Omega} \frac{d\tilde{u}}{dx} \frac{d\psi}{dx} d\Omega + d \int_{\Omega} \frac{d\tilde{u}}{dx} \psi d\Omega + s\int_{\Omega} \tilde{u} \psi d\Omega \\
  &+ \int_{\Omega} \tau_{\text{ADR}} \left( \Lu \psi \right) \left( \Lu \tilde{u} \right) d\Omega \\
  L(\psi) = &+ \int_{\Omega} f \hat{\psi} d\Omega + \int_{\Omega} \tau_{\text{ADR}} \Lu \psi f d\Omega.
\end{align*}

Note if linear Lagrange elements are used as a basis for $\tilde{u}$ and $\psi$, the diffusive terms with coefficient $\kappa$ in $\Lu \tilde{u}$ and $\Lu \psi$ will be zero, and if $s = 0$, the GLS and SUPG operators given by (\ref{bubble_gls_operator}) and (\ref{bubble_supg_operator}), respectively, would in this case be identical \citep{hughes_VIII}.

\begin{figure}
  \centering
    \includegraphics[width=\linewidth]{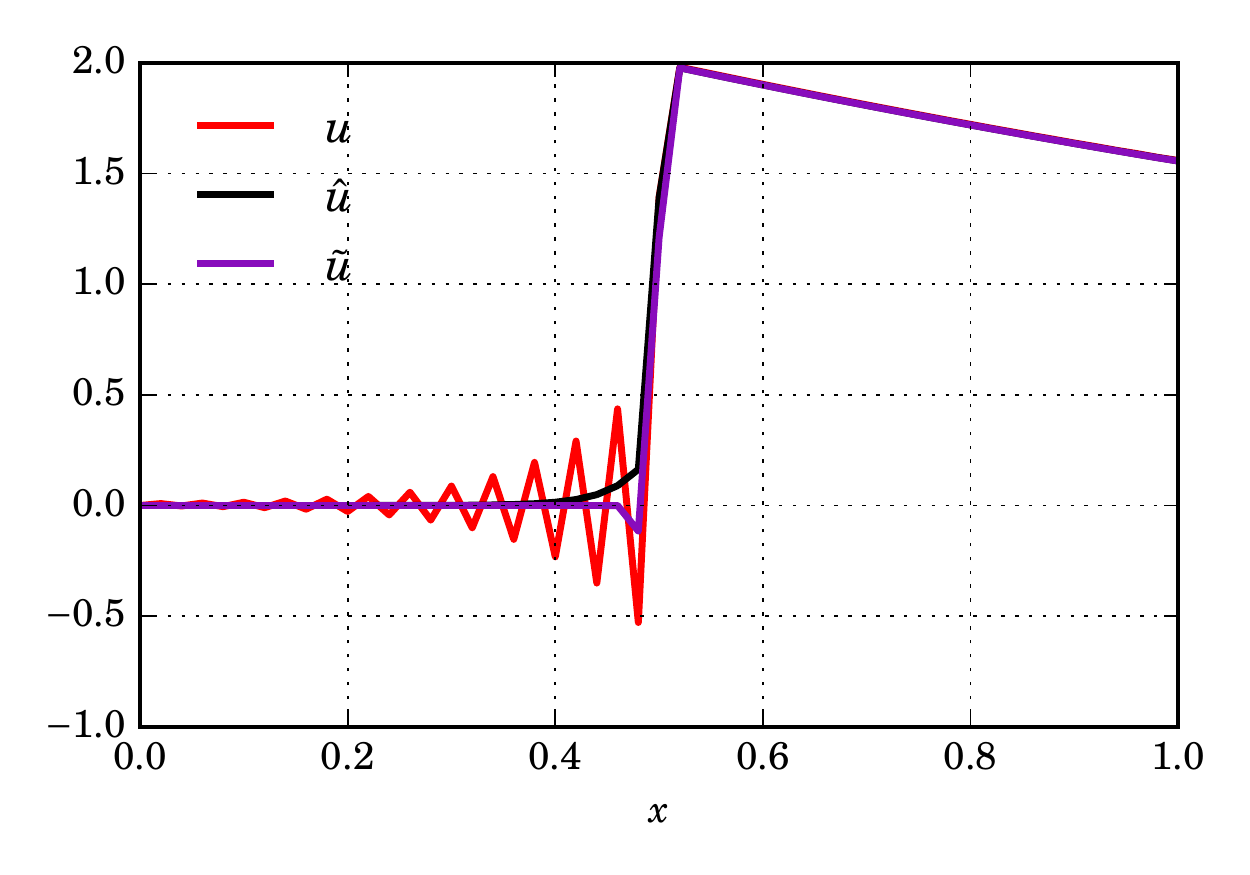}
  \caption[Advection-diffusion-reaction stabilization example]{The unstabilized method solution $u$ (red), bubble-enriched solution $\hat{u}$ (black), and GLS-stabilized solution $\tilde{u}$ (purple) with identical mesh spacing.}
  \label{dr_extreme_image}
\end{figure}

For an extreme example, we take
\begin{align*}
  \kappa = \frac{1}{100}, \hspace{4mm} d = 10, \hspace{4mm} s = 5,
\end{align*}
and
\begin{align*}
  f = \begin{cases}
        1000 & \text{ if } x = 0.5 \\
        0 & \text{ otherwise }
      \end{cases},
\end{align*}
resulting in an equation with low diffusivity that is heavily dominated by gradients of $u$ while advecting $u$ in the $+x$ direction.  Solutions determined with the standard Galerkin method with linear Lagrange elements $\psi$, quadratic-bubble-enriched linear Lagrange elements $\hat{\psi}$, and the GLS-stabilized formulation are depicted in Figure \ref{dr_extreme_image} and generated by Code Listing \ref{dr_extreme_code}.

\pythonexternal[label=dr_extreme_code, caption={FEniCS code used solve advection-diffusion-reaction problem (\ref{bubble_example_2}).}]{scripts/bubbles/ADR.py}

%===============================================================================

\section{Stabilized Stokes equations} \label{ssn_intro_stokes_2d_slip_stab}

\begin{figure*}
  \centering
  \begin{minipage}[b]{0.60\linewidth}
    \includegraphics[width=\linewidth]{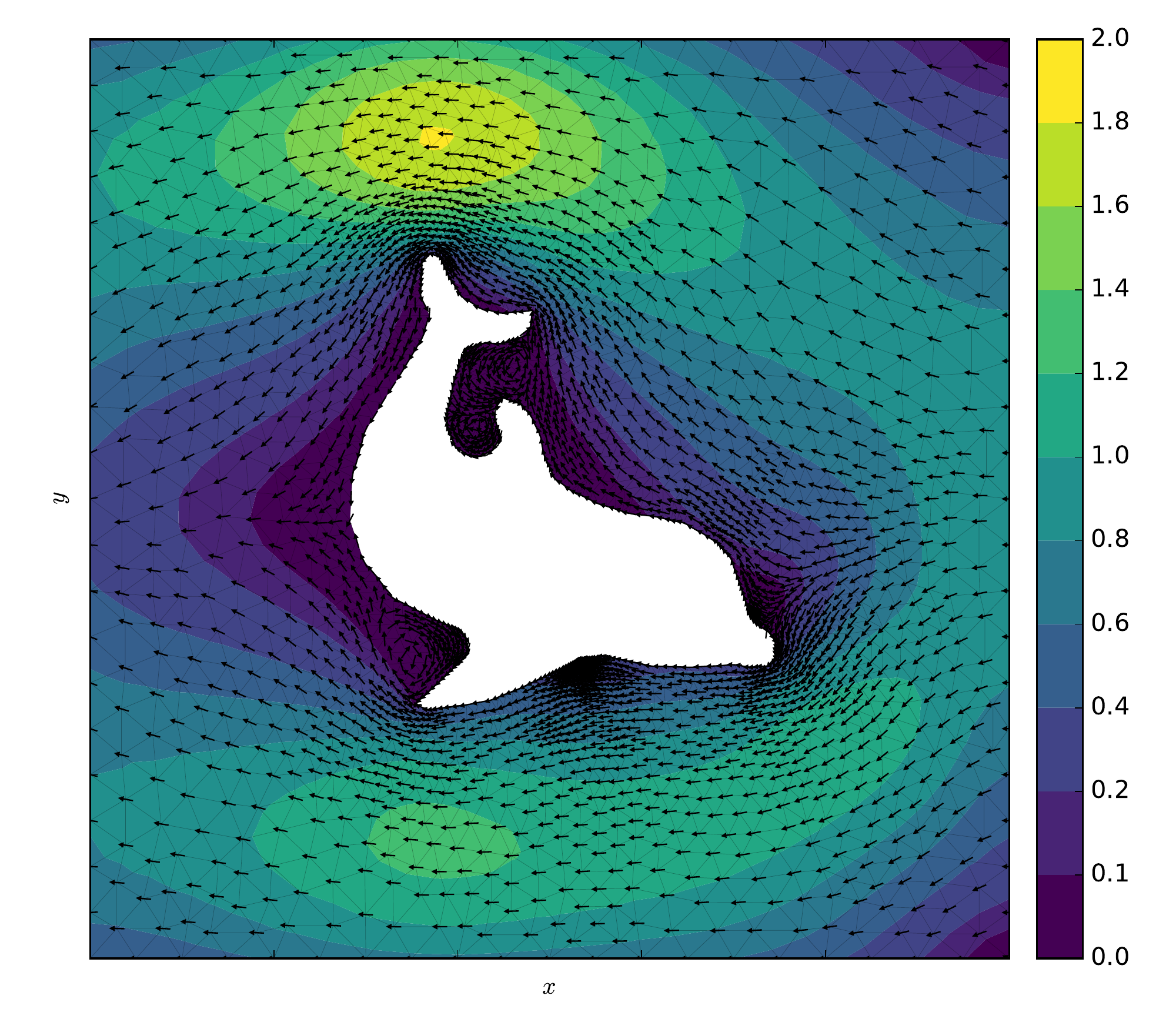}
  \end{minipage}
  \quad
  \begin{minipage}[b]{0.60\linewidth}
    \includegraphics[width=\linewidth]{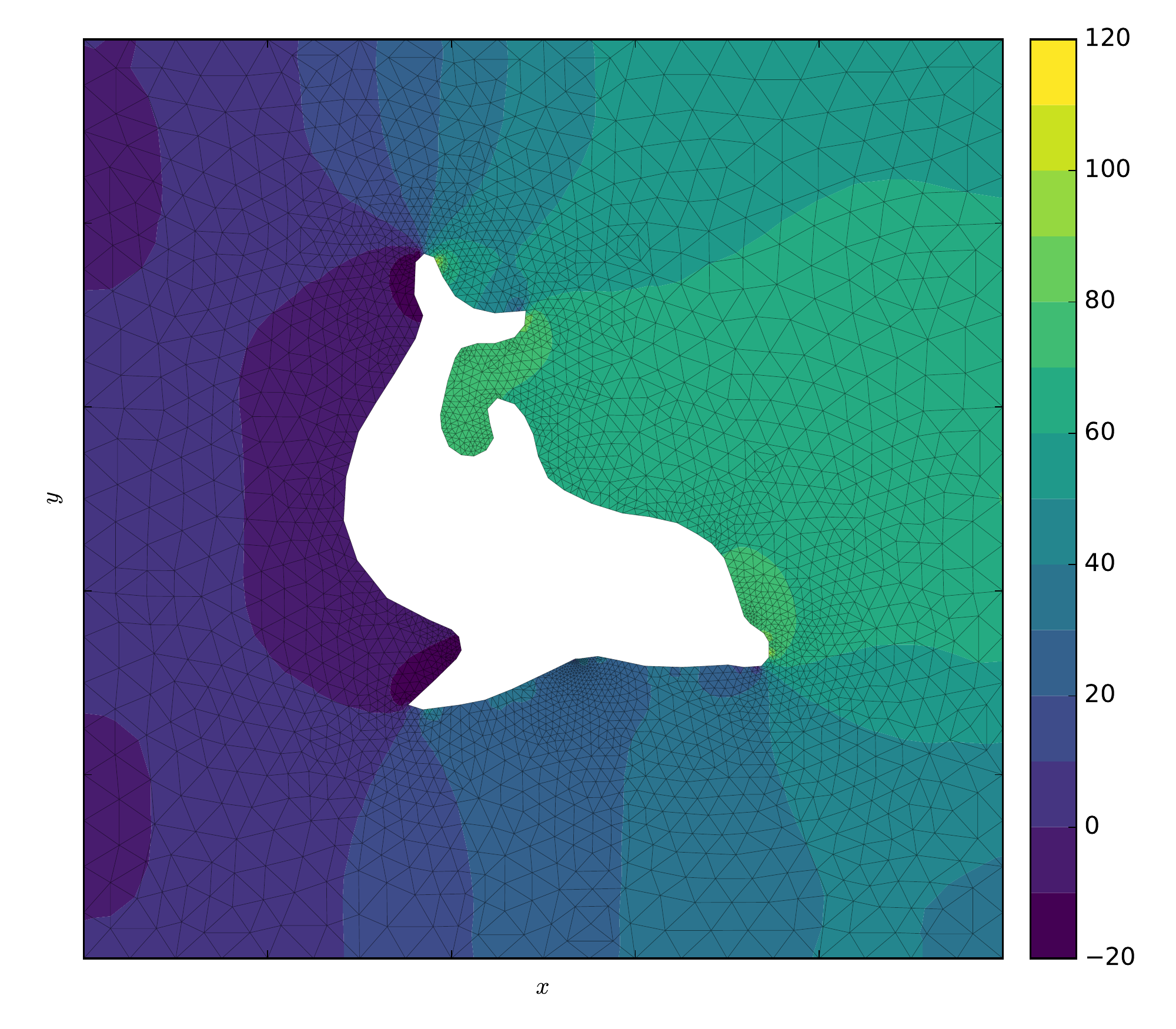}
  \end{minipage}
  \caption[GLS stabilized slip-friction Stokes example]{Galerkin/least-squares stabilized velocity field $\mathbf{u}$ (top) and pressure $p$ (bottom) with $\alpha = 0.1$, $\zeta = 0$, $\beta = 10$, and $\gamma = 100$, utilizing continuous linear Lagrange elements for both $\mathbf{u}$ and $p$ (referred to as P1 -- P1 approximation).}
  \label{intro_stokes_2d_nitsche_stab}
\end{figure*}

\index{Linear differential equations!2D}
\index{Stokes equations!Slip-friction}
\index{Nitsche method}
In this section we formulate a stabilized version of the slip-friction Stokes example described in \S \ref{ssn_intro_stokes_2d_slip} that circumvents inf-sup condition (\ref{inf_sup_condition}).  Recall that the Stokes equations for incompressible fluid over the domain $\Omega = [0,1] \times [0,1]$ are
\begin{align}
  \label{intro_stokes_stab_momentum}
  \Lu (\mathbf{u}, p) = -\nabla \cdot \sigma(\mathbf{u}, p) &= \mathbf{f} &&\text{ in } \Omega \\
  \label{intro_stokes_stab_mass}
  \nabla \cdot \mathbf{u} &= 0 &&\text{ in } \Omega,
\end{align}
where $\sigma(\mathbf{u}, p) = 2\eta\dot{\epsilon} - pI$ is the Cauchy-stress tensor.  The boundary conditions considered here are of type Dirichlet and traction (Neumann),
\begin{align}
  \label{intro_stokes_stab_gD_N_S_D_slip}
  \mathbf{u} \cdot \mathbf{n} &= g_D = 0 &&\text{ on } \Gamma_N, \Gamma_S, \Gamma_D \\
  \label{intro_stokes_stab_gN_N_S_D_fric}
  \left( \sigma \cdot \mathbf{n} \right)_{\Vert} &= \bm{g_N} = -\beta \mathbf{u} &&\text{ on } \Gamma_N, \Gamma_S, \Gamma_D \\
  \label{intro_stokes_stab_gD_E}
  \mathbf{u} &= \bm{g_D} = [-\sin(\pi y)\ 0]\T &&\text{ on } \Gamma_E \\
  \label{intro_stokes_stab_gN}
  \sigma \cdot \mathbf{n} &= \bm{g_N} = [g_{N_x}\ g_{N_y}]\T = \mathbf{0} &&\text{ on } \Gamma_W,
\end{align}
where $\Gamma_E$, $\Gamma_W$, $\Gamma_N$, and $\Gamma_S$ are the East, West, North, and South boundaries, $\Gamma_D$ is the dolphin boundary (Figure \ref{intro_stokes_2d_nitsche_stab}) and $\mathbf{n}$ is the outward-pointing normal vector to these faces.

It has been shown by \citet{hughes_V} that the stabilized Galerkin approximate solution $(\mathbf{u}, p)$ to Stokes system (\ref{intro_stokes_stab_momentum}, \ref{intro_stokes_stab_mass}) is given by solving the system
\begin{align}
  \label{stokes_stab_one}
  \begin{cases}
    \left( \bm{\Phi}, \Lu (\mathbf{u}, p) \right) = (\bm{\Phi}, \mathbf{f}), & \bm{\Phi} \in \mathbf{S_0^h}, \\
    \left( \xi, \nabla \cdot \mathbf{u} \right) + \left( \nabla \xi, \tau_{\text{S}} \left( \Lu (\mathbf{u}, p) - f \right) \right) = 0, & \xi \in M^h, 
  \end{cases}
\end{align}
where the coefficient $\alpha$ in $\tau_{\text{S}}$ given by (\ref{tau_stokes}) is constrained to obey $0 < \alpha < \alpha_0$, where the upper bound $\alpha_0$ depends on the basis used (the shape functions) for $\mathbf{u}$ and $\bm{\Phi}$.

A modification was made to (\ref{stokes_stab_one}) by \citet{hughes_VII} that allowed for discontinuous pressure spaces to be used.  The form for this model is given by
\begin{align}
  \label{stokes_stab_hughes_VII}
  B_{\text{VII}}(\mathbf{u},p,\bm{\Phi}, \xi) = L_{\text{VII}}(\bm{\Phi}, \xi),
\end{align}
where
\begin{align}
  \label{stokes_stab_hughes_VII_B}
  B_{\text{VII}}(\mathbf{u},p,\bm{\Phi}, \xi) = &+ \left( \bm{\Phi}, \Lu (\mathbf{u}, p) \right) - \left( \xi, \nabla \cdot \mathbf{u} \right) \notag \\
  &- \left( \Lu (\bm{\Phi}, \xi), \tau_{\text{S}_{\Omega}} \Lu (\mathbf{u}, p) \right) - \left( \llbracket \xi \rrbracket, \tau_{\text{S}_{\Gamma'}} \llbracket p \rrbracket \right)_{\Gamma'}, \\
  \label{stokes_stab_hughes_VII_L}
  L_{\text{VII}}(\bm{\Phi}, \xi) = &+ ( \bm{\Phi}, \mathbf{f} ) - \left( \Lu (\bm{\Phi}, \xi), \tau_{\text{S}_{\Omega}} f \right),
\end{align}
and 
\begin{align}
  \label{tau_stokes_VII}
  \tau_{\text{S}_{\Omega}} = \tau_{\text{S}} = \alpha \frac{h^2}{2\eta}, \hspace{10mm}
  \tau_{\text{S}_{\Gamma'}} = \zeta \frac{h^2}{2\eta}.
\end{align}
with constants $\alpha, \zeta \geq 0$ are dependent on the basis used for $\mathbf{u}$ and $\bm{\Phi}$.  Note that the notation $\llbracket \cdot \rrbracket$ denotes jump across interior edges, i.e. across the $+$ and $-$ sides of an edge,
\begin{align*}
  \llbracket \psi \rrbracket = \psi^+ - \psi^-.
\end{align*}
Therefore, if a continuous basis is used for $p$ and $\xi$, $\zeta$ can be taken to be zero due to the fact that the jump terms in (\ref{stokes_stab_hughes_VII}) will have no effect.

An independent analysis from \citet{hughes_VII} was presented by \citet{douglas} possessing a remarkably similar form, but where $\alpha$ and $\zeta$ were shown to be shape-independent.  This \emph{absolutely stabilized} model possesses the form
\begin{align}
  \label{stokes_stab_douglas}
  B_{\text{AS}}(\mathbf{u},p,\bm{\Phi}, \xi) = L_{\text{AS}}(\bm{\Phi}, \xi),
\end{align}
where
\begin{align}
  \label{stokes_stab_douglas_B}
  B_{\text{AS}}(\mathbf{u},p,\bm{\Phi}, \xi) = &+ \left( \bm{\Phi}, \Lu (\mathbf{u}, p) \right) + \left( \xi, \nabla \cdot \mathbf{u} \right) \notag \\
  &+ \left( \Lu (\bm{\Phi}, \xi), \tau_{\text{S}_{\Omega}} \Lu (\mathbf{u}, p) \right) + \left( \llbracket \xi \rrbracket, \tau_{\text{S}_{\Gamma'}} \llbracket p \rrbracket \right)_{\Gamma'}, \\
  \label{stokes_stab_douglas_L}
  L_{\text{AS}}(\bm{\Phi}, \xi) = &+ ( \bm{\Phi}, \mathbf{f} ) + \left( \Lu (\bm{\Phi}, \xi), \tau_{\text{S}_{\Omega}} f \right),
\end{align}
and utilizes the same coefficients $\tau_{\text{S}_{\Omega}}$ and $\tau_{\text{S}_{\Gamma'}}$ defined by (\ref{tau_stokes_VII}) with the difference that they possess only the single positivity constraint $\alpha, \zeta \geq 0$.  Note that the only difference between (\ref{stokes_stab_hughes_VII}) and (\ref{stokes_stab_douglas}) is the sign of the last two terms of bilinear forms (\ref{stokes_stab_hughes_VII_B}, \ref{stokes_stab_douglas_B}) and the last term of linear forms (\ref{stokes_stab_hughes_VII_L}, \ref{stokes_stab_douglas_L}).
  
The Galerkin/least-squares stabilized bilinear form for Dirichlet-traction-Stokes system (\ref{intro_stokes_stab_momentum}, \ref{intro_stokes_stab_mass}, \ref{intro_stokes_stab_gD_N_S_D_slip} -- \ref{intro_stokes_stab_gN}) are found identically to the formation of Nitsche variational form (\ref{intro_stokes_slip_var_form}); integration by parts of $(\bm{\Phi}, \Lu (\mathbf{u},p))$ and the addition of symmetric Nitsche terms, with the incorporation of the extra GLS terms of (\ref{stokes_stab_douglas}),

\begin{align}
  \label{intro_stokes_slip_stab_var_form}
  \mathcal{B}_{\Omega} + \mathcal{B}_{\Gamma_G} + \mathcal{B}_{\Gamma_G}^W + \mathcal{B}_{\Gamma_E} + \mathcal{B}_{\Gamma_E}^W + \mathcal{B}_{\Omega}^B + \mathcal{B}_{\Gamma'}^B = \mathcal{F} + \mathcal{F}^W + \mathcal{F}^B,
\end{align}
with individual terms
\begin{align*}
  \mathcal{B}_{\Omega} = &+ \int_{\Omega} \sigma(\mathbf{u},p) : \nabla \bm{\Phi}\ d\Omega + \int_{\Omega} \left( \nabla \cdot \mathbf{u} \right) \xi\ d\Omega \\
  \mathcal{B}_{\Gamma_G} = &- \int_{\Gamma_G} \left( \mathbf{n} \cdot \sigma(\mathbf{u},p) \cdot \mathbf{n} \right) \mathbf{n} \cdot \bm{\Phi}\ d\Gamma_G + \int_{\Gamma_G} \beta \mathbf{u} \cdot \bm{\Phi}\ d\Gamma_G \\
  \mathcal{B}_{\Gamma_G}^W = &- \int_{\Gamma_G} \left( \mathbf{n} \cdot \sigma(\bm{\Phi},\xi) \cdot \mathbf{n} \right) \mathbf{n} \cdot \mathbf{u} \ d\Gamma_G + \gamma \int_{\Gamma_G} \frac{1}{h} \left( \mathbf{u} \cdot \mathbf{n} \right) \left( \bm{\Phi} \cdot \mathbf{n} \right)\ d\Gamma_G \\
  \mathcal{B}_{\Gamma_E} = &- \int_{\Gamma_E} \sigma(\mathbf{u},p) \cdot \mathbf{n} \cdot \bm{\Phi}\ d\Gamma_E \\
  \mathcal{B}_{\Gamma_E}^W = &- \int_{\Gamma_E} \sigma(\bm{\Phi},\xi) \cdot \mathbf{n} \cdot \mathbf{u}\ d\Gamma_E + \gamma \int_{\Gamma_E} \frac{1}{h}  \left( \bm{\Phi} \cdot \mathbf{u} \right)\ d\Gamma_E \\
  \mathcal{B}_{\Omega}^B = &+ \frac{\alpha}{2} \int_{\Omega} \frac{h^2}{\eta} \Lu(\bm{\Phi},\xi) \Lu(\mathbf{u},p)\ d\Omega \\
  \mathcal{B}_{\Gamma'}^B = &+ \frac{\zeta}{2} \int_{\Gamma'} \frac{h^2}{\eta} \llbracket \xi \rrbracket \llbracket p \rrbracket\ d\Gamma'
\end{align*}
and
\begin{align*}
  \mathcal{F} = &+ \int_{\Omega} \mathbf{f} \cdot \bm{\Phi}\ d\Omega \\ 
  \mathcal{F}^W = &- \int_{\Gamma_G} \left( \mathbf{n} \cdot \sigma(\bm{\Phi},\xi) \cdot \mathbf{n} \right) g_D\ d\Gamma_G + \gamma \int_{\Gamma_G} \frac{1}{h} g_D \bm{\Phi} \cdot \mathbf{n}\ d\Gamma_G \\
  &- \int_{\Gamma_E} \sigma(\bm{\Phi},\xi) \cdot \mathbf{n} \cdot \bm{g_D}\ d\Gamma_E + \gamma \int_{\Gamma_E} \frac{1}{h} \left( \bm{\Phi} \cdot \bm{g_D} \right)\ d\Gamma_E \\
  \mathcal{F}^B = & +\frac{\alpha}{2} \int_{\Omega} \frac{h^2}{\eta} \Lu(\bm{\Phi},\xi) f\ d\Omega
\end{align*}
where $\Gamma_G = \Gamma_N \cup \Gamma_S \cup \Gamma_D$ is the entire slip-friction boundary, $h$ is the element diameter, and $\gamma > 0$ is an application-specific parameter normally derived by experimentation (see \S \ref{ssn_intro_stokes_2d_slip}).

The mixed variational formulation consistent with problem (\ref{intro_stokes_stab_momentum}, \ref{intro_stokes_stab_mass}, \ref{intro_stokes_stab_gD_N_S_D_slip} -- \ref{intro_stokes_stab_gN}) reads: find mixed approximation $\mathbf{u}, p \in \left( \mathbf{S_E^h} \subset \left( \mathcal{H}_E^1(\Omega) \right)^2 \right) \times \left( M^h \subset L^2(\Omega) \right)$ subject to (\ref{intro_stokes_slip_stab_var_form}) for all $\bm{\Phi}, \xi \in \left(\mathbf{S_0^h} \subset \left( \mathcal{H}_{E_0}^1(\Omega) \right)^2 \right) \times \left( M^h \subset L^2(\Omega) \right)$.

The velocity and pressure solutions to this problem using linear Lagrange elements for both $\mathbf{u}$ and $p$ are depicted in Figure \ref{intro_stokes_2d_nitsche_stab}, and were generated by Code Listing \ref{intro_stokes_2d_nitsche_stab_code}.

\pythonexternal[label=intro_stokes_2d_nitsche_stab_code, caption={FEniCS solution to the 2D-Nitsche-stabilized-Stokes-slip-friction problem of \S \ref{ssn_intro_stokes_2d_slip_stab}.}]{scripts/fenics_intro/2D_stokes_nitsche_stabilized.py}

%===============================================================================
%===============================================================================

\chapter{Nonlinear solution process} \label{ssn_nonlinear_solution_process}

All of the examples presented thus far have been linear equations.  For non-linear systems, the solution method described in \S \ref{ssn_galerkin_solve} no longer apply.  For these problems the $n$ unknown degrees of freedom of $u$ -- with vector representation $\mathbf{u}$ --  may be uniquely determined by solving for the down gradient direction of a quadratic model of the functional \index{Residual} \emph{residual}
\begin{align}
  \label{nonlin_resid}
  \mathscr{R}(u) = 0,
\end{align}
formed by moving all terms of a variational form to one side of the equation.  The following sections explain two ways of solving problems of this nature.

%===============================================================================

\section{Newton-Raphson method} \label{ssn_newton_raphson}

\begin{figure}
  \centering
    \def\svgwidth{\linewidth}
    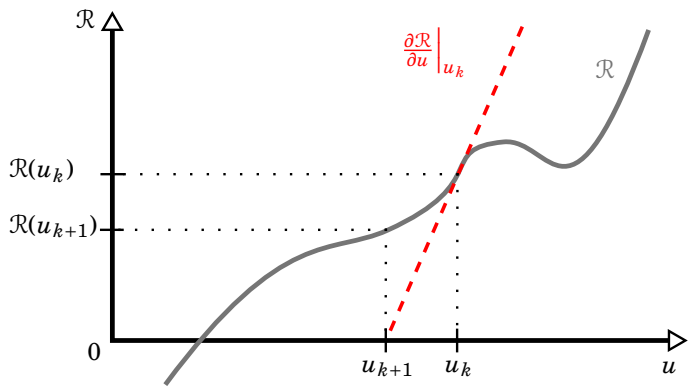
  \caption[Newton-Raphson diagram]{Illustration of the Newton-Raphson method for solving (\ref{nonlin_resid}) for $u$.}
  \label{nr_image}
\end{figure}

One way to solve system (\ref{nonlin_resid}) is the \index{Newton-Raphson method} \emph{Newton-Raphson} method \citep{nocedal}.  This method effectively linearizes the problem by first assuming an initial guess of the minimizer, $\mathbf{u}_k$, and the functional desired to be minimized, $\mathscr{R}(\mathbf{u}_k) = \mathscr{R}_k$, then uses the $\mathscr{R}$-intercept of the tangent line to this guess as a subsequent guess, $\mathbf{u}_{k+1}$.  This procedure is repeated until either the absolute value of $\mathscr{R}$ is below a desired \emph{absolute tolerance} or the relative change of $\mathscr{R}$ between guesses $\mathbf{u}_k$ and $\mathbf{u}_{k+1}$ is below a desired \emph{relative tolerance} (Figure \ref{nr_image}).

\subsection{Procedure}

First, as elaborated upon by \citet{nocedal}, the Taylor-series approximation of the residual $\mathscr{R}(\mathbf{u}_k)$ perturbed in a direction $\mathbf{p}_k$ provides a quadratic model $m_k(\mathbf{p}_k)$,
\begin{align}
  \label{newton_quadradic_model}
  \mathscr{R}(\mathbf{u}_k + \mathbf{p}_k) &\approx \mathscr{R}_k + \mathbf{p}_k\T \nabla \mathscr{R}_k + \frac{1}{2} \mathbf{p}_k\T \nabla^2 \mathscr{R}_k \mathbf{p}_k =: m_k(\mathbf{p}_k).
\end{align}
Because we wish to find the search direction $\mathbf{p}_k$ which minimizes $m_k(\mathbf{p}_k)$, we set the gradient of this function equal to zero and solve
\begin{align}
  0 &= \nabla m_k(\mathbf{p}_k) \notag \\
    &= \nabla \mathscr{R}_k + \nabla \left( \mathbf{p}_k\T \nabla \mathscr{R}_k \right) + \nabla \left( \frac{1}{2} \mathbf{p}_k\T \nabla^2 \mathscr{R}_k \mathbf{p}_k \right) \notag \\
    \label{newton_zero_grad_condition}
    &= \nabla \mathscr{R}_k + \nabla^2 \mathscr{R}_k \mathbf{p}_k,
\end{align}
giving us the \emph{Newton direction}
\begin{align}
  \label{newton_direction}
  \mathbf{p}_k &= -\left( \nabla^2 \mathscr{R}_k \right)^{-1} \nabla \mathscr{R}_k.
\end{align}

Quadratic model (\ref{newton_quadradic_model}) is simplified to a linear model if the second derivative is eliminated.  Integrating (\ref{newton_zero_grad_condition}) over $\mathbf{u}_k$ produces
\begin{align}
  \label{search_direction_equation}
  \nabla^2 \mathscr{R}_k \mathbf{p}_k &= - \nabla \mathscr{R}_k \notag \\
  \int_{\mathbf{u}_k} \nabla^2 \mathscr{R}_k(\mathbf{u}_k) \mathbf{p}_k\ d\mathbf{u}_k &= - \int_{\mathbf{u}_k} \nabla \mathscr{R}_k(\mathbf{u}_k)\ d\mathbf{u}_k \notag \\
  \nabla \mathscr{R}_k \mathbf{p}_k &= - \mathscr{R}_k,
\end{align}
a discrete system of equations representing a first-order differential equation for Newton direction $\mathbf{p}_k$.  To complete system (\ref{search_direction_equation}), one boundary condition may be specified.  For all Dirichlet boundaries, $u$ is known and can therefore not be improved.  Hence search direction $p = 0$ over essential boundaries.  Boundaries corresponding to natural conditions of $u$ require no specific treatment.

The iteration with \emph{step length} or \emph{relaxation parameter} $\alpha \in (0,1]$ is defined as
\begin{align}
  \label{newton_iteration}
  \mathbf{u}_{k+1} &= \mathbf{u}_k + \alpha \mathbf{p}_k.
\end{align}
Provided that the curve $\mathscr{R}$ is relatively smooth and that the Hessian matrix $\nabla^2 \mathscr{R}(\mathbf{u}_k)$ in (\ref{newton_direction}) is positive definite, (\ref{newton_iteration}) describes an iterative procedure for calculating the minimum of $\mathscr{R}$ and corresponding optimal value of $\mathbf{u}$.  See Algorithm \ref{newton_raphson_alg} and CSLVR source code \ref{home_rolled_method} for details.

\subsection{G\^{a}teaux derivatives} \label{ssn_gateaux}

Because residual (\ref{nonlin_resid}) is a functional, \index{G\^{a}teaux derivative} \index{Directional derivative} \emph{G\^{a}teaux derivatives} are used to calculate the directional derivatives in (\ref{search_direction_equation}).  To illustrate this derivative, consider the second-order boundary-value problem
\begin{align*}
  \nabla^2 u - u &= 0 &&\text{ in } \Omega \\
  \nabla u \cdot \mathbf{n} &= g &&\text{ on } \Gamma.
\end{align*}
where $\Omega$ is the interior domain with boundary $\Gamma$, and $\mathbf{n}$ is the outward-pointing normal vector.  The associated weak form is
\begin{align}
  \label{gateaux_example_form}
  \mathscr{R}(u) &= - \int_{\Omega} \nabla u \cdot \nabla \phi d\Omega - \int_{\Omega} \phi u d\Omega + \int_{\Gamma} \phi g d\Gamma = 0,
\end{align}

The G\^{a}teaux derivative or first variation of $\mathscr{R}$ with respect to $u$ in the direction $p$ -- with vector notation $\delta_{\mathbf{u}}^{\mathbf{p}} \mathscr{R} (\mathbf{u})$ -- is defined as
\begin{align*}
 \delta_{u}^{p}\mathscr{R}\left( u \right) = \frac{\delta}{\delta u} \mathscr{R}\left( u,p \right) = &\lim_{\epsilon \rightarrow 0} \frac{d}{d \epsilon} \mathscr{R} \left(u + \epsilon p \right) \\
  = &- \frac{d}{d \epsilon} \left[ \int_{\Omega} \nabla (u + \epsilon p) \cdot \nabla \phi d\Omega \right]_{\epsilon = 0} \\
  &- \frac{d}{d \epsilon} \left[ \int_{\Omega} \phi (u + \epsilon p) d\Omega \right]_{\epsilon = 0} \\
  &+ \frac{d}{d \epsilon} \left[ \int_{\Gamma} \phi g d\Gamma \right]_{\epsilon = 0} \\
  = &- \int_{\Omega} \nabla p \cdot \nabla \phi d\Omega - \int_{\Omega} \phi p d\Omega.
\end{align*}
It is important to recognize that if $p$ is a function with known values, i.e.~data interpolated onto a finite-element mesh, the finite-element assembly -- described in \S \ref{ssn_global_galerkin_assembly} -- will result in a vector of length $n$.  However, if $p$ is an unknown quantity and thus a member of the trial space associated with the finite-element approximation of $\mathbf{u}$, as is the case with the Newton-Raphson method here, the assembly process will result in a matrix with properties identical to the associated stiffness matrix of (\ref{gateaux_example_form}).  In order to clearly differentiate between these circumstances, the G\^{a}teaux derivative operator notation $\delta_{\mathbf{u}}^{\mathbf{p}}$ is used when $\mathbf{p}$ is a member of the trial space, and $\delta_{\mathbf{u}}$ otherwise.

These derivatives may be calculated with FEniCS using the process of \index{Automatic differentiation} \emph{automatic differentiation} \citep{nocedal}, as illustrated by the nonlinear problem example in Code Listing \ref{river_cross_code}.

\begin{algorithm}
  \normalsize
  \caption[Newton-Raphson]{ - Newton-Raphson method}
  \label{newton_raphson_alg}
  \begin{algorithmic}[1]
    \State \textbf{INPUTS}:
    \State \ \ \ $\phantom{a_{tol}}\mathllap{\mathscr{R}}$ - residual variational form
    \State \ \ \ $\phantom{a_{tol}}\mathllap{\mathbf{u}}$ - initial state parameter vector
    \State \ \ \ $\phantom{a_{tol}}\mathllap{\hat{\mathbf{p}}}$ - trial function in same space as $\mathbf{u}$
    \State \ \ \ $\phantom{a_{tol}}\mathllap{\alpha}$ - relaxation parameter
    \State \ \ \ $a_{tol}$ - absolute tolerance to stop iterating
    \State \ \ \ $r_{tol}$ - relative tolerance to stop iterating
    \State \ \ \ $\phantom{a_{tol}}\mathllap{n_{max}}$ - maximum iterations
    \State \textbf{OUTPUT}:
    \State \ \ \ $\phantom{a_{tol}}\mathllap{\mathbf{u}^*}$ - optimized state parameter vector
    \\
    \hrulefill
    \Function{NR}{$\mathscr{R},\ \mathbf{u},\ \hat{\mathbf{p}},\ \alpha,\ a_{tol},\ r_{tol},\ n_{max}$}
      \State $\phantom{\bm{\vartheta}}\mathllap{r} := \infty$
      \State $\phantom{\bm{\vartheta}}\mathllap{a} := \infty$
      \While{$(a > a_{tol}$ \textbf{or} $r > r_{tol})$ \textbf{and} $n < n_{max}$}
        \State $\phantom{\mathbf{p}}\mathllap{J} :=$ \textbf{assemble} $\delta_{\mathbf{u}}^{\hat{\mathbf{p}}} \mathscr{R}(\mathbf{u})$
        \State $\phantom{\mathbf{p}}\mathllap{\mathbf{b}} :=$ \textbf{assemble} $\mathscr{R}(\mathbf{u})$
        \State $\phantom{\mathbf{p}}\mathllap{\mathbf{p}} \leftarrow$ \textbf{solve} $J \mathbf{p} = -\mathbf{b}$
        \State $\phantom{\mathbf{p}}\mathllap{\mathbf{u}} := \mathbf{u} + \alpha \mathbf{p}$
        \State $\phantom{\mathbf{p}}\mathllap{a} := \Vert \mathbf{b} \Vert_{2}$
        \If{$n = 0$}
          \State $a_0 := a$
        \EndIf
        \State $\phantom{\mathbf{p}}\mathllap{r} := a / a_0$
        \State $\phantom{\mathbf{p}}\mathllap{n} := n + 1$
      \EndWhile
    \State \Return $\mathbf{u}^* := \mathbf{u}$ 
    \EndFunction
  \end{algorithmic}
\end{algorithm}

\pythonexternal[caption={Newton-Raphson method as implemented in CSLVR's \texttt{model} class}, label = home_rolled_method, firstline=2073, lastline=2145]{cslvr_src/model.py}

%===============================================================================

\section{Nonlinear problem example}

\index{Non-linear differential equations!1D}
Suppose we would like to minimize the time for a boat to cross a river.  The time for this boat to cross, when steered directly perpendicular to the river's parallel banks is given by
\begin{align*}
  T(y) &= \int_0^T dt = \int_0^S \frac{dt}{ds} ds = \int_0^S \frac{1}{\Vert \mathbf{u}\Vert} ds \\
  T(y) &= \int_0^{\ell} \frac{\sqrt{1 + (y')^2}}{\Vert \mathbf{u} \Vert} dx \\
  T(y) &= \int_0^{\ell} \frac{\sqrt{1 + (y')^2}}{\sqrt{v^2 + u^2}} dx = \int_0^{\ell} L(x,y') dx,
\end{align*}
where $S$ is the length of the boat's path; $\mathbf{u}$ is the velocity of the boat with components river current speed $v(x)$ and boat speed $u$ in the $x$ direction as a result of its motor; and $\ell$ is the width of the river.  For \emph{Lagrangian} $L$, the first variation of $T$ in the direction of the test function $\phi \in \testspace$ (see test space (\ref{test_space})) is
\begin{align*}
  \delta T(y,\phi) &= \frac{d}{d\epsilon} \int_0^{\ell} L(x, y + \epsilon \phi, y' + \epsilon \phi') dx \Bigg|_{\epsilon = 0} \\
                   &= \int_0^{\ell} \left( L_y \phi + L_{y'} \phi' \right) dx \\
                   &= \int_0^{\ell} \left( L_y - \frac{d}{dx} L_{y'} \right) \phi dx + L_{y'} \phi \Big|_{x=0}^{x=\ell} \\
                   &= \int_0^{\ell} \left( L_y - \left[ L_{y'x} + L_{y'y} y' + L_{y'y'} y'' \right] \right) \phi dx + L_{y'} \phi \Big|_{x=0}^{x=\ell}.
\end{align*}
Evaluating each individual term,
\begin{align*}
  L_y          &= 0 \\
  L_{y'}(y')   &= \frac{y' \left(1 + (y')^2\right)^{-1/2}}{\sqrt{v^2 + u^2}} \\
  L_{y'x}(y')  &= \frac{\left(2(y')^2 + 1\right) y''}{\sqrt{u^2 + v^2} \sqrt{1 + (y')^2}} \\
  L_{y'y}      &= 0 \\
  L_{y'y'}(y') &= \frac{1}{\sqrt{u^2 + v^2} \sqrt{1 + (y')^2}} \left( 1 - \frac{(y')^2}{\left(1 + (y')^2\right)^3} \right),
\end{align*}
and so the first variation of $T$ is reduced to
\begin{align}
  \label{boat_T_fv}
  \delta T(y,\phi) &= - \int_0^{\ell} \left( L_{y'x} + L_{y'y'} y'' \right) \phi dx + L_{y'} \phi \Big|_{x=0}^{x=\ell}.
\end{align}
In order to find the minimal time for the boat to cross, this first variation of $T$ must to be equal to zero.  First, defining
\begin{align*}
  M &= \frac{L_{y'x}}{y''},
\end{align*}
relation (\ref{boat_T_fv}) can be rewritten followed by integrating by parts of the the second-derivative term once again,
\begin{align*}
  0 &= - \int_0^{\ell} \left( M y'' + L_{y'y'} y'' \right) \phi dx + L_{y'} \phi \Big|_{x=0}^{x=\ell} \\
    &= - \int_0^{\ell} \left( M + L_{y'y'} \right) y'' \phi dx + L_{y'} \phi \Big|_{x=0}^{x=\ell} \\
    &= \int_0^{\ell} \left( M + L_{y'y'} \right) y' \phi' dx - \left[ M + L_{y'y'} \right] y' \phi \Big|_{x=0}^{x=\ell} + L_{y'} \phi \Big|_{x=0}^{x=\ell}.
\end{align*}
If the left essential boundary condition is set to $y(0) = 0$, and the right natural boundary condition is set to be equal to the trajectory of the boat at the opposite bank given by
$$y'(\ell) = g = \frac{v(\ell)}{u},$$
the final variational problem therefore consists of finding $y \in \trialspace$ (see trial space (\ref{trial_space})) such that
\footnotesize
\begin{align*}
  0 &= \int_0^{\ell} \left( M(y') + L_{y'y'}(y') \right) y' \phi' dx - \left[ M(g) + L_{y'y'}(g) \right] g \phi \Big|_{x=\ell} + L_{y'}(g) \phi \Big|_{x=\ell}.
\end{align*}
\normalsize
for all $\phi \in \testspace$.

\begin{figure}
  \centering
    \includegraphics[width=\linewidth]{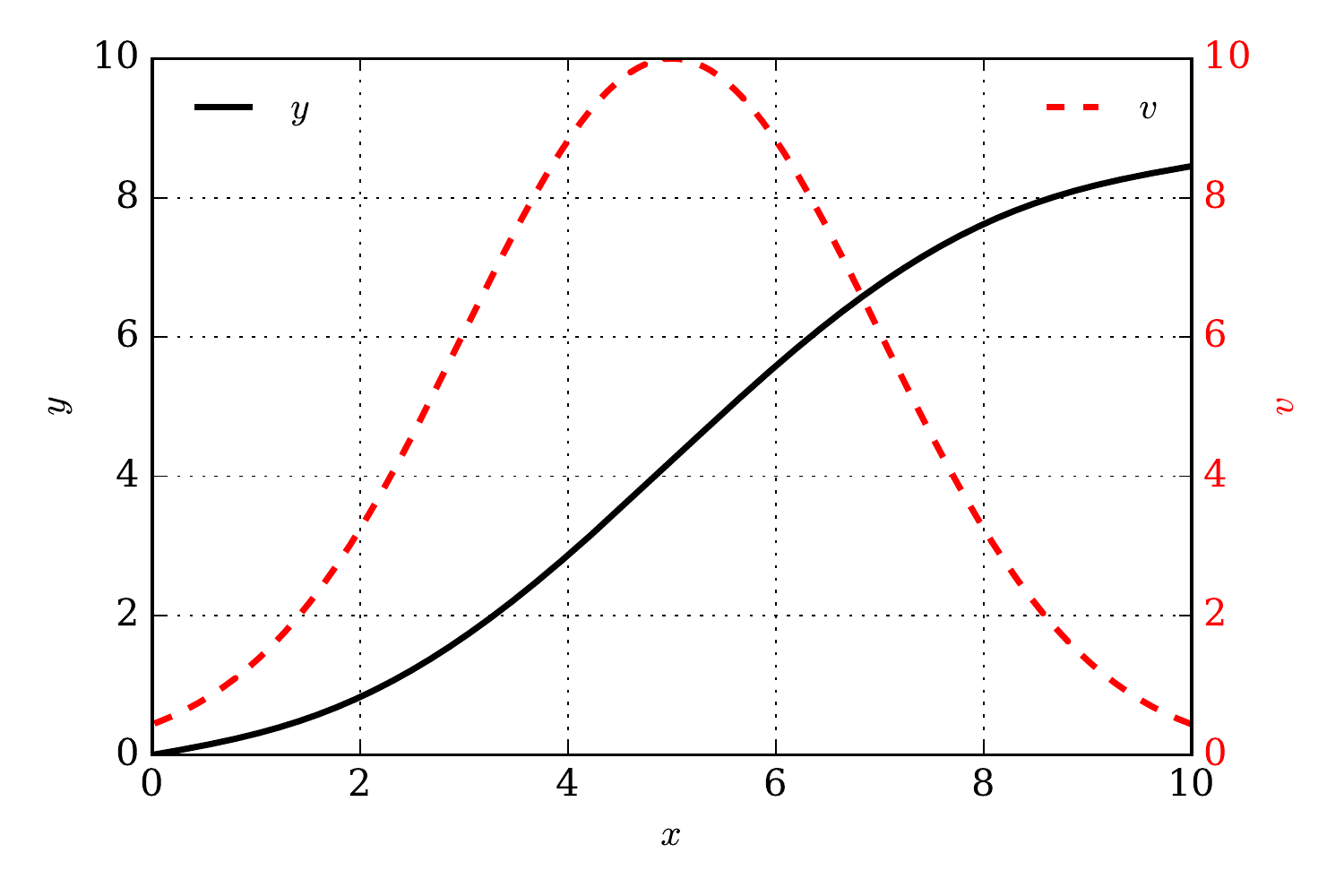}
  \caption[Nonlinear problem example]{Path taken for a boat to cross a river, $y(x)$ (solid black), with a motor speed in the $x$-direction of $u = 1$, and river velocity in the $y$-direction $v(x)$ (dashed red).}
  \label{river_cross_image}
\end{figure}

The weak solution to this equation using linear-Lagrange shape functions (\ref{linear_lagrange_functions}) using the built-in FEniCS Newton solver is shown in Figure \ref{river_cross_image} and generated from Code Listing \ref{river_cross_code}.

\pythonexternal[label=river_cross_code, caption={FEniCS source code for the river crossing example.}]{scripts/fenics_intro/river_cross.py}

%===============================================================================
%===============================================================================

\section{Quasi-Newton solution process}

\index{Quasi-Newton methods}
In some situations, full quadratic model (\ref{newton_quadradic_model}) may be preferred over linear model (\ref{search_direction_equation}) for solving a non-linear system.  When the inverse of Hessian matrix $\nabla^2 \mathscr{R}_k$ is difficult to calculate by hand, it may be approximated by using curvature information at a current guess $\mathbf{u}_k$.  Algorithms utilizing the Hessian approximation are referred to as \emph{quasi-Newton} methods.

\index{BFGS Algorithm}
As described by \citet{nocedal}, the most modern and efficient of such Hessian approximation techniques is that proposed by Broyden, Fletcher, Goldfarb, and Shanno, aptly referred to as the BFGS method (Algorithm \ref{BFGS_alg}).  This method uses the iteration
\begin{align}
  \label{qn_iteration}
  \mathbf{u}_{k+1} = \mathbf{u}_k + \alpha_k \mathbf{p}_k
\end{align}
and a quadratic model similar to (\ref{newton_quadradic_model}) with the addition of a subsequent iteration $k+1$ quadratic model:
\begin{align}
  \label{qn_k_quadradic_model}
  m_{k}(\mathbf{p}) &= \mathscr{R}_{k+1} + \mathbf{p}\T \nabla \mathscr{R}_{k} + \frac{1}{2} \mathbf{p}\T B_{k} \mathbf{p} \\
  \label{qn_k+1_quadradic_model}
  m_{k+1}(\mathbf{p}) &= \mathscr{R}_{k+1} + \mathbf{p}\T \nabla \mathscr{R}_{k+1} + \frac{1}{2} \mathbf{p}\T B_{k+1} \mathbf{p}.
\end{align}
The minimizer of (\ref{qn_k_quadradic_model}), search direction $\mathbf{p}_k$, is found identically to Newton direction (\ref{newton_direction}):
\begin{align*}
  \mathbf{p}_k &= - B_k^{-1} \nabla \mathscr{R}_k,
\end{align*}
where the approximate Hessian matrix $B_k \approx \nabla_{\mathbf{u} \mathbf{u}}^2 \mathscr{R}_k$ must be symmetric and positive definite.  Furthermore, it is required that $\nabla m_{k+1} = \nabla \mathscr{R}_j$ for $j = k, k+1$, the last two iterates.  For the last iterate $k+1$, $\mathscr{R}_{k+1}$ is evaluated at $\mathbf{u}_{k+1}$ and therefore the gradient of $m_{k+1}$ at $\mathbf{p} = \mathbf{0}$ is evaluated,
\begin{align*}
  \nabla m_{k+1}(\mathbf{0}) &= \nabla \mathscr{R}_{k+1},
\end{align*}
implying that the second condition is satisfied automatically.  Using the same reasoning and iteration (\ref{qn_iteration}), to evaluate $m_{k+1}$ at $\mathbf{u}_{k}$ the gradient of $m_{k+1}$ at $\mathbf{p} = -\alpha_k \mathbf{p}_k$ is evaluated,
\begin{align*}
  \nabla m_{k+1}(-\alpha_k \mathbf{p}_{k}) &= \nabla \mathscr{R}_{k+1} - \alpha_k \mathbf{p}_{k} B_{k+1} = \nabla \mathscr{R}_k,
\end{align*}
thus requiring that
\begin{align*}
  \alpha_k \mathbf{p}_k B_{k+1} &= \nabla \mathscr{R}_{k+1} - \nabla \mathscr{R}_k.
\end{align*}
Using iteration (\ref{qn_iteration}), the \index{Secant equation} \emph{secant equation} is
\begin{align}
  \label{secant_equation}
  B_{k+1} \mathbf{s}_k &= \mathbf{y}_k,
\end{align}
where
\begin{align*}
  \mathbf{s}_k = \mathbf{u}_{k+1} - \mathbf{u}_k \hspace{10mm} \mathbf{y}_k &= \nabla \mathscr{R}_{k+1} - \nabla \mathscr{R}_k.
\end{align*}
Equivalently, the \emph{inverse secant equation} is
\begin{align}
  \label{inverse_secant_equation}
  H_{k+1} \mathbf{y}_k &= \mathbf{s}_k,
\end{align}
where $H_{k+1} = B_{k+1}^{-1}$.

As described by \citet{nocedal}, inverse secant equation (\ref{inverse_secant_equation}) will be satisfied if the \emph{curvature condition}
\begin{align*}
  \mathbf{s}_k\T \mathbf{y}_k > 0
\end{align*} 
holds.  This is explicitly enforced by choosing step length $\alpha_k$ in (\ref{qn_iteration}) such that the \index{Armijo condition} \emph{Armijo condition}
\begin{align}
  \label{armijo_condition}
  \mathscr{R}(\mathbf{u}_k + \alpha_k \mathbf{p}_k ) \leq \mathscr{R}_k + c_1 \alpha_k \mathbf{p}\T \nabla \mathscr{R}_k
\end{align}
and the curvature condition 
\begin{align}
  \label{curvature_condition}
  \mathbf{p}\T \nabla \mathscr{R}(\mathbf{u}_k + \alpha_k \mathbf{p}_k ) \geq c_2 \mathbf{p}\T \nabla \mathscr{R}_k,
\end{align}
collectively referred to as the \emph{Wolfe conditions} \emph{Wolfe conditions}, hold for some pair of constants $c_1,c_2 \in (0,1)$.  One way of enforcing (\ref{armijo_condition}) and (\ref{curvature_condition}) is described by the \index{Backtracking line search} \emph{backtracking line search}, Algorithm \ref{qn_bls_alg} \citep{nocedal}.

In order to derive a unique $H_{k+1}$, the additional constraint that $H_{k+1}$ be close to the current matrix $H_k$ is imposed.  Thus $H_{k+1}$ is the solution to the problem
\begin{align}
  \label{qn_H_problem}
  \min_{H} \Vert H - H_k \Vert
\end{align}
\begin{align}
  \label{qn_H_problem_constraints}
  \text{subject to} \hspace{5mm} H = H\T, \hspace{5mm} H \mathbf{y}_k = \mathbf{s}_k.
\end{align}
Using the weighted Frobenius norm
\begin{align*}
  \Vert A \Vert_W = \left\Vert W^{\nicefrac{1}{2}} A W^{\nicefrac{1}{2}} \right\Vert_F
\end{align*}
with average Hessian inverse
\begin{align*}
  W = \left[ \int_0^1 \nabla^2 \mathscr{R}(\mathbf{u} + \tau \alpha_k \mathbf{p}_k ) d\tau \right]^{-1}
\end{align*}
satisfying $W\mathbf{y}_k = \mathbf{s}_k$ in (\ref{qn_H_problem}) gives the unique solution to (\ref{qn_H_problem}) and (\ref{qn_H_problem_constraints})
\begin{align*}
  H_{k+1} &= \left( I - \mathbf{\rho}_k \mathbf{s}_k \mathbf{y}_k\T \right) H_k \left( I - \rho_k \mathbf{y}_k \mathbf{s}_k\T \right) + \rho_k \mathbf{s}_k \mathbf{s}_k\T, \hspace{5mm} \rho = \left( \mathbf{y}\T \mathbf{s} \right)^{-1}.
\end{align*}

The iterative process for this method is described by Algorithm \ref{BFGS_alg}.

\begin{algorithm}
  \normalsize
  \caption[BFGS]{ - BFGS quasi-Newton method}
  \label{BFGS_alg}
  \begin{algorithmic}[1]
    \State \textbf{INPUTS}:
    \State \ \ \ $\phantom{a_{tol}}\mathllap{\mathbf{u}}$ - initial state parameter vector
    \State \ \ \ $\phantom{a_{tol}}\mathllap{H}$ - inverse Hessian approximation
    \State \ \ \ $a_{tol}$ - absolute tolerance to stop iterating
    \State \ \ \ $r_{tol}$ - relative tolerance to stop iterating
    \State \textbf{OUTPUT}:
    \State \ \ \ $\phantom{a_{tol}}\mathllap{\mathbf{u}^*}$ - optimized state parameter vector
    \\
    \hrulefill
    \Function{BFGS}{$\mathbf{u},\ H,\ a_{tol}$}
      \State $\phantom{\bm{\vartheta}}\mathllap{r} := \infty$
      \State $\phantom{\bm{\vartheta}}\mathllap{a} := \infty$
      \While{$a > a_{tol}$ \textbf{or} $r > r_{tol}$}
        \State $\phantom{\mathbf{p}}\mathllap{\mathbf{g}} :=$ \textbf{assemble} $\delta_{\mathbf{u}} \mathscr{R}(\mathbf{u})$
        \State $\mathbf{p} := - H \mathbf{g}$
        \State $\phantom{\mathbf{p}}\mathllap{\alpha} :=$ BLS$(\mathbf{p}, \mathbf{g}, \mathbf{u})$
        \State $\phantom{\mathbf{p}}\mathllap{\mathbf{u}_k} := \mathbf{u} + \alpha \mathbf{p}$
        \State $\phantom{\mathbf{p}}\mathllap{\mathbf{g}_k} :=$ \textbf{assemble} $\delta_{\mathbf{u}} \mathscr{R}(\mathbf{u}_k)$
        \State $\phantom{\mathbf{p}}\mathllap{\mathbf{s}} := \mathbf{u}_k - \mathbf{u}$
        \State $\phantom{\mathbf{p}}\mathllap{\mathbf{y}} := \mathbf{g}_k - \mathbf{g}$
        \State $\phantom{\mathbf{p}}\mathllap{\rho} := \left( \mathbf{y}\T \mathbf{s} \right)^{-1}$
        \State $\phantom{\mathbf{p}}\mathllap{H} := \left( I - \mathbf{\rho} \mathbf{s} \mathbf{y}\T \right) H \left( I - \rho \mathbf{y} \mathbf{s}\T \right) + \rho \mathbf{s} \mathbf{s}\T$
        %\State $\phantom{\mathbf{p}}\mathllap{B} := B - \frac{B \mathbf{s} \mathbf{s}\T B}{\mathbf{s}\T B \mathbf{s}} + \frac{\mathbf{y} \mathbf{y}\T}{\mathbf{y}\T \mathbf{s}}$
        \State $\phantom{\mathbf{p}}\mathllap{a} := \Vert \mathbf{g}_k \Vert_{\infty}$
        \State $\phantom{\mathbf{p}}\mathllap{r} := \Vert \mathbf{u} - \mathbf{u}_k \Vert_{\infty}$
        \State $\phantom{\mathbf{p}}\mathllap{\mathbf{u}} := \mathbf{u}_k$
      \EndWhile
    \State \Return $\mathbf{u}^* := \mathbf{u}$ 
    \EndFunction
  \end{algorithmic}
\end{algorithm}

\begin{algorithm}
  \normalsize
  \caption[Backtracking line-search]{ - Backtracking line search}
  \label{qn_bls_alg}
  \begin{algorithmic}[1] 
    \State \textbf{INPUTS}:
    \State \ \ \ $\mathbf{p}$ - search direction
    \State \ \ \ $\mathbf{g}$ - vector assembly of $\delta_{\mathbf{u}} \mathscr{R}$
    \State \ \ \ $\phantom{\mathbf{p}}\mathllap{\mathbf{u}}$ - state parameter vector
    \State \textbf{OUTPUT}: 
    \State \ \ \ $\phantom{\mathbf{p}}\mathllap{\alpha}$ - Step length.
    \\
    \hrulefill
    \Function{BLS}{$\mathbf{p},\ \mathbf{g},\ \mathbf{u}$}
      \State $\phantom{\ell_k}\mathllap{\alpha} := 1, \hspace{3mm} c_1 := 10^{-4}, \hspace{3mm} c_2 := \nicefrac{9}{10}$
      \State $\phantom{\ell_k}\mathllap{\ell_0} := $ \textbf{assemble} $\mathscr{R}\left( \mathbf{u} \right)$
      \State $\ell_k := $ \textbf{assemble} $\mathscr{R}\left( \mathbf{u} + \alpha \mathbf{p} \right)$
      \While{$\ell_k \geq \ell_0 + c_1 \alpha \mathbf{g}\T \mathbf{p}$}
        \State $\phantom{\ell_k}\mathllap{\alpha} := c_2 \alpha$
        \State $\ell_k := $ \textbf{assemble} $\mathscr{R}\left( \mathbf{u} + \alpha \mathbf{p} \right)$
      \EndWhile
      \State \Return $\alpha$
    \EndFunction
  \end{algorithmic}
\end{algorithm}

%===============================================================================
%===============================================================================

\chapter{Optimization with constraints} \label{ssn_optimization_with_constraints}

When the solution space for a problem is restricted by equality or inequality constraints, new theory is required to derive solutions.  These problems can be stated in the form \citep{nocedal}
\begin{align}
  \label{canonical_opt}
  \min_{\vartheta \in \R^n}\ \mathscr{F}(\vartheta) \hspace{10mm} \text{subject to  }
  \begin{cases}
    \mathscr{R}(\vartheta) = 0, \\
    c(x) \geq 0,
  \end{cases}
\end{align}
with parameter vector $\vartheta = [u\ x]^\intercal$.  The real-valued functions $\mathscr{F}(\vartheta)$, $\mathscr{R}(\vartheta)$, and $c(x)$ are all smooth and defined on a subset of $\mathbb{R}^n$ created from a finite-element discretization in one, two, or three dimensions.  The function to be minimized, $\mathscr{F}(\vartheta)$, is referred to as the \index{Constrained optimization!Objective function} \emph{objective} function with dependent \index{Constrained optimization!State parameter} \emph{state} parameter $u \in \R^n$ and dependent \emph{control} parameter $x \in \R^n$.

One method of solving problems of form (\ref{canonical_opt}) is through the use of a bounded version of Algorithm \ref{BFGS_alg} referred to as L\_BFGS\_B \citep{l_bfgs_b}; however, a more modern and efficient class of constrained optimization algorithms known as \index{Constrained optimization!Interior point methods} \emph{interior point} (IP) methods have been shown to perform quite well for problems of this type \citep{nocedal}.  Because CSLVR utilizes an IP method implemented by the FEniCS optimization software Dolfin-Adjoint \citep{farrell}, this is the method described here.

For the applications presented in Part II of this manuscript, objective $\mathscr{F}(\vartheta)$ and constraint $\mathscr{R}(\vartheta)$ are \index{Functionals} \emph{functionals}: the mapping from the space of functions to the space of real numbers, and so the following theory will be presented in this context. For examples of functionals, examine Chapters \ref{ssn_subgrid_scale_effects} and \ref{ssn_nonlinear_solution_process}.

%===============================================================================

\section{The control method}

\index{Control theory}
A stationary point for $\vartheta$-optimization problem (\ref{canonical_opt}) is defined as one where an arbitrary change $\delta \mathscr{F}$ of objective $\mathscr{F}$ caused by perturbations $\delta u$ or $\delta x$ in state and control parameter, respectively, lead to an increase in $\mathscr{F}$ \citep{bryson}.  Thus it is necessary that 
\begin{align}
  \label{objective_perturbations}
  \frac{\delta \mathscr{F}}{\delta u} = 0, \hspace{5mm} \text{and} \hspace{5mm}
  \frac{\delta \mathscr{F}}{\delta x} = 0.
\end{align}

Using the chain rule of variations, the perturbations of $\mathscr{F}$ and $\mathscr{R}$ in an arbitrary direction $\phi$ are
\begin{align}
  \label{delta_F}
  \frac{\delta \mathscr{F}}{\delta \phi} &= \frac{\delta \mathscr{F}}{\delta u} \frac{\delta u}{\delta \phi} + \frac{\delta \mathscr{F}}{\delta x} \frac{\delta x}{\delta \phi}, \\
  \label{delta_R}
  \frac{\delta \mathscr{R}}{\delta \phi} &= \frac{\delta \mathscr{R}}{\delta u} \frac{\delta u}{\delta \phi} + \frac{\delta \mathscr{R}}{\delta x} \frac{\delta x}{\delta \phi}.
\end{align}
Because it is desired that $\delta_{\phi} \mathscr{R} = 0$, and with non-singular $\delta_{u} \mathscr{R}$, we can solve for $\delta_{\phi} u$ in (\ref{delta_R}),
\begin{align}
  \label{delta_u}
  \frac{\delta u}{\delta \phi} = - \frac{\delta u}{\delta \mathscr{R}} \frac{\delta \mathscr{R}}{\delta x} \frac{\delta x}{\delta \phi}.
\end{align}
We then insert (\ref{delta_u}) into (\ref{delta_F}),
\begin{align}
  \frac{\delta \mathscr{F}}{\delta \phi} &= - \frac{\delta \mathscr{F}}{\delta u} \left( \frac{\delta u}{\delta \mathscr{R}} \frac{\delta \mathscr{R}}{\delta x} \frac{\delta x}{\delta \phi} \right) + \frac{\delta \mathscr{F}}{\delta x} \frac{\delta x}{\delta \phi} \notag \\
  &= \left(\frac{\delta \mathscr{F}}{\delta x}  - \frac{\delta \mathscr{F}}{\delta u} \frac{\delta u}{\delta \mathscr{R}} \frac{\delta \mathscr{R}}{\delta x} \right) \frac{\delta x}{\delta \phi},
\end{align}
and thus because we require $\delta_{\phi} \mathscr{F} = 0$ for any non-zero $\delta_{\phi} x$,
\begin{align}
  \frac{\delta \mathscr{F}}{\delta x}  - \frac{\delta \mathscr{F}}{\delta u} \frac{\delta u}{\delta \mathscr{R}} \frac{\delta \mathscr{R}}{\delta x} &=0 \notag \\
  \label{condition_two}
  \frac{\delta \mathscr{F}}{\delta x} + \lambda \frac{\delta \mathscr{R}}{\delta x} &=0,
\end{align}
where \index{Lagrange multiplier} \emph{Lagrange multiplier} or \index{Constrained optimization!Adjoint variable} \emph{adjoint variable} $\lambda$ \index{Adjoint method|seealso{Control theory}} adjoins constraint functional $\mathscr{R}$ to objective functional $\mathscr{F}$, and is given by
\begin{align}
  -\lambda &= \frac{\delta \mathscr{F}}{\delta u} \frac{\delta u}{\delta \mathscr{R}} = \frac{\delta \mathscr{F}}{\delta \mathscr{R}} \Bigg|_{u}.
\end{align}
Therefore, $\lambda$ is the direction of decent of objective $\mathscr{F}$ with respect to constraint $\mathscr{R}$ at a given energy state $u$.

It is now convenient to define the \index{Constrained optimization!Lagrangian} \emph{Lagrangian}
\begin{align}
  \label{lagrangian}
  \mathscr{L}(u, x, \lambda) &= \mathscr{F}(u, x) + \big( \lambda, \mathscr{R}(u, x) \big),
\end{align}
where the notation $(f,g) = \int_{\Omega} f g d\Omega$ is the inner product.  Using Lagrangian (\ref{lagrangian}), the first necessary condition in (\ref{objective_perturbations}) is satisfied when $\lambda$ is chosen -- say $\lambda = \lambda^*$ -- such that for a given state $u$ and control parameter $x$, 
\begin{align}
  \label{adjoint}
  \lambda^* = \argminl_{\lambda} \left\Vert \frac{\delta}{\delta u} \mathscr{L} \left( u, x; \lambda \right) \right\Vert.
\end{align}
This $\lambda^*$ may then be used in condition (\ref{condition_two}) to calculate the direction of decent of Lagrangian (\ref{lagrangian}) with respect to the control variable $x$ for a given state $u$ and adjoint variable $\lambda^*$,
\begin{align}
  G = \frac{\delta}{\delta x} \mathscr{L} (u, x, \lambda^*).
\end{align}
This \emph{G\^{a}teaux derivative}, or first variation of Lagrangian $\mathscr{L}$ with respect to $x$ (see \S \ref{ssn_gateaux}), provides a direction which control parameter $x$ may follow in order to satisfy the second condition in (\ref{objective_perturbations}) and thus minimize objective functional $\mathscr{F}$.

%===============================================================================

\section{Log-barrier solution process} \label{ssn_log_barrier}

\index{Log-barrier method}
To determine a locally optimal value of $u$, a variation of a primal-dual-interior-point algorithm with a filter-line-search method may be used, as implemented by the IPOPT framework \citep{wachter}.  Briefly, the algorithm implemented by IPOPT computes approximate solutions to a sequence of barrier problems
\begin{align}
\begin{aligned}
  &\min_{x \in \R^n}\ \left\{ \varphi_{\mu}(u, x) = \mathscr{F}(u,x) - \mu \sum_{i=1}^n \ln \left( c^i\left( x \right) \right) \right\}
  %&\text{subject to }\  \mathbf{r}(\vartheta) = 0
\end{aligned}
\label{barrier}
\end{align}
for a decreasing sequence of barrier parameters $\mu$ converging to zero, and $n$ is the number of degrees of freedom of the mesh.  Neglecting equality constraints on the control variables, the first-order \index{Constrained optimization!Necessary conditions} \emph{necessary} conditions -- known as the Karush-Kuhn-Tucker (KKT) conditions --  for barrier problem (\ref{barrier}) are
\begin{align}
\begin{aligned}
  G(\vartheta,\lambda) - \lambda_b &= 0, \\
  %\mathbf{r}(\vartheta) &= 0, \\
  \varTheta Z e - \mu e &= 0, \\
  \left\{c(\vartheta),\ \lambda_b \right\} & \geq 0,
\end{aligned}
\label{barrier_kkt}
\end{align}
where $G(\vartheta,\lambda) = \delta_x \mathscr{L}$, $\lambda_b$ is the Lagrange multiplier for the bound constraint $c(\vartheta) \geq 0$ in (\ref{canonical_opt}), $\varTheta = \mathrm{diag}(c(\vartheta))$, $Z = \mathrm{diag}(\lambda_b)$, and $e = \mathrm{ones}(n)$.  The so-called `optimality error' for barrier problem (\ref{barrier}) is
\begin{align*}
  E_{\mu}(\vartheta, \lambda_b) = \max
  \left\{
     \frac{\left\Vert G(\vartheta,\lambda) - \lambda_b \right\Vert_{\infty}}{s_d}, 
     \frac{\left\Vert \varTheta Z e - \mu e \right\Vert_{\infty}}{s_c}
  \right\},
\end{align*}
with scaling parameters $s_d,s_c \geq 1$.  This error defines the algorithm termination criteria with $\mu=0$,
\begin{align}
  \label{term_criteria}
  E_{0}(\vartheta^*, \lambda_b^*) \leq \epsilon_{\mathrm{tol}},
\end{align}
for approximate solution $(\vartheta^*, \lambda_b^*)$ and user-provided error tolerance $\epsilon_{\mathrm{tol}}$.

The solution to (\ref{barrier_kkt}) for a given $\mu$ is attained by applying a damped version of \index{Newton-Raphson method} Newton's method, whereby the sequence of iterates $(\vartheta^k, \lambda_b^k)$ for iterate $k \leq k_{max}$ solves the system
\begin{align}
  \begin{bmatrix}
   W^k   & - I \\
   Z^k   &   \varTheta^k
  \end{bmatrix}
  \begin{bmatrix}
    d_{x}^k \\
    d_{\lambda_b}^k
  \end{bmatrix} = - 
  \begin{bmatrix}
    G(\vartheta^k,\lambda) - \lambda_b^k \\
    \varTheta^k Z^k e - \mu e
  \end{bmatrix},
\end{align}
with Hessian matrix 
\begin{align}
  W^k = \frac{\delta^2}{\delta x^k \delta x^k} \mathscr{L}(u^k, x^k, \lambda) = \frac{\delta}{\delta x^k} G(\vartheta^k,\lambda).
\end{align}
Once search directions $(d_{x}^k, d_{\lambda_b}^k)$ have been found, the subsequent iterate is computed from
\begin{align}
  \label{subsequent_LB_iterate}
  x^{k+1} &= x^k + \ell^k d_{x}^k \\
  \lambda_b^{k+1} &= \lambda_b^k + \ell_z^k d_{\lambda_b}^k,
\end{align}
with step sizes $\ell$ determined by a backtracking-line-search procedure similar to Algorithm \ref{qn_bls_alg} to enforce an analogous set of Wolfe conditions as (\ref{armijo_condition}, \ref{curvature_condition}), while also requiring that a sufficient decrease in $\varphi_{\mu}$ in (\ref{barrier}) be attained.

Finally, because $\varphi_{\mu}$ is dependent on objective $\mathscr{F}$, objective $\mathscr{F}(u^{k+1},x^{k+1})$ must be evaluated for a series of potential control parameter $x^{k+1}$ values in (\ref{subsequent_LB_iterate}).  Hence multiple solutions of constraint relation $\mathscr{R}(u^{k+1},x^{k+1}) = 0$ in (\ref{canonical_opt}) are required, one for each potential state parameter $u^{k+1}$ for a given $x^{k+1}$.  Finally, at the end of each iteration, adjoint variable $\lambda$ is determined by solving (\ref{adjoint}) and used to compute the next iteration's G\^{a}teaux derivative $G(x^{k+1}, \lambda)$.  For further details, examine \citet{wachter}.

%===============================================================================
%===============================================================================

\part{Dynamics of ice-sheets and glaciers}

\chapter{Fundamentals of flowing ice}

\begin{figure*}
  \centering
    \def\svgwidth{\linewidth}
    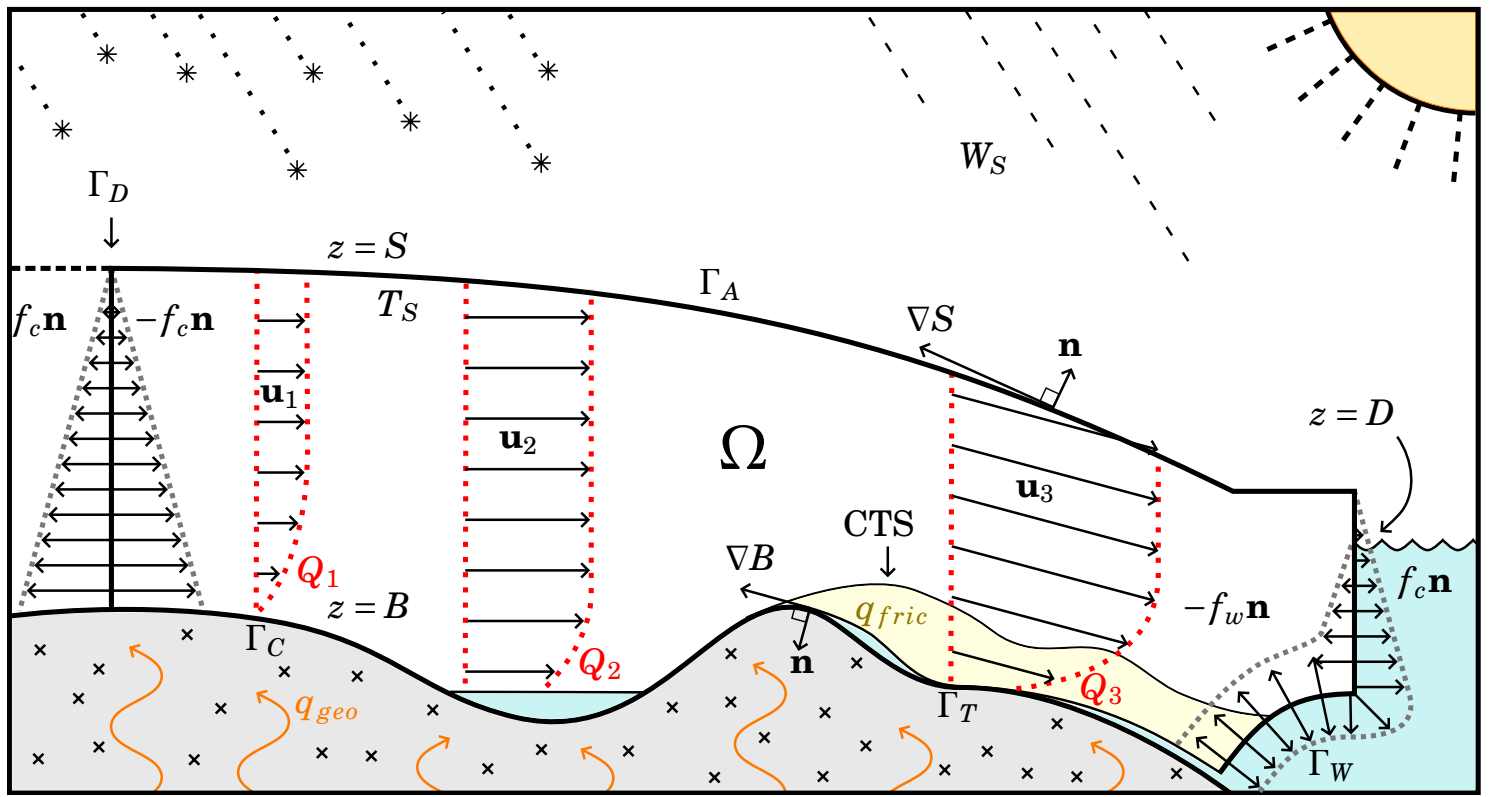
	\caption[Ice-sheet diagram]{Illustration of thermo-dynamic processes for an ice-sheet with surface of height $S$ on boundary $\Gamma_A$, grounded bed of height $B$ with cold boundary $\Gamma_C$ and temperate boundary $\Gamma_T$, ocean boundary $\Gamma_W$ with ocean height $D$, interior ice lateral boundary $\Gamma_D$, interior volume $\Omega$, outward-pointing normal vector $\mathbf{n}$, interior ice pressure normal to the boundary $f_c \mathbf{n}$, and water pressure exerted on the ice $f_w \mathbf{n}$.  Both the 2-meter depth average temperature $T_S$ and water input $W_S$ are used as surface boundary conditions, while the basal boundary is dependent upon energy-fluxes from geothermal sources {\color[rgb]{1,0.4745098,0}$q_{geo}$} and friction heat {\color[rgb]{0.56862745,0.47058824,0}$q_{fric}$}.  The velocity profiles ({\color[rgb]{1,0,0}dashed red}) depend heavily on basal traction; for example, the basal traction associated with velocity profile $\mathbf{u}_1$ is very high.  However, because the surface ice speed is small in regions far from the periphery of the ice-sheet, the gradient in velocity near the bed -- and hence strain-heat {\color{red}$Q_1$} -- is very low.  Moreover, strain-heating may also be low if the basal traction is low, as is the case for strain heat {\color{red}$Q_2$} associated with velocity profile $\mathbf{u}_2$ flowing over a lake.  Finally, observe that profile $\mathbf{u}_3$ flows over a temperate region formed from both increased friction and strain-heat {\color{red}$Q_3$}.}
  \label{ice_profile_domain}
\end{figure*}

Large bodies of ice behave as a highly viscous and thermally-dependent system.  The primary variables associated with an ice-sheet or glacier defined over a domain $\Omega$ with boundary $\Gamma$ (see Figure \ref{ice_profile_domain}) are velocity $\mathbf{u}$ with components $u$, $v$, and $w$ in the $x$, $y$, and $z$ directions; pressure $p$; and internal energy $\theta$.  These variables are inextricably linked by the fundamental conservation equations 
\begin{align}
  \label{cons_momentum}
  -\nabla \cdot \sigma &= \rho\mathbf{g} &&\text{ in } \Omega &&\leftarrow \text{ momentum} \\
  \label{cons_mass}
  \nabla \cdot \mathbf{u} &= 0 &&\text{ in } \Omega &&\leftarrow \text{ mass}  \\
  \label{cons_energy}
  \rho \dot{\theta} &= -\nabla \cdot \mathbf{q} + Q &&\text{ in } \Omega &&\leftarrow \text{ energy.}
\end{align}
These relations are in turn defined with gravitational acceleration vector $\mathbf{g}=[0\ 0\ \text{-}g]^\intercal$, ice density $\rho$, energy flux $\mathbf{q}$, strain-heat $Q$, and \index{Tensor!Cauchy-stress} \index{Tensor!Deviatoric stress} Cauchy-stress tensor
\begin{align}
  \label{stress_tensor}
  \sigma &= \tau - p I, \hspace{10mm} \tau = 2\eta\dot{\epsilon}
\end{align}
further defined with rank-two identity tensor $I$, shear viscosity $\eta$, and \index{Tensor!Strain-rate} strain-rate tensor
\begin{align}
  \label{strain_rate_tensor}
  \dot{\epsilon} 
  &= \nicefrac{1}{2} \left[ \nabla \mathbf{u} + \left(\nabla \mathbf{u} \right)^\intercal \right] \notag \\
  &= \begin{bmatrix}
       \dot{\epsilon}_{xx} & \dot{\epsilon}_{xy} & \dot{\epsilon}_{xz} \\
       \dot{\epsilon}_{yx} & \dot{\epsilon}_{yy} & \dot{\epsilon}_{yz} \\
       \dot{\epsilon}_{zx} & \dot{\epsilon}_{zy} & \dot{\epsilon}_{zz} \\
     \end{bmatrix} \notag \\
  &= \begin{bmatrix}
       \frac{\partial u}{\partial x} & \frac{1}{2}\left( \frac{\partial u}{\partial y} + \frac{\partial v}{\partial x} \right) & \frac{1}{2}\left( \frac{\partial u}{\partial z} + \frac{\partial w}{\partial x} \right) \\
       \frac{1}{2}\left( \frac{\partial v}{\partial x} + \frac{\partial u}{\partial y} \right) & \frac{\partial v}{\partial y} & \frac{1}{2}\left( \frac{\partial v}{\partial z} + \frac{\partial w}{\partial y} \right) \\
       \frac{1}{2}\left( \frac{\partial w}{\partial x} + \frac{\partial u}{\partial z} \right) & \frac{1}{2}\left( \frac{\partial w}{\partial y} + \frac{\partial v}{\partial z} \right) & \frac{\partial w}{\partial z}
     \end{bmatrix}.
\end{align}

Shear viscosity $\eta$ is derived from \index{Constitutive ice-flow relation} \emph{Nye's generalization of Glen's flow law} \citep{glen, nye} 
\begin{align}
  \label{nye}
  \dot{\epsilon} = A(\theta) \tau_e^{n-1} \tau,
\end{align}
defined with Glen's flow parameter $n$, the deviatoric part of Cauchy-stress tensor (\ref{stress_tensor}) $\tau = 2\eta \dot{\epsilon}$, and Arrhenius-type energy-dependent flow-rate factor $A(\theta)$.

The second invariant of full-stress-tensor (\ref{stress_tensor}) -- referred to as the \index{Tensor!Effective stress} \emph{effective stress} -- is given by
\begin{align}
  \tau_e^2 = & \frac{1}{2} \mathrm{tr}\left( \tau^2 \right) = \frac{1}{2} \left[ \tau_{ij} \tau_{ij} \right] \notag \\
  \label{effectivstress}
  = &\frac{1}{2} \left[ \tau_{xx}^2 + \tau_{yy}^2 + \tau_{zz}^2 + 2\tau_{xy}^2 + 2\tau_{xz}^2 + 2\tau_{yz}^2 \right].
\end{align}
Likewise, the second invariant of strain-rate tensor (\ref{strain_rate_tensor}) -- known as the \index{Tensor!Effective strain-rate} \emph{effective strain-rate} -- is given by
\begin{align}
  \dot{\varepsilon}_e^2 = & \frac{1}{2} \tr\left( \dot{\epsilon}^2 \right) = \frac{1}{2} \Bigg[ \dot{\epsilon}_{ij} \dot{\epsilon}_{ij} \Bigg] \notag \\
  \label{effective_strain_rate}
  = &\frac{1}{2} \Bigg[ \dot{\epsilon}_{xx}^2 + \dot{\epsilon}_{yy}^2 + \dot{\epsilon}_{zz}^2 + 2\dot{\epsilon}_{xy}^2 + 2\dot{\epsilon}_{xz}^2 + 2\dot{\epsilon}_{yz}^2 \Bigg].
\end{align}

Due to the fact that the viscosity of ice $\eta$ is a scalar field, the strain-rate and stress-deviator tensors in (\ref{nye}) may be set equal to their invariants.  Their relationship with \index{Viscosity} viscosity $\eta$ is then evaluated,
\begin{align}
  \label{tau_e_con}
  \dot{\varepsilon}_e = A \tau_e^{n-1} \tau_e = A \tau_e^n \hspace{5mm}
  \implies \hspace{5mm} \tau_e = A^{-\frac{1}{n}} \dot{\varepsilon}_e^{\frac{1}{n}}.
\end{align}
Inserting (\ref{tau_e_con}) into (\ref{nye}) and solving for $\tau$ results in
\begin{align*}
  \tau &= A^{-1} \tau_e^{1-n} \dot{\epsilon} 
       = A^{-1} \left( A^{-\frac{1}{n}} \dot{\varepsilon}_e^{\frac{1}{n}} \right)^{1-n} \dot{\epsilon} \\
       &= A^{-1} A^{\frac{n - 1}{n}} \dot{\varepsilon}_e^{\frac{1-n}{n}} \dot{\epsilon} 
       = A^{-\frac{1}{n}} \dot{\varepsilon}_e^{\frac{1-n}{n}} \dot{\epsilon}.
\end{align*}
Next, using deviatoric-stress-tensor definition (\ref{stress_tensor}),
\begin{align*}
  \eta = \frac{1}{2} \tau \dot{\epsilon}^{-1} 
       = \frac{1}{2} \left( A^{-\frac{1}{n}} \dot{\varepsilon}_e^{\frac{1-n}{n}} \dot{\epsilon} \right) \dot{\epsilon}^{-1} 
       = \frac{1}{2} A^{-\frac{1}{n}} \dot{\varepsilon}_e^{\frac{1-n}{n}},
\end{align*}
When solving discrete systems, a strain-regularization term $\dot{\varepsilon}_0 \ll 1$ may be introduced to eliminate singularities in areas of low strain-rate \citep{pattyn}; the resulting thermally-dependent viscosity is given by
\begin{align}
  \label{viscosity}
  \eta(\theta, \mathbf{u}) &= \frac{1}{2}A(\theta)^{-\nicefrac{1}{n}} (\dot{\varepsilon}_e(\mathbf{u}) + \dot{\varepsilon}_0)^{\frac{1-n}{n}}.
\end{align} 

Finally, the strain-heating \index{Strain heat} term $Q$ in (\ref{cons_energy}) is defined as the third invariant (the trace) of the tensor product of strain-rate tensor (\ref{strain_rate_tensor}) and the deviatoric component of Cauchy-stress tensor (\ref{stress_tensor}), $\tau = 2\eta \dot{\epsilon}$ \citep{greve}
\begin{align}
  \label{strain_heat}
  Q(\theta, \mathbf{u}) &= \mathrm{tr}\left( \dot{\epsilon} \cdot \tau \right) = \ 2 \eta \mathrm{tr}\left(\dot{\epsilon}^2\right) = 4 \eta \dot{\varepsilon}_e^2.
\end{align}

Equations (\ref{cons_momentum} -- \ref{cons_energy}) and corresponding boundary conditions are described in the following chapters.  FEniCS source code will be provided whenever possible, and are available through the open-source software \emph{Cryospheric Problem Solver} (CSLVR), an expansion of the FEniCS software \emph{Variational Glacier Simulator} (VarGlaS) developed by \citet{brinkerhoff}.

%===============================================================================

\section{List of symbols}

\begin{tabular}{lll}
$\theta$ & J kg\sups{-1} & internal energy (\ref{energy}) \\
$\theta_m$ & J kg\sups{-1} & pressure-melting energy (\ref{energy_melting}) \\
$\theta_c$ & J kg\sups{-1} & maximum energy (\ref{energy_objective}) \\
$\tilde{\theta}$ & J kg\sups{-1} & enthalpy (\ref{enthalpy}) \\
$T$ & K & temperature (\ref{temperature}) \\
$T_m$ & K & pressure-melting temp.~(\ref{temperature_melting}) \\
$T_S$ & K & 2-meter depth surface temp.~(\ref{surface_temperature}) \\
$W$ & -- & water content (\ref{water_content}) \\ 
$W_c$ & -- & maximum water content (\ref{water_demand}) \\
$W_S$ & -- & surface water content (\ref{surface_water}) \\
$\mathbf{q}$ & kg s\sups{-3} & energy flux (\ref{flux}, \ref{individual_flux}, \ref{enthalpy_grad}) \\
$\mathbf{q}_s$ & kg s\sups{-3} & sensible heat flux (\ref{individual_flux}) \\
$\mathbf{q}_l$ & kg s\sups{-3} & latent heat flux (\ref{individual_flux}) \\
$\rho$ & kg m\sups{-3} & density (\ref{mixture_density}) \\
$k$  & J s\sups{-1}m\sups{-1}K\sups{-1} & mixture thermal conductivity (\ref{mixture_thermal_conductivity}) \\
$k_i$  & J s\sups{-1}m\sups{-1}K\sups{-1} & thermal conductivity of ice (\ref{thermal_conductivity}) \\
$k_0$  & -- & non-advective transport coef.~(\ref{enthalpy_grad}) \\
$c$  & J kg\sups{-1}K\sups{-1} & mixture heat capacity (\ref{mixture_heat_capacity}) \\
$c_i$  & J kg\sups{-1}K\sups{-1} & heat capacity of ice (\ref{heat_capacity}) \\
$\kappa$ & J s\sups{-1}m\sups{-1}K\sups{-1} & enthalpy-gradient cond'v'ty (\ref{enthalpy_grad}) \\
$\nu$ & J m\sups{-1}s\sups{-1} & non-advective water-flux coef.~(\ref{individual_flux}) \\
$p$ & Pa & pressure (\ref{stress_tensor}) \\
$\mathbf{f}$ & Pa m\sups{-1} & volumetric body forces (\ref{cons_momentum}) \\
$\mathbf{g}$ & m s\sups{-2} & gravitational acceleration vector \\
$\mathbf{u}$ & m s\sups{-1} & velocity vector \\
$\mathbf{n}$ & -- & outward-normal vector \\
$Q$ & J m\sups{-3}s\sups{-1} & internal friction (\ref{strain_heat}) \\
$\Xi$ & m\sups{2}s\sups{-1} & mixture diffusivity (\ref{diffusivity}) \\
$\sigma$ & Pa & Cauchy-stress tensor (\ref{stress_tensor}) \\
$\sigma_{BP}$ & Pa & first-order stress tensor (\ref{bp_stress_tensor}) \\
$\sigma_{PS}$ & Pa & plane-strain stress tensor (\ref{ps_tensors}) \\
$\sigma_{RS}$ & Pa & reform.-Stokes stress tensor (\ref{rs_stress_tensor}) \\
$\tau$ & Pa & deviatoric-stress tensor (\ref{stress_tensor}) \\
$\dot{\epsilon}$ & s\sups{-1} & rate-of-strain tensor (\ref{strain_rate_tensor}) \\
$\dot{\varepsilon}_e$ & s\sups{-1} & effective strain-rate (\ref{effective_strain_rate}) \\
$\dot{\varepsilon}_{\text{BP}}$ & s\sups{-1} & first-order eff.~strain-rate (\ref{bp_effective_strain_rate}) \\
$\dot{\varepsilon}_{\text{PS}}$ & s\sups{-1} & plane-strain eff.~strain-rate (\ref{ps_effective_strain_rate}) \\
$\dot{\varepsilon}_{\text{RS}}$ & s\sups{-1} & reform.-Stokes eff.~strain-rate (\ref{rs_effective_strain_rate}) \\
$\eta$ & Pa s & shear viscosity (\ref{viscosity}) \\
$\eta_{\text{BP}}$ & Pa s & first-order shear viscosity (\ref{bp_viscosity}) \\
%$\eta^L_{\text{BP}}$ & Pa s & linear first-order shear visc.~(\ref{linear_bp_viscosity}) \\
$\eta_{\text{PS}}$ & Pa s & plane-strain shear viscosity (\ref{ps_viscosity}) \\
$\eta_{\text{RS}}$ & Pa s & reform.-Stokes shear viscosity (\ref{rs_viscosity}) \\
$f_w$ & Pa & hydrostatic pressure (\ref{water_pressure}) \\
$f_e$ & Pa & exterior pressure (\ref{exterior_pressure}) \\
$f_c$ & Pa & cryostatic pressure (\ref{exterior_pressure}) \\
$\beta$ & kg m\sups{-2}s\sups{-1} & basal-sliding coefficient (\ref{basal_drag}) \\
$A$ & Pa\sups{-3}s\sups{-1} & flow-rate factor with $n=3$ (\ref{rate_factor}) \\ 
$q_{geo}$ & J s\sups{-1}m\sups{-2} & geothermal heat flux \\
$q_{fric}$ & J s\sups{-1}m\sups{-2} & frictional heating (\ref{basal_friction_heat}) \\
$g_N$ & J s\sups{-1}m\sups{-2} & basal energy source (\ref{basal_energy_source}) \\
$M_b$ & m s\sups{-1} & basal melting rate (\ref{basal_melt_rate}) \\
$F_b$ & m s\sups{-1} & basal water discharge (\ref{basal_water_discharge}) \\
$S$ & m & atmospheric surface height \\
$B$ & m & basal surface height \\
$H$ & m & ice thickness \\
\end{tabular}

\begin{tabular}{lll}
$h$ & m & element diameter \\
$\tau_{\text{IE}}$ & s m\sups{3} kg\sups{-3} & energy intr'sic-time par.~(\ref{tau_ie}) \\
$\tau_{\text{BV}}$ & -- & balance vel.~intr'sic-time par.~(\ref{tau_bv}) \\
$\tau_{\text{age}}$ & s & age intrinsic-time parameter (\ref{tau_age}) \\
$P_{\'e}$ & -- & element P\'{e}clet number (\ref{tau_ie}) \\
$\xi$ & -- & energy intr'sic-time coef.~(\ref{intrinsic_time_ftn}, \ref{intrinsic_time_ftn_quad}) \\
$\alpha$ & -- & temperate zone coefficient (\ref{temperate_marker}) \\
$\mathbf{r}$ & J s\sups{-1} & energy residual vector (\ref{component_var_form}) \\
$\bm{\mathcal{C}}$ & J s\sups{-1} & energy advection matrix (\ref{advective_form}) \\
$\bm{\mathcal{K}}$ & J s\sups{-1} & conducive gradient matrix (\ref{conductive_gradient_form}) \\
$\bm{\mathcal{D}}$ & J s\sups{-1} & energy diffusion matrix (\ref{diffusion_form}) \\
$\bm{\mathcal{S}}$ & J s\sups{-1} & energy stabilization matrix (\ref{stabilization_form_a}) \\
$\mathbf{f}^{\text{ext}}$ & J s\sups{-1} & ext.~basal energy flux vec.~(\ref{basal_energy_gradient_form}) \\
$\mathbf{f}^{\text{int}}$ & J s\sups{-1} & internal strain heat vector (\ref{internal_friction_form}) \\
$\mathbf{f}^{\text{stz}}$ & J s\sups{-1} & stabilization vector (\ref{stabilization_form_l}) \\
$\Omega$ & m\sups{3} & domain volume \\
$\Gamma$ & m\sups{2} & domain outer surface \\
$\Gamma_A$ & m\sups{2} & atmospheric surface \\
$\Gamma_S$ & m\sups{2} & complete upper surface \\
$\Gamma_C$ & m\sups{2} & cold grounded basal surface \\
$\Gamma_T$ & m\sups{2} & temperate grounded basal surface \\
$\Gamma_G$ & m\sups{2} & complete grounded basal surface \\
$\Gamma_W$ & m\sups{2} & surface in contact with ocean \\
$\Gamma_E$ & m\sups{2} & non-grounded surface \\
$\Gamma_D$ & m\sups{2} & interior lateral surface \\
$\mathcal{A}$ & J s\sups{-1} & momentum variational princ. (\ref{action}) \\
$\mathcal{A}_{\text{BP}}$ & J s\sups{-1} & first-order momentum principle (\ref{bp_action}) \\
$\mathcal{A}_{\text{PS}}$ & J s\sups{-1} & plane-strain momentum pr'c'p.~(\ref{ps_action}) \\
$\mathcal{A}_{\text{RS}}$ & J s\sups{-1} & ref.-Stokes momentum princ.~(\ref{rs_action}) \\
$\Lambda$ & Pa  & impen'bil'ty Lagrange mult. (\ref{dukowicz_lambda}) \\
$\Lambda_{\text{BP}}$ & Pa  & BP impen'bil'ty Lagrange mult. (\ref{bp_dukowicz_lambda}) \\
$\Lambda_{\text{PS}}$ & Pa  & PS impen'bil'ty Lagrange mult. (\ref{ps_dukowicz_lambda}) \\
$V$ & Pa  & viscous dissipation (\ref{viscous_dissipation}) \\
$V_{\text{BP}}$ & Pa  & first-order viscous dissipation (\ref{bp_viscous_dissipation}) \\
%$V^L_{\text{BP}}$ & Pa  & linear first-order visc.~dis'pat'n (\ref{linear_viscous_dissipation}) \\
$V_{\text{PS}}$ & Pa  & plane-strain viscous dissipation (\ref{ps_viscous_dissipation}) \\
$V_{\text{RS}}$ & Pa  & reform.-Stokes viscous diss.~(\ref{rs_viscous_dissipation}) \\
$\mathscr{R}$ & Pa s\sups{-1} & energy balance residual (\ref{energy_forward_model}) \\
$\mathscr{J}$ & m\sups{6}s\sups{-4} & energy objective functional (\ref{energy_objective}) \\
$\mathscr{L}$ & m\sups{6}s\sups{-4} & energy Lagrangian functional (\ref{energy_lagrangian}) \\
$\lambda_b$ & m\sups{5}s\sups{-1} & $F_b$ inequality const. Lagrange mult. \\
$\mathscr{D}$ & J kg\sups{-1} & optimal water energy discrepancy \\
$\lambda$ & m\sups{4}kg\sups{-1}s\sups{-1} & energy adjoint variable (\ref{energy_adjoint}) \\
$\mathscr{H}$ & J s\sups{-1} & momentum Lagrangian (\ref{momentum_lagrangian}, \ref{momentum_expanded_lagrangian}) \\
$\bm{\lambda}$ & m s\sups{-1} & momentum adjoint variable (\ref{momentum_lagrangian}) \\
$\mathscr{I}$ & J s\sups{-1} & momentum objective functional (\ref{momentum_objective}) \\
$\gamma_1$ & kg m\sups{-2}s\sups{-1} & $L^2$ cost coefficient (\ref{momentum_objective}) \\
$\gamma_2$ & J s\sups{-1} & logarithmic cost coefficient (\ref{momentum_objective}) \\
$\gamma_3$ & m\sups{6}kg\sups{-1}s\sups{-1} & Tikhonov regularization coef. (\ref{momentum_objective}) \\
$\gamma_4$ & m\sups{6}kg\sups{-1}s\sups{-1} & TV regularization coeff. (\ref{momentum_objective}) \\
$\varphi_{\mu}$ & m\sups{6}s\sups{-4} & energy barrier problem (\ref{energy_barrier}) \\
$\varphi_{\omega}$ & J s\sups{-1} & momentum barrier problem (\ref{momentum_barrier}) \\
$\mathbf{d}$ & -- & imposed dir.~of balance velocity (\ref{balance_velocity_direction}) \\
$\bar{\mathbf{u}}$ & m s\sups{-1} & balance velocity (\ref{balance_velocity}) \\
$\bar{u}$ & m s\sups{-1} & balance velocity magnitude (\ref{balance_velocity_dir_and_mag}) \\
$\hat{\mathbf{u}}$ & -- & balance velocity direction (\ref{balance_velocity_dir_and_mag}) \\
$N$   & Pa m & membrane-stress tensor (\ref{membrane_stress_tensor}) \\
$M$   & Pa & membrane-stress bal.~tensor (\ref{membrane_stress_balance_tensor}) \\
%$N_e$ & -- & number of elements in mesh \\
%$N_n$ & -- & number of vertices in mesh \\
\end{tabular}

\begin{table*}[t]
\centering
\caption[Empirically-derived-ice-sheet constants]{Empirically-derived constants}
\label{constants}
\begin{tabular}{llll}
$g$ & $9.81$ & m s\sups{-2} & gravitational acceleration \\
$n$ & $3$ & -- & Glen's flow exponent\\
$R$ & $8.3144621$ & J mol\sups{-1} K\sups{-1} & universal gas constant\\
a  & $31556926$ & s a\sups{-1} & seconds per year\\
$\rho_i$ & $910$ & kg m\sups{-3} & density of ice\\
$\rho_w$ & $1000$ & kg m\sups{-3} & density of water\\
$\rho_{sw}$ & $1028$ & kg m\sups{-3} & density of seawater\\
$k_w$  & $0.561$ & J s\sups{-1}m\sups{-1}K\sups{-1} & thermal conductivity of water \\
$c_w$  & $4217.6$ & J kg\sups{-1}K\sups{-1} & heat capacity of water \\
$L_f$ & $3.34 \times 10^{5}$ & J kg\sups{-1} & latent heat of fusion \\
$T_w$  & $273.15$ & K & triple point of water\\
$\gamma$ & $9.8 \times 10\sups{-8}$ & K Pa\sups{-1} & pressure-melting coefficient \\
\end{tabular}
\end{table*}

%===============================================================================
%===============================================================================

\chapter{Momentum and mass balance} \label{ssn_momentum_and_mass_balance}

\index{Balance equations!Momentum}
\index{Balance equations!Mass}
\index{Impenetrability}
\index{Incompressibility|seealso{Balance equations!Mass}}
\index{Stokes equations!Slip-friction}
\index{Hydrostatic stress}
Momentum-balance equation (\ref{cons_momentum}) and mass-balance equation (\ref{cons_mass}), collectively referred to as the Stokes equations (see \S \ref{ssn_intro_stokes_2d}, \S \ref{ssn_intro_stokes_2d_slip}, \S \ref{ssn_intro_stokes_3d}, \S \ref{ssn_intro_stokes_2d_slip_stab}, and the introductory chapter of \citet{elman}), are completed with boundary conditions encompassing the entire outer surface $\Gamma = \Gamma_A \cup \Gamma_W \cup \Gamma_G$, with atmospheric boundary $\Gamma_A$, boundary in contact with ocean $\Gamma_W$, basal boundary in contact with bedrock $\Gamma_G$, and complete basal boundary including floating ice $\Gamma_B$ (see Figure \ref{ice_profile_domain}),
\begin{align}
  \label{surface_stress}
  \sigma \cdot \mathbf{n} &= \mathbf{0} &&\text{ on } \Gamma_A &&\leftarrow \text{ stress-free surface} \\
  \label{water_stress}
  \sigma \cdot \mathbf{n} &= -f_w \mathbf{n} &&\text{ on } \Gamma_W &&\leftarrow \text{ water pressure} \\
  \label{basal_drag}
  \big( \sigma \cdot \mathbf{n} \big)_{\Vert} &= -\beta \mathbf{u} &&\text{ on } \Gamma_G &&\leftarrow \text{ basal traction } \\
  \label{impenetrability}
  \mathbf{u} \cdot \mathbf{n} &= 0 &&\text{ on } \Gamma_B &&\leftarrow \text{ impenetrability,}
\end{align}
with outward-pointing normal vector to the boundary $\mathbf{n} = [n_x\ n_y\ n_z]^\intercal$, and hydrostatic pressure
\begin{align}
  \label{water_pressure}
  f_w = \rho_{sw} g (D - z), \hspace{5mm} z < D,
\end{align}
with seawater density $\rho_{sw}$ and ocean height $D$.  Tangential component of stress (\ref{basal_drag}) -- with tangential components denoted $(\mathbf{v})_{\Vert} = \mathbf{v} - \left( \mathbf{v} \cdot \mathbf{n} \right) \mathbf{n}$ -- is proportional to the basal velocity $\mathbf{u} |_{\Gamma_G}$ and basal-traction coefficient $\beta \geq 0$.  Notice that traction boundary (\ref{basal_drag}) and impenetrability boundary (\ref{impenetrability}) are identical to slip-friction boundary conditions (\ref{intro_stokes_gN_N_S_D_fric}) and (\ref{intro_stokes_gD_N_S_D_slip}) explored previously in \S \ref{ssn_intro_stokes_2d_slip} and \S \ref{ssn_intro_stokes_2d_slip_stab}.

Throughout the following sections, Python source code associated with these fundamental equations will be provided.  For example, viscosity (\ref{viscosity}) is created using FEniCS in Code Listing \ref{viscosity_code}. 

\begin{python}[label=viscosity_code, caption={FEniCS code used to generate viscosity $\eta$ as defined in the \texttt{Momentum} class, from which all of the momentum models of this chapter inherit.}]
def viscosity(self, U):
  """
  calculates the viscosity eta.  Uses velocity vector <U> with
  components u,v,w.  If <linear> == True, form viscosity from model.U3.
  """
  s  = "::: forming viscosity :::"
  print_text(s, self.color())
  model    = self.model
  n        = model.n
  A_f      = model.A_f
  eps_reg  = model.eps_reg
  epsdot   = self.effective_strain_rate(U)
  eta      = 0.5 * A_f**(-1/n) * (epsdot + eps_reg)**((1-n)/(2*n))
  return eta
\end{python}
%===============================================================================

\section{Full-Stokes equations} \label{ssn_full_stokes}

The expanded Stokes equations follow identically to the derivation of (\ref{3d_stokes_expanded}).  These equations are \index{Stokes equations!Applied to ice, full-Stokes} \index{Stokes equations!Slip-friction}
\footnotesize
\begin{subequations}
  \label{stokes_exp}
  \begin{eqnarray}
  \label{first_stokes_exp}
  \frac{\partial}{\partial x} \left[2\eta \frac{\partial u}{\partial x} \right]  - \frac{\partial p}{\partial x} + \frac{\partial}{\partial y} \left[\eta \left( \frac{\partial u}{\partial y} + \frac{\partial v}{\partial x} \right) \right] + \frac{\partial}{\partial z} \left[\eta \left( \frac{\partial u}{\partial z} + \frac{\partial w}{\partial x} \right) \right] = &0 \\ 
  \label{second_stokes_exp}
  \frac{\partial}{\partial x} \left[\eta \left( \frac{\partial v}{\partial x} + \frac{\partial u}{\partial y} \right) \right]  + \frac{\partial}{\partial y} \left[2\eta \frac{\partial v}{\partial y} \right] - \frac{\partial p}{\partial y} + \frac{\partial}{\partial z} \left[\eta \left( \frac{\partial v}{\partial z} + \frac{\partial w}{\partial y} \right) \right] = &0 \\ 
  \label{third_stokes_exp}
  \frac{\partial}{\partial x} \left[\eta \left( \frac{\partial w}{\partial x} + \frac{\partial u}{\partial z} \right) \right]  + \frac{\partial}{\partial y} \left[\eta \left( \frac{\partial w}{\partial y} + \frac{\partial v}{\partial z} \right) \right] + \frac{\partial}{\partial z} \left[2\eta \frac{\partial w}{\partial z} \right] - \frac{\partial p}{\partial z} = &\rho g,
  \end{eqnarray}
\end{subequations}
\normalsize
and conservation of mass relation (\ref{cons_mass}),
\begin{align}
  \label{fourth_stokes_exp}
  \frac{\partial u}{\partial x} + \frac{\partial v}{\partial y} + \frac{\partial w}{\partial z} = 0. 
\end{align}

Equations (\ref{first_stokes_exp}, \ref{second_stokes_exp}, \ref{third_stokes_exp}, and \ref{fourth_stokes_exp}) comprise a system of four equations and four unknowns $u$, $v$, $w$, and $p$.  The complexity of solving this system and associated boundary conditions (\ref{surface_stress} -- \ref{impenetrability}) has already been explored in \S \ref{ssn_intro_stokes_2d_slip} and \S \ref{ssn_intro_stokes_2d_slip_stab}; namely, the satisfaction or circumvention of inf-sup condition (\ref{inf_sup_condition}) and the correct imposition of Dirichlet condition (\ref{impenetrability}).  An elegant method satisfying these requirements is presented in the next section.

\subsection{Variational principle} \label{ssn_full_stokes_var_prin}

\index{Variational principle!Full-Stokes}
To solve system (\ref{cons_momentum}, \ref{cons_mass}, \ref{surface_stress} -- \ref{impenetrability}),  the method described in \citet{dukowicz_2010} is used.  This method makes use of a variational principle that uniquely determines velocity $\mathbf{u}$ and pressure $p$ by finding the extremum of the action

\begin{align}
  \label{action}
  \pazocal{A} \left(\mathbf{u}, p\right) = &+ \int_{\Omega} \left( V\left( \dot{\varepsilon}_e^2 \right) - \rho \mathbf{g} \cdot \mathbf{u} - p \nabla \cdot \mathbf{u} \right)\ d\Omega \notag \\
  &+ \int_{\Gamma_B} \left( \Lambda \mathbf{u} \cdot \mathbf{n} + \frac{1}{2} \beta \mathbf{u} \cdot \mathbf{u} \right)\ d\Gamma_B \notag \\
  &+ \int_{\Gamma_L} f_w \mathbf{n} \cdot \mathbf{u}\ d\Gamma_L,
\end{align}
with viscous-dissipation term
\begin{align}
  \label{viscous_dissipation}
  V\left( \dot{\varepsilon}_e^2 \right) &= \int_0^{\dot{\varepsilon}_e^2} \eta(s)\ ds \notag \\
  &= \frac{1}{2} A^{-\nicefrac{1}{n}} \int_0^{\dot{\varepsilon}_e^2} s^{\frac{1-n}{2n}}\ ds \notag \\
  &= \frac{2n}{n+1} \left( \frac{1}{2} A^{-\nicefrac{1}{n}} \right) \left( \dot{\varepsilon}_e^2 \right)^{\frac{n+1}{2n}} \notag \\
  &= \frac{2n}{n+1} \left( \frac{1}{2} A^{-\nicefrac{1}{n}} \left( \dot{\varepsilon}_e^2 \right)^{\frac{1-n}{2n}} \right) \dot{\varepsilon}_e^2 \notag \\
  &= \frac{2n}{n+1} \eta\left(\theta, \mathbf{u} \right) \dot{\varepsilon}_e^2,
\end{align}
where shear viscosity $\eta$ is given by (\ref{viscosity}).  Lagrange multiplier $\Lambda$ enforces basal-surface impenetrability condition (\ref{impenetrability}), while pressure $p$ -- defined as the mean compressive stress $p = -\sigma_{kk} / 3$ -- also takes on the role of a Lagrange multiplier to enforce incompressibility condition (\ref{cons_mass}).

This extremum is defined as the solution to 
\begin{align}
  \label{extremum_intermediate}
  \frac{\delta \pazocal{A}}{\delta \mathbf{u}} = 0, \hspace{5mm} \frac{\delta \pazocal{A}}{\delta p} = 0, \hspace{5mm} \frac{\delta \pazocal{A}}{\delta \Lambda} = 0,
\end{align}
and has been shown to be equivalent to the Stokes system (\ref{cons_momentum}, \ref{cons_mass}) by \citet{dukowicz_2010} and boundary conditions (\ref{surface_stress} -- \ref{impenetrability}) by \citet{dukowicz_2011}.  It was later explained by \citet{dukowicz_2011} how the basal stress arising from Euler-Lagrange equations (\ref{extremum_intermediate}) is constrained to obey
\begin{align}
  \label{dukowicz_basal_stress}
  \sigma \cdot \mathbf{n} \big|_{\Gamma_B} &= - \beta \mathbf{u} - \Lambda \mathbf{n}.
\end{align}
The magnitude of stress normal to the bed is determined by taking the dot product of (\ref{dukowicz_basal_stress}) with $\mathbf{n}$, making use of bed-impenetrability condition (\ref{impenetrability}), and the definition of a unit vector, resulting in
\begin{align}
  \label{dukowicz_lambda}
  \Lambda &= - \mathbf{n} \cdot \sigma \cdot \mathbf{n}.
\end{align}
Therefore, $\Lambda$ is equivalent to the magnitude of stress presented by the ice on the supporting bedrock.  Relation (\ref{dukowicz_lambda}) may be used to eliminate Lagrange multiplier $\Lambda$ in (\ref{action}); hence extremum conditions (\ref{extremum_intermediate}) are reduced to
\begin{align}
  \label{extremum}
  \frac{\delta \pazocal{A}}{\delta \mathbf{u}} = 0, \hspace{10mm} \frac{\delta \pazocal{A}}{\delta p} = 0.
\end{align}
Additionally, by assuming that the magnitude of the normal component of deviatoric-stress tensor (\ref{stress_tensor}), $\mathbf{n} \cdot \tau \cdot \mathbf{n}$, is much less than pressure $p$ along the entire basal surface $\Gamma_B$, (\ref{dukowicz_lambda}) simplifies to $\Lambda \approx p$.  This approximation has in our experience lead to improved convergence characteristics of the discrete system when the topography includes steep basal gradients.  Additionally, using both (\ref{dukowicz_basal_stress}) and (\ref{dukowicz_lambda}), observe that the tangential component of stress is
\begin{align}
  \left( \sigma \cdot \mathbf{n} \right)_{\Vert} &= \sigma \cdot \mathbf{n} - \left( \mathbf{n} \cdot \sigma \cdot \mathbf{n}\right) \mathbf{n} \notag \\
  &= - \beta \mathbf{u} - \Lambda \mathbf{n} - \left( -\Lambda \right) \mathbf{n} \notag \\
  &= - \beta \mathbf{u},
\end{align}
and is thus consistent with traction-boundary-condition (\ref{basal_drag}).

The source code of CSLVR uses an implementation similar to Code Listing \ref{cslvr_full_stokes}.

\begin{python}[label=cslvr_full_stokes, caption={CSLVR source code contained in the \texttt{MomentumDukowiczStokes} class.}]
# define variational problem :
U                    = Function(model.Q4, name = 'G')
dU                   = TrialFunction(model.Q4)
Phi                  = TestFunction(model.Q4)
phi, psi, xi,  kappa = Phi
du,  dv,  dw,  dP    = dU
u,   v,   w,   p     = U

# create velocity vector :
U3      = as_vector([u,v,w])

# viscous dissipation :
epsdot  = self.effective_strain_rate(U3)
if linear:
  s   = "    - using linear form of momentum using model.U3 in epsdot -"
  Uc  = model.U3.copy(True)
  eta = self.viscosity(Uc)
  Vd  = 2 * eta * epsdot
else:
  s   = "    - using nonlinear form of momentum -"
  eta = self.viscosity(U3)
  Vd  = (2*n)/(n+1) * A_f**(-1/n) * (epsdot + eps_reg)**((n+1)/(2*n))
print_text(s, self.color())

# potential energy :
Pe     = - rhoi * g * w

# dissipation by sliding :
Ut     = U3 - dot(U3,N)*N
Sl     = - 0.5 * beta * dot(Ut, Ut)

# incompressibility constraint :
Pc     = p * div(U3)

# impenetrability constraint :
sig    = self.stress_tensor(U3, p, eta)
lam    = - dot(N, dot(sig, N))
Nc     = - lam * (dot(U3, N) - Fb)

# pressure boundary :
Pb_w   = - rhosw*g*D * dot(U3, N)

# action :
A      = + (Vd - Pe - Pc)*dx - Nc*dBed \
         - Sl*dBed_g - Pb_w*dBed_f - Pb_w*dLat_t

# the first variation of the action in the direction of a 
# test function; the extremum :
self.mom_F = derivative(A, U, Phi)

# the first variation of the extremum in the direction 
# a trial function; the Jacobian :
self.mom_Jac = derivative(self.mom_F, U, dU)

def stress_tensor(self, U, p, eta):
  """
  return the Cauchy stress tensor.
  """
  s   = "::: forming the Cauchy stress tensor :::"
  print_text(s, self.color())

  I     = Identity(3)
  tau   = self.deviatoric_stress_tensor(U, eta)

  sigma = tau - p*I
  return sigma

def deviatoric_stress_tensor(self, U, eta):
  """
  return the deviatoric stress tensor.
  """
  s   = "::: forming the deviatoric part of the Cauchy stress tensor :::"
  print_text(s, self.color())

  epi = self.strain_rate_tensor(U)
  tau = 2 * eta * epi
  return tau

def strain_rate_tensor(self, U):
  """
  return the strain-rate tensor of <U>.
  """
  epsdot = 0.5 * (grad(U) + grad(U).T)
  return epsdot

def effective_strain_rate(self, U):
  """
  return the effective strain rate squared.
  """
  epi    = self.strain_rate_tensor(U)
  ep_xx  = epi[0,0]
  ep_yy  = epi[1,1]
  ep_zz  = epi[2,2]
  ep_xy  = epi[0,1]
  ep_xz  = epi[0,2]
  ep_yz  = epi[1,2]
  
  # Second invariant of the strain rate tensor squared
  epsdot = 0.5 * (+ ep_xx**2 + ep_yy**2 + ep_zz**2) \
                  + ep_xy**2 + ep_xz**2 + ep_yz**2
  return epsdot

def default_solve_params(self):
  """ 
  Returns a set of default solver parameters that yield good performance
  """
  nparams = {'newton_solver' :
            {
              'linear_solver'            : 'mumps',
              'relative_tolerance'       : 1e-5,
              'relaxation_parameter'     : 0.7,
              'maximum_iterations'       : 25,
              'error_on_nonconvergence'  : False,
            }}
  m_params  = {'solver'      : nparams}
  return m_params

def solve(self, annotate=False):
  """ 
  Perform the Newton solve of the full-Stokes equations 
  """
  # zero out self.velocity for good convergence for any subsequent solves,
  # e.g. model.L_curve() :
  model.assign_variable(self.get_U(), DOLFIN_EPS, cls=self)
  
  # compute solution :
  solve(self.mom_F == 0, self.U, J = self.mom_Jac, bcs = self.mom_bcs,
        annotate = annotate, solver_parameters = params['solver'])
  u, v, w, p = self.U.split()
  
\end{python}

\section{First-order approximation} \label{ssn_first_order}

\index{Stokes equations!Applied to ice, first-order}
\index{Stokes equations!Slip-friction}
Assumptions pertaining to both the state of stress and strain are appropriate over a large proportion of ice-sheets, and lead to considerable simplifications of full-Stokes equations (\ref{stokes_exp}).  These simplifications and associated variational principle are described here.

\subsection{Stress tensor simplification}

The Stokes equations with four equations and four unknowns $u$, $v$, $w$, and $p$ may be reduced to a system of three equations for the velocity components alone, as given by \citet{blatter} and \citet{pattyn}.  This is accomplished by first assuming that the shear stress components in the $z$-coordinate plane are negligible when compared to the $z$-coordinate normal stress, i.e.\ $\partial_x \sigma_{zx}, \partial_y \sigma_{zy} \ll \partial_z \sigma_{zz}$.  Using this assumption, the final equation arising from the expansion of momentum-conservation relation (\ref{cons_momentum}), Equation (\ref{third_stokes_exp}), is reduced to
\begin{align}
  \label{stress_tensor_simplification}
  \frac{\partial \sigma_{zz}}{\partial z} &\approx \rho g,
\end{align}
which may be integrated from the surface to an arbitrary $z$-coordinate,
\begin{align*}
  \int_z^S \frac{\partial \sigma_{zz}}{\partial z} dz' &\approx \int_z^S \rho g dz' \\
  \sigma_{zz}(S) - \sigma_{zz}(z) &\approx \rho g (S - z).
\end{align*}
Using surface-stress condition (\ref{surface_stress}), $\sigma_{zz}(S) = 0$, and applying Cauchy-stress tensor definition (\ref{stress_tensor}),
\begin{align}
  \label{bp_pressure}
  p(z) &\approx \rho g(S - z) + 2\eta \frac{\partial w}{\partial z}(z).
\end{align}
This pressure approximation may then be used to eliminate $p$ from the remaining pressure derivative terms in momentum-balance (\ref{cons_momentum}) with
\begin{align}
  \label{bp_pressure_gradient}
  \frac{\partial p}{\partial i} &\approx
  \rho g\frac{\partial S}{\partial i} + \frac{\partial}{\partial i} \left[ 2\eta \frac{\partial w}{\partial z} \right], \hspace{5mm} i \in x,y.
\end{align}
allowing the simplification of (\ref{first_stokes_exp}) to
\footnotesize
\begin{align}
  \frac{\partial}{\partial x} \left[2\eta \frac{\partial u}{\partial x} \right]  - \frac{\partial p}{\partial x} + \frac{\partial}{\partial y} \left[\eta \left( \frac{\partial u}{\partial y} + \frac{\partial v}{\partial x} \right) \right] + \frac{\partial}{\partial z} \left[\eta \left( \frac{\partial u}{\partial z} + \frac{\partial w}{\partial x} \right) \right] &= 0 \notag \\ 
  \frac{\partial}{\partial x} \left[2\eta \frac{\partial u}{\partial x} \right]  - \rho g\frac{\partial S}{\partial x} + \frac{\partial}{\partial x} \left[ 2\eta \frac{\partial w}{\partial z} \right] + \frac{\partial}{\partial y} \left[\eta \left( \frac{\partial u}{\partial y} + \frac{\partial v}{\partial x} \right) \right] & \notag \\
  + \frac{\partial}{\partial z} \left[\eta \left( \frac{\partial u}{\partial z} + \frac{\partial w}{\partial x} \right) \right] &= 0 \notag \\ 
  \label{first_stokes_exp_stress_simp}
  \frac{\partial}{\partial x} \left[2\eta \left( \frac{\partial u}{\partial x} - \frac{\partial w}{\partial z} \right) \right]  + \frac{\partial}{\partial y} \left[\eta \left( \frac{\partial u}{\partial y} + \frac{\partial v}{\partial x} \right) \right] + \frac{\partial}{\partial z} \left[\eta \left( \frac{\partial u}{\partial z} + \frac{\partial w}{\partial x} \right) \right] &= \rho g\frac{\partial S}{\partial x},
\end{align}
\normalsize
and (\ref{second_stokes_exp}) to
\footnotesize
\begin{align}
  \frac{\partial}{\partial x} \left[\eta \left( \frac{\partial v}{\partial x} + \frac{\partial u}{\partial y} \right) \right]  + \frac{\partial}{\partial y} \left[2\eta \frac{\partial v}{\partial y} \right] - \frac{\partial p}{\partial y} + \frac{\partial}{\partial z} \left[\eta \left( \frac{\partial v}{\partial z} + \frac{\partial w}{\partial y} \right) \right] &= 0 \notag \\ 
  \frac{\partial}{\partial x} \left[\eta \left( \frac{\partial v}{\partial x} + \frac{\partial u}{\partial y} \right) \right]  + \frac{\partial}{\partial y} \left[2\eta \frac{\partial v}{\partial y} \right] - \rho g\frac{\partial S}{\partial y} + \frac{\partial}{\partial y} \left[ 2\eta \frac{\partial w}{\partial z} \right] &\notag \\
  + \frac{\partial}{\partial z} \left[\eta \left( \frac{\partial v}{\partial z} + \frac{\partial w}{\partial y} \right) \right] &= 0 \notag \\ 
  \label{second_stokes_exp_stress_simp}
  \frac{\partial}{\partial x} \left[\eta \left( \frac{\partial v}{\partial x} + \frac{\partial u}{\partial y} \right) \right]  + \frac{\partial}{\partial y} \left[2\eta \left( \frac{\partial v}{\partial y} - \frac{\partial w}{\partial z} \right) \right] + \frac{\partial}{\partial z} \left[\eta \left( \frac{\partial v}{\partial z} + \frac{\partial w}{\partial y} \right) \right] &= \rho g \frac{\partial S}{\partial y},
\end{align}
\normalsize
which combined with conservation of mass relation (\ref{fourth_stokes_exp}),
\begin{align*}
  \frac{\partial u}{\partial x} + \frac{\partial v}{\partial y} + \frac{\partial w}{\partial z} &= 0 && \text{in } \Omega 
\end{align*}
gives three equations and three unknowns $u$, $v$, and $w$.

\subsection{Strain tensor simplification} \label{ssn_strain_tensor_simplification}

Next, assuming the horizontal gradients of $w$ are much less than the vertical gradient of the horizontal components of velocity, i.e.~$\partial_x w \ll \partial_z u$ and $\partial_y w \ll \partial_z v$, and using (\ref{stress_tensor_simplification}), strain-rate tensor (\ref{strain_rate_tensor}) is decoupled from vertical velocity $w$, resulting in the strain-rate quasi-tensor
\begin{align}
  \label{bp_strain_rate_tensor}
  \tilde{\dot{\epsilon}}
  &= \begin{bmatrix}
       \frac{\partial u}{\partial x} & \frac{1}{2}\left( \frac{\partial u}{\partial y} + \frac{\partial v}{\partial x} \right) & \frac{1}{2} \frac{\partial u}{\partial z} \\
       \frac{1}{2}\left( \frac{\partial v}{\partial x} + \frac{\partial u}{\partial y} \right) & \frac{\partial v}{\partial y} & \frac{1}{2} \frac{\partial v}{\partial z} \\
     \end{bmatrix}.
\end{align}
Effective strain-rate (\ref{effective_strain_rate}) is also decoupled from $w$ using the equivalent relation to conservation of mass relation (\ref{cons_mass}), $\dot{\epsilon}_{zz} = -\left(\dot{\epsilon}_{xx} + \dot{\epsilon}_{yy} \right)$,
\begin{align}
  \label{bp_effective_strain_rate}
  \dot{\varepsilon}_{\text{BP}}^2 = \tilde{\dot{\epsilon}}_{xx}^2 + \tilde{\dot{\epsilon}}_{yy}^2 + \tilde{\dot{\epsilon}}_{xx} \tilde{\dot{\epsilon}}_{yy} + \tilde{\dot{\epsilon}}_{xy}^2 + \tilde{\dot{\epsilon}}_{xz}^2 + \tilde{\dot{\epsilon}}_{yz}^2.
\end{align}
Using first-order effective strain-rate (\ref{bp_effective_strain_rate}), the associated first-order shear viscosity is
\begin{align}
  \label{bp_viscosity}
  \eta_{\text{BP}}(\theta, \mathbf{u}_h) &= \frac{1}{2}A(\theta)^{-\nicefrac{1}{n}} (\dot{\varepsilon}_{\text{BP}} + \dot{\varepsilon}_0)^{\frac{1-n}{n}},
\end{align} 
where horizontal vector components are denoted $\mathbf{g}_h = [g_x\ g_y]^\intercal$.

Finally, inserting pressure derivative approximation (\ref{bp_pressure_gradient}) and first-order strain-rate quasi-tensor (\ref{bp_strain_rate_tensor}) into conservation of momentum relation (\ref{cons_momentum}), simplification (\ref{first_stokes_exp_stress_simp}) becomes
\footnotesize
\begin{align}
  \frac{\partial}{\partial x} \left[2\eta \left( \frac{\partial u}{\partial x} - \frac{\partial w}{\partial z} \right) \right]  + \frac{\partial}{\partial y} \left[\eta \left( \frac{\partial u}{\partial y} + \frac{\partial v}{\partial x} \right) \right] + \frac{\partial}{\partial z} \left[\eta \left( \frac{\partial u}{\partial z} + \frac{\partial w}{\partial x} \right) \right] &= \rho g\frac{\partial S}{\partial x} \notag \\
  \frac{\partial}{\partial x} \left[2\eta \left( \frac{\partial u}{\partial x} + \left( \frac{\partial u}{\partial x} + \frac{\partial v}{\partial y} \right)\right) \right]  + \frac{\partial}{\partial y} \left[\eta \left( \frac{\partial u}{\partial y} + \frac{\partial v}{\partial x} \right) \right] + \frac{\partial}{\partial z} \left[\eta \frac{\partial u}{\partial z} \right] &= \rho g\frac{\partial S}{\partial x} \notag \\
  \label{first_stokes_exp_strain_simp}
  \frac{\partial}{\partial x} \left[2\eta \left( 2\frac{\partial u}{\partial x} + \frac{\partial v}{\partial y} \right) \right]  + \frac{\partial}{\partial y} \left[\eta \left( \frac{\partial u}{\partial y} + \frac{\partial v}{\partial x} \right) \right] + \frac{\partial}{\partial z} \left[\eta \frac{\partial u}{\partial z} \right] &= \rho g\frac{\partial S}{\partial x}
\end{align}
\normalsize
and simplification (\ref{second_stokes_exp_stress_simp}) becomes 
\footnotesize
\begin{align}
  \frac{\partial}{\partial x} \left[\eta \left( \frac{\partial v}{\partial x} + \frac{\partial u}{\partial y} \right) \right]  + \frac{\partial}{\partial y} \left[2\eta \left( \frac{\partial v}{\partial y} - \frac{\partial w}{\partial z} \right) \right] + \frac{\partial}{\partial z} \left[\eta \left( \frac{\partial v}{\partial z} + \frac{\partial w}{\partial y} \right) \right] &= \rho g \frac{\partial S}{\partial y} \notag \\
  \frac{\partial}{\partial x} \left[\eta \left( \frac{\partial v}{\partial x} + \frac{\partial u}{\partial y} \right) \right]  + \frac{\partial}{\partial y} \left[2\eta \left( \frac{\partial v}{\partial y} + \left(\frac{\partial u}{\partial x} + \frac{\partial v}{\partial y} \right)\right) \right] + \frac{\partial}{\partial z} \left[\eta \frac{\partial v}{\partial z} \right] &= \rho g \frac{\partial S}{\partial y} \notag \\
  \label{second_stokes_exp_strain_simp}
  \frac{\partial}{\partial x} \left[\eta \left( \frac{\partial v}{\partial x} + \frac{\partial u}{\partial y} \right) \right]  + \frac{\partial}{\partial y} \left[2\eta \left( 2\frac{\partial v}{\partial y} + \frac{\partial u}{\partial x} \right) \right] + \frac{\partial}{\partial z} \left[\eta \frac{\partial v}{\partial z} \right] &= \rho g \frac{\partial S}{\partial y},
\end{align}
\normalsize
two equations and two unknowns $u$ and $v$.  The first-order momentum balance is therefore
\begin{align}
  \label{bp_cons_momentum}
  \nabla \cdot \sigma_{\text{BP}} &= \rho g (\nabla S)_h &&\text{ in } \Omega,
\end{align}
with Blatter-Pattyn stress quasi-tensor
\begin{align}
  \label{bp_stress_tensor}
  \sigma_{\text{BP}} &=
  2 \eta_{\text{BP}} 
  \begin{bmatrix}
       \left( 2\frac{\partial u}{\partial x} + \frac{\partial v}{\partial y} \right) & \frac{1}{2}\left( \frac{\partial u}{\partial y} + \frac{\partial v}{\partial x} \right) & \frac{1}{2} \frac{\partial u}{\partial z} \\
       \frac{1}{2}\left( \frac{\partial v}{\partial x} + \frac{\partial u}{\partial y} \right) & \left( 2\frac{\partial v}{\partial y} + \frac{\partial u}{\partial x} \right) & \frac{1}{2} \frac{\partial v}{\partial z}
     \end{bmatrix}.
\end{align}

\subsection{First-order vertical velocity and boundary conditions}

Because vertical velocity $w$ has been eliminated from conservation of momentum (\ref{cons_momentum}) through the creation of first-order momentum balance (\ref{bp_cons_momentum}), this component of velocity may be computed directly by integrating conservation of mass (\ref{cons_mass}) vertically, resulting in
\begin{align}
  \label{bp_vertical_velocity}
  w(z) &= w(B) - \int_B^z \left( \frac{\partial u}{\partial x} + \frac{\partial v}{\partial y} \right)\ dz',
\end{align}
where the basal vertical velocity is determined directly from impenetrability condition \index{Impenetrability} (\ref{impenetrability}),
\begin{align}
  \label{bp_bed_vertical_velocity}
  w(B) &= - \frac{u(B) n_x + v(B) n_y}{n_z}. 
\end{align}
Finally, because both incompressibility (\ref{cons_mass}) and impenetrability (\ref{impenetrability}) are enforced by vertical velocity relation (\ref{bp_vertical_velocity}, \ref{bp_bed_vertical_velocity}), the remaining first-order boundary conditions are
\begin{align}
  \label{bp_exterior_stress}
  \sigma_{\text{BP}} \cdot \mathbf{n} &= f_e \mathbf{n}_h &&\text{ on } \Gamma_E &&\leftarrow \text{ exterior stress} \\
  \label{bp_basal_drag}
  \sigma_{\text{BP}} \cdot \mathbf{n} &= -\beta \mathbf{u}_h &&\text{ on } \Gamma_G &&\leftarrow \text{ basal traction},
\end{align}
where exterior stress condition (\ref{bp_exterior_stress}) is defined over exterior boundary $\Gamma_E = \Gamma_A \cup \Gamma_W$ with pressure $f_e$ derived by combining pressure approximation (\ref{bp_pressure}) and first-order stress tensor (\ref{bp_stress_tensor}) with water pressure (\ref{water_pressure}), \index{Cryostatic stress}
\begin{align}
  \label{exterior_pressure}
  f_e = f_c - f_w, \hspace{10mm} f_c = \rho g (S - z).
\end{align}
Note that boundaries located on the upper surface of the ice correspond with $f_e = 0$ and are thus stress-free, while cliff faces have $z \neq S$ and hence $f_c \neq 0$ (see Figure \ref{ice_profile_domain}).

\subsection{First-order variational principle} \label{ssn_first_order_var_prin}

\index{Variational principle!First-order}
Also compiled by \citet{dukowicz_2011} is a first-order variational principle for first-order momentum balance (\ref{bp_cons_momentum}) and associated boundary conditions (\ref{bp_exterior_stress}, \ref{bp_basal_drag}),
\begin{align}
  \label{bp_extremum}
  \frac{\delta \pazocal{A}_{\text{BP}}}{\delta \mathbf{u}_h} = 0, \hspace{10mm} \frac{\delta \pazocal{A}_{\text{BP}}}{\delta \Lambda_{\text{BP}}} = 0.
\end{align}
where
\begin{align}
  \label{bp_action}
  \pazocal{A}_{\text{BP}} \left(\mathbf{u}_h \right) = &+ \int_{\Omega} \left( V\left( \dot{\varepsilon}_{\text{BP}}^2 \right) + \rho g \mathbf{u}_h \cdot (\nabla S)_h \right)\ d\Omega \notag \\
  &+ \int_{\Gamma_B} \left( \Lambda_{\text{BP}} \mathbf{u}_h \cdot \mathbf{n}_h + \frac{1}{2} \beta \mathbf{u}_h \cdot \mathbf{u}_h \right)\ d\Gamma_B \notag \\
  &+ \int_{\Gamma_E} f_e \mathbf{n}_h \cdot \mathbf{u}_h\ d\Gamma_E, \\
  \label{bp_viscous_dissipation}
  V\left( \dot{\varepsilon}_{\text{BP}}^2 \right) = &\int_0^{\dot{\varepsilon}_{\text{BP}}^2} \eta_{\text{BP}}(s)\ ds = \frac{2n}{n+1} \eta_{\text{BP}}\left(\theta, \mathbf{u}_h \right) \dot{\varepsilon}_{\text{BP}}^2,
\end{align}
and $\Lambda_{\text{BP}}$ defined similarly to (\ref{dukowicz_lambda}),
\begin{align}
  \label{bp_dukowicz_lambda}
  \Lambda_{\text{BP}} = - \mathbf{n} \cdot \sigma_{\text{BP}} \cdot \mathbf{n} \approx 0.
\end{align}

The source code of CSLVR includes an implementation similar to Code Listing \ref{cslvr_first_order}.

\begin{python}[label=cslvr_first_order, caption={CSLVR source code contained in the \texttt{MomentumDukowiczBP} class.}]
# define variational problem :
U        = Function(model.Q2, name = 'G')
dU       = TrialFunction(model.Q2)
Phi      = TestFunction(model.Q2)
phi, psi = Phi
du,  dv  = dU
u,   v   = U

# vertical velocity :
dw     = TrialFunction(model.Q)
chi    = TestFunction(model.Q)
w      = Function(model.Q, name='w_f')

# viscous dissipation :
U3     = as_vector([u,v,0])
epsdot = self.effective_strain_rate(U3)
if linear:
  s  = "    - using linear form of momentum using model.U3 in epsdot -"
  U3_c = model.U3.copy(True)
  eta  = self.viscosity(U3_c)
  Vd   = 2 * eta * epsdot
else:
  s  = "    - using nonlinear form of momentum -"
  eta  = self.viscosity(U3)
  Vd   = (2*n)/(n+1) * A_f**(-1/n) * (epsdot + eps_reg)**((n+1)/(2*n))
print_text(s, self.color())
  
# potential energy :
Pe     = - rhoi * g * (u*S.dx(0) + v*S.dx(1))

# dissipation by sliding :
Sl     = - 0.5 * beta * (u**2 + v**2)

# pressure boundary :
Pb     = (rhoi*g*(S - z) - rhosw*g*D) * (u*N[0] + v*N[1])

# action :
A      = + (Vd_gnd - Pe)*dx - Sl*dBed_g - Pb*dBed_f - Pb*dLat_t

# the first variation of the action in the direction of a 
# test function; the extremum :
self.mom_F = derivative(A, U, Phi)

# the first variation of the extremum in the direction 
# a tril function; the Jacobian :
self.mom_Jac = derivative(self.mom_F, U, dU)

self.w_F = + (u.dx(0) + v.dx(1) + dw.dx(2))*chi*dx \
           + (u*N[0] + v*N[1] + dw*N[2] - Fb)*chi*dBed
 
def strain_rate_tensor(self, U):
  """
  return the Dukowicz 'Blatter-Pattyn' simplified strain-rate tensor of <U>.
  """
  u,v,w  = U
  epi    = 0.5 * (grad(U) + grad(U).T)
  epi02  = 0.5*u.dx(2)
  epi12  = 0.5*v.dx(2)
  epi22  = -u.dx(0) - v.dx(1)  # incompressibility
  epsdot = as_matrix([[epi[0,0],  epi[0,1],  epi02],
                      [epi[1,0],  epi[1,1],  epi12],
                      [epi02,     epi12,     epi22]])
  return epsdot
  
def effective_strain_rate(self, U):
  """
  return the Dukowicz BP effective strain rate squared.
  """
  epi    = self.strain_rate_tensor(U)
  ep_xx  = epi[0,0]
  ep_yy  = epi[1,1]
  ep_zz  = epi[2,2]
  ep_xy  = epi[0,1]
  ep_xz  = epi[0,2]
  ep_yz  = epi[1,2]
  
  # Second invariant of the strain rate tensor squared
  epsdot = + ep_xx**2 + ep_yy**2 + ep_xx*ep_yy \
           + ep_xy**2 + ep_xz**2 + ep_yz**2
  return epsdot

def default_solve_params(self):
  """ 
  Returns a set of default solver parameters that yield good performance
  """
  nparams = {'newton_solver' :
            {
              'linear_solver'            : 'cg',
              'preconditioner'           : 'hypre_amg',
              'relative_tolerance'       : 1e-5,
              'relaxation_parameter'     : 0.7,
              'maximum_iterations'       : 25,
              'error_on_nonconvergence'  : False,
              'krylov_solver'            :
              {
                'monitor_convergence'   : False,
              }
            }}
  return m_params

def solve_pressure(self, annotate=False):
  """
  Solve for the Dukowicz BP pressure to model.p.
  """
  p   = project(rhoi*g*(S - z) + 2*eta*w.dx(2),
                annotate=annotate)


def solve_vert_velocity(self, annotate=False):
  """

  Perform the Newton solve of the first-order equations 
  """
  aw       = assemble(lhs(self.w_F))
  Lw       = assemble(rhs(self.w_F))
  w_solver = LUSolver(self.solve_params['vert_solve_method'])
  w_solver.solve(aw, self.w.vector(), Lw, annotate=annotate)
  
def solve(self, annotate=False):
  """ 
  Perform the Newton solve of the first-order equations 
  """
  # zero out self.velocity for good convergence for any subsequent solves,
  # e.g. model.L_curve() :
  model.assign_variable(self.get_U(), DOLFIN_EPS, cls=self)
  
  # compute solution :
  solve(self.mom_F == 0, self.U, J = self.mom_Jac,
        annotate = annotate, solver_parameters = params['solver'])
  u, v = self.U.split()
\end{python}

\section{Plane-strain approximation} \label{ssn_plane_strain}

\index{Stokes equations!Applied to ice, plane-strain}
\index{Stokes equations!Slip-friction}
Many observations of the ice lie along $x,y$-coordinate transects.  In order to explain these observations, the \emph{plane-strain} momentum balance model \citep{hill} has been formulated for ice, and is based on the assumption that longitudinal stress and lateral shear are present only in the direction of velocity $\mathbf{u}$.  Using this model with flow specified in the $x$-direction, all $y$-component terms of stress tensor (\ref{stress_tensor}) and strain-rate tensor (\ref{strain_rate_tensor}) are eliminated, producing the two-dimensional model tensors \index{Tensor!Plane-strain stress} \index{Tensor!Plane-strain stain rate}
\begin{align}
  \label{ps_tensors}
  \sigma_{\text{PS}} = \begin{bmatrix}
             \sigma_{xx} & \sigma_{xz} \\
             \sigma_{zx} & \sigma_{zz}
           \end{bmatrix}, \hspace{5mm} \text{and} \hspace{5mm}
  \tilde{\dot{\epsilon}} = \begin{bmatrix}
                             \dot{\epsilon}_{xx} & \dot{\epsilon}_{xz} \\
                             \dot{\epsilon}_{zx} & \dot{\epsilon}_{zz}
                           \end{bmatrix}.
\end{align}
Effective strain-rate (\ref{effective_strain_rate}) is therefore reduced to
\begin{align}
  \label{ps_effective_strain_rate}
  \dot{\varepsilon}_{\text{PS}}^2 = &\frac{1}{2} \Bigg[ \tilde{\dot{\epsilon}}_{xx}^2 + \tilde{\dot{\epsilon}}_{zz}^2 + 2\tilde{\dot{\epsilon}}_{xz}^2 \Bigg],
\end{align}
which is used within the plane-strain viscosity
\begin{align}
  \label{ps_viscosity}
  \eta_{\text{PS}}(\theta, \mathbf{u}_p) &= \frac{1}{2}A(\theta)^{-\nicefrac{1}{n}} (\dot{\varepsilon}_{\text{PS}} + \dot{\varepsilon}_0)^{\frac{1-n}{n}},
\end{align} 
with $xz$-plane velocity $\mathbf{u}_p = [u\ w]\T$.  The plane-strain Stokes system analogous to full-Stokes system (\ref{cons_momentum}, \ref{cons_mass}, \ref{surface_stress} -- \ref{impenetrability}) with stress tensor $\sigma_{\text{PS}} = 2\eta_{\text{PS}} \tilde{\dot{\epsilon}} - p I$ consists of
\begin{align}
  \label{ps_cons_momentum}
  -\nabla \cdot \sigma_{\text{PS}} &= \rho\mathbf{g}_p &&\text{ in } \Omega &&\leftarrow \text{ momentum} \\
  \label{ps_cons_mass}
  \nabla \cdot \mathbf{u}_p &= 0 &&\text{ in } \Omega &&\leftarrow \text{ mass}  \\
  \label{ps_surface_stress}
  \sigma_{\text{PS}} \cdot \mathbf{n}_p &= \mathbf{0}_p &&\text{ on } \Gamma_A &&\leftarrow \text{ stress-free surface} \\
  \label{ps_water_stress}
  \sigma_{\text{PS}} \cdot \mathbf{n}_p &= -f_w \mathbf{n}_p &&\text{ on } \Gamma_W &&\leftarrow \text{ water pressure} \\
  \label{ps_basal_drag}
  \big( \sigma_{\text{PS}} \cdot \mathbf{n}_p \big)_{\Vert} &= -\beta \mathbf{u}_p &&\text{ on } \Gamma_G &&\leftarrow \text{ basal traction } \\
  \label{ps_impenetrability}
  \mathbf{u}_p \cdot \mathbf{n}_p &= 0 &&\text{ on } \Gamma_B &&\leftarrow \text{ impenetrability,}
\end{align}
with outward-pointing normal vector to the boundary $\mathbf{n}_p = [n_x\ n_z]^\intercal$, gravitational acceleration vector $\mathbf{g}_p = [0\ \text{-}g]\T$, water pressure $f_w$ as defined by (\ref{water_pressure}), and basal-traction coefficient $\beta \geq 0$.

\subsection{Plane-strain variational principle} \label{ssn_plane_strain_var_prin}

\index{Variational principle!Plane-strain}
Proceeding in an identical fashion as \S \ref{ssn_full_stokes_var_prin} and \S \ref{ssn_first_order_var_prin}, the associated action for plane-strain momentum balance (\ref{ps_cons_momentum} -- \ref{ps_impenetrability}) is
\begin{align}
  \label{ps_action}
  \pazocal{A}_{\text{PS}} \left(\mathbf{u}_p, p\right) = &+ \int_{\Omega} \left( V\left( \dot{\varepsilon}_{\text{PS}}^2 \right) - \rho \mathbf{g}_p \cdot \mathbf{u}_p - p \nabla \cdot \mathbf{u}_p \right)\ d\Omega \notag \\
  &+ \int_{\Gamma_B} \left( \Lambda_{\text{PS}} \mathbf{u}_p \cdot \mathbf{n}_p + \frac{1}{2} \beta \mathbf{u}_p \cdot \mathbf{u}_p \right)\ d\Gamma_B \notag \\
  &+ \int_{\Gamma_L} f_w \mathbf{n}_p \cdot \mathbf{u}_p\ d\Gamma_L,
\end{align}
where $\Lambda_{\text{PS}}$ is defined similarly to (\ref{dukowicz_lambda}) and (\ref{bp_dukowicz_lambda}),
\begin{align}
  \label{ps_dukowicz_lambda}
  \Lambda_{\text{PS}} = - \mathbf{n}_p \cdot \sigma_{\text{PS}} \cdot \mathbf{n}_p \approx p.
\end{align}
and with viscous dissipation term $V\left( \dot{\varepsilon}_{\text{PS}}^2 \right)$ defined from the same process leading to (\ref{viscous_dissipation}) and (\ref{bp_viscous_dissipation}), 
\begin{align}
  \label{ps_viscous_dissipation}
  V\left( \dot{\varepsilon}_{\text{PS}}^2 \right) &= \int_0^{\dot{\varepsilon}_{\text{PS}}^2} \eta_{\text{PS}}(s)\ ds = \frac{2n}{n+1} \eta_{\text{PS}}\left(\theta, \mathbf{u}_p \right) \dot{\varepsilon}_{\text{PS}}^2.
\end{align}
Finally, the extremum of action (\ref{ps_action}) is given by the solution $(\mathbf{u}_p, p)$ of
\begin{align}
  \label{ps_extremum}
  \frac{\delta \pazocal{A}_{\text{PS}}}{\delta \mathbf{u}_p} = 0, \hspace{10mm} \frac{\delta \pazocal{A}_{\text{PS}}}{\delta p} = 0,
\end{align}
and are equivalent to Euler-Lagrange plane-strain Stokes equations and boundary conditions (\ref{ps_cons_momentum} -- \ref{ps_impenetrability}).

The source code of CSLVR uses an implementation similar to Code Listing \ref{cslvr_plane_strain}.

\begin{python}[label=cslvr_plane_strain, caption={CSLVR source code contained in the \texttt{MomentumDukowiczPlaneStrain} class.}]
# define variational problem :
U               = Function(model.Q3, name = 'G')
dU              = TrialFunction(model.Q3)
Phi             = TestFunction(model.Q3)
phi, xsi, kappa = Phi
du,  dw,  dP    = dU
u,   w,   p     = U

# create velocity vector :
U2     = as_vector([u,w])

# viscous dissipation :
epsdot  = self.effective_strain_rate(U2)
if linear:
  s  = "    - using linear form of momentum using model.U3 in epsdot -"
  U3_c = model.U3.copy(True)
  U3_2 = as_vector([U3_c[0], U3_c[1]])
  eta  = self.viscosity(U3_2)
  Vd   = 2 * eta * epsdot
else:
  s  = "    - using nonlinear form of momentum -"
  eta  = self.viscosity(U2)
  Vd   = (2*n)/(n+1) * A_f**(-1/n) * (epsdot + eps_reg)**((n+1)/(2*n))
print_text(s, self.color())

# potential energy :
Pe     = - rhoi * g * w

# dissipation by sliding :
Ut     = U2 - dot(U2,N)*N
Sl     = - 0.5 * beta * dot(Ut, Ut)

# incompressibility constraint :
Pc     = p * div(U2) 

# impenetrability constraint :
sig    = self.stress_tensor(U2, p, eta)
lam    = - dot(N, dot(sig, N))
Nc     = -lam * (dot(U2,N) - Fb)

# pressure boundary :
Pb_w   = - rhosw*g*D * dot(U2,N)
Pb_l   = - rhoi*g*(S - z) * dot(U2,N)

# action :
A      = + (Vd - Pe - Pc)*dx - Nc*dBed \
         - Sl*dBed_g - Pb_w*dBed_f - Pb_w*dLat_t

# add lateral boundary conditions :  
if use_lat_bcs:
  s = "    - using internal divide lateral stress natural boundary" + \
      " conditions -"
  print_text(s, self.color())
  U3_c     = model.U3.copy(True)
  U3_2     = as_vector([U3_c[0], U3_c[1]])
  eta_l    = self.viscosity(U3_2)
  sig_l    = self.stress_tensor(U3_2, model.p, eta_l)
  #sig_l   = self.stress_tensor(U2, p, eta)
  A -= dot(dot(sig_l, N), U2) * dLat_d

# the first variation of the action in the direction of a 
# test function; the extremum :
self.mom_F = derivative(A, U, Phi)

# the first variation of the extremum in the direction 
# a tril function; the Jacobian :
self.mom_Jac = derivative(self.mom_F, U, dU)
  
def strain_rate_tensor(self, U):
  """
  return the strain-rate tensor of self.U.
  """
  epsdot = 0.5 * (grad(U) + grad(U).T)
  return epsdot

def effective_strain_rate(self, U):
  """
  return the effective strain rate squared.
  """
  epi    = self.strain_rate_tensor(U)
  ep_xx  = epi[0,0]
  ep_zz  = epi[1,1]
  ep_xz  = epi[0,1]
  
  # Second invariant of the strain rate tensor squared
  epsdot = 0.5 * (ep_xx**2 + ep_zz**2) + ep_xz**2
  return epsdot

def default_solve_params(self):
  """ 
  Returns a set of default solver parameters that yield good performance
  """
  nparams = {'newton_solver' : {'linear_solver'            : 'mumps',
                                'relative_tolerance'       : 1e-5,
                                'relaxation_parameter'     : 0.7,
                                'maximum_iterations'       : 25,
                                'error_on_nonconvergence'  : False}}
  m_params  = {'solver'      : nparams}
  return m_params

def solve(self, annotate=False):
  """ 
  Perform the Newton solve of the full-Stokes equations 
  """
  # zero out self.velocity for good convergence for any subsequent solves,
  # e.g. model.L_curve() :
  model.assign_variable(self.get_U(), DOLFIN_EPS, cls=self)
  
  # compute solution :
  solve(self.mom_F == 0, self.U, J = self.mom_Jac, bcs = self.mom_bcs,
        annotate = annotate, solver_parameters = params['solver'])
  u, w, p = self.U.split()
\end{python}

\section{Reformulated full-Stokes} \label{ssn_reformulated_stokes}

\index{Stokes equations!Applied to ice, reformulated-Stokes}
\index{Stokes equations!Slip-friction}
A novel method introduced by \citet{dukowicz_2012} utilized the foundation built by the action principles presented in \citet{dukowicz_2010} and \citet{dukowicz_2011}.  This method specifies the use of a velocity trial function that satisfies continuity equation (\ref{cons_mass}) and impenetrability condition (\ref{impenetrability}), and results in the elimination of Lagrange multipliers $p$ and $\Lambda$ in action (\ref{action}).  A version of this method has been incorporated into the CSLVR code, and varies only slightly from that presented by \citet{dukowicz_2012}.

The first step in generating the velocity trial space is to express vertical velocity component $w$ is terms of the horizontal velocity components $u$ and $v$, in a fashion similar to (\ref{bp_vertical_velocity}).  To this end, we solve the first-order BVP for the vertical velocity component $w^h$
\begin{subequations}
  \label{rs_vertical_velocity_pde}
  \begin{eqnarray}
  \label{rs_cons_mass}
  \nabla \cdot \mathbf{u}_w = 0 &\text{ in } \Omega \\
  \label{rs_impenetrability}
  \mathbf{u}_w \cdot \mathbf{n} = 0 &\text{ on } \Gamma_B,
  \end{eqnarray}
\end{subequations}
where $\mathbf{u}_w = [u\ v\ w^h]\T$ is reformulated velocity vector with previously computed horizontal velocity components $u$ and $v$.  The associated variational problem for (\ref{rs_vertical_velocity_pde}) reads: find $w^h \in S_E^h \subset \mathcal{H}^1(\Omega)$ (see trial space (\ref{trial_space})) such that
\begin{align}
  \label{rs_vertical_velocity_var_prob}
  \int_{\Omega} \left( \frac{\partial u}{\partial x} + \frac{\partial v}{\partial y} + \frac{\partial w^h}{\partial z} \right)\chi\ d\Omega + \int_{\Gamma_B} \left( u n_x + v n_y + w^h n_z \right) \chi\ d\Gamma_B &= 0,
\end{align}
for all $\chi \in S_0^h \subset \mathcal{H}^1(\Omega)$ (see test space (\ref{test_space})).  This system must be numerically calculated in tandem with the process determining the horizontal velocity components via a fixed-point or Picard iteration.

Next, strain-rate tensor (\ref{strain_rate_tensor}) is expressed in terms of reformulated velocity $\mathbf{u}_w$, \index{Tensor!Reformulated-Stokes strain-rate}
\begin{align}
  \label{rs_strain_rate_tensor}
  \tilde{\dot{\epsilon}}
  &= \begin{bmatrix}
       \frac{\partial u}{\partial x} & \frac{1}{2}\left( \frac{\partial u}{\partial y} + \frac{\partial v}{\partial x} \right) & \frac{1}{2}\left( \frac{\partial u}{\partial z} + \frac{\partial w^h}{\partial x} \right) \\
       \frac{1}{2}\left( \frac{\partial v}{\partial x} + \frac{\partial u}{\partial y} \right) & \frac{\partial v}{\partial y} & \frac{1}{2}\left( \frac{\partial v}{\partial z} + \frac{\partial w^h}{\partial y} \right) \\
       \frac{1}{2}\left( \frac{\partial w^h}{\partial x} + \frac{\partial u}{\partial z} \right) & \frac{1}{2}\left( \frac{\partial w^h}{\partial y} + \frac{\partial v}{\partial z} \right) & -\left( \frac{\partial u}{\partial x} + \frac{\partial v}{\partial y} \right)
     \end{bmatrix},
\end{align}
where incompressibility constraint (\ref{cons_mass}) has been used to express the $zz$-component.  The second invariant of this tensor provides the reformulated-Stokes effective strain-rate
\begin{align}
  \dot{\varepsilon}_{\text{RS}}^2 = & \frac{1}{2} \tr\left( \tilde{\dot{\epsilon}}^2 \right) = \frac{1}{2} \Bigg[ \tilde{\dot{\epsilon}}_{ij} \tilde{\dot{\epsilon}}_{ij} \Bigg] \notag \\
  \label{rs_effective_strain_rate}
  = &\frac{1}{2} \Bigg[ \tilde{\dot{\epsilon}}_{xx}^2 + \tilde{\dot{\epsilon}}_{yy}^2 + \tilde{\dot{\epsilon}}_{zz}^2 + 2\tilde{\dot{\epsilon}}_{xy}^2 + 2\tilde{\dot{\epsilon}}_{xz}^2 + 2\tilde{\dot{\epsilon}}_{yz}^2 \Bigg],
\end{align}
and reformulated-Stokes shear viscosity derived identically to viscosity (\ref{viscosity}),
\begin{align}
  \label{rs_viscosity}
  \eta_{\text{RS}}(\theta, \mathbf{u}_w) &= \frac{1}{2}A(\theta)^{-\nicefrac{1}{n}} (\dot{\varepsilon}_{\text{RS}} + \dot{\varepsilon}_0)^{\frac{1-n}{n}}.
\end{align} 

To eliminate the pressure dependence on the momentum balance we assume that the pressure is entirely cryostatic, such that $p = f_c = \rho g (S - z)$.  It follows from the same procedure used to derive first-order quasi-stress tensor (\ref{bp_stress_tensor}) that the reformulated-Stokes stress tensor under these assumptions is \index{Tensor!Reformulated-Stokes stress}
\begin{align}
  \label{rs_stress_tensor}
  \sigma_{\text{RS}} &=
  2 \eta_{\text{RS}} 
  \begin{bmatrix}
       \left( 2\frac{\partial u}{\partial x} + \frac{\partial v}{\partial y} \right) & \frac{1}{2}\left( \frac{\partial u}{\partial y} + \frac{\partial v}{\partial x} \right) & \frac{1}{2} \left( \frac{\partial u}{\partial z} + \frac{\partial w^h}{\partial x} \right) \\
       \frac{1}{2}\left( \frac{\partial v}{\partial x} + \frac{\partial u}{\partial y} \right) & \left( 2\frac{\partial v}{\partial y} + \frac{\partial u}{\partial x} \right) & \frac{1}{2} \left( \frac{\partial v}{\partial z} + \frac{\partial w^h}{\partial y} \right) \\
       \frac{1}{2}\left( \frac{\partial w^h}{\partial x} + \frac{\partial u}{\partial z} \right) & \frac{1}{2}\left( \frac{\partial w^h}{\partial y} + \frac{\partial v}{\partial z} \right) & -\left( \frac{\partial u}{\partial x} + \frac{\partial v}{\partial y} \right) - f_c
     \end{bmatrix}.
\end{align}
Using this stress-tensor definition in place of $\sigma$ in momentum-balance (\ref{cons_momentum}), while making use of the facts that $\partial_z f_c = -\rho g$ and $\partial_z S = 0$, result in the reformulated momentum balance $\nabla \cdot \sigma_{\text{RS}} = \rho g \nabla S$.  Therefore, the complete reformulated-Stokes momentum balance analogous to full-Stokes system (\ref{cons_momentum}, \ref{cons_mass}, \ref{surface_stress} -- \ref{impenetrability}) and first-order system (\ref{bp_cons_momentum}, \ref{bp_exterior_stress}, \ref{bp_basal_drag}) consists of
\begin{align}
  \label{rs_cons_momentum}
  \nabla \cdot \sigma_{\text{RS}} &= \rho g \nabla S &&\text{ in } \Omega &&\leftarrow \text{ momentum} \\
  \label{rs_exterior_stress}
  \sigma_{\text{RS}} \cdot \mathbf{n} &= \left(f_c - f_w \right) \mathbf{n} &&\text{ on } \Gamma_E &&\leftarrow \text{ exterior pressure} \\
  \label{rs_basal_drag}
  \big( \sigma_{\text{RS}} \cdot \mathbf{n} \big)_{\Vert} &= -\beta \mathbf{u}_w &&\text{ on } \Gamma_G &&\leftarrow \text{ basal traction }
\end{align}
with exterior boundary $\Gamma_E = \Gamma_A \cup \Gamma_W$, outward-pointing normal vector to the boundary $\mathbf{n} = [n_x\ n_y\ n_z]^\intercal$, gravitational acceleration vector $\mathbf{g}_p = [0\ 0\ \text{-}g]\T$, water pressure $f_w$ as defined by (\ref{water_pressure}), cryostatic pressure \index{Cryostatic stress} $f_c(z) = p(z) = \rho g (S - z)$, and basal-traction coefficient $\beta \geq 0$.

Note once again that the solution of reformulated system (\ref{rs_vertical_velocity_pde}, \ref{rs_cons_momentum} -- \ref{rs_basal_drag}) requires a fixed-point iteration whereby at iterate $k$, vertical velocity $w_k^h$ is coupled to a given horizontal velocity solution $u_{k-1},v_{k-1}$ via variational problem (\ref{rs_vertical_velocity_var_prob}).  See \S \ref{ssn_stokes_variational_forms} for details of the implementation used by CSLVR to accomplish this coupling.

\subsection{Reformulated-Stokes variational principle} \label{ssn_reformulated_stokes_var_prin}

\index{Variational principle!Reformulated-Stokes}
Proceeding in an identical fashion as \S \ref{ssn_full_stokes_var_prin}, \S \ref{ssn_first_order_var_prin}, and \S \ref{ssn_plane_strain_var_prin}, the associated action for reformulated-Stokes system (\ref{ps_cons_momentum} -- \ref{ps_impenetrability}) is
\begin{align}
  \label{rs_action_intermediate}
  \pazocal{A}_{\text{RS}} \left(\mathbf{u}_w, p\right) = &+ \int_{\Omega} \left( V\left( \dot{\varepsilon}_{\text{RS}}^2 \right) - \rho \mathbf{g} \cdot \mathbf{u}_w \right)\ d\Omega \notag \\
  &+ \int_{\Gamma_B} \frac{1}{2} \beta \mathbf{u}_w \cdot \mathbf{u}_w\ d\Gamma_B \notag \\
  &+ \int_{\Gamma_L} f_w \mathbf{n} \cdot \mathbf{u}_w\ d\Gamma_L,
\end{align}
with viscous dissipation term $V\left( \dot{\varepsilon}_{\text{RS}}^2 \right)$ defined from the same process leading to (\ref{viscous_dissipation}), (\ref{bp_viscous_dissipation}), and (\ref{ps_viscous_dissipation}), 
\begin{align}
  \label{rs_viscous_dissipation}
  V\left( \dot{\varepsilon}_{\text{RS}}^2 \right) &= \int_0^{\dot{\varepsilon}_{\text{RS}}^2} \eta_{\text{RS}}(s)\ ds = \frac{2n}{n+1} \eta_{\text{RS}}\left(\theta, \mathbf{u}_w \right) \dot{\varepsilon}_{\text{RS}}^2.
\end{align}

Reformulated-Stokes action (\ref{rs_action_intermediate}) was simplified in Appendix A of \citet{dukowicz_2012} by forming the expression for the gravitational work term
\begin{align}
  \label{rs_gravity_work}
  \int_{\Omega} \rho \mathbf{g} \cdot \mathbf{u}_w\ d\Omega = \int_{\Omega} \rho g \mathbf{u}_h \cdot \left( \nabla S \right)_h\ d\Omega,
\end{align}
where horizontal vector components are denoted $\mathbf{g}_h = [g_x\ g_y]^\intercal$.  Additionally, a basal vertical velocity term analogous to expression (\ref{bp_bed_vertical_velocity}) derived from impenetrability condition (\ref{rs_impenetrability}) is used to reduce the basal-traction term in (\ref{rs_action_intermediate}) to the expression
\begin{align}
  \label{rs_basal_traction}
  \int_{\Gamma_B} \frac{1}{2} \beta \mathbf{u}_w \cdot \mathbf{u}_w\ d\Gamma_B = \int_{\Gamma_B} \frac{1}{2} \beta \left( u^2 + v^2 + \left( \frac{u n_x + v n_y}{n_z} \right)^2 \right)\ d\Gamma_B.
\end{align}
Inserting (\ref{rs_gravity_work}) and (\ref{rs_basal_traction}) into (\ref{rs_action_intermediate}) results in the final reformulated-Stokes action
\begin{align}
  \label{rs_action}
  \pazocal{A}_{\text{RS}} \left(\mathbf{u}_w, p\right) = &+ \int_{\Omega} \left( V\left( \dot{\varepsilon}_{\text{RS}}^2 \right) - \rho g \mathbf{u}_h \cdot \left( \nabla S \right)_h \right)\ d\Omega \notag \\
  &+ \int_{\Gamma_B} \frac{1}{2} \beta \left( \mathbf{u}_h \cdot \mathbf{u}_h + \left( \frac{\mathbf{u}_h \cdot \mathbf{n}_h}{n_z} \right)^2 \right)\ d\Gamma_B \notag \\
  &+ \int_{\Gamma_L} f_w \mathbf{n} \cdot \mathbf{u}_w\ d\Gamma_L,
\end{align}
with extremum given by
\begin{align}
  \label{rs_extremum}
  \frac{\delta \pazocal{A}_{\text{RS}}}{\delta \mathbf{u}_h} = 0,
\end{align}
and produces the unique minimizer $(\mathbf{u}_h, p)$.  It was also shown by \citet{dukowicz_2012} that (\ref{rs_extremum}) is equivalent to reformulated-Stokes Euler-Lagrange momentum equations and boundary conditions (\ref{ps_cons_momentum} -- \ref{ps_basal_drag}).

The source code of CSLVR uses an implementation similar to Code Listing \ref{cslvr_reformulated_stokes}; note the use of Newton-Raphson Code Listing \ref{home_rolled_method} in the \texttt{solve} method.

\begin{python}[label=cslvr_reformulated_stokes, caption={CSLVR source code contained in the \texttt{MomentumDukowiczStokesReduced} class.}]
# define variational problem :
U        = Function(model.Q2, name = 'G')
dU       = TrialFunction(model.Q2)
Phi      = TestFunction(model.Q2)
phi, psi = Phi
du,  dv  = dU
u,   v   = U

# vertical velocity :
dw       = TrialFunction(model.Q)
chi      = TestFunction(model.Q)
w        = Function(model.Q, name='w_f')
self.w_F = + (u.dx(0) + v.dx(1) + dw.dx(2))*chi*dx \
           + (u*N[0] + v*N[1] + dw*N[2] - Fb)*chi*dBed

# viscous dissipation :
U3       = as_vector([u,v,model.w])
epsdot   = self.effective_strain_rate(U3)
if linear:
  s      = "    - using linear form of momentum using model.U3 in epsdot -"
  Uc     = model.U3.copy(True)
  eta    = self.viscosity(Uc)
  Vd     = 2 * eta * epsdot
else:
  s      = "    - using nonlinear form of momentum -"
  eta    = self.viscosity(U3)
  Vd     = (2*n)/(n+1) * A_f**(-1/n) * (epsdot + eps_reg)**((n+1)/(2*n))
print_text(s, self.color())

# potential energy :
Pe       = - rhoi * g * (u*S.dx(0) + v*S.dx(1))

# dissipation by sliding :
w_b      = (Fb - u*N[0] - v*N[1]) / N[2]
Sl       = - 0.5 * beta * (u**2 + v**2 + w_b**2)

# pressure boundary :
Pb       = (rhoi*g*(S - z) - rhosw*g*D) * dot(U3, N)

# action :
A        = + Vd*dx - Pe*dx - Sl*dBed_g - Pb*dBed_f - Pb_w*dLat_t

# the first variation of the action in the direction of a 
# test function; the extremum :
self.mom_F = derivative(A, U, Phi)

# the first variation of the extremum in the direction 
# a tril function; the Jacobian :
self.mom_Jac = derivative(self.mom_F, U, dU)

def strain_rate_tensor(self, U):
  """
  return the strain-rate tensor for the velocity <U>.
  """
  u,v,w  = U
  epi    = 0.5 * (grad(U) + grad(U).T)
  epi22  = -u.dx(0) - v.dx(1)          # incompressibility
  epsdot = as_matrix([[epi[0,0],  epi[0,1],  epi[0,2]],
                      [epi[1,0],  epi[1,1],  epi[1,2]],
                      [epi[2,0],  epi[2,1],  epi22]])
  return epsdot

def effective_strain_rate(self, U):
  """
  return the effective strain rate squared.
  """
  epi    = self.strain_rate_tensor(U)
  ep_xx  = epi[0,0]
  ep_yy  = epi[1,1]
  ep_zz  = epi[2,2]
  ep_xy  = epi[0,1]
  ep_xz  = epi[0,2]
  ep_yz  = epi[1,2]
  
  # Second invariant of the strain rate tensor squared
  epsdot = 0.5 * (+ ep_xx**2 + ep_yy**2 + ep_zz**2) \
                  + ep_xy**2 + ep_xz**2 + ep_yz**2
  return epsdot

def default_solve_params(self):
  """ 
  Returns a set of default solver parameters that yield good performance
  """
  nparams = {'newton_solver' :
            {
              'linear_solver'            : 'cg',
              'preconditioner'           : 'hypre_amg',
              'relative_tolerance'       : 1e-5,
              'relaxation_parameter'     : 0.7,
              'maximum_iterations'       : 25,
              'error_on_nonconvergence'  : False,
              'krylov_solver'            :
              {
                'monitor_convergence'   : False,
                #'preconditioner' :
                #{
                #  'structure' : 'same'
                #}
              }
            }}
  m_params  = {'solver'               : nparams,
               'solve_vert_velocity'  : True,
               'solve_pressure'       : True,
               'vert_solve_method'    : 'mumps'}
  return m_params

def solve_vert_velocity(self, annotate=annotate):
  """ 
  Solve for vertical velocity w.
  """
  s    = "::: solving Dukowicz reduced vertical velocity :::"
  print_text(s, self.color())
  
  aw       = assemble(lhs(self.w_F))
  Lw       = assemble(rhs(self.w_F))
  
  w_solver = LUSolver(self.solve_params['vert_solve_method'])
  w_solver.solve(aw, self.w.vector(), Lw, annotate=annotate)

def solve(self, annotate=False):
  """ 
  Perform the Newton solve of the reduced full-Stokes equations 
  """
  # zero out self.velocity for good convergence for any subsequent solves,
  # e.g. model.L_curve() :
  model.assign_variable(self.get_U(), DOLFIN_EPS, cls=self)

  def cb_ftn():
    self.solve_vert_velocity(annotate)
  
  # compute solution :
  model.home_rolled_newton_method(self.mom_F, self.U, self.mom_Jac, 
                                  self.mom_bcs, atol=1e-6, rtol=rtol,
                                  relaxation_param=alpha, max_iter=maxit,
                                  method=lin_slv, preconditioner=precon,
                                  cb_ftn=cb_ftn)
  u, v = self.U.split()
\end{python}

%===============================================================================
%===============================================================================

\section{Mass loss due to basal melting} \label{ssn_mass_loss_due_to_melting}

\index{Impenetrability}
The mass loss due to melt-water flowing from the base of the ice due to internal and external friction has the effect of lowering the ice-sheet surface.  In terms of velocity, the water discharge from the ice $F_b$ -- in units of meters of ice equivalent per second -- transforms impenetrability conditions (\ref{impenetrability}), (\ref{ps_impenetrability}) and (\ref{rs_impenetrability}) to
\begin{align}
  \label{impenetrability_Fb}
  \mathbf{u} \cdot \mathbf{n} &= F_b  &&\text{ on } \Gamma_B \\
  \label{ps_impenetrability_Fb}
  \mathbf{u}_p \cdot \mathbf{n}_p &= F_b  &&\text{ on } \Gamma_B \\
  \label{rs_impenetrability_Fb}
  \mathbf{u}_w \cdot \mathbf{n} &= F_b  &&\text{ on } \Gamma_B.
\end{align}
Furthermore, because the ice velocity may no longer be tangential to the basal surface, basal-traction conditions (\ref{basal_drag}), (\ref{ps_basal_drag}) and (\ref{rs_basal_drag}) are transformed to 
\begin{align}
  \label{basal_drag_Fb}
  \big( \sigma \cdot \mathbf{n} \big)_{\Vert} &= -\beta \mathbf{u}_{\Vert} &&\text{ on } \Gamma_G \\
  \label{ps_basal_drag_Fb}
  \big( \sigma_{\text{PS}} \cdot \mathbf{n}_p \big)_{\Vert} &= -\beta \mathbf{u}_{p\Vert} &&\text{ on } \Gamma_G \\
  \label{rs_basal_drag_Fb}
  \big( \sigma_{\text{RS}} \cdot \mathbf{n} \big)_{\Vert} &= -\beta \mathbf{u}_{w\Vert} &&\text{ on } \Gamma_G.
\end{align}
Reformulation of the full-Stokes variational principle of \S \ref{ssn_full_stokes} utilizing basal-melt-adjusted boundary conditions (\ref{impenetrability_Fb}, \ref{basal_drag_Fb}) leads to
\begin{align}
  \label{action_Fb}
  \pazocal{A} \left(\mathbf{u}, p\right) = &+ \int_{\Omega} \left( V\left( \dot{\varepsilon}_e^2 \right) - \rho \mathbf{g} \cdot \mathbf{u} - p \nabla \cdot \mathbf{u} \right)\ d\Omega \notag \\
  &+ \int_{\Gamma_B} \left( \Lambda \left( \mathbf{u} \cdot \mathbf{n} - F_b \right) + \frac{1}{2} \beta \mathbf{u}_{\Vert} \cdot \mathbf{u}_{\Vert} \right)\ d\Gamma_B \notag \\
  &+ \int_{\Gamma_L} f_w \mathbf{n} \cdot \mathbf{u}\ d\Gamma_L,
\end{align}
where $\mathbf{u}_{\Vert} = \mathbf{u} - \left( \mathbf{u} \cdot \mathbf{n} \right) \mathbf{n}$.  Reformulation of the plane-Strain variational principle of \S \ref{ssn_plane_strain} utilizing basal-melt-adjusted boundary conditions (\ref{ps_impenetrability_Fb}, \ref{ps_basal_drag_Fb}) leads to
\begin{align}
  \label{ps_action_Fb}
  \pazocal{A}_{\text{PS}} \left(\mathbf{u}_p, p\right) = &+ \int_{\Omega} \left( V\left( \dot{\varepsilon}_{\text{PS}}^2 \right) - \rho \mathbf{g}_p \cdot \mathbf{u}_p - p \nabla \cdot \mathbf{u}_p \right)\ d\Omega \notag \\
  &+ \int_{\Gamma_B} \left( \Lambda_{\text{PS}} \left( \mathbf{u}_p \cdot \mathbf{n}_p - F_b \right) + \frac{1}{2} \beta \mathbf{u}_{p\Vert} \cdot \mathbf{u}_{p\Vert} \right)\ d\Gamma_B \notag \\
  &+ \int_{\Gamma_L} f_w \mathbf{n}_p \cdot \mathbf{u}_p\ d\Gamma_L,
\end{align}
where $\mathbf{u}_{p\Vert} = \mathbf{u}_p - \left( \mathbf{u}_p \cdot \mathbf{n}_p \right) \mathbf{n}_p$.

Closer examination of melt-adjusted impenetrability condition (\ref{impenetrability_Fb}),
\begin{align*}
  u n_x + v n_y + w n_z &= F_b  &&\text{ on } \Gamma_B,
\end{align*}
suggests that given $n_z \neq 0$, the melt-adjusted basal vertical velocity is given by
\begin{align}
  \label{bp_bed_vertical_velocity_Fb}
  w(B) &= \frac{F_b - u(B)n_x - v(B)n_y}{n_z}.
\end{align}
This expression is then be used in place of Equation (\ref{bp_bed_vertical_velocity}) to solve first-order vertical-velocity-component relation (\ref{bp_vertical_velocity}).  Finally, reformulation of the reformulated-Stokes variational principle of \S \ref{ssn_reformulated_stokes} utilizing basal-melt-adjusted boundary conditions (\ref{rs_impenetrability_Fb}, \ref{rs_basal_drag_Fb}) leads to
\begin{align}
  \label{rs_action_Fb}
  \pazocal{A}_{\text{RS}} \left(\mathbf{u}_w, p\right) = &+ \int_{\Omega} \left( V\left( \dot{\varepsilon}_{\text{RS}}^2 \right) - \rho g \mathbf{u}_h \cdot \left( \nabla S \right)_h \right)\ d\Omega \notag \\
  &+ \int_{\Gamma_B} \frac{1}{2} \beta \left( \mathbf{u}_h \cdot \mathbf{u}_h + \left( \frac{F_b - \mathbf{u}_h \cdot \mathbf{n}_h}{n_z} \right)^2 \right)\ d\Gamma_B \notag \\
  &+ \int_{\Gamma_L} f_w \mathbf{n} \cdot \mathbf{u}_w\ d\Gamma_L,
\end{align}
and reformulated variational problem (\ref{rs_vertical_velocity_var_prob}) governing the solution of $w^h$,
\begin{align}
  \label{rs_vertical_velocity_var_prob_Fb}
  + \int_{\Omega} \left( \frac{\partial u}{\partial x} + \frac{\partial v}{\partial y} + \frac{\partial w^h}{\partial z} \right)\chi\ d\Omega & \\
  + \int_{\Gamma_B} \left( u n_x + v n_y + w^h n_z - F_b \right) \chi\ d\Gamma_B &= 0.
\end{align}

These are the forms of the action principles solved by CSLVR, demonstrated by Code Listings \ref{cslvr_full_stokes}, \ref{cslvr_first_order}, \ref{cslvr_plane_strain}, and \ref{cslvr_reformulated_stokes}.

\section{Stokes variational forms} \label{ssn_stokes_variational_forms}
  
Each of momentum model equations (\ref{extremum}), (\ref{bp_extremum}), (\ref{ps_extremum}), and (\ref{rs_extremum}) is nonlinear due to the velocity dependence of the viscosity $\eta$, and as such cannot be solved directly.  Instead, we determine the $k$ unknown field variables by solving for the direction of decent of these actions using the method described in \S \ref{ssn_newton_raphson}.  This method forms the G\^{a}teaux derivative of each of model extremums (\ref{extremum}), (\ref{bp_extremum}), (\ref{ps_extremum}), and (\ref{rs_extremum}) with respect to a test function.

\index{Variational principle!Euler-Lagrange equations arising from}
\index{Variational principle!Variational forms}
The variational problem associated with full-Stokes action extremum (\ref{extremum}), first-order action extremum (\ref{bp_extremum}), reformulated-Stokes action extremum (\ref{rs_extremum}), and plane-strain action extremum (\ref{ps_extremum}) consists of finding (see trial space (\ref{trial_space})) $\mathbf{U} = [u\ v\ w\ p]\T \in \mathbf{S_E^h} \subset \left( \mathscr{H}^1(\Omega) \right)^4$, $\mathbf{U}_h = [u\ v]\T \in \mathbf{S_E^h} \subset \left( \mathscr{H}^1(\Omega) \right)^2$, or $\mathbf{U}_p = [u\ w\ p]\T \in \mathbf{S_E^h} \subset \left( \mathscr{H}^1(\Omega) \right)^3$ such that
\begin{align}
  \label{full_stokes_var_form}
  \lim_{\epsilon \rightarrow 0} \left\{ \frac{\delta}{\delta \mathbf{U}} \mathscr{A}(\mathbf{U} + \epsilon \bm{\Phi}) \right\} &= 0 \\
  \label{first_order_var_form}
  \lim_{\epsilon \rightarrow 0} \left\{ \frac{\delta}{\delta \mathbf{U}_h} \mathscr{A}_{\text{BP}}(\mathbf{U}_h + \epsilon \bm{\Phi}_h) \right\} &= 0 \\
  \label{reformulated_var_form}
  \lim_{\epsilon \rightarrow 0} \left\{ \frac{\delta}{\delta \mathbf{U}_h} \mathscr{A}_{\text{RS}}(\mathbf{U}_h + \epsilon \bm{\Phi}_h) \right\} &= 0 \\
  \label{plane_strain_var_form}
  \lim_{\epsilon \rightarrow 0} \left\{ \frac{\delta}{\delta \mathbf{U}_p} \mathscr{A}_{\text{BP}}(\mathbf{U}_p + \epsilon \bm{\Phi}_p) \right\} &= 0
\end{align}
for all test functions (see test space (\ref{test_space})) $\bm{\Phi} \in \mathbf{S_0^h} \subset \left( \mathscr{H}^1(\Omega) \right)^4$, $\bm{\Phi}_h \in \mathbf{S_0^h} \subset \left( \mathscr{H}^1(\Omega) \right)^2$, and $\bm{\Phi}_p \in \mathbf{S_0^h} \subset \left( \mathscr{H}^1(\Omega) \right)^3$.

Note that when using the method described in \S \ref{ssn_newton_raphson} to solve variational forms (\ref{full_stokes_var_form}), (\ref{first_order_var_form}), (\ref{reformulated_var_form}), and (\ref{plane_strain_var_form}), $\mathbf{U}$, $\mathbf{U}_h$, and $\mathbf{U}_p$ are not trial functions, but are instead containers for the current solution approximation.  In this case trial functions enter into the equations as the down-gradient direction of G\^{a}teaux derivatives (\ref{full_stokes_var_form}), (\ref{first_order_var_form}), (\ref{reformulated_var_form}), or (\ref{plane_strain_var_form}).  Additionally, while in the process of solving for the decent direction of the reformulated Stokes model $\delta_{\mathbf{U}_h} \mathscr{A}_{\text{RS}}$ (\ref{reformulated_var_form}), discrete variational form (\ref{rs_vertical_velocity_var_prob}) must be solved for each horizontal velocity approximation $\mathbf{U}_h$.

\section{ISMIP-HOM test simulations} \label{ssn_ismip_hom_test_simulations}

\begin{figure}
  \centering
    \includegraphics[width=\linewidth]{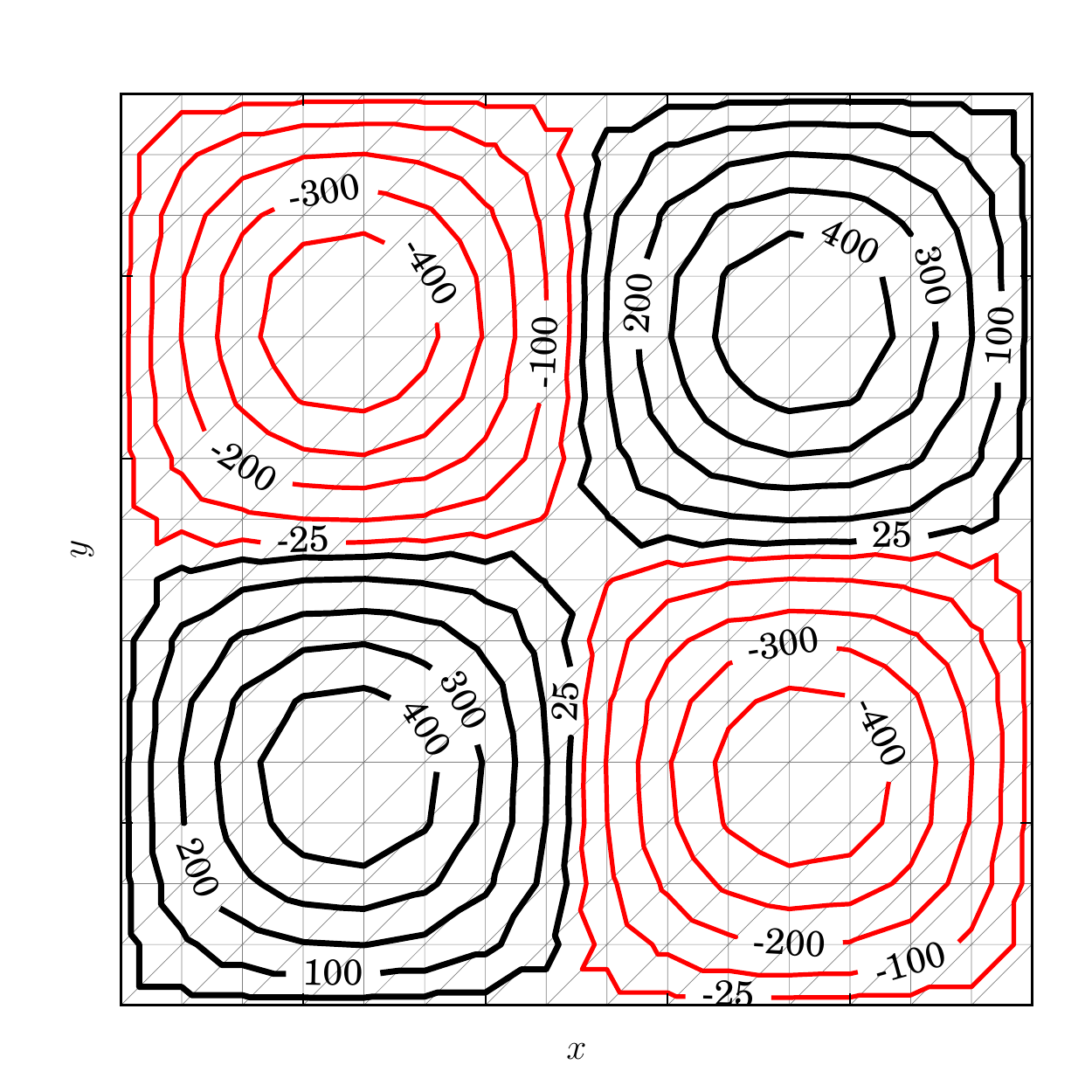}
  \caption[ISMIP-HOM bedrock topography]{ISMIP-HOM A bedrock topography deviation from the average $\hat{B} = B + \bar{B} - S$ over the $\ell \times \ell$ square km grid.  The CSLVR script used to generate the figure is shown in Code Listing \ref{ismip_hom_bed_code}.}
  \label{ismip_hom_a_B}
\end{figure}

\index{Linear differential equations!3D}
\index{ISMIP-HOM simulations}
A suitable test for the three-dimensional models defined by \S \ref{ssn_full_stokes}, \S \ref{ssn_first_order}, and \S \ref{ssn_reformulated_stokes} is the higher-order-ice-sheet-model-intercomparison project presented by \citet{ismip_hom}.  This test is defined over the domain $\Omega \in [0,\ell] \times [0,\ell] \times [B,S] \subset \R^3$ with $k_x \times k_y \times k_z$ node discretization, and specifies the use of a surface height with uniform slope $\Vert \nabla S \Vert = a$
\begin{align*}
  S(x) = - x \tan\left( a \right),
\end{align*}
and the sinusoidially-varying basal topography
\begin{align*}
  B(x,y) = S(x) - \bar{B} + b \sin\left( \frac{2 \pi}{\ell} x \right) \sin\left( \frac{2 \pi}{\ell} y \right),
\end{align*}
with average basal depth $\bar{B}$, and basal height amplitude $b$ (Figure \ref{ismip_hom_a_B}).  To enforce continuity, the periodic $\mathbf{u},p$ boundary conditions
\begin{align*}
  \mathbf{u}(0,0)    &= \mathbf{u}(\ell,\ell) & p(0,0)    &= p(\ell,\ell)\\
  \mathbf{u}(0,\ell) &= \mathbf{u}(\ell,0)    & p(0,\ell) &= p(\ell,0)   \\
  \mathbf{u}(x,0)    &= \mathbf{u}(x,\ell)    & p(x,0)    &= p(x,\ell)   \\
  \mathbf{u}(0,y)    &= \mathbf{u}(\ell,y)    & p(0,y)    &= p(\ell,y)
\end{align*}
were used.  Lastly, the basal traction coefficient is set to $\beta = 1000$ which has the effect of creating a no-slip boundary condition along the basal surface, while $A = 10\sups{-16}$ is used as an isothermal rate factor for viscosity $\eta$.  Table \ref{ismip_hom_values} lists these coefficients and values, and the CSLVR script used to solve this problem is shown in Code Listing \ref{ismip_hom_momentum_code}.

\begin{table}
\centering
\caption[ISMIP-HOM momemtum variables]{Variable values for ISMIP-HOM simulations.}
\label{ismip_hom_values}
\begin{tabular}{llll}
\hline
\textbf{Variable} & \textbf{Value} & \textbf{Units} & \textbf{Description} \\
\hline
$\dot{\varepsilon}_0$ & $10\sups{-15}$ & a\sups{-1}   & strain regularization \\
$\beta$   & $1000$          & kg m\sups{-2}a\sups{-1} & basal friction coef. \\
$A$       & $10\sups{-16}$  & Pa\sups{-3}a\sups{-1}   & flow-rate factor \\
$F_b$     & $0$             & m a\sups{-1}            & basal water discharge \\
%$\dot{a}$ & $0$             & m a\sups{-1}            & surface accumulation \\
$a$       & $0.5$           & $\circ$                 & surface gradient mag. \\
$\bar{B}$ & $1000$          & m & average basal depth \\
$b$       & $500$           & m & basal height amp.\\
$k_x$     & $15$            & -- & number of $x$ divisions \\
$k_y$     & $15$            & -- & number of $y$ divisions \\
$k_z$     & $5$             & -- & number of $z$ divisions \\
$N_e$     & $6750$          & -- & number of cells \\
$N_n$     & $1536$          & -- & number of vertices \\
\hline
\end{tabular}
\end{table}

\begin{figure*}
  \centering
    
    \includegraphics[width=0.319\linewidth]{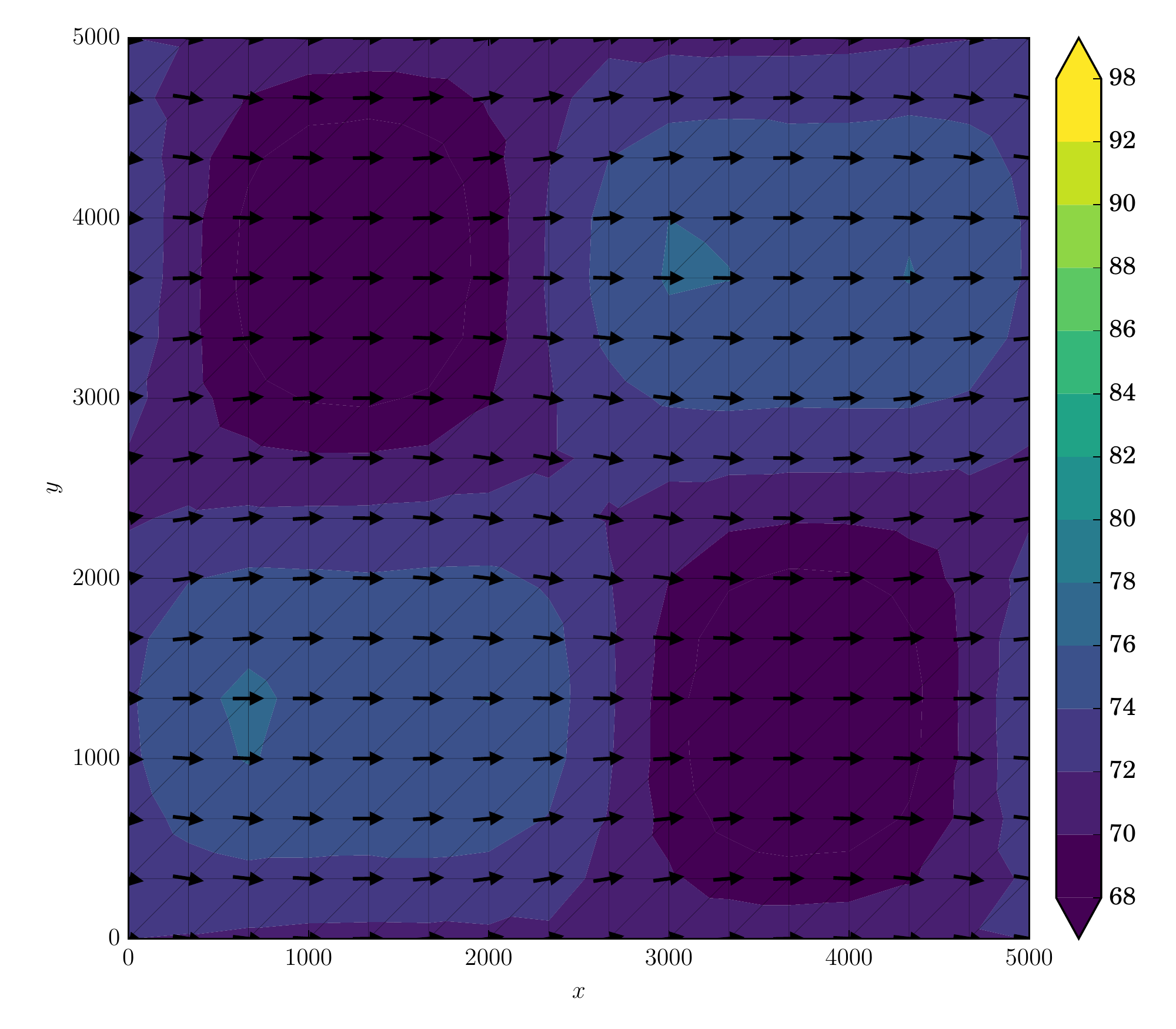}
    \quad
    \includegraphics[width=0.319\linewidth]{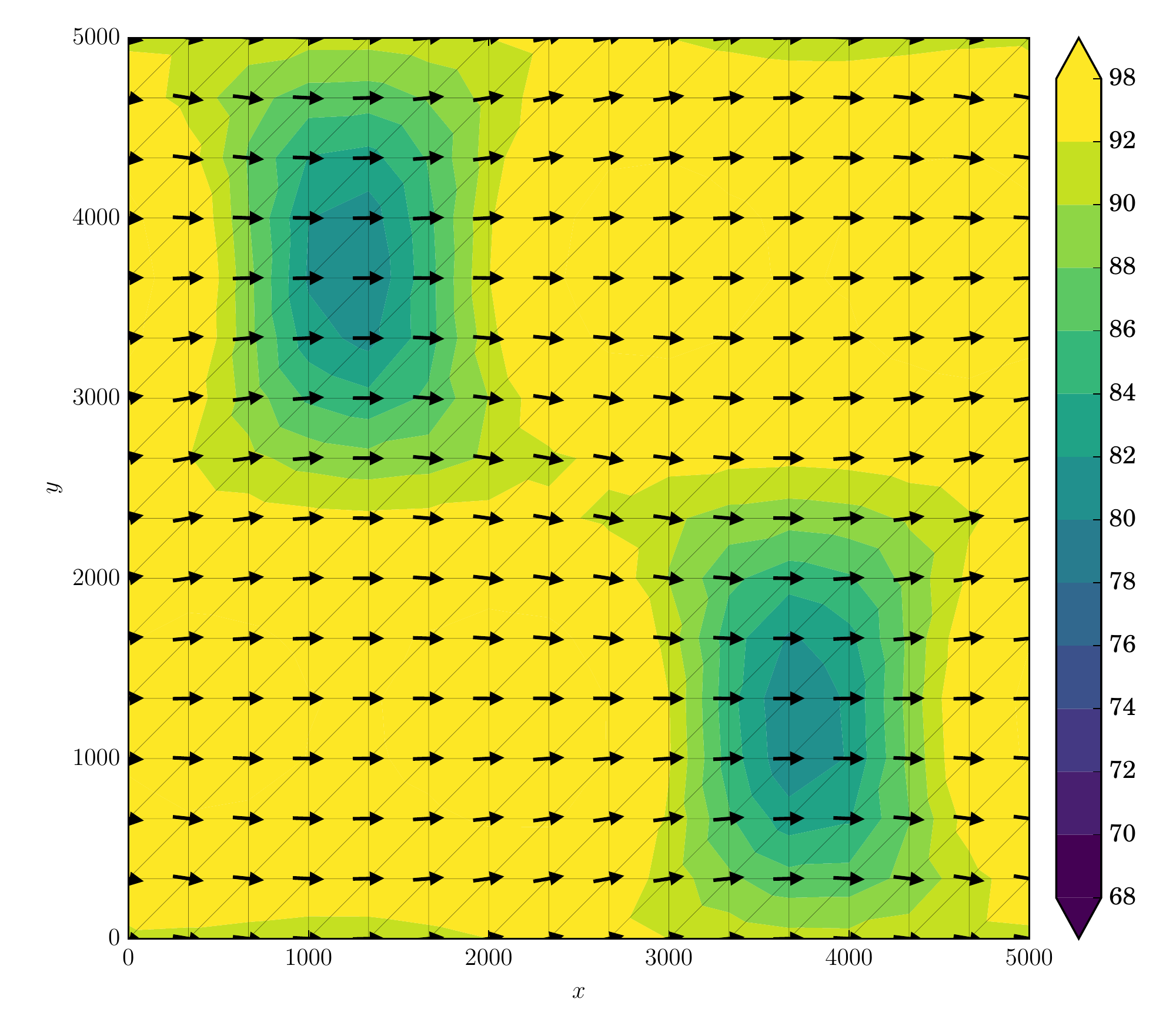}
    \quad
    \includegraphics[width=0.319\linewidth]{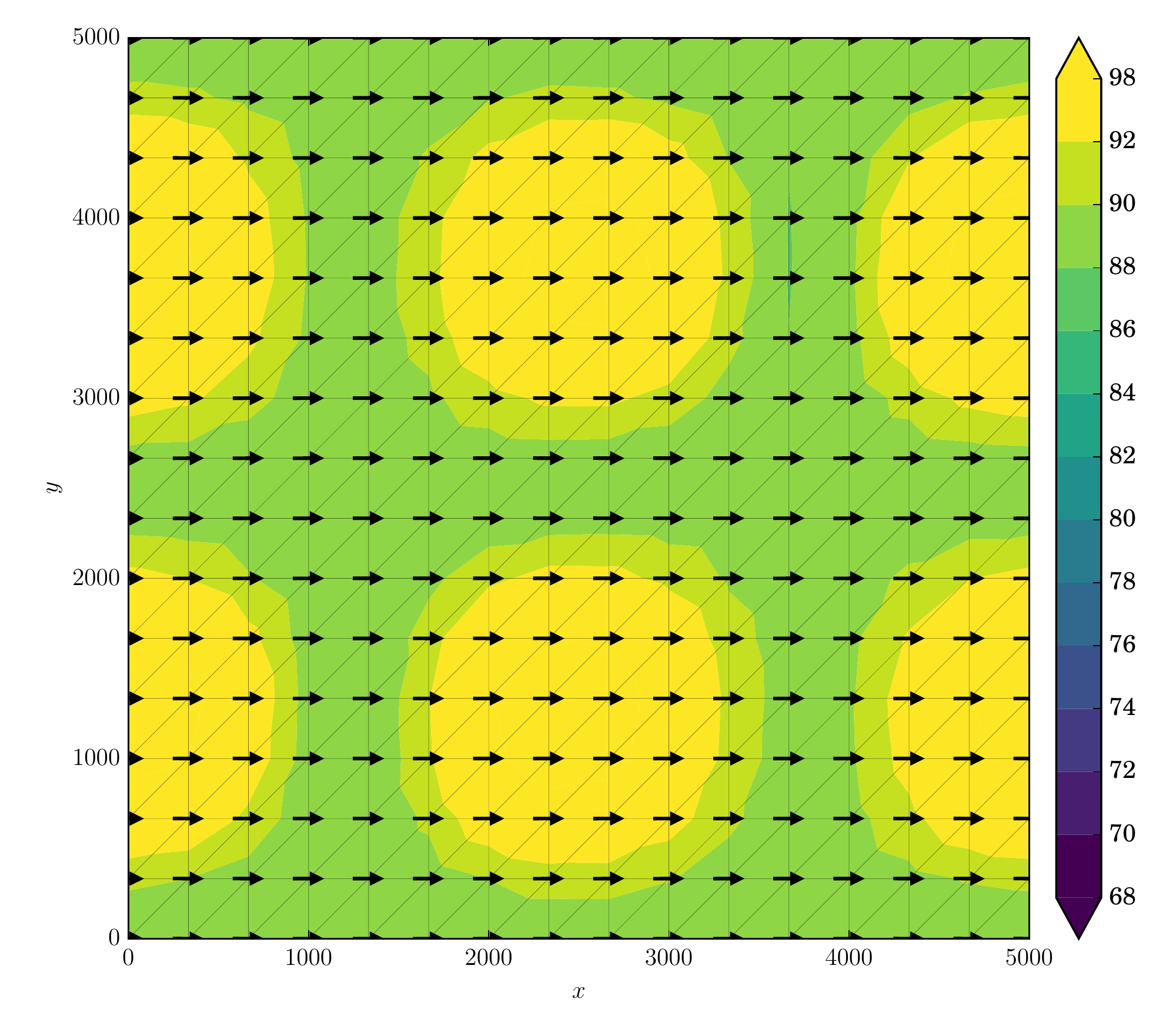}

    \includegraphics[width=0.319\linewidth]{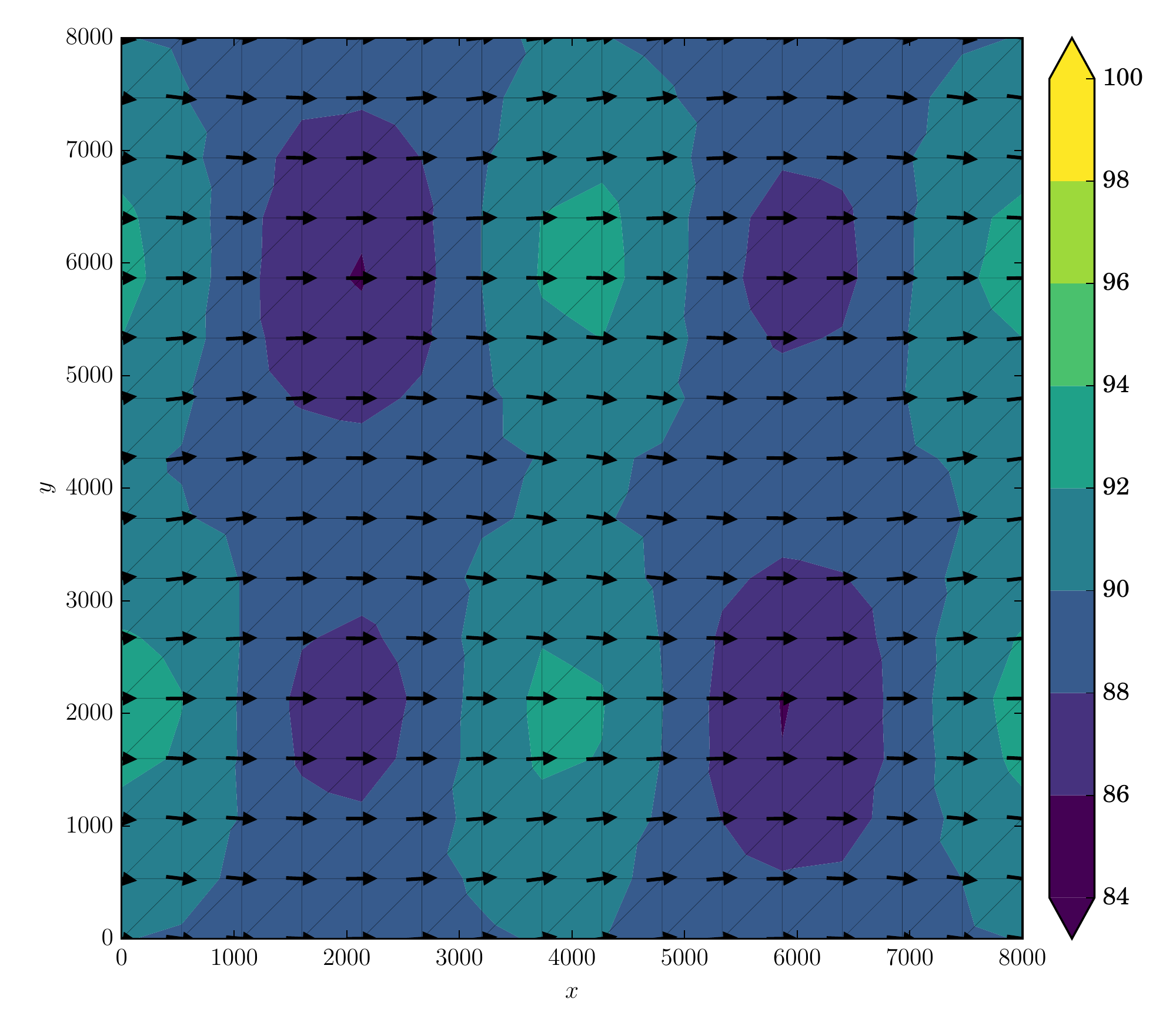}
    \quad
    \includegraphics[width=0.319\linewidth]{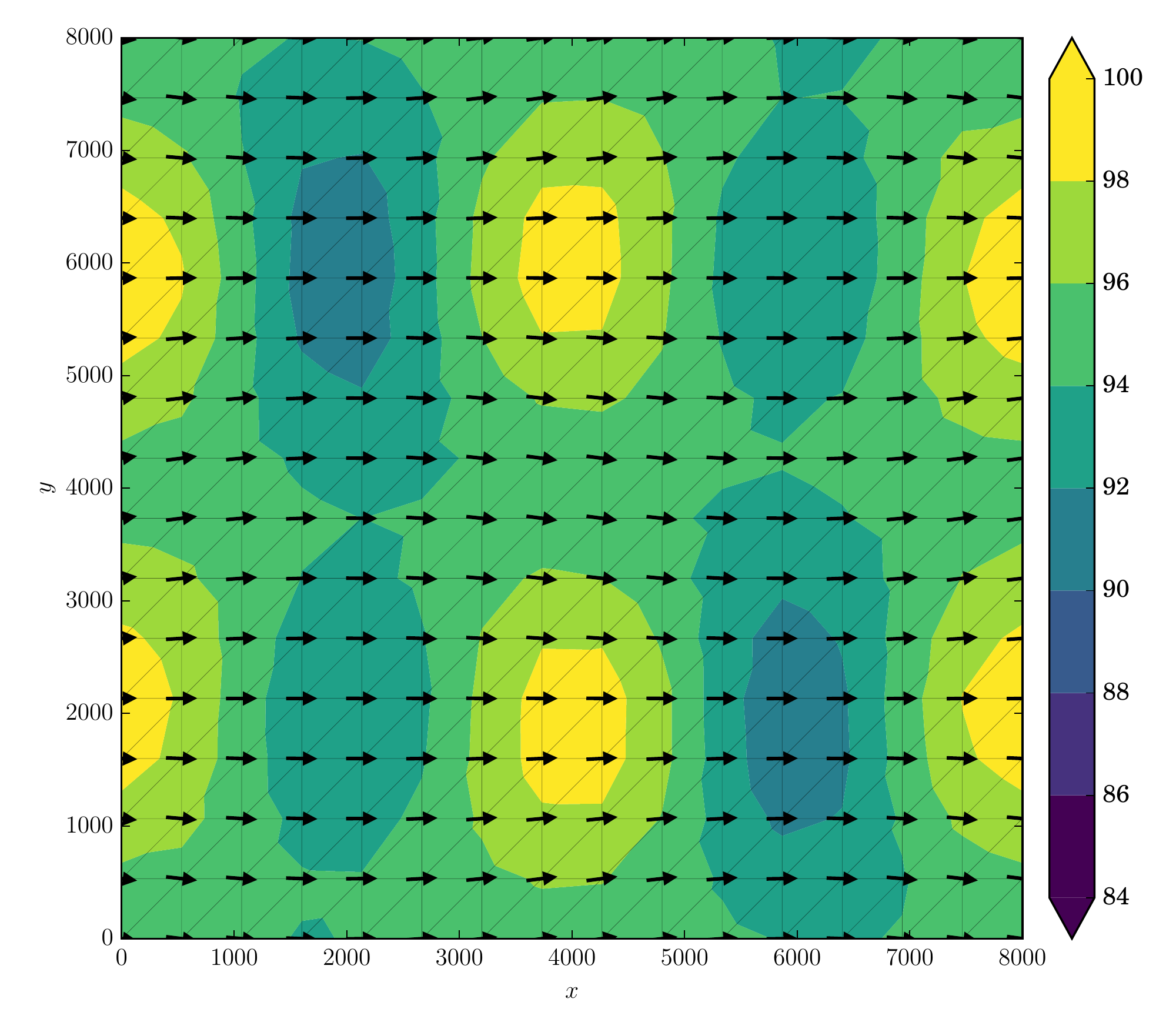}
    \quad
    \includegraphics[width=0.319\linewidth]{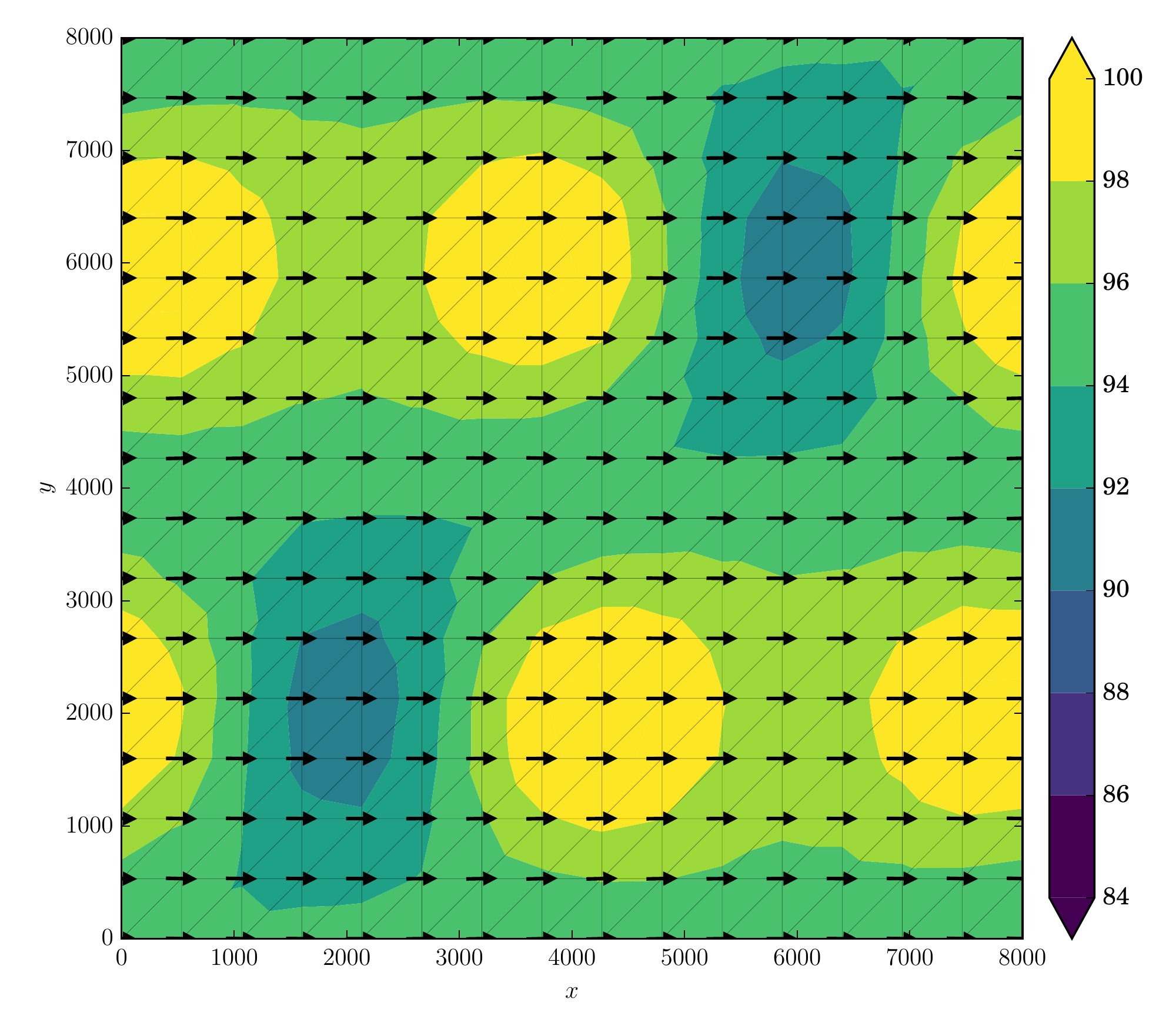}

    \includegraphics[width=0.319\linewidth]{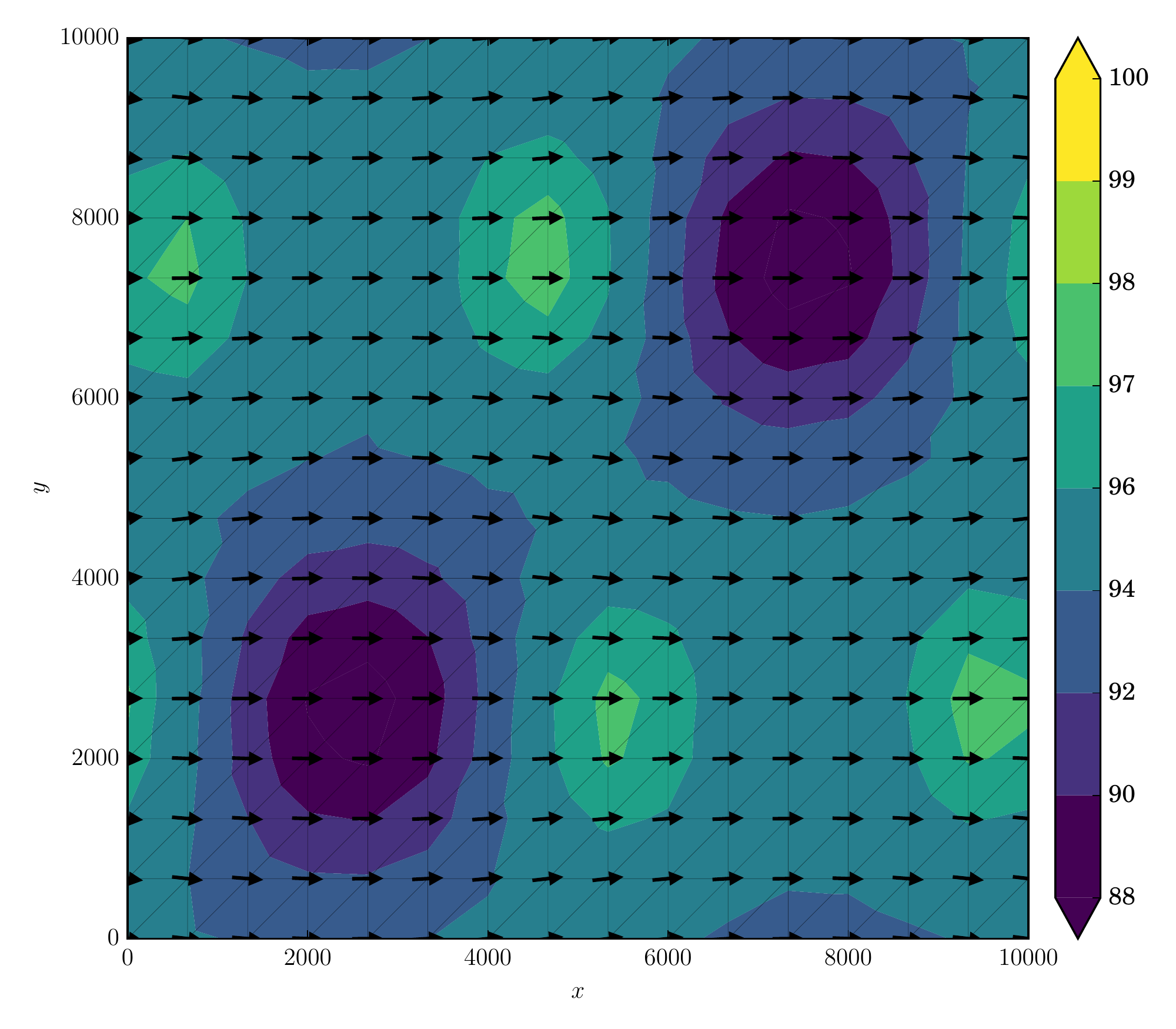}
    \quad
    \includegraphics[width=0.319\linewidth]{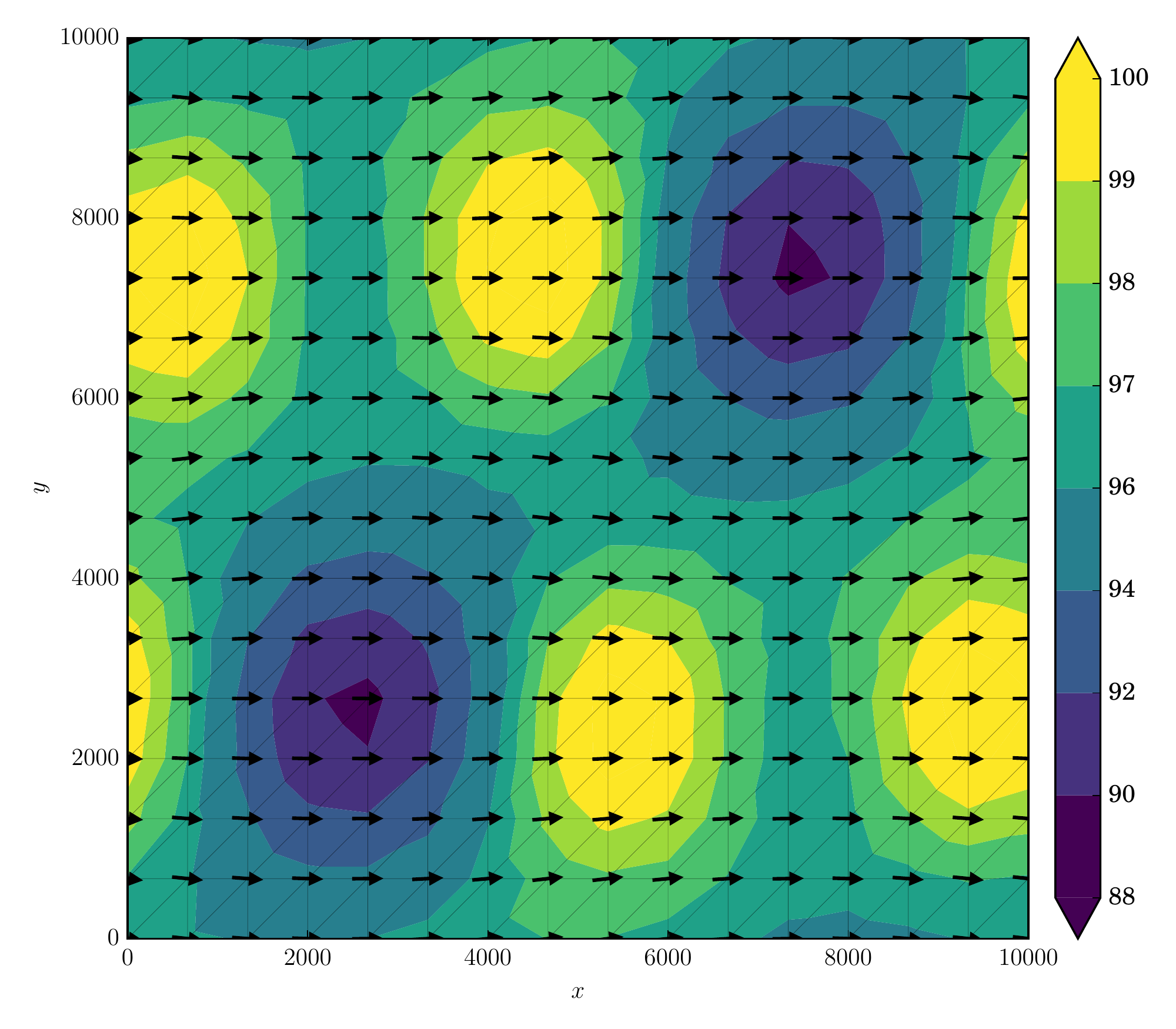}
    \quad
    \includegraphics[width=0.319\linewidth]{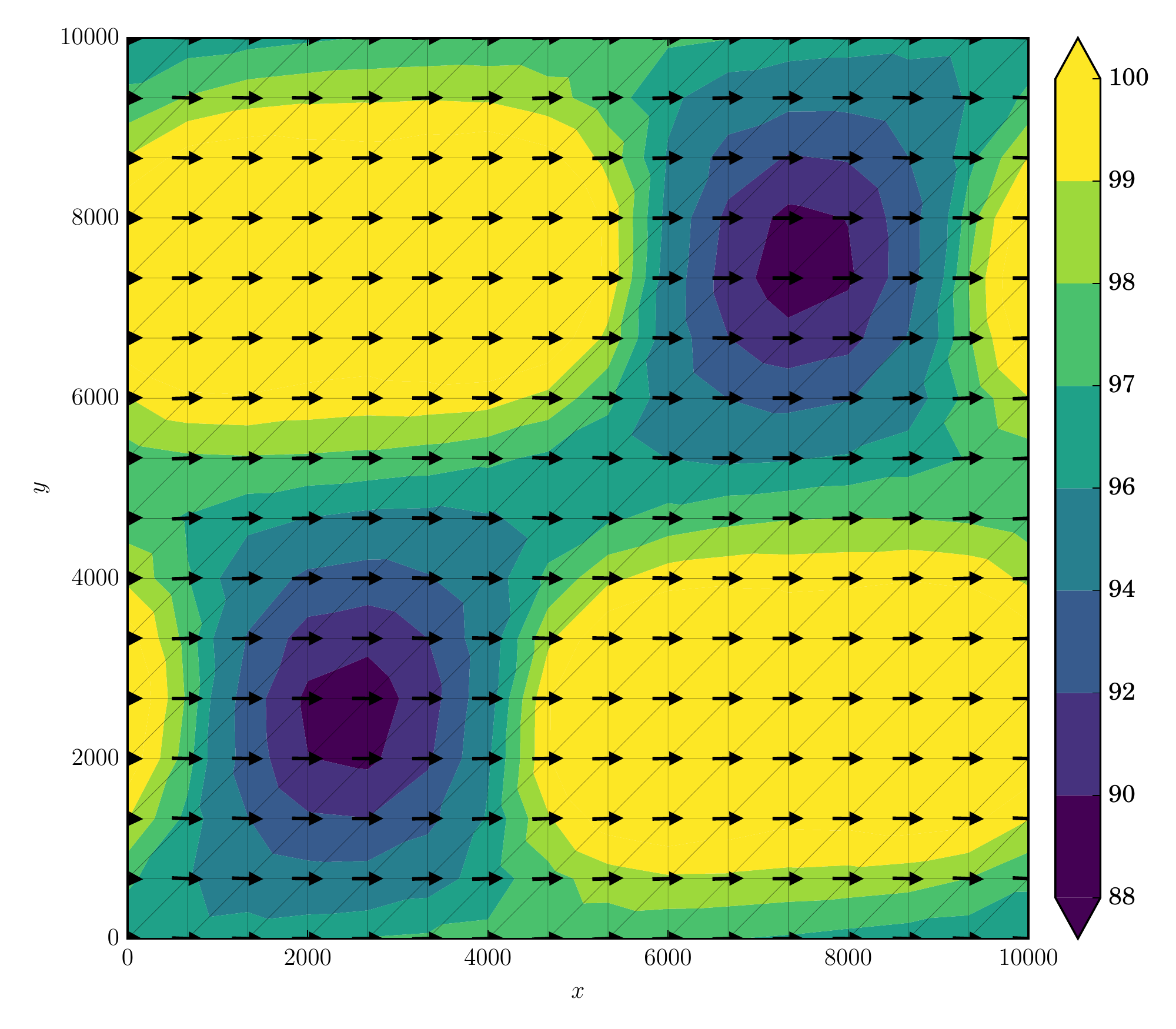}

    \includegraphics[width=0.319\linewidth]{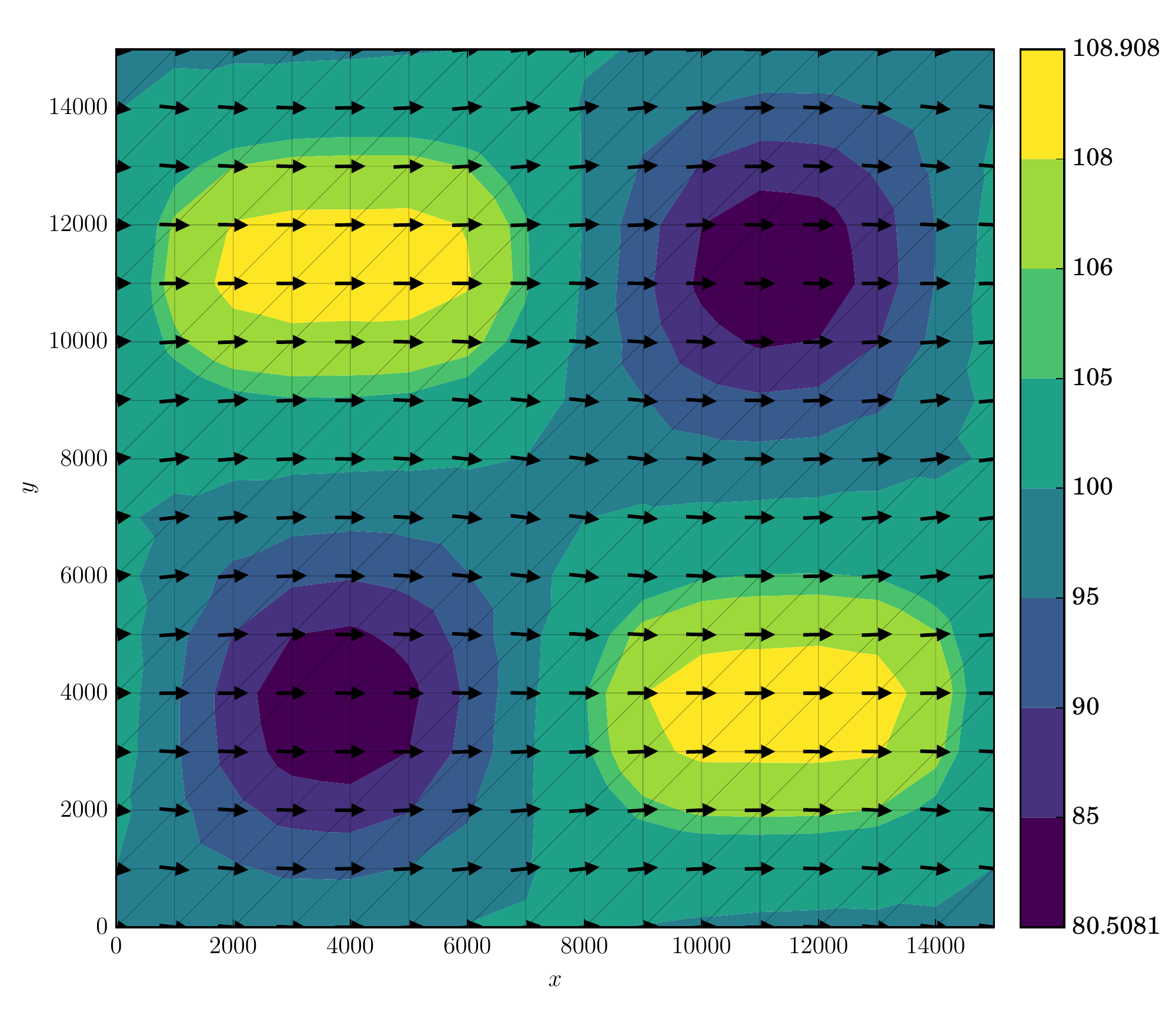}
    \quad
    \includegraphics[width=0.319\linewidth]{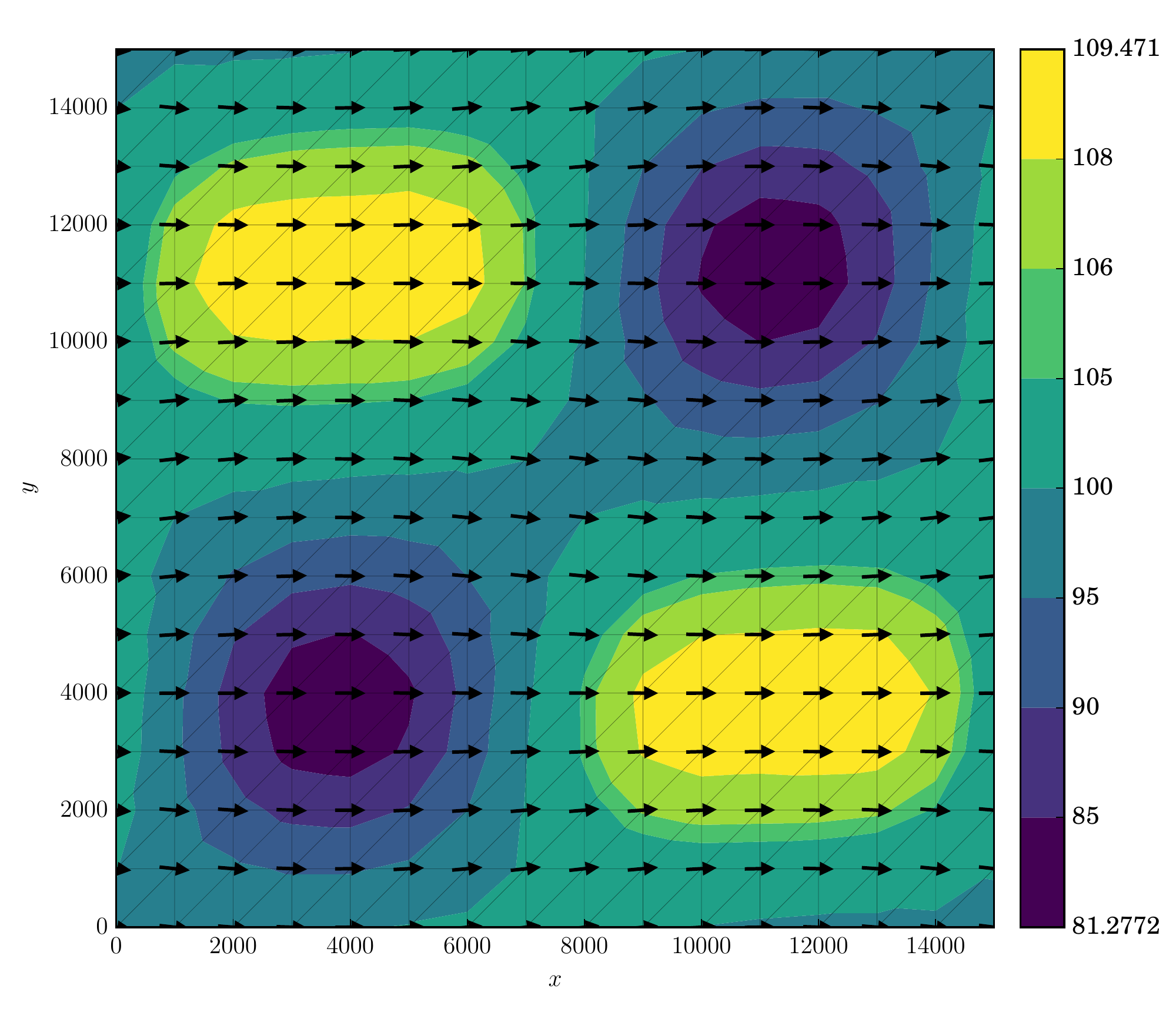}
    \quad
    \includegraphics[width=0.319\linewidth]{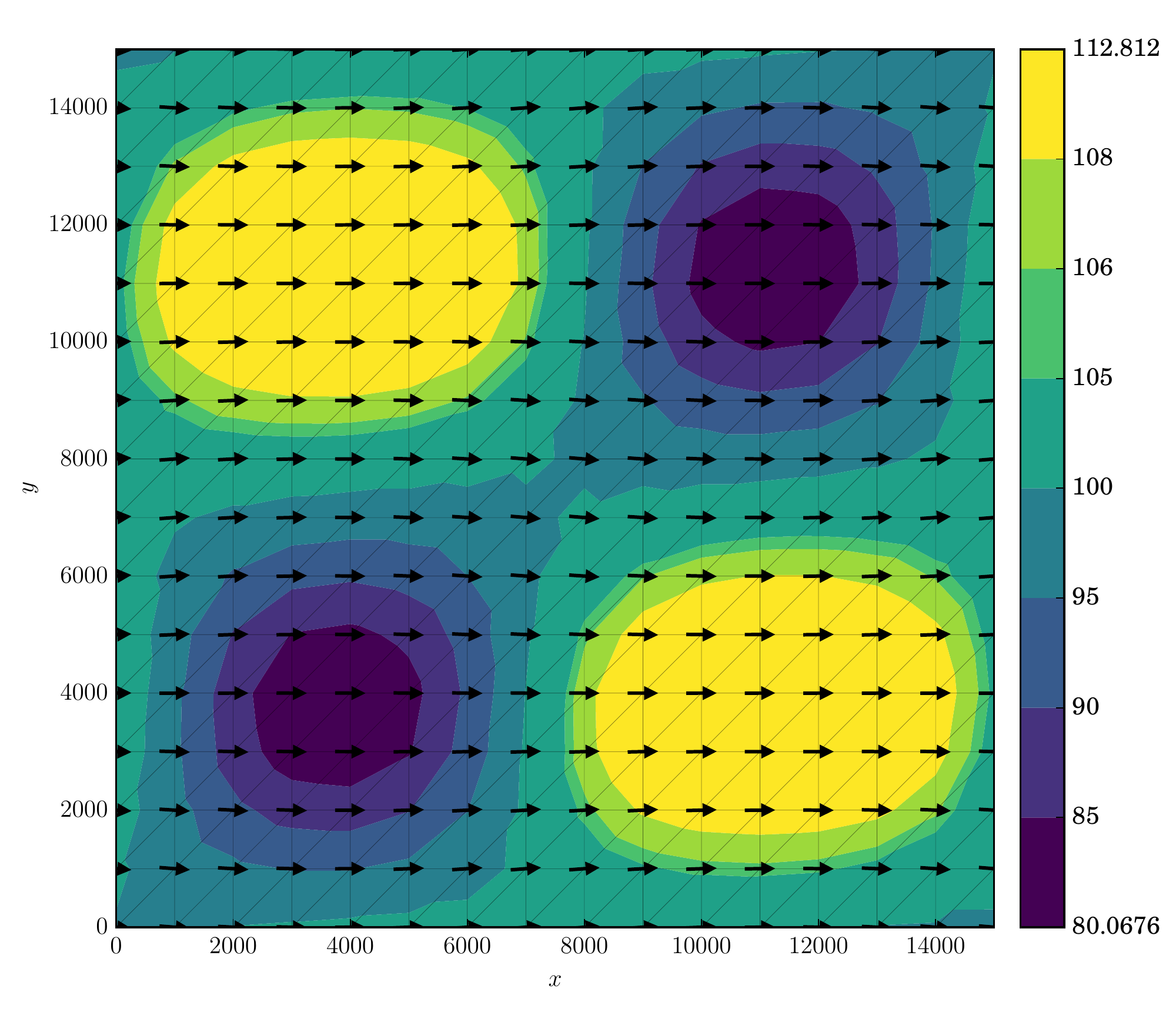}

  \caption[ISMIP-HOM momentum experiment velocities]{Surface velocity $\mathbf{u}_S$ for the ISMIP-HOM test experiment using a $5 \times 5$ square km grid (first row), $8 \times 8$ km$^2$ grid (second row), $10 \times 10$ km$^2$ grid (third row), and $15 \times 15$ km$^2$ grid (fourth row).  The first column are results attained using the full-Stokes model from \S \ref{ssn_full_stokes}, the second column using the first-order model from \S \ref{ssn_first_order}, and the third column was generated using the reformulated-Stokes model of \S \ref{ssn_reformulated_stokes}.}
  \label{ismip_hom_a_velocity}
\end{figure*}

The applicability of each of the \S \ref{ssn_full_stokes}, \S \ref{ssn_first_order}, and \S \ref{ssn_reformulated_stokes} momentum models for given basal gradient is tested by performing simulations over the range of maximum domain widths $\ell = $ 5 km, 8 km, 10 km, and 15 km.  Results indicate that the velocity approximated by the first-order and reformulated-Stokes models approach the full-Stokes solution as $\ell$ increases (Figure \ref{ismip_hom_a_velocity}).  The reformulated-Stokes model produces a quantitatively closer result to the full-Stokes model than the first-order model for all of the experiments performed.

\begin{figure*}
  \centering

    \includegraphics[width=0.319\linewidth]{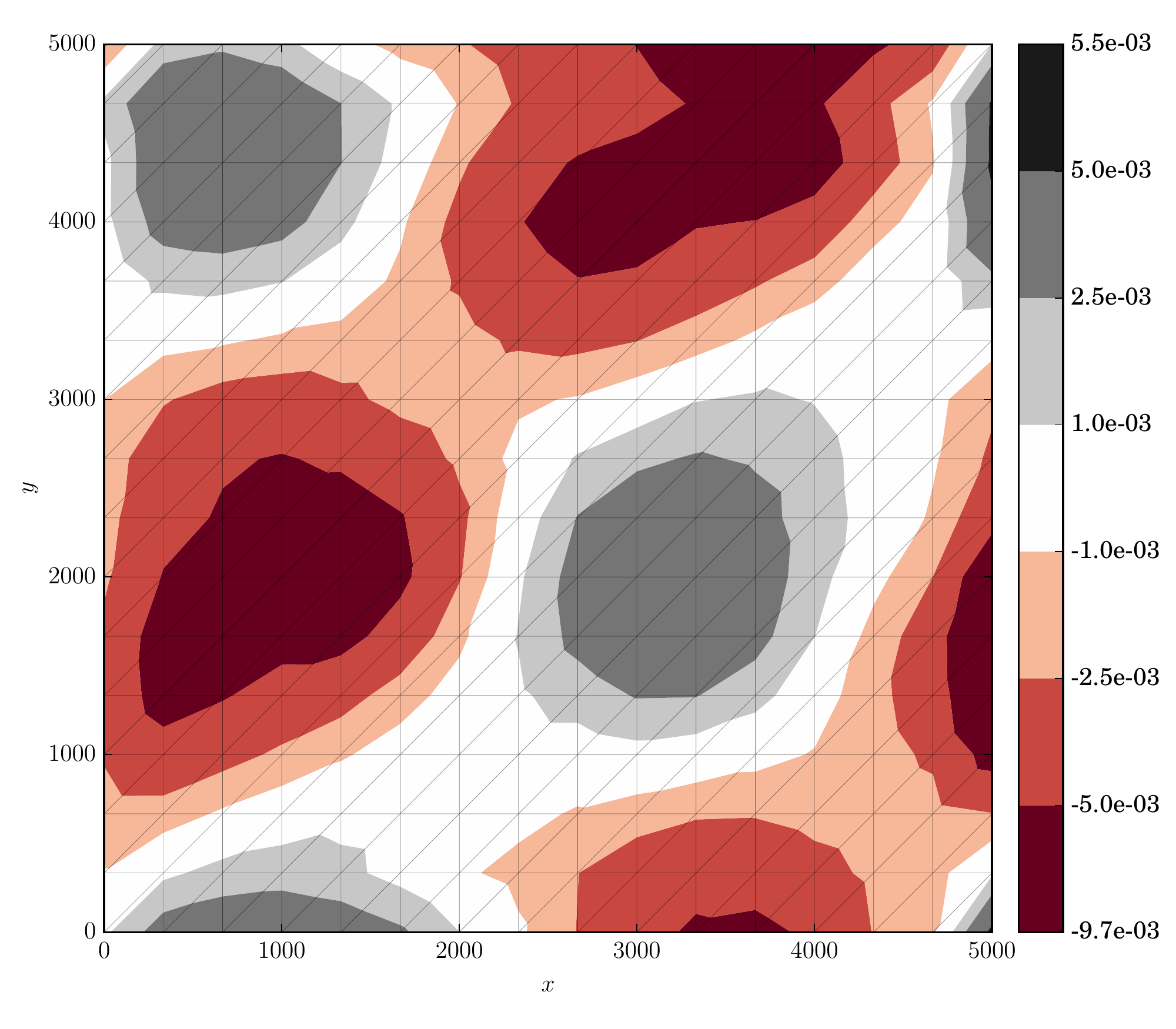}
    \quad                                             
    \includegraphics[width=0.319\linewidth]{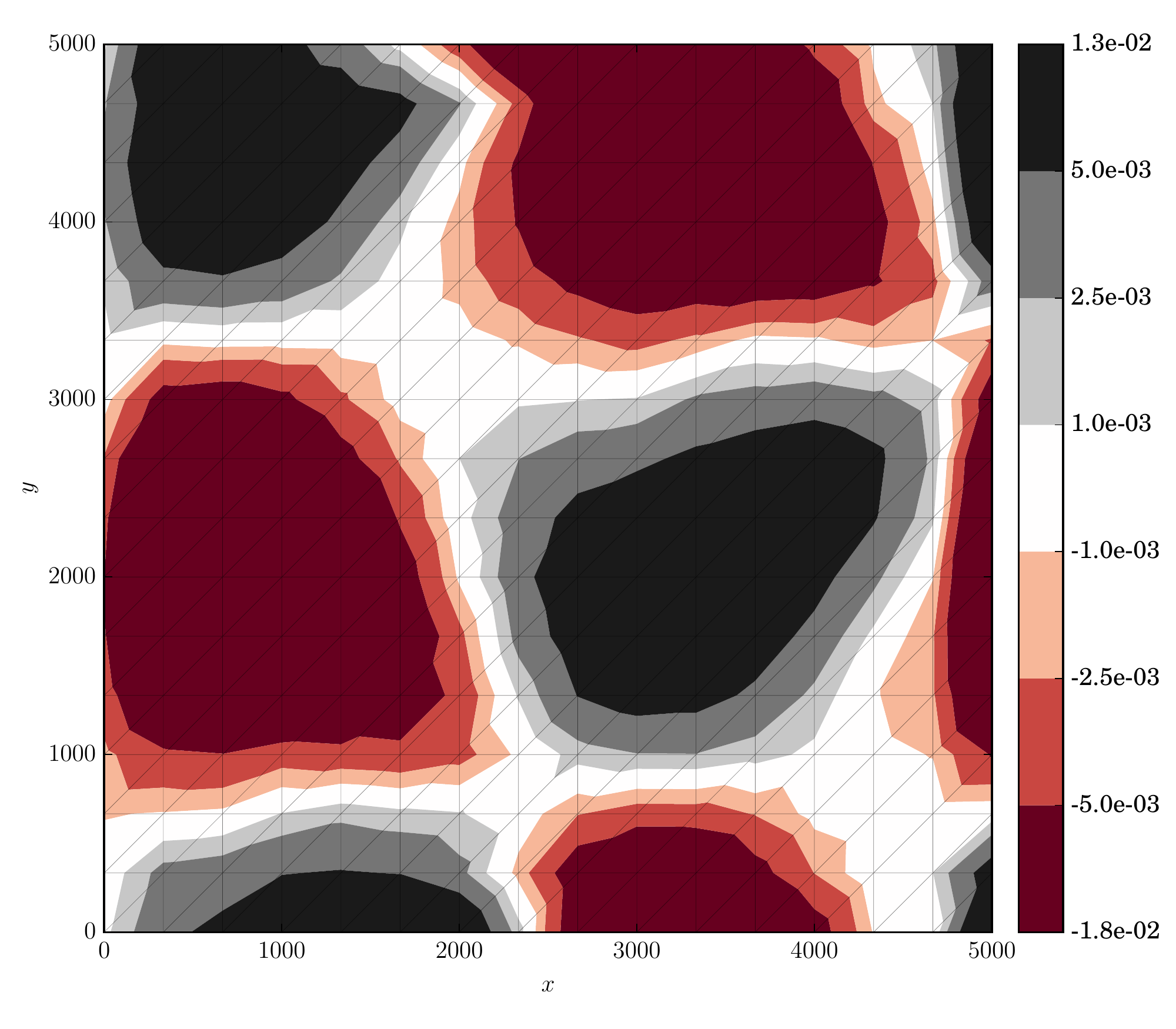}
    \quad                                             
    \includegraphics[width=0.319\linewidth]{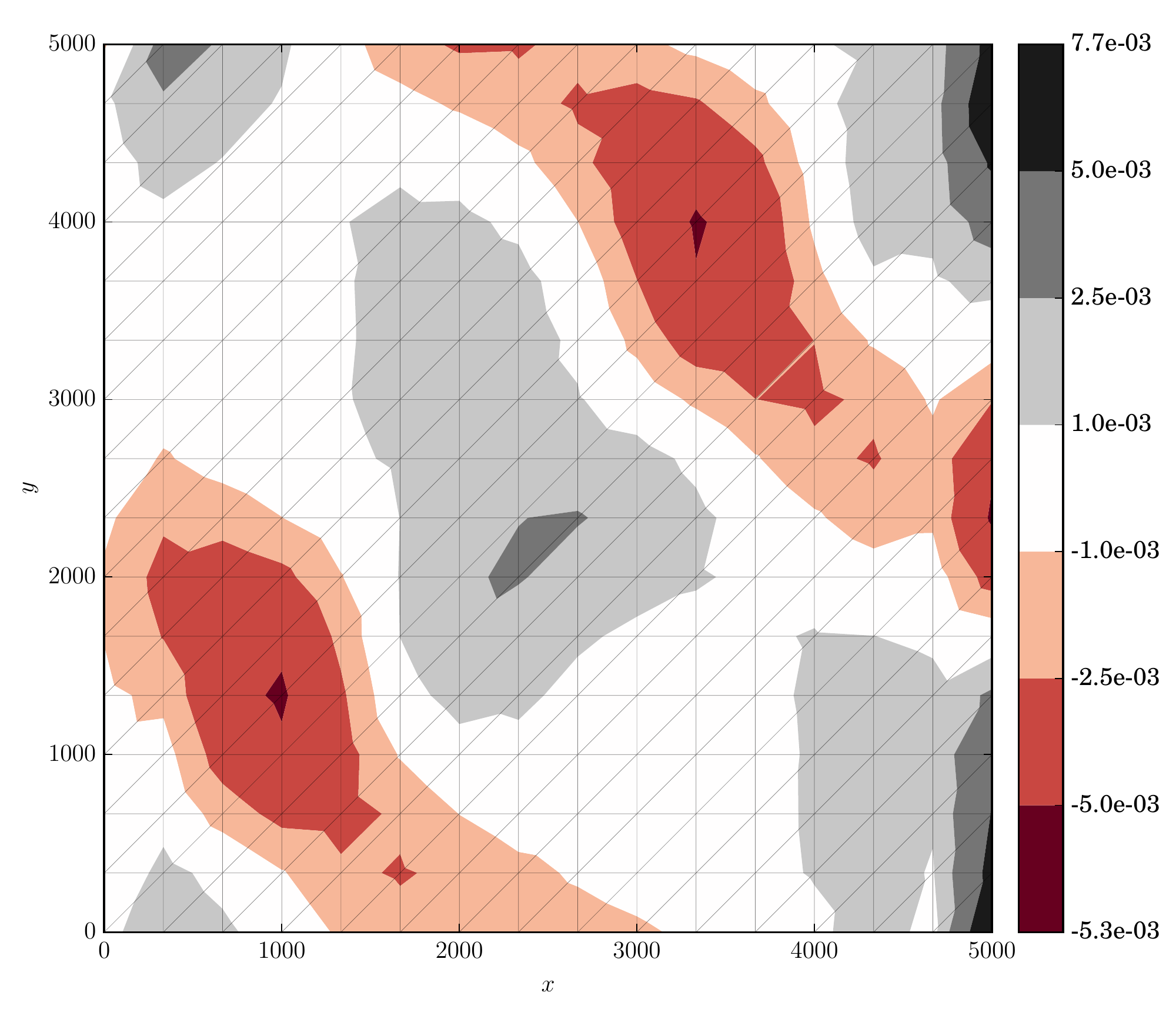}
                                                      
    \includegraphics[width=0.319\linewidth]{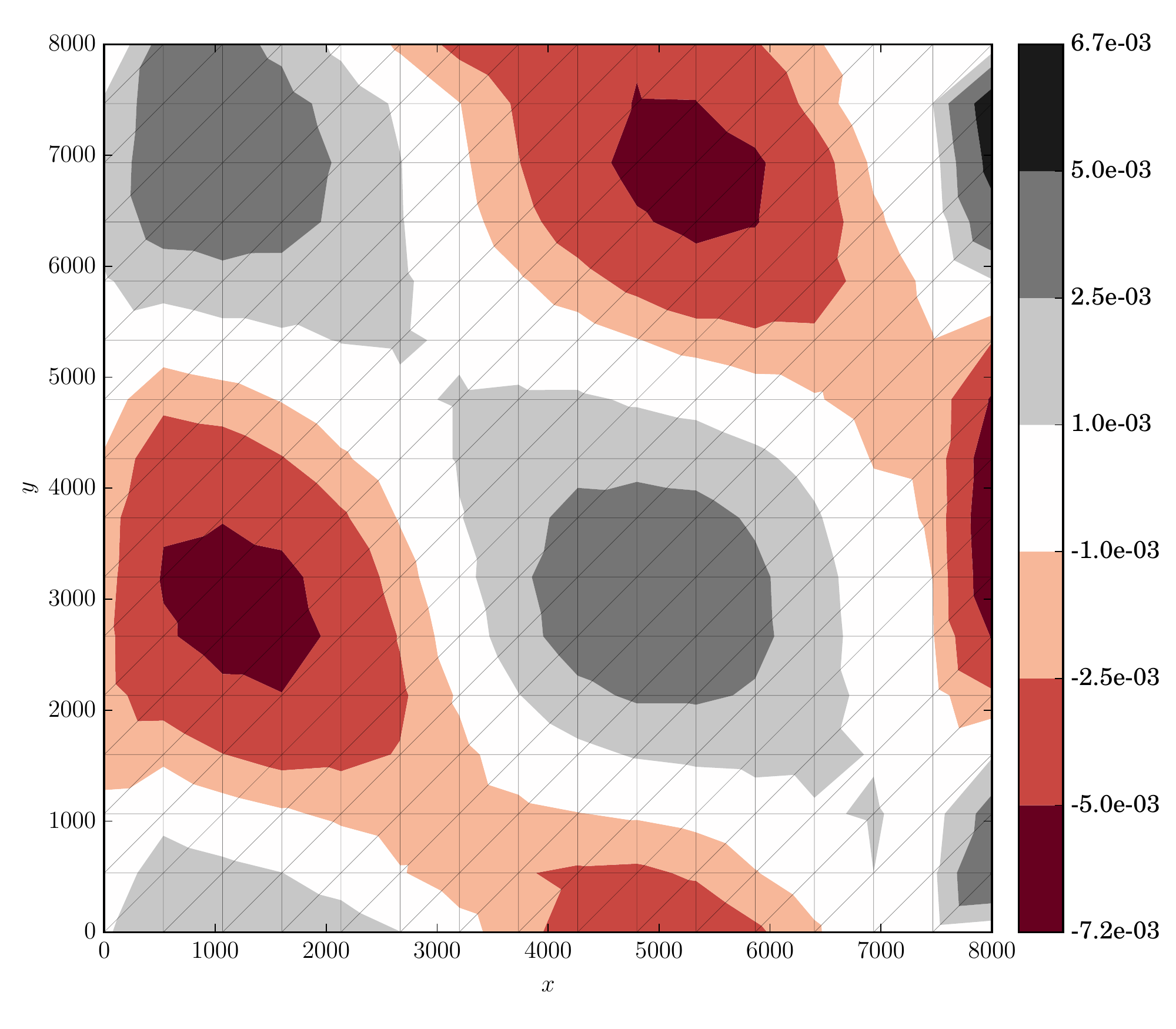}
    \quad                                             
    \includegraphics[width=0.319\linewidth]{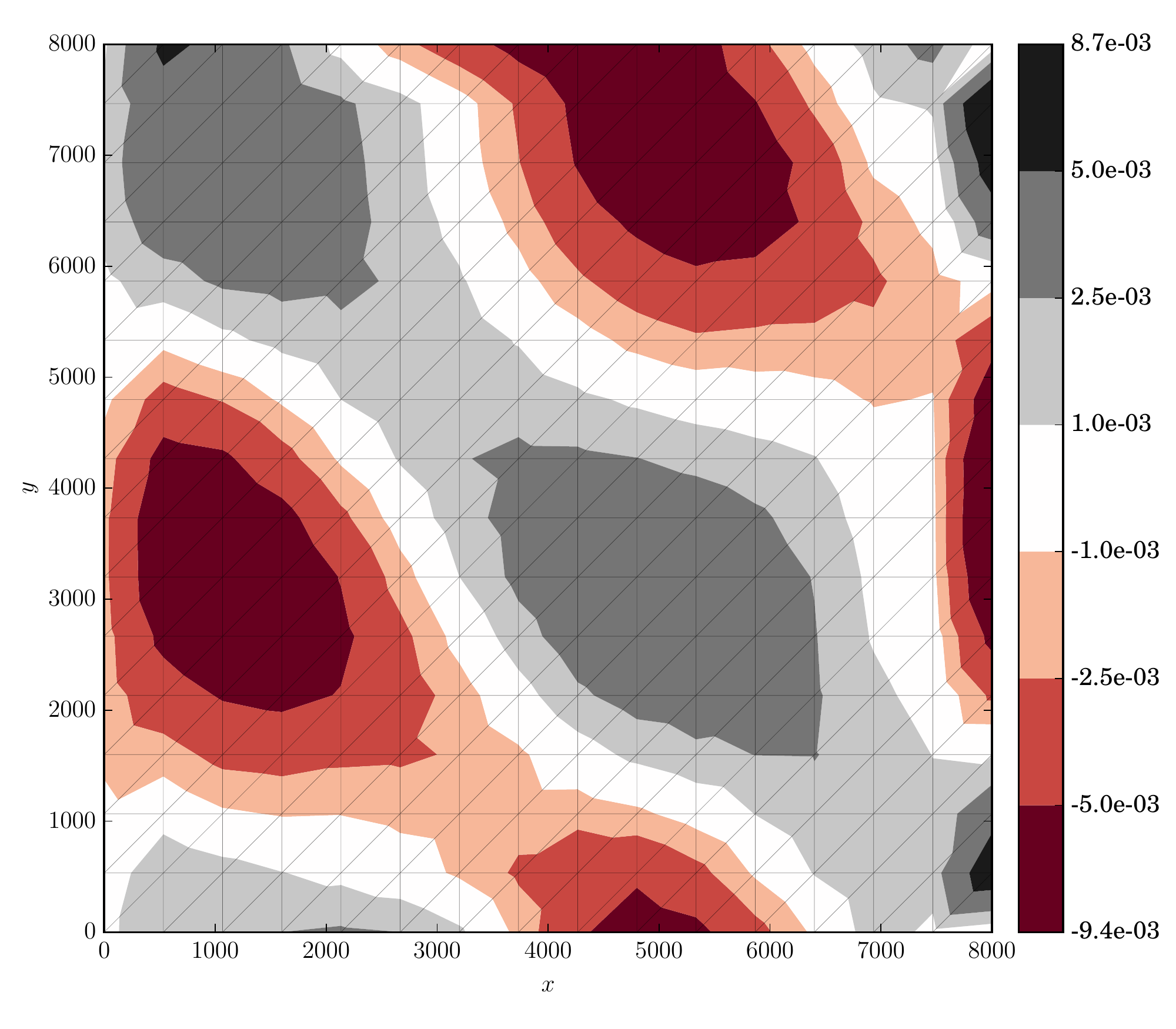}
    \quad                                             
    \includegraphics[width=0.319\linewidth]{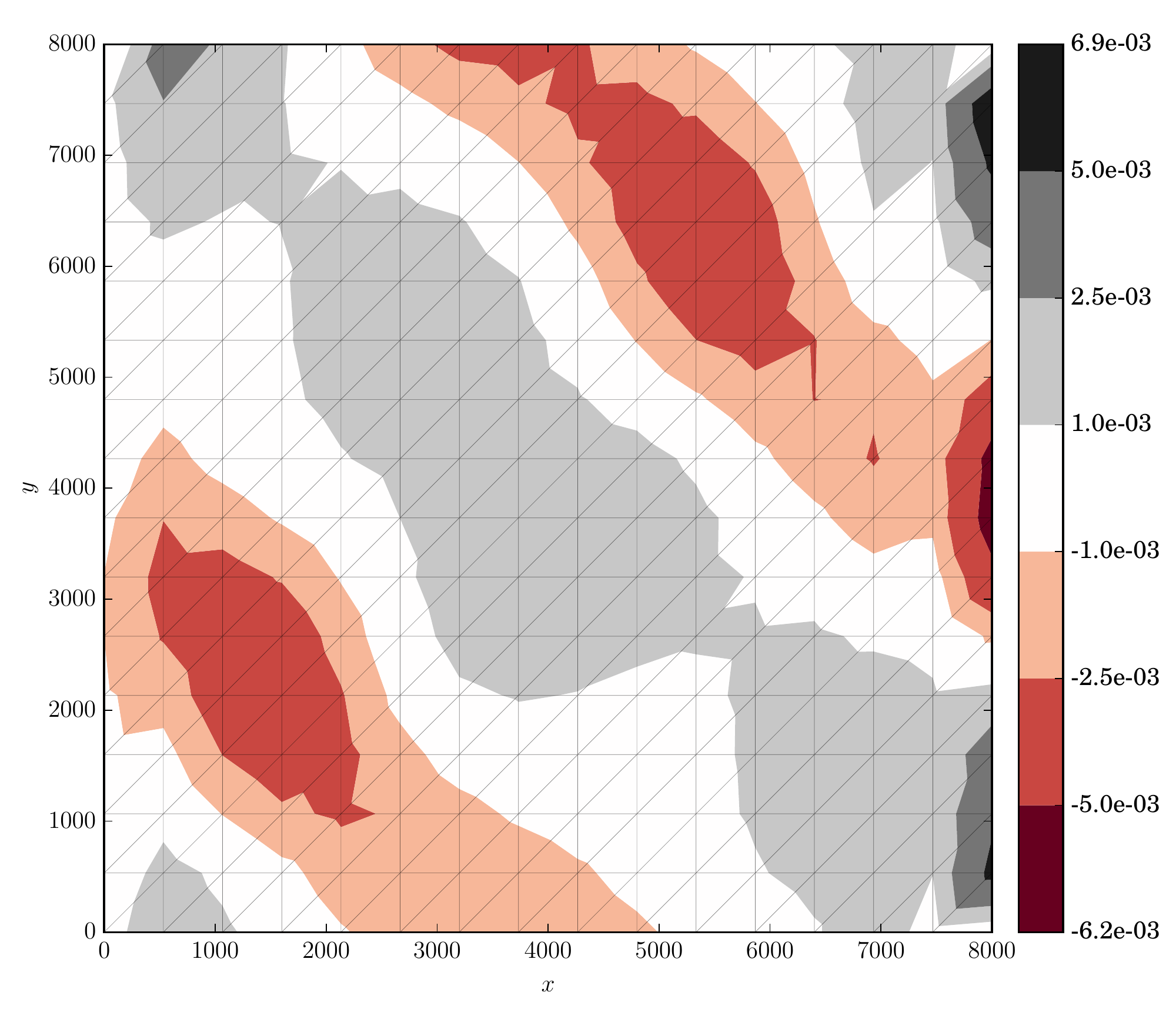}
                                                      
    \includegraphics[width=0.319\linewidth]{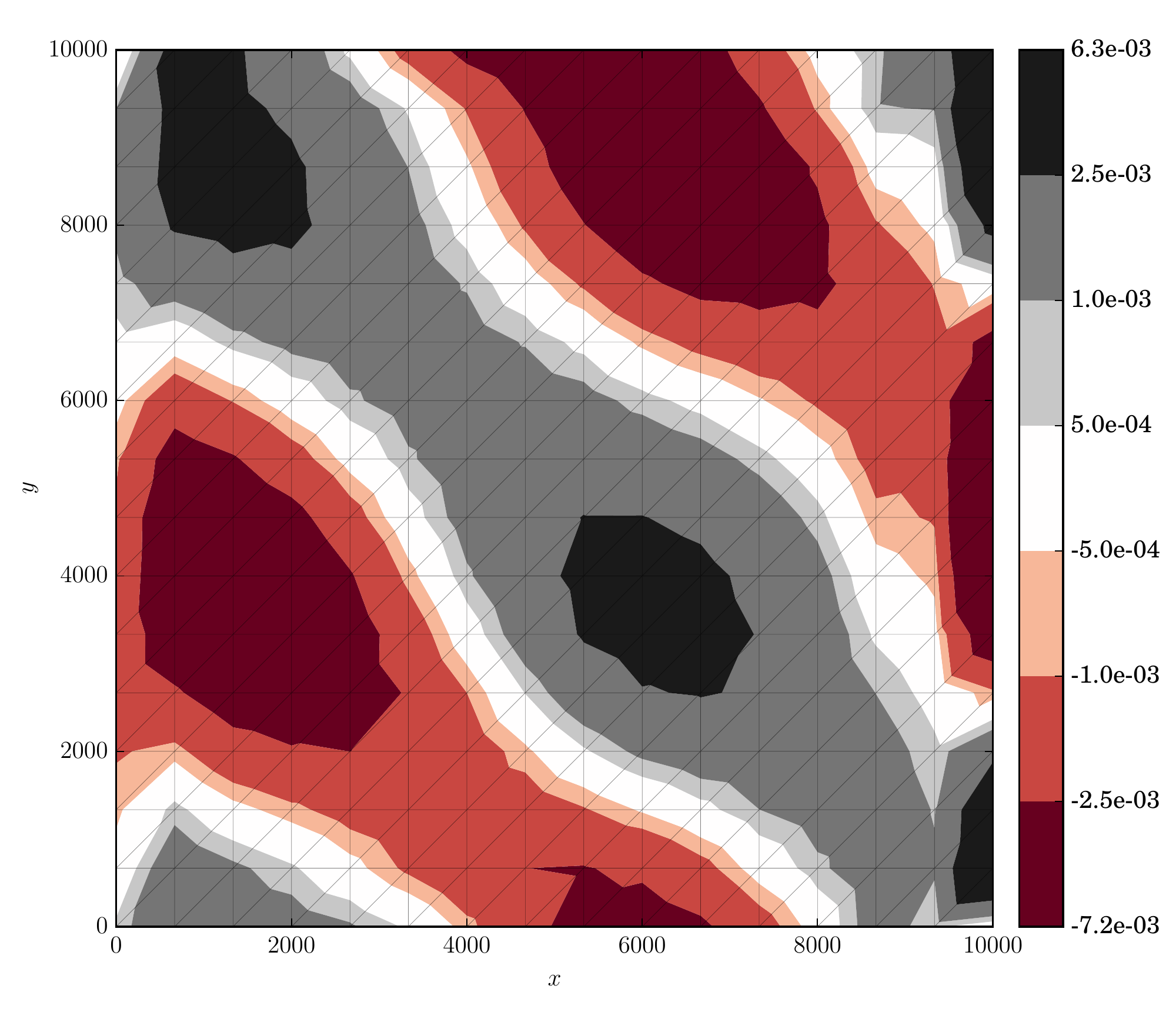}
    \quad                                             
    \includegraphics[width=0.319\linewidth]{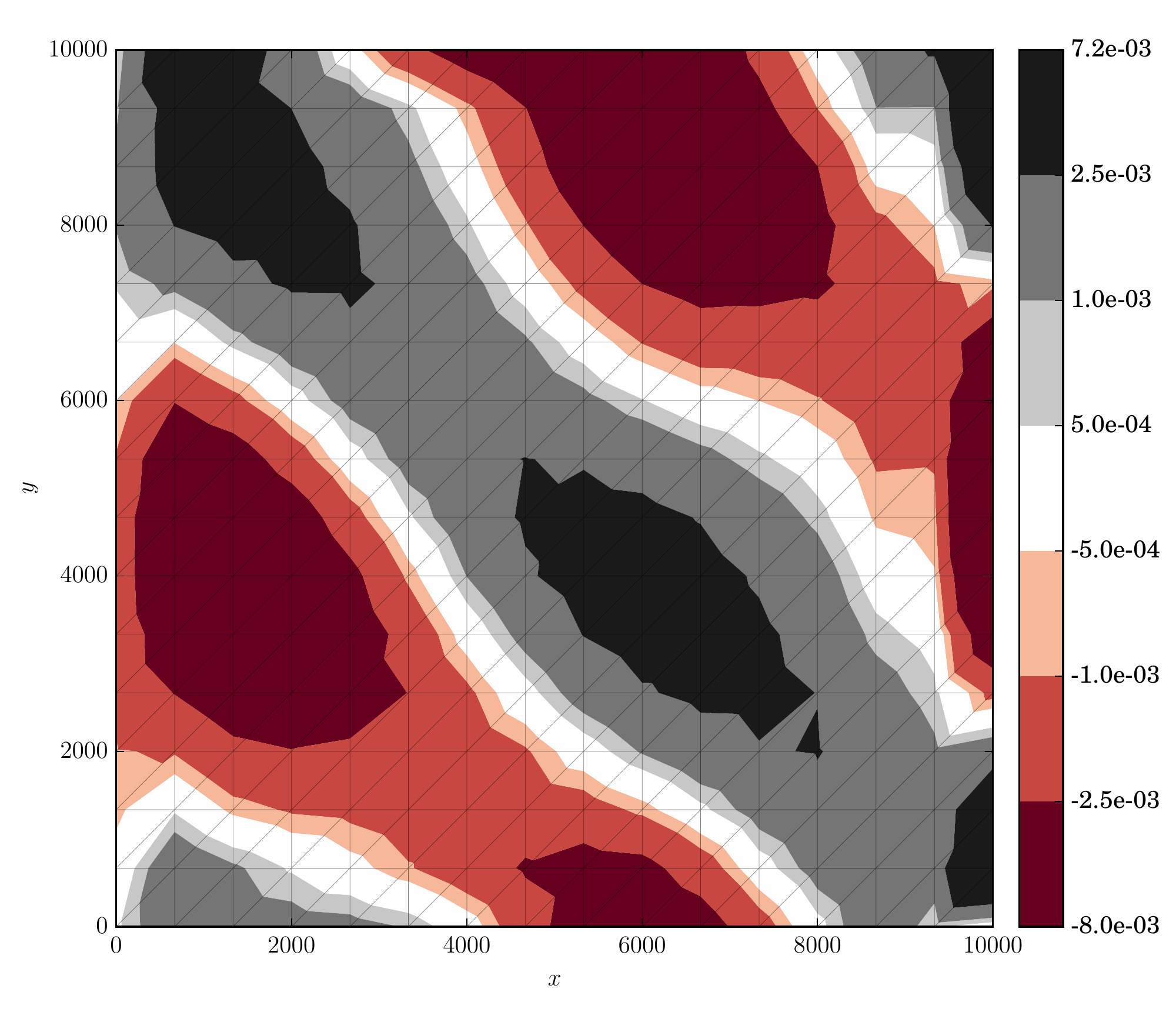}
    \quad                                             
    \includegraphics[width=0.319\linewidth]{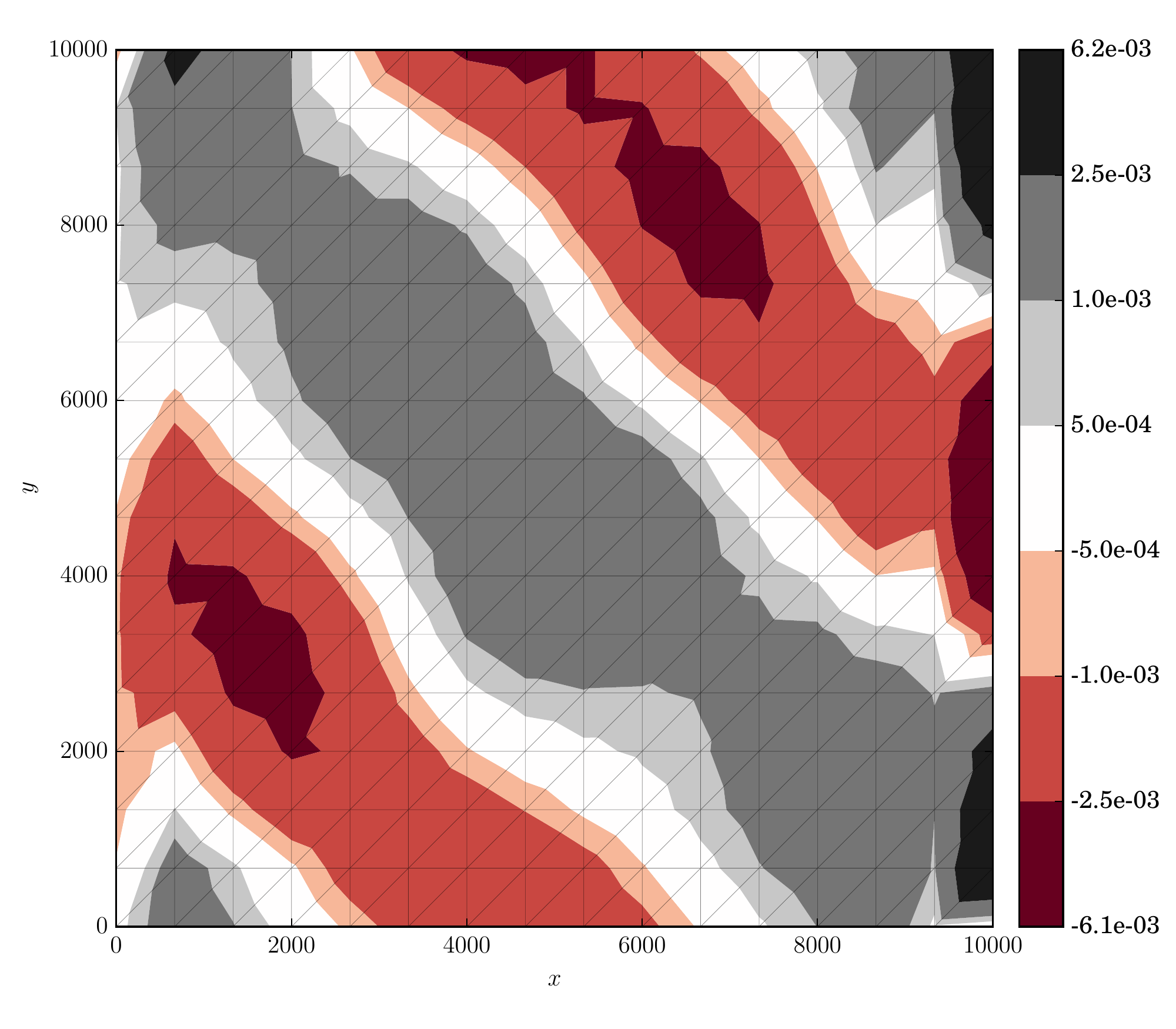}
                                                      
    \includegraphics[width=0.319\linewidth]{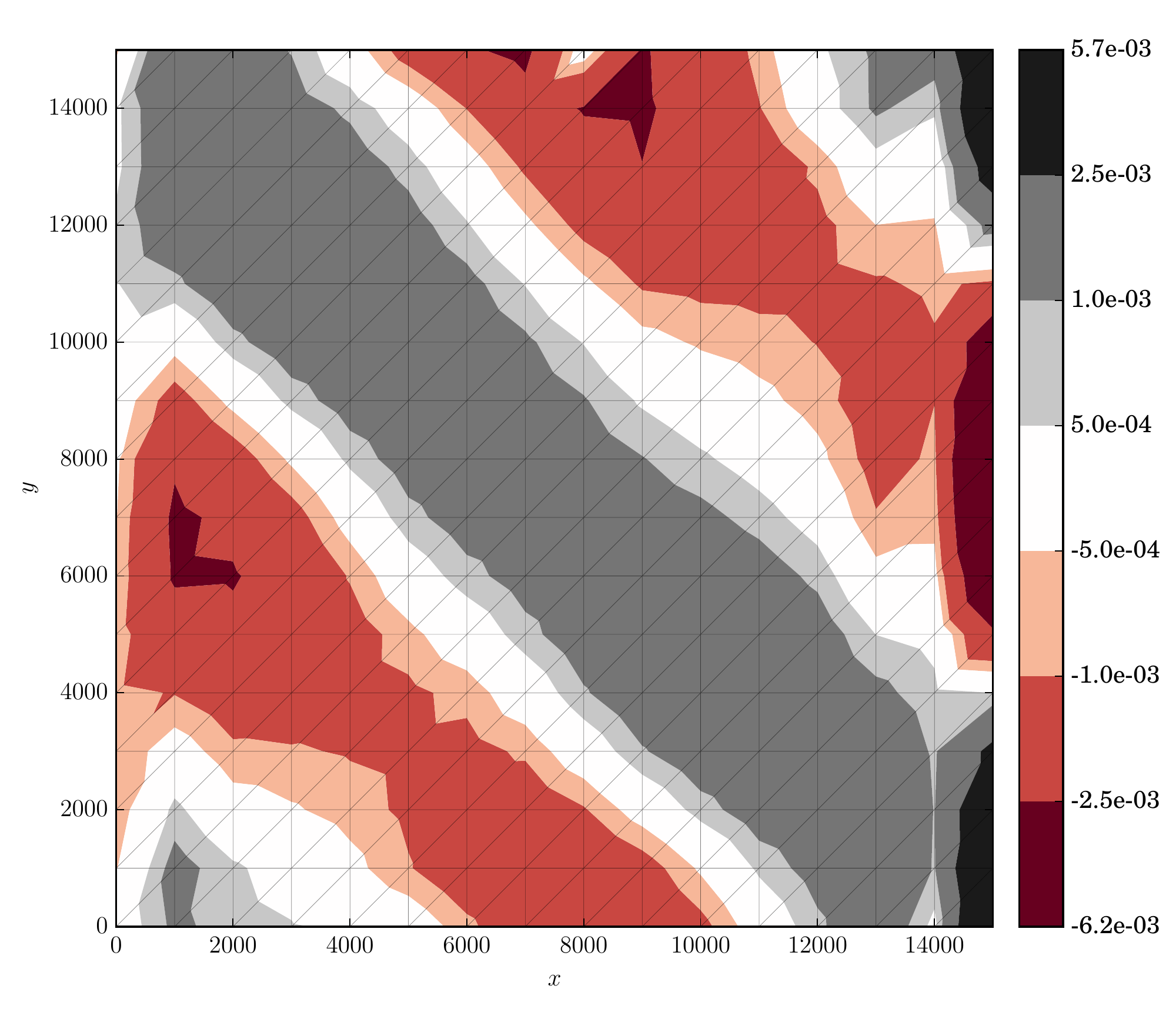}
    \quad                                             
    \includegraphics[width=0.319\linewidth]{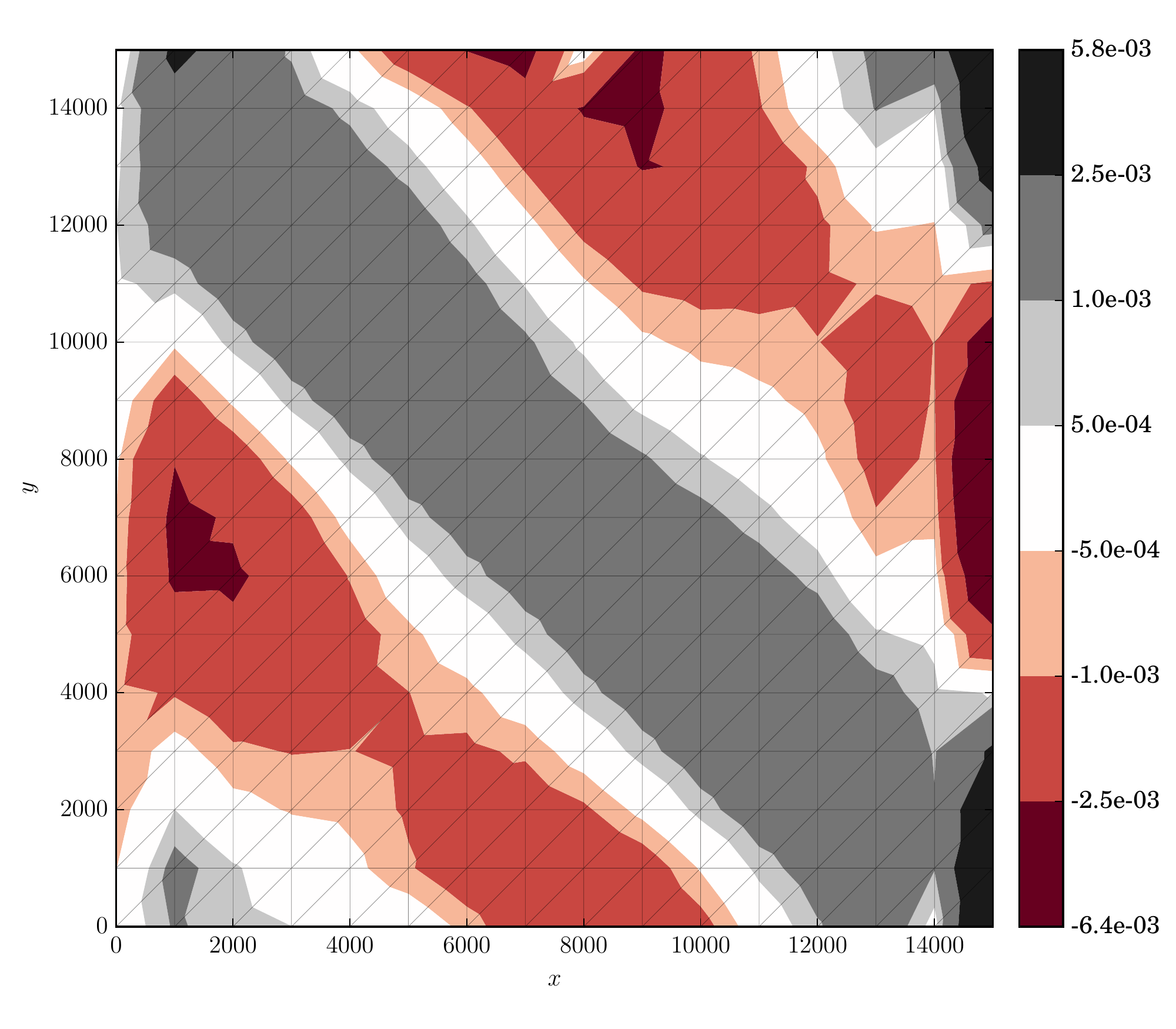}
    \quad                                             
    \includegraphics[width=0.319\linewidth]{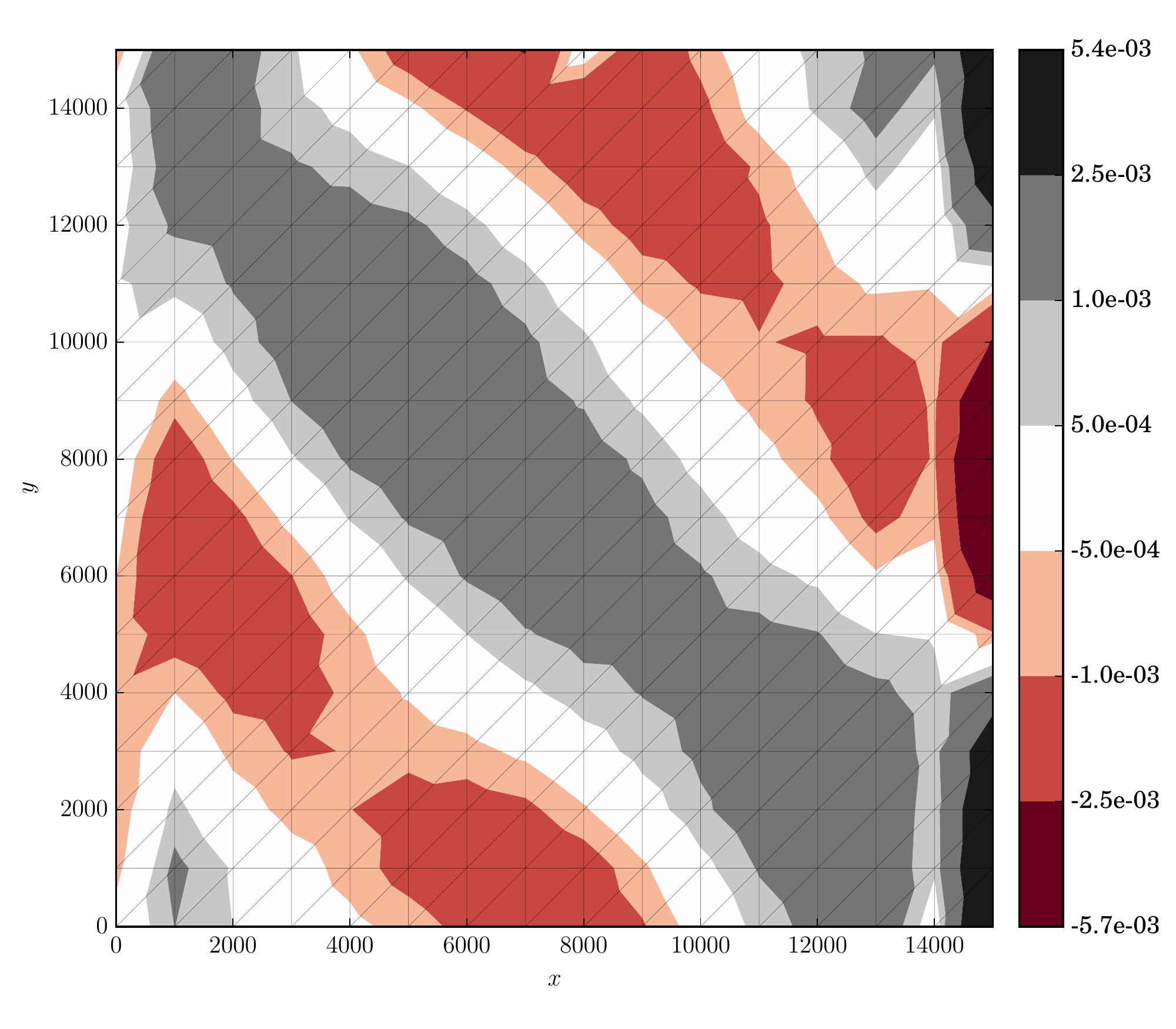}

  \caption[ISMIP-HOM momentum experiment velocity divergence]{Basal velocity divergence $\nabla \cdot \mathbf{u} |_B$ for the ISMIP-HOM test experiment using a $5 \times 5$ square km grid (first row), $8 \times 8$ km$^2$ grid (second row), $10 \times 10$ km$^2$ grid (third row), and $15 \times 15$ km$^2$ grid (fourth row).  The first column are results attained using the full-Stokes model from \S \ref{ssn_full_stokes}, the second column using the first-order model from \S \ref{ssn_first_order}, and the third column was generated using the reformulated-Stokes model of \S \ref{ssn_reformulated_stokes}.}
  \label{ismip_hom_a_divergence}
\end{figure*}

It was also interesting to observe that the velocity divergence $\nabla \cdot \mathbf{u}$ was largest near the bed, and is most likely due to the fact that inclusion of the impenetrability constraint $\mathbf{u} \cdot \mathbf{n} = 0$ demands more from the velocity approximation than was possible from the model formulations.  Similar to the surface velocity, there was observed significant variation of basal velocity divergence between each model (Figure \ref{ismip_hom_a_divergence}), with decreasing observed variance as $\ell$ increases.

\section{Plane-strain simulation} \label{ssn_plane_strain_simulation}

\begin{figure*}
  \centering

    \includegraphics[width=\linewidth]{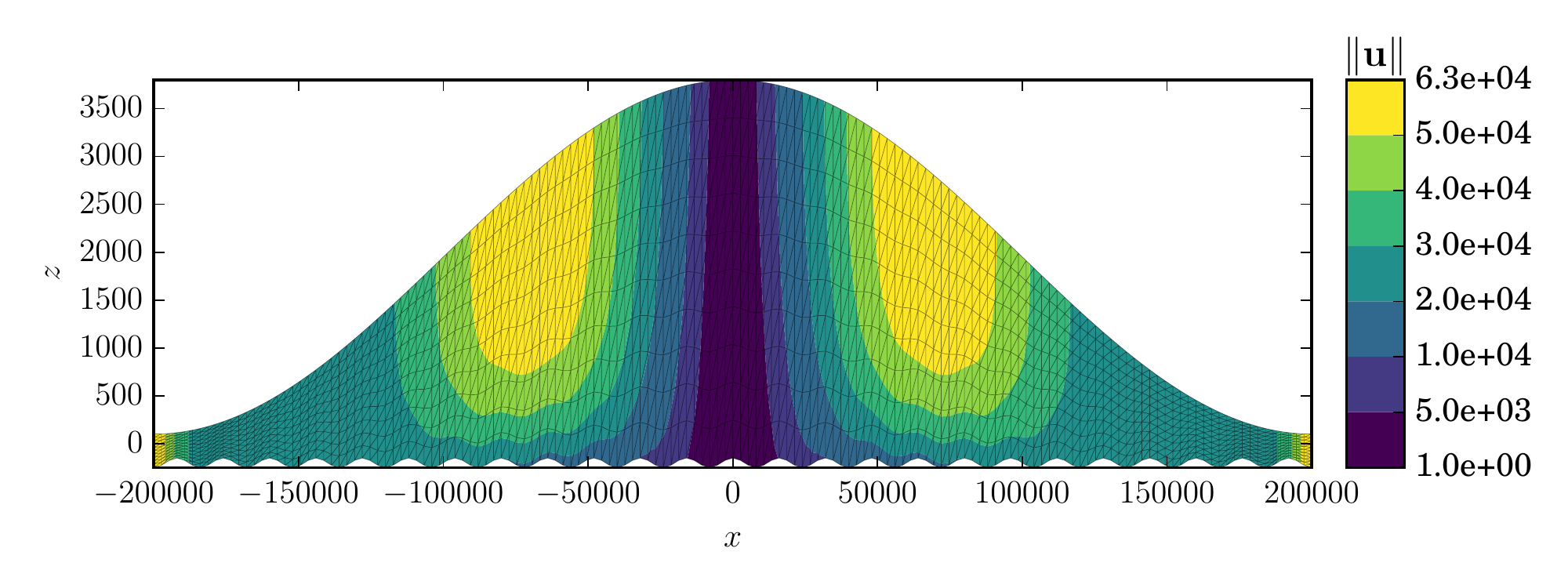}
    \includegraphics[width=\linewidth]{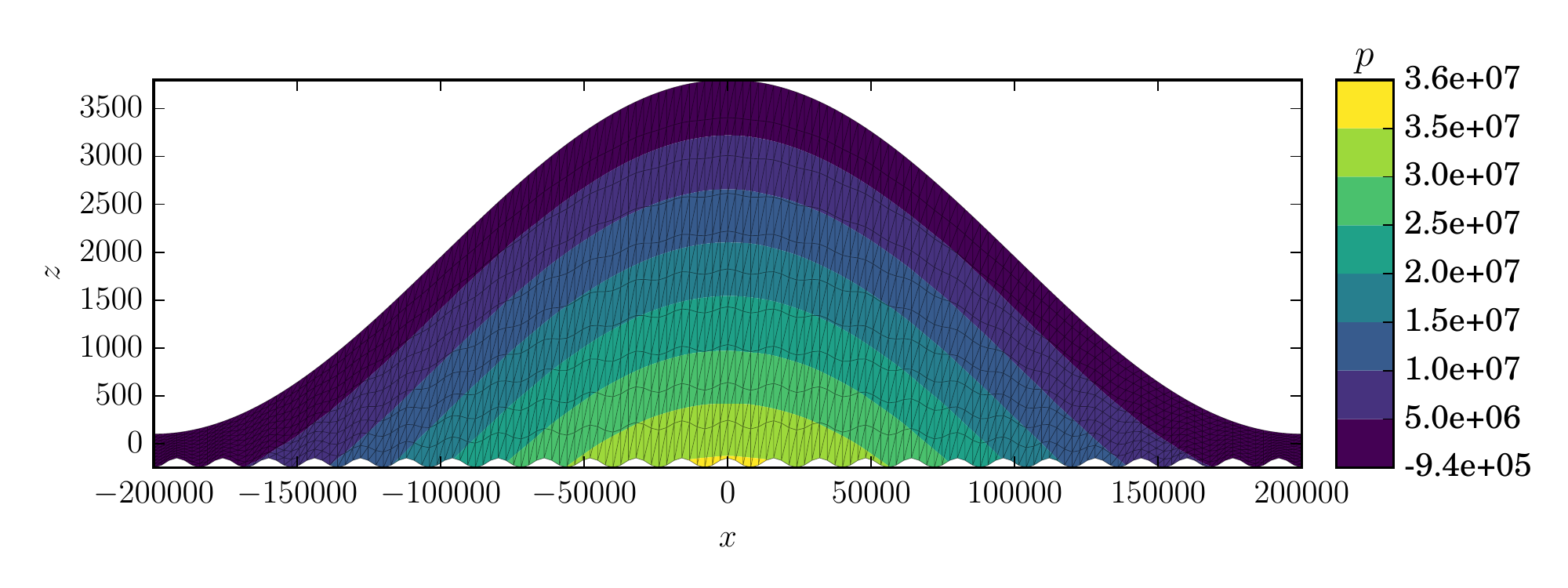}
    \includegraphics[width=\linewidth]{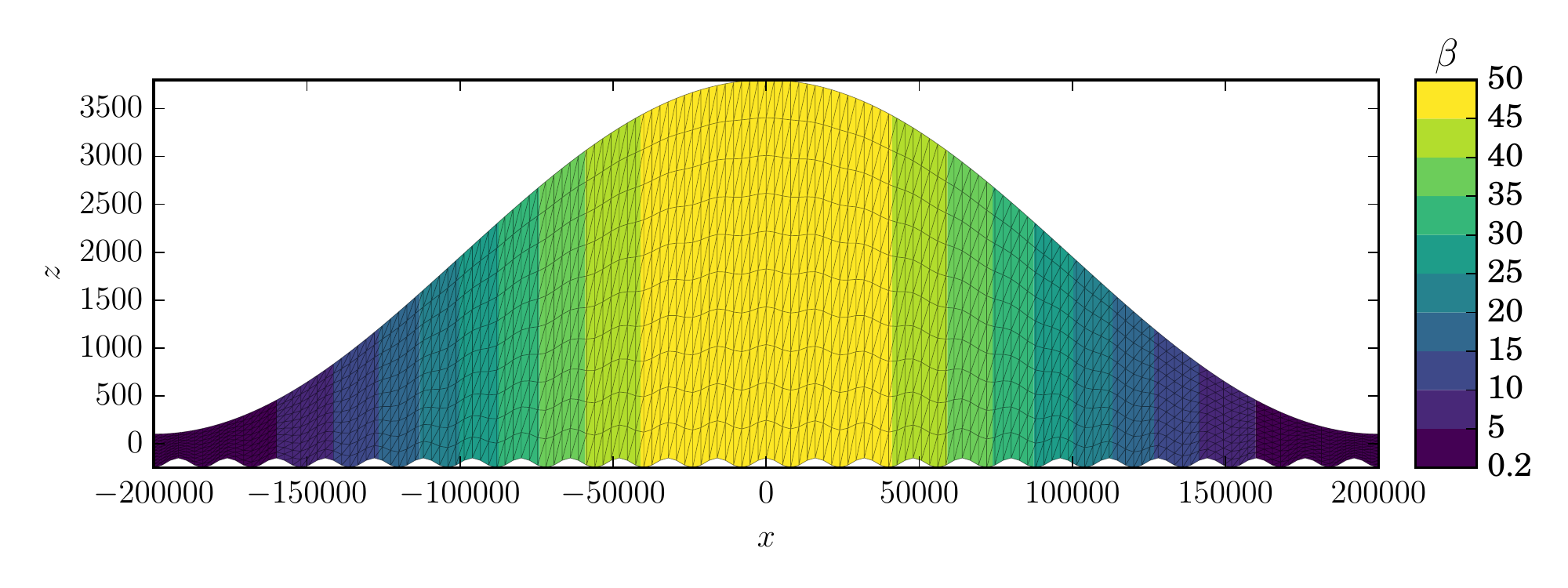}

  \caption[Plane-strain momentum experiment solution]{The plane-strain profile results for the simulation outlined by \S \ref{ssn_plane_strain_simulation}; velocity magnitude $\Vert \mathbf{u} \Vert$ (top), pressure $p$ (middle), and prescribed basal traction $\beta$ (bottom).}
  \label{plane_strain_image}
\end{figure*}

\index{Linear differential equations!2D}
\index{Plane-strain simulations}
For an example simulation of the plane-strain model of \S \ref{ssn_plane_strain}, we create a two-dimensional ice-sheet with surface height 
\begin{align*}
  S(x) = \left( \frac{ H_{max} + B_0 - S_0 }{2} \right) \cos\left( \frac{2\pi}{\ell} x \right) + \left( \frac{H_{max} + B_0 + S_0}{2} \right),
\end{align*}
with thickness at the divide $H_{max}$, height of terminus above water $S_0$, depth of ice terminus below water $B_0$, and width $\ell$.  We prescribe the sinusoidally-varying basal topography
\begin{align*}
  B(x) = b \cos\left( n_b \frac{2\pi}{\ell} x \right) + B_0,
\end{align*}
with amplitude $b$ and number of bumps $n_b$.  The basal traction field used followed the same sinusoidal variation as the surface,
\begin{align*}
  \beta(x) = \left( \frac{\beta_{max} - \beta_{min}}{2} \right) \cos\left( \frac{2\pi}{\ell} x \right) + \left( \frac{\beta_{max} + \beta_{min}}{2} \right)
\end{align*}
with maximum value $\beta_{max}$ and minimum value $\beta_{min}$.  The specific values used by the simulation are listed in Table \ref{plane_strain_values} with results depicted in Figure \ref{plane_strain_image} generated by Code Listing \ref{plane_strain_momentum_code}.

\begin{table}[H]
\centering
\caption[Plane-strain momentum variables]{Variable values for plane-strain simulation.}
\label{plane_strain_values}
\begin{tabular}{llll}
\hline
\textbf{Variable} & \textbf{Value} & \textbf{Units} & \textbf{Description} \\
\hline
$\dot{\varepsilon}_0$ & $10\sups{-15}$ & a\sups{-1}   & strain regularization \\
$A$       & $10\sups{-16}$  & Pa\sups{-3}a\sups{-1}   & flow-rate factor \\
$F_b$     & $0$             & m a\sups{-1}            & basal water discharge \\
$k_x$     & $150$           & -- & number of $x$ divisions \\
$k_z$     & $10$            & -- & number of $z$ divisions \\
$N_e$     & $3000$          & -- & number of cells \\
$N_n$     & $1661$          & -- & number of vertices \\
$\ell$    & $400$           & km & width of domain \\
$H_{max}$ & $4000$          & m  & thickness at divide \\
$S_0$     & $100$           & m  & terminus height \\
$B_0$     & $-200$          & m  & terminus depth \\
$n_b$     & $25$            & -- & number of bed bumps \\
$b$       & $50$            & m  & bed bump amplitude \\
$\beta_{max}$ & $50$        & kg m\sups{-2}a\sups{-1} & max basal traction \\ 
$\beta_{min}$ & $0.2$       & kg m\sups{-2}a\sups{-1} & min basal traction \\ 
\hline
\end{tabular}
\end{table}

\pythonexternal[label=ismip_hom_bed_code, caption={CSLVR script used to generate ISMIP-HOM basal topography Figure \ref{ismip_hom_a_B}.}]{scripts/momentum/ISMIP_HOM_A/plot_bed.py}

\pythonexternal[label=ismip_hom_momentum_code, caption={CSLVR script which solves the ISMIP-HOM experiment of \S \ref{ssn_ismip_hom_test_simulations}.}]{scripts/momentum/ISMIP_HOM_A/xsmall/ISMIP_HOM_A_BP.py}

\pythonexternal[label=plane_strain_momentum_code, caption={CSLVR source code used to solve the plane-strain problem of \S \ref{ssn_plane_strain_simulation}.}]{scripts/momentum/plane_strain/plane_strain.py}

\begin{figure*}
  \centering

    \includegraphics[width=0.7\linewidth]{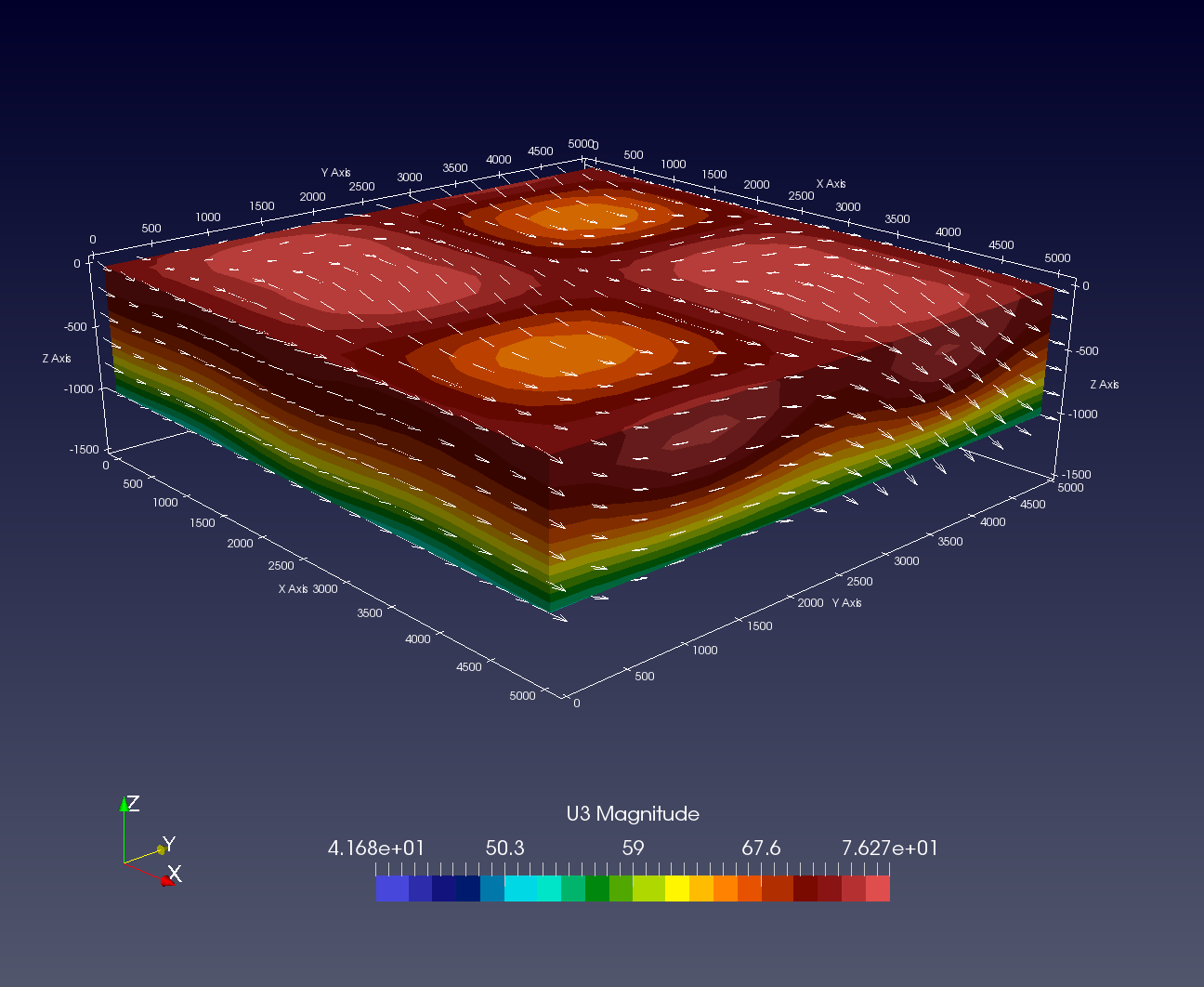}

    \vspace{10mm}

    \includegraphics[width=0.7\linewidth]{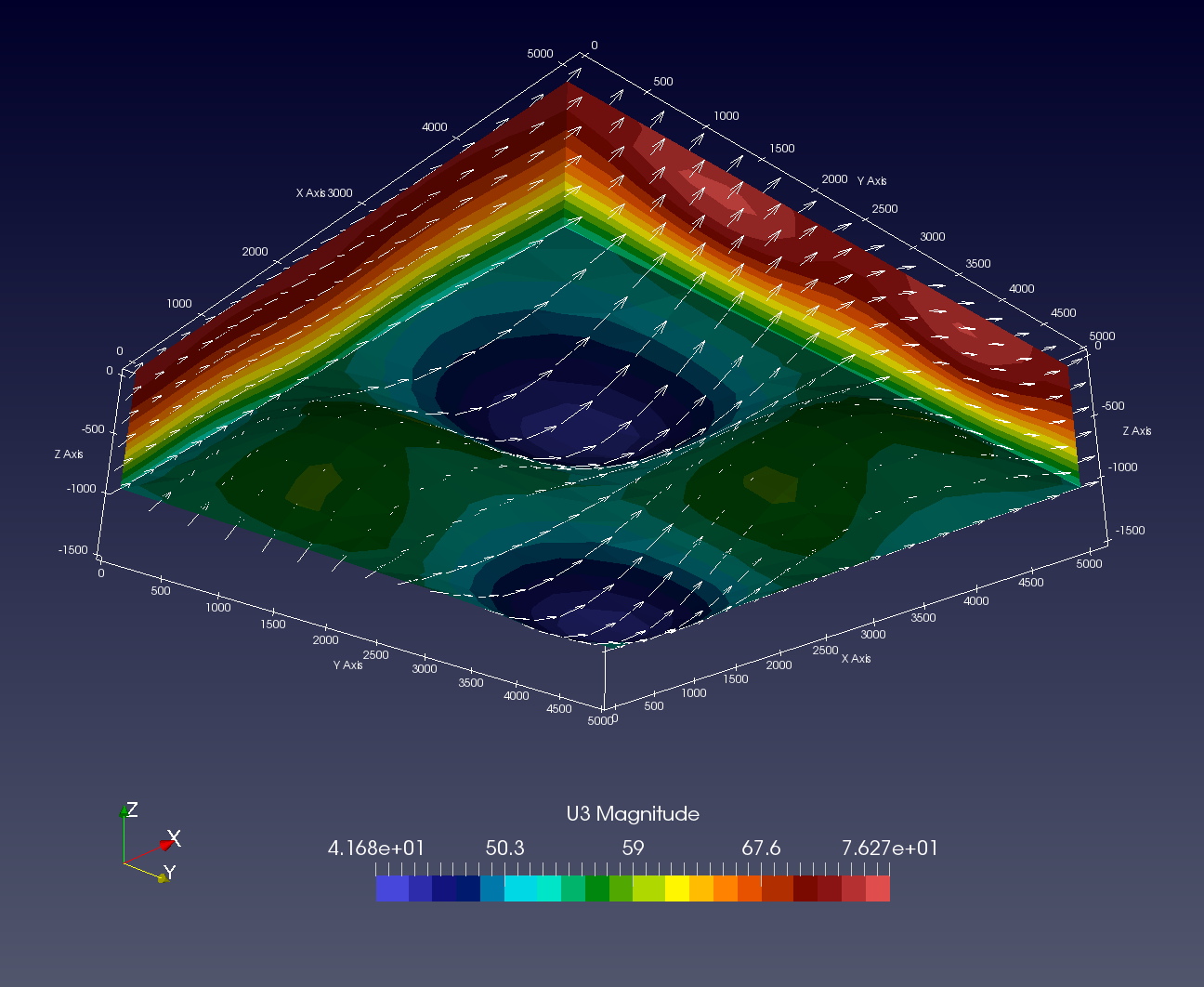}

    \caption[Three-dimensional ISMIP-HOM momentum solution]{Full-Stokes velocity $\mathbf{u}$ view from above (top) and below (bottom) with a domain of $5 \times 5$ square km.}

  \label{ismip_hom_a_paraview}
\end{figure*}

%===============================================================================
%===============================================================================

\chapter{Internal energy balance} \label{ssn_internal_energy_balance}

The internal energy boundary conditions appropriate to the base of an ice-sheet or glacier are complicated by the fact that both essential and natural types have been specified.  Proposed here is a unification of these conditions in a single natural form, presented as a water content minimization problem whereby an observed maximum value of moisture retention within ice is enforced.  Using this method, previous constraints on the basal energy flux are no longer required, allowing abnormally high intra-ice water contents resulting from internal friction to be drained from the ice using established energy transport equations pertinent to polythermal glaciers.  An algorithm is presented in \S \ref{ssn_dual_optimization} which couples this method with a surface-velocity data-assimilation procedure for basal traction (described in \S \ref{ssn_momentum_optimization_procedure}), resulting in a set fully thermo-mechanically coupled basal traction, velocity, internal energy, and basal water discharge.

\section{Introduction}

Within a polythermal glacier or ice-sheet, both liquid and solid phases are present.  It follows that any mathematical description of polythermal ice must account for the role of liquid water in terms of both rheology and energy.  First, ice is defined as cold if a change in energy leads to a change in temperature alone, and temperate if a change in energy leads to a change in water content alone.  Methods from mixture theory have been incorporated and models proposed which either track the transition surface from cold to temperate \index{Cold/temperate surface} (CTS) explicitly \citep{hutter_1982, greve97, greve, blatter_ArXiv}, or solve the internal energy as a single continuous variable \citep{aschwanden_blatter, aschwanden}.  Irrespective of the implementation, these models yield a potentially time-varying distribution of internal energy, which via bijection provides temperature and water content.

For cold ice, defined as ice with zero liquid water content, an advection-diffusion equation with a strain-heating source describes heat flow \citep{patterson}. The energy distribution in temperate ice is more complex.  Instead of raising the temperature of ice, strain-heating in these areas generates water.  This water, once produced, is advected along the trajectory of ice flow.  Additionally, the water is thought to either diffuse in the same way as heat \citep{hutter_1982}, as considered here, or move in a similar fashion to groundwater in a Darcy-type pattern whereby pockets of water drain under pressure through micro-connecting veins \citep{fowler}.  The precise mechanism governing the non-advective transport of water remains unclear.

In order to solve equations governing the transport of energy defining these models, appropriate boundary conditions must be prescribed.  Over the surface in contact with atmosphere, estimated water content and temperatures are readily applied as essential-type conditions.  Over the basal surface in contact with bedrock, both geothermal and frictional heat is presented to the ice as a natural-type boundary.  Further complexity arises in imposing this natural boundary condition.

For cold ice, the entirety of geothermal and frictional heat flows into the interior and raises the temperature.  The temperature of ice becomes fixed once it reaches its pressure-melting point, and the energy flux into the ice then becomes proportional to its basal melting rate adjusted by water discharge \citep{greve97}.  Additionally, for interior ice with temperature at its pressure-melting point and possessing a non-zero strain-heat source, a \index{Temperate zone} \emph{temperate zone} is created.  While there is no reason a temperate zone could not be formed within the ice interior, the strain-rate is greatest near the bed for ice-sheets and Canadian-type glaciers considered here \citep{aschwanden}, and thus temperate zones typically lie near the basal surface at these localities (Figure \ref{ice_profile_domain}).

As there is no existing constitutive relation explaining the water flux either out of or into the ice in temperate zones, the flux of energy into the ice has been previously specified to be in balance with the flux of water out of the ice, such that a homogeneous natural boundary condition exists \citep{greve97, greve, aschwanden_blatter, aschwanden, kleiner, blatter_ArXiv}.  In the process of applying this homogeneous energy boundary condition to observed glacial geometries and surface boundary conditions, complications arise when strain-heating creates unreasonably high water content within the ice.  To address this issue, models heretofore have either moved all internally-generated water above a threshold directly into a basal-water storage-layer \citep{greve97}, or eliminated some fraction of water at a rate proportional to its magnitude \citep[][section 4.6]{aschwanden}.  In so doing, these models abandon their mathematical formulations in order to ensure that the ice does not deviate from an expected maximum water content. 

The work presented here addresses this shortcoming by expressing the basal boundary condition as a control optimization problem \citep{bryson, nocedal}, whereby an observed maximum basal water content is enforced.  This method is coupled with the basal traction inversion procedure outlined by \citet{macayeal} in order to incorporate satellite- and radar-observed ice surface velocities.  The result of this optimization process is consistent set of of energy, basal water discharge, velocity, pressure, and basal traction fields.  The method is applicable to all diffusive-type energy balance models, requiring only the re-specification of the basal boundary condition and that a method of water transport is available; i.e.~a non-zero non-advective water flux coefficient is assumed.

%===============================================================================
%===============================================================================

\section{Mathematical foundation}

To begin, the internal energy \index{Internal energy} of ice is defined from the \emph{constitutive equation for internal energy}, as found in \citet{greve},
\begin{align}
  \label{energy}
  \theta(T,W) = \int_{T_0}^T c_i(T')\ dT' + W L_f,
\end{align}
with absolute temperature $T$, reference temperature $T_0$, and \index{Water content of ice} water content or moisture density $W \in [0,1]$.  Water content $W$ is defined as the ratio of the mass of water contained within a unit-volume of water-ice mixture to the mass of the mixture, such that $W=0$ and $W=1$ corresponds with 100\% ice and 100\% water, respectively.  Energy definition (\ref{energy}) is in turn characterized by the definitions of \index{Heat capacity!Latent} \index{Heat capacity!Sensible} \emph{sensible} heat capacity $c_i$ and \emph{latent} heat capacity $L_f$; the amount of energy required to raise one unit mass of ice one unit of temperature and completely melt one unit mass of ice, respectively.

The \index{Balance equations!Internal energy} \emph{energy balance} takes the form as presented by \citet{greve97},
\begin{align}
  \label{energy_balance}
  \rho \frac{d \theta}{dt} = - \nabla \cdot \mathbf{q} + Q,
\end{align}
with mixture density $\rho$, strain-heat $Q$, and energy-flux composed of sensible and latent terms
\begin{align}
  \label{flux}
  \mathbf{q} = \mathbf{q}_s + \mathbf{q}_l,
\end{align}
derived from \emph{Fourier's Law} of conduction \citep{davis}, with thermal conductivity $k$ associated with temperature $T$ and non-advective water flux coefficient $\nu$ associated with water content $W$,
\begin{align}
  \label{individual_flux}
  \mathbf{q}_s = - k(\theta) \nabla T \hspace{10mm} 
  \mathbf{q}_l = - \nu(\theta) \nabla W.
\end{align}
Note here that the latent heat flux coefficient $\nu$ differs from previous formulations such as \citet{greve97} and \citet{aschwanden} in that it has multiplicatively absorbed latent heat capacity $L_f$.

Bulk water-ice mixture properties apply to thermal conductivity $k$, heat capacity $c$, and density $\rho$,
\begin{align}
  \label{mixture_thermal_conductivity}
  k &= (1-W) k_i + W k_w \\
  \label{mixture_heat_capacity}
  c &= (1-W) c_i + W c_w \\
  \label{mixture_density}
  \rho &= (1-W) \rho_i + W \rho_w,
\end{align}
where the subscripts $i$ and $w$ respectively refer to ice and water.  Thermal conductivity \index{Thermal conductivity} $k_i$ in (\ref{mixture_thermal_conductivity}) has been shown to relate to temperature \citep{yen, ritz, greve},
\begin{align}
  \label{thermal_conductivity}
  k_i = 9.828 \exp\left( -5.7 \times 10^{-3} T \right),
\end{align}
as well as heat capacity $c_i$ in (\ref{mixture_heat_capacity}),
\begin{align}
  \label{heat_capacity}
  c_i = a + b T, \hspace{5mm} a = 146.3, \hspace{5mm} b = 7.253,
\end{align}
while densities $\rho_i$, $\rho_w$ and latent heat properties $k_w$, $c_w$, and $L_f$ are taken as constant.
  
As stated in \citet{hutter_1982, greve97, greve, aschwanden}, because the maximum water retention of ice has mostly been observed to be less than 5\% of the total mass, the maximum change in density is less than 0.5\%.  Hence it is reasonable to abandon separate momentum balances for disparate masses of ice and water and instead treat the mixture as a single homogeneous and incompressible fluid.  Thus it is \emph{demanded} that
\begin{align}
  \label{water_demand}
  W \leq W_c \leq 0.05,
\end{align}
where $W_c$ is an observed maximum water content.

Using definition (\ref{heat_capacity}) for heat capacity $c_i$, the integral in energy definition (\ref{energy}) is evaluated using for simplicity $T_0=0$, providing the quadratic equation for energy
\begin{align}
  \label{energy_quad}
  \theta(T,W) &= a T + \frac{b}{2} T^2 + W L_f.
\end{align}

The temperature of ice is also constrained by its melting point, shown to be dependent on pressure $p$ by the \index{Clausius-Clapeyron relationship} \emph{Clausius-Clapeyron} relationship \citep{patterson}
\begin{align}
  \label{temperature_melting}
  T_m = T_w - \gamma p,
\end{align} 
with triple point of water $T_w = 273.15$ and empirically-derived coefficient $\gamma = 9.8 \times 10\sups{-8}$ (represented by the dashed red line in Figure \ref{temperate_zone_revised_image}).  Using this relation, the internal energy of pure ice raised to its pressure-melting point is
\begin{align}
  \label{energy_melting}
  \theta_m = \theta(T_m,0) &= a T_m + \frac{b}{2} T_m^2,
\end{align}
while the internal energy of ice that has been $(W \times 100$)\% melted is
\begin{align}
  \label{temperate_energy}
  \theta(T_m,W) &= \theta_m + W L_f.
\end{align}
Note here that because $0 \leq W \leq 1$, temperatures above the pressure melting point are explicitly forbidden.  This is acceptable, given the very low allowable percentage of water as demanded by (\ref{water_demand}); water internal to the ice will be in close contact with ice and should thus not exceed the melting temperature of ice.

Using (\ref{temperate_energy}), water content $W$ of the mixture is defined as the fraction of internal energy above that of pure ice at the pressure melting point to the specific latent heat of fusion $L_f$,
\begin{align}
  \label{water_content}
  W(\theta) = &\left\{% 
    \begin{array}{ll}
      \frac{\theta - \theta_m}{L_f}, & \theta > \theta_m \\
      0,                             & \theta \leq \theta_m
    \end{array} \right. ,
\end{align}
while temperature $T$ is derived using the quadratic formula with $W=0$ in Equation (\ref{energy_quad}),
\begin{align}
  \label{temperature}
  T(\theta) = &\left\{%
    \begin{array}{ll}
      T_m,                                               & \theta > \theta_m \\
      \frac{-a + \sqrt{a^2 + 2b \theta}}{b}, & \theta \leq \theta_m
    \end{array} \right. .
\end{align}

If a different lower integration bound $T_0$ in (\ref{energy}) were used to calculate (\ref{energy_quad}) and (\ref{energy_melting}) -- say $T_0 = T_m$ at pressure-melting temperature (\ref{temperature_melting}) -- a simple calculation will show that (\ref{energy}) will be negative in areas with temperature below $T_m$, while temperature (\ref{temperature}) and water content (\ref{water_content}) will produce identical values as when taking $T_0 = 0$.  If a more accurate estimate of internal energy is required, one may be attained by using a heat capacity appropriate for the entire range of $T$ from zero to $T_m$, as compiled by \citet{yen}.

Combining energy flux definitions (\ref{flux}) and (\ref{individual_flux}) with definitions of water content (\ref{water_content}) and temperature (\ref{temperature}), the flow of energy may be stated
\begin{align}
  \label{energy_flux_cases}
  \mathbf{q} = 
  \begin{cases}
    \mathbf{q}_s + \mathbf{q}_l = - k \nabla T_m - \nu \nabla W, & \theta > \theta_m \\
    \mathbf{q}_s + \mathbf{q}_l = - k \nabla T, & \theta \le \theta_m
  \end{cases}.
\end{align}
Because no universally agreed upon constitutive relation exists between latent-energy non-advective flux coefficient $\nu$ and energy $\theta$, the use of the Fickian-type \emph{regularizing} choice of $\nu \approx k / k_0$ for some constant $k_0$ as suggested by \citet{aschwanden_blatter} is used here.  The \index{Enthalpy} \index{Enthalpy-gradient method} \emph{enthalpy-gradient} method \citep{pham, aschwanden_blatter, aschwanden} simplifies energy-flux term (\ref{energy_flux_cases}) further by using the unified energy flux
\begin{align}
  \label{enthalpy_grad}
  \mathbf{q} = - \left( \frac{\kappa}{c} \right) \nabla \theta, \hspace{10mm}
  \kappa =
  \begin{cases}
    \frac{k}{k_0}, & \theta > \theta_m \\
    k,             & \theta \leq \theta_m
  \end{cases}.
\end{align}
Note that for enthalpy $\tilde{\theta}$, the relationship to internal energy $\theta$, pressure $p$, and volume $V$ is \citep{yen}
\begin{align}
  \label{enthalpy}
  d \tilde{\theta} &= d \theta + p d V,
\end{align}
and using the same reasoning leading to demand (\ref{water_demand}), for incompressible ice $dV \approx 0$, and therefore enthalpy $\tilde{\theta}$ can be taken synonymously with internal energy $\theta$ in the context of glaciers and ice-sheets.

While the enthalpy-gradient method has been shown to reproduce a nearly identical CTS as models which track the CTS explicitly \citep{kleiner}, great care must be taken when discritizing the system of equations \citep{blatter_ArXiv}.  However, the study here only requires alteration of the basal boundary conditions appropriate to these models, and hence any variability in the CTS caused by either the choice of $k_0$ or the method of discretization of enthalpy-gradient flux (\ref{enthalpy_grad}) may be safely ignored, without losing generality.

\begin{figure}
  \centering
    \includegraphics[width=\linewidth]{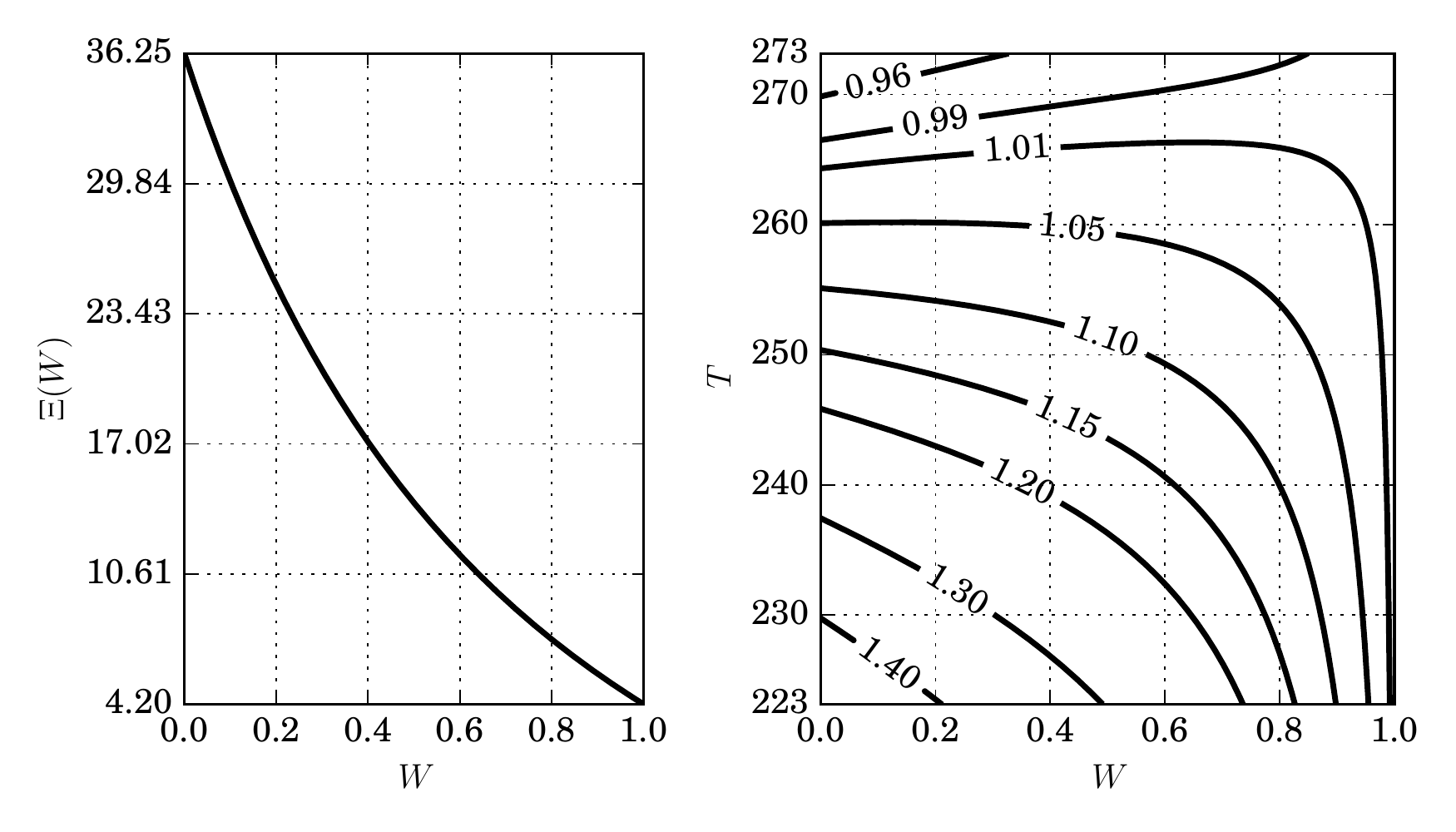}
  \caption[Energy diffusivity diagram]{Energy diffusivity $\Xi(W, T=268.15)$ in units of m\sups{2}a\sups{-1} (left) as a function of water content $W$ only, and the ratio of $\Xi(W,T=268.15)$ to $\Xi(W,T)$ (right) utilizing the temperature-dependent heat capacity (\ref{heat_capacity}) and thermal conductivity (\ref{thermal_conductivity}) in diffusivity (\ref{diffusivity}).}
  \label{bulk_thermal_image}
\end{figure}

The material (convective) derivative and divergence terms in energy-balance (\ref{energy_balance}) are expanded using enthalpy-gradient flux (\ref{enthalpy_grad}),
\begin{align}
  \label{energy_euler_lagrange}
  \rho \left( \frac{\partial \theta}{\partial t} + \mathbf{u} \cdot \nabla \theta \right) &= \nabla \left( \frac{\kappa}{c} \right) \cdot \nabla \theta + \frac{\kappa}{c} \nabla \cdot \nabla \theta + Q,
\end{align}
where $\mathbf{u} = [u\ v\ w]^\intercal$ is the mixture velocity in Cartesian coordinates with respective horizontal axes $(x,y)$ and vertical $z$ (Figure \ref{ice_profile_domain}).  Note that if a constant heat capacity and thermal conductivity were used, as is commonly done, the first term on the right-hand-side of this equation would be zero and could hence be eliminated.  I prefer to include this temperature relationship, as it accounts for up to 50\% extra variation in diffusive capability (Figure \ref{bulk_thermal_image}).  In the presence of water the ice mixture will be less able to conduct energy due to decreased bulk thermal conductivity (\ref{mixture_thermal_conductivity}), increased bulk heat capacity (\ref{mixture_heat_capacity}), and increased bulk density (\ref{mixture_density}) used with energy-balance (\ref{energy_euler_lagrange}).  This variation in diffusion is measured by dividing both sides of energy balance (\ref{energy_euler_lagrange}) by density $\rho$, resulting in the \index{Diffusivity} \emph{diffusivity}
\begin{align}
  \label{diffusivity}
  \Xi=\kappa/(\rho c),
\end{align}
so named because it is attached to the \emph{diffusive} term $\nabla \cdot \nabla \theta$.

%===============================================================================
%===============================================================================

\subsection{Momentum interdependence}

Coupling between energy balance (\ref{energy_euler_lagrange}) and momentum balance (\ref{cons_momentum}, \ref{cons_mass}) occurs through pressure-melting relation (\ref{temperature_melting}, \ref{energy_melting}) and friction, both internally with strain-heat (\ref{strain_heat}) and externally with basal fiction heat generated by sliding over rough terrain.

\begin{figure}
  \centering
    \includegraphics[width=\linewidth]{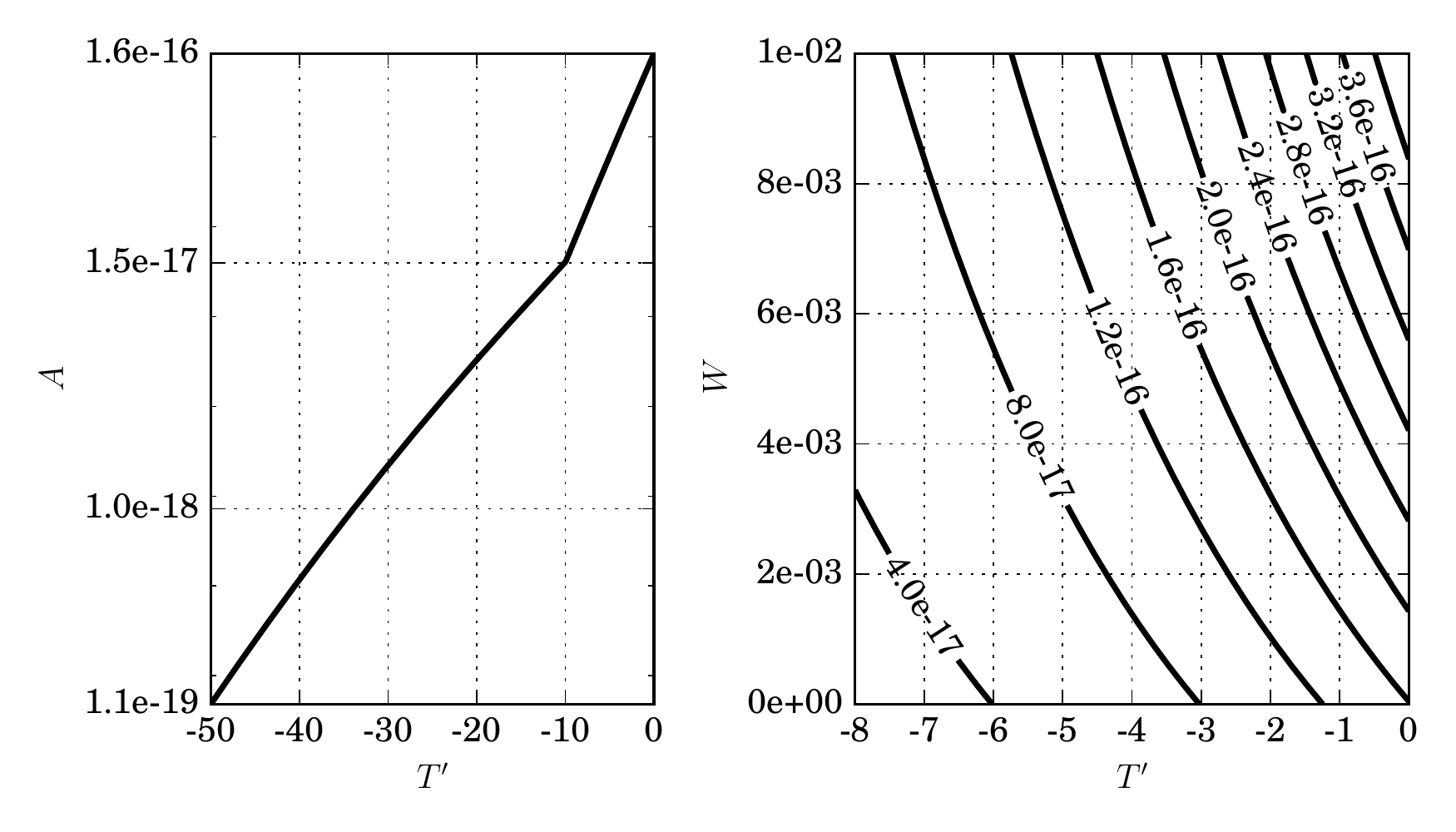}
  \caption[Flow-rate-factor diagram]{Flow rate factor (\ref{rate_factor}) in Pa\sups{-3}a\sups{-1} used in viscosity (\ref{viscosity}) with $W=0$ (left), and a range of water contents (right).  The change in slope in the left figure at $-10^\circ$ C is due the discontinuity of parameters (\ref{flow_parameter}) and (\ref{activation_energy}).}
  \label{rate_factor_image}
\end{figure}

First, shear viscosity (\ref{viscosity}) and internal friction (\ref{strain_heat}) are defined with the Arrhenius-type, energy-dependent flow-rate factor \index{Flow-rate factor} \index{Flow enhancement factor}
\begin{align}
  \label{rate_factor}
  A(\theta) = a_T E \left( 1 + 181.5 W_f \right) \exp\left(-\frac{Q_T}{RT'} \right),
\end{align}
with enhancement factor $E = 1$ unless otherwise specified (see \S \ref{ssn_shelf_inversion}), universal gas constant $R$, empirically-constrained water content $W_f = \min\{W, 0.01\}$ \citep{patterson}, energy-dependent flow-parameter \citep{pattersonBudd}
\begin{align}
  \label{flow_parameter}
  a_T &= \begin{cases}
           3.985 \times 10^{-13} \hspace{3mm} \text{s\sups{-1}Pa\sups{-3}} & T' < 263.15 \\
           1.916 \times 10^{3\hphantom{-1}} \hspace{3mm} \text{s\sups{-1}Pa\sups{-3}} & T' \geq 263.15 \\
         \end{cases},
\end{align}
temperature-dependent creep activation energy 
\begin{align}
  \label{activation_energy}
  Q_T & = \begin{cases}
            6.00 \times 10^{4} \hspace{3mm} \text{J mol\sups{-1}} & T' < 263.15 \\
            1.39 \times 10^{5} \hspace{3mm} \text{J mol\sups{-1}} & T' \geq 263.15 \\
          \end{cases},
\end{align}
with pressure-melting adjusted temperature $T' = T + \gamma p$ \citep{greveP}.  Note that rate factor (\ref{rate_factor}) will decrease with decreasing temperature, and increase with increasing water content.  This dependence on energy is thus also expressed by shear viscosity (\ref{viscosity}) and internal friction (\ref{strain_heat}).  Therefore, a decrease in energy $\theta$ produces stiffer ice that is more resistance to deformation, while an increase in $\theta$ produces softer ice which is easier to deform (Figure \ref{rate_factor_image}).

The second coupling between energy and momentum is by external friction heat flowing into the ice -- defined analogously to internal friction (\ref{strain_heat}) -- formed from the negative product of \emph{tangential stress} with \emph{tangential velocity} \citep{greve}
\begin{align}
  \label{basal_friction_heat}
  q_{fric} = -\big( \sigma \cdot \mathbf{n} \big)_{\Vert} \cdot \mathbf{u}_{\Vert} = \beta \mathbf{u} \cdot \mathbf{u} = \beta \Vert \mathbf{u} \Vert^2,
\end{align}
where impenetrable bed condition (\ref{impenetrability}) was used.

%===============================================================================
%===============================================================================

\section{Energy boundary conditions} \label{ssn_energy_boundary_conditions}

The flow of energy present on the base of the ice, when neglecting sub-glacial water transport, is a combination of geothermal $q_{geo}$ and friction energy from sliding $q_{fric}$ sources
\begin{align}
  \label{basal_energy_source}
  g_N = q_{geo} + q_{fric},
\end{align}
in units of Wm\sups{-2}.  For cold regions this energy flux can only raise the temperature of ice, whereas for ice at its pressure melting point, the energy flux will create water along the basal surface by melting.  Once generated, this water is available for transport via a sub-glacial hydraulic network.  If basal water transport is prohibited, the energy flux will raise the water content along the basal surface, which may then flow under pressure through veins located between ice three-grain boundaries \citep{nye_and_frank, shreve, raymond_and_harrison, lliboutry}.  The specification of sub-glacial water transport is therefore critical for determining the correct distribution of water -- and thus energy -- both interior and exterior to the ice sheet.  While beyond the scope of study here, basal hydraulic models may be easily incorporated into the solution method presented in the following sections.

The boundary conditions over exterior ice-sheet surface $\Gamma = \Gamma_A \cup \Gamma_W \cup \Gamma_C \cup \Gamma_T$ with atmosphere boundary $\Gamma_A$, boundary in contact with ocean $\Gamma_W$, cold or temperate basal surface without overlying temperate ice $\Gamma_C$, and temperate basal surface with overlying temperate ice $\Gamma_T$ (Figure \ref{ice_profile_domain}) for temperature $T$ are
\begin{align}
  \label{surface_temperature}
  T &= T_S &&\text{ on } \Gamma_A, \\
  \label{sea_temperature}
  T &= T_{sea} &&\text{ on } \Gamma_W, \\
  \label{temperature_flux}
  \big( k \nabla T \big) \cdot \mathbf{n} &= g_N &&\text{ on } \Gamma_C, \\
  \label{temperate_temperature}
  T &= T_m &&\text{ on } \Gamma_T,
\end{align}
where seawater temperature $T_{sea}$ may possibly be unequal to pressure melting temperature $T_m$.  A similar set of conditions exist for water content $W$,
\begin{align}
  \label{surface_water}
  W &= W_S      &&\text{ on } \Gamma_A, \\
  \label{sea_water_content}
  W &= W_{sea}  &&\text{ on } \Gamma_W, \\
  \label{cold_water}
  W &= 0        &&\text{ on } \Gamma_C, \\
  \label{latent_flux}
  \big( \nu \nabla W \big) \cdot \mathbf{n} &= \rho L_f M_b - \rho_w L_f F_b  &&\text{ on } \Gamma_T,
\end{align}
with basal melting rate $M_b$, basal water discharge \index{Basal water discharge} from the ice $F_b$, and water content on ocean boundaries $W_{sea}$.  Note that latent energy flux \index{Latent energy flux} (\ref{latent_flux}) has been defined previously by \citet{greve97}; make the substitution $\rho_{\text{w}}^+ = \rho_w$, $\mathcal{P}_{\text{b}}^{\text{w}} = \rho M_b$, and $\dot{m}_{\text{b}}^{\text{w}} = \rho_w F_b$ in Equation (2.44) and $\omega^- \approx 0$ in Equation (2.51) of this work (see Appendix A).

The relationship between basal water content and the above stated basal boundary conditions may be put into perspective by considering the \index{Balance equations!Basal energy} \emph{basal energy balance}, defined as a combination of energy flowing into the mixture, sensible energy flux out of the mixture, and energy fluctuations caused by latent heat of fusion transitions \citep[][section 9.3.4]{greve}:
\begin{align}
  \label{basal_energy_balance}
  M_b L_f \rho = q_{geo} + q_{fric} - \big( k \nabla T \big) \cdot \mathbf{n} &&\text{ on } \Gamma_G,
\end{align}
where $\Gamma_G = \Gamma_C \cup \Gamma_T$ is the entire grounded basal surface.  Note that because $L_f$ and $\rho$ are both positive and non-zero, if $M_b > 0$, mass is able to be accumulated by the basal hydraulic network in the form of melting ice.  Likewise, if $M_b < 0$, the ice is able to accumulate mass on its basal surface in the form of freezing water, if available from the hydraulic network.  Furthermore, note that for $T |_{\Gamma_G} < T_m$, the flux of temperature from the ice (\ref{temperature_flux}) inserted into basal energy balance (\ref{basal_energy_balance}) implies that $M_b = 0$.  Finally, at temperate basal regions, essential temperature condition (\ref{temperate_temperature}) applies and the basal melt rate \index{Basal melting rate} becomes quantifiable from basal energy balance (\ref{basal_energy_balance}),
\begin{align}
  \label{basal_melt_rate}
  M_b = \frac{q_{geo} + q_{fric} - \big( k \nabla T \big) \cdot \mathbf{n}}{L_f \rho} &&\text{ on } \Gamma_{CT}.
\end{align}
where $\Gamma_{CT} = \left( \Gamma_C \cap \Gamma_T \right) \cup \Gamma_T$ is the entire temperate basal surface.  Therefore, basal melt rate $M_b$ is a means to quantify both the interaction of the ice with sub-glacial water by way of basal melting and accretion by freezing (referred to as \index{Basal freeze-on} \emph{basal freeze-on}), and the flux of water into the base of the ice.  

Similarly, solving for basal water discharge $F_b$ in latent energy flux (\ref{latent_flux}) results in
\begin{align}
  \label{basal_water_discharge}
  F_b &= \frac{\rho L_f M_b - \left( \nu \nabla W \right) \cdot \mathbf{n}}{L_f \rho_w} &&\text{ on } \Gamma_T.
\end{align}
This is the total rate of water flowing from the ice in units of m s\sups{-1}.  Note that if $F_b = 0$ no amount of water is able to flow from the ice.  In this case, latent energy flux (\ref{latent_flux}) is solely determined by basal melt rate (\ref{basal_melt_rate}) and all basally-generated melt water is available for transport to the interior of the ice as governed by enthalpy gradient flux (\ref{enthalpy_grad}).  Basal water discharge $F_b$ may also be negative, corresponding with water flowing into the ice from the basal hydraulic network, and positive, corresponding with water flowing out; this will respectively increase and decrease latent energy flux (\ref{latent_flux}).  For the purposes of this paper, water is not allowed to flow into the interior of the ice from the basal hydraulic network, corresponding to the requirement $F_b \geq 0$.

Next, temperature boundary conditions (\ref{surface_temperature}, \ref{sea_temperature}) and water boundary conditions (\ref{surface_water}, \ref{sea_water_content}) are combined using energy constitutive relation (\ref{energy}),
\begin{align}
  \label{gamma_a_ebc}
  \theta &= \int_{T_0}^{T_S} c_i(T')\ dT' + W_S L_f &&\text{ on } \Gamma_A \\
  \label{gamma_w_ebc}
  \theta &= \int_{T_0}^{T_{sea}} c_i(T')\ dT' + W_{sea} L_f &&\text{ on } \Gamma_W,
\end{align}
which may be evaluated using Equation (\ref{energy_quad}) if $T_0 = 0$.

Basal regions containing overhead temperature at the temperature melting point $T_m$ are defined by the coefficient
\begin{align}
  \label{weak_temperate_marker}
  \alpha_w = \begin{cases}
             0, & \nabla T \cdot \mathbf{n} \neq \nabla T_m \cdot \mathbf{n} \hspace{2.5mm} \text{ on } \Gamma_G \\
             1, & \nabla T \cdot \mathbf{n} = \nabla T_m \cdot \mathbf{n} \hspace{2.5mm} \text{ on } \Gamma_G 
           \end{cases},
\end{align}
or, using the continuity of internal water content $W$, by the coefficient
\begin{align}
  \label{temperate_marker}
  \alpha = \begin{cases}
             0, & W = 0 \hspace{2.5mm} \text{ on } \Gamma_G \\
             1, & W > 0 \hspace{2.5mm} \text{ on } \Gamma_G 
           \end{cases}.
\end{align}
Coefficient (\ref{temperate_marker}) is stronger than (\ref{weak_temperate_marker}), as it does not require calculation of derivatives and is thus not affected by low-resolution approximation errors arising from the discretization.

Coefficient (\ref{temperate_marker}) used in conjunction with enthalpy-gradient flux (\ref{enthalpy_grad}) and basal melt rate (\ref{basal_melt_rate}) combines sensible energy flux (\ref{temperature_flux}) and latent energy flux (\ref{latent_flux}) over the entire grounded surface $\Gamma_G$:
\begin{align}
  \label{temperate_flux}
  \left( \frac{\kappa}{c} \nabla \theta \right) \cdot \mathbf{n} &= \begin{cases}
             g_N, & \alpha = 0  \\
             g_N - (k \nabla T_m) \cdot \mathbf{n} - \rho_w L_f F_b, & \alpha = 1
           \end{cases},
\end{align}
producing finally the basal energy flux
\begin{align}
  \label{energy_flux}
  \left( \frac{\kappa}{c} \nabla \theta \right) \cdot \mathbf{n}
  &= g_N - \alpha g_W &&\text{ on } \Gamma_G,
\end{align}
where $g_W = (k \nabla T_m) \cdot \mathbf{n} + \rho_w L_f F_b$.  By stating basal energy flux boundary (\ref{energy_flux}) in this manner, a continuous range of basal energy flux values across the entire grounded basal surface is allowed.

In areas with overlying temperate ice, it has been previously assumed \citep{aschwanden, kleiner} that the flux of water -- and therefore energy -- out of the ice is in balance with the water gradient caused by basal melt, corresponding with the condition $F_b=M_b \rho / \rho_w$ in energy-flux (\ref{energy_flux}) and
\begin{align}
  \label{zero_basal_water_discharge}
  \left( \frac{\kappa}{c} \nabla \theta \right) \cdot \mathbf{n} &= g_N - \alpha g_N &&\text{ on } \Gamma_G.
\end{align}
By the strict use of this boundary condition, strain-heating may increase the water content of interior ice to abnormally high levels \citep{aschwanden}.  In contrast, by using generalized basal energy flux (\ref{energy_flux}) while allowing the possibility for $F_b > M_b \rho / \rho_w$, the energy flux across the basal surface is able to adapt to large quantities of internally-generated water produced by internal friction (\ref{strain_heat}).

Therefore, provided that the non-advective water-flux coefficient $\nu$ in water flux (\ref{latent_flux}) is greater than zero -- corresponding to an enthalpy-gradient coefficient (\ref{enthalpy_grad}) with $k_0 < \infty$ in temperate regions -- a procedure for choosing an appropriate value of $F_b$ in (\ref{energy_flux}) defines a mechanism for enforcing maximum water retention demand (\ref{water_demand}).

%===============================================================================
%===============================================================================

\section{Exploring basal-melting-rate}

\index{Basal melting rate}
As discussed in \S \ref{ssn_energy_boundary_conditions}, basal melting rate (\ref{basal_melt_rate}) is only valid in regions where $\theta \ge \theta_m$ and is positive only when
\begin{align}
  q_{geo} + q_{fric} > \big( k_i \nabla T_m \big) \cdot \mathbf{n}.
\end{align}
Digging deeper into the pressure-melting gradient,
\begin{align}
  \big( k_i \nabla T_m \big) \cdot \mathbf{n} &= \big( k_i \gamma \nabla p \big) \cdot \mathbf{n} \notag \\
  &= k_i \gamma \nabla \big( \delta p \big) \cdot \mathbf{n} + k_i \gamma \big( \nabla (\rho g (S - z)) \big) \cdot \mathbf{n} \notag \\
  &= k_i \gamma \nabla \big( \delta p \big) \cdot \mathbf{n} + k_i \gamma \rho g \left( \frac{\partial S}{\partial x} n_x + \frac{\partial S}{\partial y} n_y  + n_z \right) \notag,
\end{align}
where Clausius-Clapeyron relationship (\ref{temperature_melting}) has been applied and basal pressure $p$ was separated into cryostatic $\rho g H$ with $z$-varying thickness $H = S - z$ for surface height $S$, and dynamic $\delta p$ terms.  Next, applying the low basal-slope requirement of first-order momentum \citet{blatter, pattyn}, $\mathbf{n} \approx [0\ 0\ \text{-}1]^\intercal$ and
\begin{align}
  \big( k_i \nabla T_m \big) \cdot \mathbf{n}
  &\approx - k_i \gamma \frac{\partial \delta p}{\partial z} - k_i \gamma \rho g = - k_i \gamma \left( \frac{\partial \delta p}{\partial z} + \rho g \right).
\end{align}
Therefore, because $q_{geo}$ and $q_{fric}$ (\ref{basal_friction_heat}) are both positive-definite functions, refreeze only occurs in regions where the dynamic pressure gradient is able to overcome the geothermal and frictional energy flux adjusted by a small scalar value,
\begin{align}
  \label{refreeze}
  q_{geo} + q_{fric} > k_i \gamma \left( \frac{\partial \delta p}{\partial z} + \rho g \right) \implies \text{ refreezing.}
\end{align}
Note also that in the case of cryostatic assumptions, $\delta p$ is zero and refreeze condition (\ref{refreeze}) simplifies to $q_{geo} + q_{fric} > k_i \gamma \rho g$.  Hence with an average geothermal flux of $\mathcal{O}(10\sups{-2})$ W m\sups{-2}, $q_{fric} \gg q_{geo}$, and $k_i \gamma \rho g = \mathcal{O}(10\sups{-3})$ W m\sups{-2}, refreeze is unlikely to occur under these assumptions.  Therefore, the full-Stokes momentum balance must be applied in order to properly identify refreezing or melt.

%===============================================================================
%===============================================================================

\section{Weak energy approximation}

The variational and weak form of the steady-state ($\partial_t \theta = 0$) version of  energy balance (\ref{energy_euler_lagrange}) and associated boundary conditions (\ref{gamma_a_ebc}, \ref{gamma_w_ebc}, \ref{energy_flux}) is constructed by taking the inner product of the residual 
\begin{align}
  \label{energy_forward_model}
  \mathscr{R}(\theta, F_b) &= \rho \mathbf{u} \cdot \nabla \theta - \nabla \left( \frac{\kappa}{c} \right) \cdot \nabla \theta - \frac{\kappa}{c} \nabla \cdot \nabla \theta - Q.
\end{align}
with the test function $\psi \in \testspace$ (see test space (\ref{test_space})), integrating over the entire ice-sheet volume $\Omega$,
\begin{align}
  - \int_{\Omega} Q \psi\ d\Omega
  + \int_{\Omega} \rho \mathbf{u} \cdot \nabla \theta \psi\ d\Omega 
  - \int_{\Omega} \nabla \left( \frac{\kappa}{c} \right) \cdot \nabla \theta \psi\ d\Omega & \notag \\
  + \int_{\Omega} \left( \frac{\kappa}{c} \right) \nabla \theta \cdot \nabla \psi\ d\Omega 
  \label{ie_intermediate_var_form}
  - \int_{\Gamma} \left( \frac{\kappa}{c} \nabla \theta \right) \cdot \mathbf{n} \psi\ d\Gamma &= 0,
\end{align}
where the diffusive term has been integrated by parts.  Because outward flux terms over essential boundaries vanish (see test space (\ref{test_space})), the boundary integral over $\Gamma$ is reduced to an integral over $\Gamma_G$ using energy flux (\ref{energy_flux}),
\begin{align}
  \label{energy_basal_boundary_form}
  \int_{\Gamma} \left( \frac{\kappa}{c} \nabla \theta \right) \cdot \mathbf{n} \psi\ d\Gamma &= \int_{\Gamma_G} \left( g_N - \alpha g_W \right)\ \psi\ d\Gamma_G.
\end{align}

%===============================================================================

\subsection{Numerical stabilization}

Numerical instabilities will manifest in areas where the transport of energy is dominated by advection (see \S \ref{ssn_stabilized_methods}).  Hence stabilization is required to reduce non-physical oscillations resulting from solving Galerkin-form (\ref{ie_intermediate_var_form}).

To begin, the linear differential operator associated with problem (\ref{energy_euler_lagrange}) is
\begin{align}
  \label{internal_energy_linear_operator}
  \Lu \theta &= \rho \mathbf{u} \cdot \nabla \theta - \nabla \left( \frac{\kappa}{c} \right) \cdot \nabla \theta - \frac{\kappa}{c} \nabla \cdot \nabla \theta.
\end{align}
On close inspection, the advective part of this operator is
\begin{align}
  \label{internal_energy_advective_operator}
  \Lu_{\text{adv}} \theta &= \tilde{\mathbf{u}} \cdot \nabla \theta,
\end{align}
with quasi-velocity
\begin{align}
  \label{internal_energy_quasi_velocity}
  \tilde{\mathbf{u}} &= \rho \mathbf{u} - \nabla \left( \frac{\kappa}{c} \right).
\end{align}
The conductive gradient term $\nabla \left( \nicefrac{\kappa}{c} \right)$ of the energy-flux therefore contributes in an advective sense to the transport of energy; energy transport increases as the magnitude of the conductive gradient increases.  For example, if the ice is taken -- without loss of generality -- as stationary, $\mathbf{u} = \mathbf{0}$ and energy transport occurs by diffusion and quasi-advection in the down-gradient direction of the conduction term $\nicefrac{\kappa}{c}$.

Next, the stabilized form of internal-energy balance (\ref{ie_intermediate_var_form}) with linear operator (\ref{internal_energy_linear_operator}) using general stabilized form (\ref{generalized_form}) with test function $\psi$ and intrinsic-time parameter $\tau_{\text{IE}}$ is 
\begin{align}
  \label{internal_energy_generalized_form}
  (\psi, \Lu \theta) + (\mathbb{L} \psi, \tau_{\text{IE}}(\Lu\theta - Q)) &= (\psi, Q),
\end{align}
where operator $\mathbb{L}$ is a differential operator chosen from 
\begin{align}
  \label{ie_gls_operator}
  \mathbb{L} &= + \Lu && \text{Galerkin/least-squares (GLS)} \\
  \label{ie_supg_operator}
  \mathbb{L} &= + \Lu_{\text{adv}} && \text{SUPG} \\
  \label{ie_ssm_operator}
  \mathbb{L} &= - \Lu^* && \text{subgrid-scale model (SSM)}
\end{align}
with adjoint of operator $\Lu$ denoted $\Lu^*$.

\begin{figure}
  \centering
    \includegraphics[width=\linewidth]{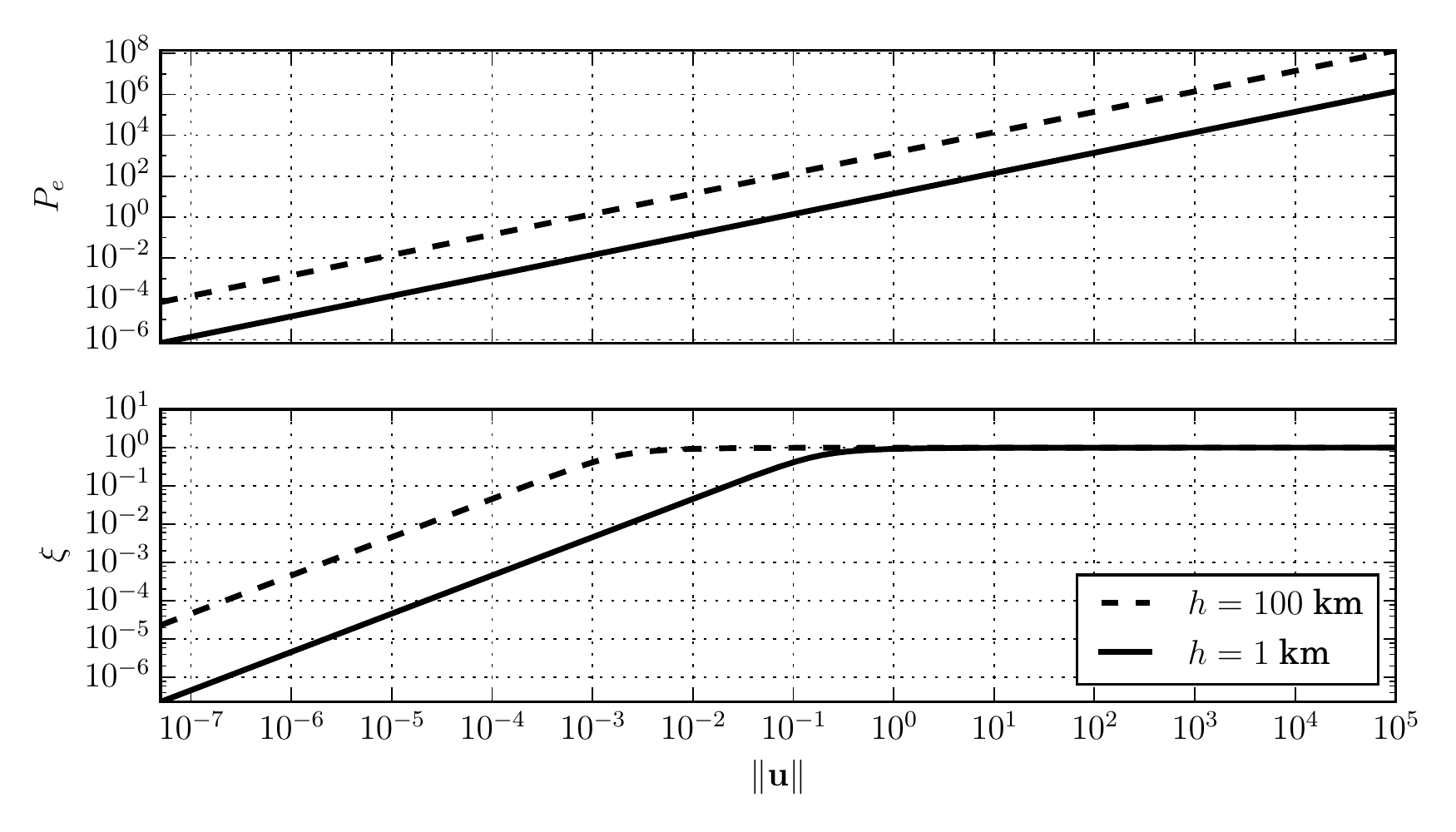}
  \caption[P\'eclet number and intrinsic-time parameter diagram]{The element P\'eclet number $P_{\'e}$ (top) with $h = 1$ km (solid) and $h=100$ km (dashed) over a range of velocity values with magnitude $\Vert \mathbf{u} \Vert$ in m a\sups{-1} appropriate to ice-sheets. The corresponding intrinsic time $\xi$ multiplicative term to the SUPG formulation for linear-Lagrange elements (\ref{intrinsic_time_ftn}) (bottom) becomes very close to unity after the ice speed gets above $\Vert \mathbf{u} \Vert \approx 2$ cm a$^{-1}$ with $h=1$ km.}
  \label{peclet_image}
\end{figure}

Making the appropriate substitutions in SUPG parameter (\ref{tau_supg}) results in the intrinsic-time parameter \index{Intrinsic-time parameter!Internal energy}
\begin{align}
  \label{tau_ie}
  \tau_{\text{IE}} &= \frac{h \xi(P_{\'e})}{2 \Vert \tilde{\mathbf{u}} \Vert}, \hspace{10mm} P_{\'e} = \frac{h \Vert \tilde{\mathbf{u}} \Vert}{2 \rho \Xi},
\end{align}
where $h$ is the cell diameter, $P_{\'e}$ is the ratio of advective to diffusive flux coefficients, referred to as the \index{Peclet@P\`eclet number!Internal energy} \emph{element P\'eclet} number, and $\xi(P_{\'e})$ is a function that is dependent on the element-shape-functions utilized.  For example, if the shape functions $\psi$ are taken as the linear Lagrange elements described in \S \ref{ssn_intro_galerkin_equations}, the accuracy-optimal function choice for $\xi$ is given by \citep{brooks}
\begin{align}
  \label{intrinsic_time_ftn}
  \xi\left( P_{\'e} \right) &= \coth \left( P_{\'e} \right) - \frac{1}{P_{\'e}}.
\end{align}
It was later determined by \citet{codina_1992} that when quadratic shape functions are used, the accuracy-optimal function for $\xi$ changes to
\begin{align}
  \xi_1(P_{\'e}) &= \frac{1}{2} \left( \coth\left( \frac{P_{\'e}}{2} \right) - \frac{2}{P_{\'e}} \right) \notag \\
  \label{intrinsic_time_ftn_quad}
  \xi\left( P_{\'e} \right) &= \frac{\big(3 + 3 P_{\'e} \xi_1\big) \tanh(P_{\'e}) - \big(3 P_{\'e} + P_{\'e}^2 \xi_1 \big)}{ \big(2 - 3 \xi_1 \tanh(P_{\'e}) \big) P_{\'e}^2}.
\end{align}
Note that for any substantial ice flow, $P_{\'e}$ will be very large and thus $\xi \approx 1$ (Figure \ref{peclet_image}).  Additionally, note that if linear-Lagrange elements are used, the application of SUPG stabilization (\ref{ie_supg_operator}) and GLS stabilization (\ref{ie_gls_operator}) in form (\ref{internal_energy_generalized_form}) are identical.

Therefore, using boundary integral (\ref{energy_basal_boundary_form}) with variational form (\ref{ie_intermediate_var_form}) in stabilized form (\ref{internal_energy_generalized_form}), the problem consists of finding $\theta \in \mathcal{H}^1(\Omega)$ such that
\begin{align}
  \label{var_form}
  - \int_{\Omega} Q \psi\ d\Omega
  + \int_{\Omega} \rho \mathbf{u} \cdot \nabla \theta \psi\ d\Omega 
  - \int_{\Omega} \nabla \left( \frac{\kappa}{c} \right) \cdot \nabla \theta \psi\ d\Omega & \notag \\
  + \int_{\Omega} \left( \frac{\kappa}{c} \right) \nabla \theta \cdot \nabla \psi\ d\Omega 
  - \int_{\Gamma_G} \left( g_N - \alpha g_W \right)\ \psi\ d\Gamma_G & \notag \\
  + \int_{\Omega} \tau_{\text{IE}} \big( \mathbb{L} \psi \big) \big( \Lu\theta - Q \big)\ d\Omega - \int_{\Omega} Q\ \psi\ d\Omega  &= 0,
\end{align}
for all $\psi \in \mathcal{H}^1(\Omega)$ subject to the remaining essential boundary conditions (\ref{gamma_a_ebc}) and (\ref{gamma_w_ebc}). 

%===============================================================================

\subsection{Energy balance discretization}

For a mesh with $N_n$ vertices and $N_{\Gamma}$ essential exterior vertices corresponding with Dirichlet boundaries (\ref{gamma_a_ebc}, \ref{gamma_w_ebc}), the approximation 
\begin{align}
  \label{galerkin_approximation}
  \theta \approx \theta_h = \sum_{j=1}^{N_n} \bm{\theta}_j \psi_j + \sum_{j=N_n+1}^{N_n+N_{\Gamma}} \bm{\theta}_j \psi_j
\end{align}
defines an expansion of $\theta$ with unknown coefficients $\bm{\theta}_j$ associated with the trial functions $\psi_j$ (see trial space (\ref{trial_space})). Inserting approximation (\ref{galerkin_approximation}) into (\ref{var_form}) results in the matrix-vector set of equations
\begin{align}
  \label{component_var_form}
  \mathbf{r} = \bm{\mathcal{C}}\bm{\theta} - \bm{\mathcal{K}}\bm{\theta} + \bm{\mathcal{D}}\bm{\theta} + \bm{\mathcal{S}} \bm{\theta} + \mathbf{f}^{\text{ext}} - \mathbf{f}^{\text{int}} - \mathbf{f}^{\text{stz}},
\end{align}
where for each test function $\psi_i$ with $i,j = 1, 2, \ldots, N_n$,
\begin{align}
  \label{advective_form}
  \bm{\mathcal{C}}_{ij} &= \int_{\Omega} \rho \mathbf{u} \cdot \nabla \psi_j\ \psi_i\ d\Omega &&\leftarrow \text{advection} \\
  \label{conductive_gradient_form}
  \bm{\mathcal{K}}_{ij} &= \int_{\Omega} \nabla \left( \frac{\kappa}{c} \right) \cdot \nabla \psi_j\ \psi_i\ d\Omega &&\leftarrow \text{cond.~grad.} \\
  \label{diffusion_form}
  \bm{\mathcal{D}}_{ij} &= \int_{\Omega} \left( \frac{\kappa}{c} \right) \nabla \psi_j \cdot \nabla \psi_i\ d\Omega &&\leftarrow \text{diffusion} \\
  \label{stabilization_form_a}
  \bm{\mathcal{S}}_{ij} &= \int_{\Omega} \tau_{\text{IE}} \big( \mathbb{L} \psi_i \big) \big( \Lu \psi_j \big)\ d\Omega &&\leftarrow \text{stab'z'tion} \\
  \label{basal_energy_gradient_form}
  \mathbf{f}_i^{\text{ext}} &= \int_{\Gamma_G} \left( g_N - \alpha g_W \right)\ \psi_i\ d\Gamma_G &&\leftarrow \text{energy flux} \\
  \label{internal_friction_form}
  \mathbf{f}_i^{\text{int}} &= \int_{\Omega} Q\ \psi_i\ d\Omega &&\leftarrow \text{strain heat} \\
  \label{stabilization_form_l}
  \mathbf{f}_i^{\text{stz}} &= \int_{\Omega} \tau_{\text{IE}} \big( \mathbb{L} \psi_i \big)\ Q\ d\Omega &&\leftarrow \text{stab'z'tion},
\end{align}
and $\mathbf{r}$ is the residual error vector, defining a \emph{weak solution} $\bm{\theta}$ to variational form (\ref{var_form}) when $\Vert \mathbf{r} \Vert = 0$.  When using approximation (\ref{galerkin_approximation}), the sum involving the $N_{\Gamma}$ exterior vertices will produce a set of matrices with identical properties as (\ref{advective_form} -- \ref{diffusion_form}), but where the unknown coefficients $\bm{\theta}_j$, $j = N_n+1,N_n+2,\ldots,N_n+N_{\Gamma}$ are  known from the surface essential boundaries (\ref{gamma_a_ebc} -- \ref{gamma_w_ebc}).  Thus, the last sum in approximation (\ref{galerkin_approximation}) interpolates the boundary data onto the finite-element basis \citep{elman}.

If an identical discrete basis is used for both $\psi_i$ and $\psi_j$, $\theta_h$ defined by (\ref{galerkin_approximation}) corresponds with \emph{Bubnov-Galerkin} approximation.  In such a case, it is easily seen that $\bm{\mathcal{D}}$ is symmetric, and as it turns out, positive-definite \citep{elman}.  The same cannot be said of $\bm{\mathcal{C}}$ and $\bm{\mathcal{K}}$.  However, symmetry is added back to linear system (\ref{component_var_form}) by the term $\bm{\mathcal{S}}$.  For example, if linear Lagrange elements are used as a basis for $\psi$ and streamline-upwind/Petrov-Galerkin stabilization operator (\ref{ie_supg_operator}) is used,
\begin{align}
  \label{stabilization_form_a_expanded}
  \bm{\mathcal{S}}_{ij} &= \int_{\Omega} \tau_{\text{IE}} \big( \Lu_{\text{adv}} \psi_i \big) \big( \Lu \psi_j \big)\ d\Omega \notag \\
  &= \int_{\Omega} \tau_{\text{IE}} \left( \left( \rho \mathbf{u} + \nabla \left( \frac{\kappa}{c} \right) \right) \cdot \nabla \psi_i \right) \left( \left( \rho \mathbf{u} + \nabla \left( \frac{\kappa}{c} \right) \right) \cdot \nabla \psi_j \right)\ d\Omega \notag \\
  &= \int_{\Omega} \tau_{\text{IE}} \left( \tilde{\mathbf{u}} \cdot \nabla \psi_i \right) \left( \tilde{\mathbf{u}} \cdot \nabla \psi_j \right)\ d\Omega,
\end{align}
where quasi-velocity (\ref{internal_energy_quasi_velocity}) has been applied and the fact that second-derivatives of linear element shape-functions are zero.  Matrix (\ref{stabilization_form_a_expanded}) is symmetric-positive-definite, and thus the addition of this term in (\ref{component_var_form}) has the effect of increasing the stability or \emph{coercivity} of the linear system.

An algorithm well suited for the solution of problems possessing non-symmetric matrices such as these is the \emph{generalized minimum residual method} (GMRES).  This procedure is one of many \emph{Krylov subspace methods}, which iteratively reduces the energy norm of the error.  However, because the non-advective flux coefficient $\kappa$ is discontinuous, depending on the unknown $\theta$, and also because thermal properties (\ref{thermal_conductivity}) and (\ref{heat_capacity}) are non-linear with respect to $\theta$, system (\ref{component_var_form}) is non-linear.  Thus, a linearization of this system is required.  One such linearization is \emph{Newton's method}, which iteratively reduces residual (\ref{component_var_form}) by solving for the direction of decent with respect to the solution space of $\theta$ (investigate \S \ref{ssn_newton_raphson}).

The source code of CSLVR uses an implementation similar to Code Listing \ref{cslvr_enthalpy}.

\begin{python}[label=cslvr_enthalpy, caption={CSLVR source code contained in the \texttt{Enthalpy} class.}]
# define test and trial functions : 
psi     = TestFunction(model.Q)
dtheta  = TrialFunction(model.Q)
theta   = Function(model.Q, name='energy.theta')
  
# momentum-dependent properties :
U       = momentum.velocity()
epsdot  = momentum.effective_strain_rate(U) + model.eps_reg
eta     = momentum.viscosity(U)
    
# internal friction (strain heat) :
Q_s     = 4 * eta * epsdot

# coefficient for non-advective water flux (enthalpy-gradient) :
k_c     = conditional( gt(W, 0.0), model.k_0, 1 )

# thermal conductivity and heat capacity (Greve and Blatter 2009) :
ki      = 9.828 * exp(-0.0057*T)
ci      = 146.3 + 7.253*T

# bulk properties :
k       =  (1 - W)*ki   + W*kw     # bulk thermal conductivity
c       =  (1 - W)*ci   + W*cw     # bulk heat capacity
rho     =  (1 - W)*rhoi + W*rhow   # bulk density
kappa   =  spy * k_c * k           # discontinuous with water, J/(a*m*K)
Xi      =  kappa / (rho*c)         # bulk enthalpy-gradient diffusivity

# basal heat-flux natural boundary condition :
q_fric  = beta * inner(U,U)
g_w     = model.gradTm_B + rhow*L*Fb
g_n     = q_geo + q_fric
g_b     = g_n - alpha*g_w

# the Peclet number : 
Ut     = rho*U - grad(kappa/c)
Unorm  = sqrt(dot(Ut, Ut) + DOLFIN_EPS)
PE     = Unorm*h/(2*kappa/c)

# for linear elements :
if model.order == 1:
  xi     = 1/tanh(PE) - 1/PE

# for quadratic elements :
if model.order == 2:
  xi_1  = 0.5*(1/tanh(PE) - 2/PE)
  xi    =     ((3 + 3*PE*xi_1)*tanh(PE) - (3*PE + PE**2*xi_1)) \
           /  ((2 - 3*xi_1*tanh(PE))*PE**2)

# intrinsic time parameter :
tau   = h*xi / (2 * Unorm)

# the linear differential operator for this problem :
def Lu(u):
  Lu  = + rho * dot(U, grad(u)) \
        - kappa/c * div(grad(u)) \
        - dot(grad(kappa/c), grad(u))
  return Lu

# the advective part of the operator : 
def L_adv(u):
  return dot(Ut, grad(u))

# the adjoint of the operator :
def L_star(u):
  Ls  = - dot(U, grad(u)) \
        - Xi * div(grad(u)) \
        + 1/rho * dot(grad(kappa/c), grad(u))
  return Ls

# use streamline-upwind/Petrov-Galerkin stabilization : 
if stabilization_method == 'SUPG':
  s      = "    - using streamline-upwind/Petrov-Galerkin stabilization -"
  LL     = lambda x: + L_adv(x)
# use Galerkin/least-squares stabilization :
elif stabilization_method == 'GLS':
  s      = "    - using Galerkin/least-squares stabilization -"
  LL     = lambda x: + Lu(x)
# use subgrid-scale-model stabilization :
elif stabilization_method == 'SSM':
  s      = "    - using subgrid-scale-model stabilization -"
  LL     = lambda x: - L_star(x)
print_text(s, cls=self)

self.theta_a = + rho * dot(U, grad(dtheta)) * psi * dx \
               + kappa/c * inner(grad(psi), grad(dtheta)) * dx \
               - dot(grad(kappa/c), grad(dtheta)) * psi * dx \
               + inner(LL(psi), tau*Lu(dtheta)) * dx

self.theta_L = + g_b * psi * dBed_g \
               + Q_s * psi * dx \
               + inner(LL(psi), tau * Q_s) * dx

# surface boundary condition : 
self.theta_bc = []
self.theta_bc.append( DirichletBC(Q, theta_surface, 
                                  model.ff, model.GAMMA_S_GND) )
self.theta_bc.append( DirichletBC(Q, theta_surface,
                                  model.ff, model.GAMMA_S_FLT) )
self.theta_bc.append( DirichletBC(Q, theta_surface, 
                                  model.ff, model.GAMMA_U_GND) )
self.theta_bc.append( DirichletBC(Q, theta_surface,
                                  model.ff, model.GAMMA_U_FLT) )

# apply T_melt conditions of portion of ice in contact with water :
self.theta_bc.append( DirichletBC(Q, theta_float, 
                                  model.ff, model.GAMMA_B_FLT) )
self.theta_bc.append( DirichletBC(Q, theta_float, 
                                  model.ff, model.GAMMA_L_UDR) )

# form the solver parameters :
self.solve_params = self.default_solve_params()

def default_solve_params(self):
  """ 
  Returns a set of default solver parameters that yield good performance
  """
  params  = {'solver' : {'linear_solver'       : 'gmres',
                         'preconditioner'      : 'amg'},
             'use_surface_climate' : False}
  return params

def solve(self, annotate=False):
  """ 
  Solve the energy equations, saving enthalpy to model.theta, temperature 
  to model.T, and water content to model.W.
  """
  solve(self.theta_a == self.theta_L, self.theta, self.theta_bc,
        solver_parameters = self.solve_params['solver'], annotate=annotate)
\end{python}

%===============================================================================

\section{Water content optimization} \label{ssn_water_content_optimization}

In the absence of a constitutive relation for basal water discharge $F_b$, system of equations (\ref{energy_euler_lagrange}, \ref{gamma_a_ebc}, \ref{gamma_w_ebc}, \ref{energy_flux}) is ill-posed.  However, this problem can be overcome using methods from control theory \citep{bryson, macayeal, nocedal}.  Notice that it is expected that for very low basal water content $W$ the discharge of water from the base of the ice-sheet be small; likewise, in areas of abnormally high basal water content it is expected that the discharge of water from the ice be high.  Thus it is desired to minimize the difference between water content $W$ and maximum water content $W_c$ as given by demand (\ref{water_demand}) over the entire basal surface.  Mathematically, this can be stated in terms of the \index{Constrained optimization!State parameter} \emph{state} parameter $\theta$ as minimizing the \index{Constrained optimization!Objective function} $L^2$ \emph{objective} functional
\begin{align}
  \label{energy_objective}
  \mathscr{J}(\theta) &= \frac{1}{2} \int_{\Gamma_G} \left( \theta - \theta_c \right)^2\ d\Gamma_G,
\end{align}
where $\theta_c = \theta_m + W_c L_f$ is the maximum energy associated with maximum water content demand (\ref{water_demand}).  This functional will be minimized in two ways: first, by minimizing the flow of water out of the ice in regions with $\theta < \theta_c$, corresponding with the lower bound $F_b = 0$ in basal energy flux (\ref{energy_flux}); and second, by maximizing the flow of water out of the ice at regions with $\theta > \theta_c$, corresponding with $F_b > M_b \rho / \rho_w$.  The role of $F_b$ in basal energy flux (\ref{energy_flux}) is is hence the \index{Constrained optimization!Control parameter} \emph{control} parameter for the minimization of objective (\ref{energy_objective}).

For additional illustration, recall that the inward-directed flow of energy-flux (\ref{energy_flux}) is maximal for lower bound $F_b = 0$.  Therefore, at regions with $T < T_m$, it is expected that $F_b = 0$.  Furthermore, because parameter $\alpha$ defined by (\ref{temperate_marker}) eliminates any basal water discharge over cold regions in energy flux (\ref{energy_flux}), this expectation is automatically satisfied and integration across the entire grounded basal domain $\Gamma_G$ by objective (\ref{energy_objective}) is justified.  Notice also that an $F_b$ which produces a minimum of objective (\ref{energy_objective}) will of course affect the distribution of water content $W$ in the mixture interior, and thus also affect the enhancement of flow as evident by the water-content dependence of rate-factor (\ref{rate_factor}).  Because of this, it is either necessary to re-compute the momentum balance for each optimization of $\theta$, or combine both energy and momentum into a single mixed formulation.

The minimization of objective (\ref{energy_objective}), solution of variational problem (\ref{var_form}, \ref{gamma_a_ebc}, \ref{gamma_w_ebc}), and satisfaction of the positivity constraint of basal water flux $F_b \geq 0$ can be stated as a constrained optimization problem analogous to that presented in Chapter \ref{ssn_optimization_with_constraints}.   Thus we state the problem in the form
\begin{align}
  \min_{\theta}\ \mathscr{J}(\theta) \hspace{10mm} \text{subject to  }
  \begin{cases}
    \mathscr{R}(\theta, F_b) = 0,\\
    F_b \geq 0,
  \end{cases}
  \label{w_opt}
\end{align}
where $\mathscr{R}$ is the residual, or in this context \index{Residual} \index{Constrained optimization!Forward model} \emph{forward model} defined by (\ref{energy_forward_model}).  The energy Lagrangian \index{Constrained optimization!Lagrangian} associated with problem (\ref{w_opt}) is (see Chapter \ref{ssn_optimization_with_constraints})
\begin{align}
  \label{energy_lagrangian}
  \mathscr{L}(\theta, F_b, \lambda) &= \mathscr{J}(\theta, F_b) + \big( \lambda, \mathscr{R}(\theta, F_b) \big),
\end{align}
where the notation $(f,g) = \int_{\Omega} f g d\Omega$ is the inner product.  Using Lagrangian (\ref{energy_lagrangian}), the first necessary condition in (\ref{objective_perturbations}) is satisfied when \index{Constrained optimization!Adjoint variable} \emph{adjoint variable} $\lambda$ is chosen -- say $\lambda = \lambda^*$ -- such that for a given energy state $\theta$ and control parameter $F_b$, 
\begin{align}
  \label{energy_adjoint}
  \lambda^* = \argminl_{\lambda} \left\Vert \frac{\delta}{\delta \theta} \mathscr{L} \left( \theta, F_b; \lambda \right) \right\Vert.
\end{align}
This $\lambda^*$ may then be used in condition (\ref{condition_two}) to calculate the direction of decent of $\mathscr{L}$ with respect to the control variable $F_b$ for a given energy state $\theta$ and adjoint variable $\lambda^*$,
\begin{align}
  G = \frac{\delta}{\delta F_b} \mathscr{L} (\theta, F_b, \lambda^*).
\end{align}
This \emph{G\^{a}teaux derivative} provides a direction which basal water discharge $F_b$ may follow in order to satisfy the second condition in (\ref{objective_perturbations}) and thus minimize objective functional (\ref{energy_objective}).

\begin{figure}
  \centering
    \def\svgwidth{\linewidth}
    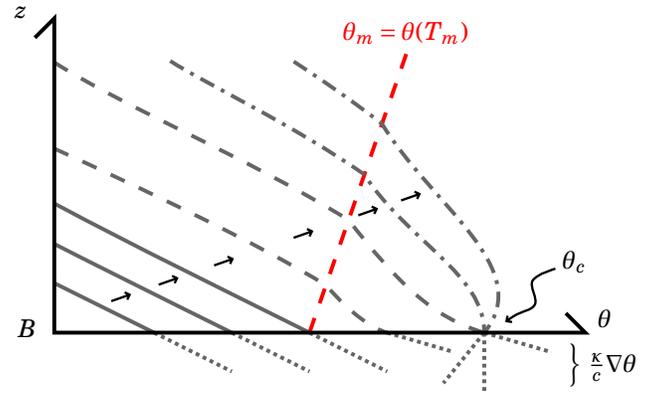
  \caption[Water-content optimization diagram]{Illustration of the transition from cold ice to temperate.  The arrows point in the direction of increasing energy profiles, with cold ice profiles ({\color[rgb]{0.39607843,0.39607843,0.39607843}solid gray}), temperate ice with basal water contents less than $W_c$ ({\color[rgb]{0.39607843,0.39607843,0.39607843}dashed gray}), and temperate ice which would have basal water contents higher than $W_c$ without some amount of basal water discharge $F_b$ ({\color[rgb]{0.39607843,0.39607843,0.39607843}dashed-dotted gray}).  Note that the gradient in $\theta$ increases once the ice becomes temperate, as required when $(k \nabla T_m) \cdot \mathbf{n} < 0$ in basal energy flux (\ref{energy_flux}).}
  \label{temperate_zone_revised_image}
\end{figure}

\subsection{Variations}

Lagrangian (\ref{energy_lagrangian}) is formed by taking the inner product of the adjoint variable $\lambda$ with forward model (\ref{energy_forward_model}), integrating over the entire domain $\Omega$, integrating the diffusive term by parts, and adding stabilization,
\begin{align}
  \label{expanded_lagrangian}
  \mathscr{L}(\theta, F_b, \lambda) =
  &+ \frac{1}{2} \int_{\Gamma_G} \left( \theta - \theta_c \right)^2\ d\Gamma_G \notag \\ 
  &+ \int_{\Omega} \rho \mathbf{u} \cdot \nabla \theta \lambda\ d\Omega - \int_{\Omega} Q \lambda\ d\Omega \notag \\
  &- \int_{\Omega} \nabla \left( \frac{\kappa}{c} \right) \cdot \nabla \theta \lambda\ d\Omega + \int_{\Omega} \left( \frac{\kappa}{c} \right) \nabla \theta \cdot \nabla \lambda\ d\Omega \notag \\
  &- \int_{\Gamma_G} \left( g_N - \alpha g_W \right)\ \lambda\ d\Gamma_G \notag \\
  &+ \int_{\Omega} \tau_{\text{IE}} \big( \mathbb{L} \lambda \big) \big( \Lu\theta - Q \big)\ d\Omega.
\end{align}
Therefore, the first variation of $\mathscr{L}$ with respect to $\theta$ in the direction $\psi$ is
\begin{align}
 \label{dLdtheta}
 \frac{\delta \mathscr{L}}{\delta \theta} = 
 &+ \int_{\Gamma_G} \left( \theta - \theta_c \right) \psi\ d\Gamma_G 
  + \int_{\Omega} \rho \mathbf{u} \cdot \nabla \psi \lambda\ d\Omega \notag \\
  &- \int_{\Omega} \nabla \left( \frac{\kappa}{c} \right) \cdot \nabla \psi \lambda\ d\Omega 
  + \int_{\Omega} \left( \frac{\kappa}{c} \right) \nabla \psi \cdot \nabla \lambda\ d\Omega \notag \\
  &+ \int_{\Omega} \tau_{\text{IE}} \big( \mathbb{L} \lambda \big) \left( \rho \mathbf{u} \cdot \nabla \psi - \nabla \left( \frac{\kappa}{c} \right) \cdot \nabla \psi - \left( \frac{\kappa}{c} \right) \nabla \cdot \nabla \psi \right) \lambda\ d\Omega.
\end{align}
while the first variation of $\mathscr{L}$ with respect to $F_b$ is
\begin{align}
  \label{dLdFb}
  G(F_b, \lambda) = \frac{\delta \mathscr{L}}{\delta F_b}
  = \int_{\Gamma_G} \psi \alpha \rho_w L_f \lambda\ d\Gamma_G.
\end{align}

\subsection{Energy optimization procedure}

\index{Log-barrier method}
To determine an optimal value of basal water discharge $F_b$, a variation of a primal-dual-interior-point algorithm with a filter-line-search method implemented by the IPOPT framework \citep{wachter} may be used (read \S \ref{ssn_log_barrier}).  In the context of problem (\ref{w_opt}), the algorithm implemented by IPOPT computes approximate solutions to a sequence of barrier problems
\begin{align}
\begin{aligned}
  &\min_{F_b}\ \left\{ \varphi_{\mu}(\theta, F_b) = \mathscr{J}(\theta) - \mu \sum_{i=1}^{N_n} \ln(F_b^i) \right\}
\end{aligned}
\label{energy_barrier}
\end{align}
for a decreasing sequence of barrier parameters $\mu$ converging to zero, and $N_n$ is the number of degrees of freedom of the finite-element mesh.  For further details of this algorithm, see \S \ref{ssn_log_barrier}.

We defer an example of the energy-optimization and solution process until Chapter \ref{ssn_thermo_mechanical_coupling} which describes the thermo-mechanical coupling of energy $\theta$ and momentum $(\mathbf{u}, p)$.  The CSLVR implementation is shown in Code Listing \ref{cslvr_water_opt}.

%===============================================================================
%===============================================================================

\section{Effect of discontinuous energy conductivity}

A cause of concern for polythermal glaciologists is to correctly calculate the position of the CTS.  While this analysis addresses a separate issue pertaining to the energy balance -- the basal boundary condition -- it is important to note that this method is compatible with alternative solution methods of the energy balance equations.

First, it is expected that for a lower non-advective water diffusion coefficient $\nu$, the basal water discharge $F_b$ must adapt to reduce the water generated from strain-heating, located interior to the ice.  Of course, if $\nu = 0$ as in \citet{greve}, no amount increase in $F_b$ will reduce the internal water content due to the fact that basal-latent-heat-flux-boundary-condition (\ref{latent_flux}) is zero.  Likewise, for large values of $\nu$, water will be very efficiently routed and $W$ will be very sensitive to perturbations in $F_b$.

For any given energy-balance formulation using $\nu > 0$ implementing the procedure of \S \ref{ssn_water_content_optimization}, as the outward flux of water increases, intra-ice water is moved from the interior to the bed until the gradient in $\theta$ reaches the point that cost functional (\ref{energy_objective}) cannot be decreased further.

\pythonexternal[label=cslvr_water_opt, caption={CSLVR source code contained in the \texttt{Energy} class for solving the water-content-optimization procedure of  \S \ref{ssn_water_content_optimization}.}, firstline=327, lastline=520]{cslvr_src/energy.py}

%===============================================================================
%===============================================================================

\chapter{Thermo-mechanical coupling} \label{ssn_thermo_mechanical_coupling}

Coupling between the momentum-balance models described in Chapter \ref{ssn_momentum_and_mass_balance} and the energy-balance model of Chapter \ref{ssn_internal_energy_balance} -- referred to as \index{Thermo-mechanical coupling} \emph{thermo-mechanical coupling} (TMC) -- is accomplished here via fixed-point iteration, a process which generates approximations of velocity $\mathbf{u}$, pressure $p$, and energy $\theta$ that eventually converge to a stationary point.  This stationary point is attained when the norm of the difference between two successive approximations of $\theta$ are below a specified tolerance.

First, as discussed in \S \ref{ssn_stokes_variational_forms}, because effective strain-rates (\ref{effective_strain_rate}), (\ref{bp_effective_strain_rate}), (\ref{ps_effective_strain_rate}), and (\ref{rs_effective_strain_rate}) are non-linear in $\mathbf{u}$, momentum systems (\ref{extremum}), (\ref{bp_extremum}), (\ref{ps_extremum}), and (\ref{rs_extremum}) are linearized using Newton's method (see \S \ref{ssn_newton_raphson}).

To begin, it has been observed that this linearization will converge consistently provided that the current velocity guess $\mathbf{u}$ is sufficiently far from a stationary point, and so we initialize $\mathbf{u}$ to zero at the start of every iteration.  Once $\mathbf{u}$ and $p$ have been obtained by solving the momentum balance, the water-content optimization procedure described in the previous section is performed.  This procedure results in an optimal distribution of water for a given friction $\beta$, and thus an improved estimate of rate factor (\ref{rate_factor}) to be used in the subsequent iteration's momentum formulation (Algorithm \ref{tmc}).

Returning to non-linear energy balance discretization (\ref{component_var_form}), note that a solution process for this non-linear system -- such as Newton's method -- requires several intermediate solutions of (\ref{component_var_form}) be solved.  Thus, the non-linearity with respect to $\theta$ present in thermal conductivity (\ref{thermal_conductivity}), heat capacity (\ref{heat_capacity}), energy-flux (\ref{enthalpy_grad}), and rate-factor (\ref{rate_factor}) may be eliminated if instead the discontinuities are evaluated with regard to the previous TMC iteration's pressure-melting point.  This simplification saves considerable time, especially when considering the many forward-model (\ref{energy_forward_model}) solutions required by energy barrier problem (\ref{energy_barrier}).

It remains to properly define the temperate zone, and thus correct values for temperate zone marker $\alpha$ defined by (\ref{temperate_marker}).  To accomplish this, $\alpha$ is initially set to zero across the entire basal surface, hence assuming that the ice is cold throughout, and basal energy source (\ref{basal_energy_source}) is universally applied.  The energy distribution resulting from the energy balance is then evaluated, and $\alpha$ is assigned a value of one along any facet containing energies above pressure-melting energy (\ref{energy_melting}).

While this method of boundary marking is approximate, note that for temperate regions where the pressure-melting temperature decreases when moving up the ice column, flux of water (\ref{latent_flux}) with zero basal water discharge $F_b$ will be larger than basal energy source (\ref{basal_energy_source}), and will thus clearly be temperate (Figure \ref{temperate_zone_revised_image}).  It follows that this marking method has the potential to incorrectly mark boundaries only in areas where the pressure-melting point \emph{increases} when moving up from the basal surface and with negligible water transported by advection to its location.  The temperate basal boundary marking process and thermal-parameter linearization scheme is outlined by Algorithm \ref{tpu}.

Note in Algorithm \ref{tmc} that we initialize basal-water discharge $F_b$ to the value $M_b \rho / \rho_w$ -- consistent with zero-energy-flux boundary condition (\ref{zero_basal_water_discharge}) -- prior to solving $F_b$-optimization problem (\ref{w_opt}).  This has the effect of starting the optimization process at the same point for each iteration of TMC Algorithm \ref{tmc}, and leads to better convergence characteristics of the algorithm.

The implementation used by CSLVR for Algorithms \ref{tmc} and \ref{tpu} are shown in Code Listing \ref{cslvr_tmc} and \ref{cslvr_tpu}, respectively.

\begin{algorithm}
  \caption[Thermo-mechanical coupling]{ -- TMC fixed-point iteration}
  \label{tmc}
  \begin{algorithmic}[1] 
    \Function{TMC}{$\beta, \theta, F_b$}
      \State $a_{tol} := 100$; $n_{\max} := 350$; $r := \infty$; $a := \infty$; $i := 1$
      \While{$(a > a_{tol}$ \textbf{or} $r > r_{tol})$ \textbf{and} $i < n_{\max}$}
        \State $\phantom{F_b^*}\mathllap{\mathbf{U}} := [\mathbf{u}\ p]^{\intercal} \in \left( \mathcal{H}^1(\Omega) \right)^4 = [\mathbf{0}, p]^{\intercal}$
        \State $\phantom{F_b^*}\mathllap{\mathbf{U}^*} := \argminl_{\mathbf{U}} \Big\Vert \delta_{\mathbf{U}} \mathcal{A}(\beta, \theta, F_b; \mathbf{U}) \Big\Vert$
        \State TPU$(\mathbf{U}^*, \beta, F_b)$
        \State $\phantom{F_b^*}\mathllap{q_{fric}} := \beta \Vert \mathbf{u} \Vert^2$
        \State $\phantom{F_b^*}\mathllap{M_b} := \frac{q_{geo} + q_{fric} - \big( k \nabla T \big) \cdot \mathbf{n}}{L_f \rho}$
        \State $\phantom{F_b^*}\mathllap{F_b} := M_b \rho / \rho_w$
        \If{we want to optimize $F_b$}
          \State $F_b^* := \argminl_{F_b} \Big\{ \varphi_{\mu}(F_b) \Big\}$
        \Else
          \State $F_b^* := F_b$
        \EndIf
        \State $\phantom{F_b^*}\mathllap{\theta^*} := \argminl_{\theta} \Big\Vert \mathscr{R}(\mathbf{U}^*, \beta, F_b^*; \theta) \Big\Vert$
        \State $\phantom{F_b^*}\mathllap{a_n} := \Vert \theta - \theta^* \Vert_{2}$
        \If{$i = 1$}
          \State $r := a_n$
        \Else
          \State $r := |a - a_n|$
        \EndIf
        \State $\phantom{F_b^*}\mathllap{\theta} := \theta^*$, $F_b := F_b^*$, $i := i + 1$; $a := a_n$
      \EndWhile
    \State \Return $\theta, F_b$
    \EndFunction
  \end{algorithmic}
\end{algorithm}

\begin{algorithm}
  \caption[Thermal-parameters update]{ -- thermal parameters update (TPU)}
  \label{tpu}
  \begin{algorithmic}[1] 
    \Function{TPU}{$\mathbf{U}, \beta, F_b$}
      \State $a_{tol} := 100$; $n_{\max} := 50$; $a := \infty$; $r := \infty$; $i := 1$; $\theta := 0$
      \State $\alpha := 0$
      \While{$(a > a_{tol}$ \textbf{or} $r > r_{tol})$ \textbf{and} $i < n_{\max}$}
        \State $\theta^* := \argminl_{\theta} \Big\Vert \mathscr{R}(\mathbf{U}, \beta, F_b, T, W, a_T, Q_T, W_f; \theta) \Big\Vert$
        \If{$i = 1$}
        \State $\alpha^k := 1$ \textbf{if} $\theta^k > \theta_m^k$, $k \in [1,n]$
        \EndIf
        \State $\theta^* \rightarrow (T,W, a_T, Q_T, W_f, \kappa)$
        \State $a_n := \Vert \theta - \theta^* \Vert_{2}$
        \If{$i = 1$}
          \State $r := a_n$
        \Else
          \State $r := |a - a_n|$
        \EndIf
        \State $\theta := \theta^*$; $\phantom{\theta}\mathllap{i} := i + 1$; $a := a_n$
      \EndWhile
    \EndFunction
  \end{algorithmic}
\end{algorithm}

\pythonexternal[label=cslvr_tmc, caption={CSLVR source code contained in the \texttt{Model} class used to perform the thermo-mechanical coupling between the \texttt{Momentum} and \texttt{Energy} classes.}, firstline=2147, lastline=2393]{cslvr_src/model.py}

\pythonexternal[label=cslvr_tpu, caption={CSLVR source code contained in the \texttt{Enthalpy} class used to update the thermal parameters.}, firstline=1208, lastline=1301]{cslvr_src/energy.py}

\section{Plane-strain simulation} \label{ssn_tmc_plane_strain_simulation}

\begin{figure}
  \centering
    \includegraphics[width=\linewidth]{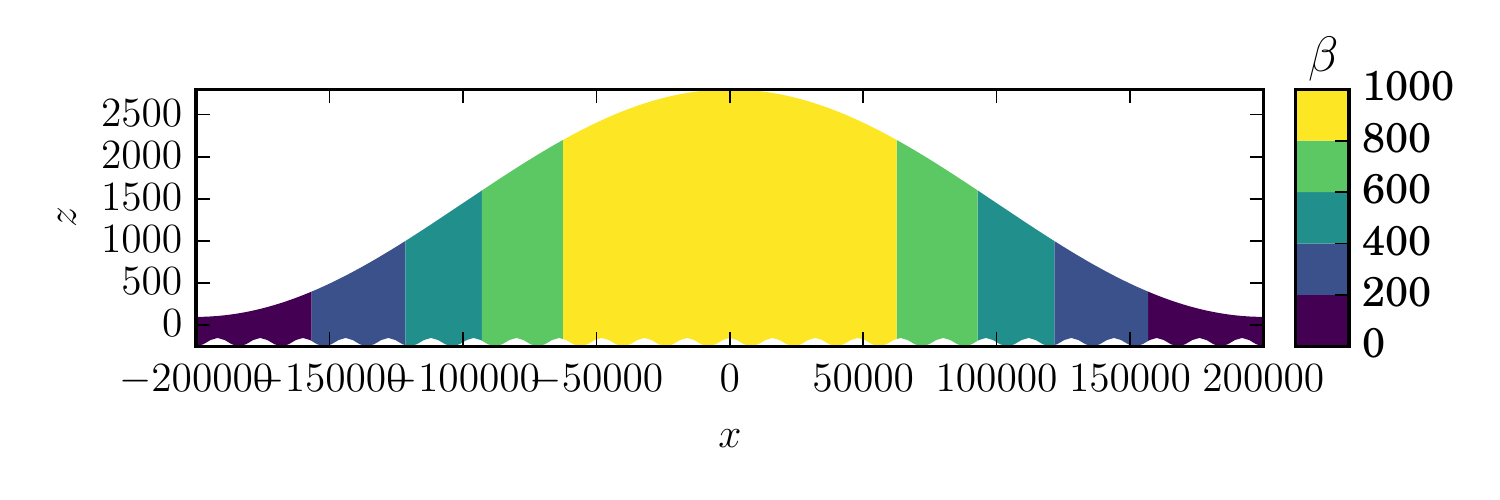}
  \caption[Plane-strain TMC example basal traction field]{The basal traction $\beta$ used for the TMC-simulation.}
  \label{tmc_beta_image}
\end{figure}

\index{Non-linear differential equations!2D}
\index{Plane-strain simulations}
For a simple example of TMC Algorithm \ref{tmc}, we use the same plane-strain model as \S \ref{ssn_plane_strain_simulation} using an altered maximum thickness and basal traction.  The two-dimensional ice-sheet model uses surface height 
\begin{align*}
  S(x) = \left( \frac{ H_{max} + B_0 - S_0 }{2} \right) \cos\left( \frac{2\pi}{\ell} x \right) + \left( \frac{H_{max} + B_0 + S_0}{2} \right),
\end{align*}
with thickness at the divide $H_{max}$, height of terminus above water $S_0$, depth of ice terminus below water $B_0$, and width $\ell$.  We prescribe the sinusoidally-varying basal topography
\begin{align*}
  B(x) = b \cos\left( n_b \frac{2\pi}{\ell} x \right) + B_0,
\end{align*}
with amplitude $b$ and number of bumps $n_b$.  The basal traction field prescribed follows the surface topography,
\begin{align*}
  \beta(x) = \left( \frac{\beta_{max} - \beta_{min}}{2} \right) \cos\left( \frac{2\pi}{\ell} x \right) + \left( \frac{\beta_{max} + \beta_{min}}{2} \right)
\end{align*}
with maximum value $\beta_{max}$ and minimum value $\beta_{min}$ (see Figure \ref{tmc_beta_image}).  The surface temperature followed the same sinusoidal pattern as the surface,
\begin{align*}
  T_S(x,z) &= T_{min} + \lambda_t (H_{max} + B_0 - S_0 - z)
\end{align*}
with minimum temperature $T_{min}$ and lapse rate $\lambda_t$.  The specific values used by the simulation are listed in Table \ref{tmc_plane_strain_values}.

\begin{figure*}
  \centering

    \includegraphics[width=\linewidth]{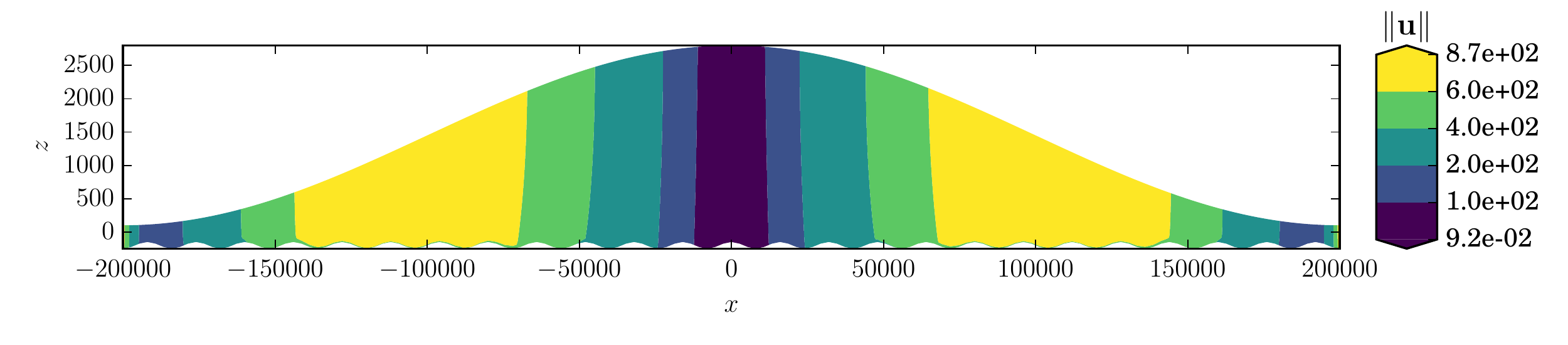}
    \includegraphics[width=\linewidth]{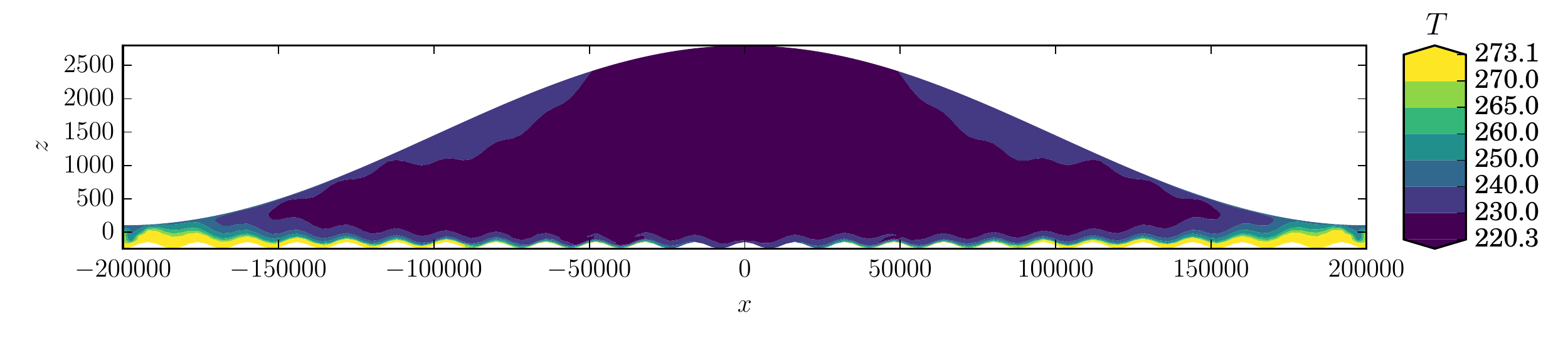}
    \includegraphics[width=\linewidth]{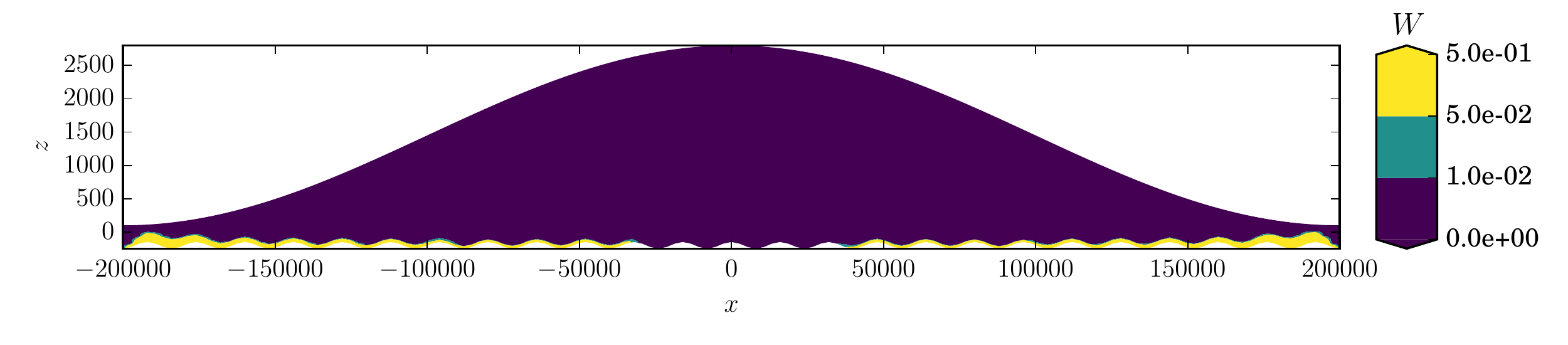}
    \includegraphics[width=\linewidth]{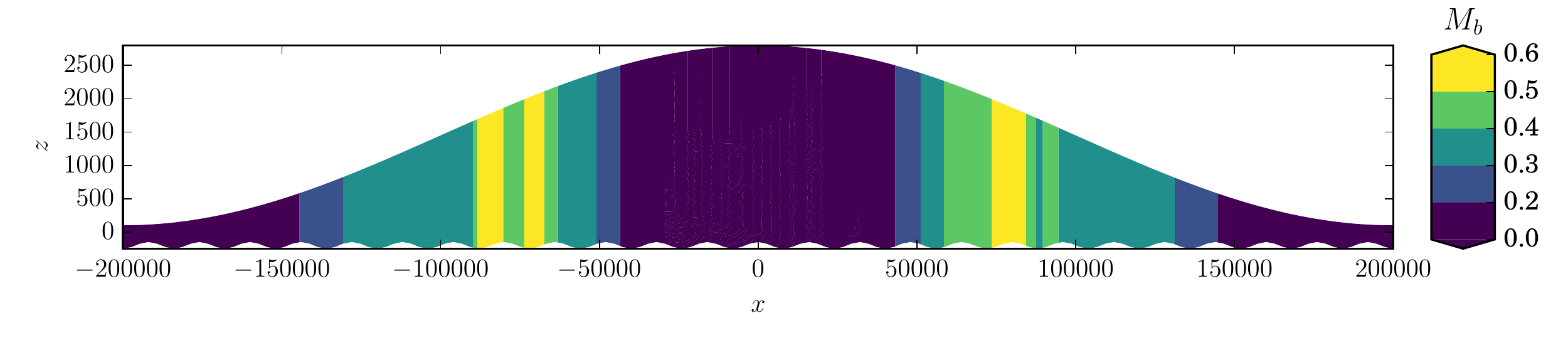}
    \includegraphics[width=\linewidth]{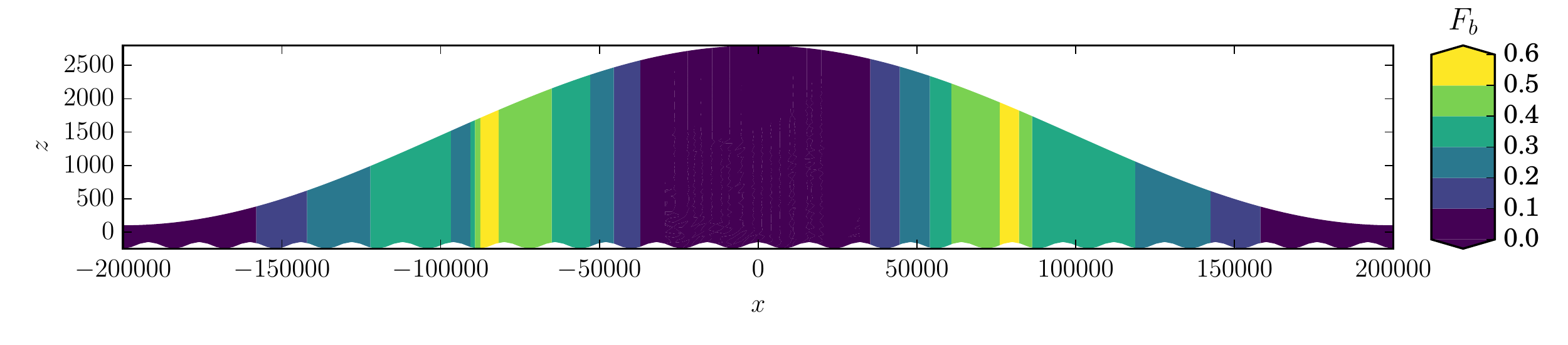}

  \caption[Plane-strain zero-energy-flux solution]{The plane-strain results attained using zero-temperate-energy-flux boundary condition (\ref{zero_basal_water_discharge}).  From top to bottom: velocity magnitude $\Vert \mathbf{u} \Vert$, temperature $T$, water content $W$, basal melt rate $M_b$, and basal water discharge $F_b$.}
  \label{tmc_zero_energy_image}
\end{figure*}

\begin{figure*}
  \centering

    \includegraphics[width=\linewidth]{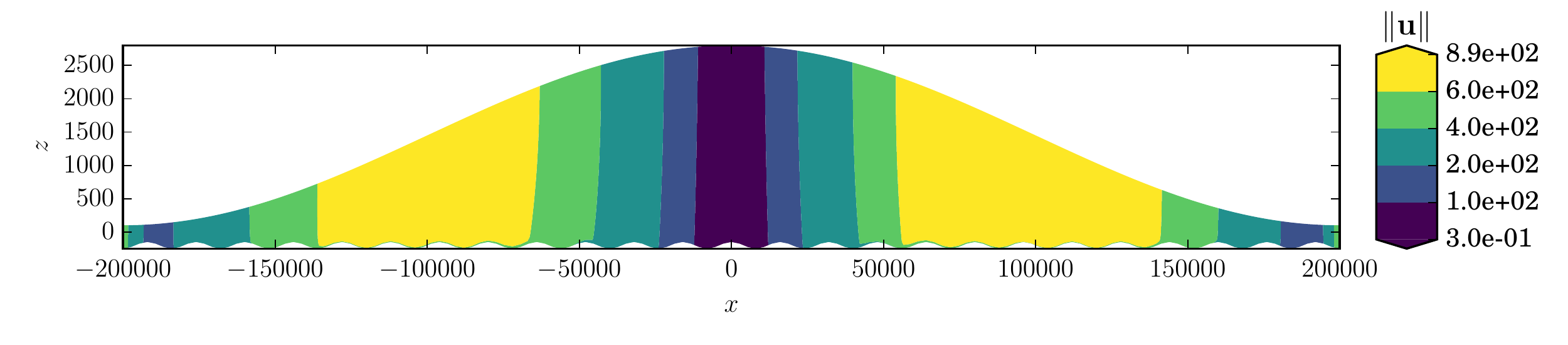}
    \includegraphics[width=\linewidth]{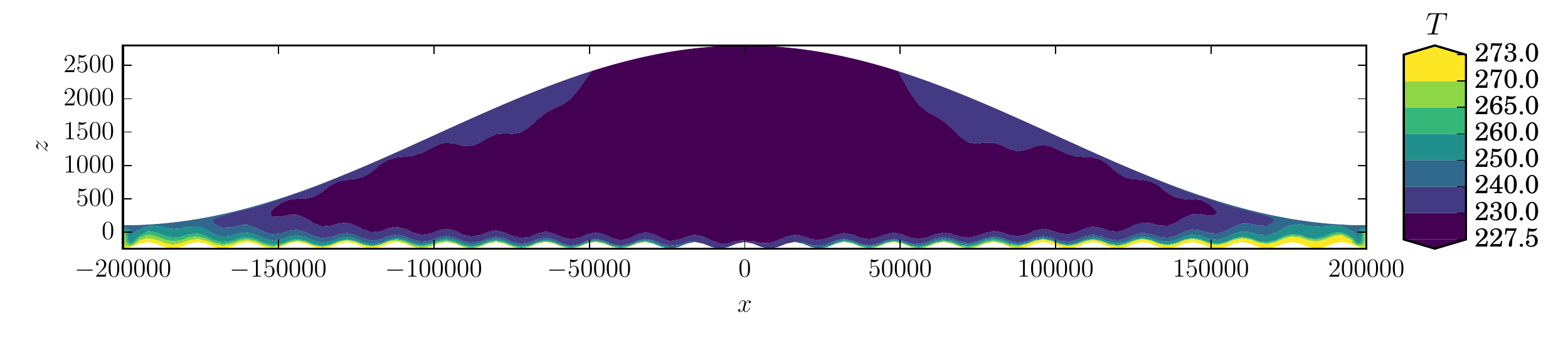}
    \includegraphics[width=\linewidth]{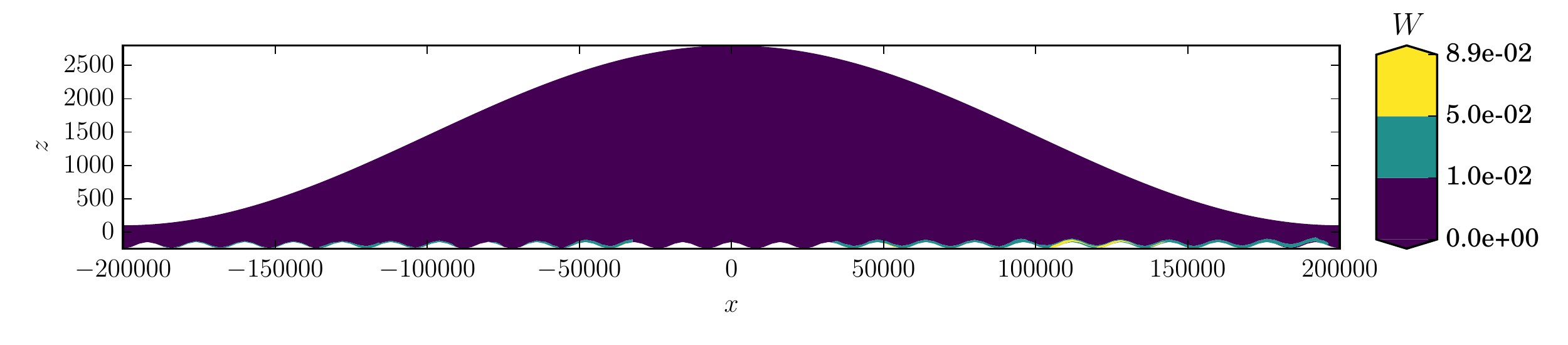}
    \includegraphics[width=\linewidth]{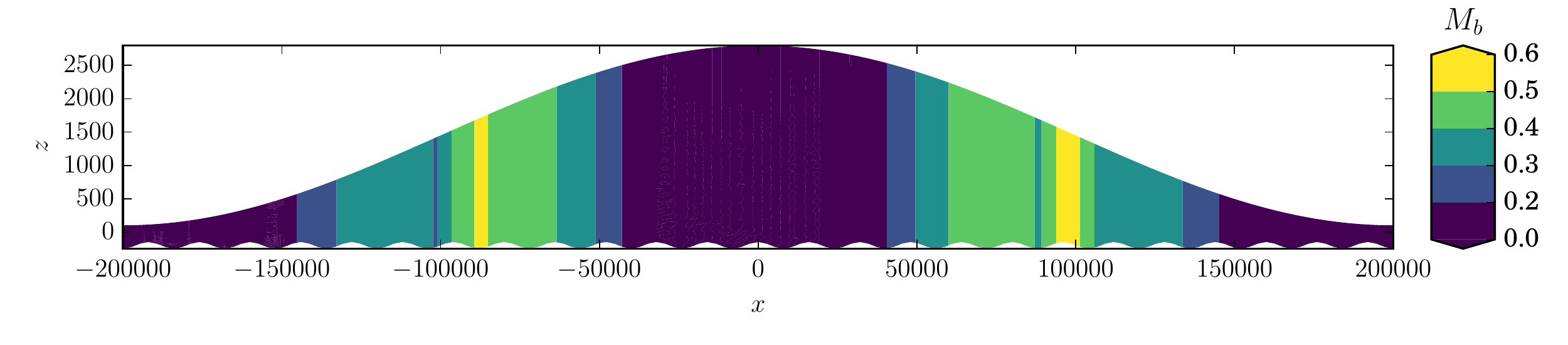}
    \includegraphics[width=\linewidth]{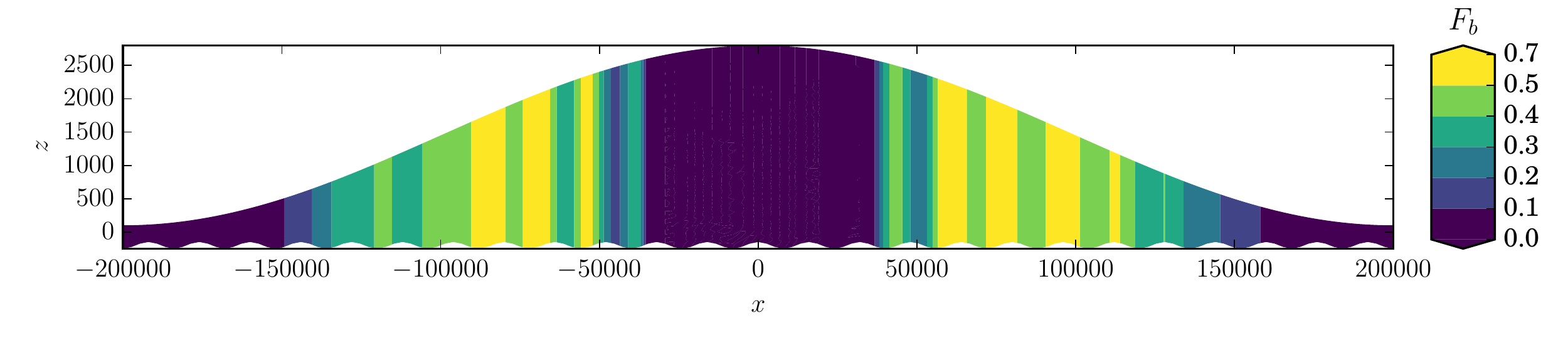}

  \caption[Plane-strain water-optimization solution]{Plane-strain results attained using the $F_b$ optimization process of \S \ref{ssn_water_content_optimization} with boundary condition (\ref{energy_flux}).  From top to bottom: velocity magnitude $\Vert \mathbf{u} \Vert$, temperature $T$, water content $W$, basal melt rate $M_b$, and basal water discharge $F_b$.}
  \label{tmc_Fb_image}
\end{figure*}

The difference between the use of zero-temperate-energy-flux condition (\ref{zero_basal_water_discharge}) and water-optimization procedure (\ref{w_opt}) is examined by performing Algorithm \ref{tmc} using both boundary conditions.  Results obtained solving the energy balance with the zero-temperate-energy-flux condition results in a water content in temperate areas reaching unreasonably high levels (Figure \ref{tmc_zero_energy_image}); while the water content field obtained using the $F_b$-optimization procedure also possessed areas with unreasonably high water content, the situation is much improved (Figure \ref{tmc_Fb_image}).

\begin{figure*}
  \centering
    \includegraphics[width=\linewidth]{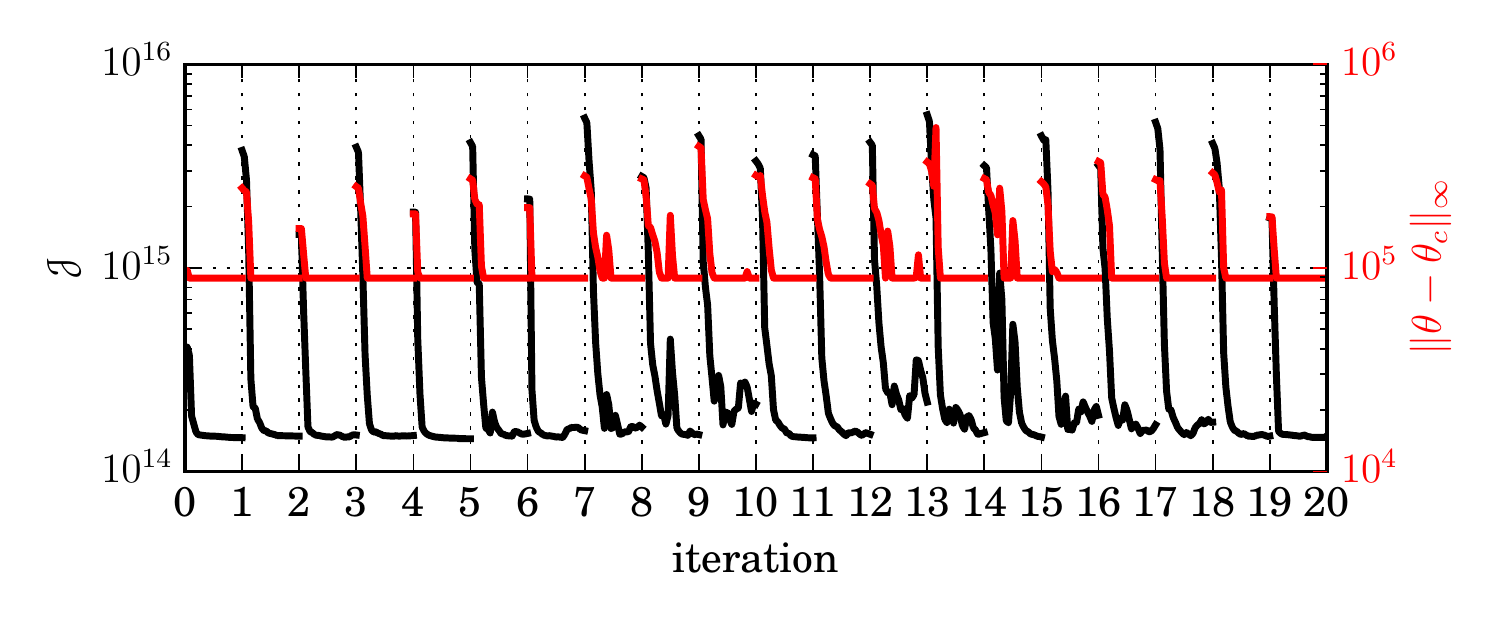}
    \caption[Plane-strain water-optimization convergence diagram]{Objective functional values (\ref{energy_objective}) for each iteration of Algorithm \ref{tmc} (left axis, black) and misfit between the critical value of current energy value $\theta$ and the critical energy value $\theta_c = \theta + W_c L$ (right axis, red) for the $F_b$-optimization procedure of \S \ref{ssn_water_content_optimization}.}
  \label{opt_convergence_image}
\end{figure*}

\begin{figure*}
  \centering
    \includegraphics[width=\linewidth]{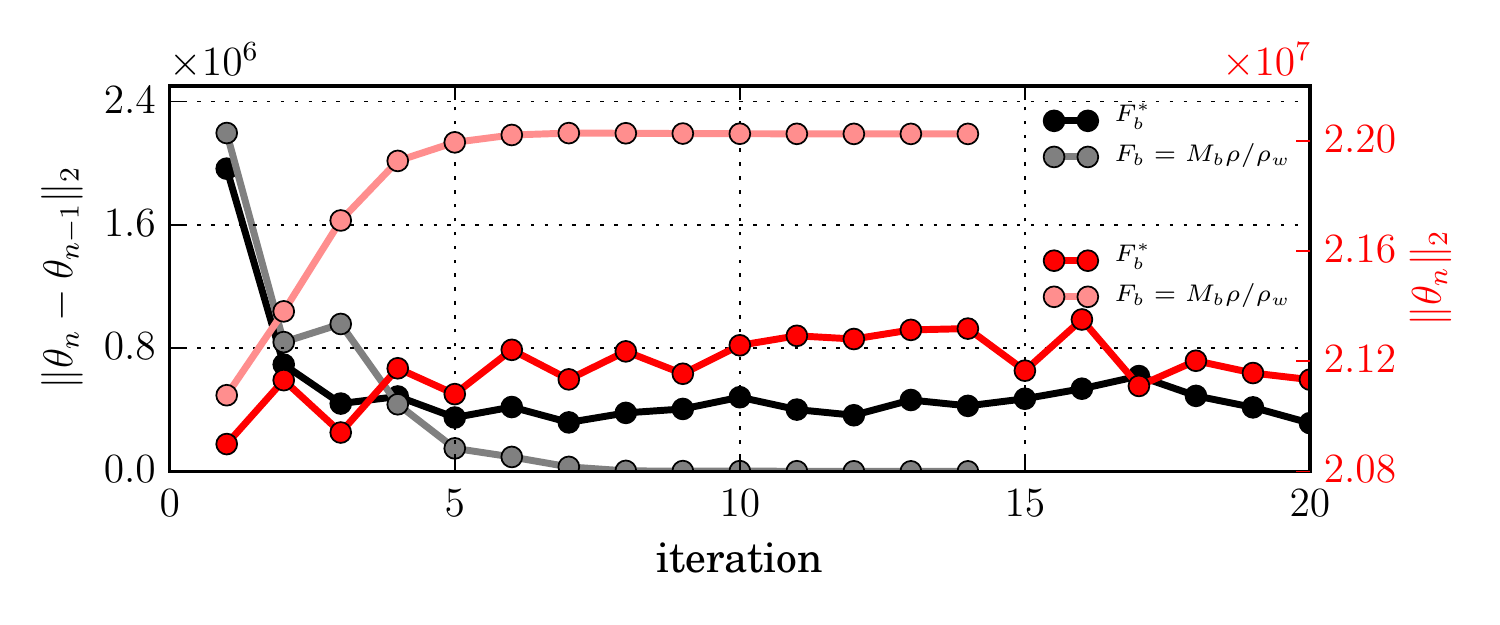}
    \caption[Plane-strain TMC convergence diagram]{Convergence plot of TMC algorithm \ref{tmc} applied to the plane-strain simulation using both the zero-energy-basal-boundary condition (\ref{zero_basal_water_discharge}) corresponding with $F_b = M_b \rho/\rho_w$ (left axis, grey), and the $F_b$-optimization procedure of \S \ref{ssn_water_content_optimization} using basal-water-discharge-boundary condition (\ref{energy_flux}) (left axis, black).  Also shown is the norm of the current energy guess $\theta_n$ for the zero-energy-basal-boundary condition simulation (right axis, pink) and the $F_b$-optimization procedure simulation (right axis, red).} 
  \label{tmc_convergence_image}
\end{figure*}

The convergence behavior associated with the optimization-procedure simulation appears to becomes unstable around iterate 200 (Figure \ref{opt_convergence_image}).  This may be due to the extreme nature of the simulation, or an indication that the optimization procedure of \S \ref{ssn_water_content_optimization} may require further constraints on basal water discharge $F_b$.  Note that the norm of the current energy guess using a zero basal-energy flux is much larger than that obtained using the $F_b$-optimization procedure (Figure \ref{tmc_convergence_image}).  This is because the water content using this method is allowed to reach values associated with ice composed of up to $\approx 50$\% water, an extremely unlikely event.  Additionally, note that the use of $W_f$ in flow-rate factor (\ref{rate_factor}) prevents rate-factor $A$ from reaching levels beyond empirical evidence, and also that if we had removed any energy above $\theta_c = \theta + W_c L$ as in \citet{greve}, or through a time-dependent function similar to that used by \citet{aschwanden}, the values of $\Vert \theta_n \Vert_2$ would be closer to that obtained by our $F_b$-optimization procedure in Figure \ref{tmc_convergence_image}.  Finally, due to the fact that zero-basal-energy flux boundary (\ref{zero_basal_water_discharge}) only removes water generated at the basal surface, the unadjusted internal water content values quantify the amount of water generated by strain-heating within the ice.

The CSLVR script used to solve this problem is shown in Code Listing \ref{cslvr_tmc_plane_strain}, and the code used to generate Figures \ref{tmc_zero_energy_image} and \ref{tmc_Fb_image} in Code Listing \ref{cslvr_tmc_plane_strain_dv}.

\begin{table}
\centering
\caption[Plane-strain TMC example variables]{Plane-strain TMC variables.}
\label{tmc_plane_strain_values}
\begin{tabular}{llll}
\hline
\textbf{Variable} & \textbf{Value} & \textbf{Units} & \textbf{Description} \\
\hline
$\dot{\varepsilon}_0$ & $10\sups{-15}$ & a\sups{-1}   & strain regularization \\
$F_b$         & $0$           & m a\sups{-1}  & basal water discharge \\
$k_x$         & $150$         & -- & number of $x$ divisions \\
$k_z$         & $50$          & -- & number of $z$ divisions \\
$N_e$         & $15000$       & -- & number of cells \\
$N_n$         & $7701$        & -- & number of vertices \\
$\ell$        & $400$         & km & width of domain \\
$H_{max}$     & $3000$        & m  & thickness at divide \\
$S_0$         & $100$         & m  & terminus height \\
$B_0$         & $-200$        & m  & terminus depth \\
$n_b$         & $25$          & -- & number of bed bumps \\
$b$           & $50$          & m  & bed bump amplitude \\
$\beta_{max}$ & $1000$        & kg m\sups{-2}a\sups{-1} & max basal traction \\ 
$\beta_{min}$ & $100$         & kg m\sups{-2}a\sups{-1} & min basal traction \\ 
$T_{min}$     & $228.15$      & K  & min.~temperature \\
$\lambda_t$   & $6.5$e$-3$    & K m\sups{-1} & lapse rate \\ 
$W_c$         & $0.03$        & -- & maximum basal $W$ \\
$k_0$         & $10\sups{-3}$ & -- & non-adv.~flux coef. \\ 
$q_{geo}$     & $4.2 \times 10\sups{-2}$ & W m\sups{-2} & geothermal heat flux \\
\hline
\end{tabular}
\end{table}

\pythonexternal[label=cslvr_tmc_plane_strain, caption={CSLVR script which performs the plane-strain TMC simulation of \S \ref{ssn_tmc_plane_strain_simulation}.}]{scripts/tmc/plane_strain_tmc.py}

\pythonexternal[label=cslvr_tmc_plane_strain_dv, caption={CSLVR script which plots the result generated by Code Listing \ref{cslvr_tmc_plane_strain}.}]{scripts/tmc/plot_ps_tmc.py}

%===============================================================================
%===============================================================================

\chapter{Inclusion of velocity data} \label{ssn_inclusion_of_velocity_data}

\index{Basal traction}
In order to partition the effects of energy-enhancement and traction-diminishment on momentum -- and thus ensure consistency between rate factor $A(T,W)$ given by (\ref{rate_factor}) and traction coefficient $\beta$ in tangential stress condition (\ref{basal_drag}), (\ref{bp_basal_drag}), (\ref{ps_basal_drag}), and (\ref{rs_basal_drag}) -- an optimization procedure for basal traction is performed for a previously attained energy $\theta$.  Because shear viscosity (\ref{viscosity}), (\ref{bp_viscosity}), (\ref{ps_viscosity}) and (\ref{rs_viscosity}) decreases for increasing energy, the ice speeds up for higher temperature $T$ and water content $W$ due to increased deformation.  Similarly, larger values of basal-traction $\beta$ decrease the vertical average of the mixture velocity while possibly generating more energy due to frictional- and strain-heating.  Indubitably, any generated heat resulting from these processes induces deformation, and therefore also increases the mixture velocity.  Hence a feedback process exists which requires extra care be taken in order to constrain both energy and traction.

An energy constraint already exists as evident by coefficient $W_f$ in rate factor (\ref{rate_factor}); due to lack of empirical evidence, enhancement is limited for ice containing water contents in excess of 1\%.  Regarding basal traction $\beta$, it is desirable to penalize abnormally high spatial gradients in order to reduce non-physical oscillations \citep{vogel}.

The process used to optimize traction coefficient $\beta$ -- referred to in this context as \index{Data assimilation} \emph{data assimilation} -- is directly analogous to the process used to optimize the basal-water content and energy previously described; a momentum objective functional \index{Constrained optimization!Objective function} $\mathscr{I}(\mathbf{u}_h, \beta) : \mathcal{H}^1(\Omega) \times \mathcal{H}^1(\Omega) \rightarrow \mathbb{R}$ is minimized over the domain of the ice-sheet $\Omega$.  This objective functional measures the misfit between the observed velocities $\mathbf{u}_{ob} = [u_{ob}\ v_{ob}]^\intercal$ and modeled velocities $\mathbf{u}_h = [u\ v]^\intercal$ over upper ice-sheet surface $\Gamma_S$,
\begin{align}
  \label{momentum_objective}
  \mathscr{I}(\mathbf{u}_h, \beta) = \gamma_1 \mathscr{I}_1(\mathbf{u}_h)
                                   + \gamma_2 \mathscr{I}_2(\mathbf{u}_h)
                                   + \gamma_3 \mathscr{I}_3(\beta)
                                   + \gamma_4 \mathscr{I}_4(\beta),
\end{align}
where
\begin{align}
  \label{l2_cost}
  \mathscr{I}_1(\mathbf{u}_h) = 
    & \frac{1}{2} \int_{\Gamma_S} \left[ (u - u_{ob})^2 + (v - v_{ob})^2 \right] d\Gamma_S \\ 
  \label{logarithmic_cost}
  \mathscr{I}_2(\mathbf{u}_h) = 
    & \frac{1}{2} \bigintsss_{\Gamma_S} \ln\left( \frac{ \left(u^2 + v^2\right)^{\nicefrac{1}{2}} + u_0}{\left(u_{ob}^2 + v_{ob}^2\right)^{\nicefrac{1}{2}} + u_0} \right)^2 d\Gamma_S \\ 
  \label{tikhonov_regularization}
  \mathscr{I}_3(\beta) = 
    & \frac{1}{2} \int_{\Gamma_G} \nabla \beta \cdot \nabla \beta\ d\Gamma_G \\
  \label{tv_regularization}
  \mathscr{I}_4(\beta) = 
    & \int_{\Gamma_G} \left( \nabla \beta \cdot \nabla \beta + \beta_0 \right)^{\nicefrac{1}{2}}\ d\Gamma_G,
\end{align}
with $L^2$ cost coefficient $\gamma_1$, \emph{logarithmic} cost coefficient $\gamma_2$, \index{Constrained optimization!Tikhonov regularization function} \emph{Tikhonov} regularization parameter $\gamma_3$, and \index{Constrained optimization!Total-variation regularization function} \emph{total variation} (TV) regularization parameter $\gamma_4$.  Here, $u_0 = 10\sups{-2}$ and $\beta_0 = 10\sups{-16}$ terms are added to avoid singularities.  Note that the functionals (\ref{l2_cost}) and (\ref{logarithmic_cost}) are referred to as \index{Constrained optimization!Cost function} \emph{cost functionals} while (\ref{tikhonov_regularization}) and (\ref{tv_regularization}) are referred to as \emph{regularization functionals}.

By forming this objective with cost functionals stated in terms of both $L^2$ and logarithmic velocity misfit terms, priority is given to either fast or slow areas of flow by adjusting the values of $\gamma_1$ and $\gamma_2$, respectively \citep{morlighem}.  Note that larger values of $\gamma_3$ and $\gamma_4$ result in increased regularity of $\beta$, at the cost of increased misfit $\left\Vert \mathbf{u}_h - \mathbf{u}_{ob} \right\Vert$, thus weighting the associated gradient penalty functionals in relation to the other functionals in (\ref{momentum_objective}).  Finally, the addition of the total variation functional (\ref{tv_regularization}) in objective (\ref{momentum_objective}) reduces short-wavelength oscillations that only marginally affect Tikhonov regularization functional (\ref{tikhonov_regularization}).

%===============================================================================
%===============================================================================

\section{Momentum optimization procedure} \label{ssn_momentum_optimization_procedure}

One of the advantages of the action principles presented by \citet{dukowicz_2010} is that actions (\ref{action}), (\ref{bp_action}), (\ref{ps_action}), and (\ref{rs_action}) are all self-adjoint.  Indeed, the momentum \index{Constrained optimization!Lagrangian} Lagrangian functional associated with momentum objective (\ref{momentum_objective}) and first-order action principle (\ref{bp_action}) -- used in an analogous set of KKT conditions for momentum as general KKT condition (\ref{barrier_kkt}) -- is defined as
\begin{align}
  \label{momentum_lagrangian}
  \mathscr{H}(\mathbf{u}_h, \beta, \bm{\lambda}) &= \mathscr{I}(\mathbf{u}_h, \beta) + \delta_{\mathbf{u}_h} \mathcal{A}_{\text{BP}}(\bm{\lambda}),
\end{align}
with momentum adjoint variable \index{Constrained optimization!Adjoint variable} $\bm{\lambda} = [\lambda_x\ \lambda_y]^\intercal \in \left( \mathcal{H}^1(\Omega) \right)^2$.

The log-barrier problem \index{Log-barrier method} (see \S \ref{ssn_log_barrier}) for momentum is of the same form as energy barrier problem (\ref{energy_barrier}).  Thus, the associated minimization problem for \index{Constrained optimization!Control parameter} control parameter $\beta$ and \index{Constrained optimization!State parameter} state parameter $\mathbf{u}_h$ is
\begin{align}
  \min_{\beta}\ \left\{ \varphi_{\omega}(\mathbf{u}_h, \beta) = \mathscr{I}(\mathbf{u}_h, \beta) - \omega \sum_{i=1}^n \ln(\beta_i) \right\}
\label{momentum_barrier}
\end{align}
for a decreasing sequence of barrier parameters $\omega$ converging to zero, and $n$ is the number of quadrature points in the discretization.  The logarithmic sum term in (\ref{momentum_barrier}) ensures that $\beta$ remains positive, as required by tangential basal stress condition (\ref{basal_drag}).
  
Note that I have used first-order action (\ref{bp_action}) in Lagrangian (\ref{momentum_lagrangian}); it has been my experience that solving full-Stokes momentum balance (\ref{extremum}) using a coarse mesh ($\approx$ 1 km minimum cell diameter) and traction $\beta^*$ resulting from the minimization of first-order momentum barrier problem (\ref{momentum_barrier}) results in a velocity field that differs only slightly from that obtained using first-order momentum balance (\ref{bp_extremum}).  This simplification provides a several-fold improvement in computation time.  Note also that the reformulated-Stokes action principle presented in \S \ref{ssn_reformulated_stokes} presents a substantial improvement in the velocity approximation from the first-order model at higher basal gradients (see \S \ref{ssn_ismip_hom_test_simulations}).  However, at its current state of development, CSLVR has not incorporated this model with the automated adjoint-optimization software it currently employs, Dolfin-Adjoint.
  
Additional computational energy is saved by using a linearization of rate-factor (\ref{rate_factor}) derived from a previously thermo-mechanically coupled $\mathbf{u}_h^{i-1}$.  This converts first-order viscosity (\ref{bp_viscosity}) into
\begin{align}
  \label{linear_bp_viscosity}
  \eta_{\text{BP}}^L(\theta, \mathbf{u}_h^{i-1}) &= \frac{1}{2}A(\theta)^{-\nicefrac{1}{n}} (\dot{\varepsilon}_{\text{BP}}(\mathbf{u}_h^{i-1}) + \dot{\varepsilon}_0)^{\frac{1-n}{n}},
\end{align} 
and first-order-viscous-dissipation term (\ref{bp_viscous_dissipation}) in (\ref{bp_action}) into 
\begin{align}
  \label{linear_viscous_dissipation}
  V^L\left( \dot{\varepsilon}_{\text{BP}}^2 \right) &= \int_0^{\dot{\varepsilon}_{\text{BP}}^2} \eta_{\text{BP}}^L(s)\ ds = \eta_{\text{BP}}^L\left(\theta, \mathbf{u}_h^{i-1} \right) \dot{\varepsilon}_{\text{BP}}^2.
\end{align}
  
Finally, $\delta_{\mathbf{u}_h} \mathcal{A}_{\text{BP}}(\bm{\lambda})$ in Lagrangian (\ref{momentum_lagrangian}) is as derived by \citet{dukowicz_2010},
\begin{align}
  \label{momentum_expanded_lagrangian}
  \mathscr{H}(\mathbf{u}_h, \beta, \bm{\lambda}) =
  &+ \mathscr{I}(\mathbf{u}_h, \beta) + \int_{\Gamma_E}\ f_e \mathbf{n}_h \cdot \bm{\lambda}\ d\Gamma_E \notag \\
  &+ \int_{\Omega} \sigma_{\text{BP}}^L : \nabla \bm{\lambda}\ d\Omega
   - \int_{\Omega} \rho g (\nabla S)_h \cdot \bm{\lambda}\ d\Omega \notag \\
  &+ \int_{\Gamma_B} \left( \Lambda_{\text{BP}}^L \bm{\lambda} \cdot \mathbf{n}_h + \beta \mathbf{u}_h \cdot \bm{\lambda} \right)\ d\Gamma_B,
\end{align}
where $\sigma_{\text{BP}}^L$ and $\Lambda_{\text{BP}}^L$ are the linearized counterparts to first-order quasi-stress tensor (\ref{bp_stress_tensor}) and impenetrability Lagrange multiplier (\ref{bp_dukowicz_lambda}) utilizing linear viscosity (\ref{linear_bp_viscosity}).

The CSLVR source code implementation of this procedure is shown in Code Listing \ref{cslvr_u_opt}.

\pythonexternal[label=cslvr_u_opt, caption={CSLVR source code of the abstract class \texttt{Momentum} for optimizing the velocity and traction.  All of the momentum models of \S \ref{ssn_momentum_and_mass_balance} inherit this method.}, firstline=565, lastline=820]{cslvr_src/momentum.py}

\section{Dual optimization for energy and momentum} \label{ssn_dual_optimization}

To begin the procedure of solving energy and momentum, the thermo-mechanical coupling process described in Algorithm \ref{tmc} is performed for an initial friction field $\beta^i$, energy $\theta^i$, and basal water discharge $F_b^i$.  After this, barrier problem (\ref{momentum_barrier}) is repeatedly solved until the difference between two subsequent traction fields are below a specified tolerance.  This procedure is outlined by Algorithm \ref{tmc_da} with CSLVR implementation shown in Code Listing \ref{cslvr_tmc_da}.

As suggested by \citet{morlighem}, a suitable initialization of the traction field $\beta^i$ may be formed by vertically integrating first-order momentum balance (\ref{bp_cons_momentum}) and eliminating any horizontal derivatives -- i.e., longitudinal stretching and lateral shearing -- from the left-hand side, resulting in
\begin{align}
  \label{SIA}
  \beta \mathbf{u}_h |_B = \rho g H (\nabla S)_h,
\end{align}
where $H = S - B$ is the ice thickness and $\beta \mathbf{u}_h |_B$ is the entire contribution of vertical shear, referred to as \emph{basal traction}.  Note that this derivation of the momentum balance is equivalent to the \index{Stokes equations!Applied to ice, shallow ice approximation} \emph{shallow-ice approximation} for glacier flow \citep{greve}.  Finally, using the observed surface velocity as an approximation for basal velocity $\mathbf{u}_h |_B$ and taking the norm of vector expression (\ref{SIA}), the shallow-ice-approximate traction field is derived,
\begin{align}
  \label{beta_SIA}
  \beta_{\text{SIA}} = \frac{\rho g H \Vert (\nabla S)_h \Vert}{\Vert \mathbf{u}_{ob} \Vert+ u_0},
\end{align}
where $u_0$ is a small positive speed to avoid singularities.  Notice that in areas without surface velocity observations, the approximate traction field (\ref{beta_SIA}) will be invalid.  Therefore, for the purposes of creating an initial traction field, we replace any areas containing missing measurements of velocity $\mathbf{u}_{ob}$ in (\ref{beta_SIA}) with the balance velocity $\bar{\mathbf{u}}$ as derived in Chapter \ref{ssn_balance_velocity}.

It has been my experience that traction $\beta$ will converge consistently if initialized to the same value prior to solving system (\ref{momentum_barrier}).  Therefore, at the start of every iteration, $\beta$ is reset to its initial value $\beta^i$.  See Algorithm \ref{tmc_da} for details, and Chapter \ref{ssn_application_jakobshavn} for an example of the full energy and momentum optimization procedure applied to the region of Greenland's Jakobshavn Glacier.  In the example that follows, we solve an isothermal momentum-optimization problem.

\begin{algorithm}
  \caption[Thermo-mechanically coupled data-assimilation]{ -- TMC basal-friction data assimilation}
  \label{tmc_da}
  \begin{algorithmic}[1]
    \Function{TMC\_DA}{$\beta^i, \theta^i, F_b^i, n_{\max}$}
      \State $\phantom{\theta,F_b}\mathllap{a_{tol}} := 100$
      \State $\phantom{\theta,F_b}\mathllap{r} := \infty$
      \State $\phantom{\theta,F_b}\mathllap{i} := 1$
      \State $\phantom{\theta,F_b}\mathllap{\beta^p} := \beta^i$
      \State $\theta, F_b := \text{ TMC} \left( \beta^i, \theta^i, F_b^i \right)$
      \While{$r > a_{tol}$ \textbf{and} $i < n_{\max}$}
        \State $\phantom{\theta,F_b}\mathllap{\beta} := \beta^i$
        \State $\phantom{\theta, F_b}\mathllap{\beta^*} := \argminl_{\beta} \Big\{ \varphi_{\omega}(\beta) \Big\}$
        \State $\theta, F_b := \text{TMC} \left( \beta^*, \theta, F_b \right)$
        \State $\phantom{\theta, F_b}\mathllap{r} := \Vert \beta^p - \beta^* \Vert_{2}$
        \State $\phantom{\theta, F_b}\mathllap{\beta^p} := \beta^*$
        \State $\phantom{\theta, F_b}\mathllap{i} := i + 1$
      \EndWhile
      \State \Return $\beta^*$
    \EndFunction
  \end{algorithmic}
\end{algorithm}

\pythonexternal[label=cslvr_tmc_da, caption={Implementation of Algorithm \ref{tmc_da} by CSLVR contained in the \texttt{Model} class for the optimization of a \texttt{Momentum} and \texttt{Energy} instance.}, firstline=2395, lastline=2586]{cslvr_src/model.py}

%===============================================================================

\section{L-curve analysis} \label{ssn_l_curve}

\index{Constrained optimization!L-curve analysis}
\index{L-curve analysis|seealso{Constrained optimization!L-curve analysis}}
In order to determine the correct values of $\gamma_3$ and $\gamma_4$, we use a process referred to as \emph{L-curve analysis} developed by \citet{hansen}.  The method relies on the fact that for many inverse problems, the shape of the curve resulting from plotting the regularization functional values against the cost functional values for a series of regularization parameters resembles the shape of an `L'.  The `corner' of this curve approximately corresponds to the point whereby increasing regularization begins to negatively impact the cost functional minimization with negligible improvement in smoothness of the control parameter.  One might say that the `cost' of regularization after this point becomes too high.

To this end, we have created the function \texttt{L\_curve} contained within the \texttt{Model} class (Code Listing \ref{cslvr_l_curve}) for automatically performing this calculation for any child class of the CSLVR \texttt{Physics} class (Code Listing \ref{cslvr_physics}).  All of the physics calculations presented in \S \ref{ssn_momentum_and_mass_balance}, \S \ref{ssn_internal_energy_balance}, \S \ref{ssn_balance_velocity}, \S \ref{ssn_stress_balance}, and \S \ref{ssn_ice_age} inherit from this class; this is because any physics calculation may have an application as an optimization problem.  In regards to momentum objective (\ref{momentum_objective}), we plot the values of the total cost functional $\mathscr{I} = \mathscr{I}_1 + \mathscr{I}_2$ versus either regularization functional $\mathscr{I}_3$ or $\mathscr{I}_4$.

In \S \ref{ssn_ismip_hom_inverse_sims} we examine this procedure for a similar problem as that presented in \S \ref{ssn_ismip_hom_test_simulations}.  Finally, this method was also employed to derive the regularization parameters utilized to generate the simulations presented in Chapter \ref{ssn_application_jakobshavn}.

\pythonexternal[label=cslvr_physics, caption={CSLVR source code for the abstract class \texttt{Physics} from which all physics calculation classes inherit.}]{cslvr_src/physics.py}

\pythonexternal[label=cslvr_l_curve, caption={CSLVR implementation of the L-curve procedure of \S \ref{ssn_l_curve}.  Both the cost and regularization functional values are saved as \texttt{txt} files; and plots of the convergence behavior and L-curve as \texttt{pdf} files.}, firstline=2588, lastline=2814]{cslvr_src/model.py}

\section{Ice-shelf inversion procedure} \label{ssn_shelf_inversion}

\index{Flow enhancement factor}
\index{Shelf inversion|seealso{Flow enhancement factor}}
Due to the fact that basal traction $\beta$ defined over floating ice-shelves is very close to zero \citep{greve}, a different control parameter must be specified in order to match the surface velocity observations in these areas.  One choice for this control is enhancement factor $E$ in flow-rate factor (\ref{rate_factor}).  The inversion for this parameter has been used to generate preliminary continent-scale inversions of Antarctica with low error velocity misfit $\Vert \mathbf{u} - \mathbf{u}_{ob} \Vert$ over ice-shelves.  We only make note of the fact that this option is easily implemented with CSLVR, and results in a depth-varying distribution of enhancement $E$.

%===============================================================================

\section{ISMIP-HOM inverse test simulation} \label{ssn_ismip_hom_inverse_sims}

\index{Linear differential equations!3D}
\index{ISMIP-HOM simulations}
For a simple test of the momentum optimization procedure described in \S \ref{ssn_momentum_optimization_procedure}, an inverse form of the ISMIP-HOM project presented previously in \S \ref{ssn_ismip_hom_test_simulations} is performed \citep{ismip_hom}.  This test is defined over the domain $\Omega \in [0,\ell] \times [0,\ell] \times [B,S] \subset \R^3$ with $k_x \times k_y \times k_z$ node discretization, and specifies the use of a surface height with uniform slope $\Vert \nabla S \Vert = a$
\begin{align*}
  S(x) = - x \tan\left( a \right)
\end{align*}
and basal topography matching the surface slope
\begin{align*}
  B(x,y) = S(x) - H,
\end{align*}
with ice thickness $H$.  As before, we enforce continuity via the periodic $\mathbf{u}$ boundary conditions (the first-order momentum model does not solve for pressure $p$)
\begin{align*}
  \mathbf{u}(0,0)    &= \mathbf{u}(\ell,\ell) \\
  \mathbf{u}(0,\ell) &= \mathbf{u}(\ell,0) \\
  \mathbf{u}(x,0)    &= \mathbf{u}(x,\ell) \\
  \mathbf{u}(0,y)    &= \mathbf{u}(\ell,y).
\end{align*}

\begin{figure}
  \centering
    \includegraphics[width=0.9\linewidth]{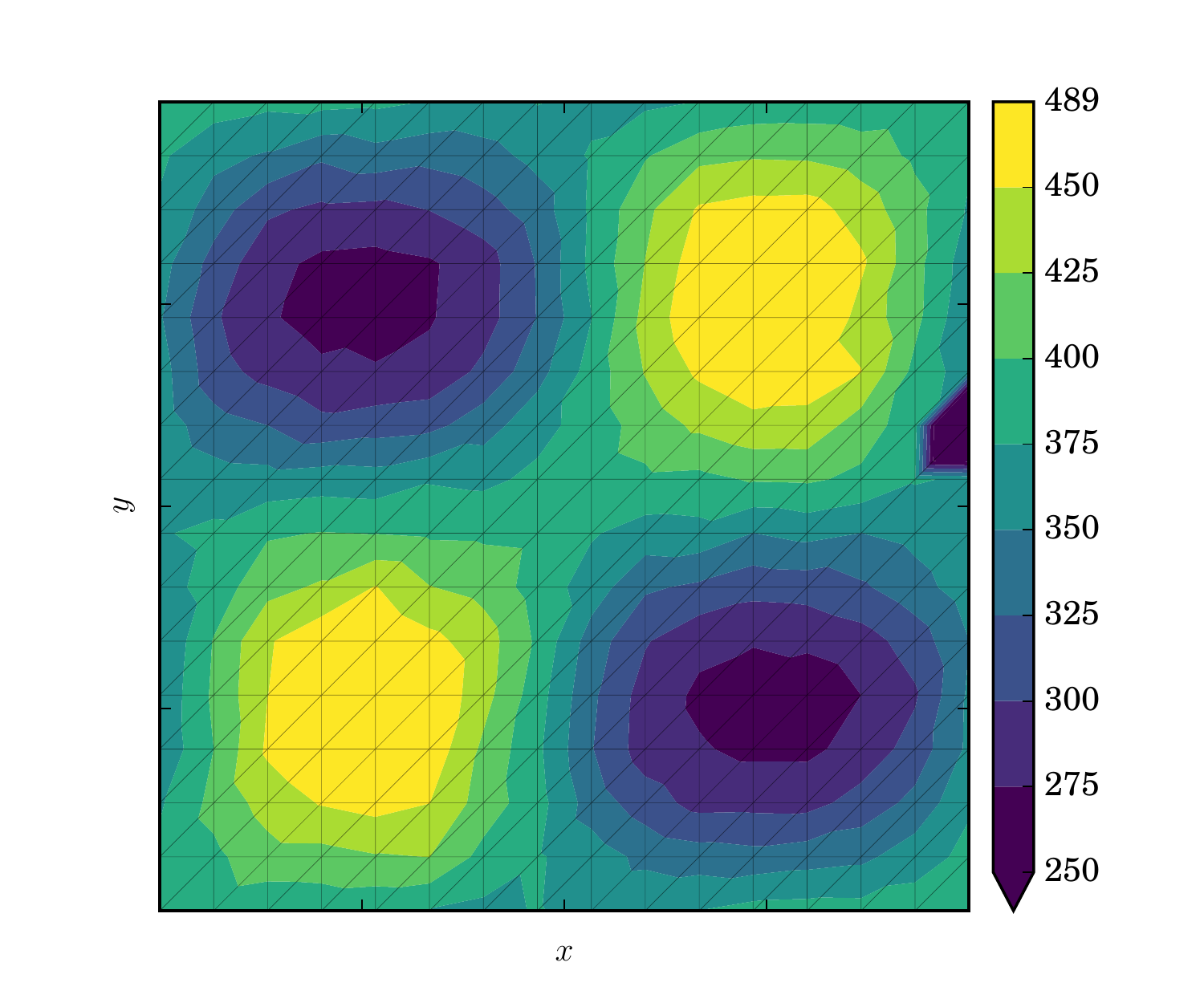}
    \includegraphics[width=0.9\linewidth]{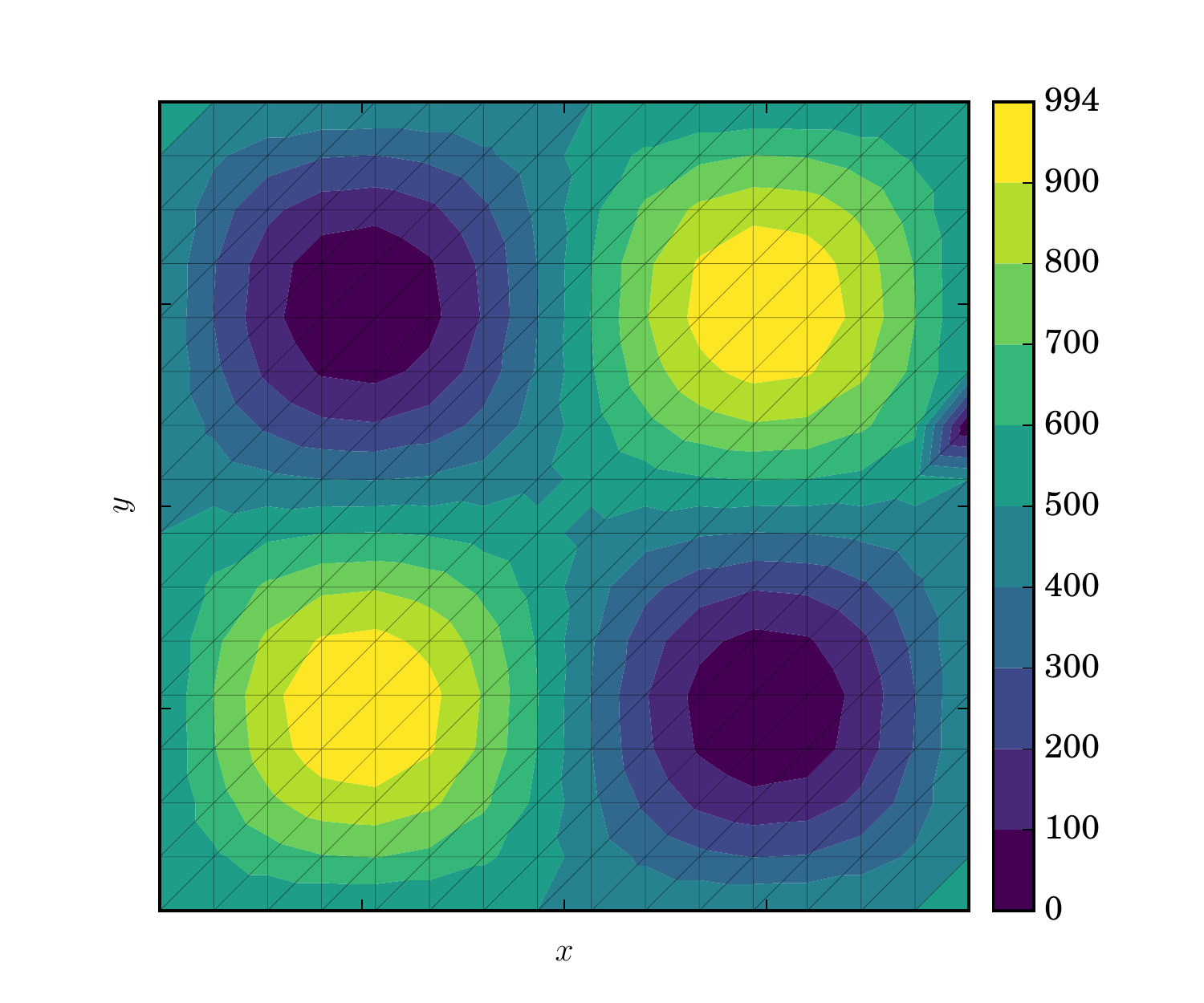}
    \includegraphics[width=0.9\linewidth]{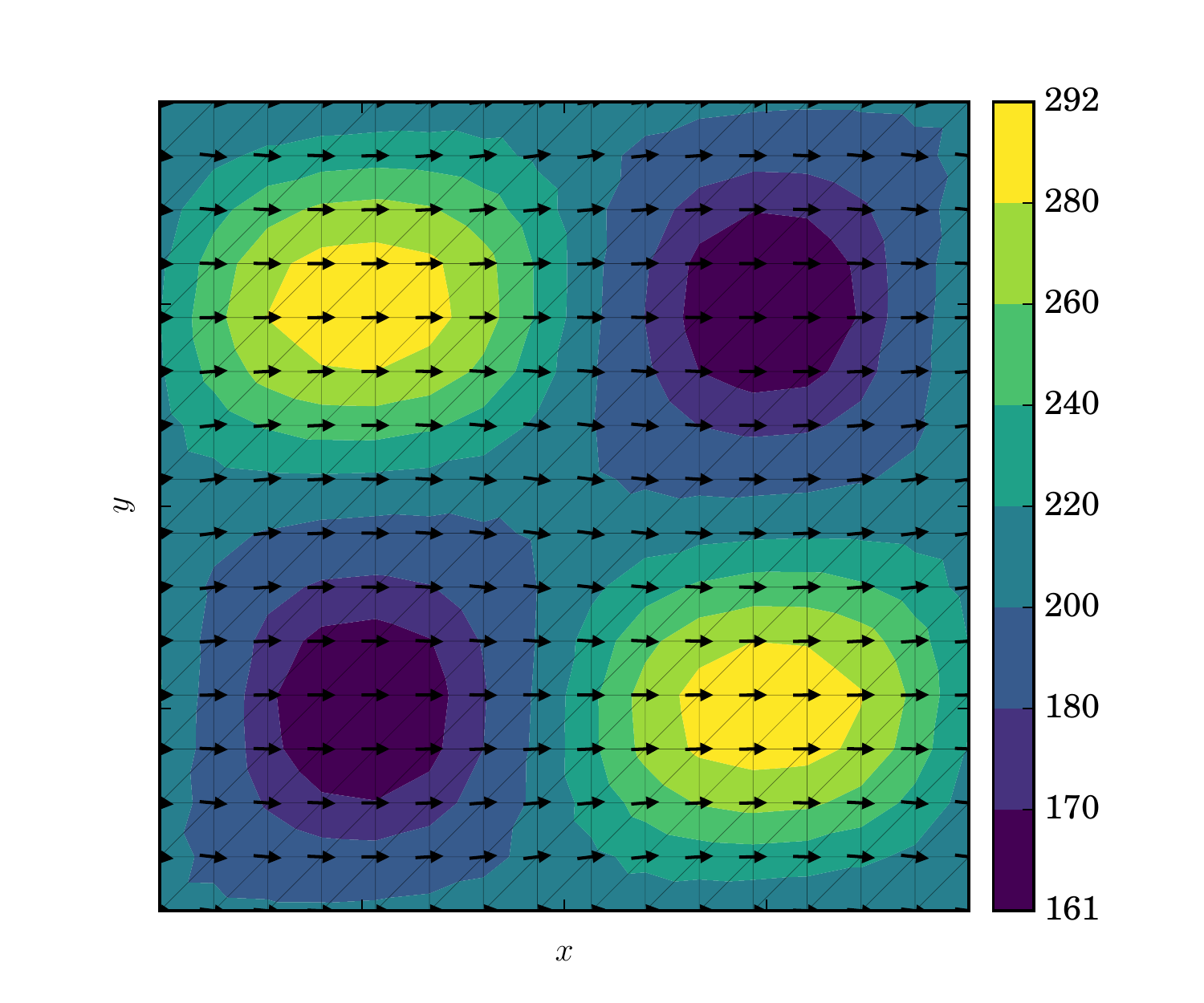}
  \caption[Inverse ISMIP-HOM `true' data fields]{ISMIP-HOM C initial value for traction $\beta$, SIA traction coefficient $\beta_{\text{SIA}}$ (top), `true' traction coefficient $\beta_{\text{true}}$ (middle), and `true' velocity $\mathbf{u}_{\text{true}}$ (bottom) with a $20 \times 20$ square km grid.}
  \label{inverse_ismip_true}
\end{figure}

To begin, first-order momentum system (\ref{bp_extremum}) is solved using the `true' basal traction field
\begin{align*}
  \beta_{\text{true}}(x,y) = \left( \frac{\beta_{max}}{2} \right) \sin\left( \frac{2\pi}{\ell} x \right)\sin\left( \frac{2\pi}{\ell} y \right) + \left( \frac{\beta_{max}}{2} \right)
\end{align*}
with maximum value $\beta_{max}$ (Figure \ref{inverse_ismip_true}).  Similar to \citet{petra}, we add normally-distributed-random noise to the resulting `true' velocity field $\mathbf{u}_{\text{true}}$ with standard deviation $\sigma$ to create the simulated `observed' velocity
\begin{align*}
  \mathbf{u}_{ob} = \mathbf{u}_{\text{true}} + \bm{\epsilon} 
\end{align*}
where 
\begin{align*}
  \bm{\epsilon} \distras{iid}\mathcal{N}\left(\bm{0},\sigma^2 I \right), \hspace{10mm} \sigma = \frac{\Vert \mathbf{u}_{\text{true}} \Vert_{\infty}}{\text{SNR}}
\end{align*}
with signal-to-noise ratio SNR (Figure \ref{inverse_ismip_true}).

For simplicity, we use the isothermal rate-factor $A = 10\sups{-16}$  for use with viscosity $\eta$, thus removing the necessity to optimize energy $\theta$.  Table \ref{ismip_hom_inverse_values} lists the coefficients and values used.

\begin{figure*}
  \centering
    \includegraphics[width=\linewidth]{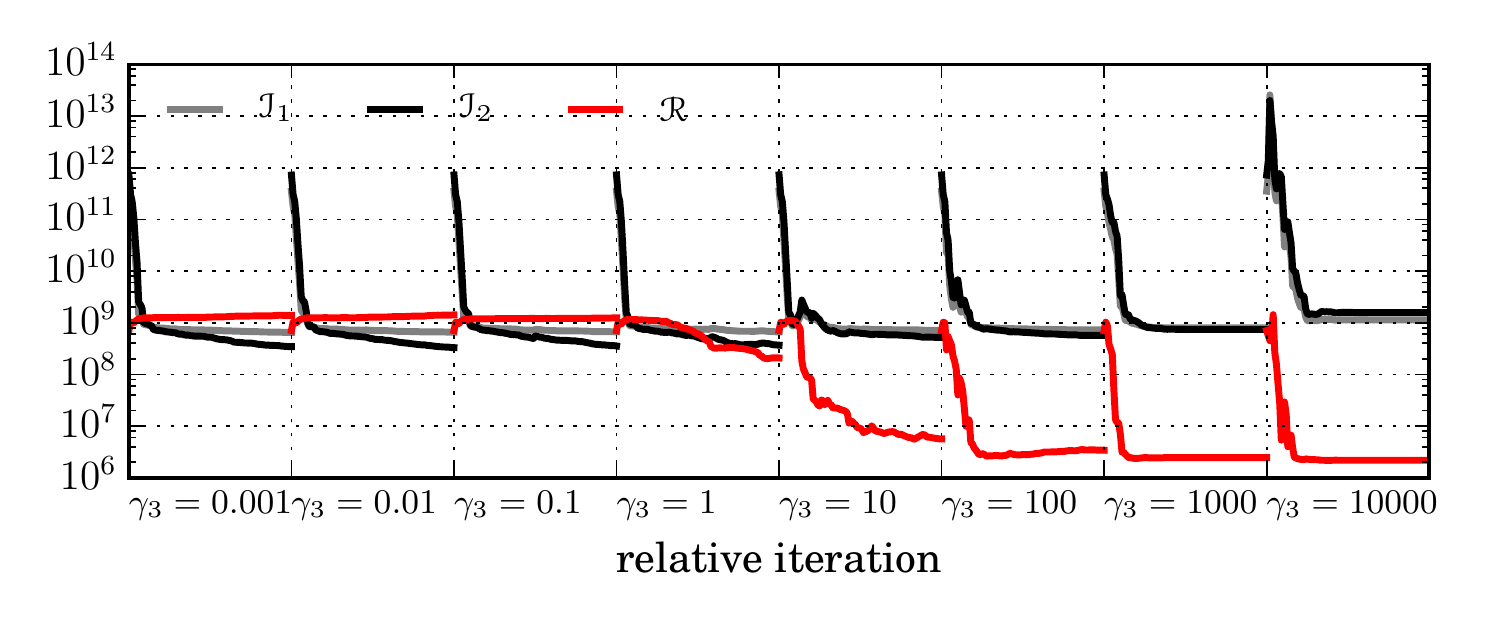}
    \caption[Tikhonov-regularized inverse ISMIP-HOM convergence diagram]{Convergence plot of the cost functional $\gamma_1 \mathscr{I}_1$ (grey) and $\gamma_2 \mathscr{I}_2$ (black), and the Tikhonov regularization functional $\mathscr{I}_3$ (red) for each of the Tikhonov-regularization parameters $\gamma_3$ shown on the $x$-axis.  For this procedure, the TV-regularization parameter $\gamma_4 = 0$, and cost functional coefficient chosen to be $\gamma_1 =1$, $\gamma_2 = 10^5$.}
  \label{ismip_l_curve_convergence_tik}
\end{figure*}

\begin{figure*}
  \centering
    \includegraphics[width=\linewidth]{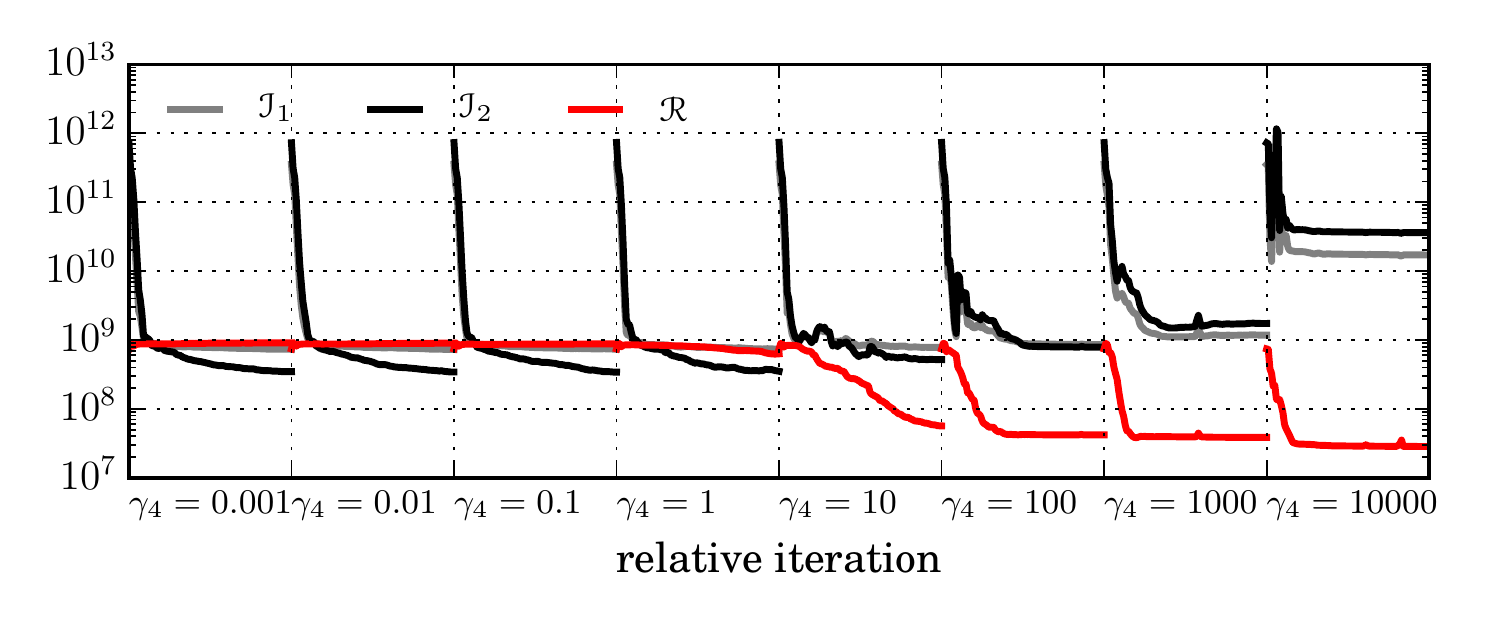}
    \caption[Total-variation regularized ISMIP-HOM convergence diagram]{Convergence plot of the cost functional $\gamma_1 \mathscr{I}_1$ (grey) and $\gamma_2 \mathscr{I}_2$ (black), and the TV regularization functional $\mathscr{I}_4$ (red) for each of the TV-regularization parameters $\gamma_4$ shown.  For this procedure, the Tikhonov-regularization parameter $\gamma_3 = 0$, and cost functional coefficient chosen to be $\gamma_1 =1$, $\gamma_2 = 10^5$.}
  \label{ismip_l_curve_convergence_tv}
\end{figure*}

To begin the inversion process, $L^2$ cost functional coefficient $\gamma_1$ and logarithmic cost functional coefficient $\gamma_2$ in (\ref{momentum_objective}) are determined by solving momentum optimization problem (\ref{momentum_barrier}) and adjusting their relative values such that at the end of the optimization their associated functionals are of approximately the same order.  Following \citet{morlighem}, we set $\gamma_1 = 1$ and derive by this process $\gamma_2 = 10\sups{5}$ (Figures \ref{ismip_l_curve_convergence_tik} and \ref{ismip_l_curve_convergence_tv}).

\begin{figure*}
  \centering
    \includegraphics[width=\linewidth]{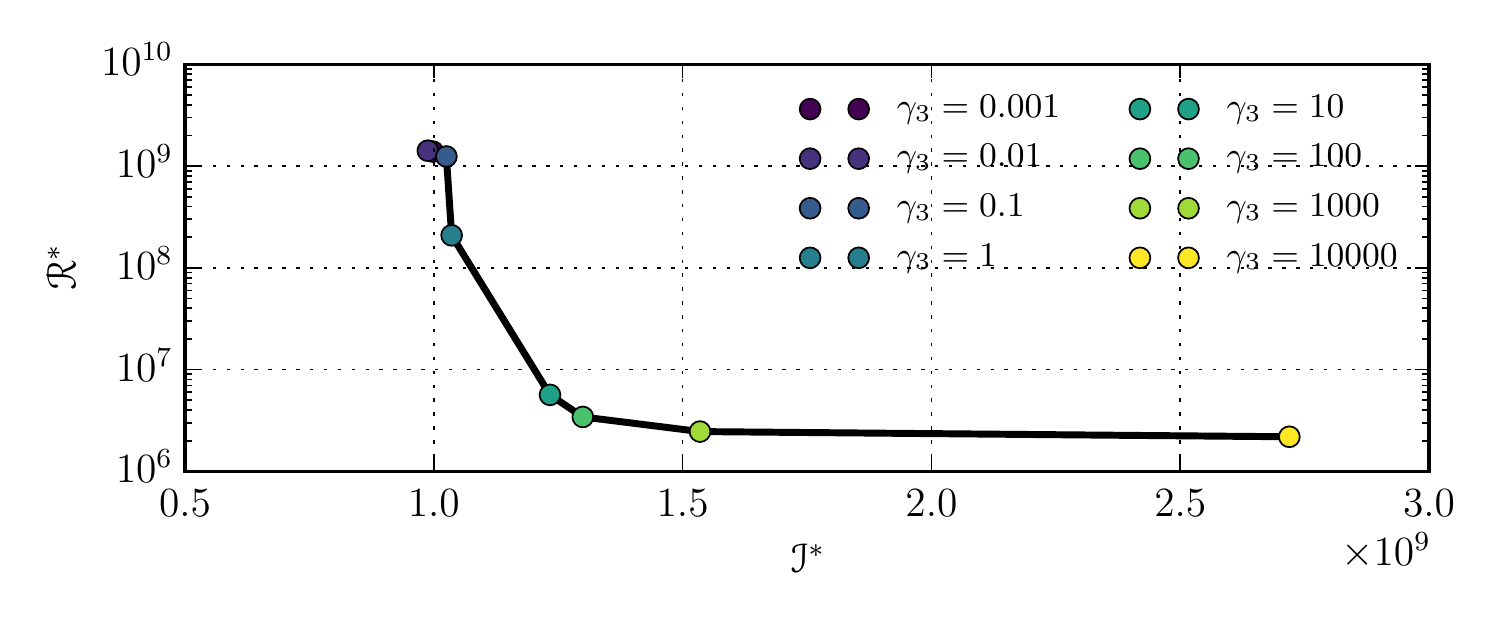}
    \caption[ISMIP-HOM Tikhonov L-curve diagram]{L-curve for Tikhonov parameter $\gamma_3$ with $\gamma_4=0$.}
  \label{ismip_l_curve_tik}
\end{figure*}

\begin{figure*}
  \centering
    \includegraphics[width=\linewidth]{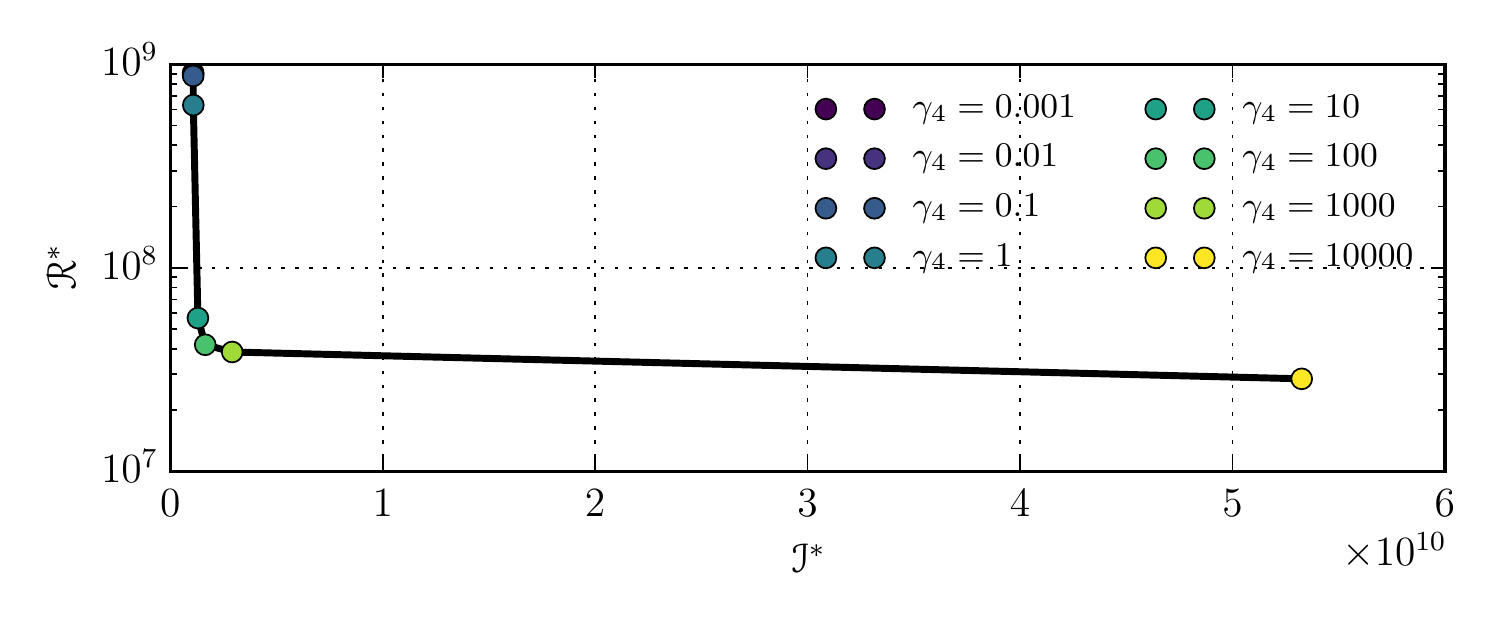}
    \caption[ISMIP-HOM total-variation L-curve diagram]{L-curve for total-variation parameter $\gamma_4$ with $\gamma_3=0$.}
  \label{ismip_l_curve_tv}
\end{figure*}

The next step is to derive a proper value for the regularization parameters $\gamma_3$ and $\gamma_4$ associated respectively with Tikhonov regularization functional (\ref{tikhonov_regularization}) and total variation regularization functional (\ref{tv_regularization}).  To this end, we perform the L-curve procedure described in \S \ref{ssn_l_curve} over a range of Tikhonov parameters $\gamma_3$ and TV parameters $\gamma_4$ in objective (\ref{momentum_objective}) (see Code Listing \ref{cslvr_l_curve_script}).  For simplicity, we vary only one of $\gamma_3$ or $\gamma_4$ and set the other to zero (Figures \ref{ismip_l_curve_tik} and \ref{ismip_l_curve_tv}).

\begin{figure}
  \centering
    \includegraphics[width=0.9\linewidth]{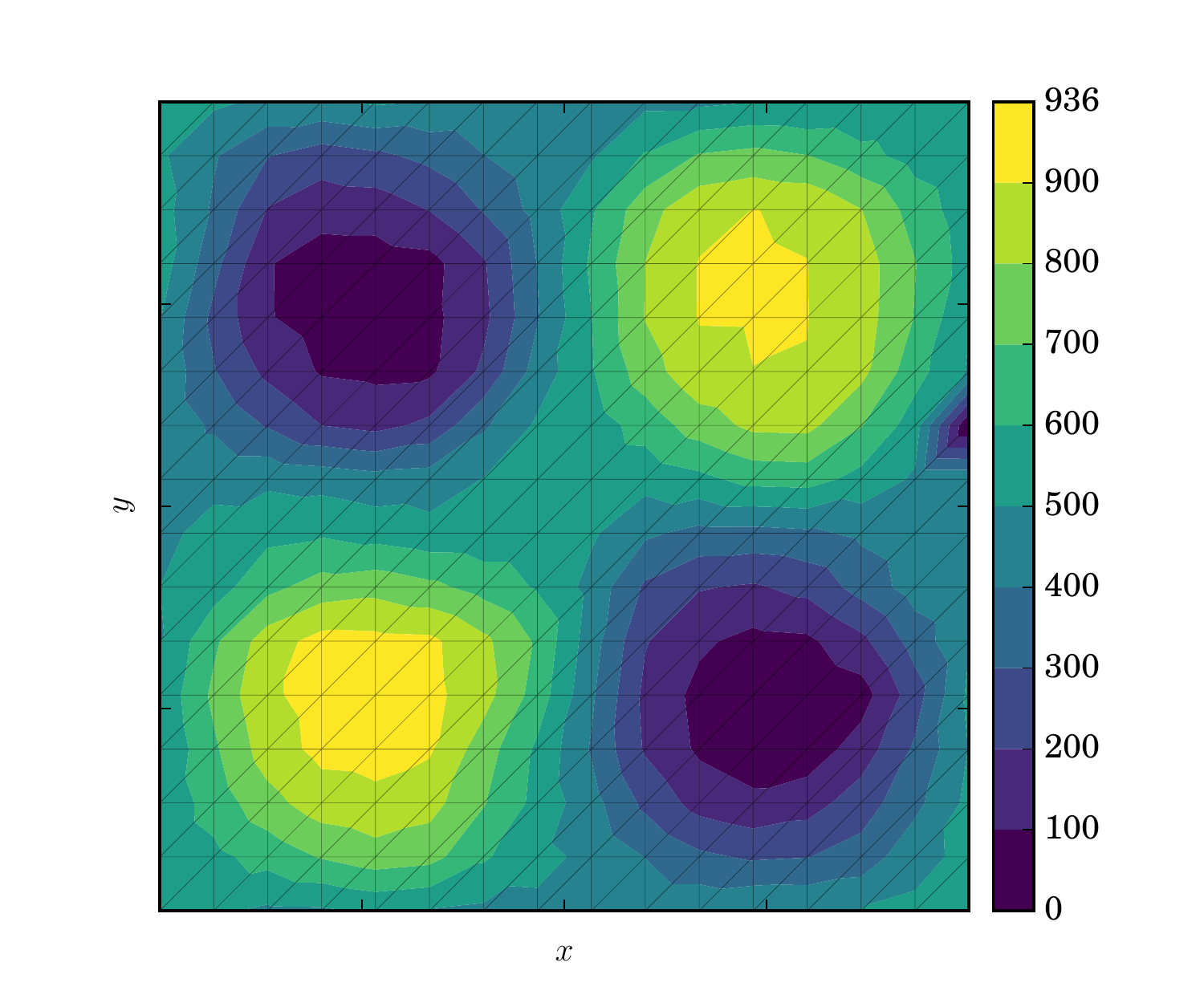}
    \includegraphics[width=0.9\linewidth]{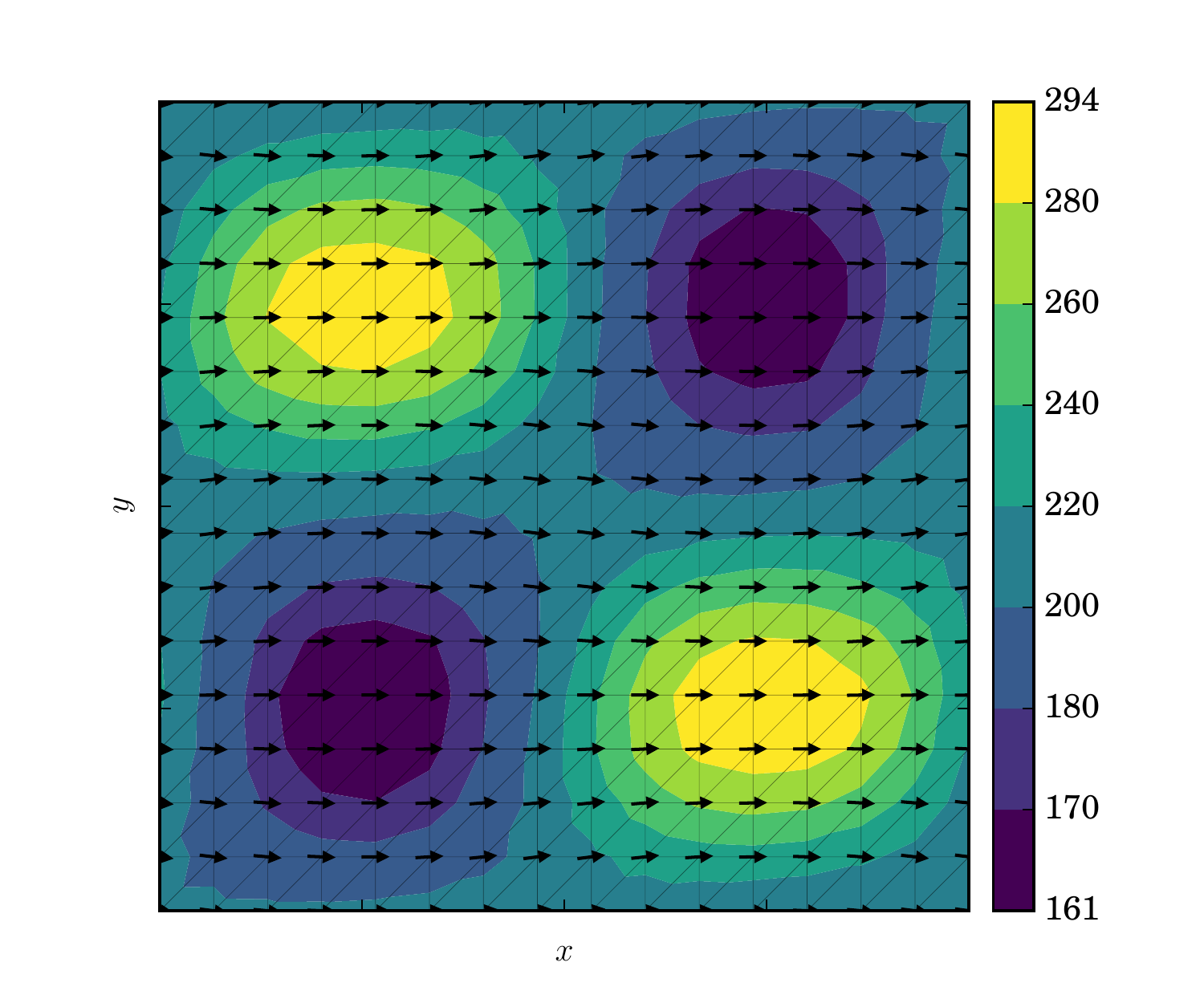}
    \includegraphics[width=0.9\linewidth]{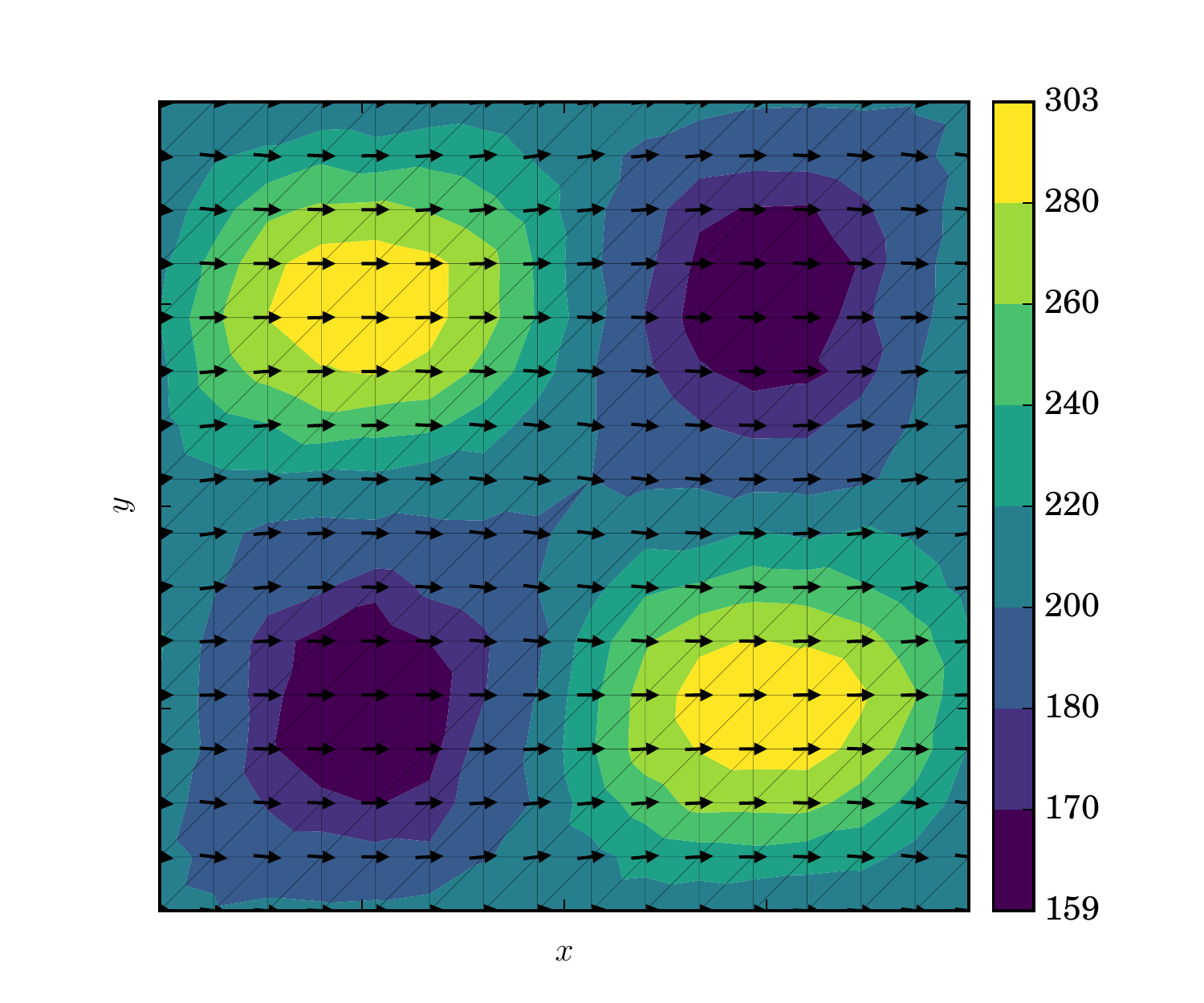}
  \caption[TV-regularized inverse ISMIP-HOM results]{Results obtained using TV-regularization with $\gamma_3=0$, $\gamma_4=100$; optimized traction coefficient $\beta^*$ (top), optimized velocity $\mathbf{u}^*$ (middle), and `observed' velocity $\mathbf{u}_{ob}$ (bottom) with a $20 \times 20$ square km grid.}
  \label{inverse_ismip_opt_tv}
\end{figure}

\begin{figure}
  \centering
    \includegraphics[width=0.9\linewidth]{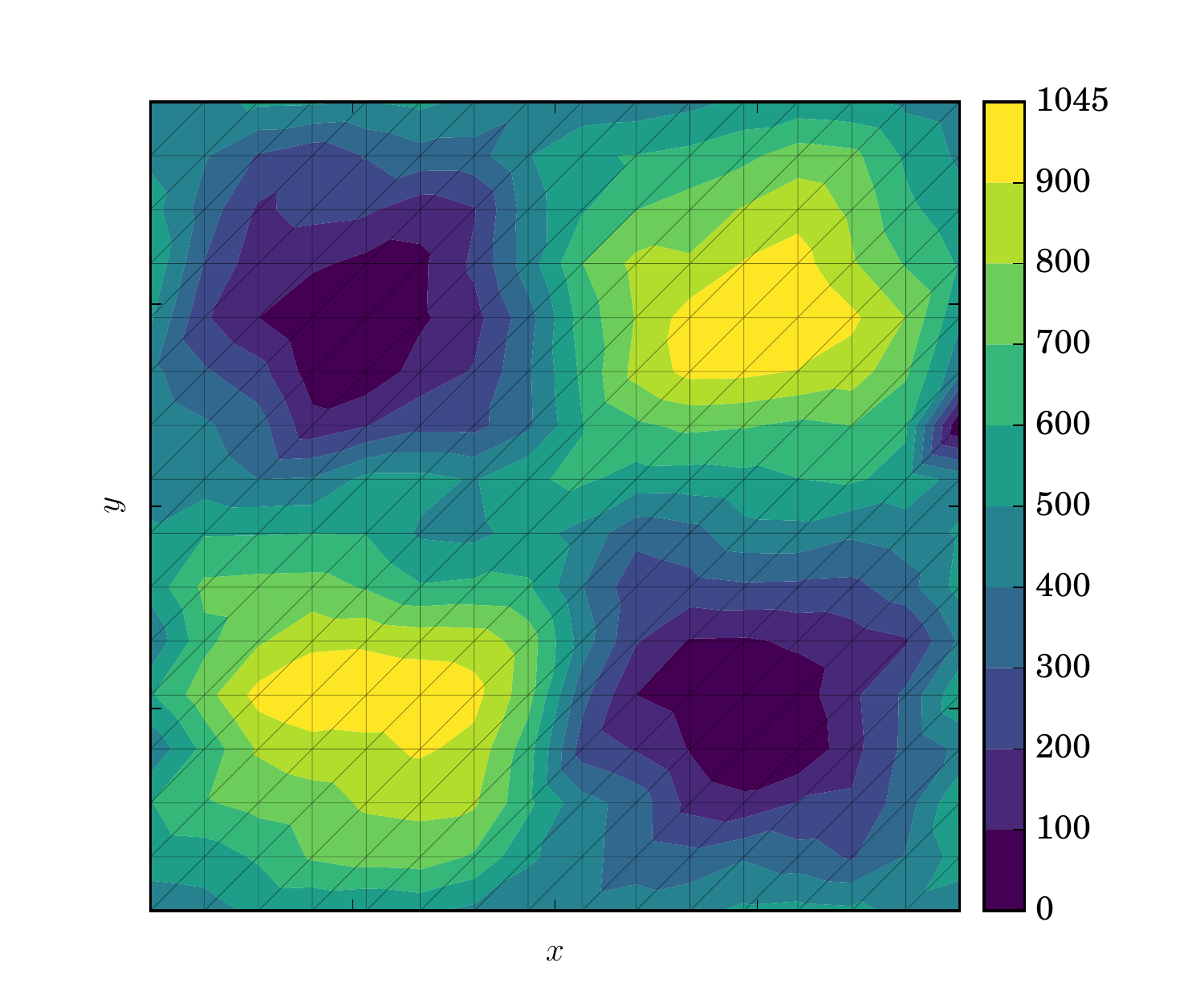}
    \includegraphics[width=0.9\linewidth]{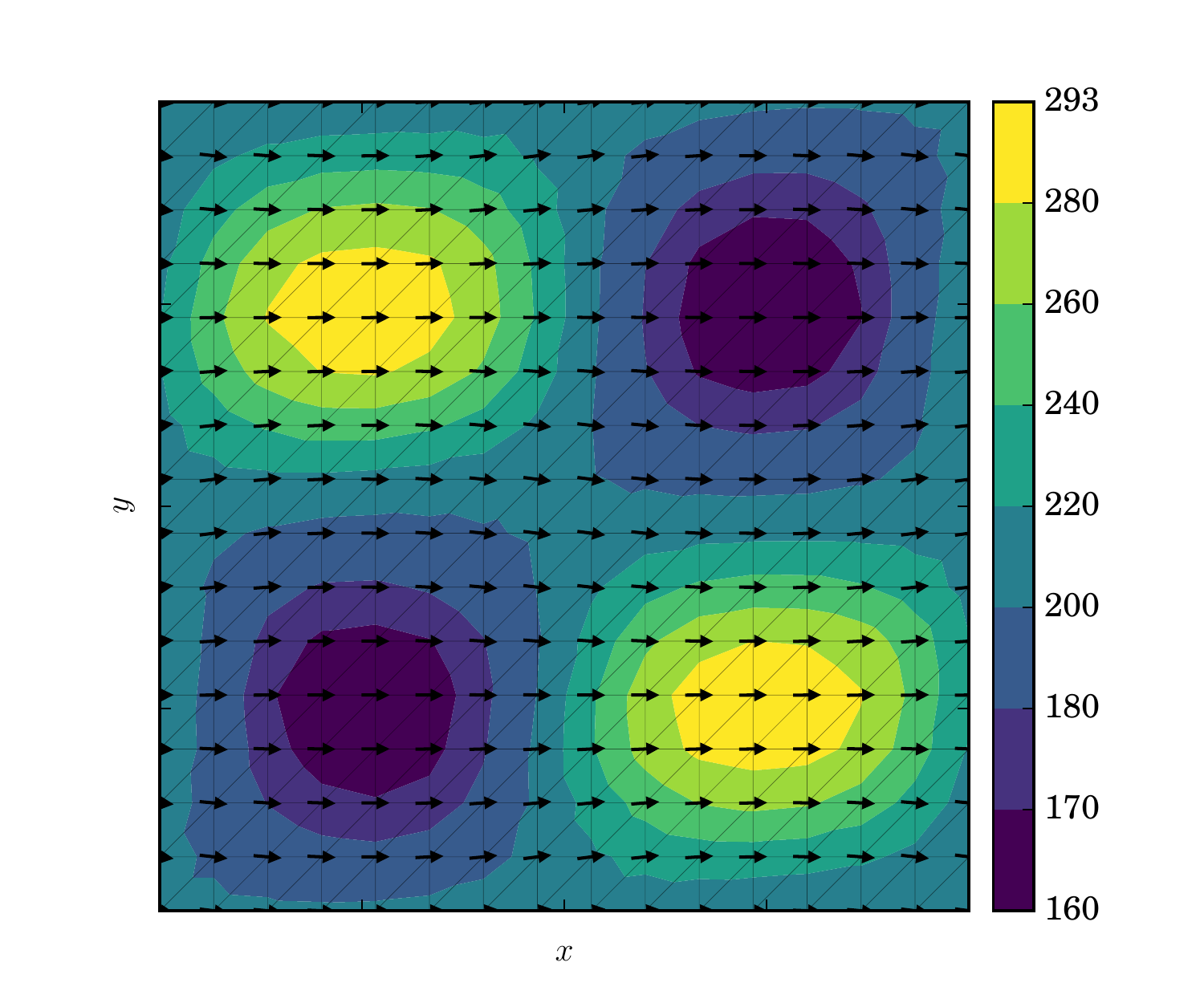}
    \includegraphics[width=0.9\linewidth]{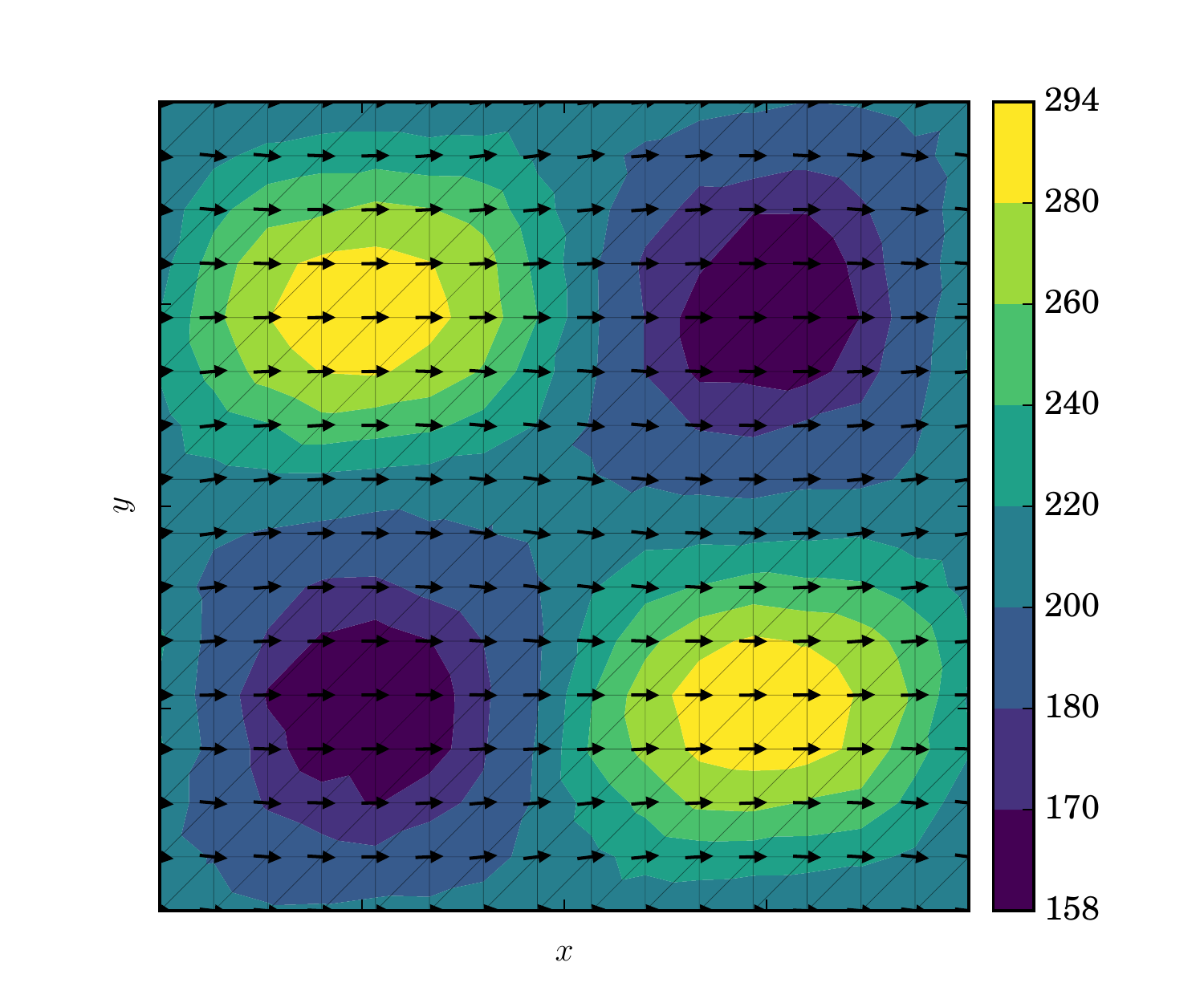}
  \caption[Tikhonov-regularized inverse ISMIP-HOM results with $\gamma_3=100$]{Results obtained using Tikhonov-regularization with $\gamma_3=100$, $\gamma_4=0$; optimized traction coefficient $\beta^*$ (top), optimized velocity $\mathbf{u}^*$ (middle), and `observed' velocity $\mathbf{u}_{ob}$ (bottom) with a $20 \times 20$ square km grid.}
  \label{inverse_ismip_opt_tikhonov_100}
\end{figure}

\begin{figure}
  \centering
    \includegraphics[width=0.9\linewidth]{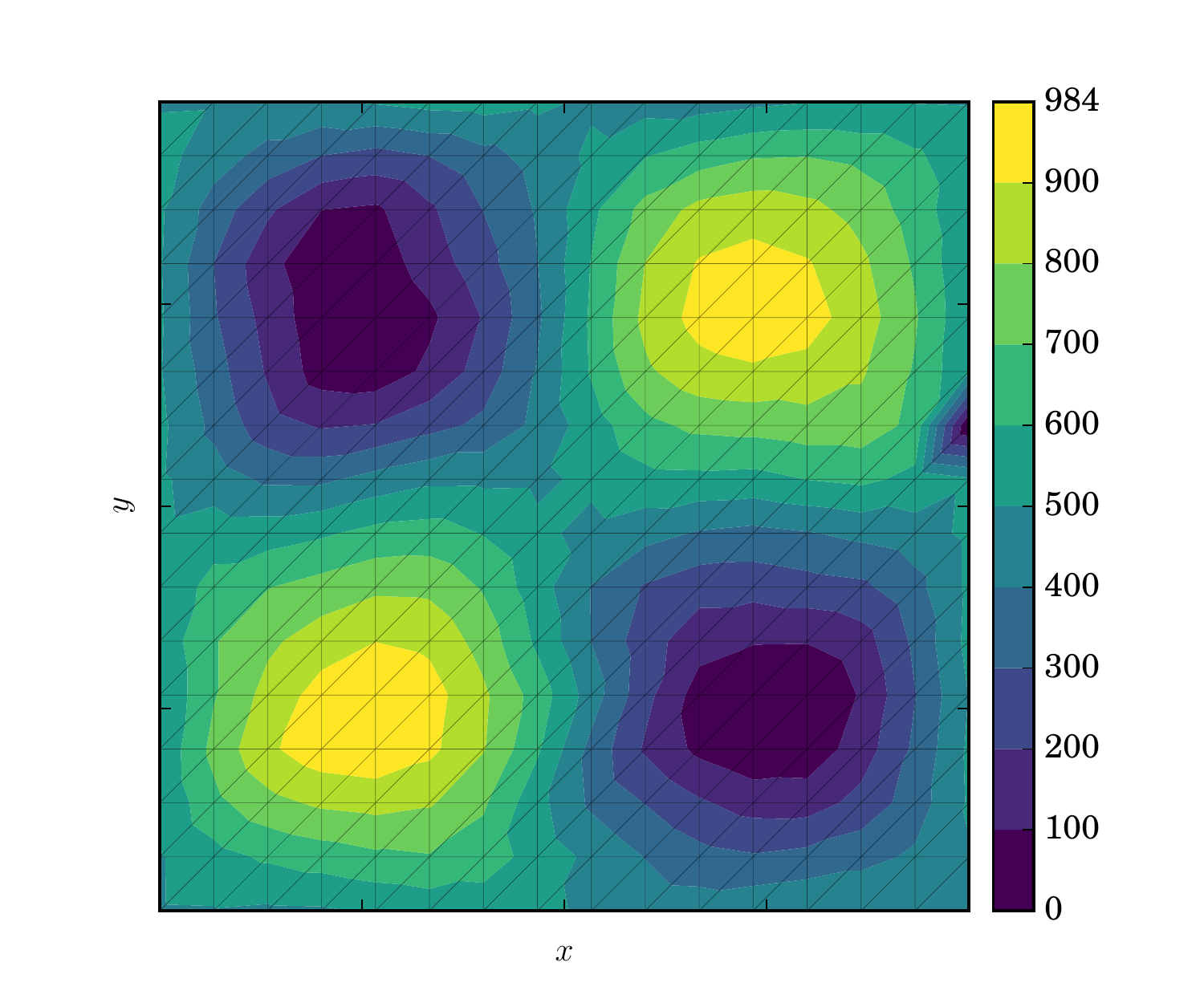}
    \includegraphics[width=0.9\linewidth]{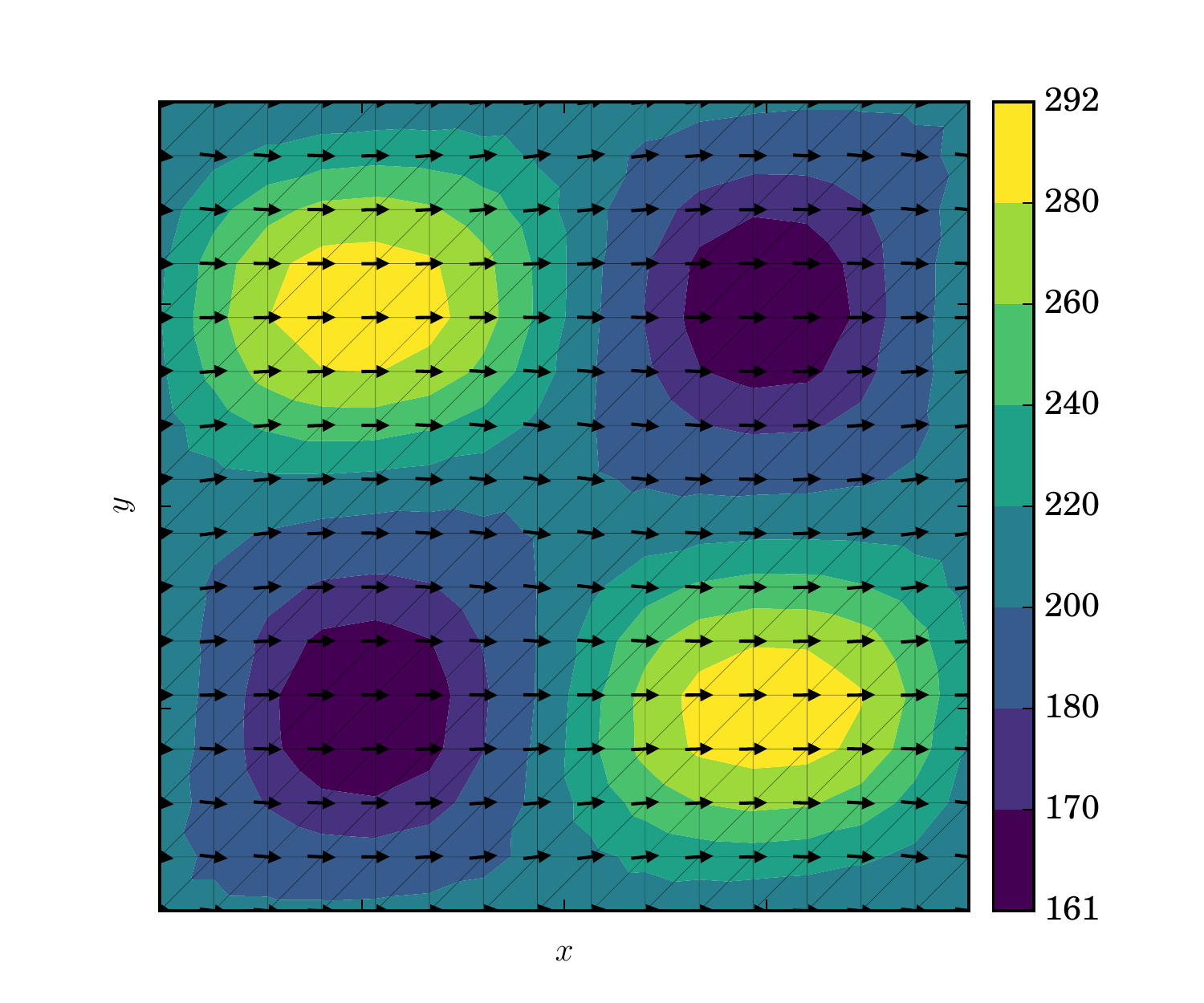}
    \includegraphics[width=0.9\linewidth]{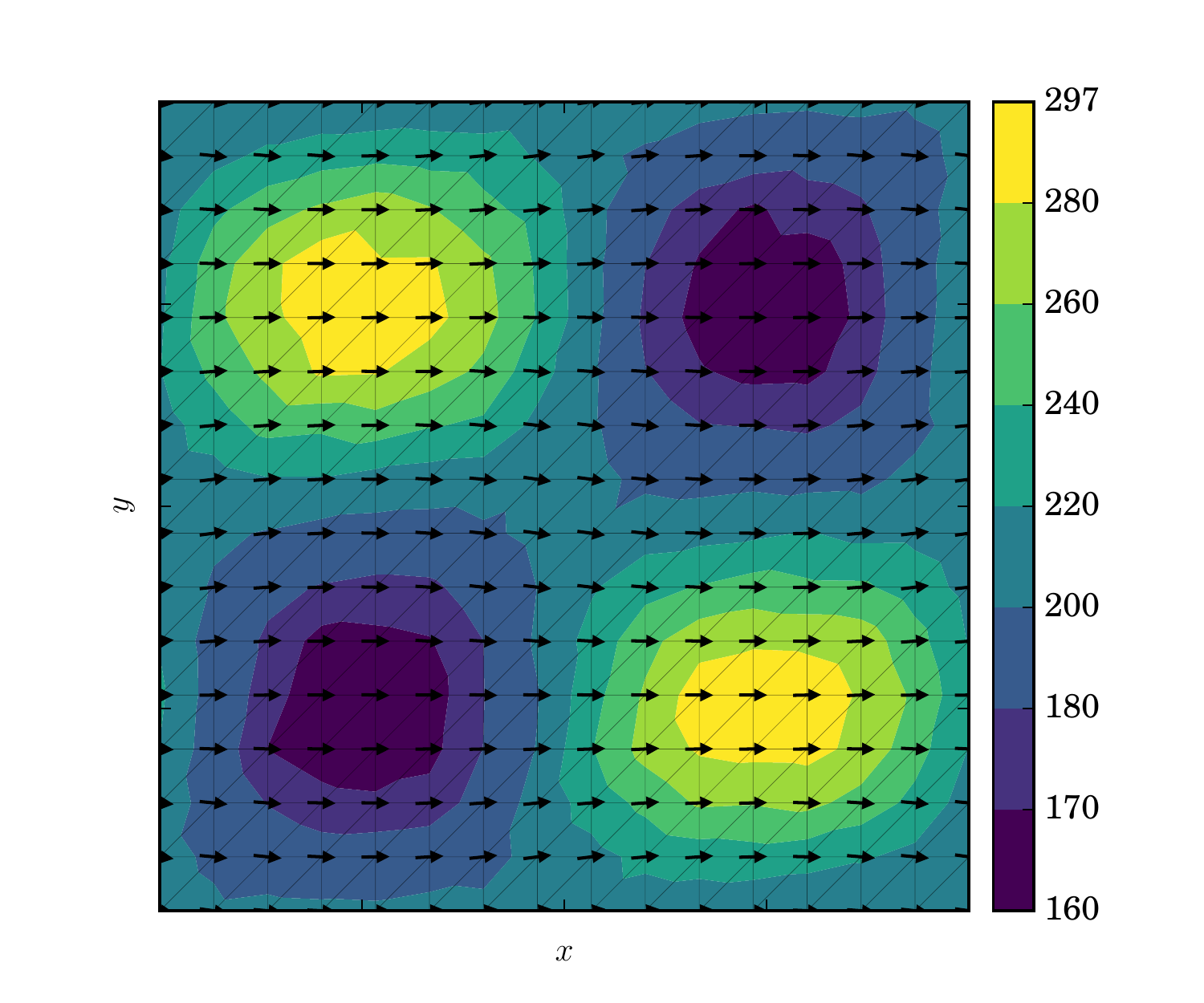}
  \caption[Tikhonov-regularized inverse ISMIP-HOM results with $\gamma_3=500$]{Results obtained using Tikhonov-regularization with $\gamma_3=500$, $\gamma_4=0$; optimized traction coefficient $\beta^*$ (top), optimized velocity $\mathbf{u}^*$ (middle), and `observed' velocity $\mathbf{u}_{ob}$ (bottom) with a $20 \times 20$ square km grid.}
  \label{inverse_ismip_opt_tikhonov_500}
\end{figure}

Results generated with Code Listing \ref{cslvr_inverse_ismip_script} indicate that an appropriate value for both parameters is $\gamma_3, \gamma_4 = 100$.  However, the traction field resulting from Tikhonov regularization with $\gamma_3 = 100$, $\gamma_4=0$ are much more irregular than the results obtained using TV-regularization with $\gamma_3=0$, $\gamma_4=100$ (compare Figures \ref{inverse_ismip_opt_tv} and \ref{inverse_ismip_opt_tikhonov_100}).  Results obtained using Tikhonov-regularization with $\gamma_3=500$, $\gamma_4=0$ produced a qualitatively-smoother result, closer to that obtained via TV-regularization with $\gamma_3=0$, $\gamma_4=100$ (Figure \ref{inverse_ismip_opt_tikhonov_500}).

We conclude by noting that the L-curve procedure described in \S \ref{ssn_l_curve} is a means to derive values for regularization parameters that are \emph{approximately} to optimal.  Thus simulation and examination of results may be required in order to derive an appropriate value for these parameters.  Additionally, when the true value of the unknown quantity is known, such as the case here, the regularization parameter may be chosen to minimize the error $\Vert \mathbf{u}^* - \mathbf{u}_{\text{true}} \Vert$.  Real-world data assimilations such as that presented in Chapter \ref{ssn_application_jakobshavn} do not include `true' values for the velocity, and so we rely on the technique of trial-and-error to derive these parameters.

\begin{table}
\centering
\caption[Inverse ISMIP-HOM variables]{ISMIP-HOM inverse variables.}
\label{ismip_hom_inverse_values}
\begin{tabular}{llll}
\hline
\textbf{Variable} & \textbf{Value} & \textbf{Units} & \textbf{Description} \\
\hline
$\dot{\varepsilon}_0$ & $10\sups{-15}$ & a\sups{-1}   & strain regularization \\
$A$       & $10\sups{-16}$  & Pa\sups{-3}a\sups{-1}   & flow-rate factor \\
$\ell$    & $20$            & km & width of domain \\
$a$       & $0.5$           & $\circ$                 & surface gradient mag. \\
$H$       & $1000$          & m & ice thickness \\
$\beta_i$ & $\beta_{\text{SIA}}$   & kg m\sups{-2}a\sups{-1} & ini.~traction coef. \\
SNR &     $100$   & -- & $\mathbf{u}_{ob}$ signal-to-noise ratio \\
$\gamma_1$ & $10^{-2}$ & kg m\sups{-2}a\sups{-1} & $L^2$ cost coefficient \\
$\gamma_2$ & $5 \times 10^3$ & J a\sups{-1} & log. cost coefficient \\
$\gamma_3$ & $10^{-1}$ & m\sups{6}kg\sups{-1}a\sups{-1} & Tikhonov reg. coef. \\
$\gamma_4$ & $10$ & m\sups{6}kg\sups{-1}a\sups{-1} & TV reg. coeff. \\
$F_b$     & $0$             & m a\sups{-1}            & basal water discharge \\
$k_x$     & $15$            & -- & number of $x$ divisions \\
$k_y$     & $15$            & -- & number of $y$ divisions \\
$k_z$     & $5$             & -- & number of $z$ divisions \\
$N_e$     & $6750$          & -- & number of cells \\
$N_n$     & $1536$          & -- & number of vertices \\
\hline
\end{tabular}
\end{table}

\pythonexternal[label=cslvr_l_curve_script, caption={CSLVR script for performing the L-curve procedure.}]{scripts/data_assimilation/L_curve.py}

\pythonexternal[label=cslvr_inverse_ismip_script, caption={CSLVR script used to solve the inverse ISMIP-HOM experiment with regularization parameters derived by Code Listing \ref{cslvr_l_curve_script}.}]{scripts/data_assimilation/ISMIP_HOM_C_inverse.py}

%===============================================================================
%===============================================================================

\chapter{Velocity balance} \label{ssn_balance_velocity}

\index{Balance equations!Velocity}
Due to the complexity of solving the higher-order Stokes models of Chapter \ref{ssn_momentum_and_mass_balance}, it is desirable to be able to run CPU-time-inexpensive computations over continent-scale regions.  One way to calculate a less-expensive momentum property is by solving the \emph{balance velocity}; the vertically-averaged velocity
\begin{align}
  \label{balance_velocity}
  \bar{\mathbf{u}} = \frac{1}{H} \int_z \mathbf{u}\ dz,
\end{align}
with components $\bar{\mathbf{u}} = [\bar{u}\ \bar{v}\ \bar{w}]\T$.  

First, assuming that a given mass within an arbitrary volume of ice $\Omega \in \R^3$ with boundary $\Gamma$ remains constant, \emph{Leibniz's rule in three dimensions} -- better known in continuum mechanics as \index{Reynold's Transport Theorem} \emph{Reynold's Transport Theorem} -- states that
\begin{align}
  \label{general_mass_balance}
  \frac{d}{dt} \int_{\Omega} \rho d\Omega = \int_{\Omega} \frac{\partial \rho}{\partial t} d\Omega + \int_{\Gamma} \rho \mathbf{u} \cdot \mathbf{n} d\Gamma = 0.
\end{align}
In the context of ice-sheets, there is a flux of ice across the upper surface $\Gamma_S$ by either accumulation or ablation $\dot{a}$, and a flux of ice lost across the basal surface $\Gamma_B$ by melting $F_b$.  Hence
\begin{align}
  \label{surface_mass_loss}
  \left( \mathbf{w}_S - \mathbf{u} \right) \cdot \mathbf{n} |_{\Gamma_S} &= \dot{a} \\
  \label{basal_mass_loss}
  \left( \mathbf{u} - \mathbf{w}_B \right) \cdot \mathbf{n} |_{\Gamma_B} &= F_b,
\end{align}
where $\mathbf{w}_B$ and $\mathbf{w}_S$ is the velocity of the free surfaces $F_S(\mathbf{x},t) = z - S(x,y,t)$ and $F_B(\mathbf{x},t) = B(x,y,t) - z$ at the coordinate $\mathbf{x} = [x\ y\ z]\T$, respectively \citep{greve97}.  Therefore, the boundary integral of general mass balance relation (\ref{general_mass_balance}) may be decomposed into
\begin{align*}
  \int_{\Gamma} \rho \mathbf{u} \cdot \mathbf{n} d\Gamma = &+ \int_{\Gamma_B} \rho \left( \mathbf{w}_B \cdot \mathbf{n} + F_b \right) d\Gamma_B + \int_{\Gamma_S} \rho \left( \mathbf{w}_S \cdot \mathbf{n} - \dot{a} \right) d\Gamma_S \\
  &+ \int_{\Gamma_L} \rho \mathbf{u} \cdot \mathbf{n} d\Gamma_L 
\end{align*}
where $\Gamma_B$ is the basal surface, $\Gamma_S$ is the upper surface, and $\Gamma_L$ is the lateral surface of the volume.

Next, if the rate of change of the free surface $F_S(\mathbf{x},t)$ and $F_B(\mathbf{x},t)$ is unchanging, the Eulerian coordinate system demands that
\begin{align}
  \label{kinematic_surface}
  \frac{dF_S}{dt} &= \frac{\partial F_S}{\partial t} + \mathbf{w}_S \cdot \nabla F_S = 0 \\
  \label{kinematic_bed}
  \frac{dF_B}{dt} &= \frac{\partial F_B}{\partial t} + \mathbf{w}_B \cdot \nabla F_B = 0.
\end{align}
Using the outward-pointing normal vector definition on the surface and bed
\begin{align}
  \label{normal_vector}
  \mathbf{n} |_{\Gamma_S} = \frac{\nabla F_S}{\Vert \nabla F_S \Vert}, \hspace{10mm}
  \mathbf{n} |_{\Gamma_B} = \frac{\nabla F_B}{\Vert \nabla F_B \Vert},
\end{align}
and relations (\ref{surface_mass_loss}, \ref{basal_mass_loss}),
\begin{align*}
  \mathbf{w}_S \cdot \nabla F_S &= \Vert \nabla F_S \Vert \left( \dot{a} + \mathbf{u} \cdot \frac{\nabla F_S}{\Vert \nabla F_S \Vert} \right) \\
  \mathbf{w}_B \cdot \nabla F_B &= \Vert \nabla F_B \Vert \left( - F_b + \mathbf{u} \cdot \frac{\nabla F_B}{\Vert \nabla F_B \Vert} \right),
\end{align*}
which used in (\ref{kinematic_surface}, \ref{kinematic_bed}) results in  
\begin{align*}
  \frac{\partial F_S}{\partial t} + \mathbf{u} \cdot \nabla F_S &= - \Vert \nabla F_S \Vert \dot{a} \\
  \frac{\partial F_B}{\partial t} + \mathbf{u} \cdot \nabla F_B &= \Vert \nabla F_B \Vert F_b.
\end{align*}
Evaluating the derivatives above,
\begin{align*}
  \mathbf{u} \cdot \nabla F_S &= \mathbf{u} \cdot \left( \hat{\mathbf{k}} - \nabla S \right) = - \Vert \hat{\mathbf{k}} - \nabla S \Vert \dot{a} + \frac{\partial S}{\partial t} \\
  \mathbf{u} \cdot \nabla F_B &= \mathbf{u} \cdot \left( \nabla B - \hat{\mathbf{k}} \right) = \Vert \nabla B - \hat{\mathbf{k}} \Vert F_b - \frac{\partial B}{\partial t},
\end{align*}
and with the use of $\hat{\mathbf{k}} \cdot \mathbf{u}(\mathbf{x},t) = w(\mathbf{x},t)$, 
\begin{align}
  \label{u_dot_gradS}
  \mathbf{u} |_{\Gamma_S} \cdot \nabla S &= w(S) + \Vert \hat{\mathbf{k}} - \nabla S \Vert \dot{a} - \frac{\partial S}{\partial t} \\
  \label{u_dot_gradB}
  \mathbf{u} |_{\Gamma_B} \cdot \nabla B &= w(B) + \Vert \nabla B - \hat{\mathbf{k}} \Vert F_b - \frac{\partial B}{\partial t}.
\end{align}

Next, employing the \index{Divergence Theorem} \emph{Divergence Theorem}
\begin{align}
  \label{divergence_theorem}
  \int_{\Gamma} \mathbf{v} \cdot \mathbf{n} d\Gamma = \int_{\Omega} \nabla \cdot \mathbf{v} d\Omega
\end{align}
to the boundary integral in general mass balance (\ref{general_mass_balance}) results in
\begin{align*}
  \frac{d}{dt} \int_{\Omega} \rho d\Omega + \int_{\Omega} \rho \nabla \cdot \mathbf{u} d\Omega = \int_{\Omega} \left( \frac{\partial \rho}{\partial t} + \rho \nabla \cdot \mathbf{u} \right) d\Omega = 0,
\end{align*}
and thus
\begin{align*}
  \frac{\partial \rho}{\partial t} + \rho \nabla \cdot \mathbf{u} = 0.
\end{align*}
Due to the fact that ice is incompressible, $\partial_t \rho = 0$ and conservation of mass relation (\ref{cons_mass}), $\nabla \cdot \mathbf{u} = 0$, has been derived.  Integrating this expression vertically and applying \index{Leibniz's Rule} \emph{Leibniz's Rule} (Appendix \ref{leibniz_formula}),
\begin{align}
  \label{integrated_cons_mass}
  \int_B^S \nabla \cdot \mathbf{u} dz &= \nabla \cdot \left( \int_B^S \mathbf{u}\ dz \right) + \mathbf{u} |_{\Gamma_B} \cdot \nabla B - \mathbf{u} |_{\Gamma_S} \cdot \nabla S = 0,
\end{align}
which using balance-velocity definition (\ref{balance_velocity}) and inserting (\ref{u_dot_gradS}, \ref{u_dot_gradB}) provide
\begin{align*}
  \nabla \cdot \big( H \bar{\mathbf{u}} \big) = &+ \mathbf{u} |_{\Gamma_S} \cdot \nabla S - \mathbf{u} |_{\Gamma_B} \cdot \nabla B \\
  =&+ \left( w(S) + \Vert \hat{\mathbf{k}} - \nabla S \Vert \dot{a} - \frac{\partial S}{\partial t}\right) \\
  &- \left( w(B) + \Vert \nabla B - \hat{\mathbf{k}} \Vert F_b - \frac{\partial B}{\partial t} \right) \\
  =& \Vert \hat{\mathbf{k}} - \nabla S \Vert \dot{a} - \Vert \nabla B - \hat{\mathbf{k}} \Vert F_b - \frac{\partial H}{\partial t} + w(S) - w(B)
\end{align*}
where in the last step, ice thickness definition $H(x,y) = S(x,y) - B(x,y)$ has been used.

Finally, we can state the \emph{balance velocity equation}
\begin{align}
  \label{balance_velocity_equation_one}
  \nabla \cdot \big( H \bar{\mathbf{u}} \big) = f,
\end{align}
with forcing term
\begin{align}
  \label{balance_velocity_equation_f}
  f = \Vert \hat{\mathbf{k}} - \nabla S \Vert \dot{a} - \Vert \nabla B - \hat{\mathbf{k}} \Vert F_b - \frac{\partial H}{\partial t} + w(S) - w(B).
\end{align}

Note that $w(S) - w(B)$ is the difference between the vertical component of velocity at the surface to that on the bed, and may be set to values obtained from one of the higher-order models of Chapter \ref{ssn_momentum_and_mass_balance}.  The rate of change of the thickness $\partial_t H$ and the surface accumulation/ablation rate $\dot{a}$ may be estimated from observations.  Lastly, the mass loss due to basal water melting $F_b$ can be attained by the methods of Chapter \ref{ssn_internal_energy_balance}.

\section{The direction of flowing ice}

Balance velocity $\bar{\mathbf{u}}$ may be decomposed into a direction $\hat{\mathbf{u}}$ and magnitude $\bar{u}$ \citep{balance} such that
\begin{align}
  \label{balance_velocity_dir_and_mag}
  \bar{\mathbf{u}} = \bar{u} \mathbf{\hat{u}},
\end{align}
with unit vector $\hat{\mathbf{u}} = [\hat{u}\ \hat{v}\ \hat{w}]\T$.  From balance equation (\ref{balance_velocity_equation_one}),
\begin{align}
  \label{balance_velocity_equation}
  \nabla \cdot \big( H \bar{u} \mathbf{\hat{u}} \big) &= f,
\end{align}
which can be expanded with terms $H$ and $\hat{u}$ grouped together,
\begin{align*}
  \bar{u} H \nabla \cdot \mathbf{\hat{u}} + \nabla \big( \bar{u} H \big) \cdot \mathbf{\hat{u}} &= f \\
  \bar{u} H \nabla \cdot \mathbf{\hat{u}} + \big( H \nabla \bar{u} + \bar{u} \nabla H \big) \cdot \mathbf{\hat{u}} &= f.
\end{align*}
After a little rearranging, this produces
\begin{align}
  H \hat{\mathbf{u}} \cdot \nabla \bar{u} + \big( H \nabla \cdot \mathbf{\hat{u}} + \nabla H \cdot \mathbf{\hat{u}} \big) \bar{u} &= f \notag \\
  \label{balance_velocity_equation_expanded}
  H \hat{\mathbf{u}} \cdot \nabla \bar{u} + \nabla \cdot \big( H \mathbf{\hat{u}} \big) \bar{u} &= f.
\end{align}

Next, because this system includes one equation and three unknowns, two are eliminated by \emph{prescribing} the direction of flow.  In the absence of surface velocity data, one choice for this direction is down the gradient of \index{Driving stress} \emph{driving stress} $\bm{\tau_d} = [\tau_x\ \tau_y\ 0]\T$, derived by integrating vertically the forcing term of first-order momentum balance (\ref{bp_cons_momentum}):
\begin{align}
  \label{driving_stress}
  \bm{\tau_d} = \int_B^S \rho g \nabla S\ dz = \rho g H \nabla S.
\end{align}

Upon closer inspection of balance velocity relation (\ref{balance_velocity_equation_one}), and using the fact that the partial derivative of the vertically integrated velocity with respect to $z$ is zero,
\begin{align*}
  \nabla \cdot \big( H \bar{\mathbf{u}} \big) &= \nabla \cdot \left( \int_B^S \mathbf{u}\ dz \right) = \frac{\partial}{\partial x} \left( \int_B^S \mathbf{u}\ dz \right)
     + \frac{\partial}{\partial y} \left( \int_B^S \mathbf{u}\ dz \right).
\end{align*}
Hence only the two horizontal components $\hat{u}$ and $\hat{v}$ of balance velocity (\ref{balance_velocity}) remain, and the domain is reduced to $\Omega \in \R^2$.  Additionally, because only the \emph{direction} of ice flow is imposed, the direction $\nabla S$ may be used in place of full driving stress (\ref{driving_stress}).

Let the direction of imposed flow be defined as
\begin{align}
  \label{balance_velocity_direction}
  \hat{\mathbf{u}} = \frac{\mathbf{d}}{\Vert \mathbf{d} \Vert}, \hspace{10mm} 
  \mathbf{d} = \begin{bmatrix}
                 d_x \\
                 d_y \\
               \end{bmatrix}.
\end{align}
Due to the fact that the solution to balance-velocity Equation (\ref{balance_velocity_equation}, \ref{balance_velocity_equation_f}) will be highly sensitive to variations in the data used to define (\ref{balance_velocity_direction}), an additional term is added to flow-direction (\ref{balance_velocity_direction}) in such a way that direction is decomposed into data $\mathbf{d}^{\text{data}}$ and Laplace-blurring $\mathbf{d}^s$ terms \citep{balance}
\begin{align*}
  \mathbf{d} &= \mathbf{d}^{\text{data}} + \mathbf{d}^s, \hspace{5mm} \mathbf{d}^s = \big( \kappa H \big)^2 \nabla \cdot \big( \nabla \mathbf{d} \big),
\end{align*}
where the constant $\kappa$ adjusts the amount of smoothing proportional to the ice thickness $H$.  The expansion of this linear system is
\begin{align*}
  \begin{bmatrix}
    d_x \\
    d_y
  \end{bmatrix} - \big( \kappa H \big)^2
  \begin{bmatrix}
    \frac{\partial^2 d_x}{\partial x^2} + \frac{\partial^2 d_x}{\partial y^2} \\ 
    \frac{\partial^2 d_y}{\partial x^2} + \frac{\partial^2 d_y}{\partial y^2}
  \end{bmatrix} &=
  \begin{bmatrix}
     d_x^{\text{data}} \\
     d_y^{\text{data}}
  \end{bmatrix},
\end{align*}
thus illuminating two equations for the unknowns $d_x$ and $d_y$,
\begin{align}
  \label{smoothed_driving_stress_x}
  d_x - \big( \kappa H \big)^2 \left( \frac{\partial^2 d_x^s}{\partial x^2} + \frac{\partial^2 d_x^s}{\partial y^2} \right) &= d_x^{\text{data}} \\
  \label{smoothed_driving_stress_y}
  d_y - \big( \kappa H \big)^2 \left( \frac{\partial^2 d_y^s}{\partial x^2} + \frac{\partial^2 d_y^s}{\partial y^2} \right) &= d_y^{\text{data}}.
\end{align}

\subsection{Variational forms for $\mathbf{d}$}

The variational problem associated with smoothed driving stress equations (\ref{smoothed_driving_stress_x}, \ref{smoothed_driving_stress_y}) reads: find $d_x \in \ltwospace$ (see $L^2$ space (\ref{l2_space})) such that
\begin{align*}
  \int_{\Omega} d_x \phi\ d\Omega - \int_{\Omega} \big( \kappa H \big)^2 \left( \frac{\partial^2 d_x}{\partial x^2} + \frac{\partial^2 d_x}{\partial y^2} \right) \phi\ d\Omega &= \int_{\Omega} d_x^{\text{data}} \phi\ d\Omega \\
  + \int_{\Omega} d_x \phi\ d\Omega + \int_{\Omega} \big( \kappa H \big)^2 \big( \nabla d_x \big)_h \cdot \big( \nabla \phi \big)_h\ d\Omega &\\
  - \int_{\Gamma} \phi \big( \kappa H \big)^2 \big( \nabla d_x \big)_h \cdot \mathbf{n}_h\ d\Gamma &= \int_{\Omega} d_x^{\text{data}} \phi\ d\Omega
\end{align*}
for all $\phi \in \testspace$ (see test space (\ref{test_space})); and $d_y \in \ltwospace$ such that
\begin{align*}
  \int_{\Omega} d_y \psi\ d\Omega - \int_{\Omega} \big( \kappa H \big)^2 \left( \frac{\partial^2 d_y}{\partial x^2} + \frac{\partial^2 d_y}{\partial y^2} \right) \psi\ d\Omega &= \int_{\Omega} d_y^{\text{data}} \psi\ d\Omega \\
  \int_{\Omega} d_y \psi\ d\Omega + \int_{\Omega} \big( \kappa H \big)^2 \big( \nabla d_y \big)_h \cdot \big( \nabla \psi \big)_h\ d\Omega &\\
  - \int_{\Gamma} \psi \big( \kappa H \big)^2 \big( \nabla d_y \big)_h \cdot \mathbf{n}_h\ d\Gamma &= \int_{\Omega} d_y^{\text{data}} \psi\ d\Omega
\end{align*}
for all $\psi \in \testspace$.  Horizontal vector components are denoted $\mathbf{v}_h = [v_x\ v_y]\T$.

\section{The magnitude of flowing ice}

On closer examination of velocity-balance relation (\ref{balance_velocity_equation_expanded}), the equation is seen to be of the same form as the advection-reaction equation described in \S \ref{ssn_stabilized_methods},
\begin{align*}
  \mathbf{a} \cdot \nabla u + s u = f,
\end{align*}
with unknown quantity $u = \bar{u}$, velocity $\mathbf{a} = H \hat{\mathbf{u}}$, and reaction coefficient $s = \nabla \cdot \big( H \mathbf{\hat{u}} \big)$.  As previously discussed in \S \ref{ssn_stabilized_methods}, equations of this type suffer from numerical oscillations requiring the use of stabilization.  Therefore, defining the linear differential operator associated with problem (\ref{balance_velocity_equation_expanded}),
\begin{align}
  \label{bv_linear_operator}
  \Lu u &= H \hat{\mathbf{u}} \cdot \nabla \bar{u} + \nabla \cdot \big( H \mathbf{\hat{u}} \big) \bar{u},
\end{align}
the stabilized form is stated using general stabilized form (\ref{generalized_form}) with test function $\phi \in \testspace$ and intrinsic-time parameter \index{Intrinsic-time parameter!Balance velocity} $\tau_{\text{BV}}$,
\begin{align}
  \label{bv_generalized_form}
  (\phi, \Lu \bar{u}) + (\mathbb{L} \phi, \tau_{\text{BV}}(\Lu\bar{u} - f)) &= (\phi, f),
\end{align}
where operator $\mathbb{L}$ is a differential operator typically chosen from
\begin{align}
  \label{bv_gls_operator}
  \mathbb{L} &= + \Lu && \text{Galerkin/least-squares (GLS)} \\
  \label{bv_supg_operator}
  \mathbb{L} &= + \Lu_{\text{adv}} && \text{SUPG} \\
  \label{bv_ssm_operator}
  \mathbb{L} &= - \Lu^* && \text{subgrid-scale model (SSM)}
\end{align}
with $\Lu_{\text{adv}} = H\hat{\mathbf{u}} \cdot \nabla \bar{u}$, the advective part of the operator $\Lu$.

An appropriate stabilization parameter for this problem is given by (\ref{tau_adr}), as derived by \cite{codina}.  After making the appropriate substitutions to ADR parameter (\ref{tau_adr}), the intrinsic-time parameter is
\begin{align}
  \label{tau_bv}
  \tau_{\text{BV}} = \frac{1}{\frac{2 H}{h} + \nabla \cdot \big( H \hat{\mathbf{u}} \big)},
\end{align}
with cell diameter $h$.  Here, the fact that $\Vert \hat{\mathbf{u}} \Vert = 1$ and that there is no diffusion present has been used.

\subsection{Variational form for $\bar{u}$}

Using intrinsic-time parameter (\ref{tau_bv}) and operator (\ref{bv_linear_operator}), the variational problem associated with (\ref{balance_velocity_equation}, \ref{balance_velocity_equation_f}) reads: find $\bar{u} \in \ltwospace$ such that
\begin{align}
  \label{balance_velocity_weak_problem}
  \mathscr{B}(\phi, \bar{u}) = \ell(\phi),
\end{align}
where
\begin{align*}
  \mathscr{B}(\phi, \bar{u} &= (\phi, \Lu \bar{u}) + (\mathbb{L} \phi, \tau_{\text{BV}}(\Lu\bar{u})) \\
  \ell(\phi) &= (\phi, f) + (\mathbb{L} \phi, \tau_{\text{BV}}f),
\end{align*}
for all $\phi \in \trialspace$ and $\mathbb{L}$ given by one of (\ref{bv_gls_operator}), (\ref{bv_supg_operator}), or (\ref{bv_ssm_operator}).

The implementation of this problem by CSLVR is shown in Code Listing \ref{cslvr_balance_velocity}.

\pythonexternal[label=cslvr_balance_velocity, caption={CSLVR implementation of the \texttt{BalanceVelocity} class.}]{cslvr_src/balancevelocity.py}

\section{Continent-wide simulations} \label{ssn_balance_velocity_simulations}

\index{Linear differential equations!2D}
Solutions to balance-velocity variational problem (\ref{balance_velocity_weak_problem}) were obtained over the entire continents of Antarctica and Greenland for each of stabilization schemes (\ref{bv_gls_operator}), (\ref{bv_supg_operator}), and (\ref{bv_ssm_operator}).  In order to complete these simulations, some assumptions had to be made regarding forcing term (\ref{balance_velocity_equation_f}).

First, we assumed that $w(S) - w(B) = 0$ everywhere.  While this is assumption is not true in general, it has been our experience from investigation of velocity solutions obtained via the methods of Chapter \ref{ssn_inclusion_of_velocity_data} that the vertical velocity remains relatively constant throughout the vertical coordinate over much of the ice-sheet domain, except areas where the velocity changes direction abruptly.  Second, we assumed that the ice sheet thickness is in equilibrium, and hence $\partial_t H = 0$.  Finally, we did not prescribe any basal melting, and thus $F_b = 0$.

To create the finite-element mesh, we incorporated the dynamic version of GMSH \citep{gmsh} into CSLVR.  This software allows us to generate finite-element meshes with cell diameters set by any function we wish.  Following the work of \citet{balance}, we generated meshes for each ice-sheet with cell diameter $h$ given by 
$$h = mH$$
for some constant $m$.

\begin{figure*}
  \centering
    \includegraphics[width=0.8\linewidth]{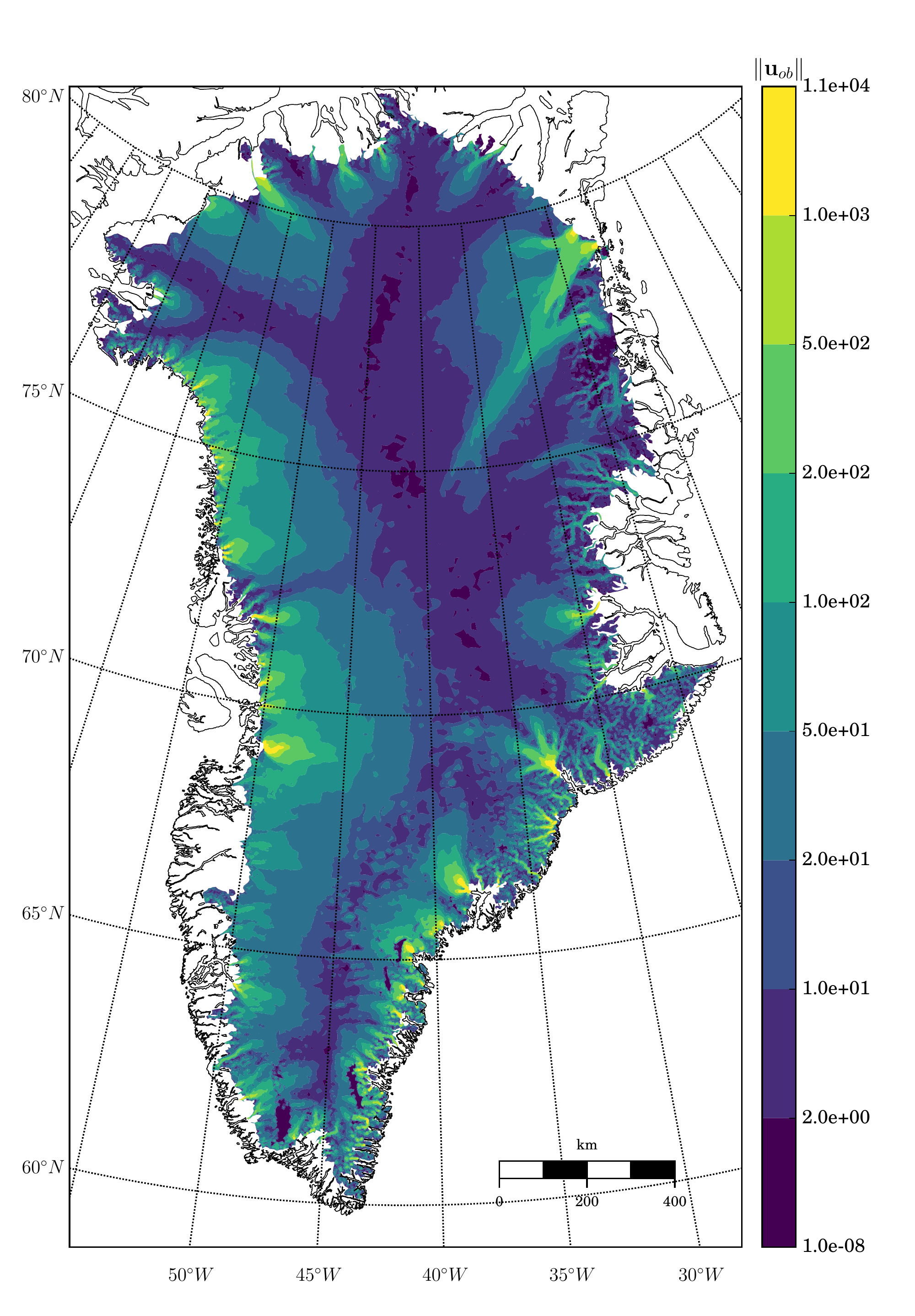}
  \caption[Greenland surface-velocity magnitude]{Surface velocity magnitude of Greenland for the polar year 2008--2009 provided by \citet{rignot_greenland}.}
  \label{greenland_u_ob_image}
\end{figure*}

\begin{figure*}
  \centering
    \includegraphics[width=\linewidth]{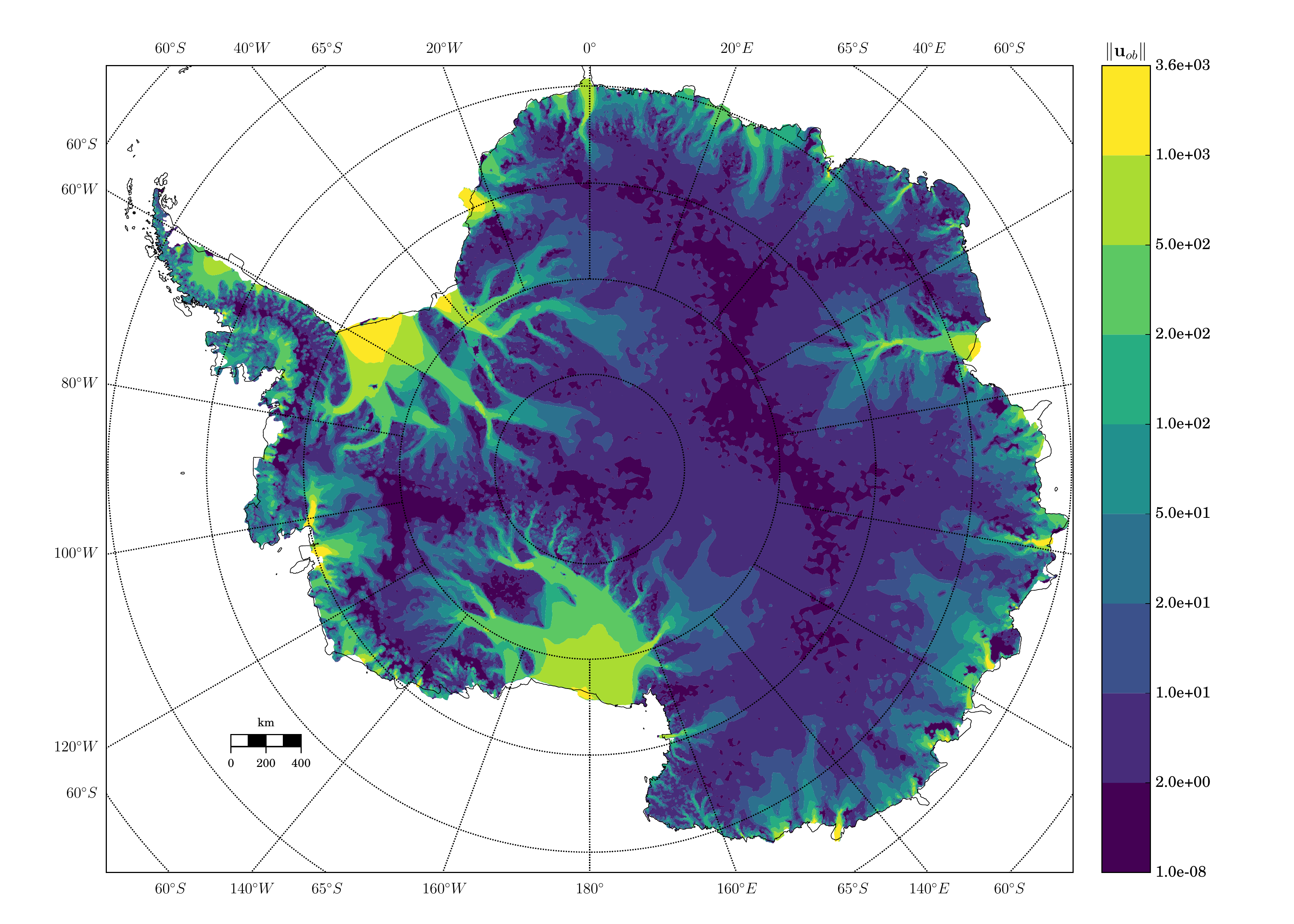}
  \caption[Antarctica surface velocity magnitude]{Surface velocity magnitude of Antarctica as provided by \citet{rignot}.}
  \label{antarctica_u_ob_image}
\end{figure*}

\begin{figure*}

  \centering

  \begin{subfigure}[b]{0.45\linewidth}
    \includegraphics[width=\linewidth]{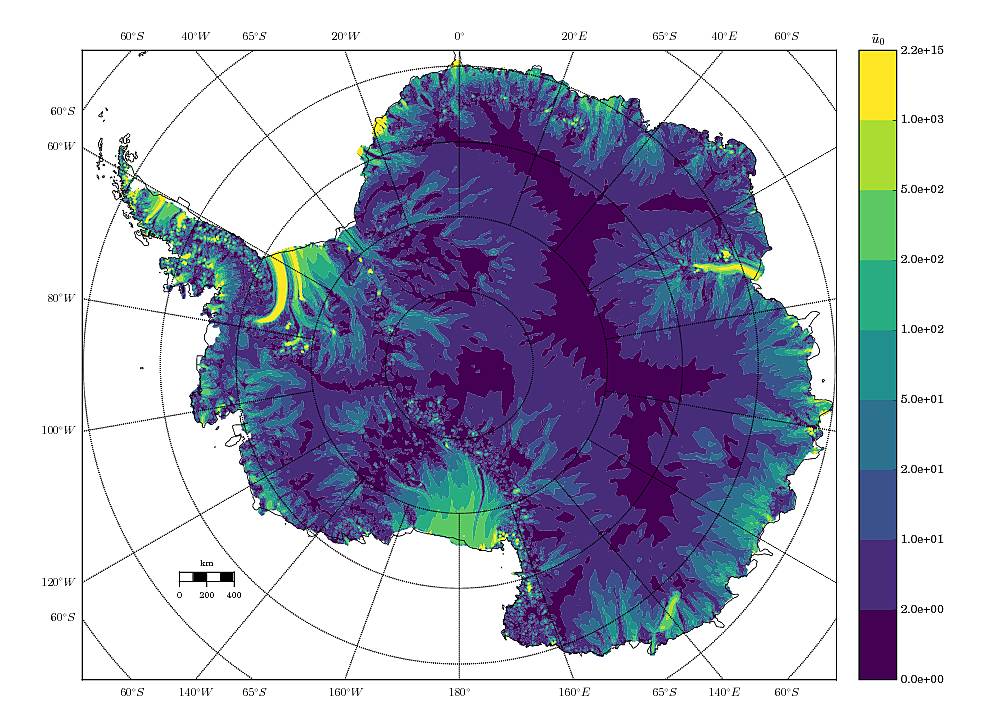}
  \caption{$\kappa = 0$, GLS.}
  \label{antarctica_bv_image_kappa_0_GLS_U_ob_S}
  \end{subfigure}
  \begin{subfigure}[b]{0.45\linewidth}
    \includegraphics[width=\linewidth]{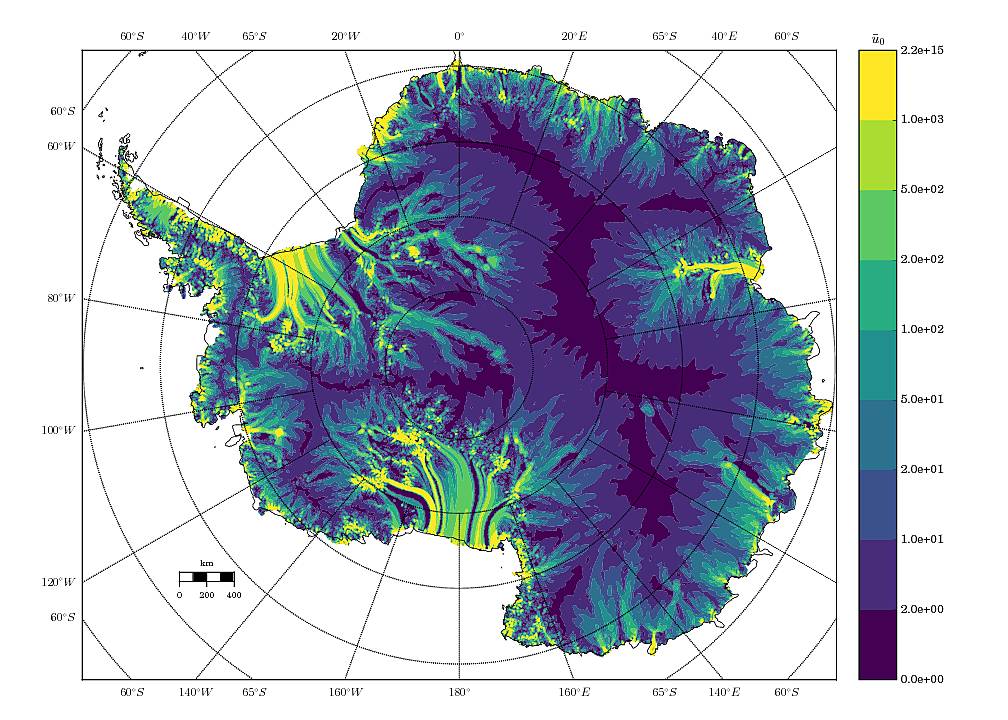}
  \caption{$\kappa = 0$, SUPG.}
  \label{antarctica_bv_image_kappa_0_SUPG_U_ob_S}
  \end{subfigure}

  \begin{subfigure}[b]{0.45\linewidth}
    \includegraphics[width=\linewidth]{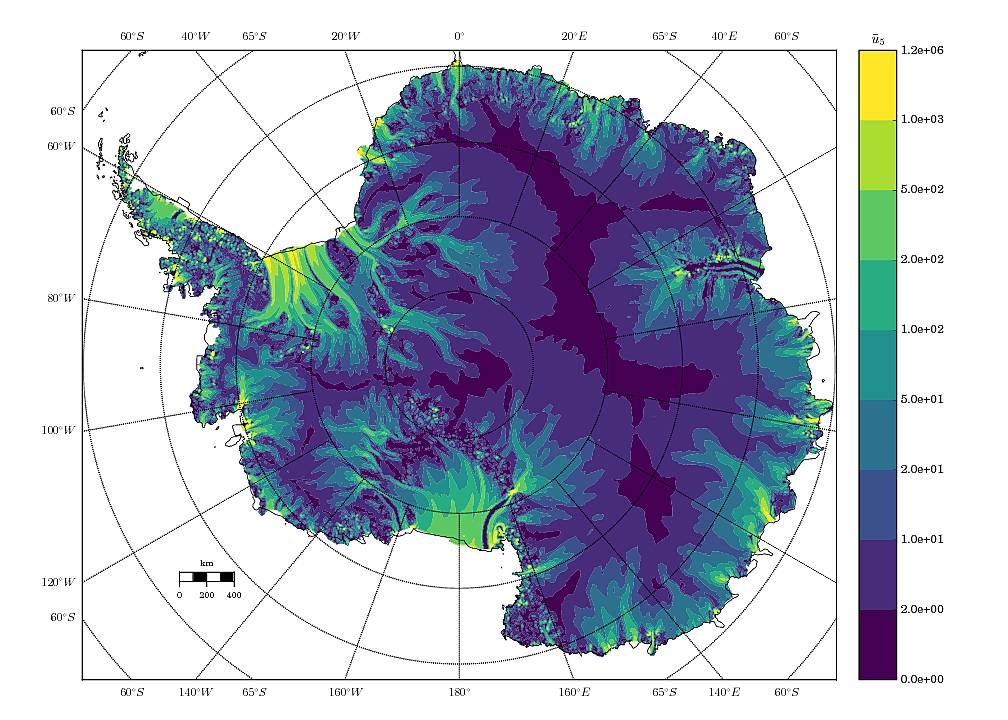}
  \caption{$\kappa = 5$, GLS.}
  \label{antarctica_bv_image_kappa_5_GLS_U_ob_S}
  \end{subfigure}
  \begin{subfigure}[b]{0.45\linewidth}
    \includegraphics[width=\linewidth]{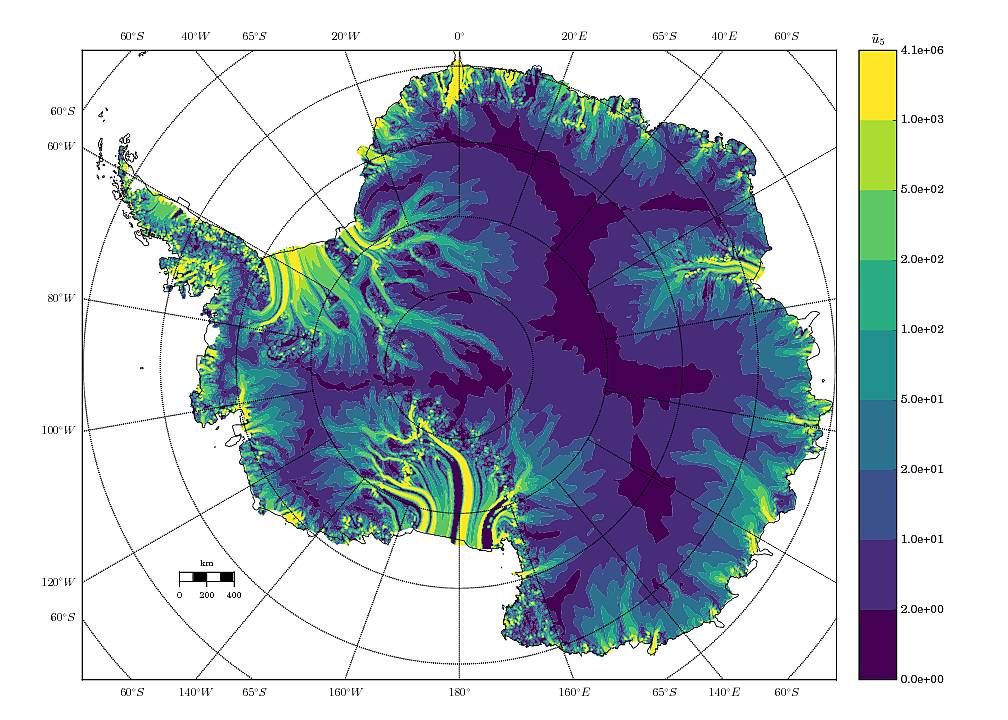}
  \caption{$\kappa = 5$, SUPG.}
  \label{antarctica_bv_image_kappa_5_SUPG_U_ob_S}
  \end{subfigure}

  \begin{subfigure}[b]{0.45\linewidth}
    \includegraphics[width=\linewidth]{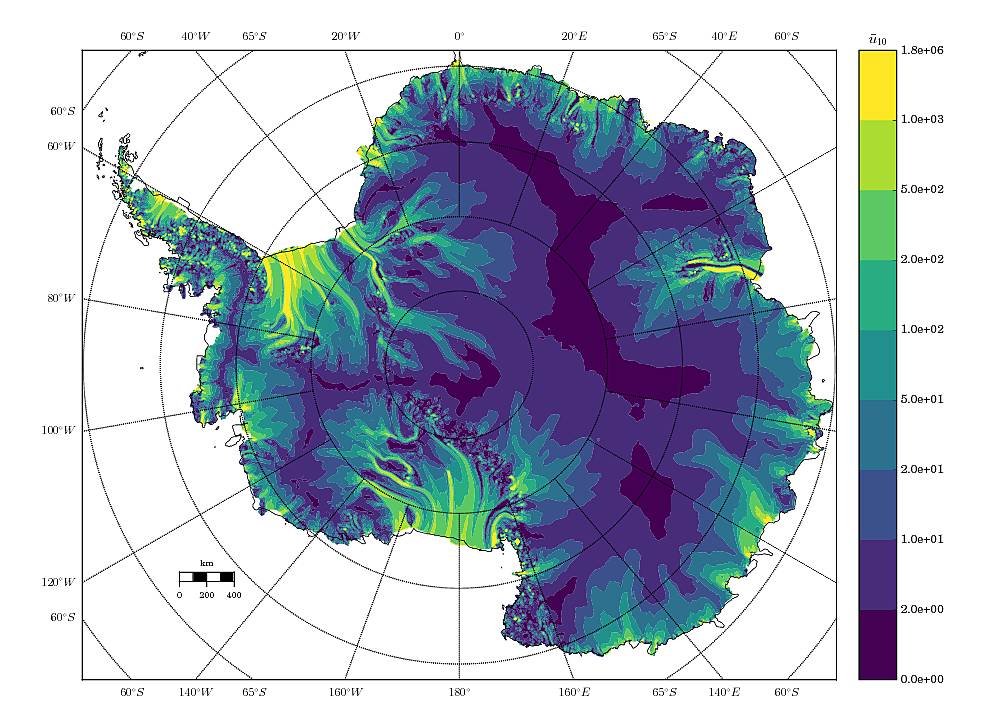}
  \caption{$\kappa = 10$, GLS.}
  \label{antarctica_bv_image_kappa_5_GLS_U_ob_S}
  \end{subfigure}
  \begin{subfigure}[b]{0.45\linewidth}
    \includegraphics[width=\linewidth]{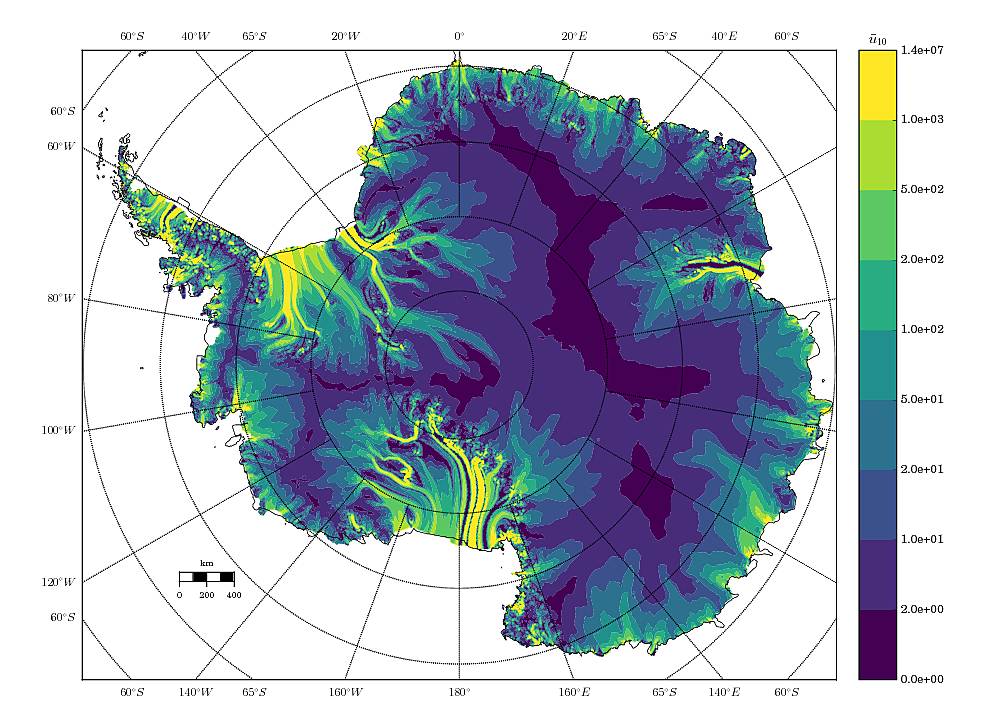}
  \caption{$\kappa = 10$, SUPG.}
  \label{antarctica_bv_image_kappa_5_SUPG_U_ob_S}
  \end{subfigure}
  
  \caption[Antarctica balance-velocity with $\mathbf{d}^{\text{data}} = \mathbf{u}_{ob}$ over shelves.]{Balance velocity $\bar{u}$ derived over Antarctica with imposed direction of flow down the surface gradient $\nabla S$ over grounded ice and in the direction of surface observations $\mathbf{u}_{ob}$ over floating ice, where smoothing radius $\kappa$ varies as indicated.  The columns vary according to stabilization used; either Galerkin/least-squares (GLS) stabilization (\ref{bv_gls_operator}) or streamline-upwind/Petrov-Galerkin (SUPG) stabilization (\ref{bv_supg_operator}) in variational form (\ref{balance_velocity_weak_problem}).  Results using subgrid-scale-model stabilization (\ref{bv_ssm_operator}) (not shown) appeared more unstable than the (SUPG) method.}

  \label{antarctica_bv_image_U_ob_S}

\end{figure*}

\begin{figure*}

  \centering

  \begin{subfigure}[b]{0.45\linewidth}
    \includegraphics[width=\linewidth]{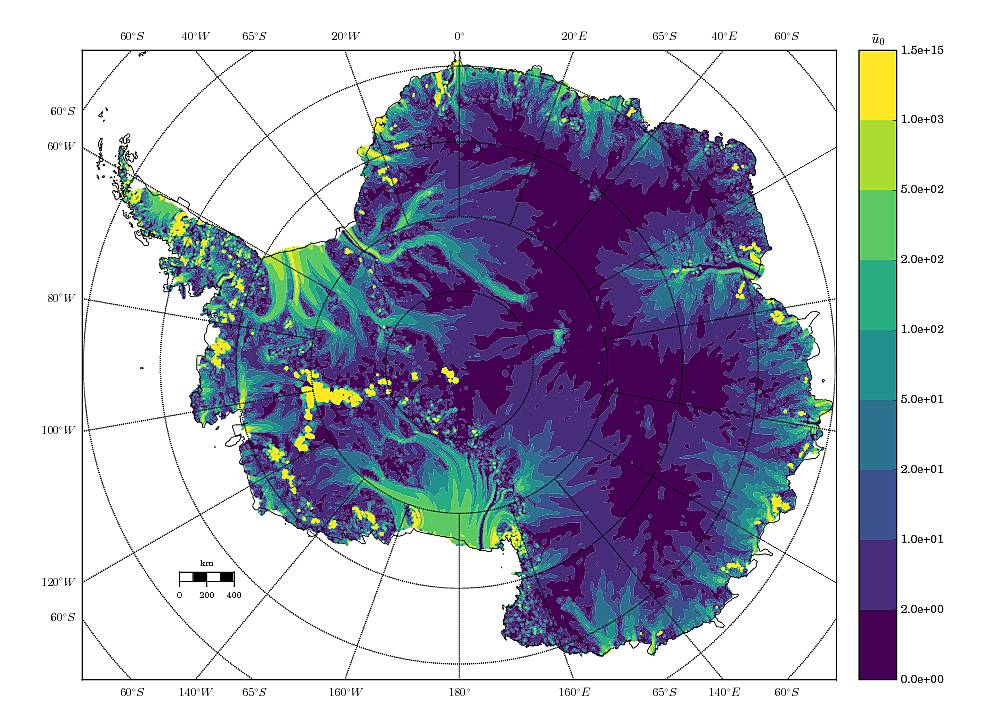}
  \caption{$\kappa = 0$, GLS.}
  \label{antarctica_bv_image_d_U_ob_kappa_0_GLS}
  \end{subfigure}
  \begin{subfigure}[b]{0.45\linewidth}
    \includegraphics[width=\linewidth]{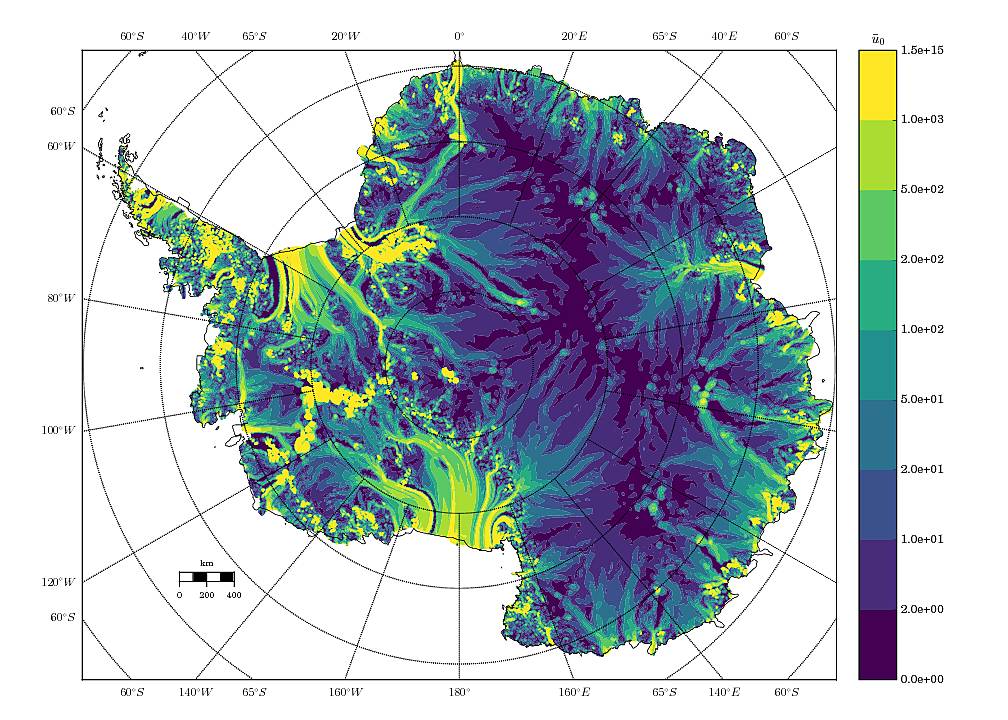}
  \caption{$\kappa = 0$, SUPG.}
  \label{antarctica_bv_image_d_U_ob_kappa_0_SUPG}
  \end{subfigure}

  \begin{subfigure}[b]{0.45\linewidth}
    \includegraphics[width=\linewidth]{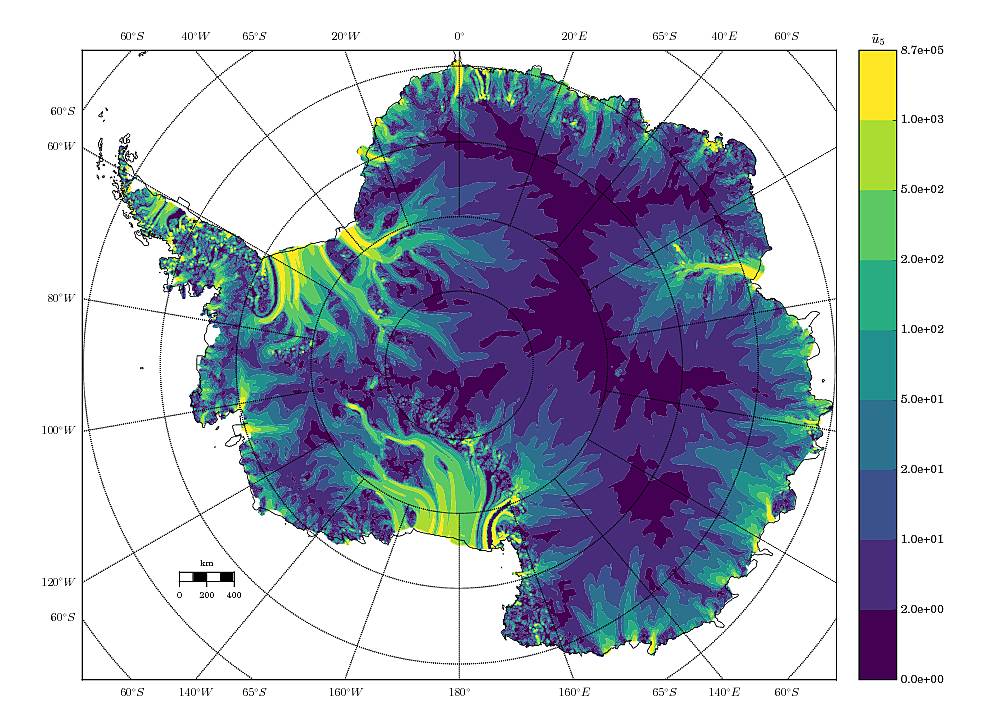}
  \caption{$\kappa = 5$, GLS.}
  \label{antarctica_bv_image_d_U_ob_kappa_5_GLS}
  \end{subfigure}
  \begin{subfigure}[b]{0.45\linewidth}
    \includegraphics[width=\linewidth]{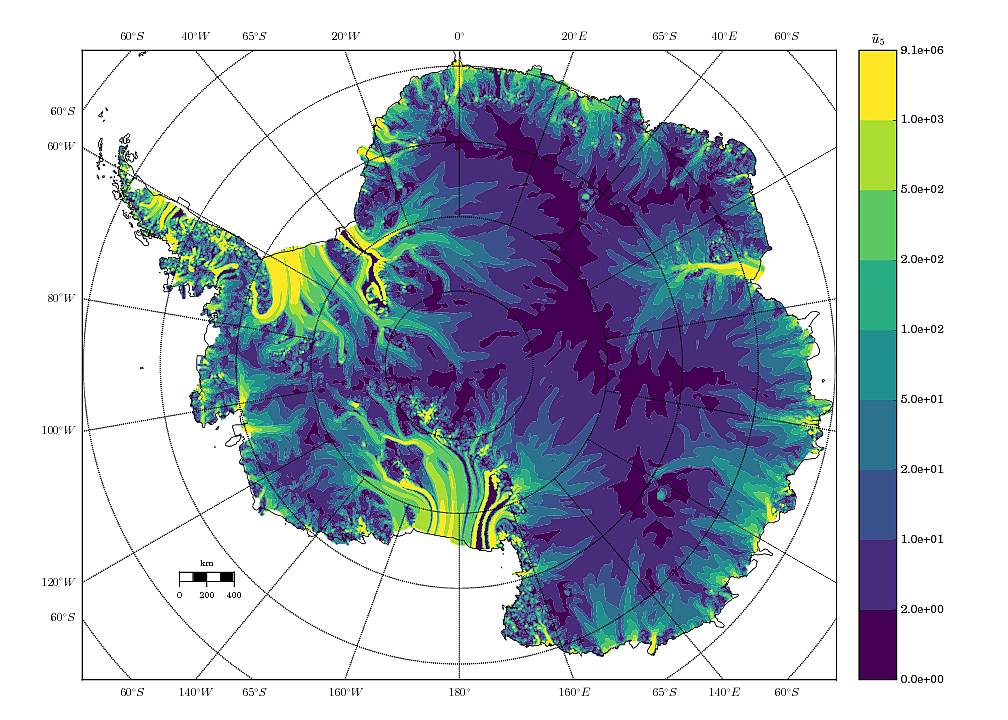}
  \caption{$\kappa = 5$, SUPG.}
  \label{antarctica_bv_image_d_U_ob_kappa_5_SUPG}
  \end{subfigure}

  \begin{subfigure}[b]{0.45\linewidth}
    \includegraphics[width=\linewidth]{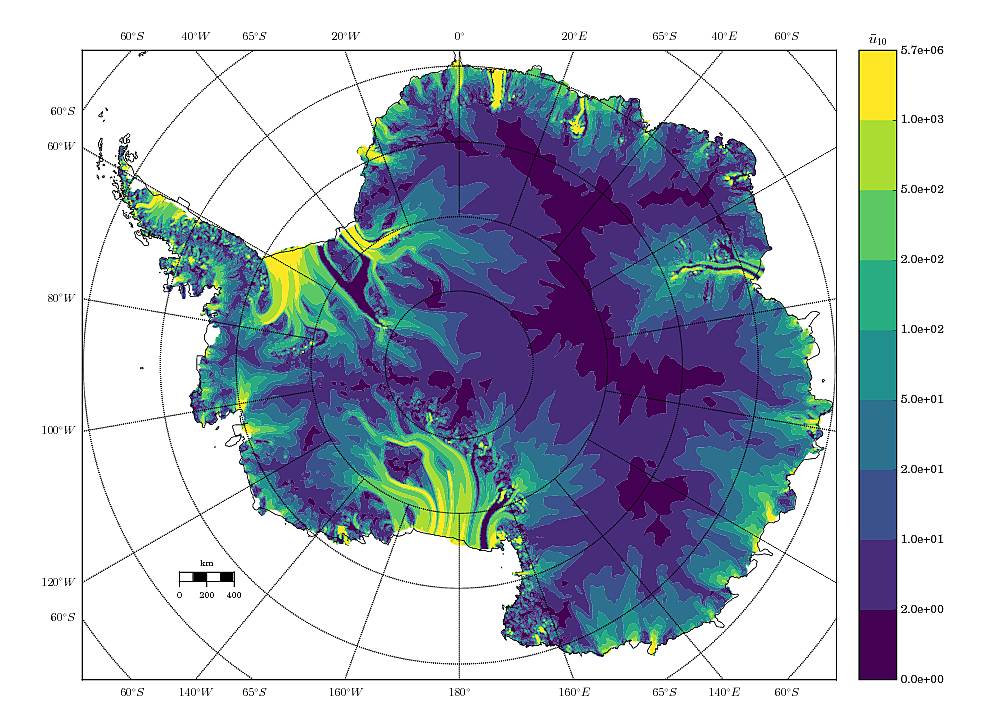}
  \caption{$\kappa = 10$, GLS.}
  \label{antarctica_bv_image_d_U_ob_kappa_10_GLS}
  \end{subfigure}
  \begin{subfigure}[b]{0.45\linewidth}
    \includegraphics[width=\linewidth]{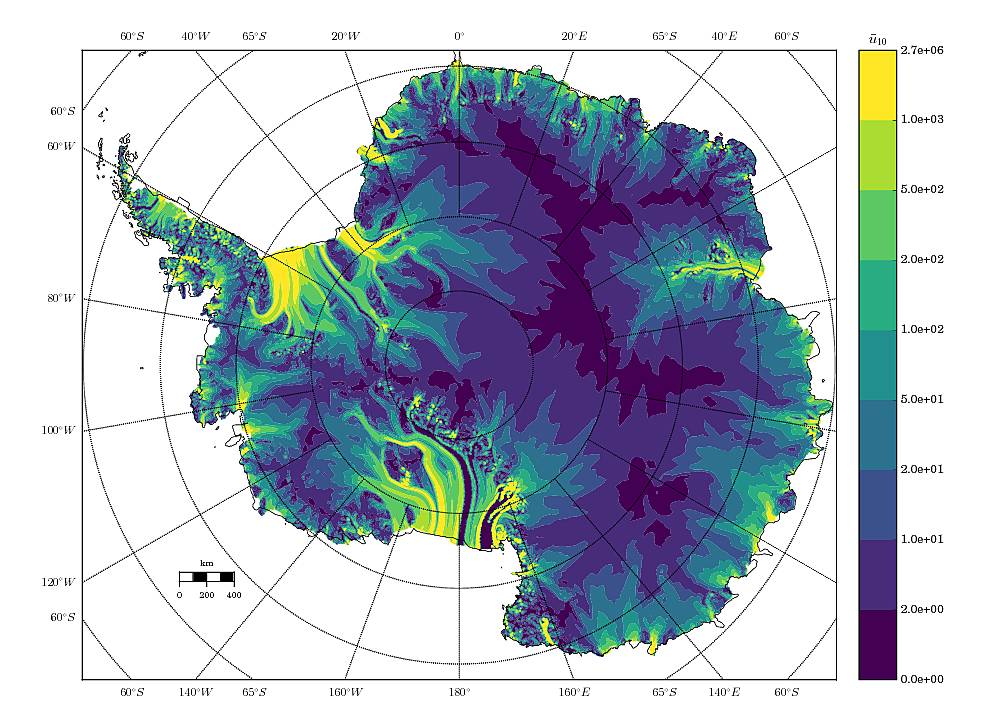}
  \caption{$\kappa = 10$, SUPG.}
  \label{antarctica_bv_image_d_U_ob_kappa_10_SUPG}
  \end{subfigure}
  
  \caption[Antarctica balance-velocity with $\mathbf{d}^{\text{data}} = \mathbf{u}_{ob}$.]{Balance velocity $\bar{u}$ derived over Antarctica with imposed direction of flow in the direction of surface observations $\mathbf{u}_{ob}$, where smoothing radius $\kappa$ varies as indicated.  The columns vary according to stabilization used; either Galerkin/least-squares (GLS) stabilization (\ref{bv_gls_operator}) or streamline-upwind/Petrov-Galerkin (SUPG) stabilization (\ref{bv_supg_operator}) in variational form (\ref{balance_velocity_weak_problem}).  Results using subgrid-scale-model stabilization (\ref{bv_ssm_operator}) (not shown) appeared more unstable than the (SUPG) method.}

  \label{antarctica_bv_image_d_U_ob}
  
\end{figure*}

\begin{figure*}

  \centering

  \begin{subfigure}[b]{0.45\linewidth}
    \includegraphics[width=\linewidth]{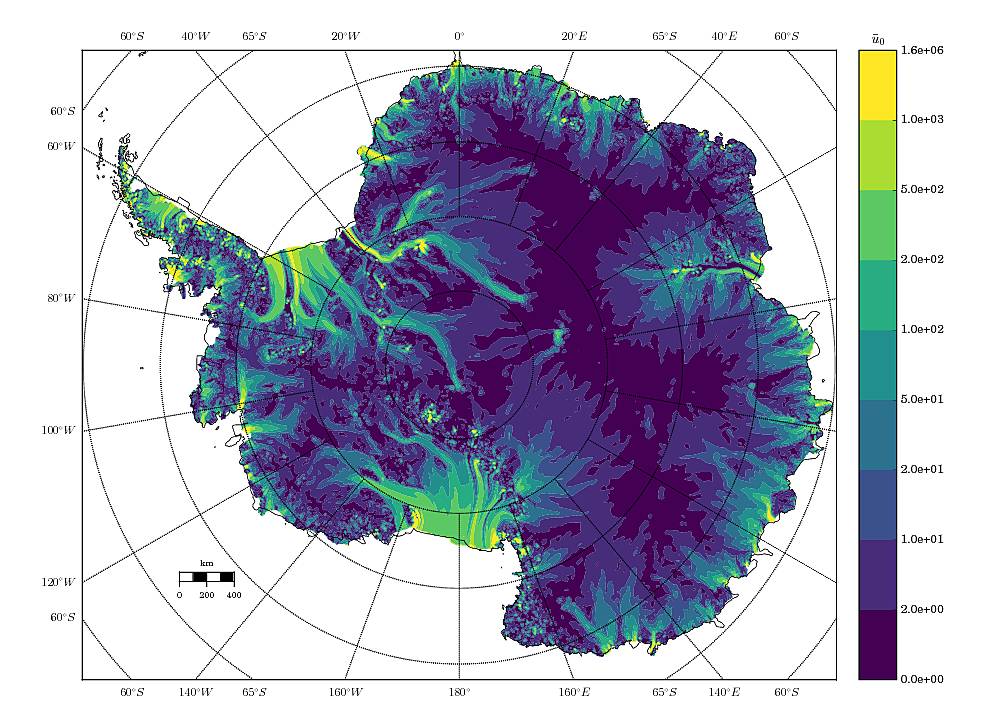}
  \caption{$\kappa = 0$, GLS.}
  \label{antarctica_bv_image_d_gS_m_U_kappa_0_GLS}
  \end{subfigure}
  \begin{subfigure}[b]{0.45\linewidth}
    \includegraphics[width=\linewidth]{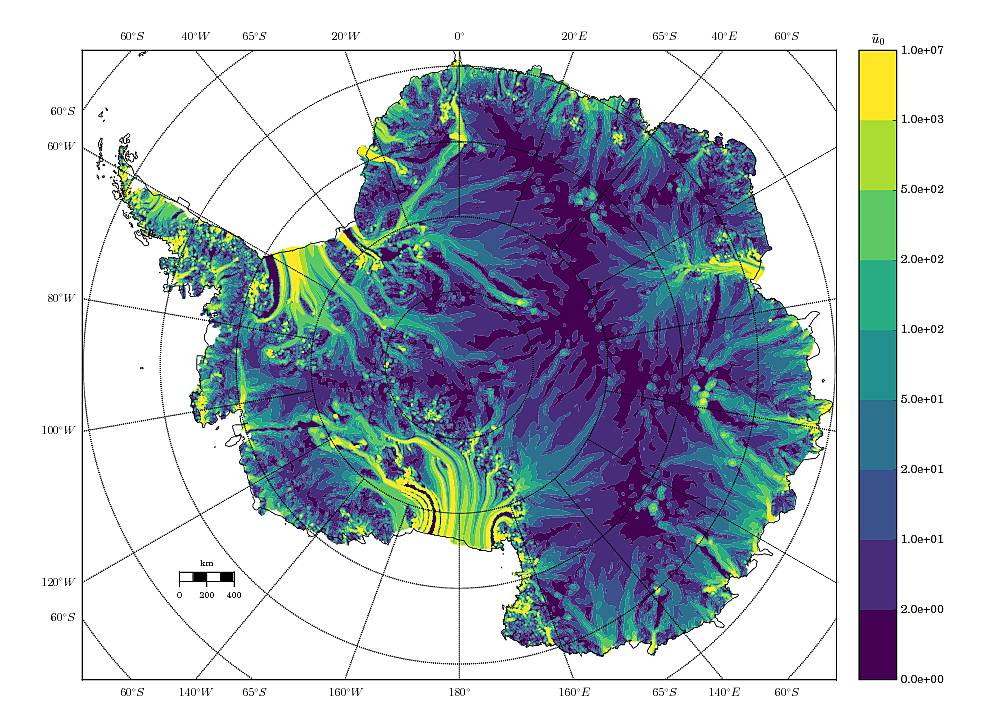}
  \caption{$\kappa = 0$, SUPG.}
  \label{antarctica_bv_image_d_gS_m_U_kappa_0_SUPG}
  \end{subfigure}

  \begin{subfigure}[b]{0.45\linewidth}
    \includegraphics[width=\linewidth]{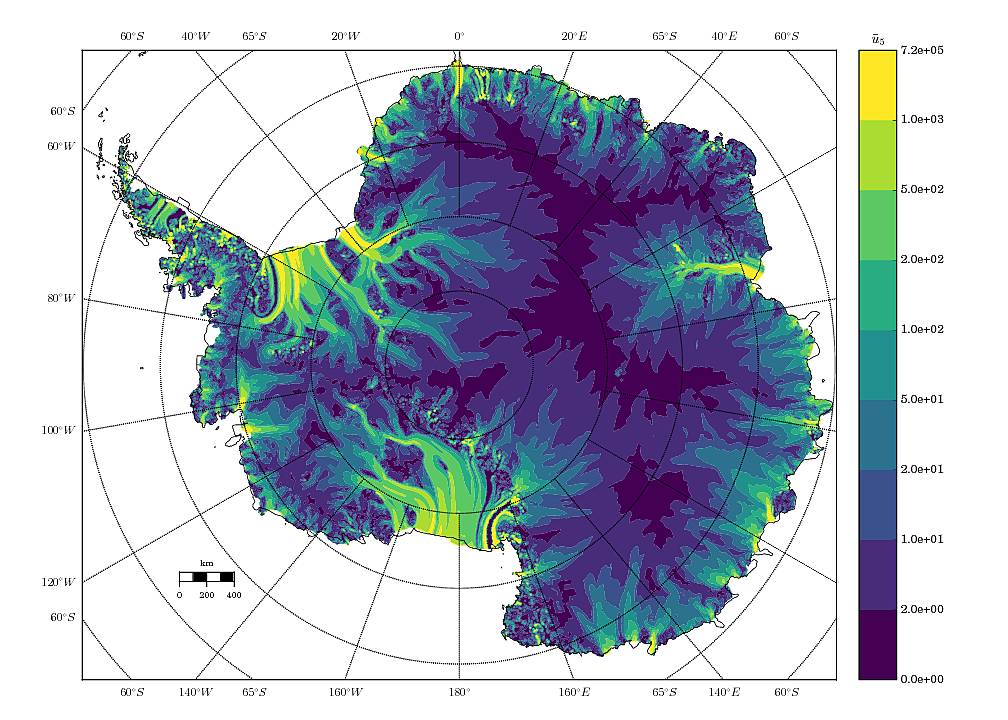}
  \caption{$\kappa = 5$, GLS.}
  \label{antarctica_bv_image_d_gS_m_U_kappa_5_GLS}
  \end{subfigure}
  \begin{subfigure}[b]{0.45\linewidth}
    \includegraphics[width=\linewidth]{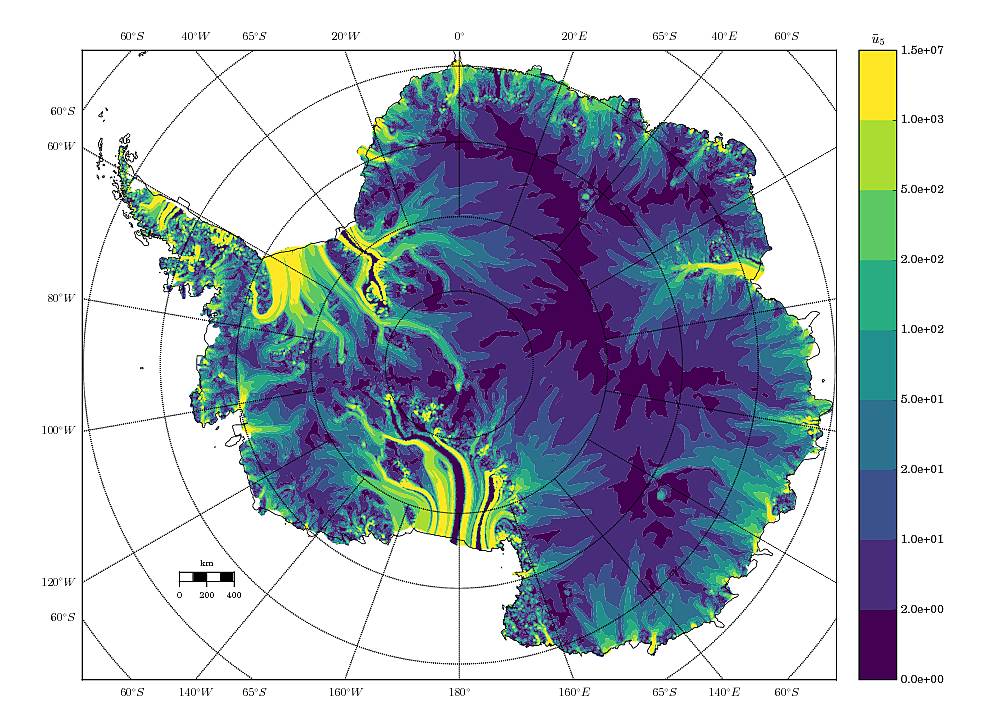}
  \caption{$\kappa = 5$, SUPG.}
  \label{antarctica_bv_image_d_gS_m_U_kappa_5_SUPG}
  \end{subfigure}

  \begin{subfigure}[b]{0.45\linewidth}
    \includegraphics[width=\linewidth]{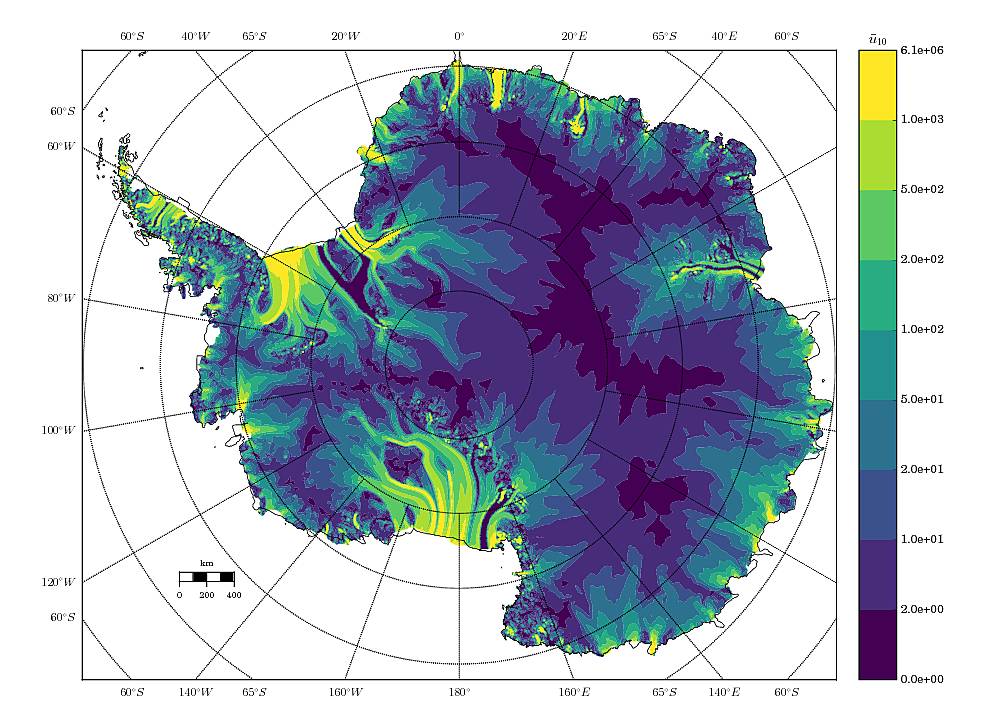}
  \caption{$\kappa = 10$, GLS.}
  \label{antarctica_bv_image_d_gS_m_U_kappa_10_GLS}
  \end{subfigure}
  \begin{subfigure}[b]{0.45\linewidth}
    \includegraphics[width=\linewidth]{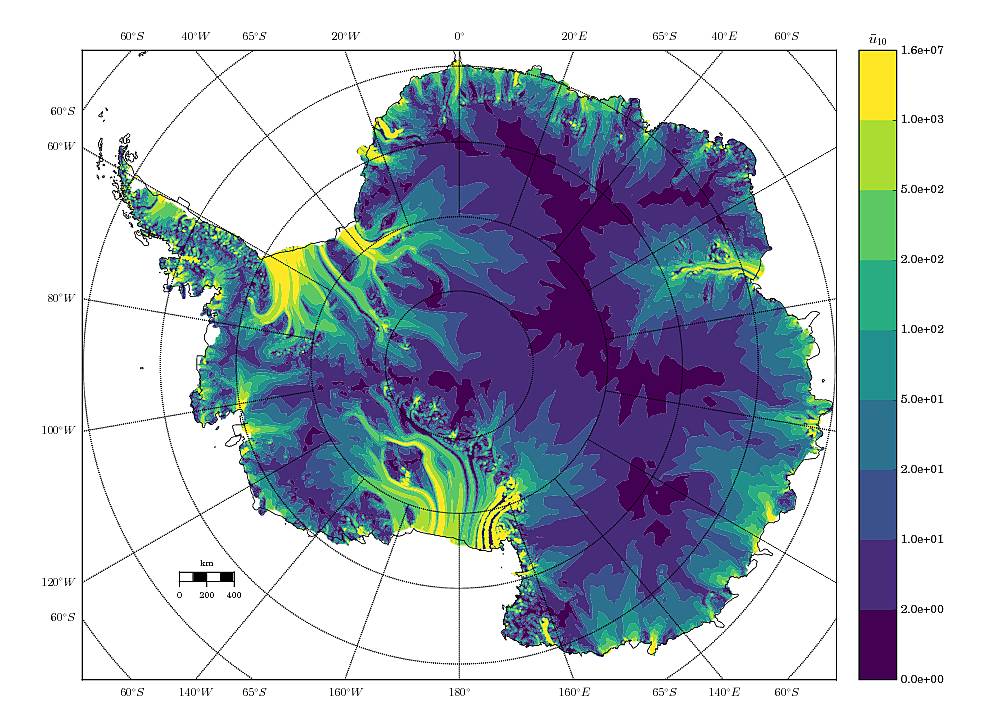}
  \caption{$\kappa = 10$, SUPG.}
  \label{antarctica_bv_image_d_gS_m_U_kappa_10_SUPG}
  \end{subfigure}
  
  \caption[Antarctica balance-velocity with $\mathbf{d}^{\text{data}} = -\nabla S$ where $\mathbf{u}_{ob}$ are missing.]{Balance velocity $\bar{u}$ derived over Antarctica with imposed direction of flow in the direction of surface observations $\mathbf{u}_{ob}$ and down the surface gradient $\nabla S$ where $\mathbf{u}_{ob}$ values are missing, where smoothing radius $\kappa$ varies as indicated.  The columns vary according to stabilization used; either Galerkin/least-squares (GLS) stabilization (\ref{bv_gls_operator}) or streamline-upwind/Petrov-Galerkin (SUPG) stabilization (\ref{bv_supg_operator}) in variational form (\ref{balance_velocity_weak_problem}).  Results using subgrid-scale-model stabilization (\ref{bv_ssm_operator}) (not shown) appeared more unstable than the (SUPG) method.}
  
  \label{antarctica_bv_image_d_gS_m_U}
  
\end{figure*}

Finally, simulations were run using flow directions $\mathbf{d}^{\text{data}}$ both in the down-surface-gradient direction (\ref{balance_velocity_direction}) and in the direction of surface observations $\mathbf{u}_{ob}$ (Figures \ref{greenland_u_ob_image} and \ref{antarctica_u_ob_image}).  Due to the fact that the surface gradient is nearly zero over the floating shelves of Antarctica, the flow direction resulting from the surface gradient is not well defined.  To investigate this problem, we applied the direction of flow to be in the direction of surface velocity observations $\mathbf{u}_{ob} = [u_{ob}\ v_{ob}]\T$ in these areas, with improved results (Figure \ref{antarctica_bv_image_U_ob_S}).  For comparison purposes, we then ran a simulation over Antarctica imposing the direction of flow entirely in the direction of observations $\mathbf{u}_{ob}$, and noticed that the results obtained without smoothing contained high error in regions without velocity observations (Figure \ref{antarctica_bv_image_d_U_ob}).  To remedy this, we ran one final test over Antarctica with direction of flow imposed in the direction of velocity observations $\mathbf{u}_{ob}$ where observations are present and down the surface gradient $\nabla S$ where they are not (Figure \ref{antarctica_bv_image_d_gS_m_U}).

\begin{figure*}
  
  \centering 

  \begin{subfigure}[b]{0.25\linewidth}
    \includegraphics[width=\linewidth]{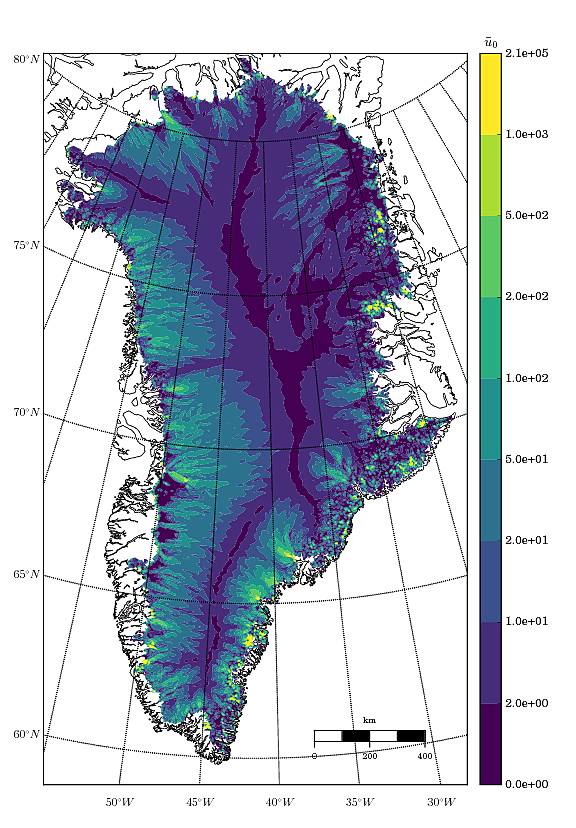}
  \caption{$\kappa = 0$, GLS.}
  \label{greenland_bv_image_kappa_0_GLS}
  \end{subfigure}
  \begin{subfigure}[b]{0.25\linewidth}
    \includegraphics[width=\linewidth]{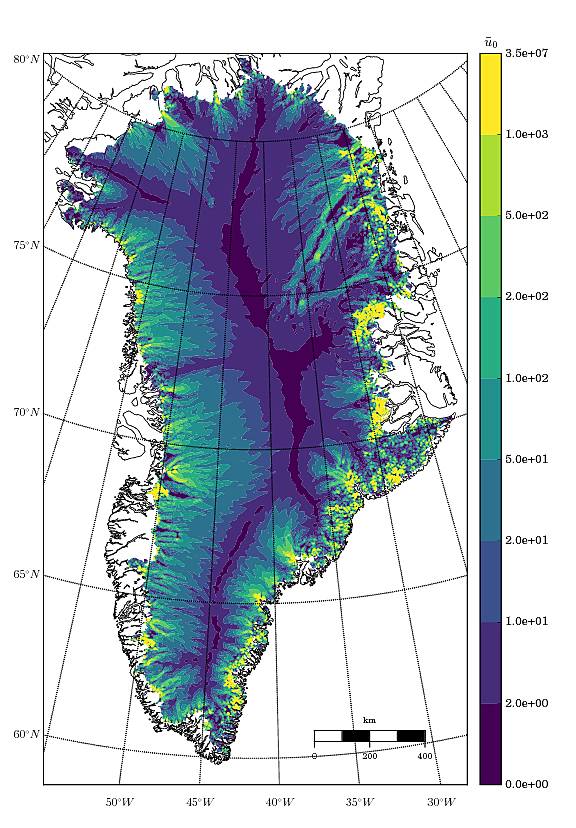}
  \caption{$\kappa = 0$, SUPG.}
  \label{greenland_bv_image_kappa_0_SUPG}
  \end{subfigure}
  \begin{subfigure}[b]{0.25\linewidth}
    \includegraphics[width=\linewidth]{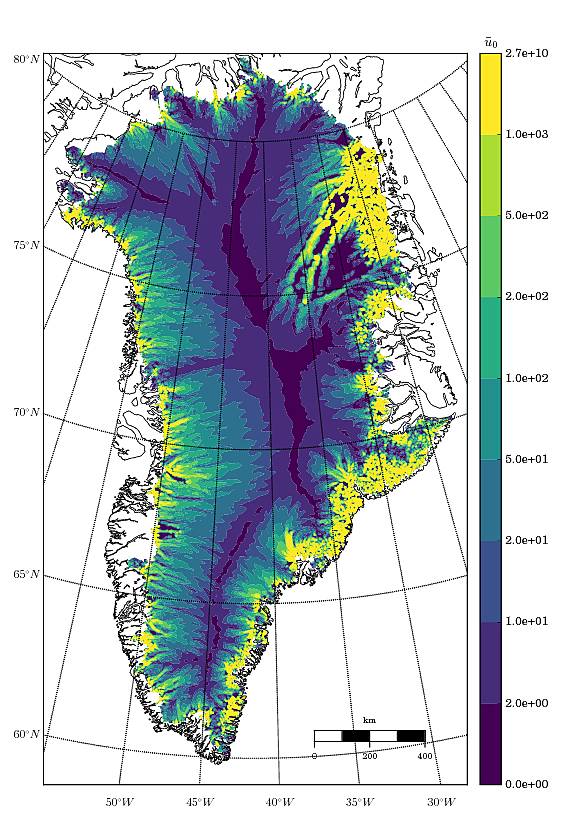}
  \caption{$\kappa = 0$, SSM.}
  \label{greenland_bv_image_kappa_5_SSM}
  \end{subfigure}

  \begin{subfigure}[b]{0.25\linewidth}
    \includegraphics[width=\linewidth]{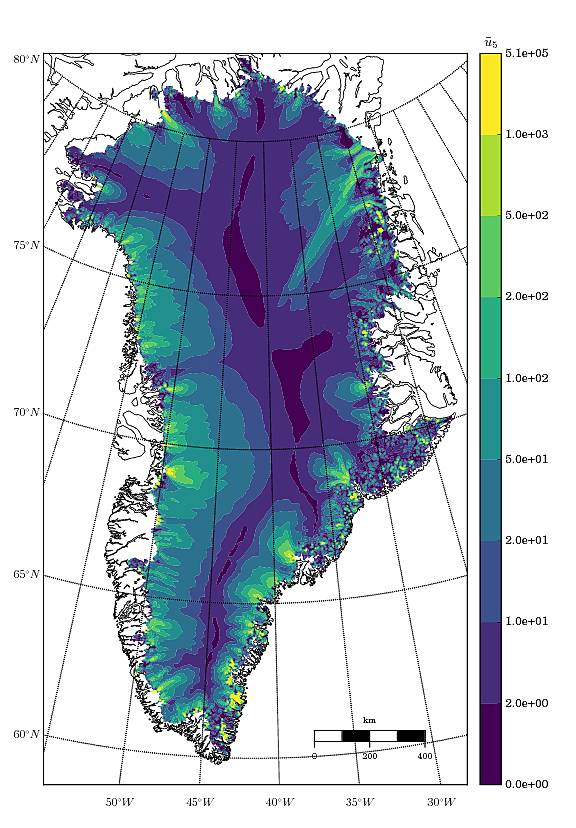}
  \caption{$\kappa = 5$, GLS.}
  \label{greenland_bv_image_kappa_5_GLS}
  \end{subfigure}
  \begin{subfigure}[b]{0.25\linewidth}
    \includegraphics[width=\linewidth]{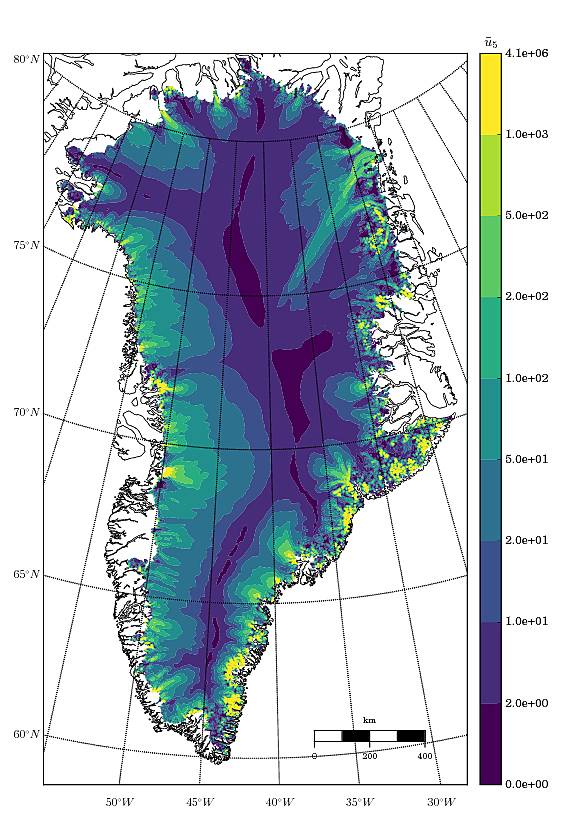}
  \caption{$\kappa = 5$, SUPG.}
  \label{greenland_bv_image_kappa_5_SUPG}
  \end{subfigure}
  \begin{subfigure}[b]{0.25\linewidth}
    \includegraphics[width=\linewidth]{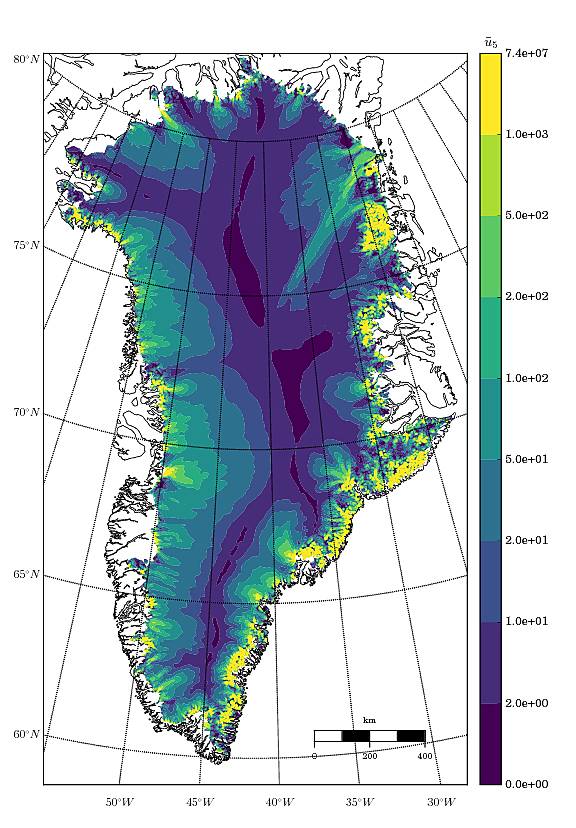}
  \caption{$\kappa = 5$, SSM.}
  \label{greenland_bv_image_kappa_5_SSM}
  \end{subfigure}

  \begin{subfigure}[b]{0.25\linewidth}
    \includegraphics[width=\linewidth]{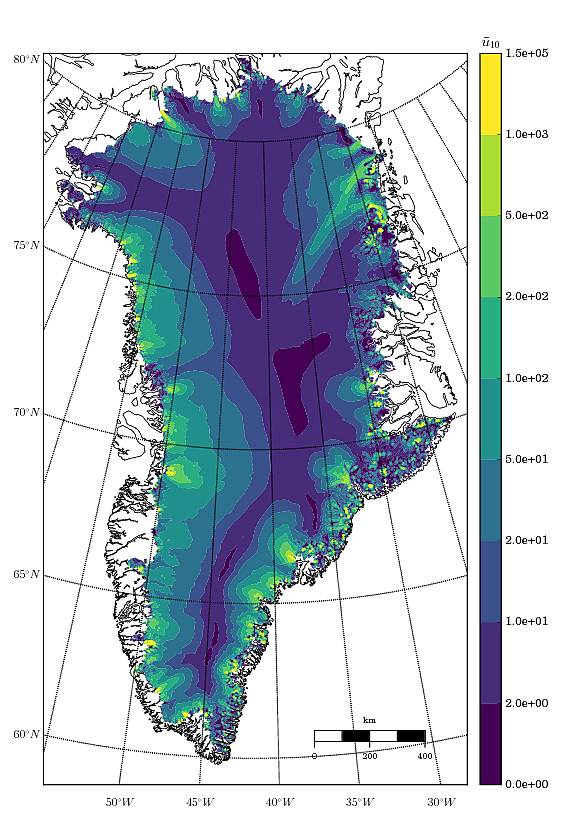}
  \caption{$\kappa = 10$, GLS.}
  \label{greenland_bv_image_kappa_10_GLS}
  \end{subfigure}
  \begin{subfigure}[b]{0.25\linewidth}
    \includegraphics[width=\linewidth]{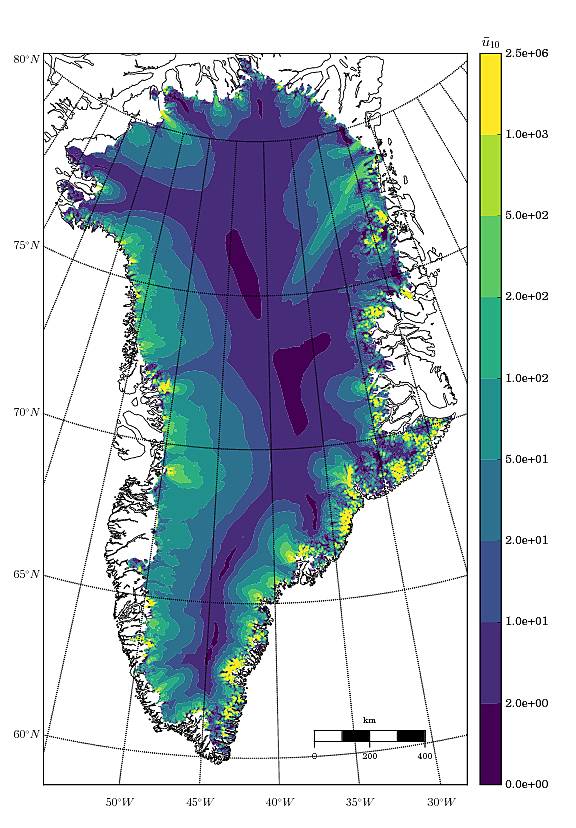}
  \caption{$\kappa = 10$, SUPG.}
  \label{greenland_bv_image_kappa_10_SUPG}
  \end{subfigure}
  \begin{subfigure}[b]{0.25\linewidth}
    \includegraphics[width=\linewidth]{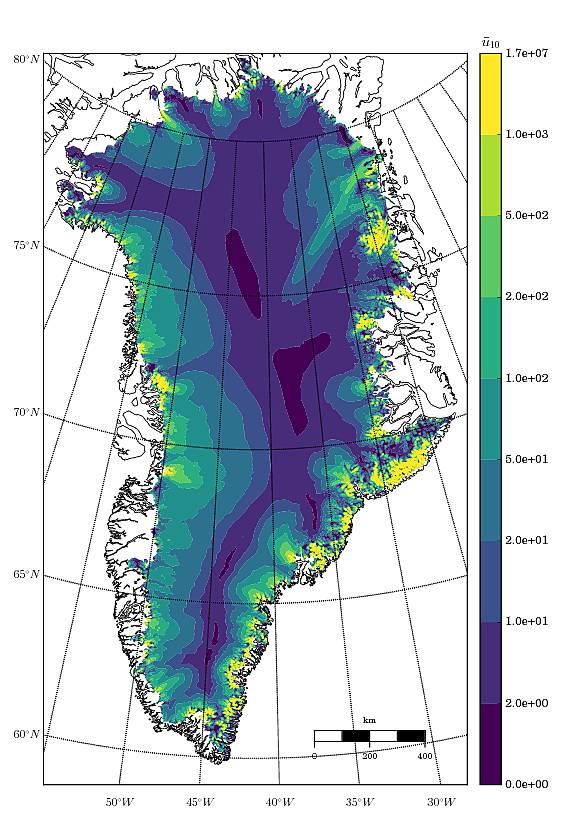}
  \caption{$\kappa = 10$, SSM.}
  \label{greenland_bv_image_kappa_10_SSM}
  \end{subfigure}
 
  \caption[Greenland balance-velocity with $\mathbf{d}^{\text{data}} = -\nabla S$.]{Balance velocity $\bar{u}$ derived over Greenland with direction of flow imposed down the surface gradient $\nabla S$, where smoothing radius $\kappa$ varies as indicated.  The columns vary according to stabilization used; either Galerkin/least-squares (GLS) stabilization (\ref{bv_gls_operator}), streamline-upwind/Petrov-Galerkin (SUPG) stabilization (\ref{bv_supg_operator}), or subgrid-scale-model (SSM) stabilization (\ref{bv_ssm_operator}) in variational form (\ref{balance_velocity_weak_problem}).}

  \label{greenland_bv_image}

\end{figure*}

\begin{figure*}
  
  \centering 

  \begin{subfigure}[b]{0.25\linewidth}
    \includegraphics[width=\linewidth]{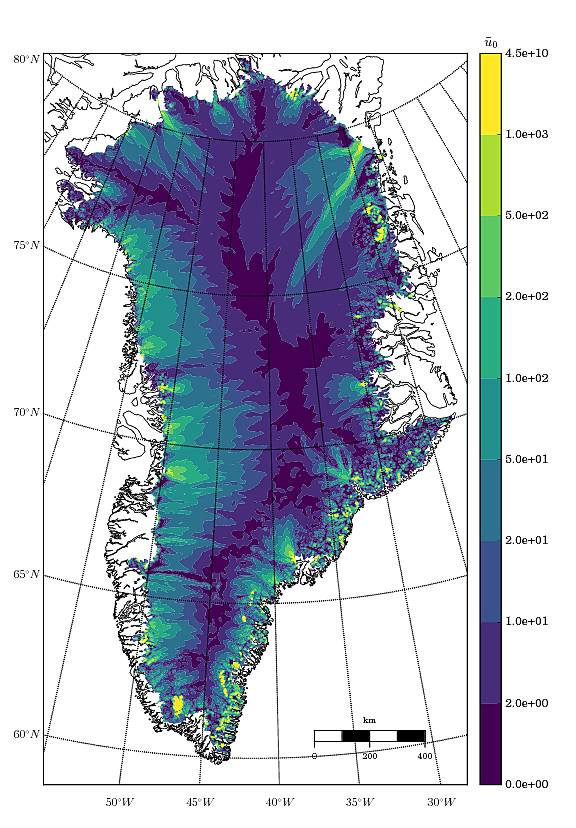}
  \caption{$\kappa = 0$, GLS.}
  \label{greenland_bv_image_d_U_ob_kappa_0_GLS}
  \end{subfigure}
  \begin{subfigure}[b]{0.25\linewidth}
    \includegraphics[width=\linewidth]{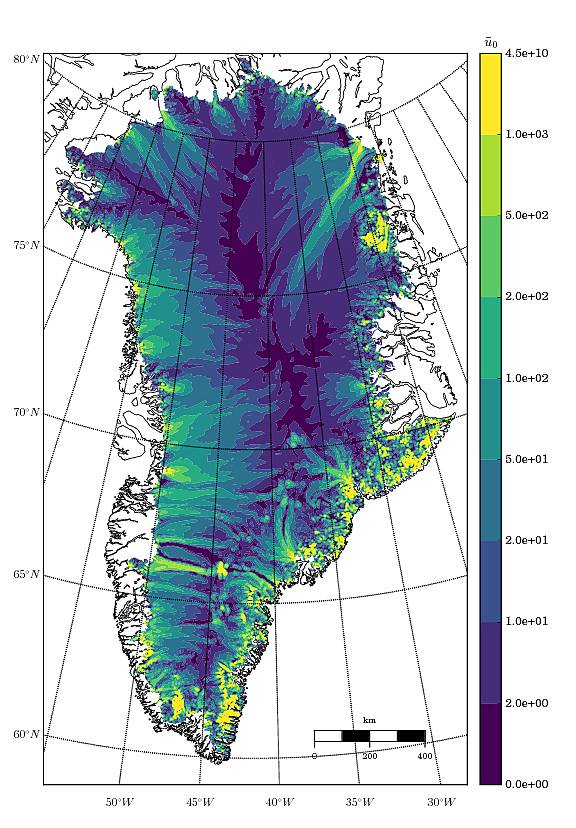}
  \caption{$\kappa = 0$, SUPG.}
  \label{greenland_bv_image_d_U_ob_kappa_0_SUPG}
  \end{subfigure}
  \begin{subfigure}[b]{0.25\linewidth}
    \includegraphics[width=\linewidth]{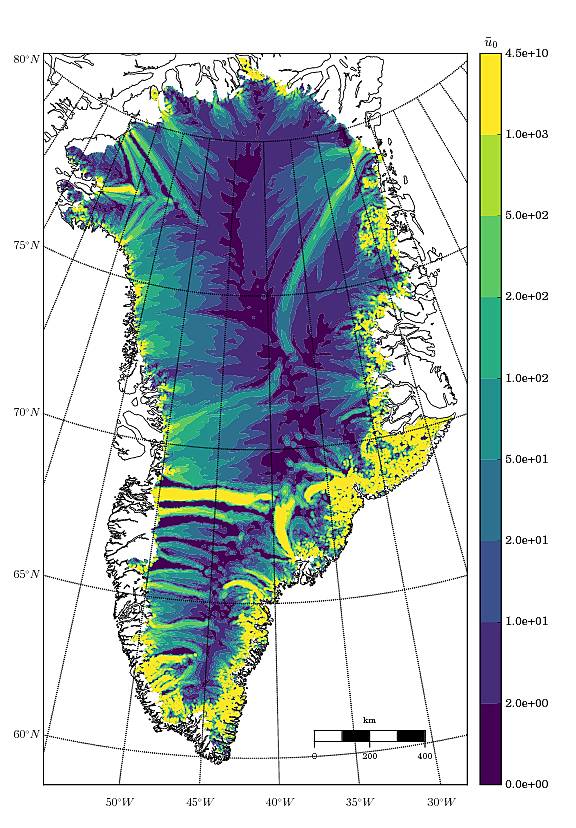}
  \caption{$\kappa = 0$, SSM.}
  \label{greenland_bv_image_d_U_ob_kappa_5_SSM}
  \end{subfigure}

  \begin{subfigure}[b]{0.25\linewidth}
    \includegraphics[width=\linewidth]{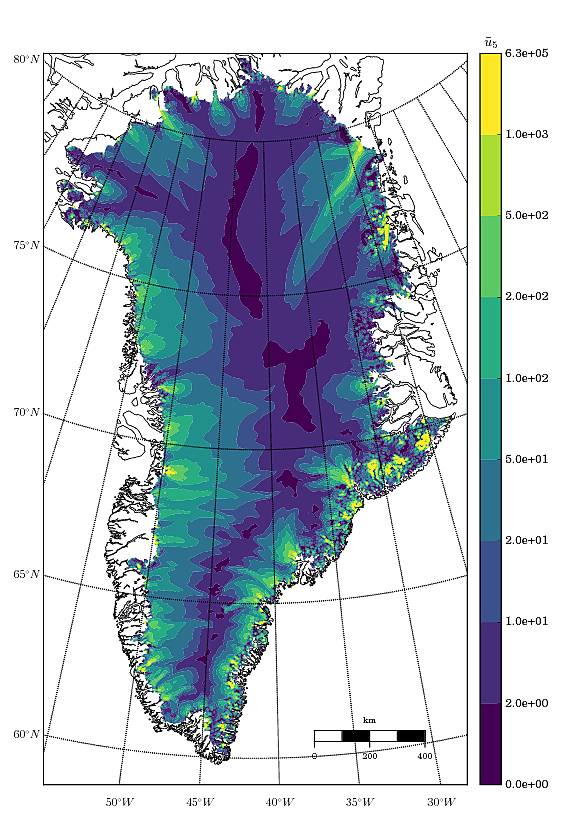}
  \caption{$\kappa = 5$, GLS.}
  \label{greenland_bv_image_d_U_ob_kappa_5_GLS}
  \end{subfigure}
  \begin{subfigure}[b]{0.25\linewidth}
    \includegraphics[width=\linewidth]{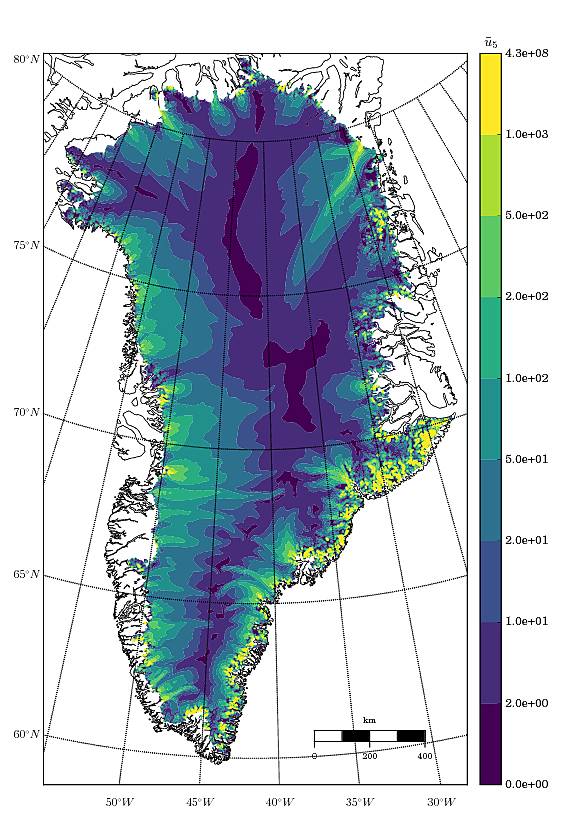}
  \caption{$\kappa = 5$, SUPG.}
  \label{greenland_bv_image_d_U_ob_kappa_5_SUPG}
  \end{subfigure}
  \begin{subfigure}[b]{0.25\linewidth}
    \includegraphics[width=\linewidth]{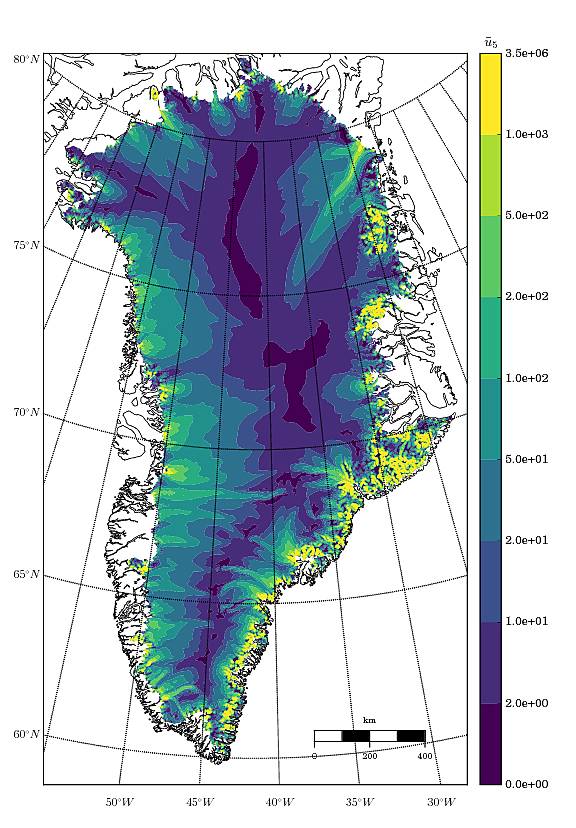}
  \caption{$\kappa = 5$, SSM.}
  \label{greenland_bv_image_d_U_ob_kappa_5_SSM}
  \end{subfigure}

  \begin{subfigure}[b]{0.25\linewidth}
    \includegraphics[width=\linewidth]{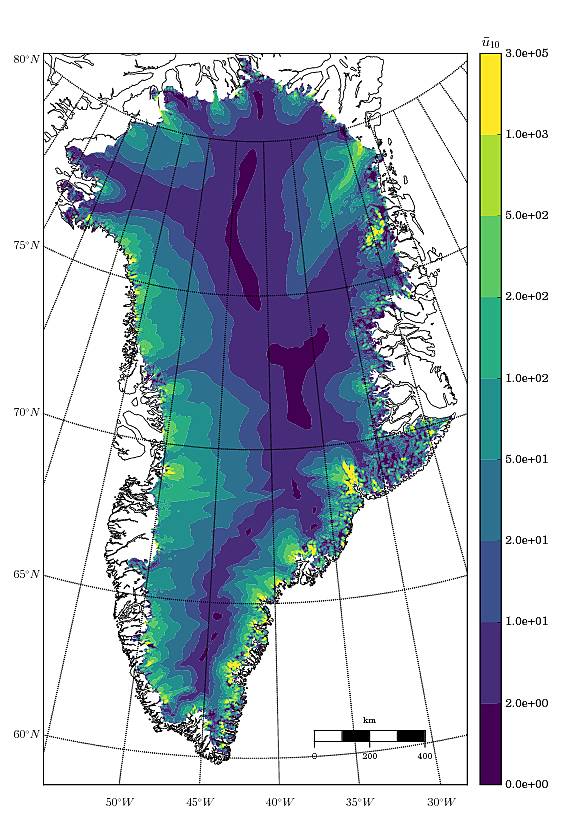}
  \caption{$\kappa = 10$, GLS.}
  \label{greenland_bv_image_d_U_ob_kappa_10_GLS}
  \end{subfigure}
  \begin{subfigure}[b]{0.25\linewidth}
    \includegraphics[width=\linewidth]{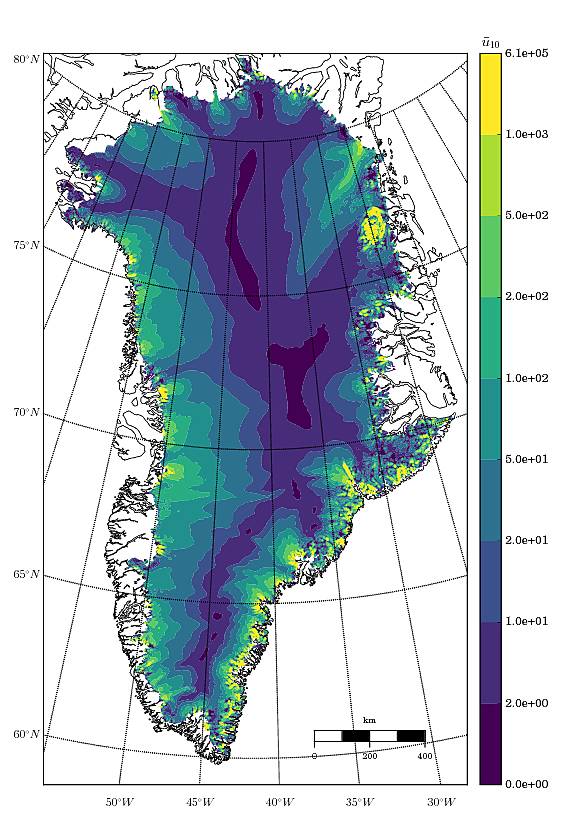}
  \caption{$\kappa = 10$, SUPG.}
  \label{greenland_bv_image_d_U_ob_kappa_10_SUPG}
  \end{subfigure}
  \begin{subfigure}[b]{0.25\linewidth}
    \includegraphics[width=\linewidth]{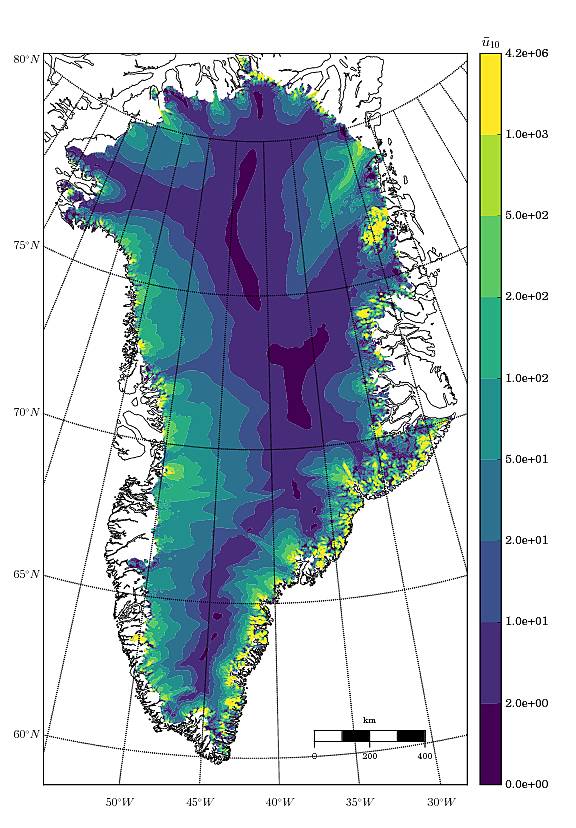}
  \caption{$\kappa = 10$, SSM.}
  \label{greenland_bv_image_d_U_ob_kappa_10_SSM}
  \end{subfigure}
 
  \caption[Greenland balance-velocity with $\mathbf{d}^{\text{data}} = \mathbf{u}_{ob}$.]{Balance velocity $\bar{u}$ derived over Greenland with direction of flow imposed in the direction of surface velocity observations $\mathbf{u}_{ob}$, where smoothing radius $\kappa$ varies as indicated.  The columns vary according to stabilization used; either Galerkin/least-squares (GLS) stabilization (\ref{bv_gls_operator}), streamline-upwind/Petrov-Galerkin (SUPG) stabilization (\ref{bv_supg_operator}), or subgrid-scale-model (SSM) stabilization (\ref{bv_ssm_operator}) in variational form (\ref{balance_velocity_weak_problem}).}

  \label{greenland_bv_image_d_U_ob}

\end{figure*}

\begin{figure*}
  
  \centering 

  \begin{subfigure}[b]{0.25\linewidth}
    \includegraphics[width=\linewidth]{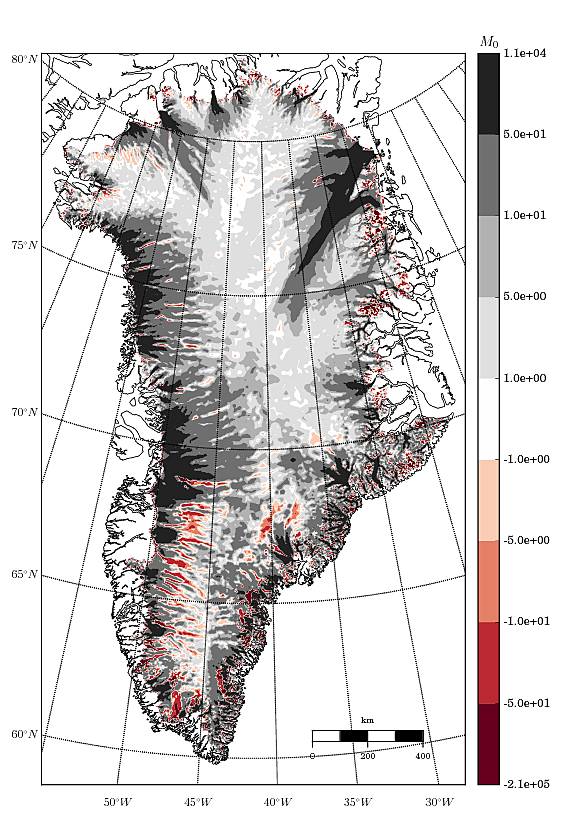}
  \caption{$\kappa = 0$, GLS.}
  \label{greenland_bv_image_kappa_0_GLS_misfit}
  \end{subfigure}
  \begin{subfigure}[b]{0.25\linewidth}
    \includegraphics[width=\linewidth]{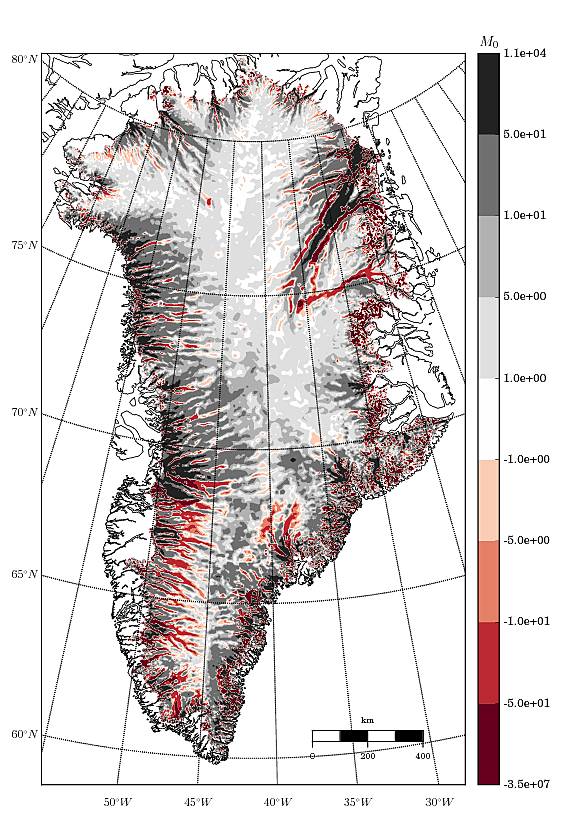}
  \caption{$\kappa = 0$, SUPG.}
  \label{greenland_bv_image_kappa_0_SUPG_misfit}
  \end{subfigure}
  \begin{subfigure}[b]{0.25\linewidth}
    \includegraphics[width=\linewidth]{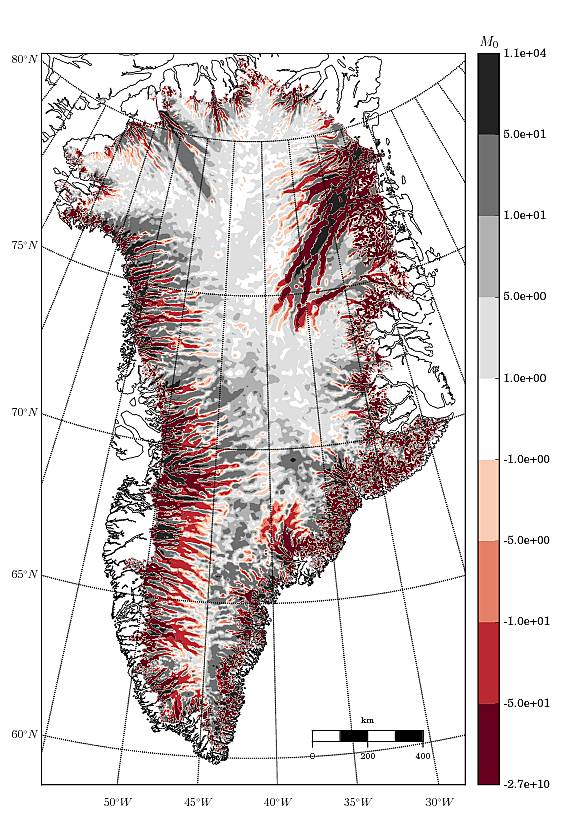}
  \caption{$\kappa = 0$, SSM.}
  \label{greenland_bv_image_kappa_5_SSM_misfit}
  \end{subfigure}

  \begin{subfigure}[b]{0.25\linewidth}
    \includegraphics[width=\linewidth]{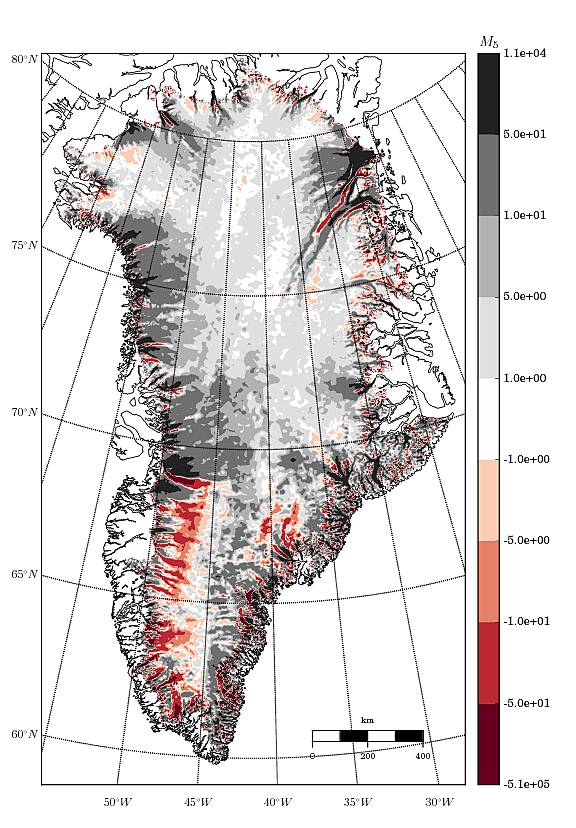}
  \caption{$\kappa = 5$, GLS.}
  \label{greenland_bv_image_kappa_5_GLS_misfit}
  \end{subfigure}
  \begin{subfigure}[b]{0.25\linewidth}
    \includegraphics[width=\linewidth]{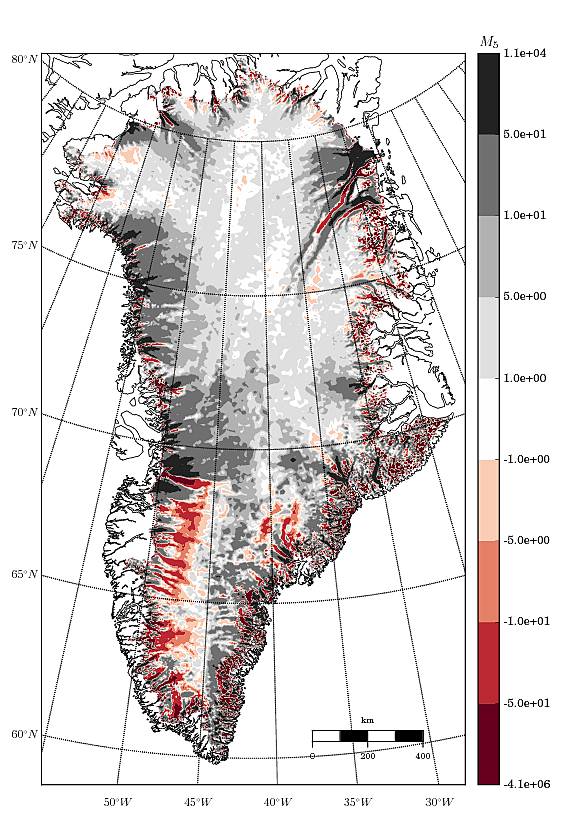}
  \caption{$\kappa = 5$, SUPG.}
  \label{greenland_bv_image_kappa_5_SUPG_misfit}
  \end{subfigure}
  \begin{subfigure}[b]{0.25\linewidth}
    \includegraphics[width=\linewidth]{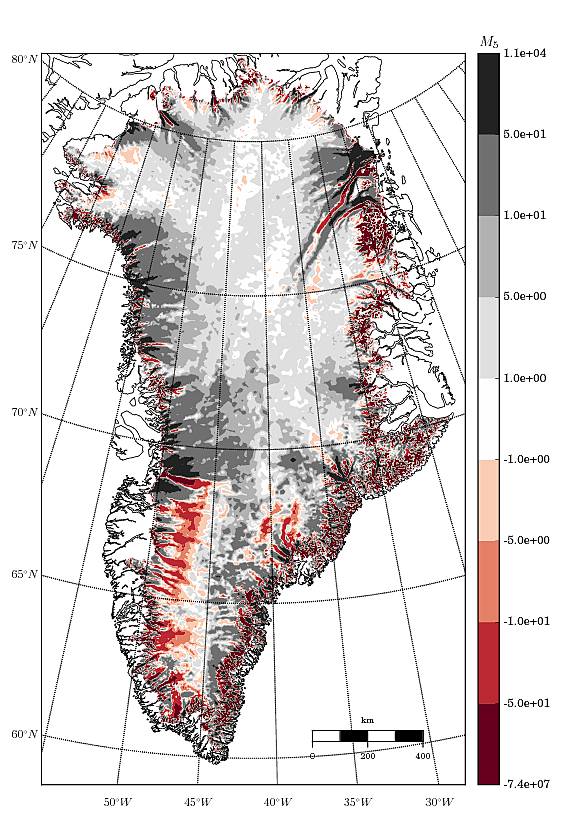}
  \caption{$\kappa = 5$, SSM.}
  \label{greenland_bv_image_kappa_5_SSM_misfit}
  \end{subfigure}

  \begin{subfigure}[b]{0.25\linewidth}
    \includegraphics[width=\linewidth]{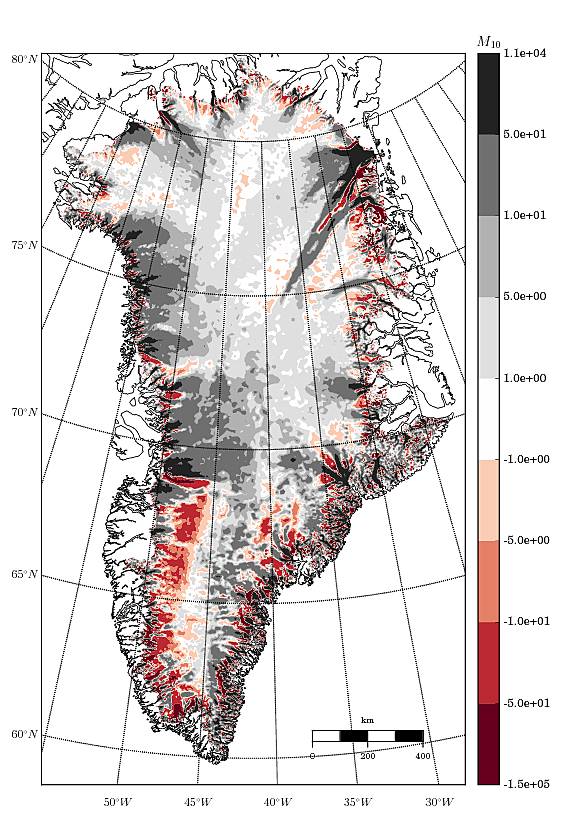}
  \caption{$\kappa = 10$, GLS.}
  \label{greenland_bv_image_kappa_10_GLS_misfit}
  \end{subfigure}
  \begin{subfigure}[b]{0.25\linewidth}
    \includegraphics[width=\linewidth]{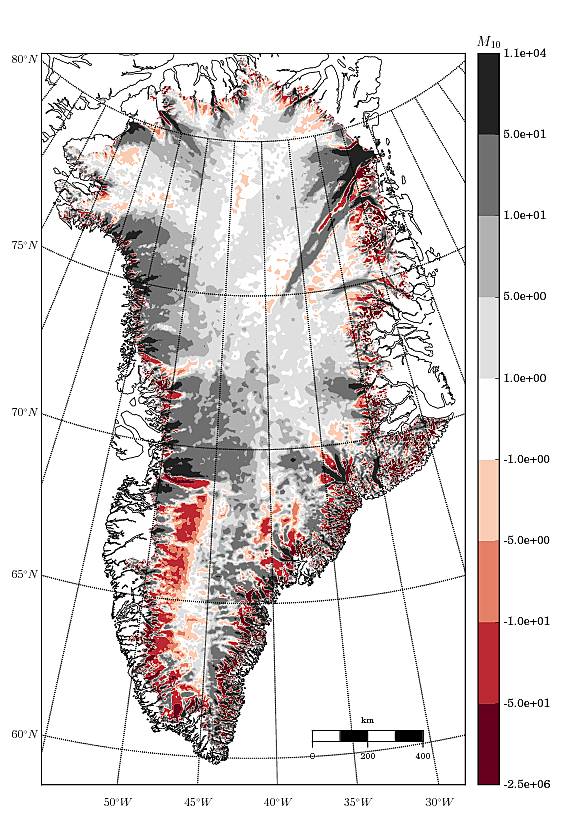}
  \caption{$\kappa = 10$, SUPG.}
  \label{greenland_bv_image_kappa_10_SUPG_misfit}
  \end{subfigure}
  \begin{subfigure}[b]{0.25\linewidth}
    \includegraphics[width=\linewidth]{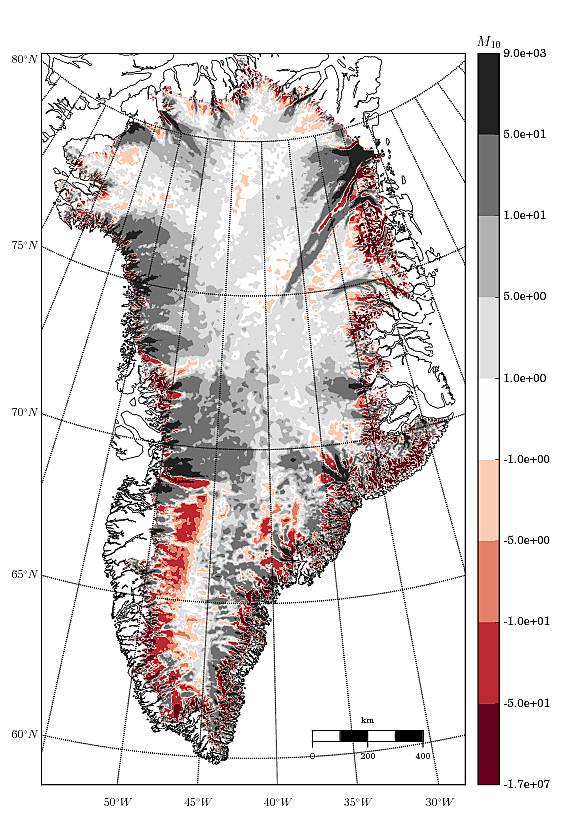}
  \caption{$\kappa = 10$, SSM.}
  \label{greenland_bv_image_kappa_10_SSM_misfit}
  \end{subfigure}
 
  \caption[Greenland balance-velocity misfit with $\mathbf{d}^{\text{data}} = -\nabla S$.]{Difference $\Vert \mathbf{u}_{ob} \Vert - \bar{u}$ between balance velocity $\bar{u}$ and the magnitude of the observed surface velocity $\mathbf{u}_{ob}$ derived over Greenland with imposed direction of flow down the surface gradient $\nabla S$, where smoothing radius $\kappa$ varies as indicated.  The columns vary according to stabilization used; either Galerkin/least-squares (GLS) stabilization (\ref{bv_gls_operator}), streamline-upwind/Petrov-Galerkin (SUPG) stabilization (\ref{bv_supg_operator}), or subgrid-scale-model (SSM) stabilization (\ref{bv_ssm_operator}) in variational form (\ref{balance_velocity_weak_problem}).}

  \label{greenland_bv_image_misfit}

\end{figure*}

\begin{figure*}
  
  \centering 

  \begin{subfigure}[b]{0.25\linewidth}
    \includegraphics[width=\linewidth]{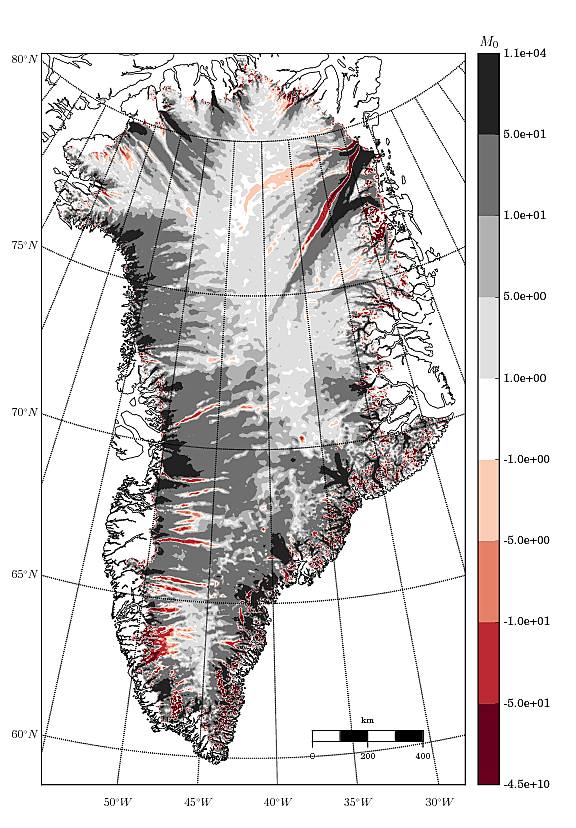}
  \caption{$\kappa = 0$, GLS.}
  \label{greenland_bv_image_d_U_ob_kappa_0_GLS_misfit}
  \end{subfigure}
  \begin{subfigure}[b]{0.25\linewidth}
    \includegraphics[width=\linewidth]{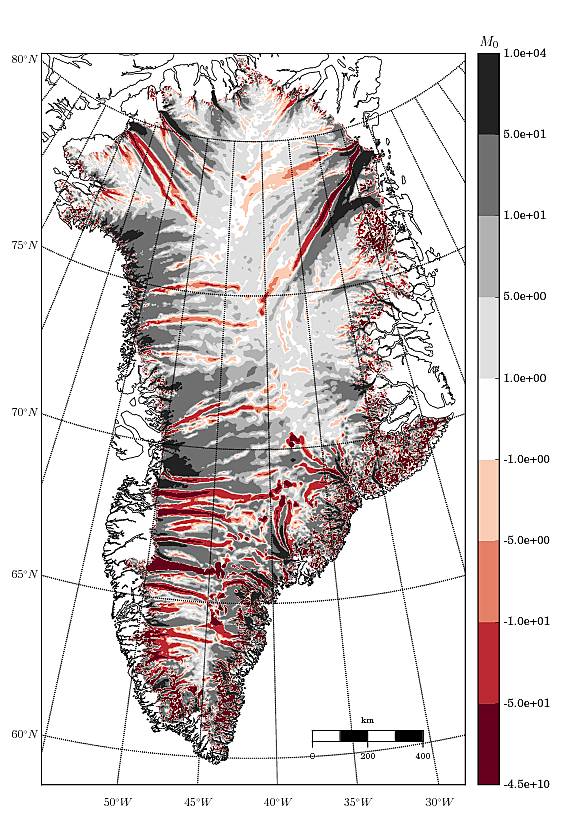}
  \caption{$\kappa = 0$, SUPG.}
  \label{greenland_bv_image_d_U_ob_kappa_0_SUPG_misfit}
  \end{subfigure}
  \begin{subfigure}[b]{0.25\linewidth}
    \includegraphics[width=\linewidth]{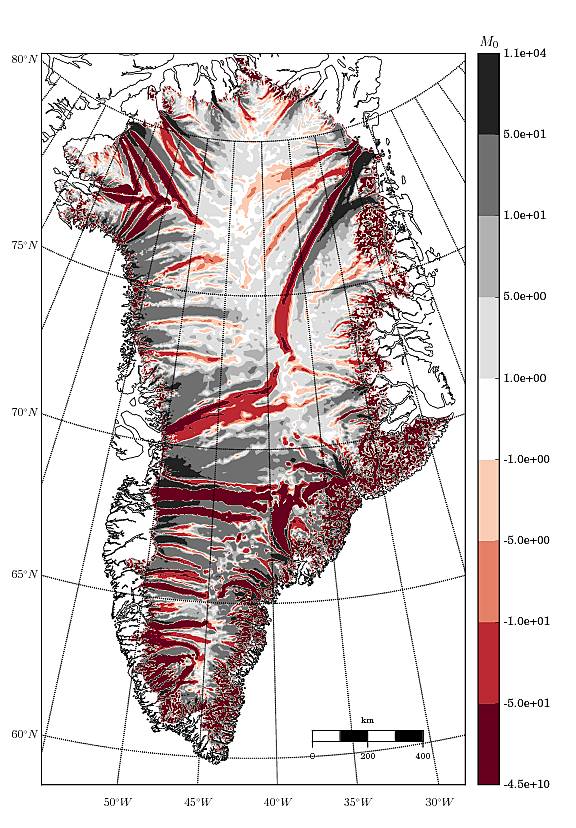}
  \caption{$\kappa = 0$, SSM.}
  \label{greenland_bv_image_d_U_ob_kappa_5_SSM_misfit}
  \end{subfigure}

  \begin{subfigure}[b]{0.25\linewidth}
    \includegraphics[width=\linewidth]{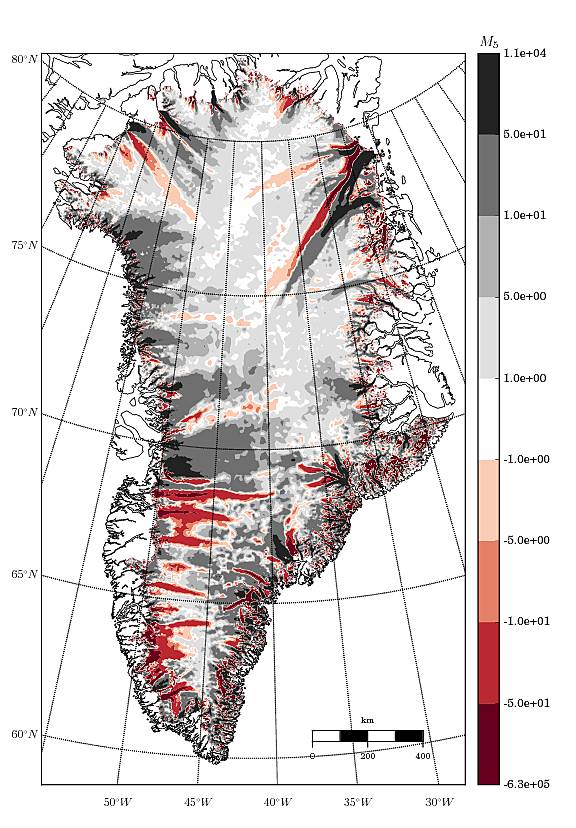}
  \caption{$\kappa = 5$, GLS.}
  \label{greenland_bv_image_d_U_ob_kappa_5_GLS_misfit}
  \end{subfigure}
  \begin{subfigure}[b]{0.25\linewidth}
    \includegraphics[width=\linewidth]{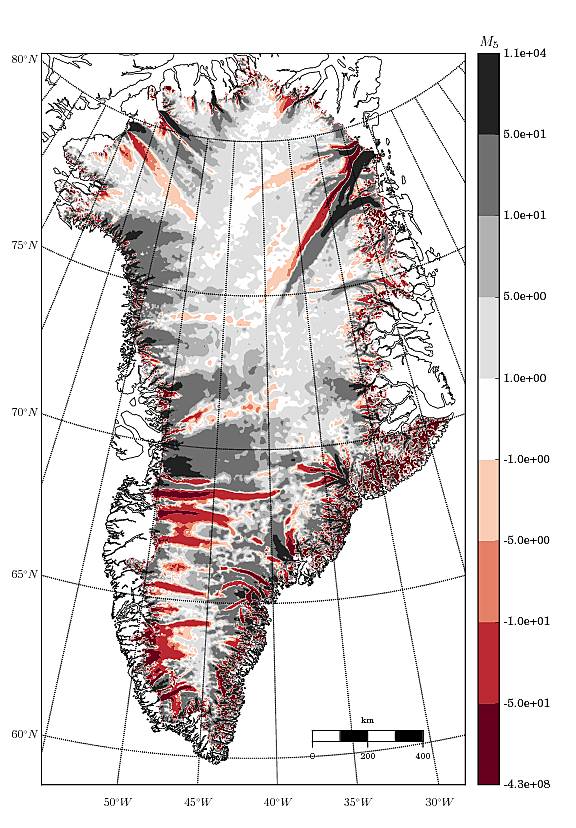}
  \caption{$\kappa = 5$, SUPG.}
  \label{greenland_bv_image_d_U_ob_kappa_5_SUPG_misfit}
  \end{subfigure}
  \begin{subfigure}[b]{0.25\linewidth}
    \includegraphics[width=\linewidth]{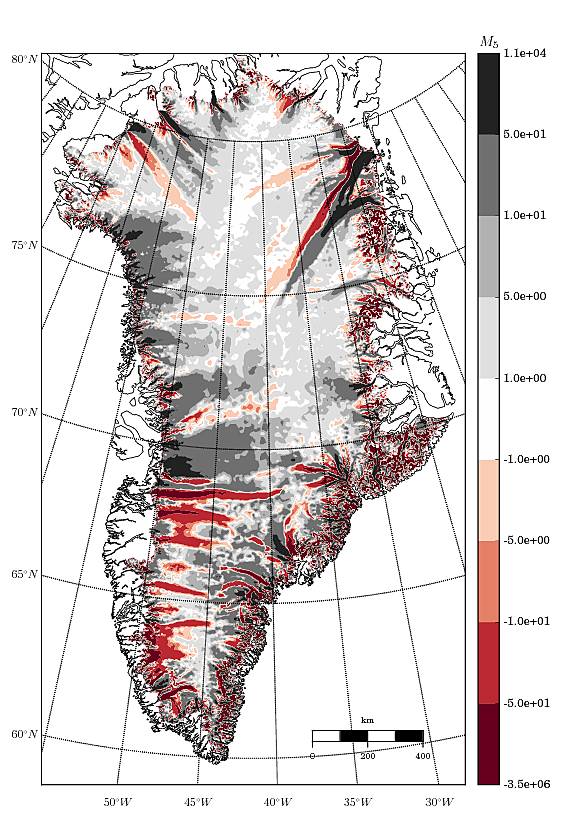}
  \caption{$\kappa = 5$, SSM.}
  \label{greenland_bv_image_d_U_ob_kappa_5_SSM_misfit}
  \end{subfigure}

  \begin{subfigure}[b]{0.25\linewidth}
    \includegraphics[width=\linewidth]{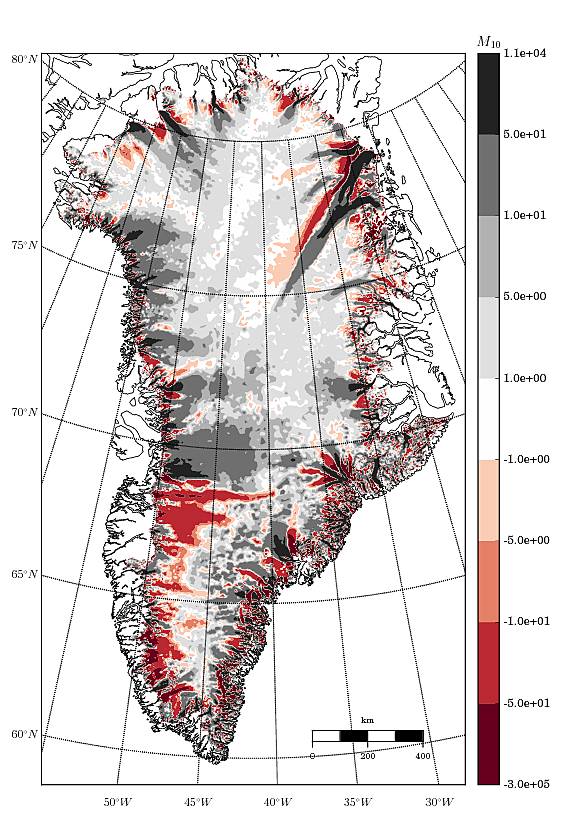}
  \caption{$\kappa = 10$, GLS.}
  \label{greenland_bv_image_d_U_ob_kappa_10_GLS_misfit}
  \end{subfigure}
  \begin{subfigure}[b]{0.25\linewidth}
    \includegraphics[width=\linewidth]{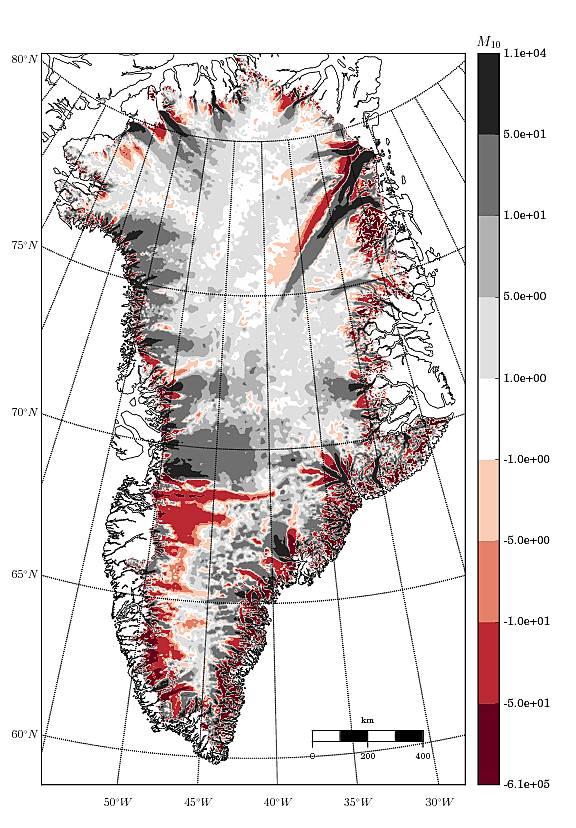}
  \caption{$\kappa = 10$, SUPG.}
  \label{greenland_bv_image_d_U_ob_kappa_10_SUPG_misfit}
  \end{subfigure}
  \begin{subfigure}[b]{0.25\linewidth}
    \includegraphics[width=\linewidth]{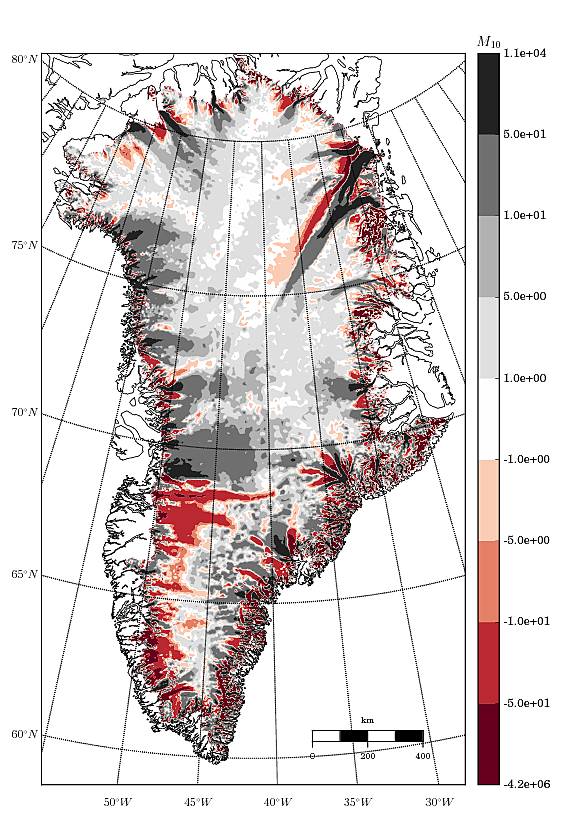}
  \caption{$\kappa = 10$, SSM.}
  \label{greenland_bv_image_d_U_ob_kappa_10_SSM_misfit}
  \end{subfigure}
 
  \caption[Greenland balance-velocity misfit with $\mathbf{d}^{\text{data}} = \mathbf{u}_{ob}$.]{Difference $\Vert \mathbf{u}_{ob} \Vert - \bar{u}$ between balance velocity $\bar{u}$ and the magnitude of the observed surface velocity $\mathbf{u}_{ob}$ derived over Greenland with imposed direction of flow in the direction of surface velocity observations $\mathbf{u}_{ob}$, where smoothing radius $\kappa$ varies as indicated.  The columns vary according to stabilization used; either Galerkin/least-squares (GLS) stabilization (\ref{bv_gls_operator}), streamline-upwind/Petrov-Galerkin (SUPG) stabilization (\ref{bv_supg_operator}), or subgrid-scale-model (SSM) stabilization (\ref{bv_ssm_operator}) in variational form (\ref{balance_velocity_weak_problem}).}

  \label{greenland_bv_image_d_U_ob_misfit}

\end{figure*}

\begin{figure*}

  \centering

  \begin{subfigure}[b]{0.45\linewidth}
    \includegraphics[width=\linewidth]{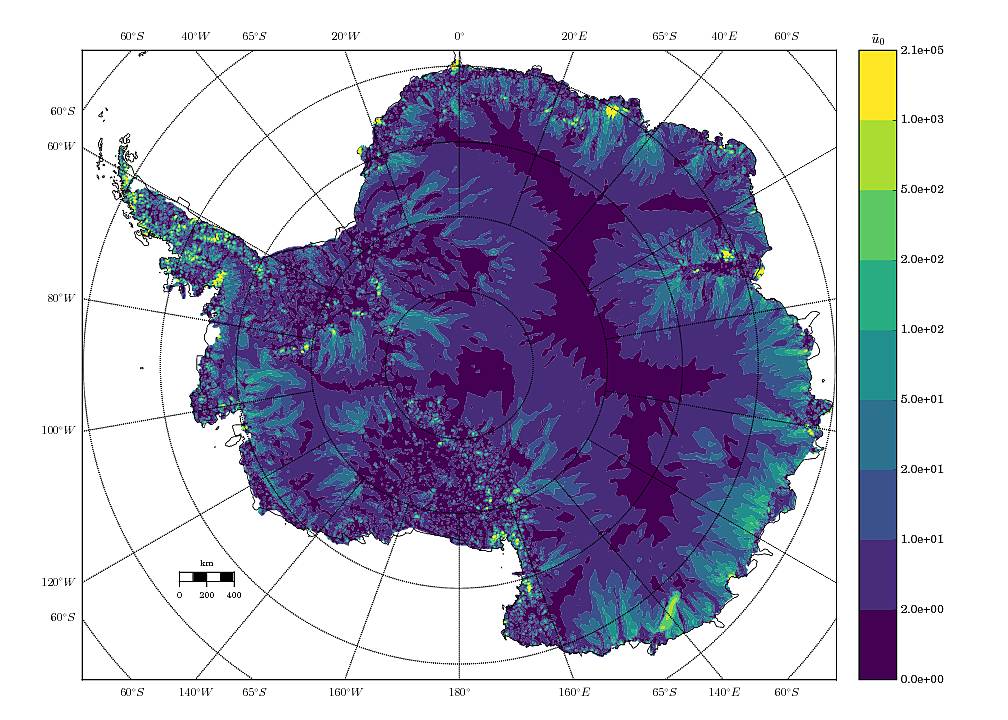}
  \caption{$\kappa = 0$, GLS.}
  \label{antarctica_bv_image_kappa_0_GLS}
  \end{subfigure}
  \begin{subfigure}[b]{0.45\linewidth}
    \includegraphics[width=\linewidth]{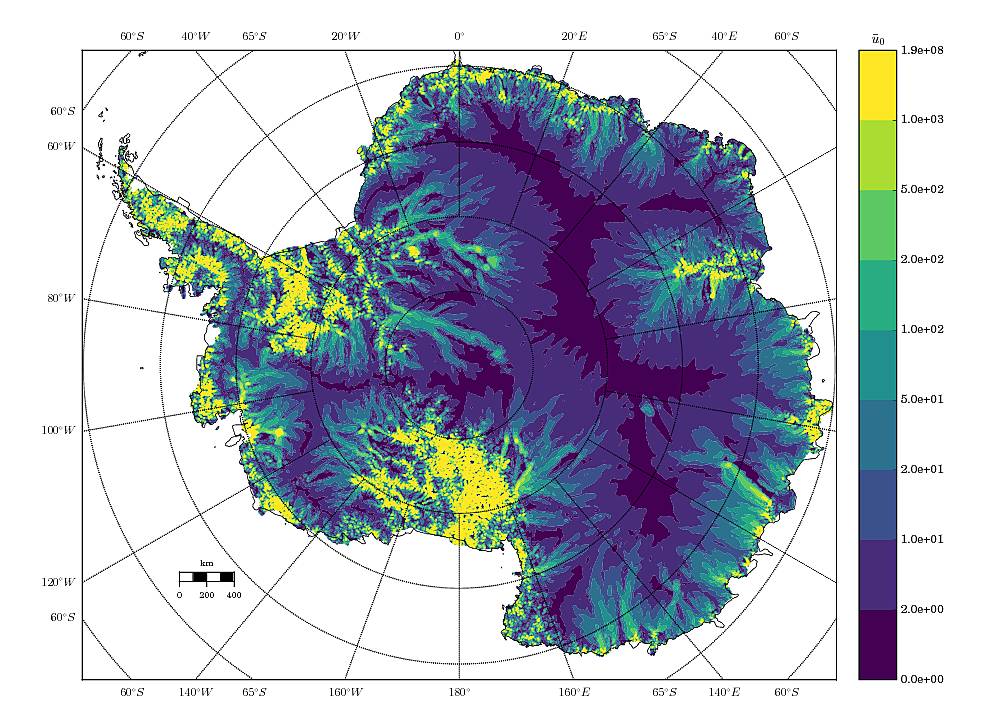}
  \caption{$\kappa = 0$, SUPG.}
  \label{antarctica_bv_image_kappa_0_SUPG}
  \end{subfigure}

  \begin{subfigure}[b]{0.45\linewidth}
    \includegraphics[width=\linewidth]{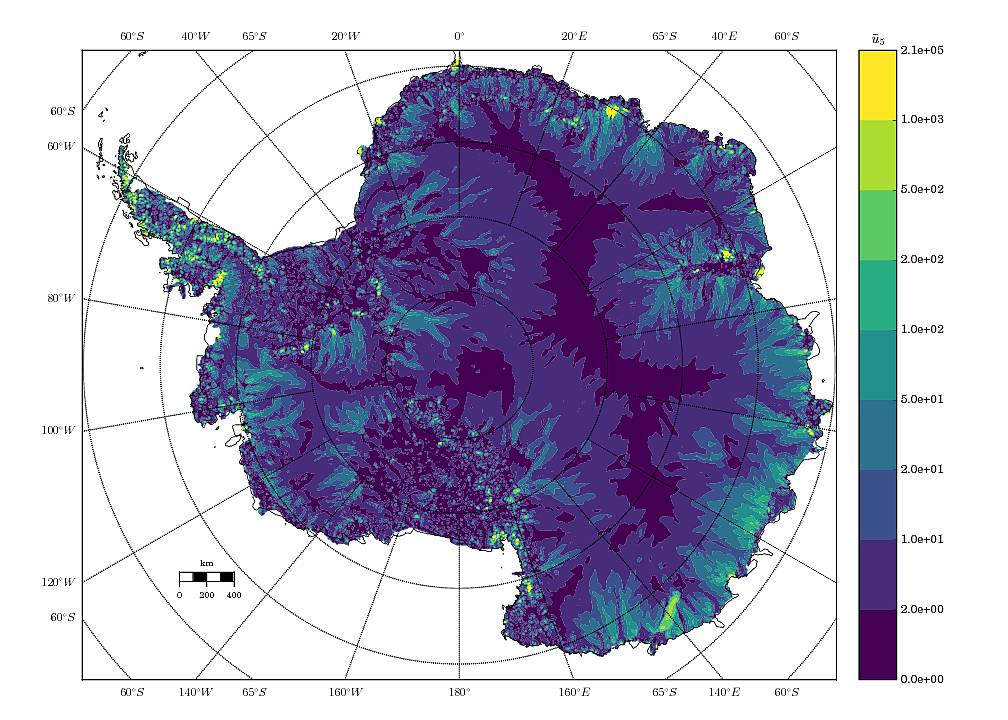}
  \caption{$\kappa = 5$, GLS.}
  \label{antarctica_bv_image_kappa_5_GLS}
  \end{subfigure}
  \begin{subfigure}[b]{0.45\linewidth}
    \includegraphics[width=\linewidth]{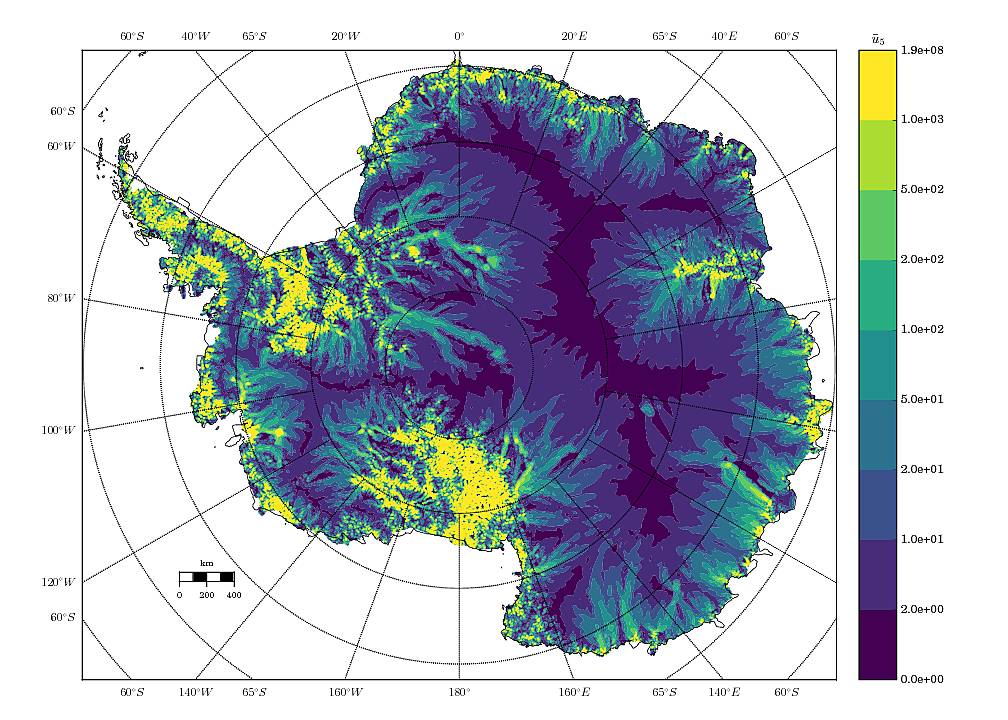}
  \caption{$\kappa = 5$, SUPG.}
  \label{antarctica_bv_image_kappa_5_SUPG}
  \end{subfigure}

  \begin{subfigure}[b]{0.45\linewidth}
    \includegraphics[width=\linewidth]{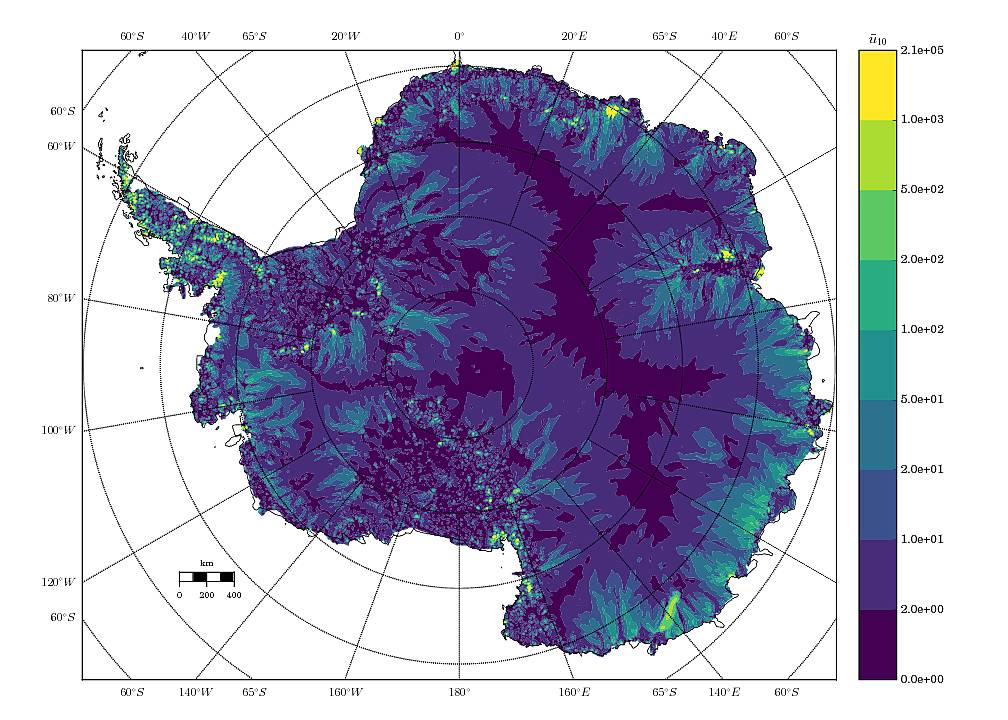}
  \caption{$\kappa = 10$, GLS.}
  \label{antarctica_bv_image_kappa_10_GLS}
  \end{subfigure}
  \begin{subfigure}[b]{0.45\linewidth}
    \includegraphics[width=\linewidth]{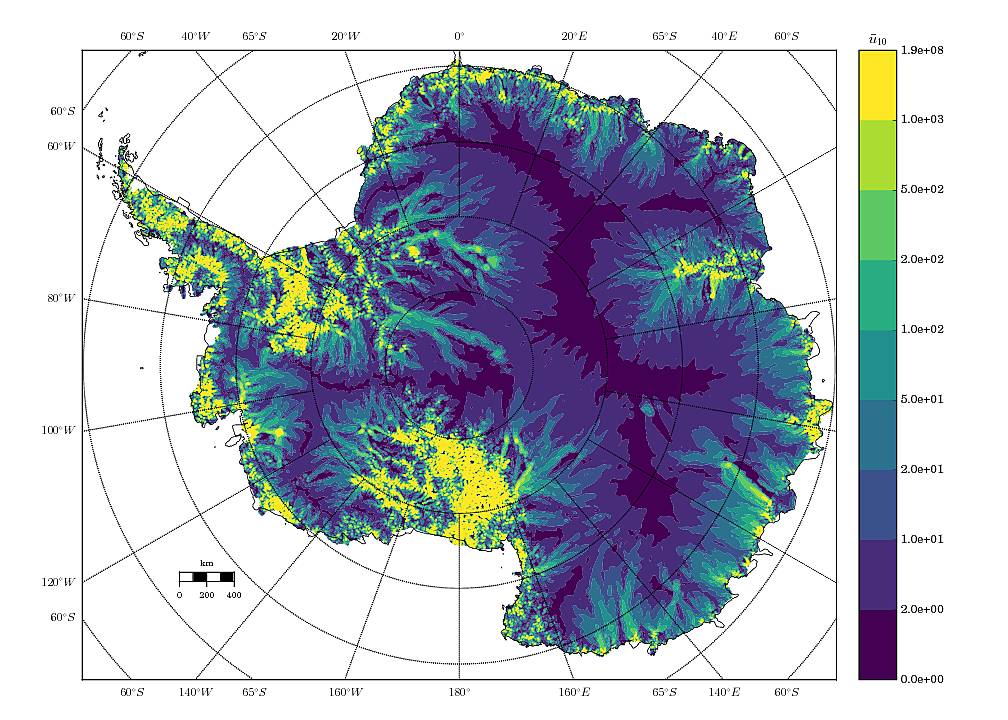}
  \caption{$\kappa = 10$, SUPG.}
  \label{antarctica_bv_image_kappa_10_SUPG}
  \end{subfigure}
  
  \caption[Antarctica balance-velocity with $\mathbf{d}^{\text{data}} = -\nabla S$.]{Balance velocity $\bar{u}$ derived over Antarctica with imposed direction of flow down the surface gradient $\nabla S$, where smoothing radius $\kappa$ varies as indicated.  The columns vary according to stabilization used; either Galerkin/least-squares (GLS) stabilization (\ref{bv_gls_operator}) or streamline-upwind/Petrov-Galerkin (SUPG) stabilization (\ref{bv_supg_operator}) in variational form (\ref{balance_velocity_weak_problem}).  Results using subgrid-scale-model stabilization (\ref{bv_ssm_operator}) (not shown) appeared more unstable than the (SUPG) method.}

  \label{antarctica_bv_image}
  
\end{figure*}

\begin{figure*}

  \centering

  \begin{subfigure}[b]{0.45\linewidth}
    \includegraphics[width=\linewidth]{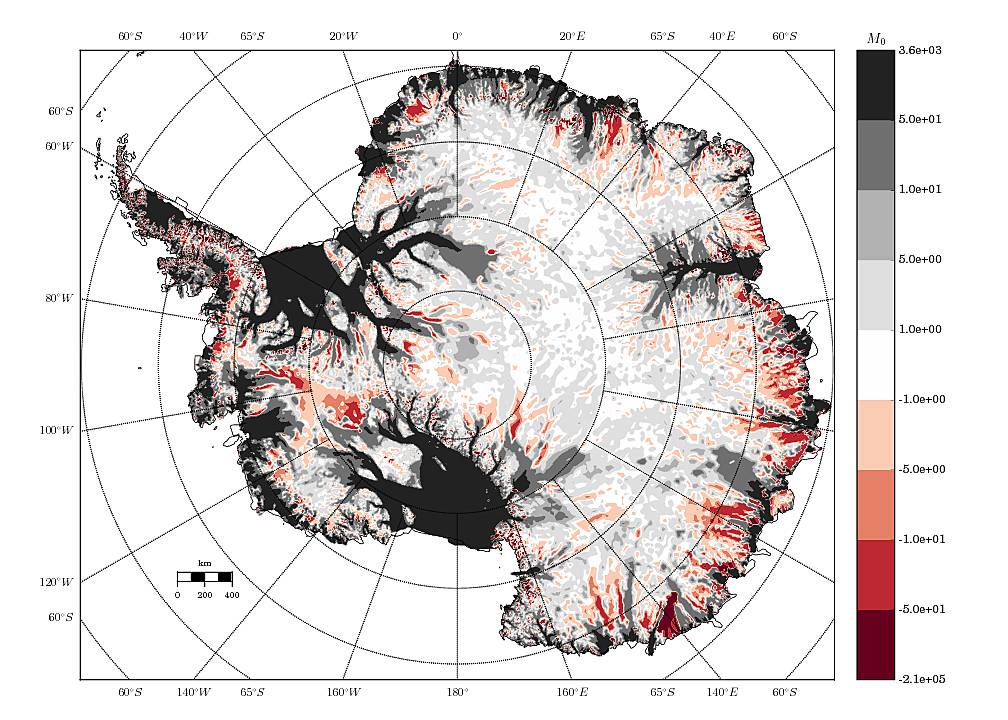}
  \caption{$\kappa = 0$, GLS.}
  \label{antarctica_bv_image_kappa_0_GLS_misfit}
  \end{subfigure}
  \begin{subfigure}[b]{0.45\linewidth}
    \includegraphics[width=\linewidth]{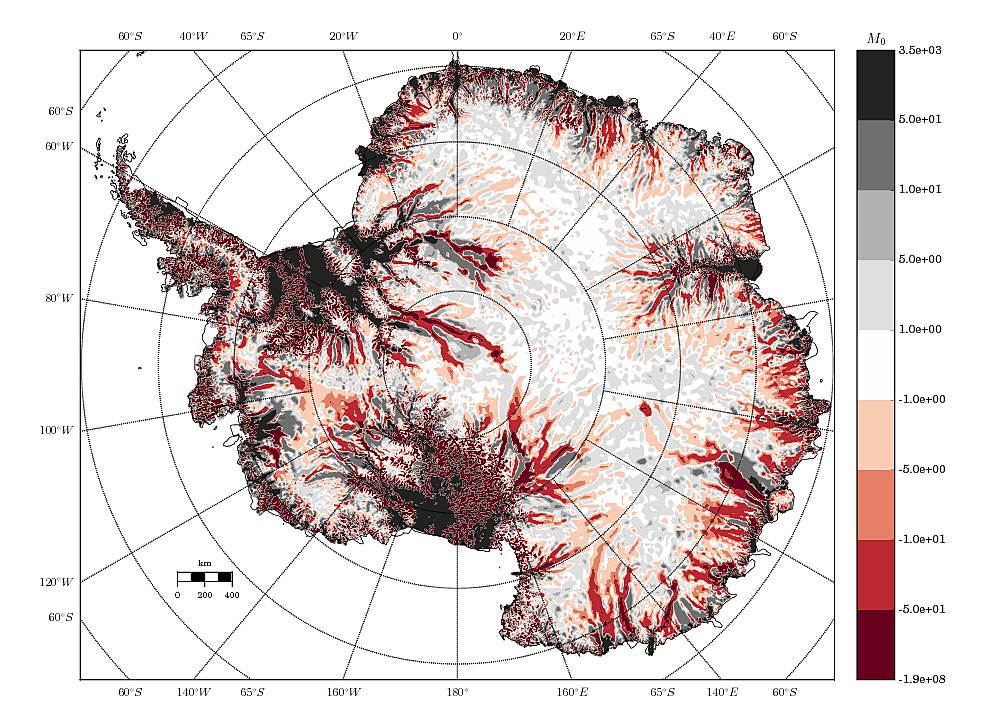}
  \caption{$\kappa = 0$, SUPG.}
  \label{antarctica_bv_image_kappa_0_SUPG_misfit}
  \end{subfigure}

  \begin{subfigure}[b]{0.45\linewidth}
    \includegraphics[width=\linewidth]{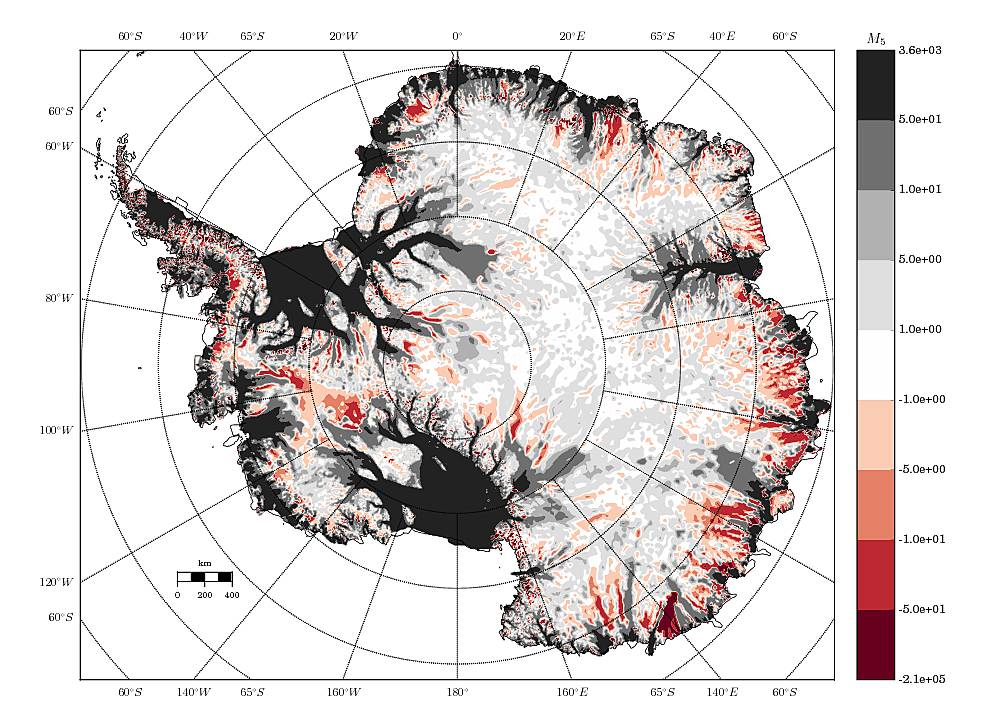}
  \caption{$\kappa = 5$, GLS.}
  \label{antarctica_bv_image_kappa_5_GLS_misfit}
  \end{subfigure}
  \begin{subfigure}[b]{0.45\linewidth}
    \includegraphics[width=\linewidth]{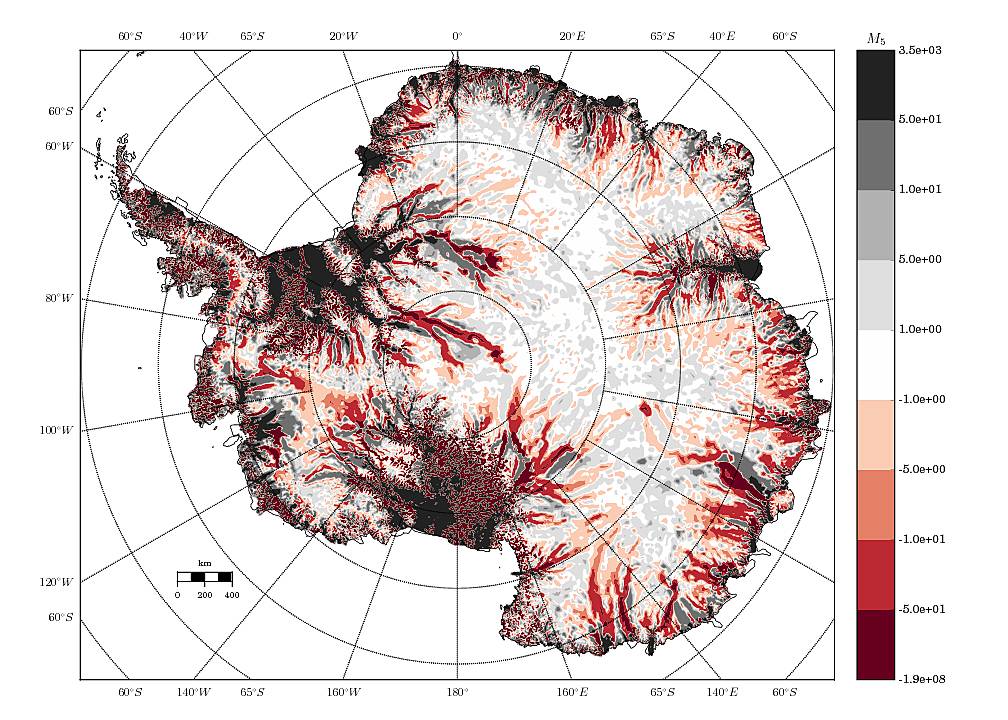}
  \caption{$\kappa = 5$, SUPG.}
  \label{antarctica_bv_image_kappa_5_SUPG_misfit}
  \end{subfigure}

  \begin{subfigure}[b]{0.45\linewidth}
    \includegraphics[width=\linewidth]{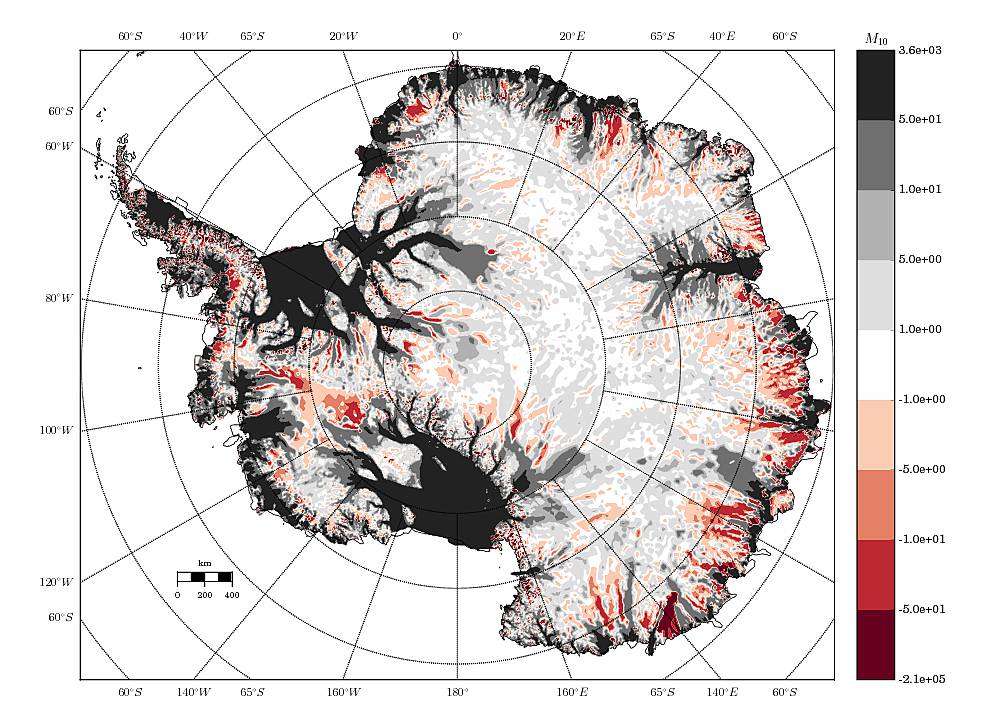}
  \caption{$\kappa = 10$, GLS.}
  \label{antarctica_bv_image_kappa_5_GLS_misfit}
  \end{subfigure}
  \begin{subfigure}[b]{0.45\linewidth}
    \includegraphics[width=\linewidth]{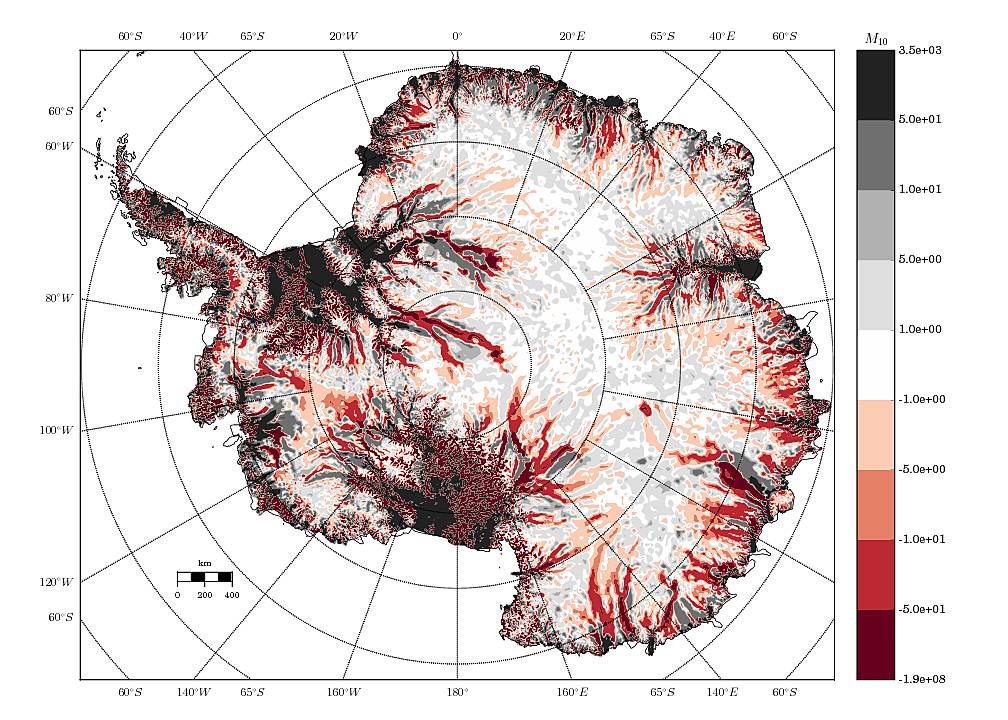}
  \caption{$\kappa = 10$, SUPG.}
  \label{antarctica_bv_image_kappa_5_SUPG_misfit}
  \end{subfigure}
  
  \caption[Antarctica balance-velocity misfit with $\mathbf{d}^{\text{data}} = - \nabla S$.]{Difference $\Vert \mathbf{u}_{ob} \Vert - \bar{u}$ between balance velocity $\bar{u}$ and the magnitude of the observed surface velocity $\mathbf{u}_{ob}$ over Antarctica with imposed direction of flow down the surface gradient $\nabla S$, where smoothing radius $\kappa$ varies as indicated.  The columns vary according to  stabilization used; either Galerkin/least-squares (GLS) stabilization (\ref{bv_gls_operator}) or streamline-upwind/Petrov-Galerkin (SUPG) stabilization (\ref{bv_supg_operator}) in variational form (\ref{balance_velocity_weak_problem}).}

  \label{antarctica_bv_image_misfit}

\end{figure*}

\begin{figure*}

  \centering

  \begin{subfigure}[b]{0.45\linewidth}
    \includegraphics[width=\linewidth]{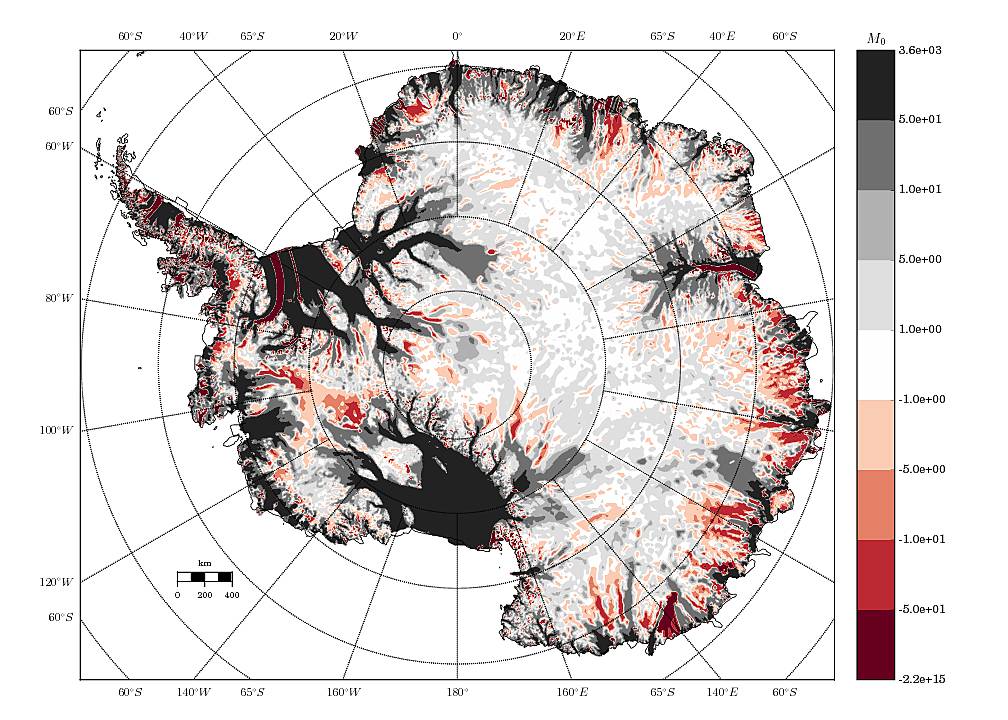}
  \caption{$\kappa = 0$, GLS.}
  \label{antarctica_bv_image_kappa_0_GLS_U_ob_S_misfit}
  \end{subfigure}
  \begin{subfigure}[b]{0.45\linewidth}
    \includegraphics[width=\linewidth]{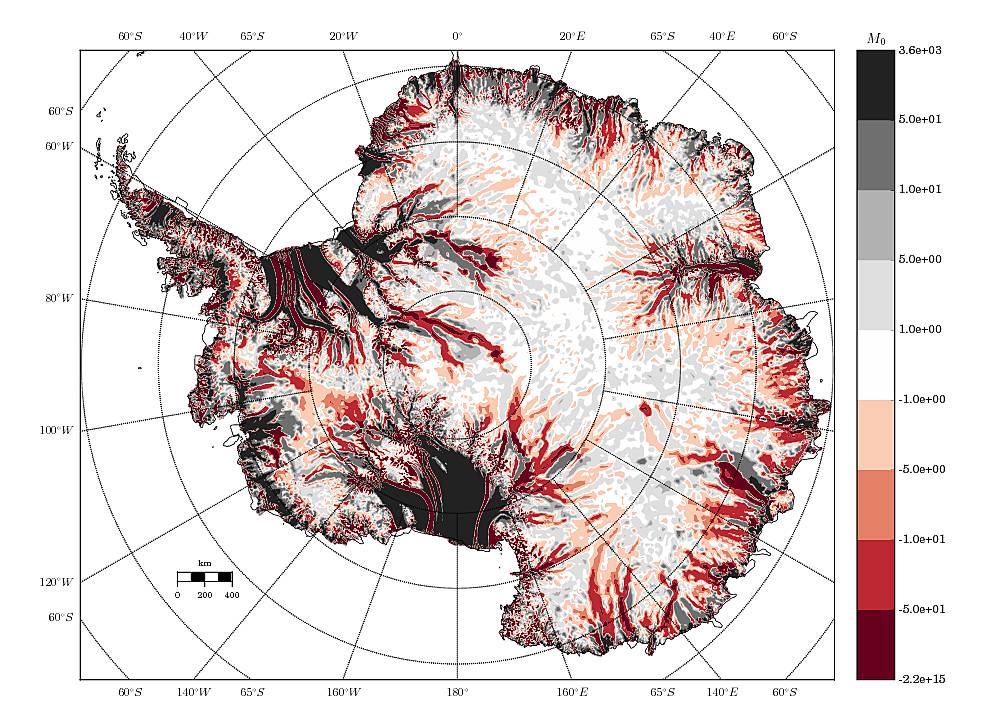}
  \caption{$\kappa = 0$, SUPG.}
  \label{antarctica_bv_image_kappa_0_SUPG_U_ob_S_misfit}
  \end{subfigure}

  \begin{subfigure}[b]{0.45\linewidth}
    \includegraphics[width=\linewidth]{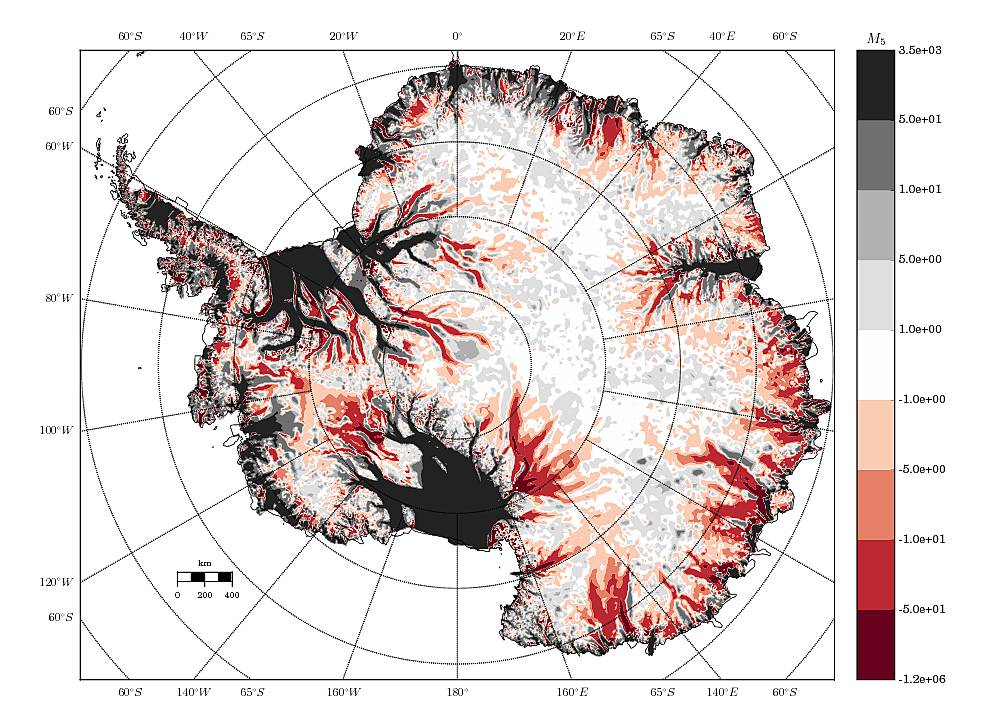}
  \caption{$\kappa = 5$, GLS.}
  \label{antarctica_bv_image_kappa_5_GLS_U_ob_S_misfit}
  \end{subfigure}
  \begin{subfigure}[b]{0.45\linewidth}
    \includegraphics[width=\linewidth]{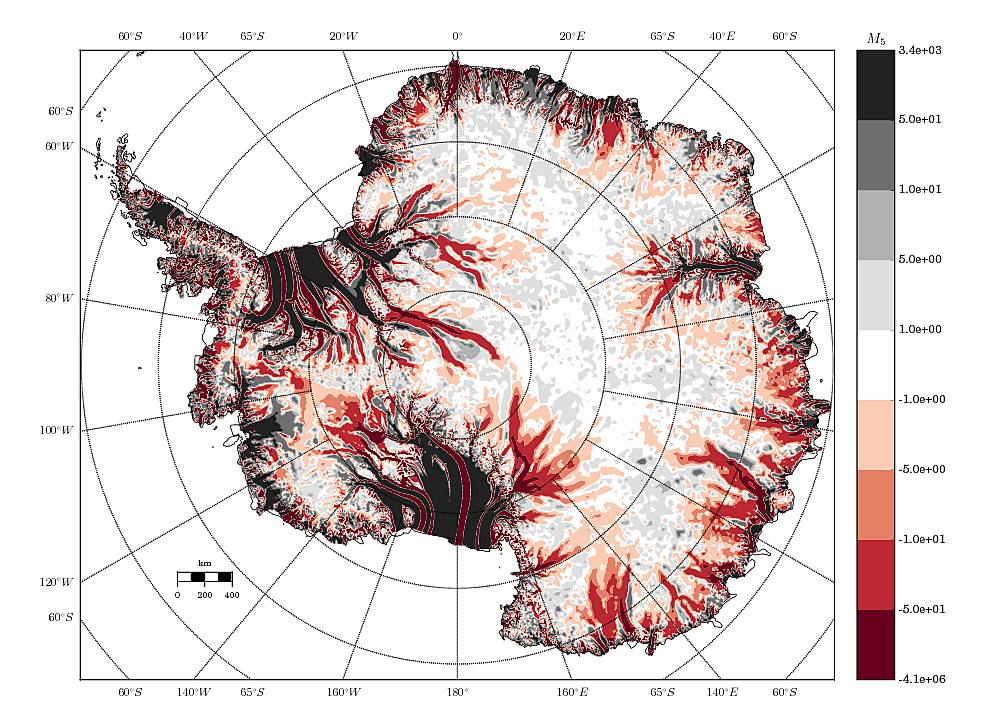}
  \caption{$\kappa = 5$, SUPG.}
  \label{antarctica_bv_image_kappa_5_SUPG_U_ob_S_misfit}
  \end{subfigure}

  \begin{subfigure}[b]{0.45\linewidth}
    \includegraphics[width=\linewidth]{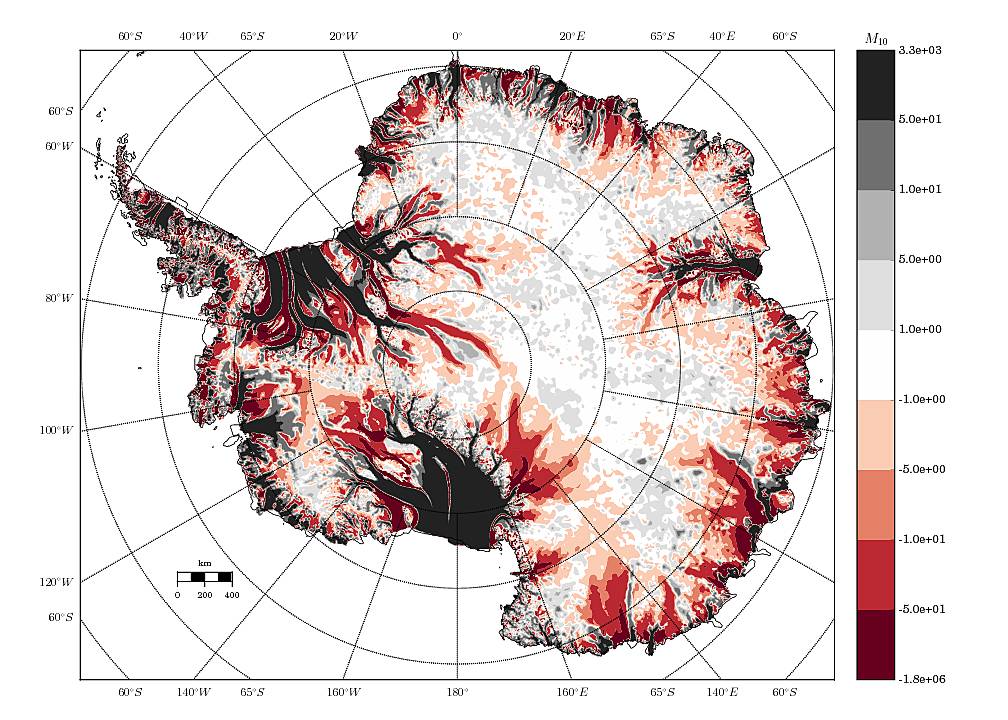}
  \caption{$\kappa = 10$, GLS.}
  \label{antarctica_bv_image_kappa_5_GLS_U_ob_S_misfit}
  \end{subfigure}
  \begin{subfigure}[b]{0.45\linewidth}
    \includegraphics[width=\linewidth]{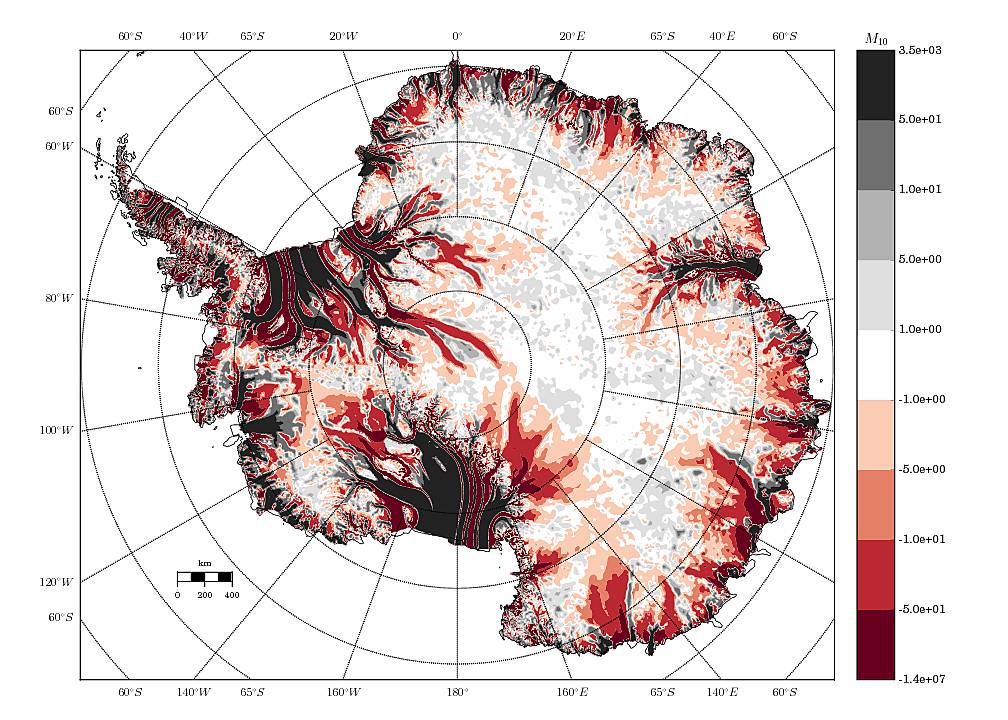}
  \caption{$\kappa = 10$, SUPG.}
  \label{antarctica_bv_image_kappa_5_SUPG_U_ob_S_misfit}
  \end{subfigure}
  
  \caption[Antarctica balance-velocity misfit with $\mathbf{d}^{\text{data}} = \mathbf{u}_{ob}$ over shelves.]{Difference $\Vert \mathbf{u}_{ob} \Vert - \bar{u}$ between balance velocity $\bar{u}$ and the magnitude of the observed surface velocity $\mathbf{u}_{ob}$ over Antarctica with imposed direction of flow down the surface gradient $\nabla S$ over grounded ice and in the direction of surface observations $\mathbf{u}_{ob}$ over floating ice, where smoothing radius $\kappa$ varies as indicated.  The columns vary according to  stabilization used; either Galerkin/least-squares (GLS) stabilization (\ref{bv_gls_operator}) or streamline-upwind/Petrov-Galerkin (SUPG) stabilization (\ref{bv_supg_operator}) in variational form (\ref{balance_velocity_weak_problem}).}
  
  \label{antarctica_bv_image_U_ob_S_misfit}

\end{figure*}

\begin{figure*}

  \centering

  \begin{subfigure}[b]{0.45\linewidth}
    \includegraphics[width=\linewidth]{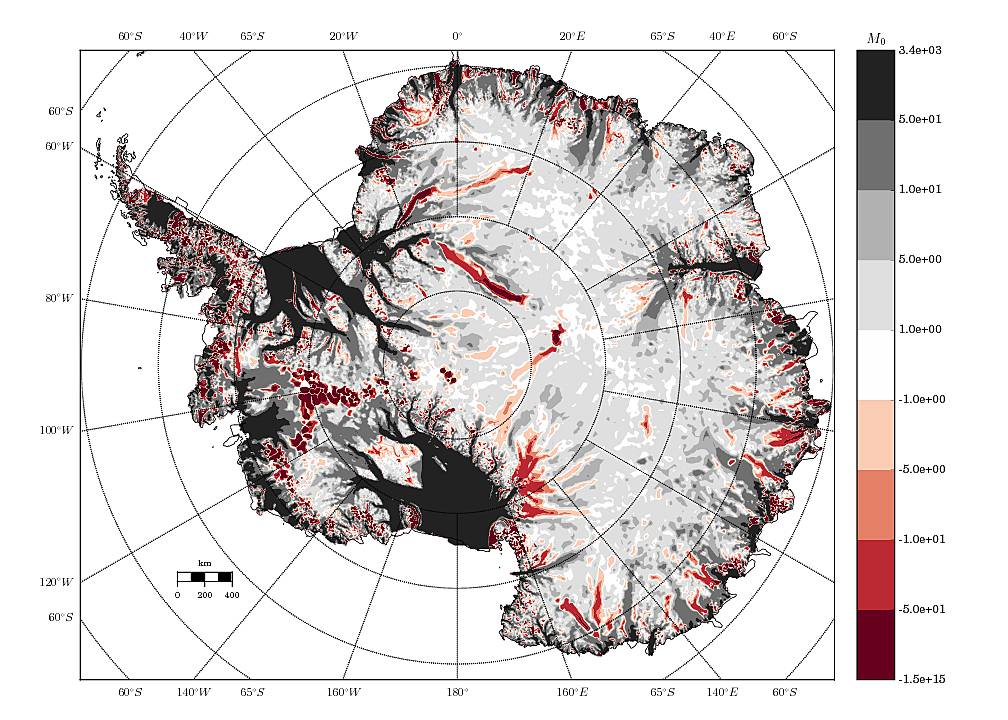}
  \caption{$\kappa = 0$, GLS.}
  \label{antarctica_bv_image_kappa_0_GLS_U_ob_misfit}
  \end{subfigure}
  \begin{subfigure}[b]{0.45\linewidth}
    \includegraphics[width=\linewidth]{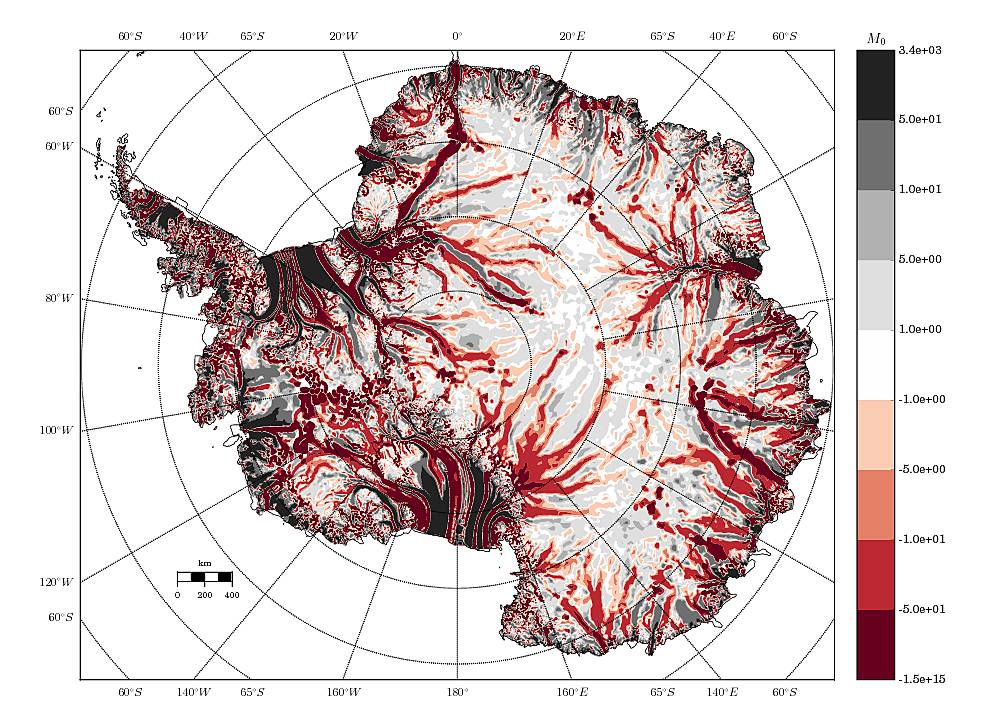}
  \caption{$\kappa = 0$, SUPG.}
  \label{antarctica_bv_image_kappa_0_SUPG_U_ob_misfit}
  \end{subfigure}

  \begin{subfigure}[b]{0.45\linewidth}
    \includegraphics[width=\linewidth]{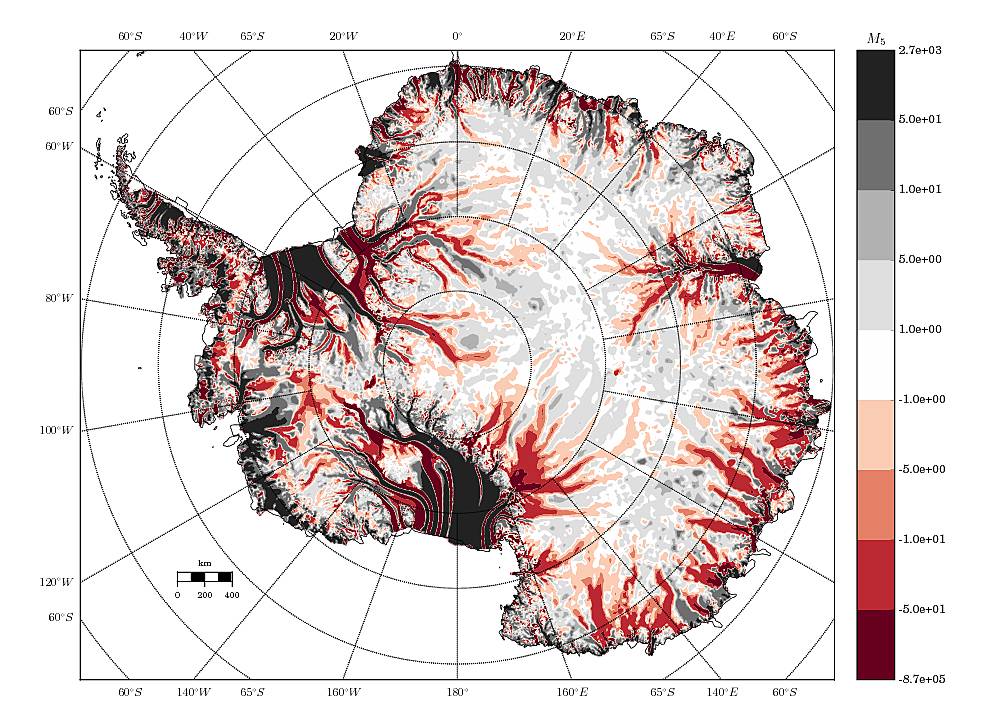}
  \caption{$\kappa = 5$, GLS.}
  \label{antarctica_bv_image_kappa_5_GLS_U_ob_misfit}
  \end{subfigure}
  \begin{subfigure}[b]{0.45\linewidth}
    \includegraphics[width=\linewidth]{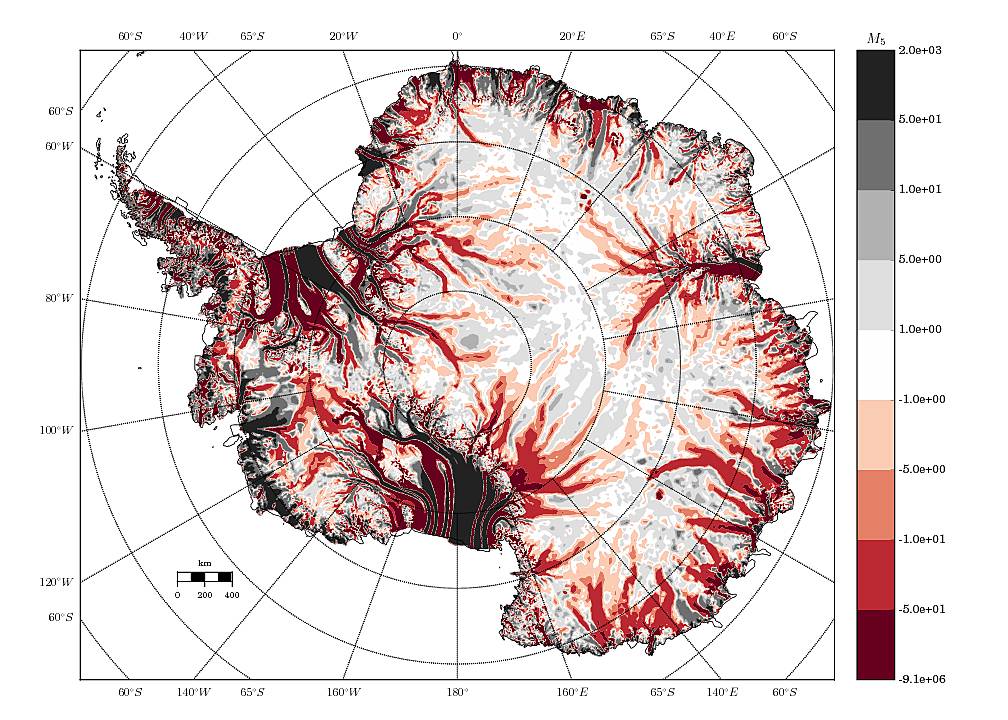}
  \caption{$\kappa = 5$, SUPG.}
  \label{antarctica_bv_image_kappa_5_SUPG_U_ob_misfit}
  \end{subfigure}

  \begin{subfigure}[b]{0.45\linewidth}
    \includegraphics[width=\linewidth]{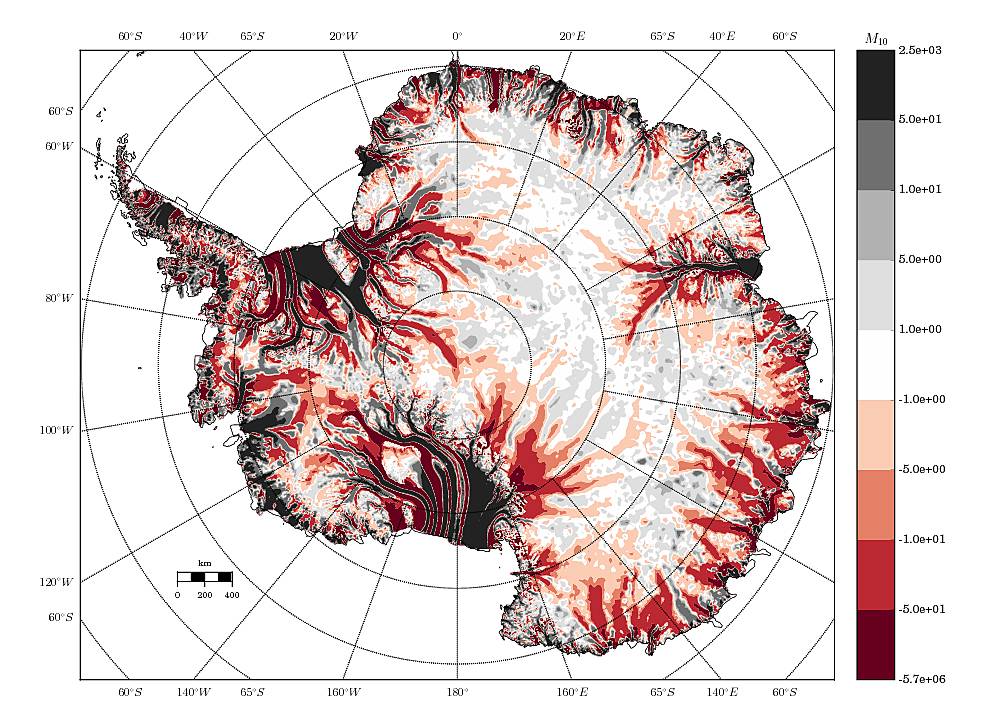}
  \caption{$\kappa = 10$, GLS.}
  \label{antarctica_bv_image_kappa_5_GLS_U_ob_misfit}
  \end{subfigure}
  \begin{subfigure}[b]{0.45\linewidth}
    \includegraphics[width=\linewidth]{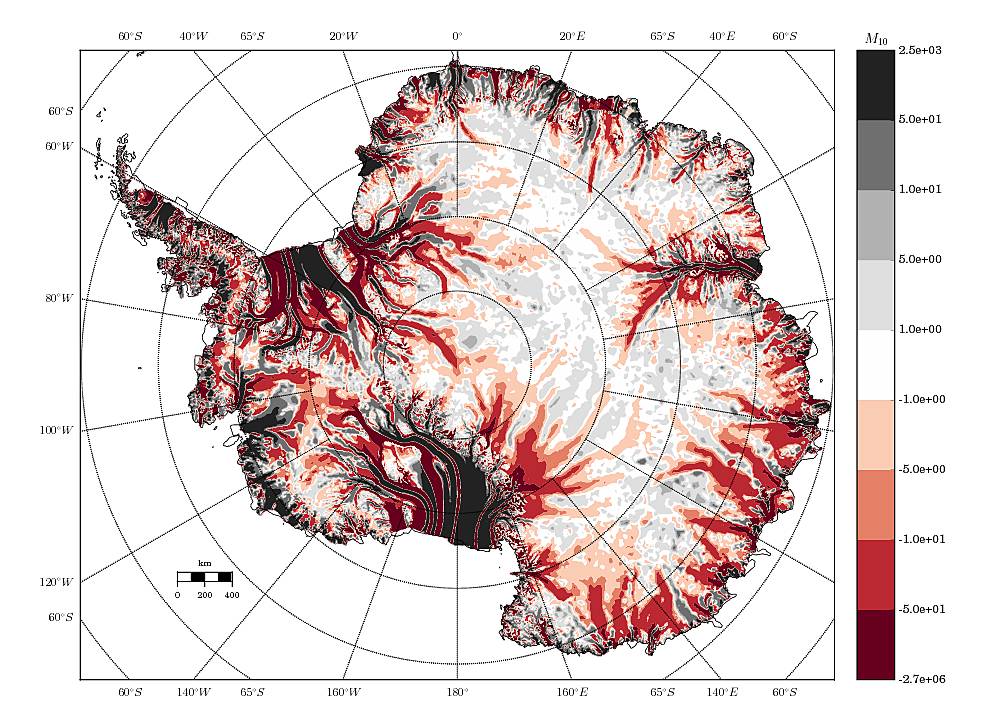}
  \caption{$\kappa = 10$, SUPG.}
  \label{antarctica_bv_image_kappa_5_SUPG_U_ob_misfit}
  \end{subfigure}
  
  \caption[Antarctica balance-velocity misfit with $\mathbf{d}^{\text{data}} = \mathbf{u}_{ob}$.]{Difference $\Vert \mathbf{u}_{ob} \Vert - \bar{u}$ between balance velocity $\bar{u}$ and the magnitude of the observed surface velocity $\mathbf{u}_{ob}$ over Antarctica with imposed direction of flow in the direction of surface observations $\mathbf{u}_{ob}$, where smoothing radius $\kappa$ varies as indicated.  The columns vary according to  stabilization used; either Galerkin/least-squares (GLS) stabilization (\ref{bv_gls_operator}) or streamline-upwind/Petrov-Galerkin (SUPG) stabilization (\ref{bv_supg_operator}) in variational form (\ref{balance_velocity_weak_problem}).}
  
  \label{antarctica_bv_image_U_ob_misfit}

\end{figure*}

\begin{figure*}

  \centering

  \begin{subfigure}[b]{0.45\linewidth}
    \includegraphics[width=\linewidth]{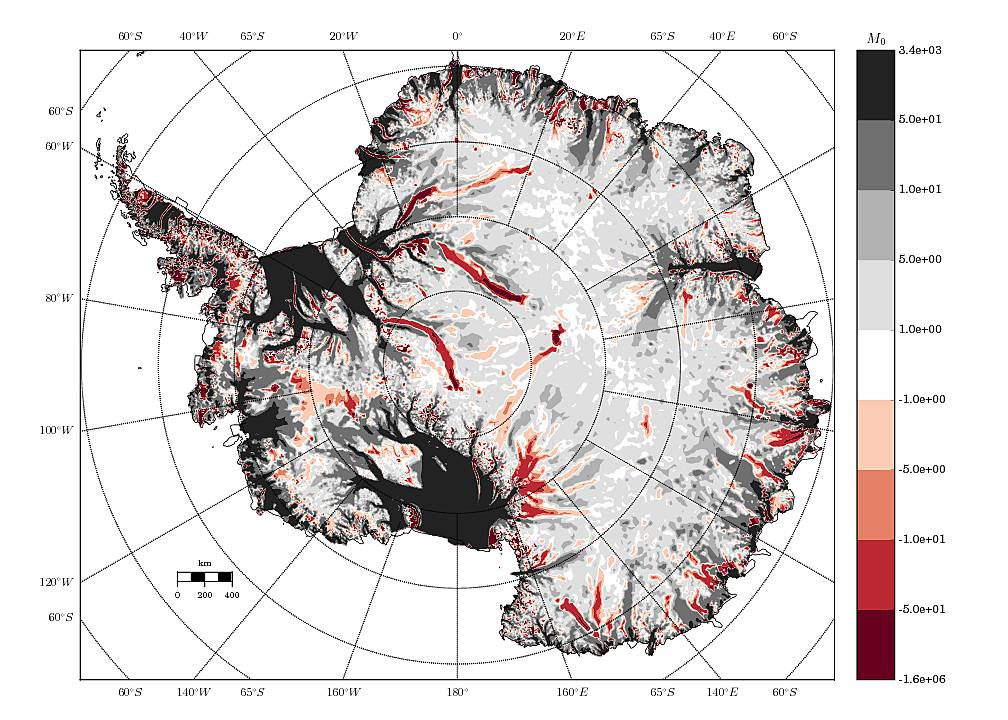}
  \caption{$\kappa = 0$, GLS.}
  \label{antarctica_bv_image_kappa_0_GLS_gS_m_U_misfit}
  \end{subfigure}
  \begin{subfigure}[b]{0.45\linewidth}
    \includegraphics[width=\linewidth]{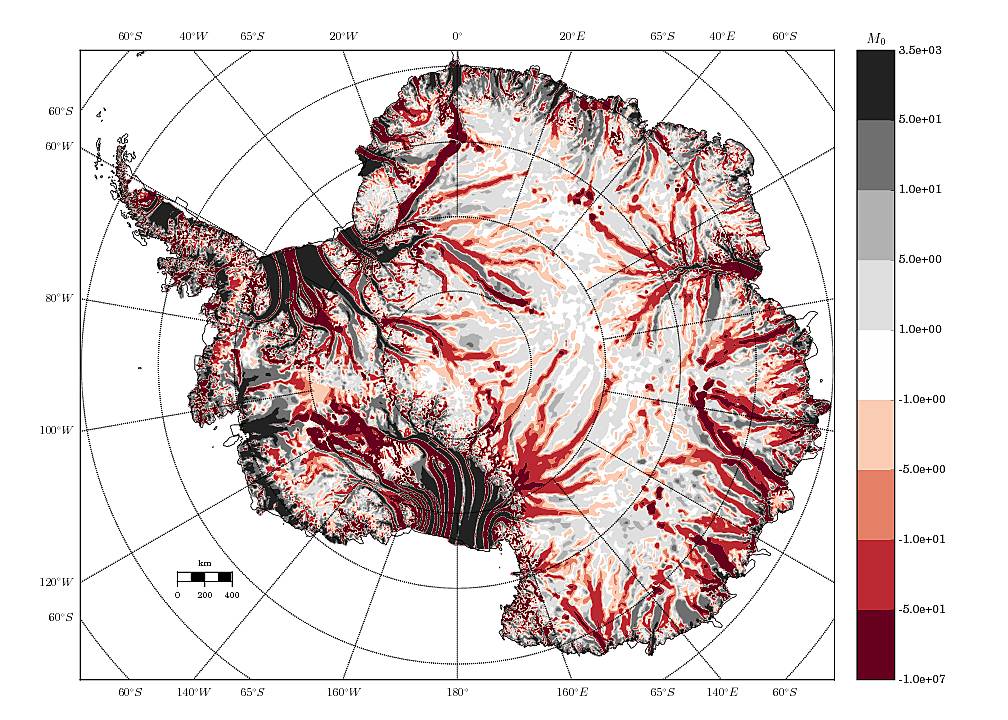}
  \caption{$\kappa = 0$, SUPG.}
  \label{antarctica_bv_image_kappa_0_SUPG_gS_m_U_misfit}
  \end{subfigure}

  \begin{subfigure}[b]{0.45\linewidth}
    \includegraphics[width=\linewidth]{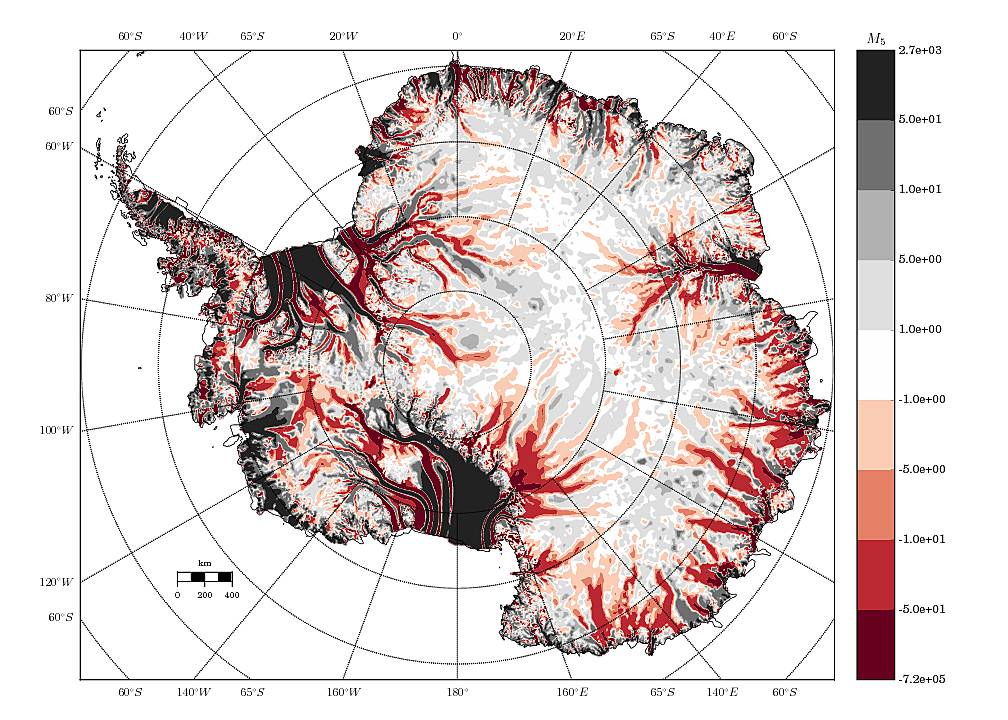}
  \caption{$\kappa = 5$, GLS.}
  \label{antarctica_bv_image_kappa_5_GLS_gS_m_U_misfit}
  \end{subfigure}
  \begin{subfigure}[b]{0.45\linewidth}
    \includegraphics[width=\linewidth]{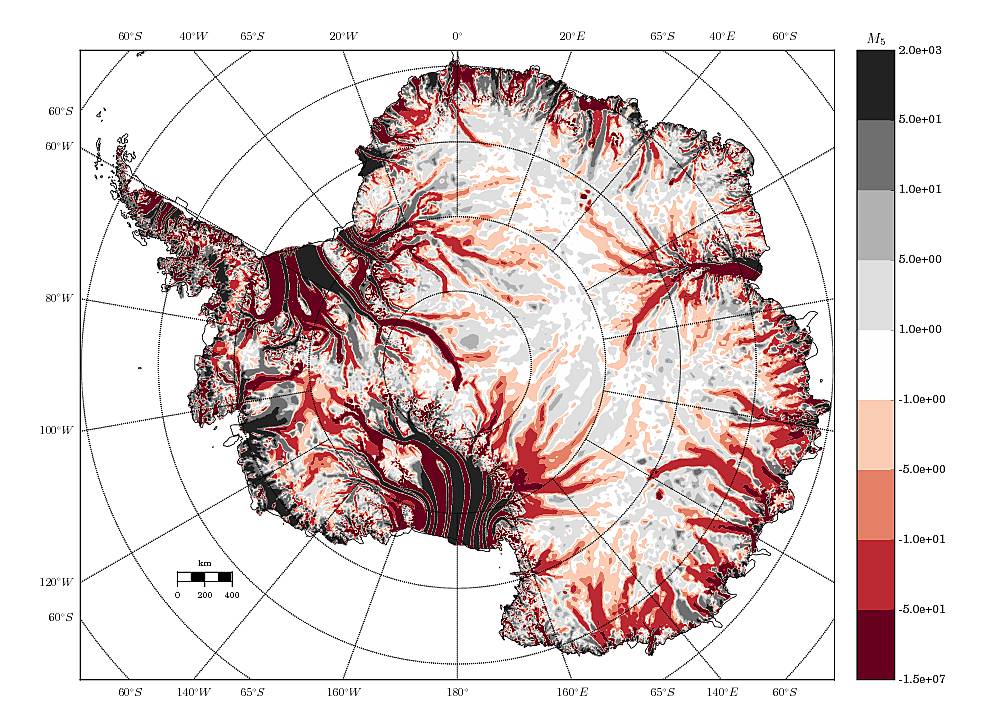}
  \caption{$\kappa = 5$, SUPG.}
  \label{antarctica_bv_image_kappa_5_SUPG_gS_m_U_misfit}
  \end{subfigure}

  \begin{subfigure}[b]{0.45\linewidth}
    \includegraphics[width=\linewidth]{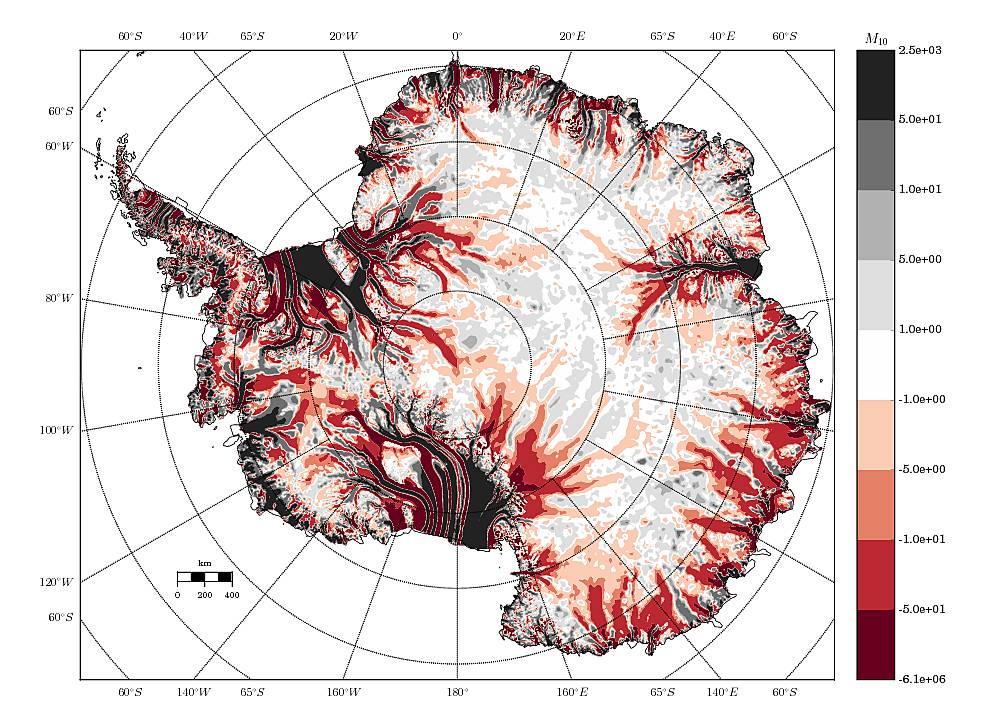}
  \caption{$\kappa = 10$, GLS.}
  \label{antarctica_bv_image_kappa_5_GLS_gS_m_U_misfit}
  \end{subfigure}
  \begin{subfigure}[b]{0.45\linewidth}
    \includegraphics[width=\linewidth]{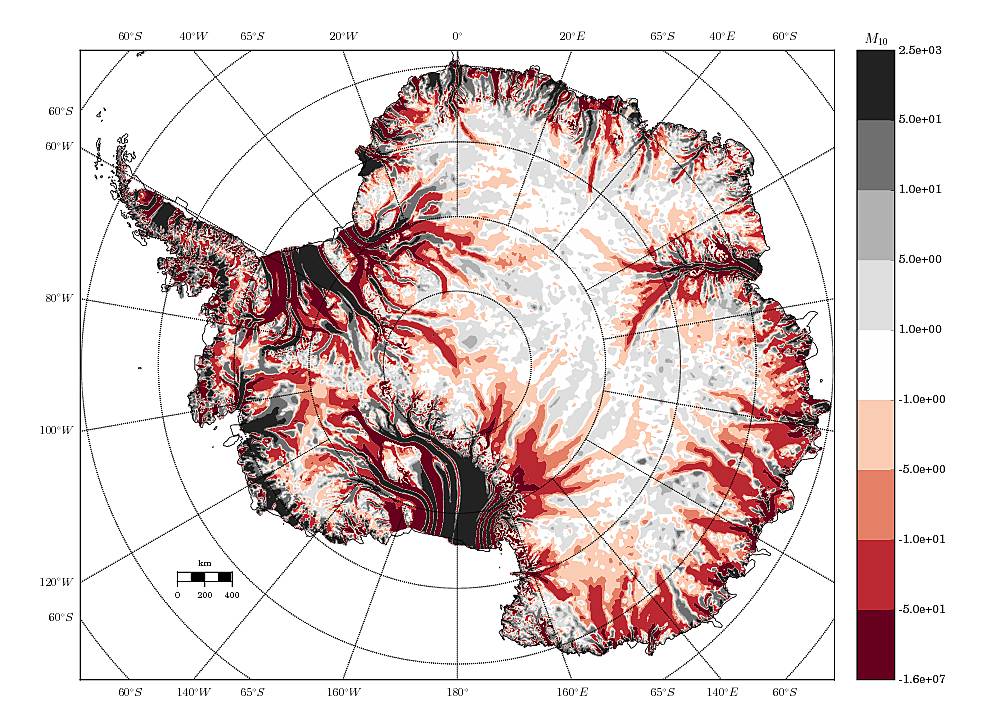}
  \caption{$\kappa = 10$, SUPG.}
  \label{antarctica_bv_image_kappa_5_SUPG_gS_m_U_misfit}
  \end{subfigure}
  
  \caption[Antarctica balance-velocity misfit with $\mathbf{d}^{\text{data}} = -\nabla S$ where $\mathbf{u}_{ob}$ are missing.]{Difference $\Vert \mathbf{u}_{ob} \Vert - \bar{u}$ between balance velocity $\bar{u}$ and the magnitude of the observed surface velocity $\mathbf{u}_{ob}$ over Antarctica with imposed direction of flow in the direction of surface observations $\mathbf{u}_{ob}$ and down the surface gradient $\nabla S$ where $\mathbf{u}_{ob}$ values are missing, where smoothing radius $\kappa$ varies as indicated.  The columns vary according to stabilization used; either Galerkin/least-squares (GLS) stabilization (\ref{bv_gls_operator}) or streamline-upwind/Petrov-Galerkin (SUPG) stabilization (\ref{bv_supg_operator}) in variational form (\ref{balance_velocity_weak_problem}).  Results using subgrid-scale-model stabilization (\ref{bv_ssm_operator}) (not shown) appeared more unstable than the (SUPG) method.}

  \label{antarctica_bv_image_gS_m_U_misfit}

\end{figure*}

Results indicate that the vertically averaged flow field $\bar{u}$ roughly reproduces the general pattern of recorded surface velocities over both Greenland (Figures \ref{greenland_bv_image} and \ref{greenland_bv_image_d_U_ob}) and Antarctica (Figures \ref{antarctica_bv_image} and \ref{antarctica_bv_image_d_U_ob}).  The difference between the vertically-averaged velocity and the recorded surface speed provides some insight into the variation of velocity with depth (Figures \ref{greenland_bv_image_misfit}, \ref{greenland_bv_image_d_U_ob_misfit}, \ref{antarctica_bv_image_misfit}, \ref{antarctica_bv_image_U_ob_S_misfit}, \ref{antarctica_bv_image_U_ob_misfit}, and \ref{antarctica_bv_image_gS_m_U_misfit}).  That is, in regions where the misfit is high, we expect significant differences between the surface and the basal velocity.  However, due to the fact that the data -- including topography, accumulation, $\partial_t H$ and $F_b$ -- contain a high degree of uncertainty, we expect that with better data the sporadic variance structure evident here will be diminished.

Defining the direction of flow in the direction of velocity observations over the shelves of Antarctica is an improvement over the down-surface-gradient direction (compare Figures \ref{antarctica_bv_image} and \ref{antarctica_bv_image_U_ob_S}).  However, there appear to remain substantial differences between these two velocities in these regions (Figure \ref{antarctica_bv_image_U_ob_S_misfit}).  This may be due to substantial basal melt or freeze-on under the shelves, implying in this case that our specification of $F_b = 0$ is entirely inappropriate.  Additionally, the balance velocity changes from under-estimation to over-estimation of the surface velocity over the fastest-moving parts of interior regions when a meaningful direction of flow is imposed over floating-ice regions (compare Figures \ref{antarctica_bv_image_misfit} and \ref{antarctica_bv_image_U_ob_S_misfit}).  This illuminates the fact that the balance velocity of the floating-ice shelves has at least some affect on the velocity deep within the interior.

Filling in the gaps of missing velocity data $\mathbf{u}_{ob}$ with $-\nabla S$ appeared to have the most effect on results derived without smoothing (Figure \ref{antarctica_bv_image_d_gS_m_U} and \ref{antarctica_bv_image_gS_m_U_misfit}).

Finally, the solutions attained with the stabilization schemes \ref{bv_gls_operator}), (\ref{bv_supg_operator}), and (\ref{bv_ssm_operator}) varied most with lower values of smoothing radius $\kappa$.  At the lowest $\kappa$-values, the GLS model appears to on-average under estimate the velocity, while the SSM model seems to on-average over estimate.  Remarkably, all calculations over estimate the velocity of South-West Greenland (Figures \ref{greenland_bv_image_misfit} and \ref{greenland_bv_image_d_U_ob_misfit}) and match the deep-interior velocity observations of Antarctica relatively well (Figures \ref{antarctica_bv_image_misfit}, \ref{antarctica_bv_image_U_ob_S_misfit}, \ref{antarctica_bv_image_U_ob_misfit}, and \ref{antarctica_bv_image_gS_m_U_misfit}). 

\subsection{Greenland}

The Greenland simulations used topography $S$ and $B$ given by \citet{bamber}, and an accumulation/ablation function $\dot{a}$ provided by \citet{veen,burgess}.

The CSLVR scripts used to generate the mesh, data, and perform the simulation are shown in Code Listings \ref{cslvr_greenland_gen_mesh}, \ref{cslvr_greenland_gen_data}, and \ref{cslvr_greenland_bv}, respectively. 

\begin{table}[H]
\centering
\caption[Greenland balance-velocity variables]{Greenland $\bar{u}$ simulation variables.}
\label{greenland_balance_velocity_values}
\begin{tabular}{llll}
\hline
\textbf{Variable} & \textbf{Value} & \textbf{Units} & \textbf{Description} \\
\hline
$\partial_t H$ & $0$     & m a\sups{-1}  & $H$ obs.~rate of change \\
$F_b$          & $0$     & m a\sups{-1}  & basal water discharge \\
$m$            & $5$     & --            & mesh refinement \\
$N_e$          & $92855$ & --            & number of cells \\
$N_n$          & $49064$ & --            & number of vertices \\
\hline
\end{tabular}
\end{table}

\pythonexternal[label=cslvr_greenland_gen_mesh, caption={CSLVR script used to generate the 2D mesh for Greenland.}]{scripts/balance_velocity/greenland/gen_mesh.py}

\pythonexternal[label=cslvr_greenland_gen_data, caption={CSLVR script used to generate the data used by the Greenland balance velocity calculation of Code Listing \ref{cslvr_greenland_bv}.}]{scripts/balance_velocity/greenland/gen_data.py}

\pythonexternal[label=cslvr_greenland_bv, caption={CSLVR script used to calculate the balance velocity for Greenland.}]{scripts/balance_velocity/greenland/balance_velocity.py}

\subsection{Antarctica}

The Antarctica simulations used topography $S$ and $B$ provided by \citet{fretwell}, and an accumulation/ablation function $\dot{a}$ provided by \citet{arthern_2006,lebrocq}.

The CSLVR scripts used to generate the mesh, data, and perform the simulation are shown in Code Listings \ref{cslvr_antarctica_gen_mesh}, \ref{cslvr_antarctica_gen_data}, and \ref{cslvr_antarctica_bv}, respectively. 

\begin{table}[H]
\centering
\caption[Antarctica balance-velocity variables]{Antarctica $\bar{u}$ simulations variables.}
\label{antarctica_balance_velocity_values}
\begin{tabular}{llll}
\hline
\textbf{Variable} & \textbf{Value} & \textbf{Units} & \textbf{Description} \\
\hline
$\partial_t H$ & $0$       & m a\sups{-1}  & $H$ obs.~rate of change \\
$F_b$          & $0$       & m a\sups{-1}  & basal water discharge \\
$m$            & $10$      & --            & mesh refinement \\
$N_e$          & $162211$  & --            & number of cells \\
$N_n$          & $82894$   & --            & number of vertices \\
\hline
\end{tabular}
\end{table}

\pythonexternal[label=cslvr_antarctica_gen_mesh, caption={CSLVR script used to generate the 2D mesh for Antarctica.}]{scripts/balance_velocity/antarctica/gen_mesh.py}

\pythonexternal[label=cslvr_antarctica_gen_data, caption={CSLVR script used to generate the data used by the Antarctica balance velocity calculation of Code Listing \ref{cslvr_antarctica_bv}.}]{scripts/balance_velocity/antarctica/gen_data.py}

\pythonexternal[label=cslvr_antarctica_bv, caption={CSLVR script used to calculate the balance velocity for Antarctica.}]{scripts/balance_velocity/antarctica/balance_velocity.py}

%===============================================================================

%===============================================================================

%===============================================================================

%===============================================================================

%===============================================================================

%===============================================================================

%===============================================================================

%===============================================================================

%===============================================================================

%===============================================================================

%===============================================================================

%===============================================================================

%===============================================================================

%===============================================================================

%===============================================================================
%===============================================================================

\chapter{Stress balance} \label{ssn_stress_balance}

\index{Balance equations!Stress}
In this chapter we decompose the state of stress with an ice-sheet into along-flow and across-flow components
\begin{align*}
  \bm{\sigma_n} &= \sigma \cdot \hat{\mathbf{u}}_n, \hspace{10mm}
  \bm{\sigma_t}  = \sigma \cdot \hat{\mathbf{u}}_t,
\end{align*}
where the unit-vectors $\hat{\mathbf{u}}_n$ and $\hat{\mathbf{u}}_t$ point in the direction of and tangential to the flow, respectively:
\begin{align}
  \label{flow_coordinate_system}
  \hat{\mathbf{u}}_n &= \frac{[u\ v\ 0]\T}{\Vert \mathbf{u}_h \Vert}, \hspace{5mm} \text{and} \hspace{5mm}
  \hat{\mathbf{u}}_t  = \frac{[v\ \text{-}u\ 0]\T}{\Vert \mathbf{u}_h \Vert},
\end{align}
where $\mathbf{f}_h = [f_x\ f_y]\T$ denotes the horizontal components of the vector $\mathbf{f}$.

\section{Membrane stress} \label{ssn_membrane_stress}

\index{Membrane stress}
Let coordinates $(i,j,z)$ be the transformed velocity-coordinate system in the direction of flow, tangential to flow, and in the positive vertical direction, respectively.  The unit vectors of this rotated coordinate system are
\begin{align}
  \label{flow_coordinate_vectors}
  \hat{\mathbf{u}} = [\hat{u}\ \hat{v}\ 0]\T \hspace{5mm} 
  \hat{\mathbf{v}} = [\hat{v}\ \text{-}\hat{u}\ 0]\T \hspace{5mm}
  \hat{\mathbf{w}} = [0\ 0\ \hat{w}]\T,
\end{align}
with normalized velocity vector components
\begin{align}
  \label{normalized_velocity_components}
  \hat{u} = \frac{u}{\Vert \mathbf{u}_h \Vert}, \hspace{8mm}
  \hat{v} = \frac{v}{\Vert \mathbf{u}_h \Vert}, \hspace{8mm}
  \hat{w} = 1.
\end{align}
The partial derivatives in this new coordinate system can be evaluated using the directional derivatives \index{Directional derivative}
\begin{align*}
  \frac{\partial f}{\partial i} &= \nabla_{\hat{\mathbf{u}}} f = \nabla f \cdot \hat{\mathbf{u}} = \frac{\partial f}{\partial x} \hat{u} + \frac{\partial f}{\partial y} \hat{v} \\
  \frac{\partial f}{\partial j} &= \nabla_{\hat{\mathbf{v}}} f = \nabla f \cdot \hat{\mathbf{v}} = \frac{\partial f}{\partial x} \hat{v} - \frac{\partial f}{\partial y} \hat{u} \\
  \frac{\partial f}{\partial z} &= \nabla_{\hat{\mathbf{w}}} f = \nabla f \cdot \hat{\mathbf{w}} = \frac{\partial f}{\partial z},
\end{align*} 
and the $z$-rotated gradient operator
\begin{align}
  \label{transformed_gradient_operator}
  \nabla_r \equiv \left[ \frac{\partial}{\partial i}\ \frac{\partial}{\partial j}\ \frac{\partial}{\partial z} \right]\T.
\end{align}

Next, let $\phi$ be the signed angle in radians between the $x$-axis and the horizontal velocity vector $\mathbf{u}_h = [u\ v]\T$.  The rotation matrix about the $z$-axis used to transform any vector in the $x$, $y$, $z$ coordinate system to the $i$, $j$, $z$ coordinate system is
\begin{align*}
  R_z &= \begin{bmatrix}[c]
           \cos(\phi)  & -\sin(\phi) & 0 \\
           \sin(\phi)  &  \cos(\phi) & 0 \\
           0           &  0          & 1
         \end{bmatrix},
\end{align*}
and the $\phi$-rotation of the rank-two Cauchy-stress tensor $\sigma$ about the $z$-axis is
\begin{align}
  \label{rotated_stress_tensor}
  \sigma_r = R_z \cdot \sigma \cdot R_z\T
           = \begin{bmatrix}
               \sigma_{ii} & \sigma_{ij} & \sigma_{iz} \\
               \sigma_{ji} & \sigma_{jj} & \sigma_{jz} \\
               \sigma_{zi} & \sigma_{zj} & \sigma_{zz}
             \end{bmatrix}.
\end{align}

Returning to momentum balance (\ref{cons_momentum}), the $\phi$-rotated momentum balance is
\begin{align}
  \label{rotated_momentum_balance}
  - \nabla_r \cdot \sigma_r &= \mathbf{f},
\end{align}
leading to an expansion similar to (\ref{stokes_exp}),
\begin{align*}
  \frac{\partial \sigma_{ii}}{\partial i} + \frac{\partial \sigma_{ij}}{\partial j} + \frac{\partial\sigma_{iz}}{\partial z} &= 0 \\ 
  \frac{\partial \sigma_{ji}}{\partial i} + \frac{\partial \sigma_{jj}}{\partial j} + \frac{\partial\sigma_{jz}}{\partial z} &= 0 \\ 
  \frac{\partial \sigma_{zi}}{\partial i} + \frac{\partial \sigma_{zj}}{\partial j} + \frac{\partial\sigma_{zz}}{\partial z} &= \rho g.
\end{align*}

Next, rotated momentum-balance (\ref{rotated_momentum_balance}) is integrated vertically, 
\begin{align}
  \label{vertically_integrated_stress_balance_one}
  - \int_B^S \nabla_r \cdot \sigma_r\ dz &= \int_B^S \mathbf{f}\ dz.
\end{align}
Similar to the derivation of vertically-integrated mass-balance (\ref{integrated_cons_mass}), Leibniz's rule is applied to the above (Appendix \ref{leibniz_formula}), resulting in the vertically-integrated stress-balance
\begin{align}
  \label{vertically_integrated_stress_balance}
  - \nabla_r \cdot \left( \int_B^S \sigma_r\ dz \right) + \sigma_r |_S \cdot \nabla_r S - \sigma_r |_B \cdot \nabla_r B &= \int_B^S \mathbf{f}\ dz.
\end{align}

It is also of interest to examine the state of stress without the contribution of the mean compressive stress $p = -\sigma_{kk} / 3$,  and thus perform a similar set of calculations as above to deviatoric stress-tensor $\tau$ in (\ref{stress_tensor}).  The rotated-stress-deviator tensor is thus
\begin{align}
  \label{rotated_stress_deviator_tensor}
  \tau_r = R_z \cdot \tau \cdot R_z\T
         = \begin{bmatrix}
             \tau_{ii} & \tau_{ij} & \tau_{iz} \\
             \tau_{ji} & \tau_{jj} & \tau_{jz} \\
             \tau_{zi} & \tau_{zj} & \tau_{zz}
           \end{bmatrix},
\end{align}
such that stress tensor (\ref{rotated_stress_tensor}) may be decomposed using (\ref{stress_tensor}) into
\begin{align*}
  \sigma_r = \tau_r - pI
           = \begin{bmatrix}
               \tau_{ii} & \tau_{ij} & \tau_{iz} \\
               \tau_{ji} & \tau_{jj} & \tau_{jz} \\
               \tau_{zi} & \tau_{zj} & \tau_{zz}
             \end{bmatrix} - 
           p \begin{bmatrix}
               1 & 0 & 0 \\
               0 & 1 & 0 \\
               0 & 0 & 1 
             \end{bmatrix}.
\end{align*}
Therefore, using the fact that $\nabla_r \cdot \big(p I \big) = \nabla_r p$, vertically-integrated stress-balance (\ref{vertically_integrated_stress_balance}) may also be written
\begin{align}
  \label{vertically_integrated_stress_balance_decomposed}
  - \nabla_r \cdot \left( \int_B^S \tau_r\ dz \right) + \tau_r |_S \cdot \nabla_r S - \tau_r |_B \cdot \nabla_r B &= \int_B^S \left( \mathbf{f} - \nabla_r p \right)\ dz.
\end{align}

Finally, the terms of the vertically-integrated deviatoric-stress tensor $\int_z \tau_r dz = N$ contains individual components \index{Tensor!Membrane stress}
\begin{align}
  \label{membrane_stress_tensor}
  N &= \begin{bmatrix}
         N_{ii} & N_{ij} & N_{iz} \\
         N_{ji} & N_{jj} & N_{jz} \\
         N_{zi} & N_{zj} & N_{zz}
       \end{bmatrix}
     = \begin{bmatrix}
         \int_z \tau_{ii} & \int_z \tau_{ij} & \int_z \tau_{iz}  \\ 
         \int_z \tau_{ji} & \int_z \tau_{jj} & \int_z \tau_{jz}  \\ 
         \int_z \tau_{zi} & \int_z \tau_{zj} & \int_z \tau_{zz}
       \end{bmatrix},
\end{align}
and are referred to as \emph{membrane stresses} \citep{greve}.  

\section{Membrane stress balance} \label{ssn_membrane_stress_balance}

Re-writing stress balance (\ref{vertically_integrated_stress_balance_one}) using (\ref{vertically_integrated_stress_balance_decomposed}), we have an equivalent form of (\ref{vertically_integrated_stress_balance_decomposed}), the \emph{membrane stress balance}
\begin{align}
  \label{membrane_stress_balance}
  - M \mathbf{1} = \mathbf{f}_{\text{int}}, 
\end{align}
where $\mathbf{1} = [1\ 1\ 1]\T$ is the rank-one tensor of ones, $\mathbf{f}_{\text{int}} = \int_z \left( \mathbf{f} - \nabla_r p \right)\ dz$ is the right-hand side of (\ref{vertically_integrated_stress_balance_decomposed}), and the tensor $M$ is defined as
\begin{align}
  \label{membrane_stress_balance_tensor}
  M &= \begin{bmatrix}
         M_{ii} & M_{ij} & M_{iz} \\
         M_{ji} & M_{jj} & M_{jz} \\
         M_{zi} & M_{zj} & M_{zz}
       \end{bmatrix}
     = \begin{bmatrix}
         \int_z \frac{\partial \tau_{ii}}{\partial i} & \int_z \frac{\partial \tau_{ij}}{\partial j} & \int_z \frac{\partial\tau_{iz}}{\partial z}  \\ 
         \int_z \frac{\partial \tau_{ji}}{\partial i} & \int_z \frac{\partial \tau_{jj}}{\partial j} & \int_z \frac{\partial\tau_{jz}}{\partial z}  \\ 
         \int_z \frac{\partial \tau_{zi}}{\partial i} & \int_z \frac{\partial \tau_{zj}}{\partial j} & \int_z \frac{\partial\tau_{zz}}{\partial z}
       \end{bmatrix}.
\end{align}
Applying Leibniz's rule to the $i$- and $j$-derivative terms, and the first fundamental theorem of calculus to the $z$-derivative terms,
\begin{align}
\label{individual_membrane_stress_balance_components}
\left.
\def\arraystretch{2}
\begin{array}{lcl}
  M_{ii} &=& \frac{\partial}{\partial i} N_{ii} + \tau_{ii}(S) \frac{\partial S}{\partial i} - \tau_{ii}(B) \frac{\partial B}{\partial i} \\
  M_{ij} &=& \frac{\partial}{\partial j} N_{ij} + \tau_{ij}(S) \frac{\partial S}{\partial j} - \tau_{ij}(B) \frac{\partial B}{\partial j} \\
  M_{iz} &=& \tau_{iz}(S) - \tau_{iz}(B) \\
  M_{ji} &=& \frac{\partial}{\partial i} N_{ji} + \tau_{ji}(S) \frac{\partial S}{\partial i} - \tau_{ji}(B) \frac{\partial B}{\partial i} \\
  M_{jj} &=& \frac{\partial}{\partial j} N_{jj} + \tau_{jj}(S) \frac{\partial S}{\partial j} - \tau_{jj}(B) \frac{\partial B}{\partial j} \\
  M_{jz} &=& \tau_{jz}(S) - \tau_{jz}(B) \\
  M_{zi} &=& \frac{\partial}{\partial i} N_{zi} + \tau_{zi}(S) \frac{\partial S}{\partial i} - \tau_{zi}(B) \frac{\partial B}{\partial i} \\
  M_{zj} &=& \frac{\partial}{\partial j} N_{zj} + \tau_{zj}(S) \frac{\partial S}{\partial j} - \tau_{zj}(B) \frac{\partial B}{\partial j} \\
  M_{zz} &=& \tau_{zz}(S) - \tau_{zz}(B)
\end{array}
\right\}.
\end{align}

Provided that the elements of tensors $\sigma$ and $\tau$ have been populated with values obtained by solving one of the three-dimensional momentum-balance formulations of Chapter \ref{ssn_momentum_and_mass_balance}, the components of membrane-stress tensor (\ref{membrane_stress_tensor}) may be calculated using numerical integration (see \S \ref{ssn_integration} for an analogous problem in one dimension).  These stresses, once derived, may then be used to calculate the individual stress terms of membrane-stress balance (\ref{membrane_stress_balance}) given by (\ref{individual_membrane_stress_balance_components}).

Finally, note that the last row of first-order strain-rate tensor (\ref{bp_strain_rate_tensor}) has been eliminated.  While it is surely possible to employ full-Cauchy stress-deviator tensor definition (\ref{stress_tensor}) to evaluate membrane-stress tensor (\ref{membrane_stress_tensor}) using a velocity field computed from first-order momentum balance (\ref{bp_extremum}), it is more instructive to examine the state of stress for this model from the point of view of its mathematical formulation.  Therefore, the first-order stain-rate tensor, \index{Tensor!First-order strain-rate} defined as
\begin{align}
  \label{bp_full_strain_rate_tensor}
  \tilde{\dot{\epsilon}}
  &= \begin{bmatrix}
       \frac{\partial u}{\partial x} & \frac{1}{2}\left( \frac{\partial u}{\partial y} + \frac{\partial v}{\partial x} \right) & \frac{1}{2} \frac{\partial u}{\partial z} \\
       \frac{1}{2}\left( \frac{\partial v}{\partial x} + \frac{\partial u}{\partial y} \right) & \frac{\partial v}{\partial y} & \frac{1}{2} \frac{\partial v}{\partial z} \\
       \frac{1}{2} \frac{\partial u}{\partial z} & \frac{1}{2} \frac{\partial v}{\partial z} & -\left( \frac{\partial u}{\partial x} + \frac{\partial v}{\partial y} \right)
     \end{bmatrix},
\end{align}
and first-order stress-deviator tensor
\begin{align}
  \label{bp_full_stress_deviator}
  \tau_{\text{BP}} = 2 \eta_{\text{BP}} \tilde{\dot{\epsilon}},
\end{align}
derived from the simplifications described in \S \ref{ssn_strain_tensor_simplification} are used to evaluate the balance of stress associated with this model.

The CSLVR implementation of this problem is shown in Code Listing \ref{cslvr_stress_balance}.

\pythonexternal[label=cslvr_stress_balance, caption={CSLVR source code for the \texttt{StressBalance} class.}]{cslvr_src/stressbalance.py}

\section{ISMIP-HOM test simulation} \label{ssn_stress_balance_ismip_hom_test_simulations}

\index{Linear differential equations!3D}
\index{ISMIP-HOM simulations}
In this section we revisit one of the higher-wavelength ISMIP-HOM experiment presented in \S \ref{ssn_ismip_hom_test_simulations} in order to examine the distributions of stress for each of the momentum models defined in \S \ref{ssn_full_stokes}, \S \ref{ssn_first_order}, and \S \ref{ssn_reformulated_stokes}.  Once again, this test is defined over the domain $\Omega \in [0,\ell] \times [0,\ell] \times [B,S] \subset \R^3$ with a $k_x \times k_y \times k_z$ element discretization, and specifies the use of a surface height with uniform slope $\Vert \nabla S \Vert = a$
\begin{align*}
  S(x) = - x \tan\left( a \right),
\end{align*}
and the sinusoidially-varying basal topography
\begin{align*}
  B(x,y) = S(x) - \bar{B} + b \sin\left( \frac{2 \pi}{\ell} x \right) \sin\left( \frac{2 \pi}{\ell} y \right),
\end{align*}
with average basal depth $\bar{B}$, and basal height amplitude $b$ (Figure \ref{ismip_hom_a_B}).  To enforce continuity, the periodic $\mathbf{u},p$ boundary conditions
\begin{align*}
  \mathbf{u}(0,0)    &= \mathbf{u}(\ell,\ell) & p(0,0)    &= p(\ell,\ell)\\
  \mathbf{u}(0,\ell) &= \mathbf{u}(\ell,0)    & p(0,\ell) &= p(\ell,0)   \\
  \mathbf{u}(x,0)    &= \mathbf{u}(x,\ell)    & p(x,0)    &= p(x,\ell)   \\
  \mathbf{u}(0,y)    &= \mathbf{u}(\ell,y)    & p(0,y)    &= p(\ell,y)
\end{align*}
were applied.  Lastly, the basal traction coefficient was set to $\beta = 1000$, having the effect of creating a no-slip boundary condition along the entire basal surface, while $A = 10\sups{-16}$ was used as an isothermal rate factor for viscosity $\eta$.  Table \ref{ismip_hom_stress_balance_values} lists these and other coefficients used, and CSLVR script used to calculate the momentum- and stress-balance is shown in Code Listing \ref{cslvr_stress_balance_script}.

\begin{figure*}
  
  \centering 

  \begin{subfigure}[b]{0.3\linewidth}
    \includegraphics[width=\linewidth]{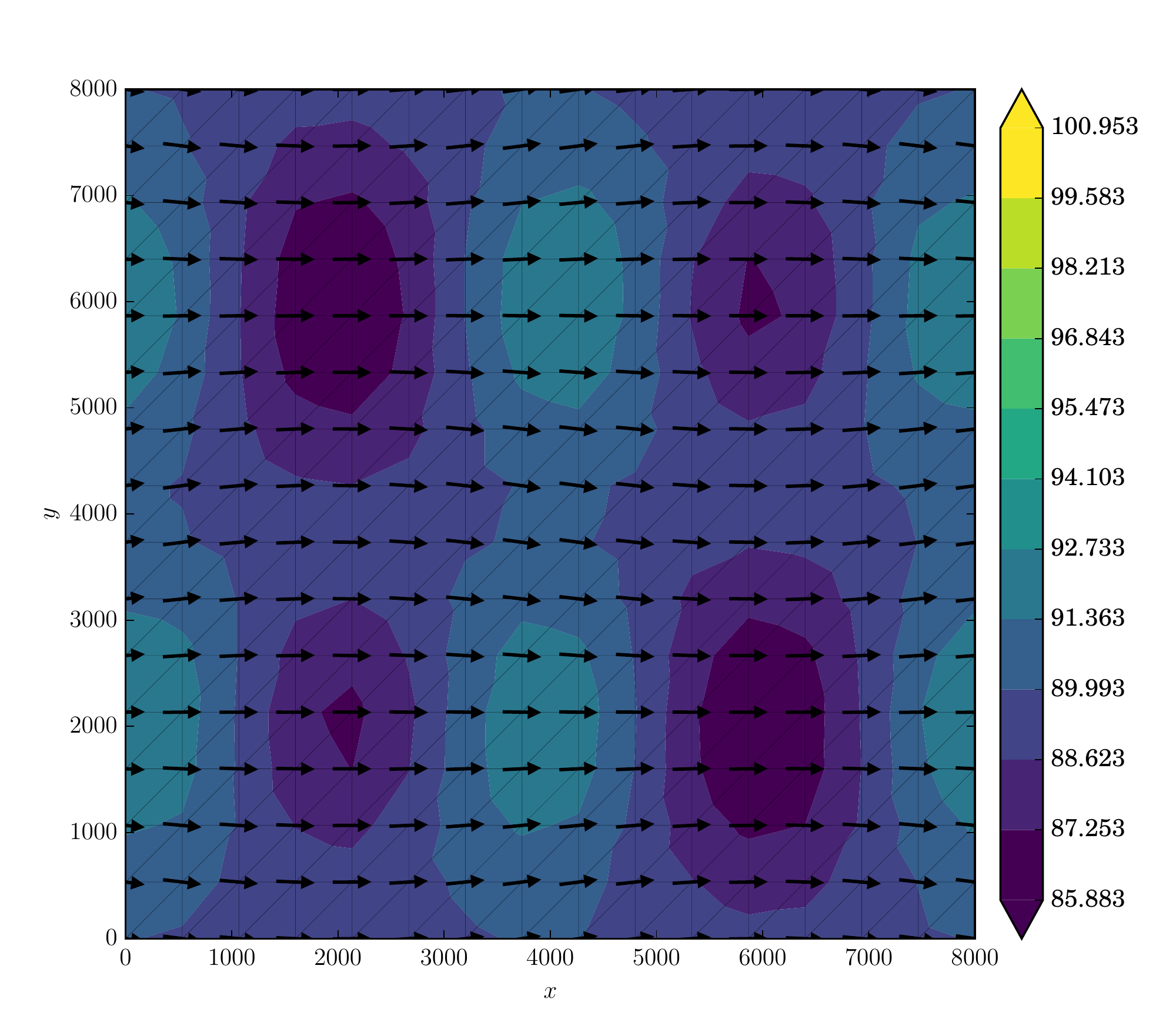}
  \caption{$\mathbf{u}_S$}
  \label{fs_ms_U}
  \end{subfigure}
  \begin{subfigure}[b]{0.3\linewidth}
    \includegraphics[width=\linewidth]{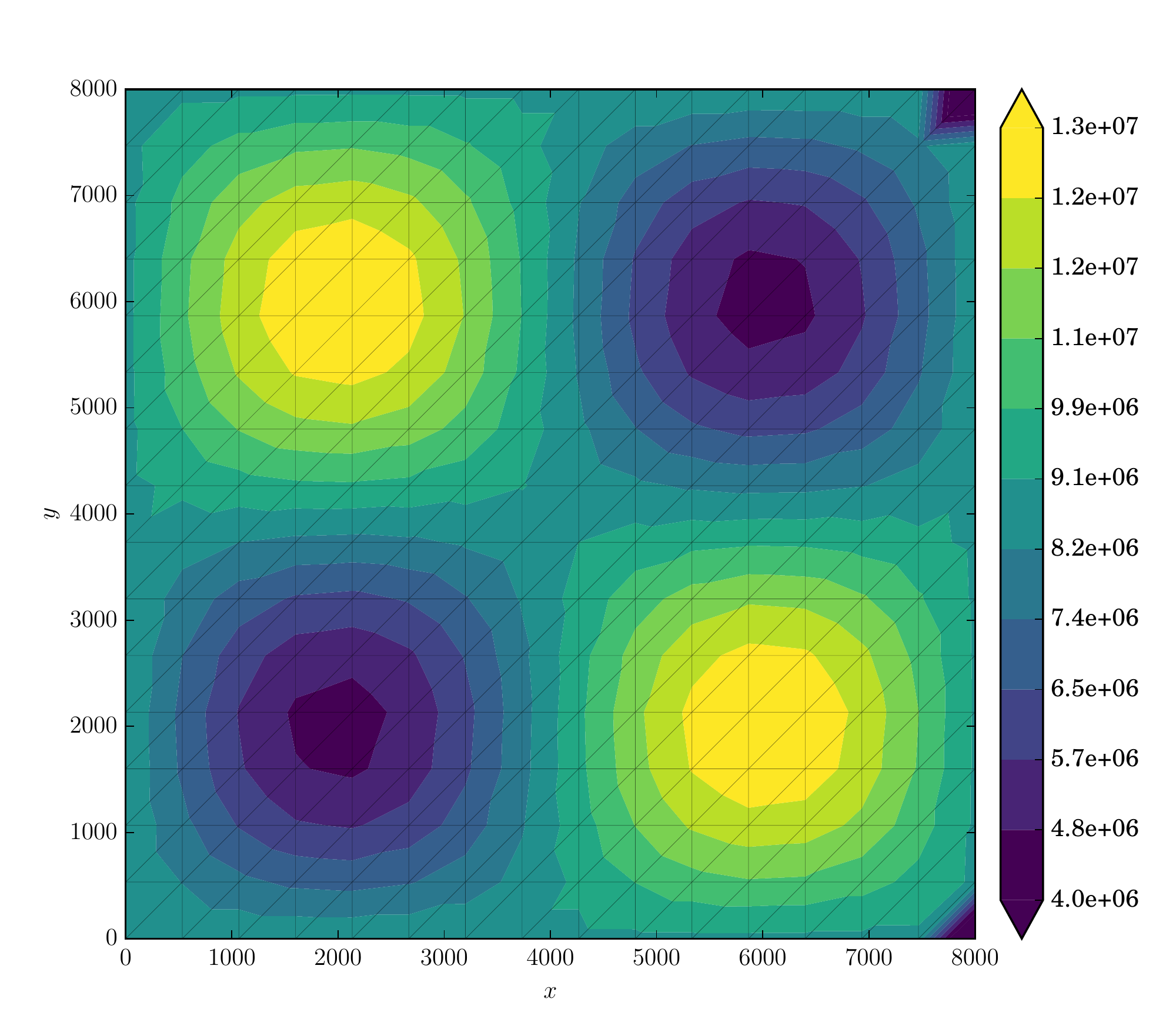}
  \caption{$p |_B$}
  \label{fs_ms_p}
  \end{subfigure}

  \begin{subfigure}[b]{0.3\linewidth}
    \includegraphics[width=\linewidth]{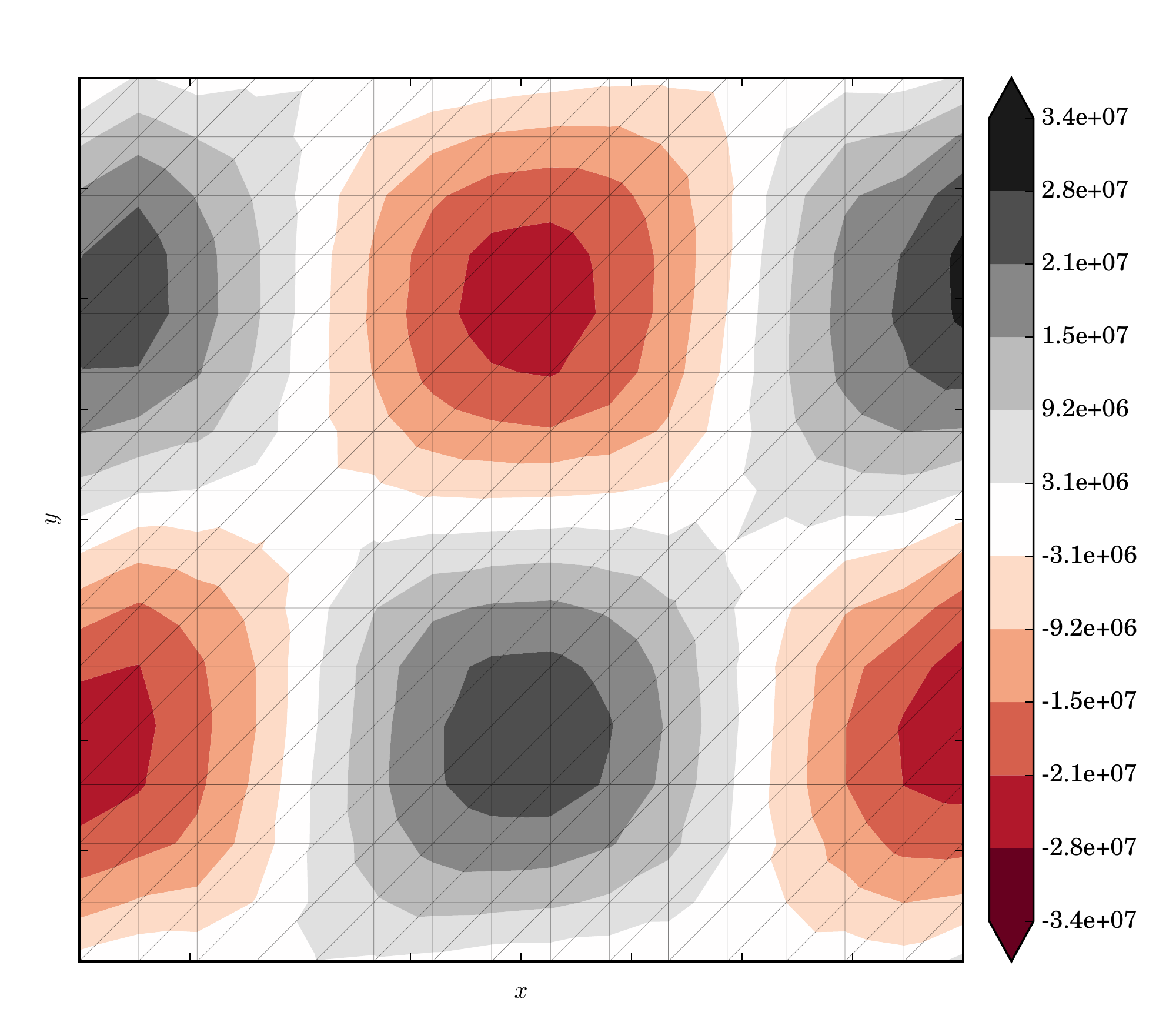}
  \caption{$N_{ii}$}
  \label{fs_N_ii}
  \end{subfigure}
  \begin{subfigure}[b]{0.3\linewidth}
    \includegraphics[width=\linewidth]{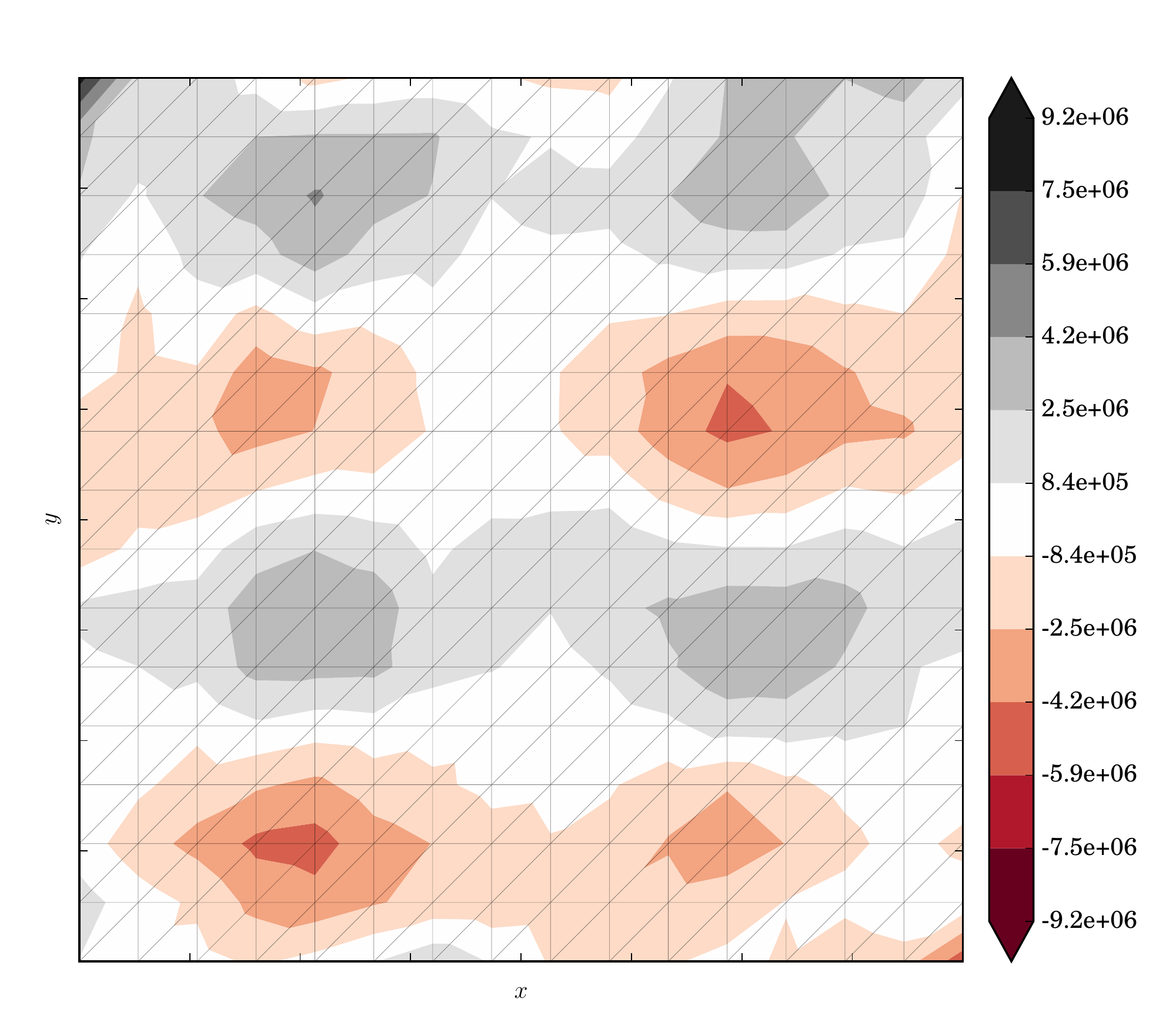}
  \caption{$N_{ij}$}
  \label{fs_N_ij}
  \end{subfigure}
  \begin{subfigure}[b]{0.3\linewidth}
    \includegraphics[width=\linewidth]{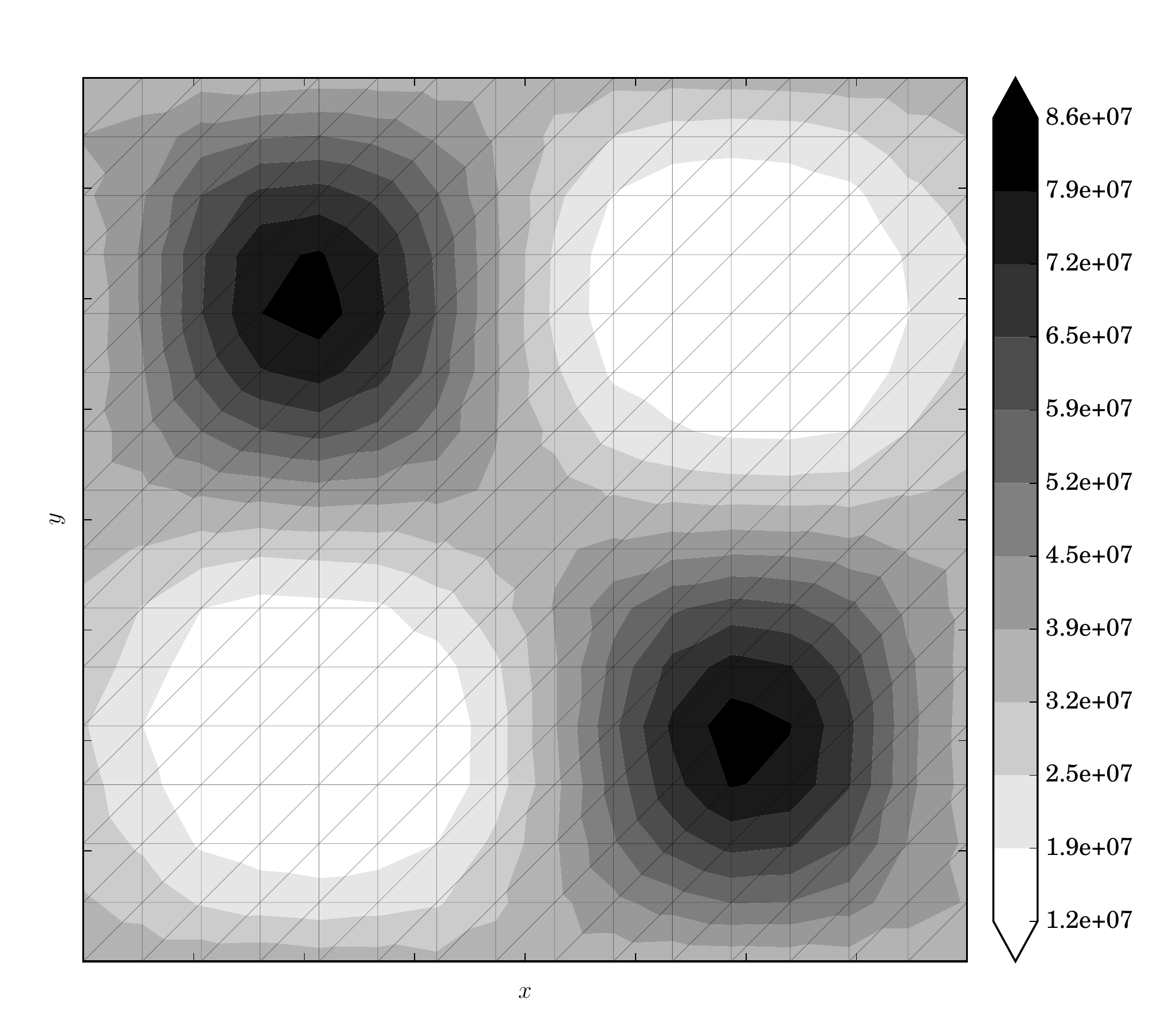}
  \caption{$N_{iz}$}
  \label{fs_N_iz}
  \end{subfigure}

  \begin{subfigure}[b]{0.3\linewidth}
    \includegraphics[width=\linewidth]{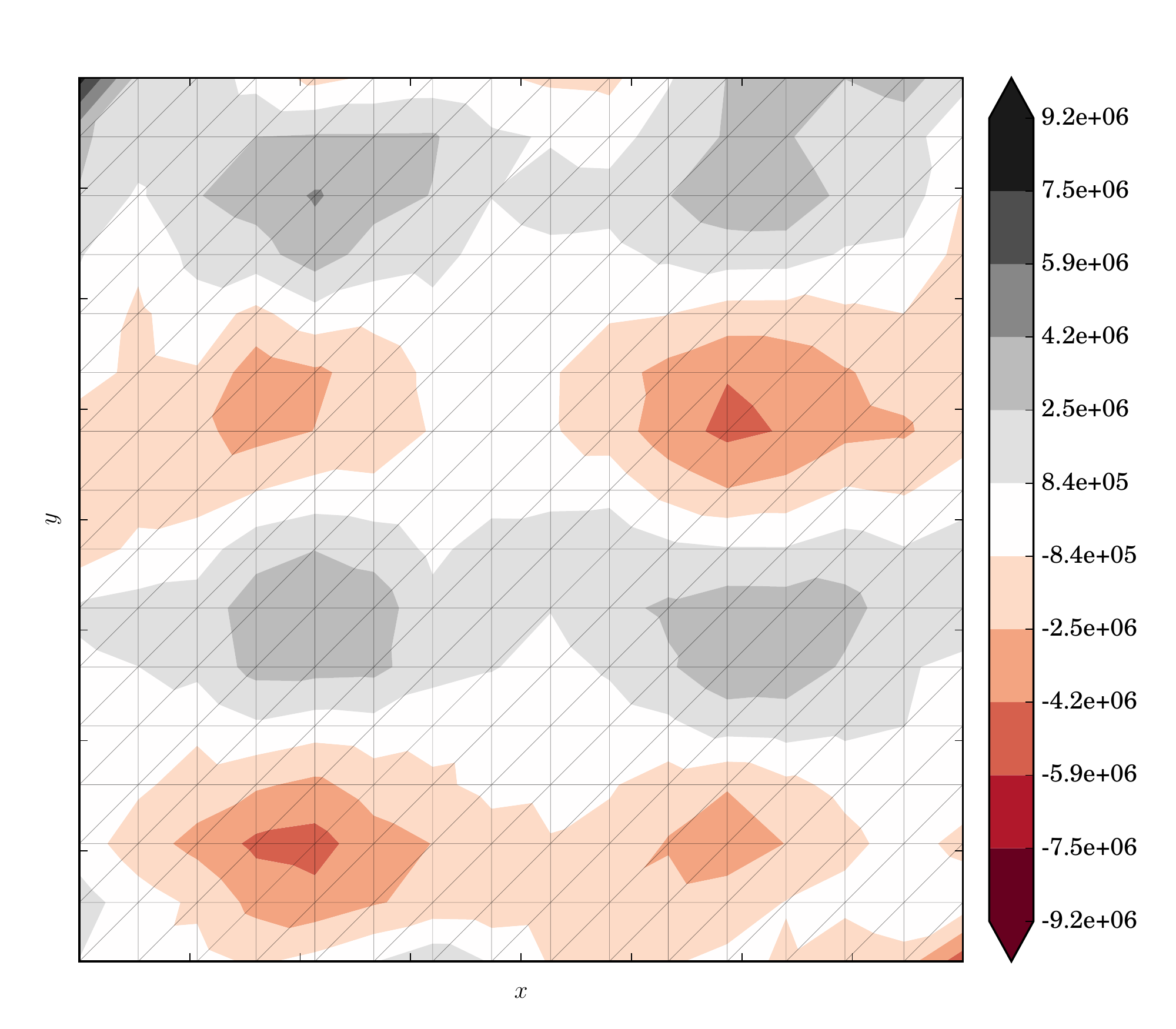}
  \caption{$N_{ji}$}
  \label{fs_N_ji}
  \end{subfigure}
  \begin{subfigure}[b]{0.3\linewidth}
    \includegraphics[width=\linewidth]{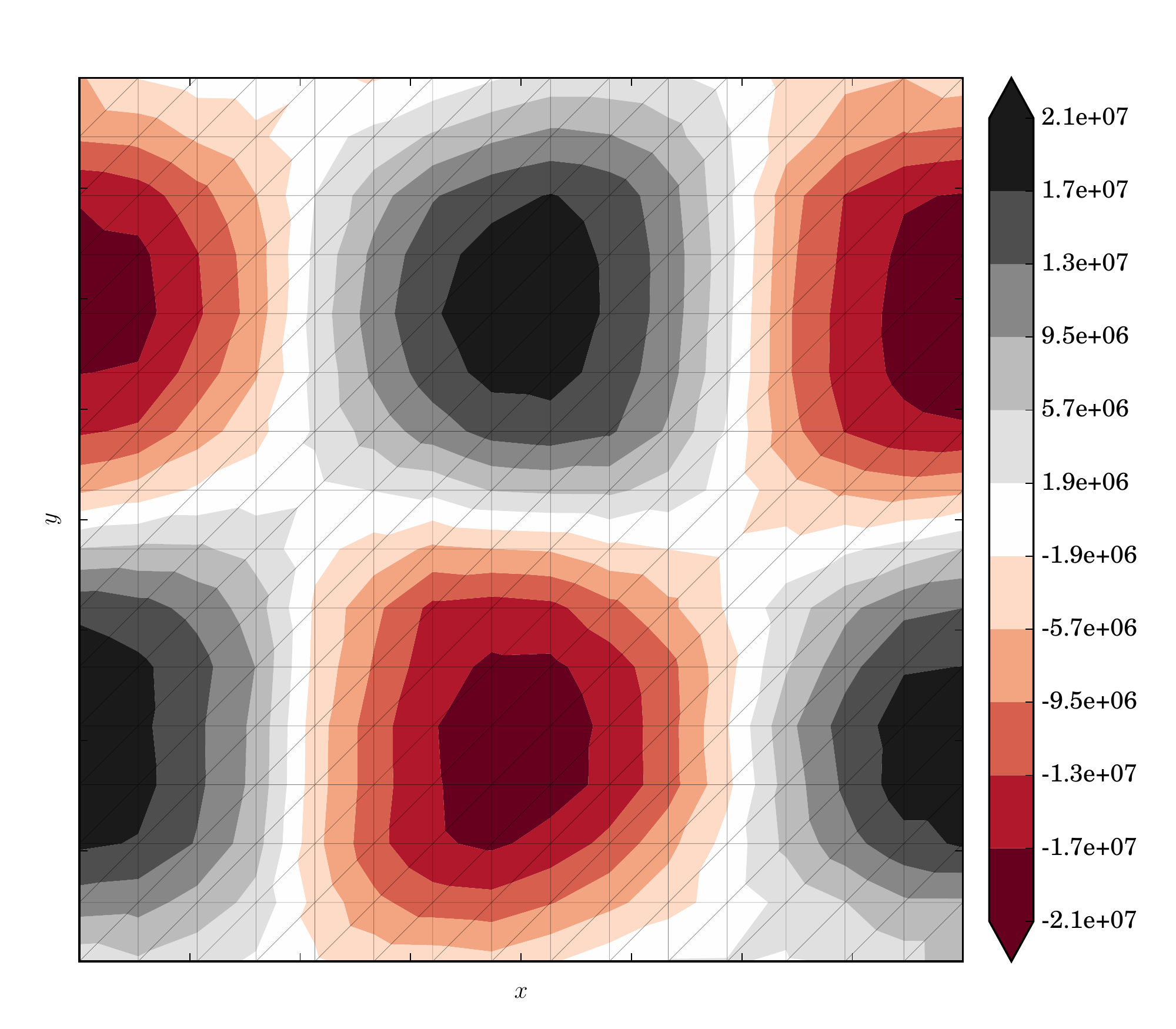}
  \caption{$N_{jj}$}
  \label{fs_N_jj}
  \end{subfigure}
  \begin{subfigure}[b]{0.3\linewidth}
    \includegraphics[width=\linewidth]{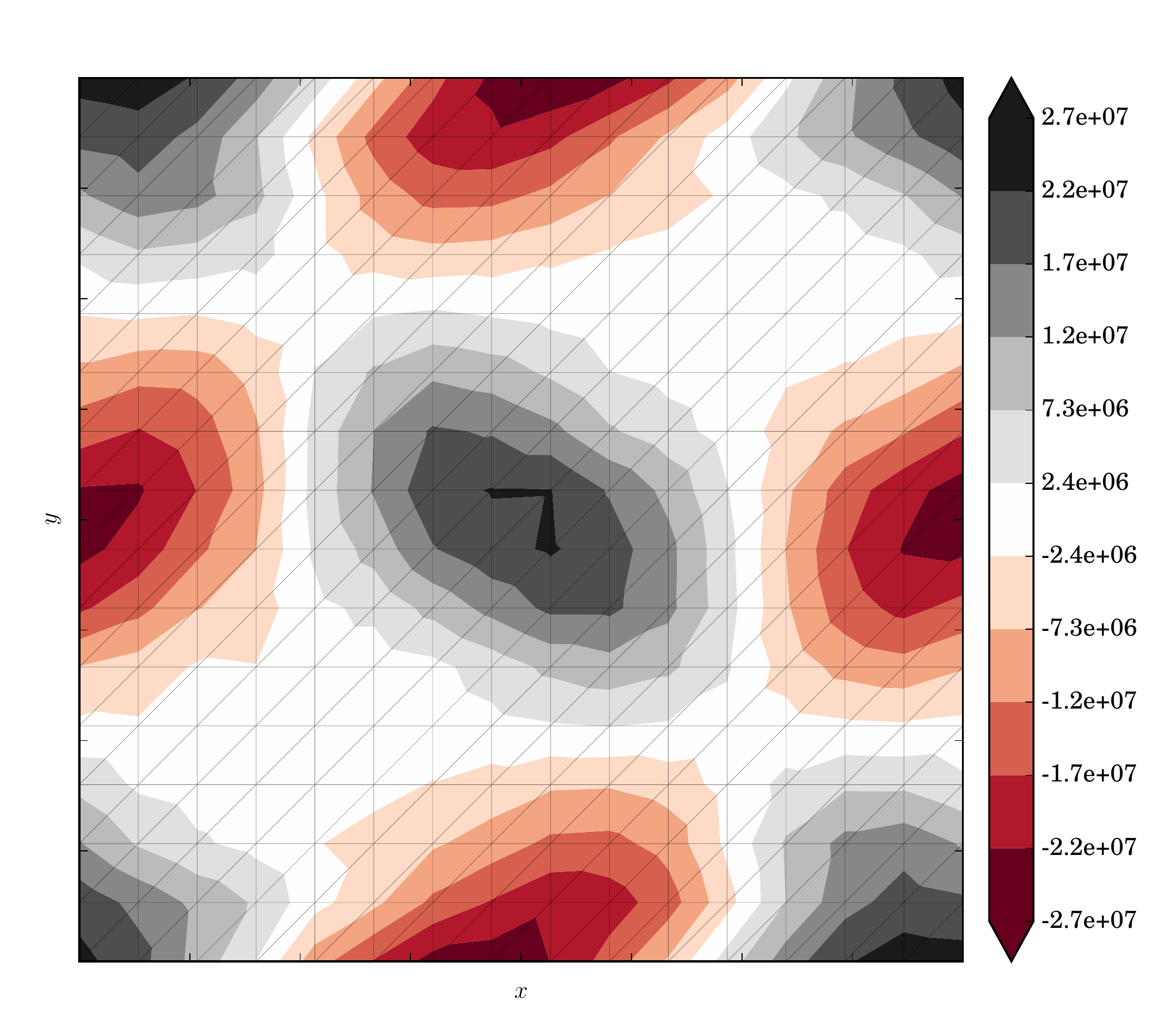}
  \caption{$N_{jz}$}
  \label{fs_N_jz}
  \end{subfigure}

  \begin{subfigure}[b]{0.3\linewidth}
    \includegraphics[width=\linewidth]{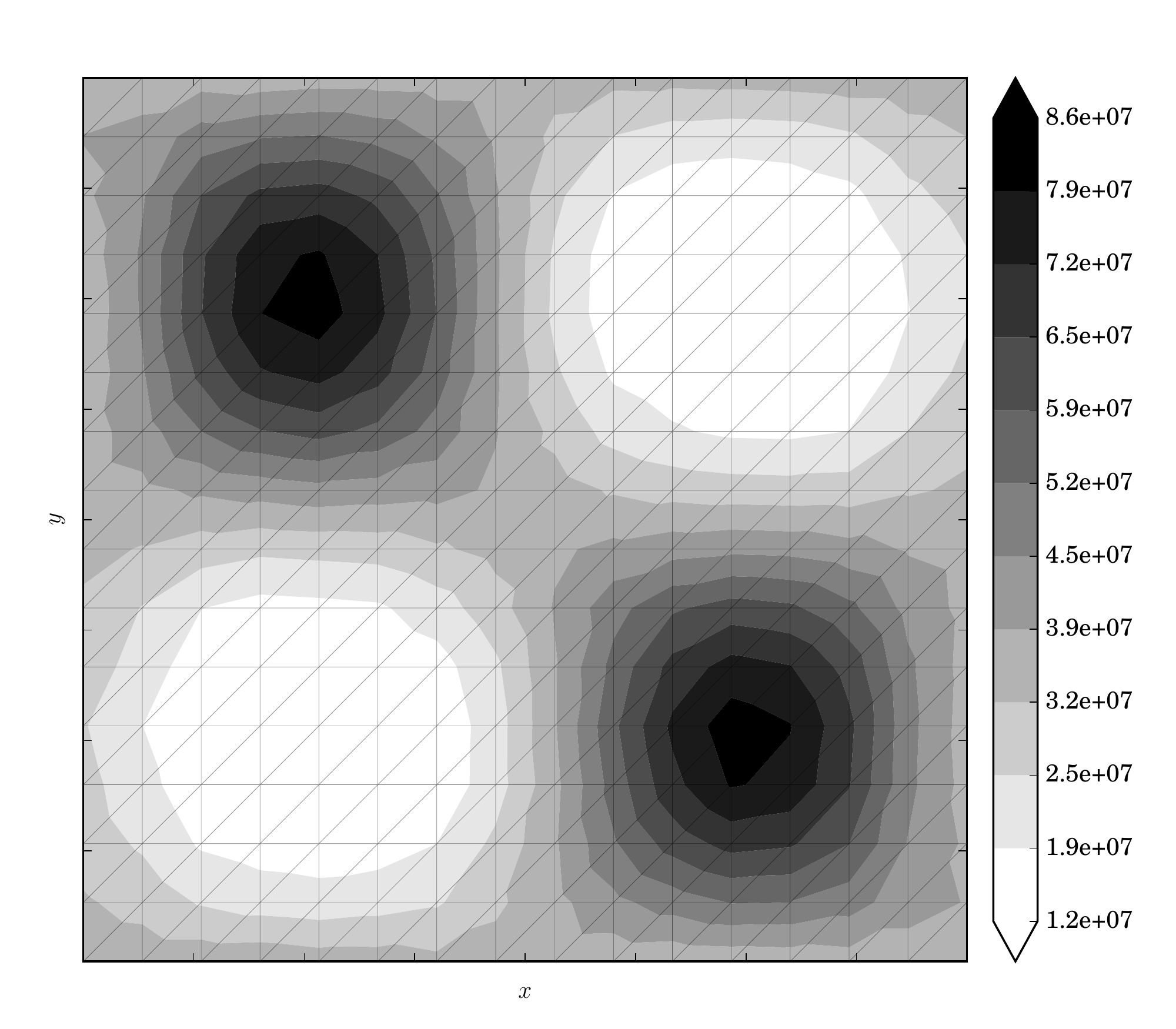}
  \caption{$N_{zi}$}
  \label{fs_N_zi}
  \end{subfigure}
  \begin{subfigure}[b]{0.3\linewidth}
    \includegraphics[width=\linewidth]{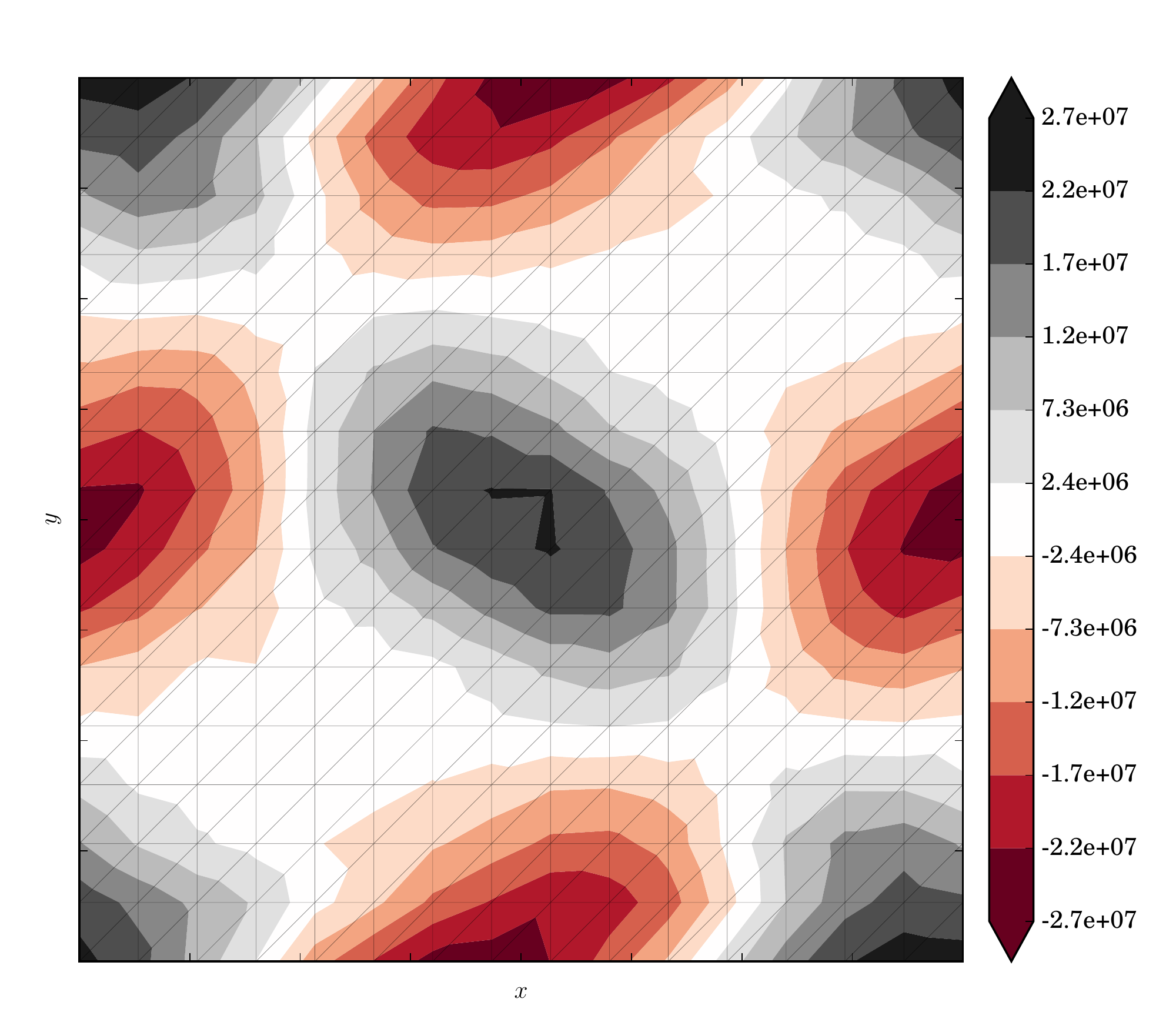}
  \caption{$N_{zj}$}
  \label{fs_N_zj}
  \end{subfigure}
  \begin{subfigure}[b]{0.3\linewidth}
    \includegraphics[width=\linewidth]{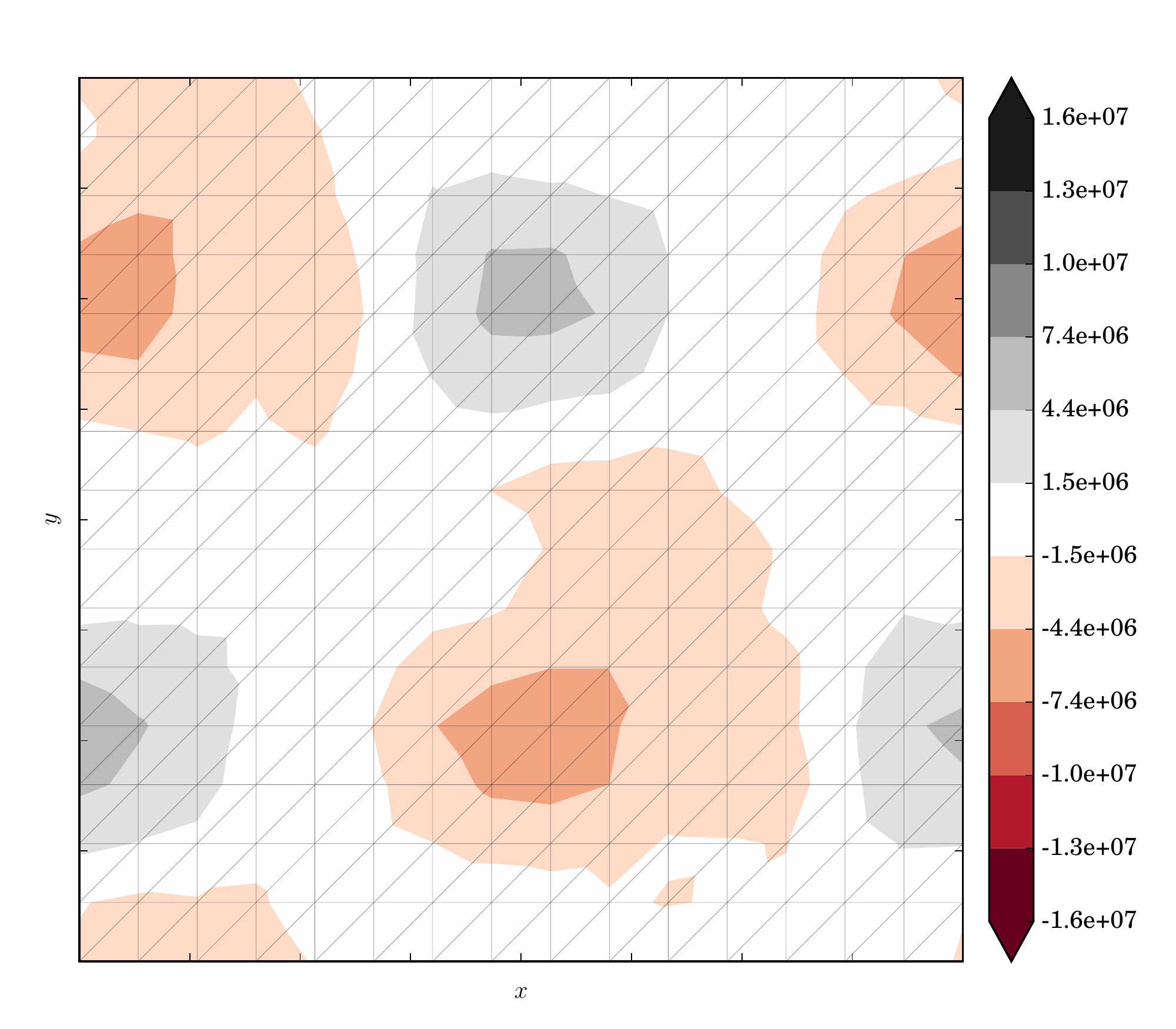}
  \caption{$N_{zz}$}
  \label{fs_N_zz}
  \end{subfigure}
 
  \caption[ISMIP-HOM full-Stokes membrane stress]{Full-Stokes membrane stress $N_{kk}$.}

  \label{fs_membrane_stress}

\end{figure*}

\begin{figure*}
  
  \centering 
  
  \begin{subfigure}[b]{0.3\linewidth}
    \includegraphics[width=\linewidth]{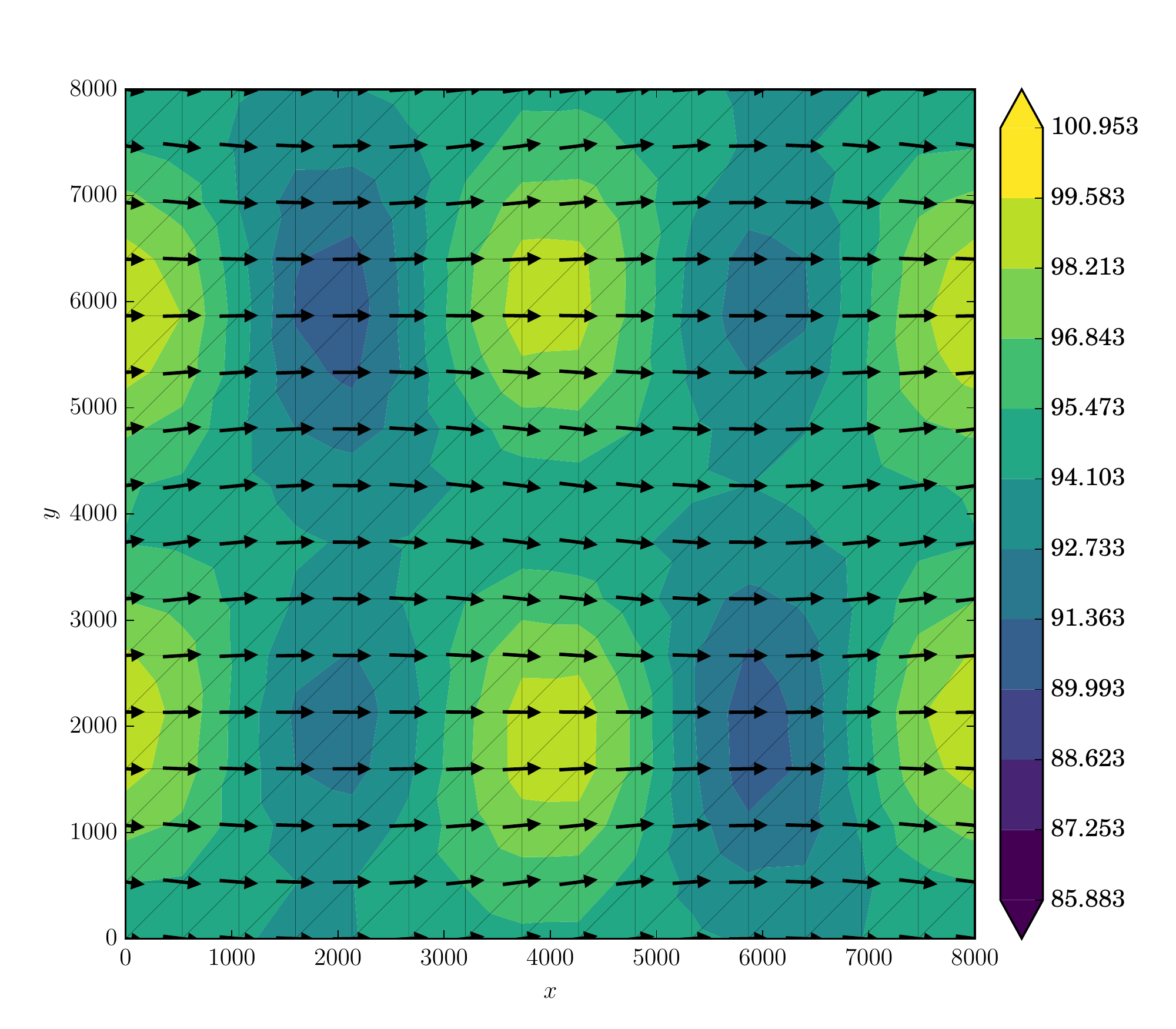}
  \caption{$\mathbf{u}_S$}
  \label{rs_ms_U}
  \end{subfigure}
  \begin{subfigure}[b]{0.3\linewidth}
    \includegraphics[width=\linewidth]{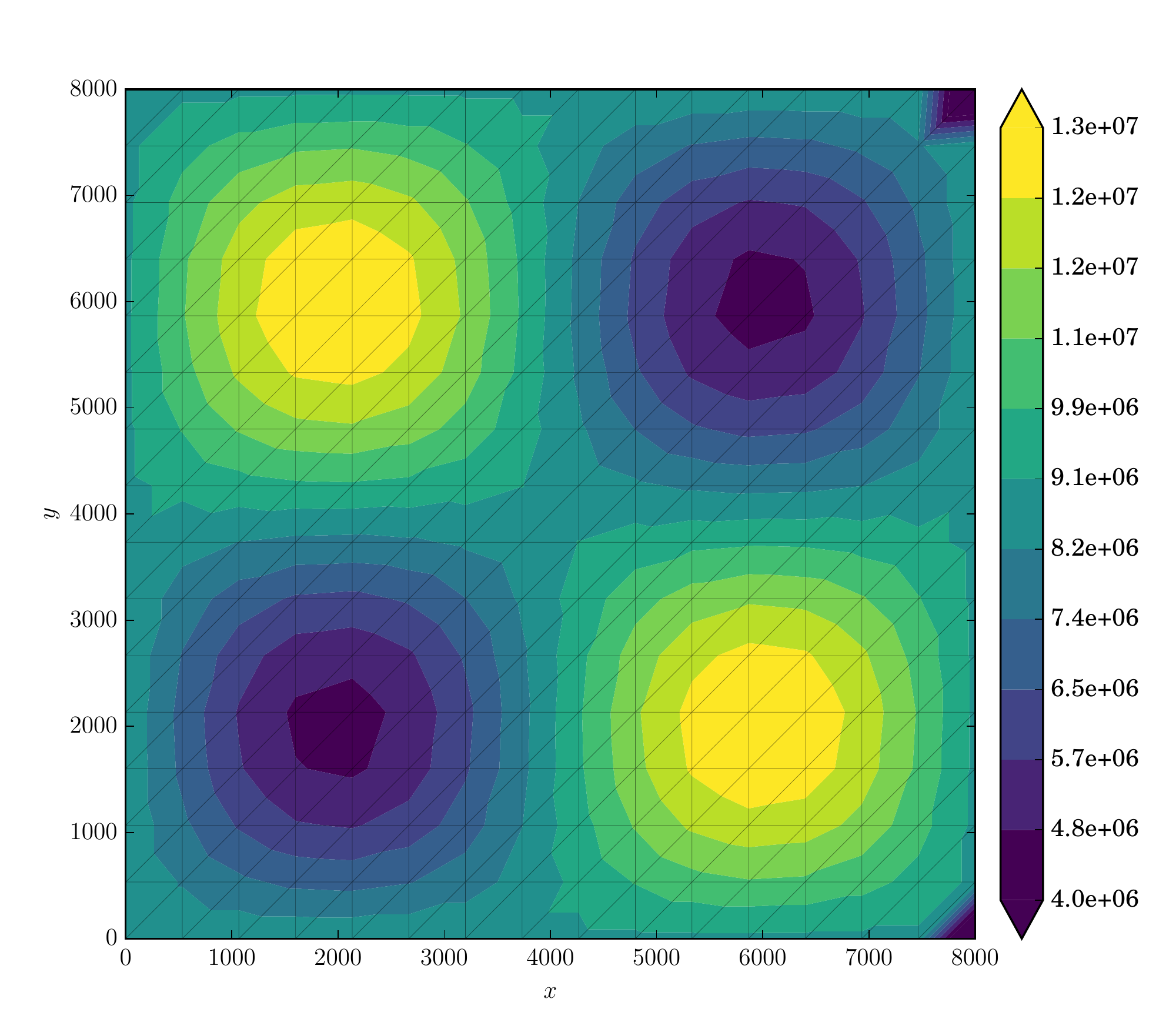}
  \caption{$p |_B$}
  \label{rs_ms_p}
  \end{subfigure}

  \begin{subfigure}[b]{0.3\linewidth}
    \includegraphics[width=\linewidth]{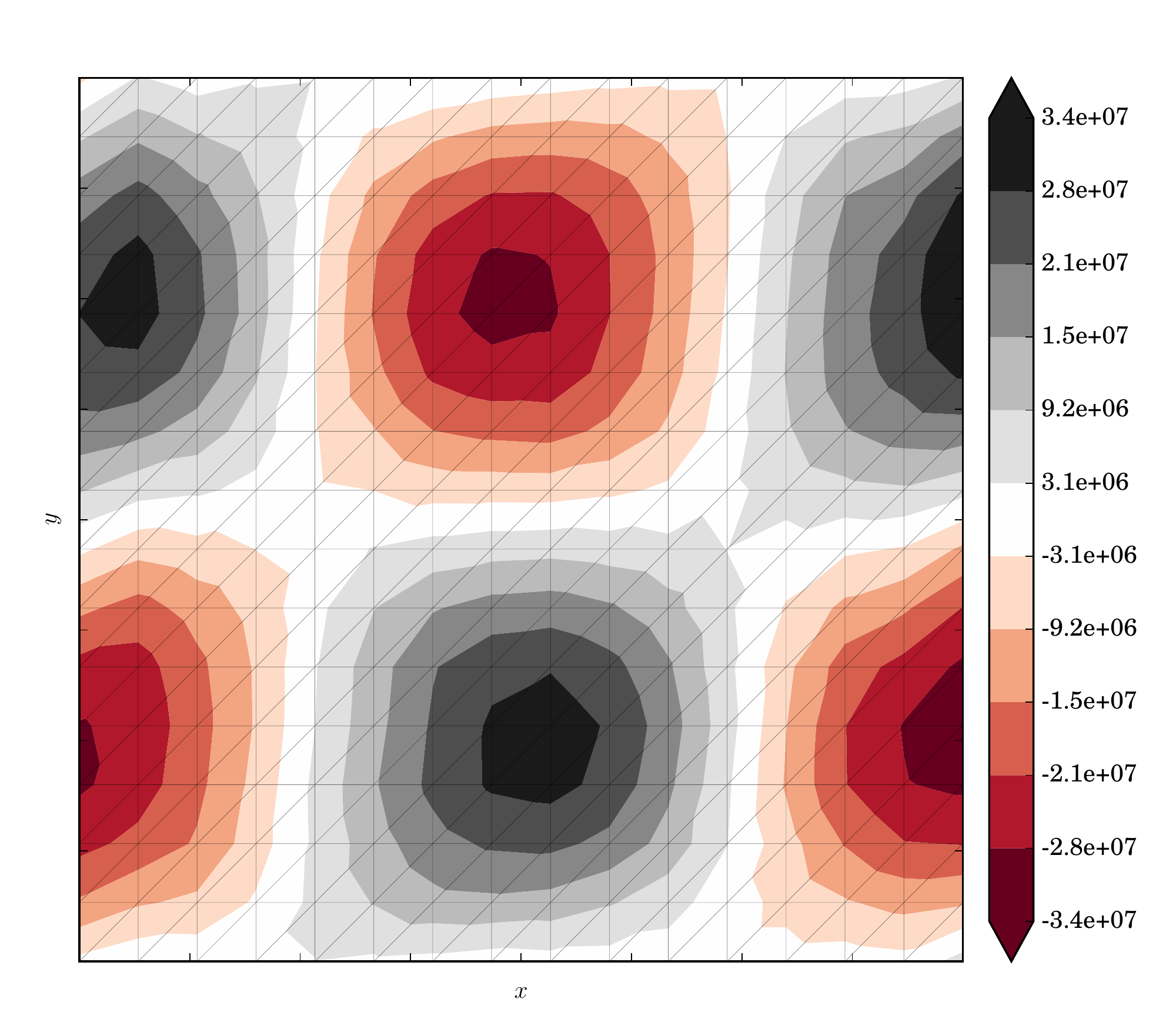}
  \caption{$N_{ii}$}
  \label{rs_N_ii}
  \end{subfigure}
  \begin{subfigure}[b]{0.3\linewidth}
    \includegraphics[width=\linewidth]{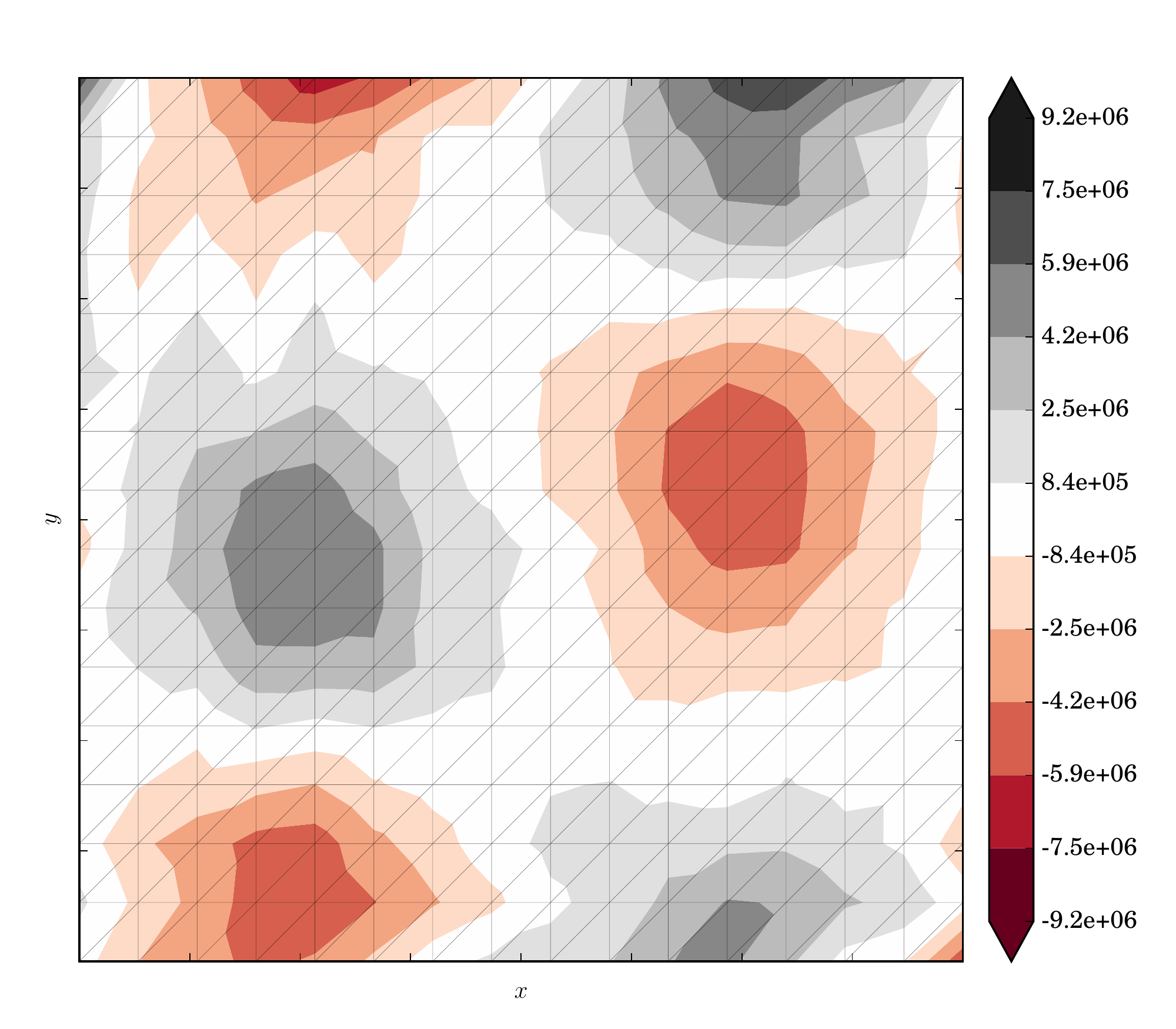}
  \caption{$N_{ij}$}
  \label{rs_N_ij}
  \end{subfigure}
  \begin{subfigure}[b]{0.3\linewidth}
    \includegraphics[width=\linewidth]{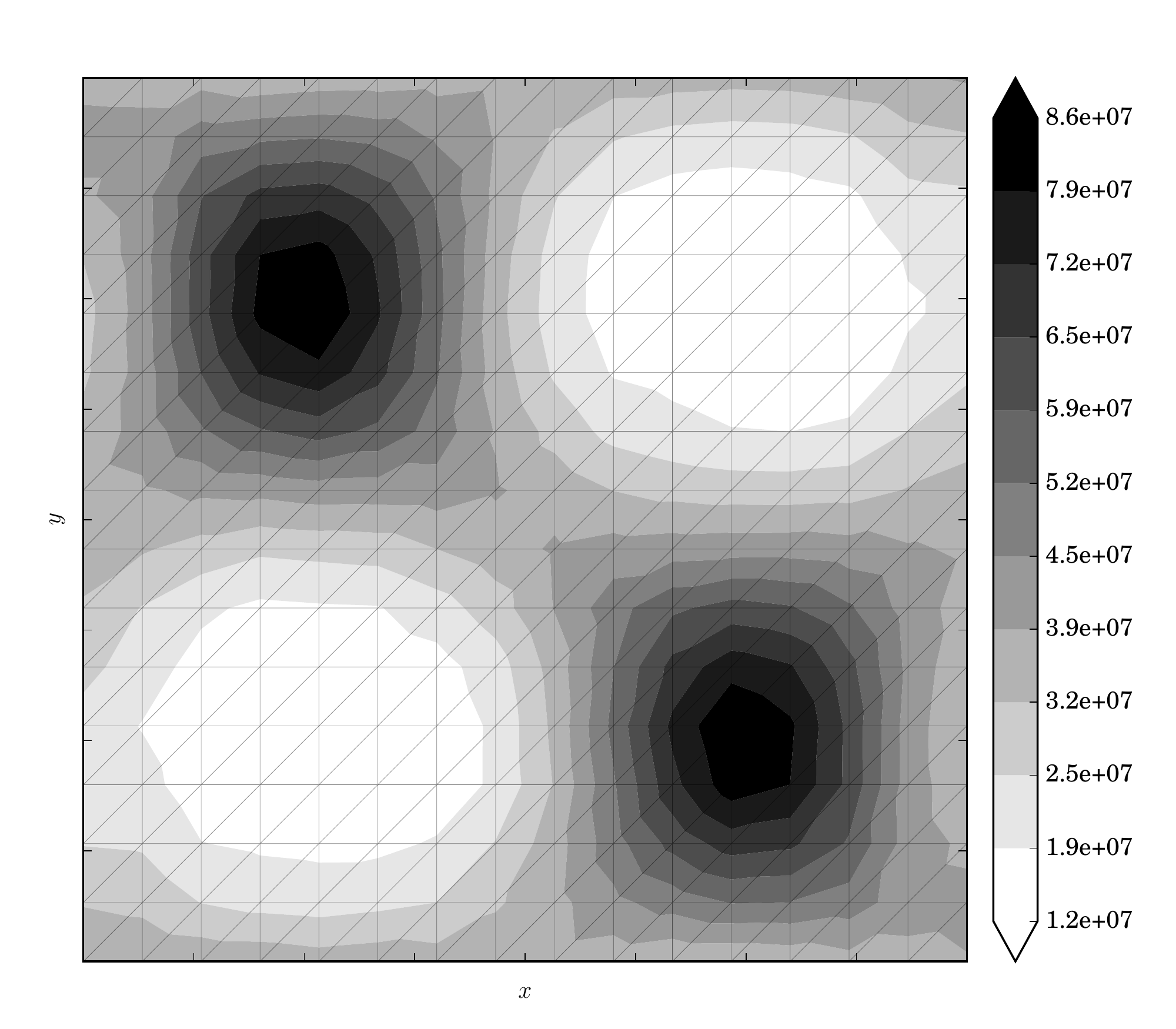}
  \caption{$N_{iz}$}
  \label{rs_N_iz}
  \end{subfigure}

  \begin{subfigure}[b]{0.3\linewidth}
    \includegraphics[width=\linewidth]{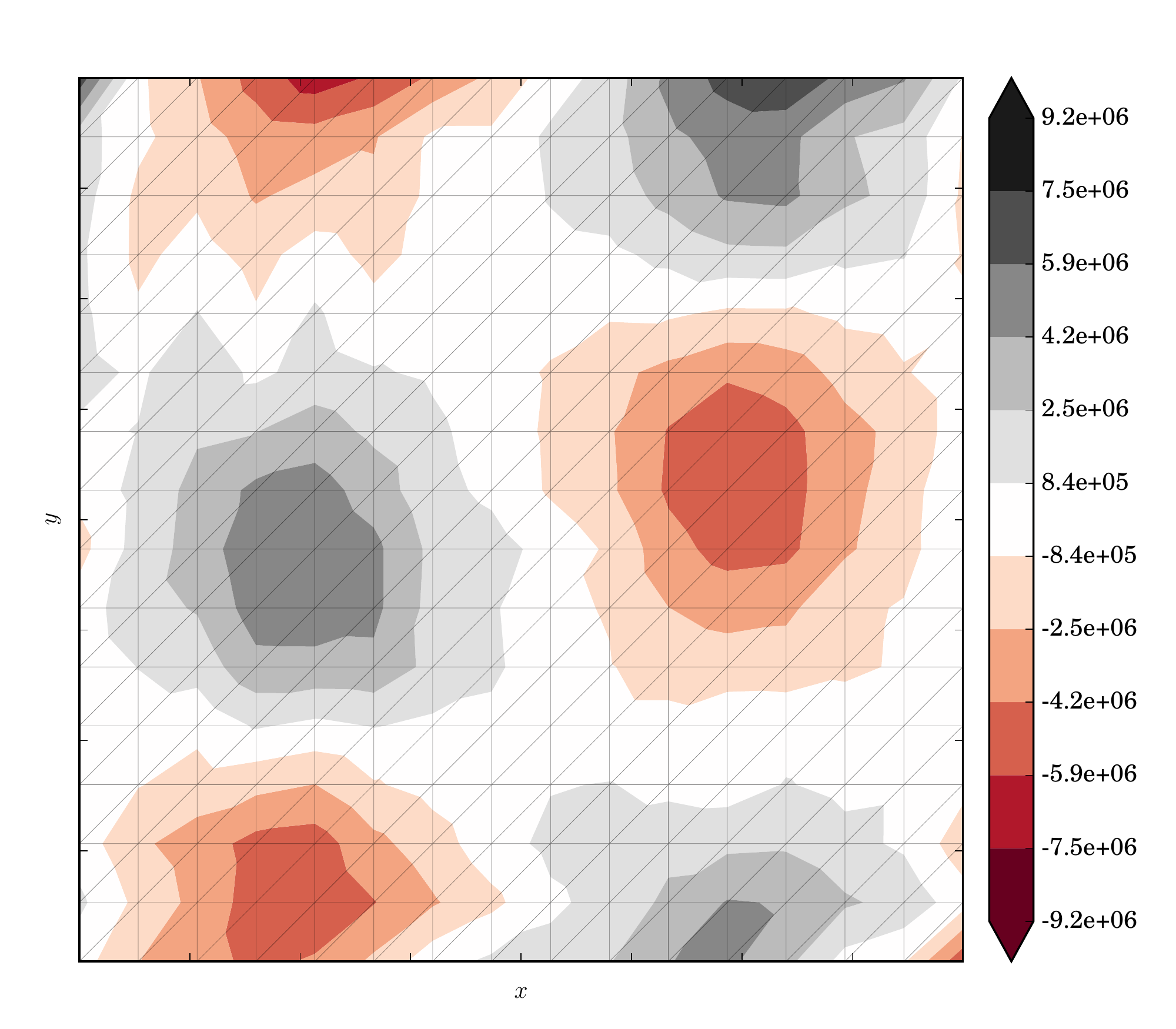}
  \caption{$N_{ji}$}
  \label{rs_N_ji}
  \end{subfigure}
  \begin{subfigure}[b]{0.3\linewidth}
    \includegraphics[width=\linewidth]{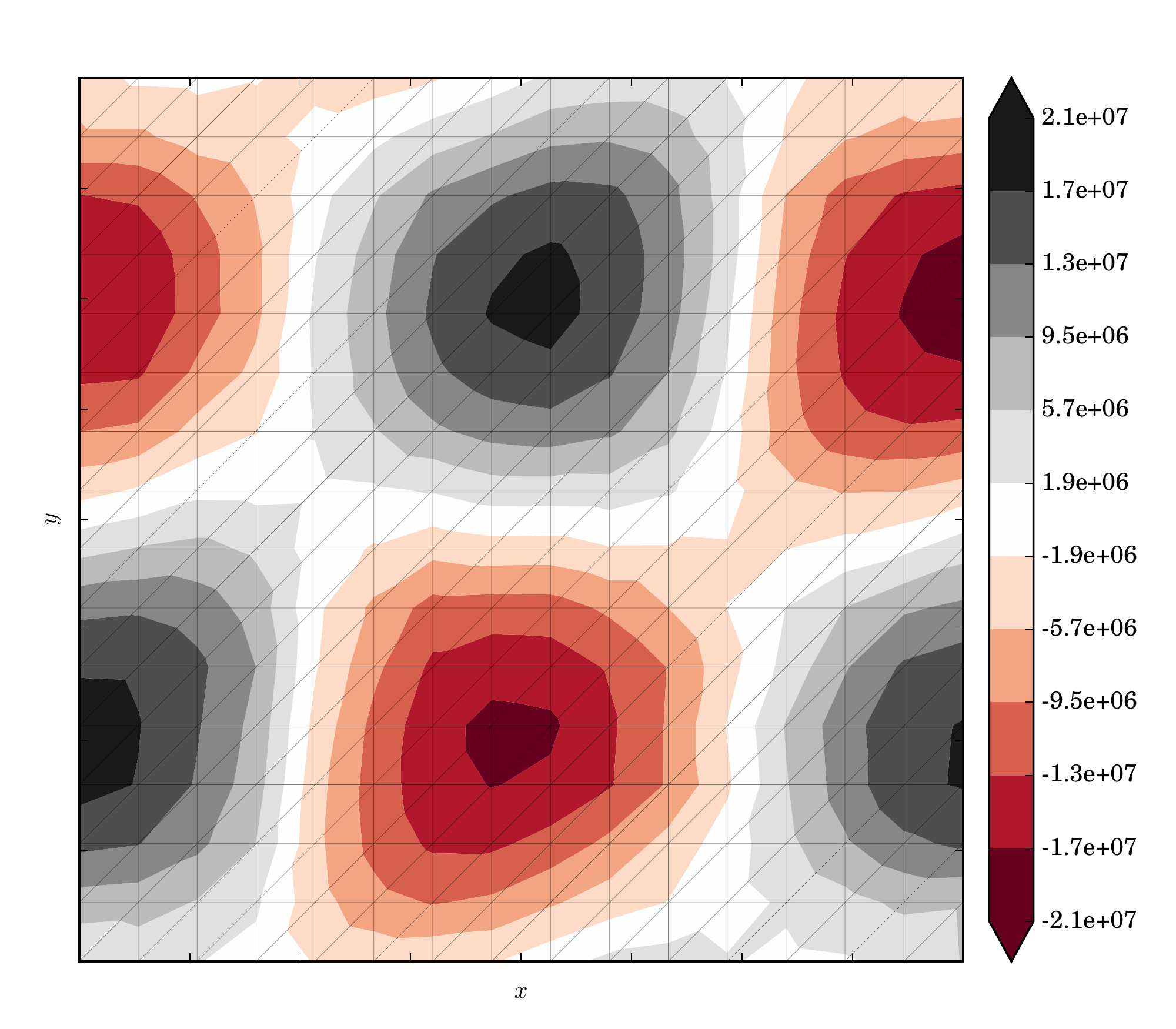}
  \caption{$N_{jj}$}
  \label{rs_N_jj}
  \end{subfigure}
  \begin{subfigure}[b]{0.3\linewidth}
    \includegraphics[width=\linewidth]{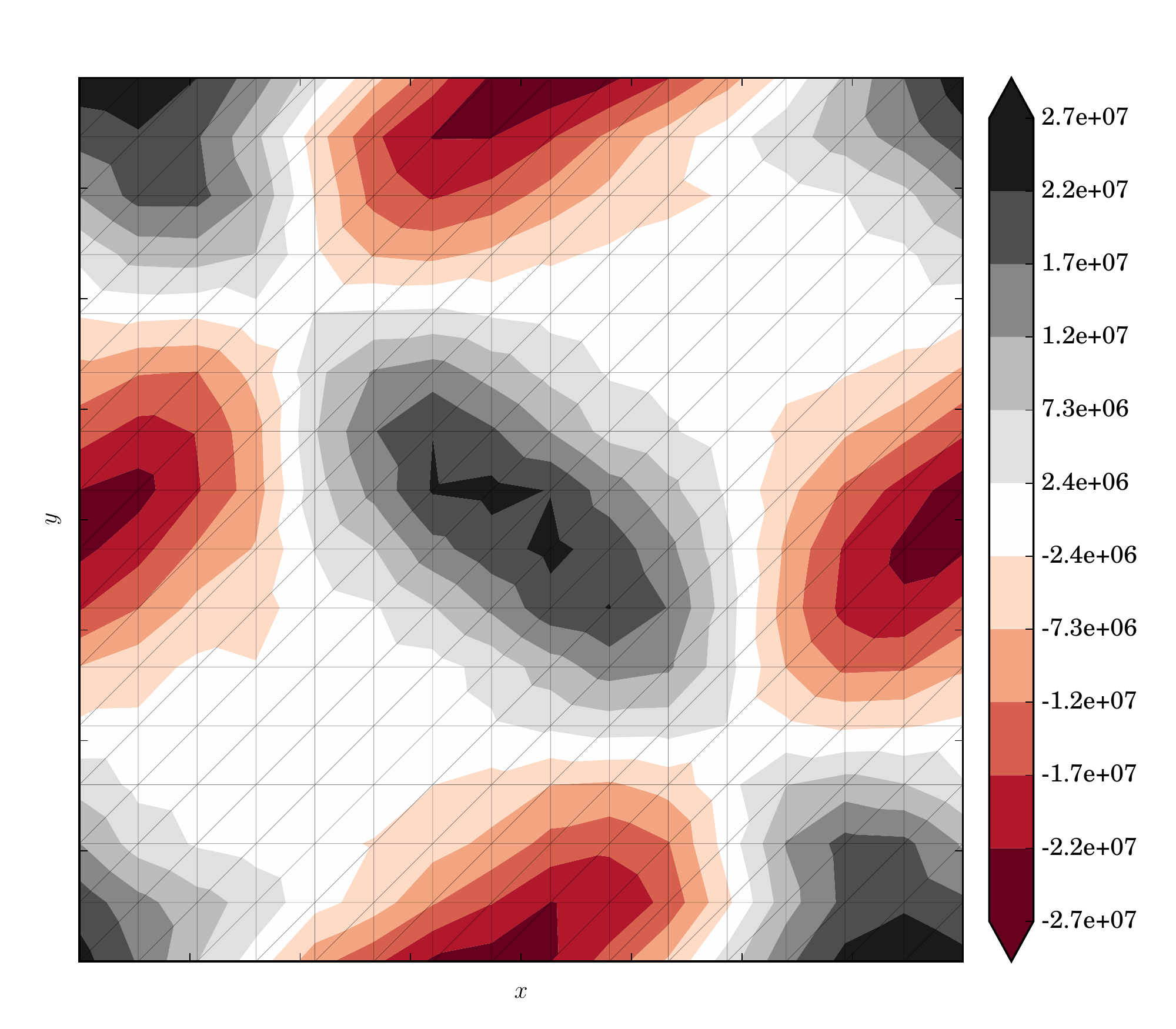}
  \caption{$N_{jz}$}
  \label{rs_N_jz}
  \end{subfigure}

  \begin{subfigure}[b]{0.3\linewidth}
    \includegraphics[width=\linewidth]{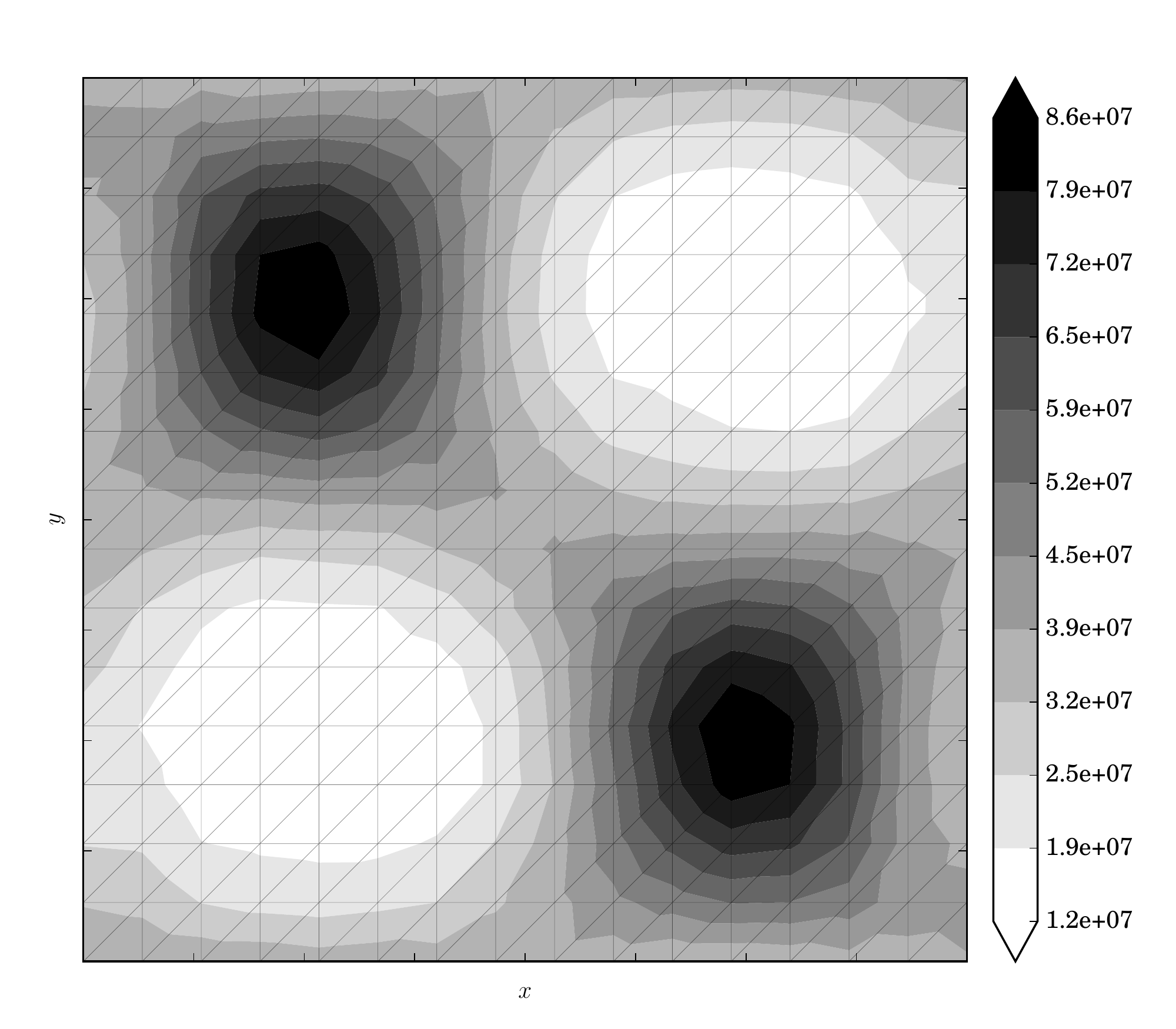}
  \caption{$N_{zi}$}
  \label{rs_N_zi}
  \end{subfigure}
  \begin{subfigure}[b]{0.3\linewidth}
    \includegraphics[width=\linewidth]{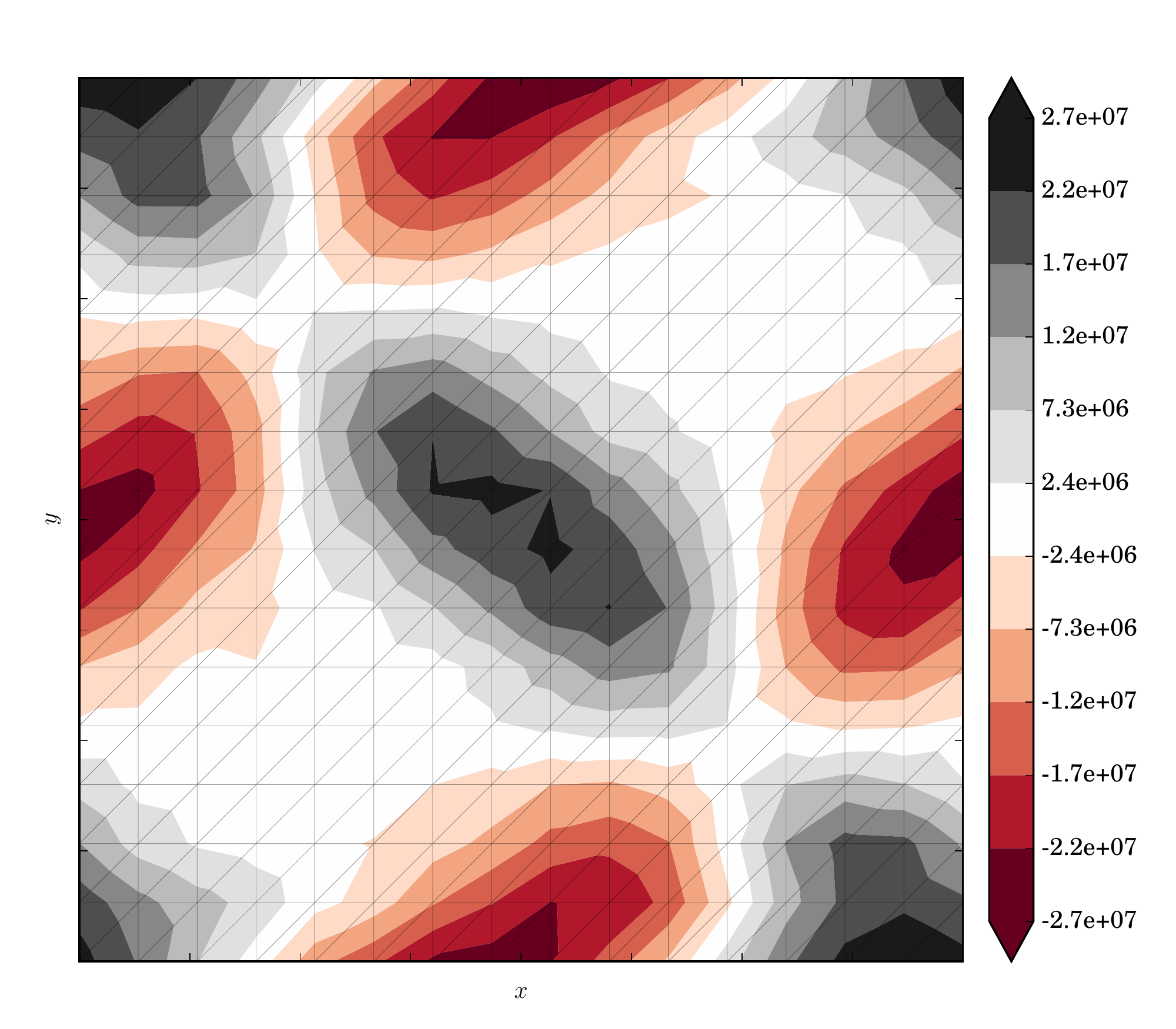}
  \caption{$N_{zj}$}
  \label{rs_N_zj}
  \end{subfigure}
  \begin{subfigure}[b]{0.3\linewidth}
    \includegraphics[width=\linewidth]{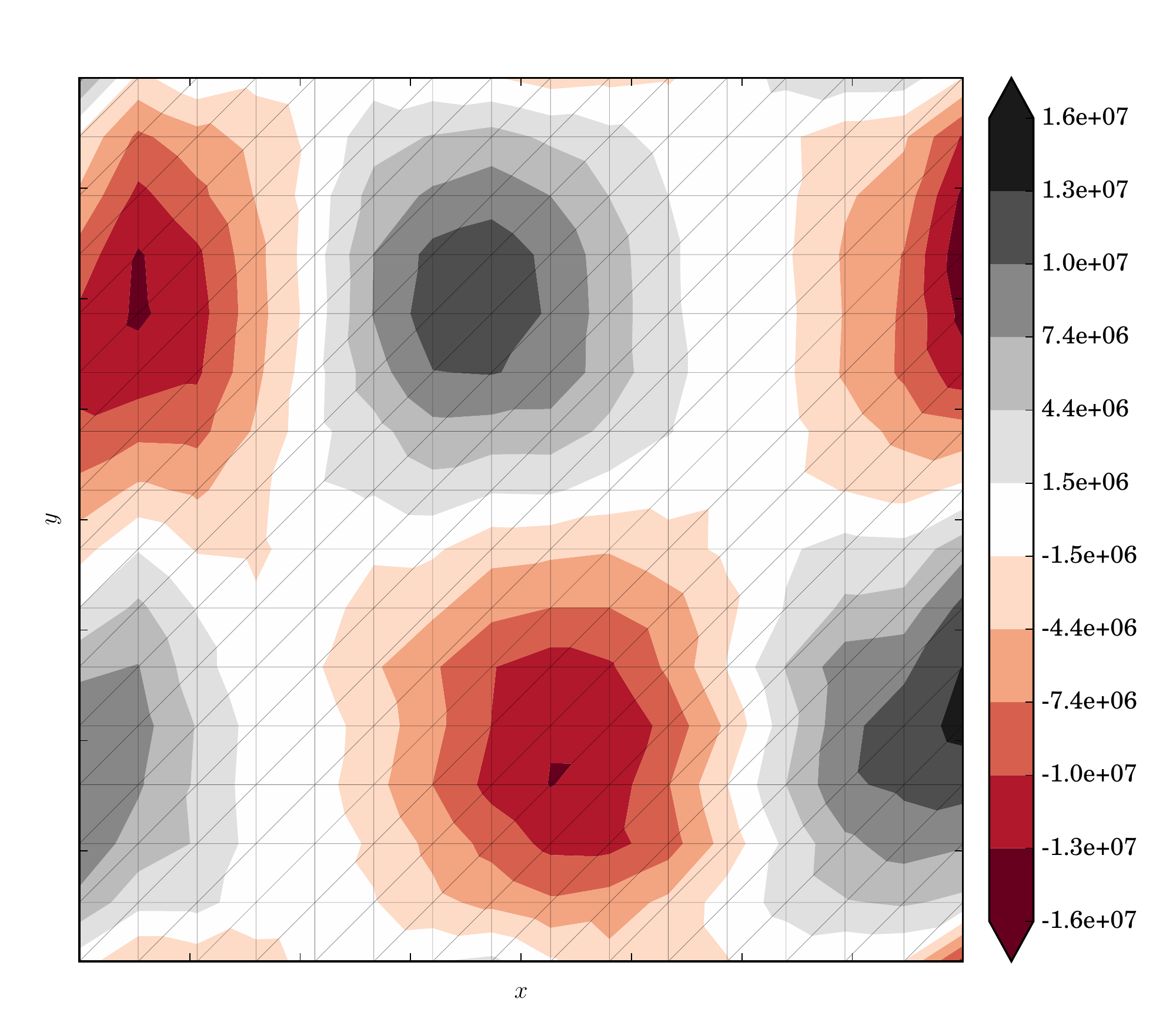}
  \caption{$N_{zz}$}
  \label{rs_N_zz}
  \end{subfigure}
 
  \caption[ISMIP-HOM reformulated-Stokes membrane stress]{Reformulated-Stokes membrane stress $N_{kk}$.}

  \label{rs_membrane_stress}

\end{figure*}

\begin{figure*}
  
  \centering 

  \begin{subfigure}[b]{0.3\linewidth}
    \includegraphics[width=\linewidth]{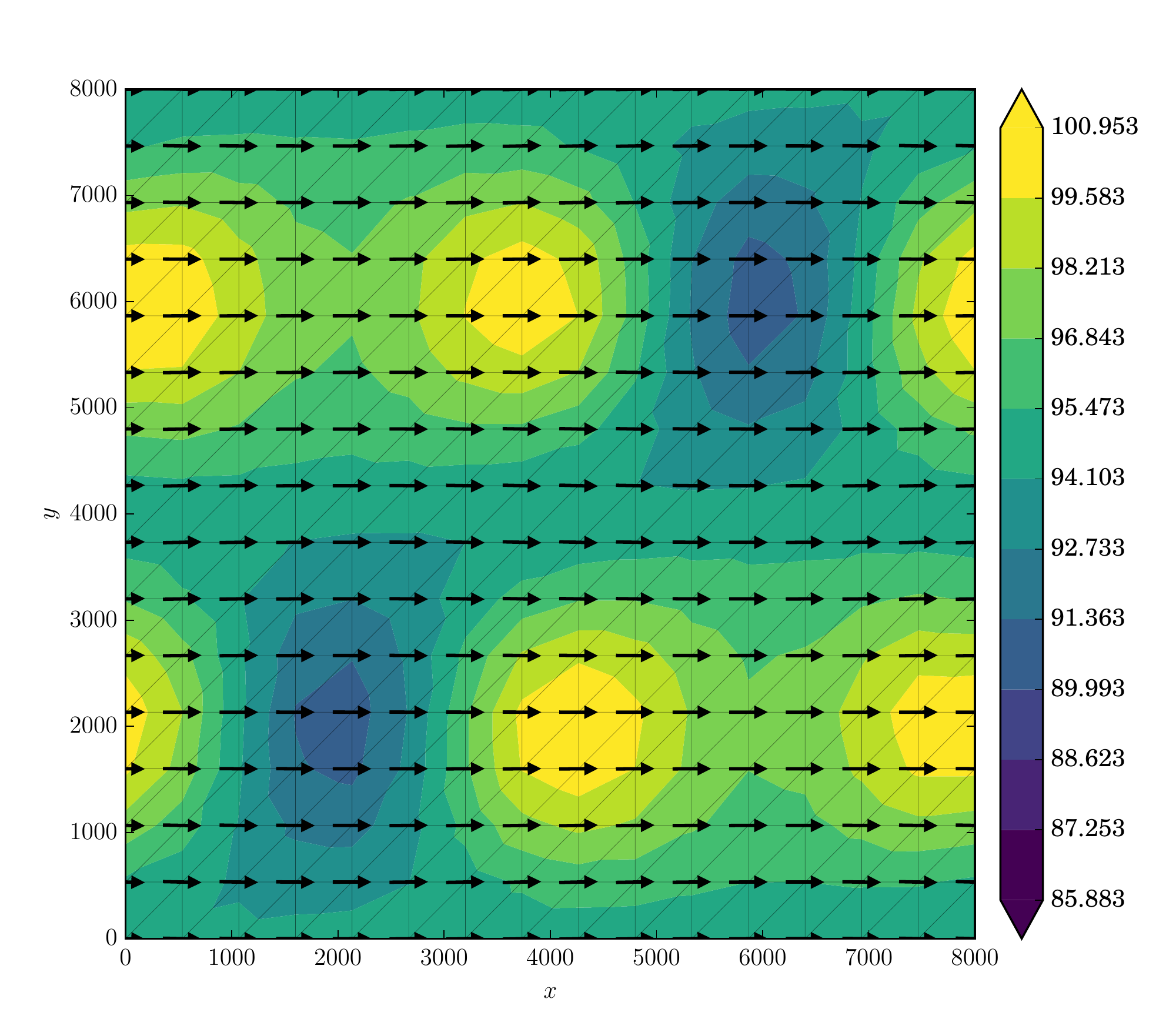}
  \caption{$\mathbf{u}_S$}
  \label{bp_ms_U}
  \end{subfigure}
  \begin{subfigure}[b]{0.3\linewidth}
    \includegraphics[width=\linewidth]{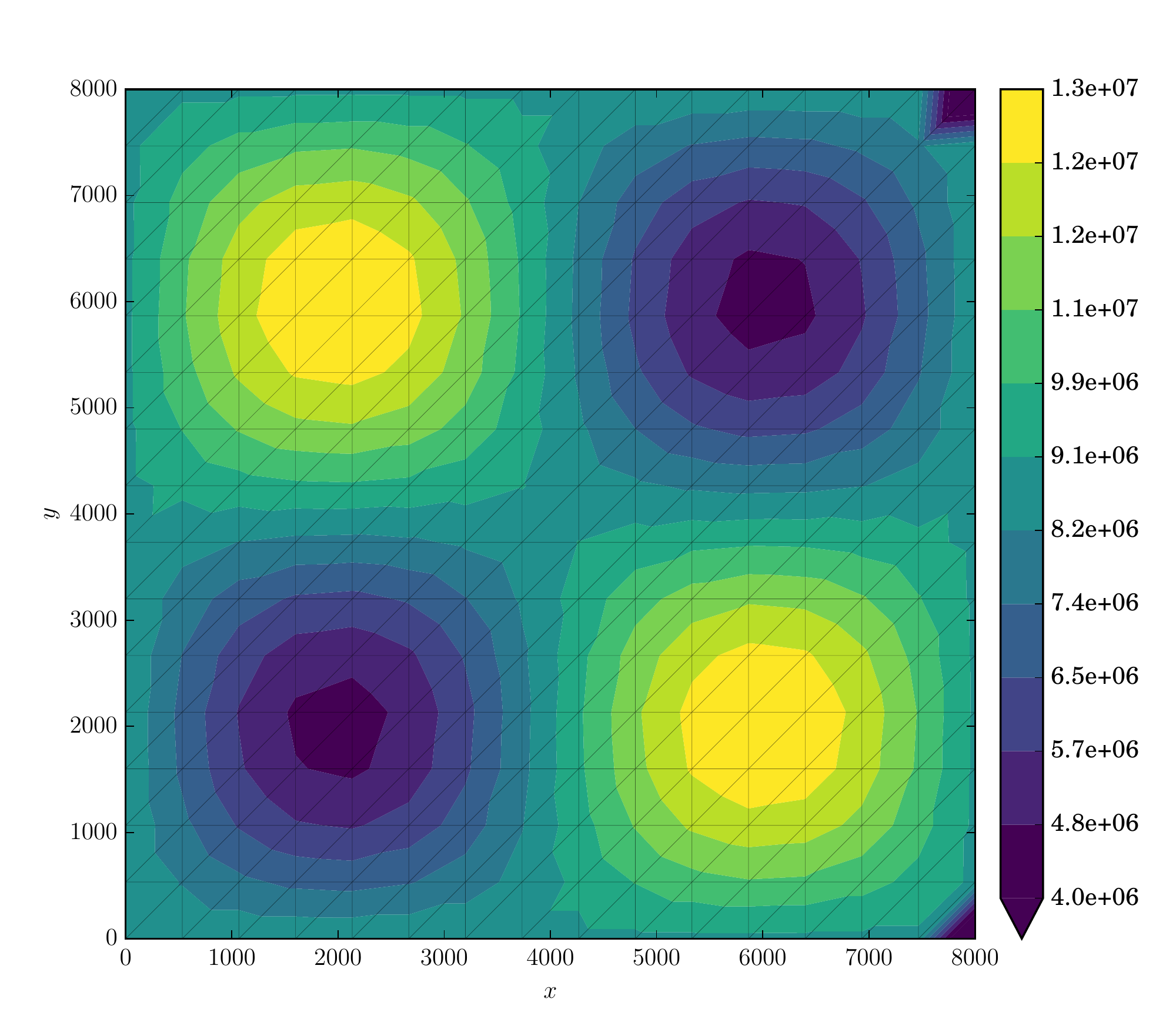}
  \caption{$p |_B$}
  \label{bp_ms_p}
  \end{subfigure}

  \begin{subfigure}[b]{0.3\linewidth}
    \includegraphics[width=\linewidth]{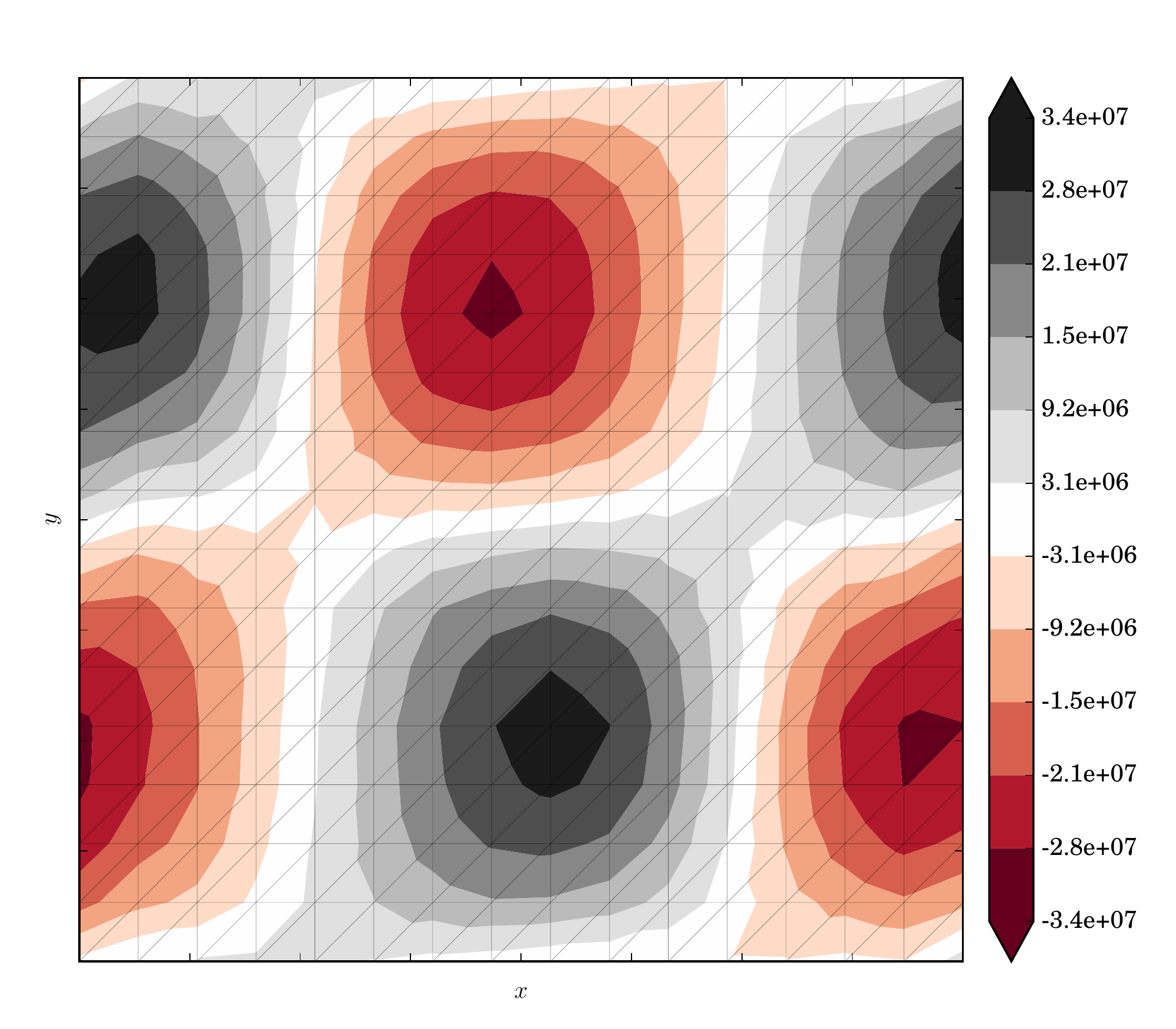}
  \caption{$N_{ii}$}
  \label{bp_N_ii}
  \end{subfigure}
  \begin{subfigure}[b]{0.3\linewidth}
    \includegraphics[width=\linewidth]{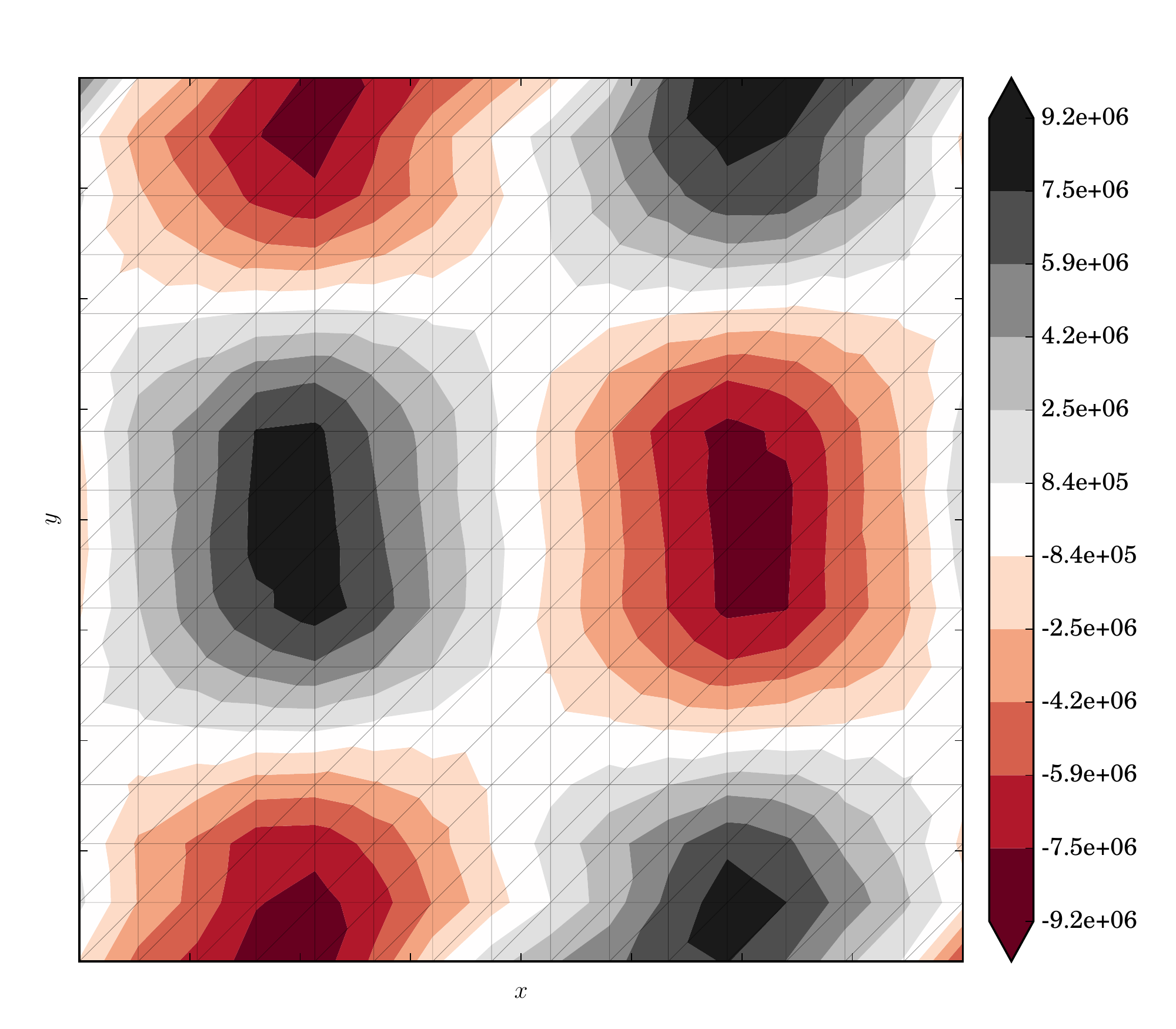}
  \caption{$N_{ij}$}
  \label{bp_N_ij}
  \end{subfigure}
  \begin{subfigure}[b]{0.3\linewidth}
    \includegraphics[width=\linewidth]{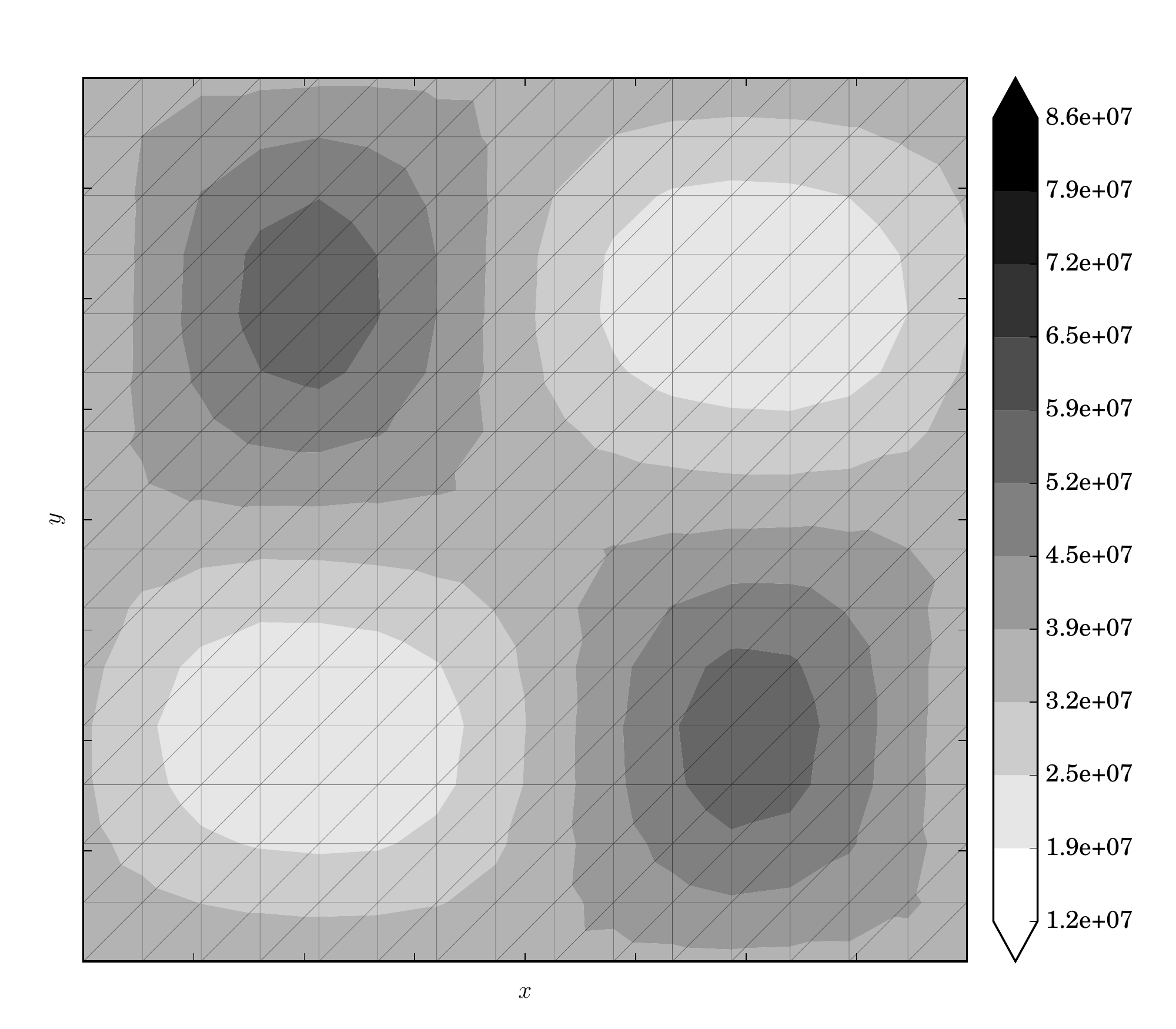}
  \caption{$N_{iz}$}
  \label{bp_N_iz}
  \end{subfigure}

  \begin{subfigure}[b]{0.3\linewidth}
    \includegraphics[width=\linewidth]{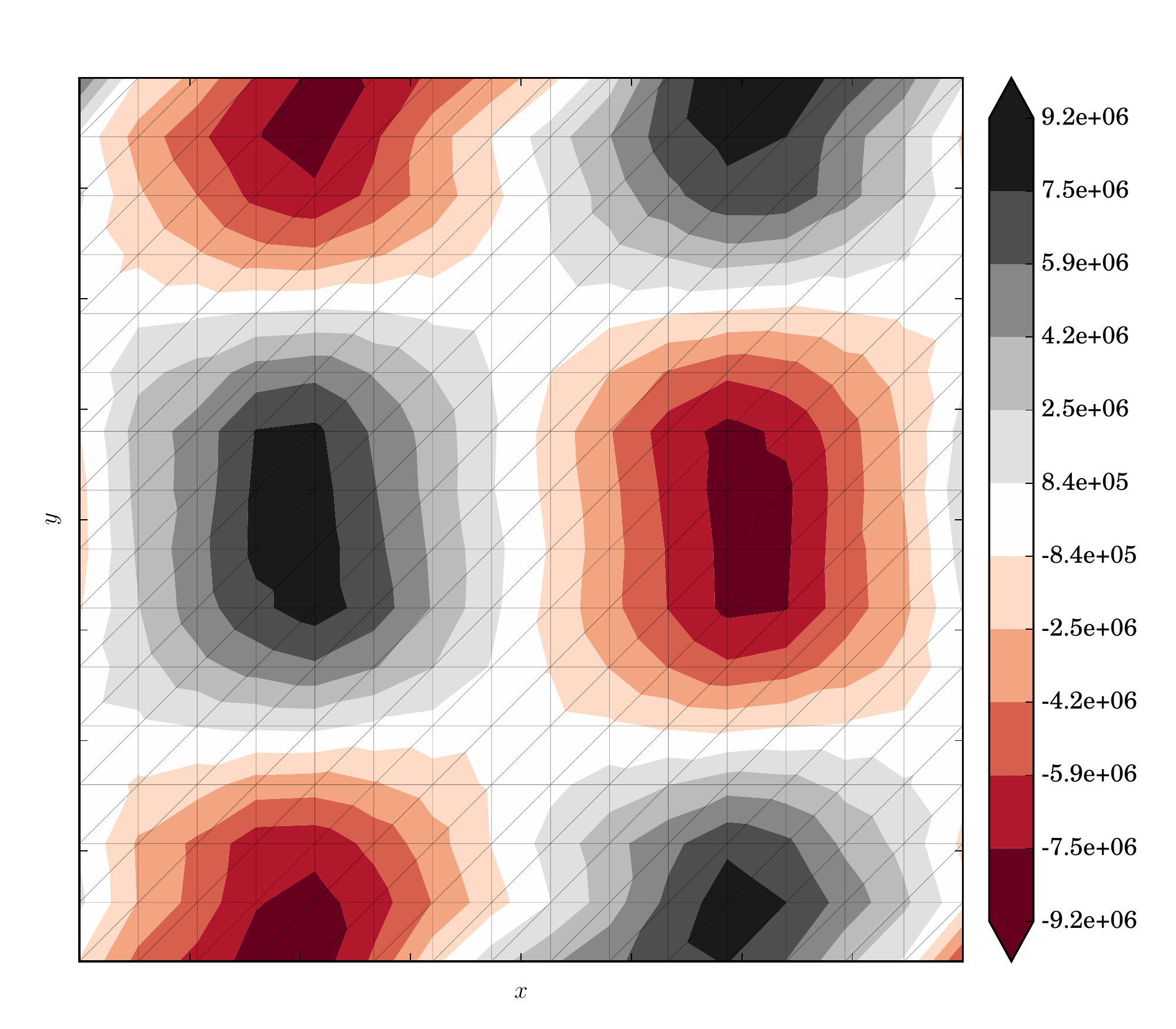}
  \caption{$N_{ji}$}
  \label{bp_N_ji}
  \end{subfigure}
  \begin{subfigure}[b]{0.3\linewidth}
    \includegraphics[width=\linewidth]{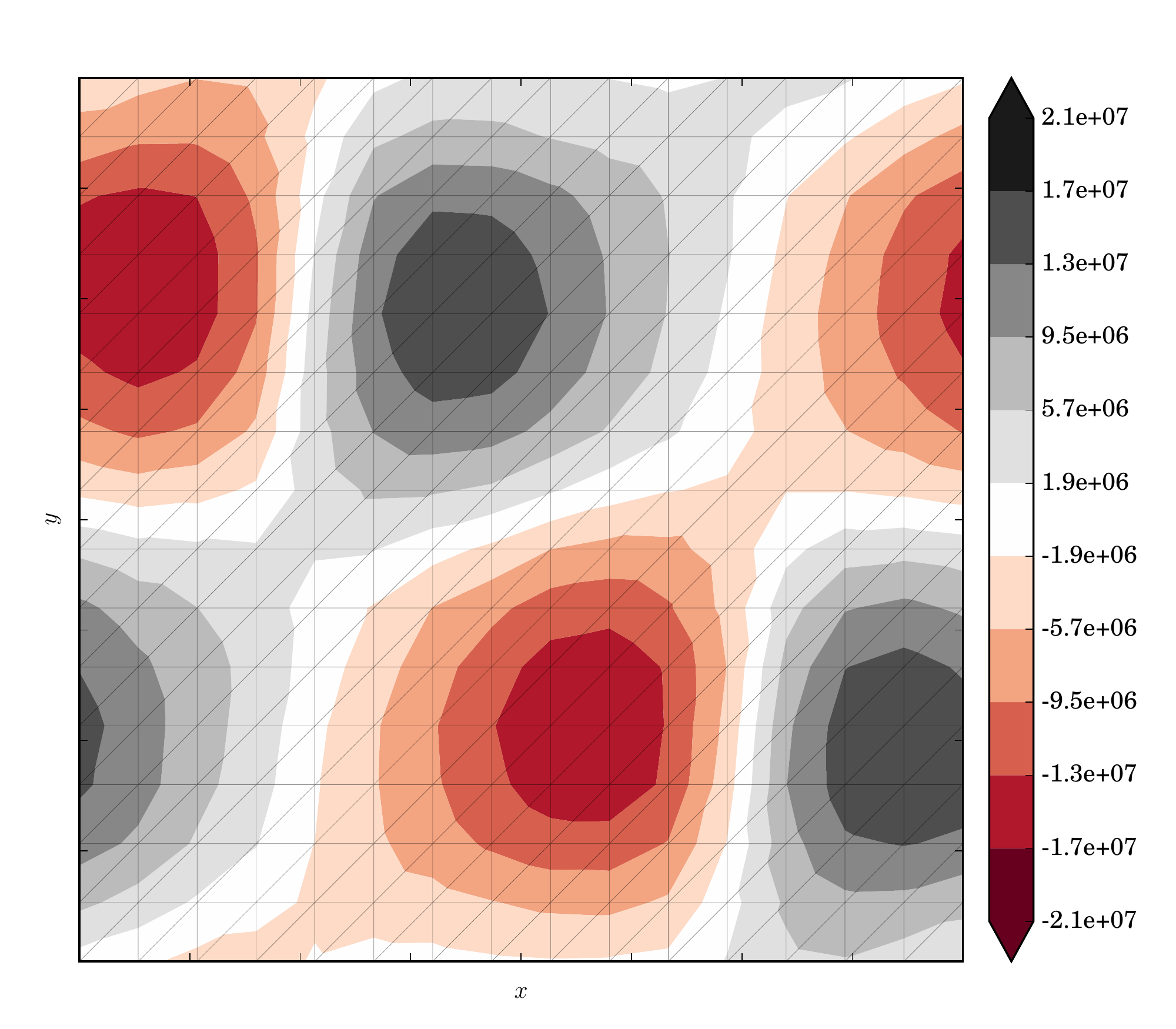}
  \caption{$N_{jj}$}
  \label{bp_N_jj}
  \end{subfigure}
  \begin{subfigure}[b]{0.3\linewidth}
    \includegraphics[width=\linewidth]{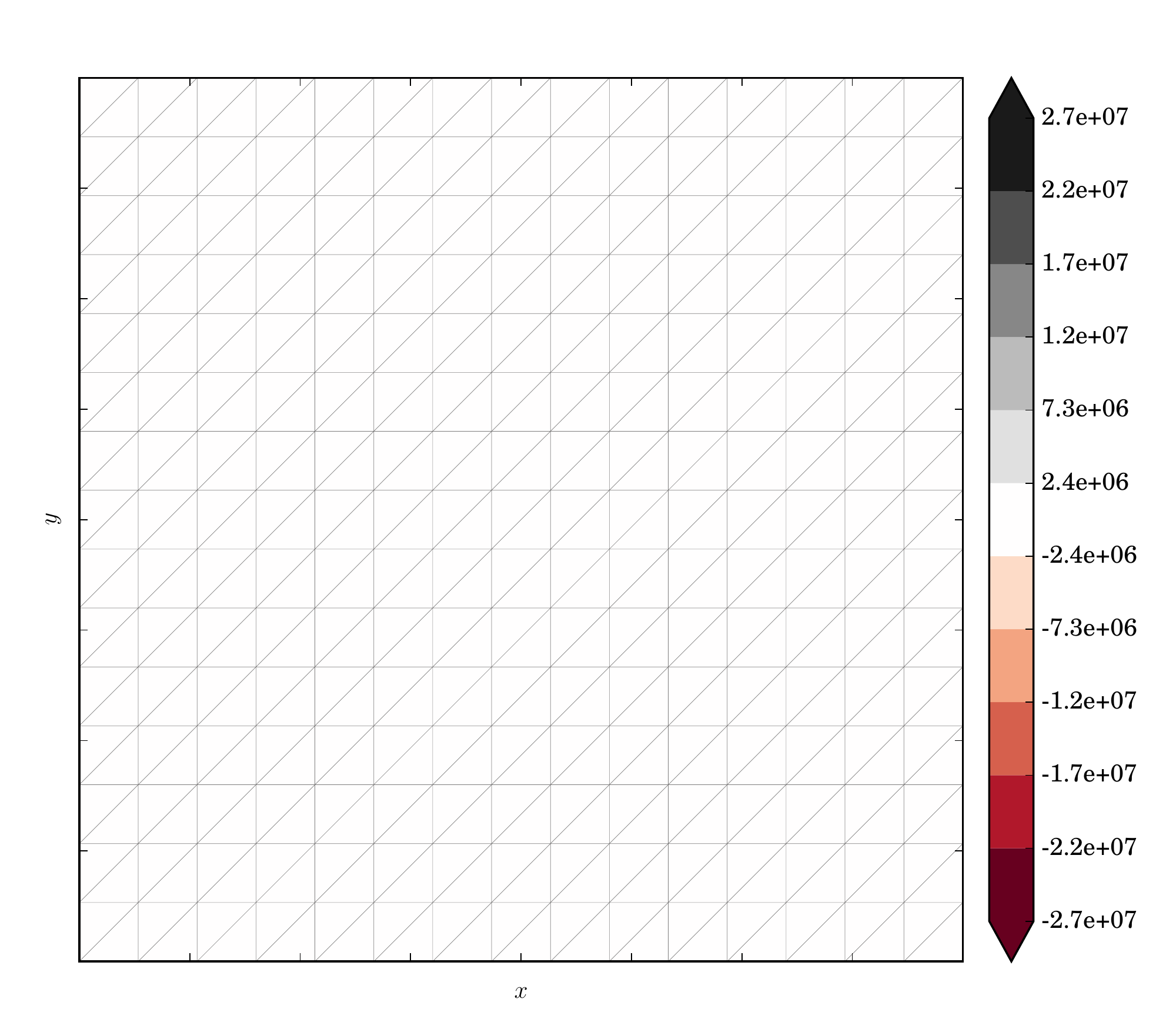}
  \caption{$N_{jz}$}
  \label{bp_N_jz}
  \end{subfigure}

  \begin{subfigure}[b]{0.3\linewidth}
    \includegraphics[width=\linewidth]{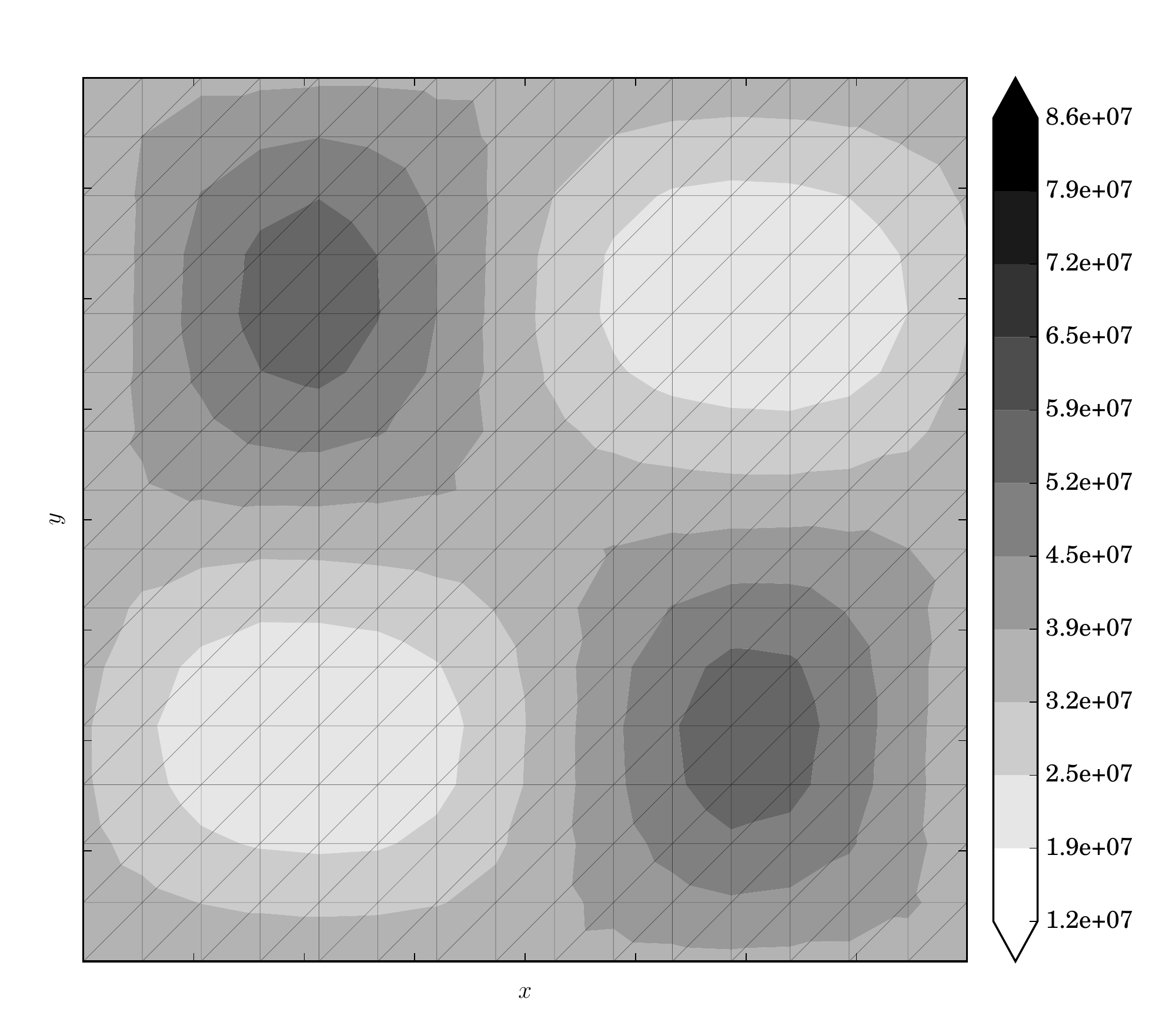}
  \caption{$N_{zi}$}
  \label{bp_N_zi}
  \end{subfigure}
  \begin{subfigure}[b]{0.3\linewidth}
    \includegraphics[width=\linewidth]{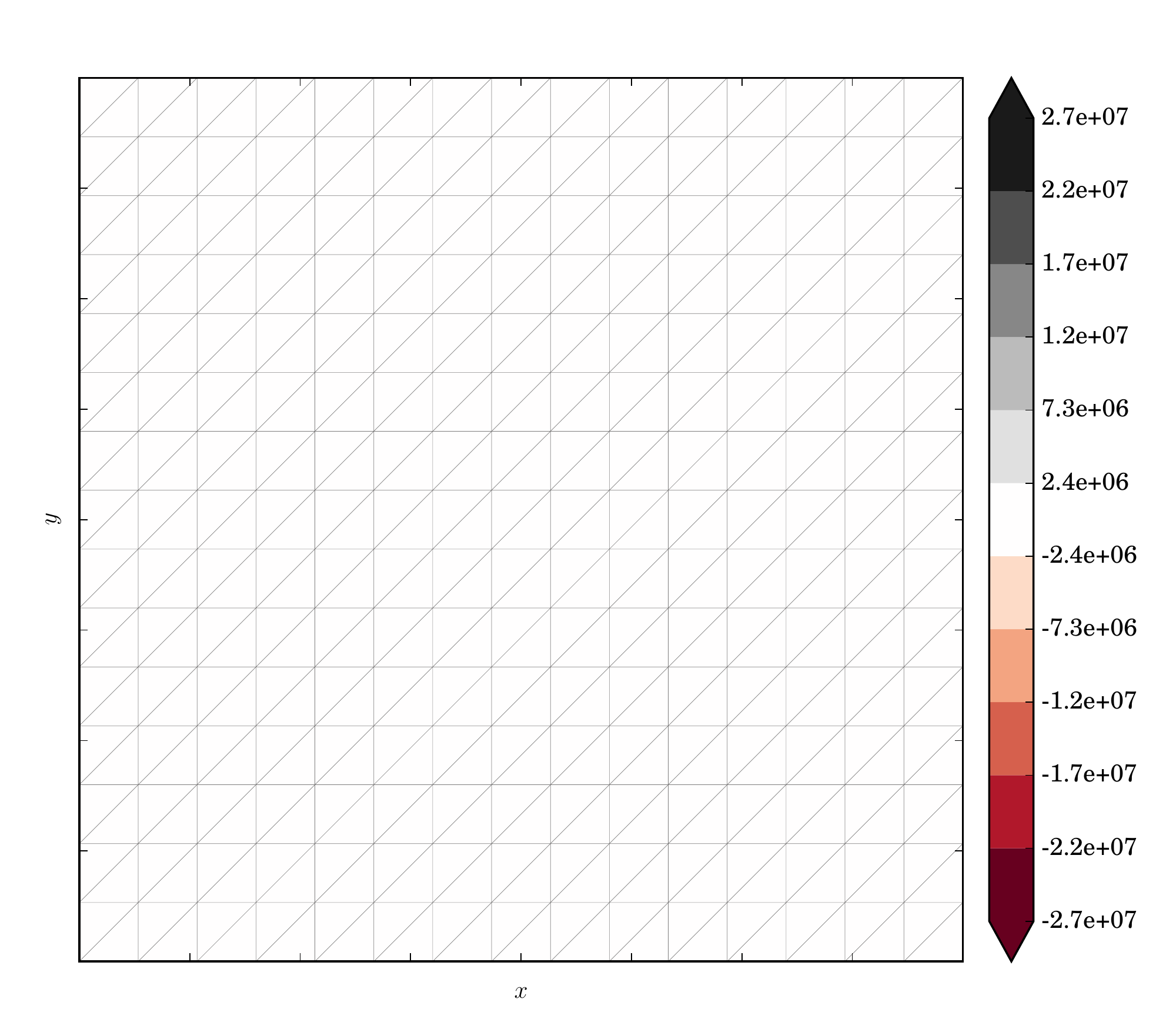}
  \caption{$N_{zj}$}
  \label{bp_N_zj}
  \end{subfigure}
  \begin{subfigure}[b]{0.3\linewidth}
    \includegraphics[width=\linewidth]{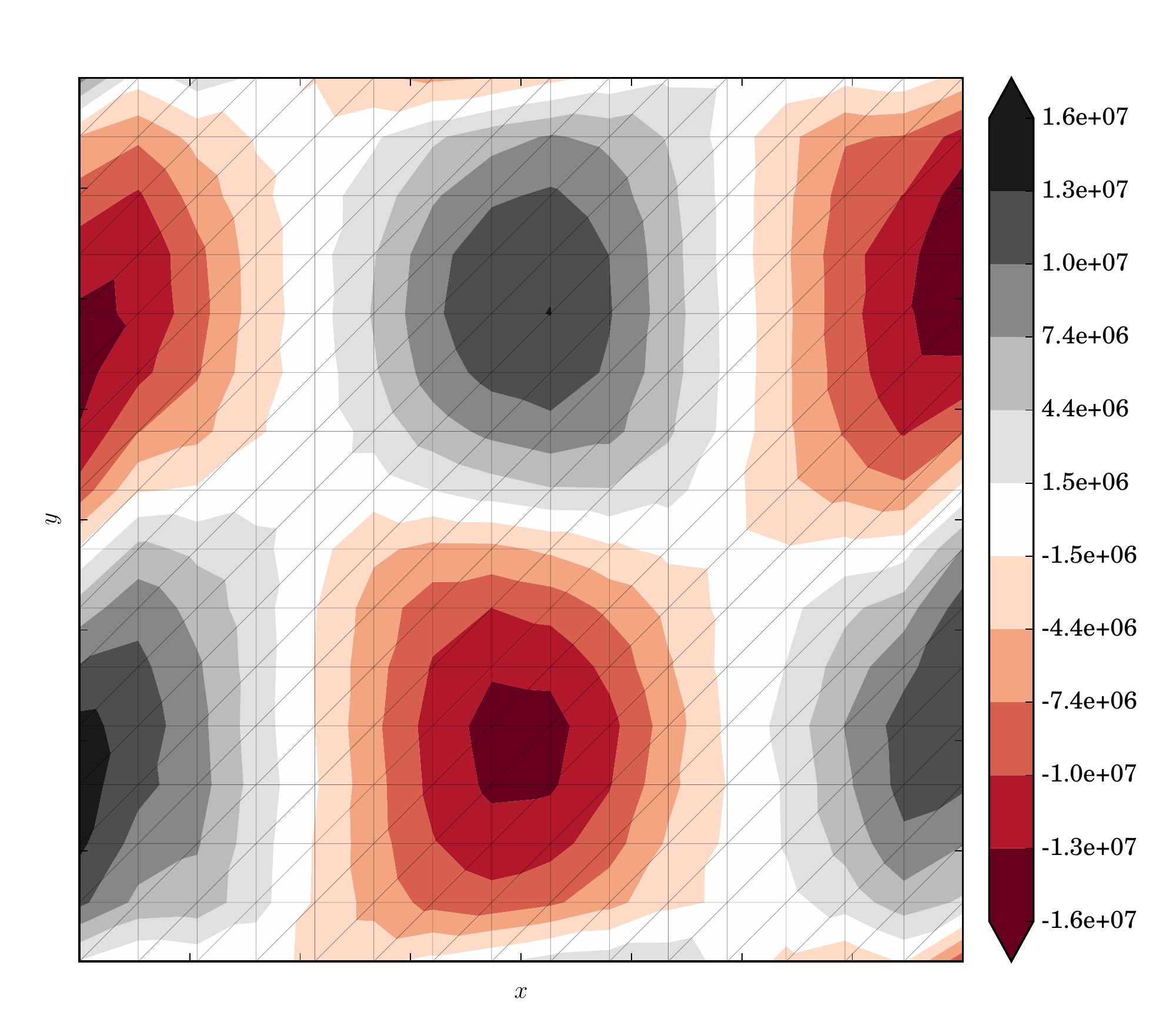}
  \caption{$N_{zz}$}
  \label{bp_N_zz}
  \end{subfigure}
 
  \caption[ISMIP-HOM first-order membrane stress]{First-order membrane stress $N_{kk}$.}

  \label{bp_membrane_stress}

\end{figure*}

The domain width $\ell = 8$ km was chosen for this analysis due to the fact that significant differences exist between solutions obtained by each of the higher-order models (see \S \ref{ssn_ismip_hom_test_simulations}).  Results indicate that the membrane-stress distributions associated with the reformulated-Stokes balance of \S \ref{ssn_reformulated_stokes} are more similar to that associated with the full-Stokes balance of \S \ref{ssn_full_stokes} than the first-order model of \S \ref{ssn_first_order} (Figures \ref{fs_membrane_stress}, \ref{rs_membrane_stress}, and \ref{bp_membrane_stress}).  The most striking difference between the reformulated-Stokes and full-Stokes balance appears to be the $z$-normal stress $N_{zz}$ involving the $z$-derivative of vertical velocity $w$; the reformulated-Stokes solution is approximately one order of magnitude larger than the full-Stokes solution for this term (Figures \ref{rs_N_zz} and \ref{fs_N_zz}).  It is also interesting to note that the oscillations in the $y$-direction of lateral-shearing term $N_{ij} = N_{ji}$ corresponding to the reformulated-Stokes model (Figures \ref{rs_N_ij} and \ref{rs_N_ji}) is one wavelength larger than the full-Stokes model (Figures \ref{fs_N_ij} and \ref{fs_N_ji}).

\begin{figure*}
  
  \centering 
  
  \begin{subfigure}[b]{0.3\linewidth}
    \includegraphics[width=\linewidth]{images/stress_balance/FS/U_mag.pdf}
  \caption{$\mathbf{u}_S$}
  \label{fs_msb_U}
  \end{subfigure}
  \begin{subfigure}[b]{0.3\linewidth}
    \includegraphics[width=\linewidth]{images/stress_balance/FS/p.pdf}
  \caption{$p |_B$}
  \label{fs_msb_p}
  \end{subfigure}

  \begin{subfigure}[b]{0.3\linewidth}
    \includegraphics[width=\linewidth]{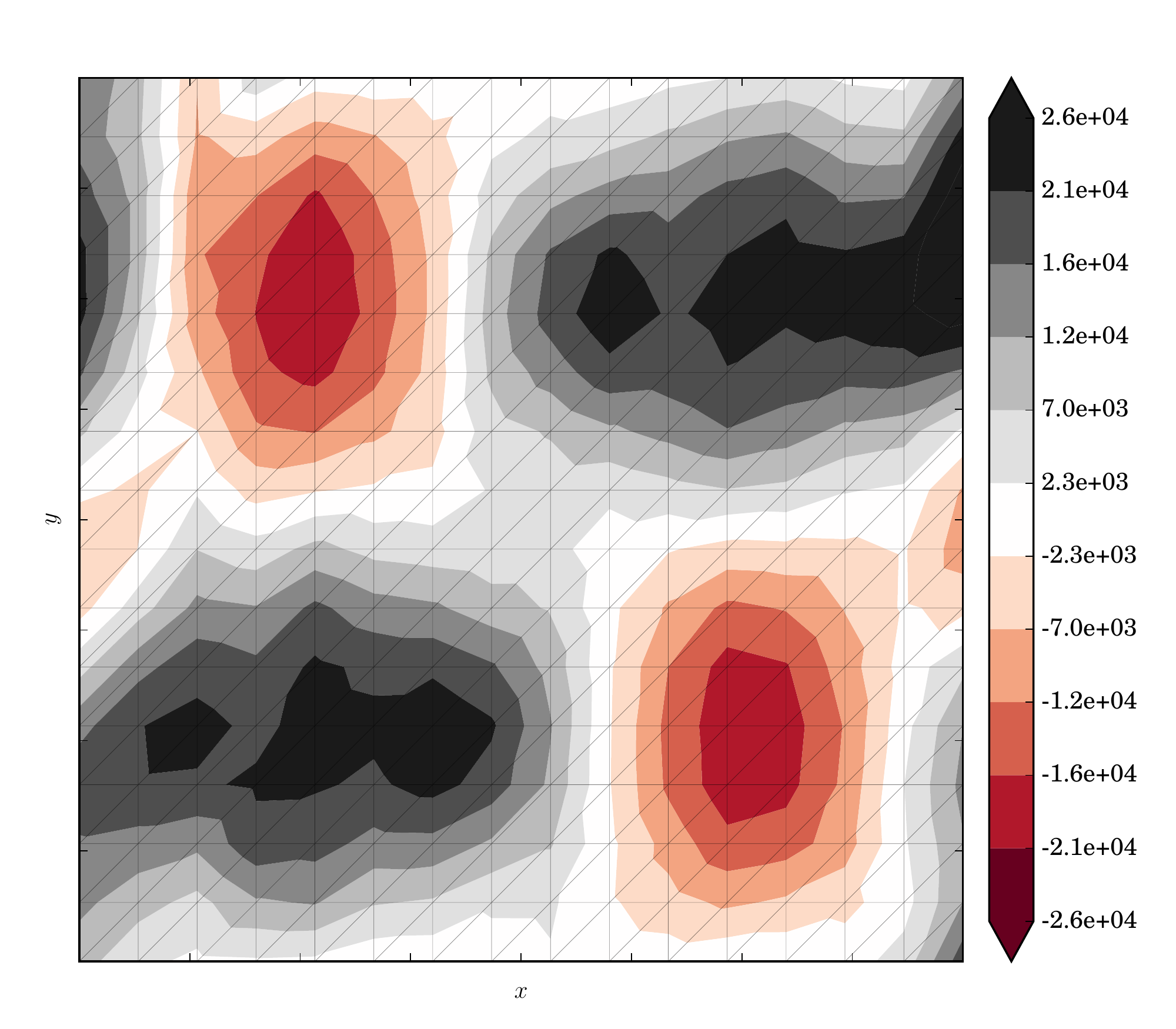}
  \caption{$M_{ii}$}
  \label{fs_M_ii}
  \end{subfigure}
  \begin{subfigure}[b]{0.3\linewidth}
    \includegraphics[width=\linewidth]{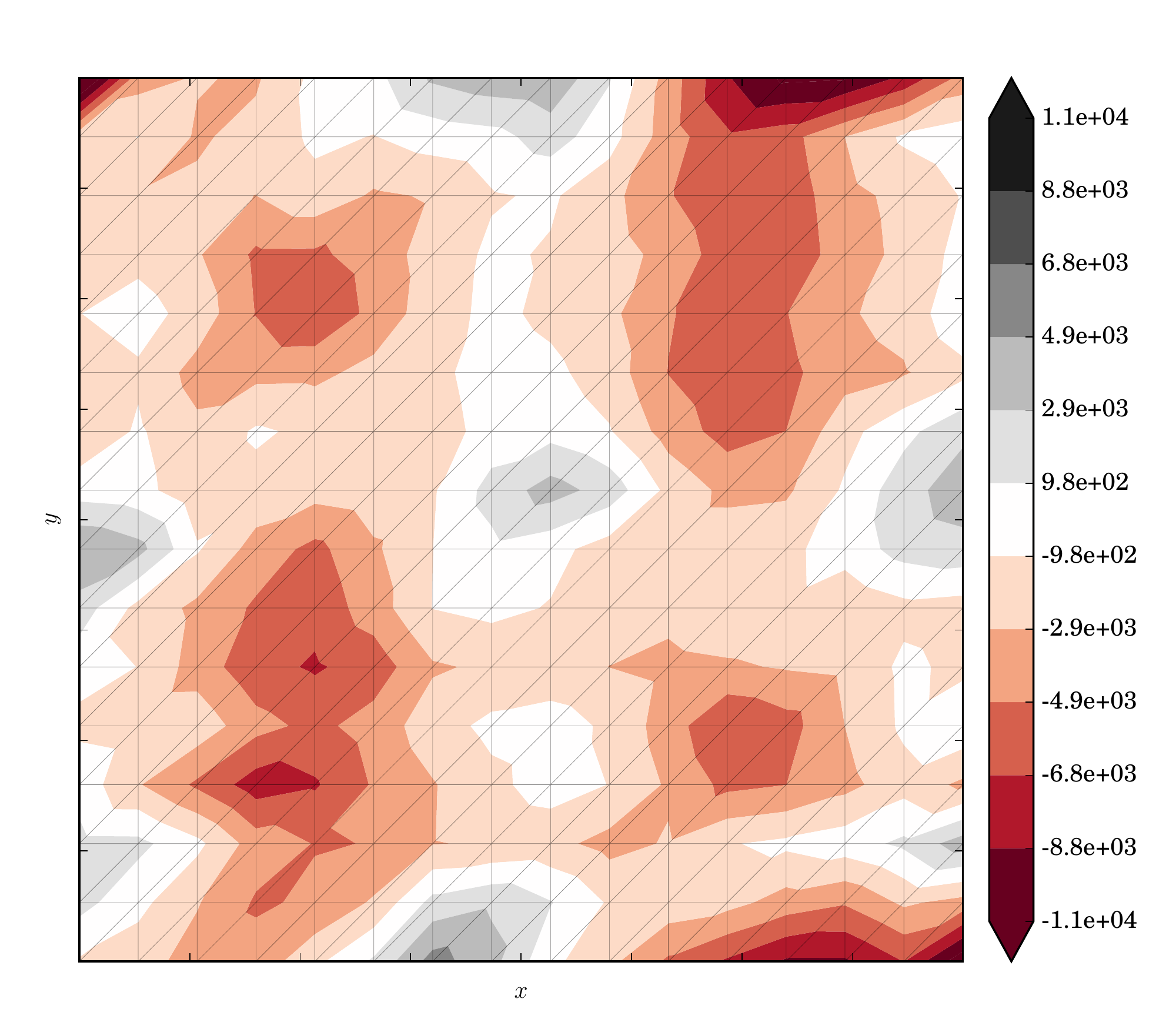}
  \caption{$M_{ij}$}
  \label{fs_M_ij}
  \end{subfigure}
  \begin{subfigure}[b]{0.3\linewidth}
    \includegraphics[width=\linewidth]{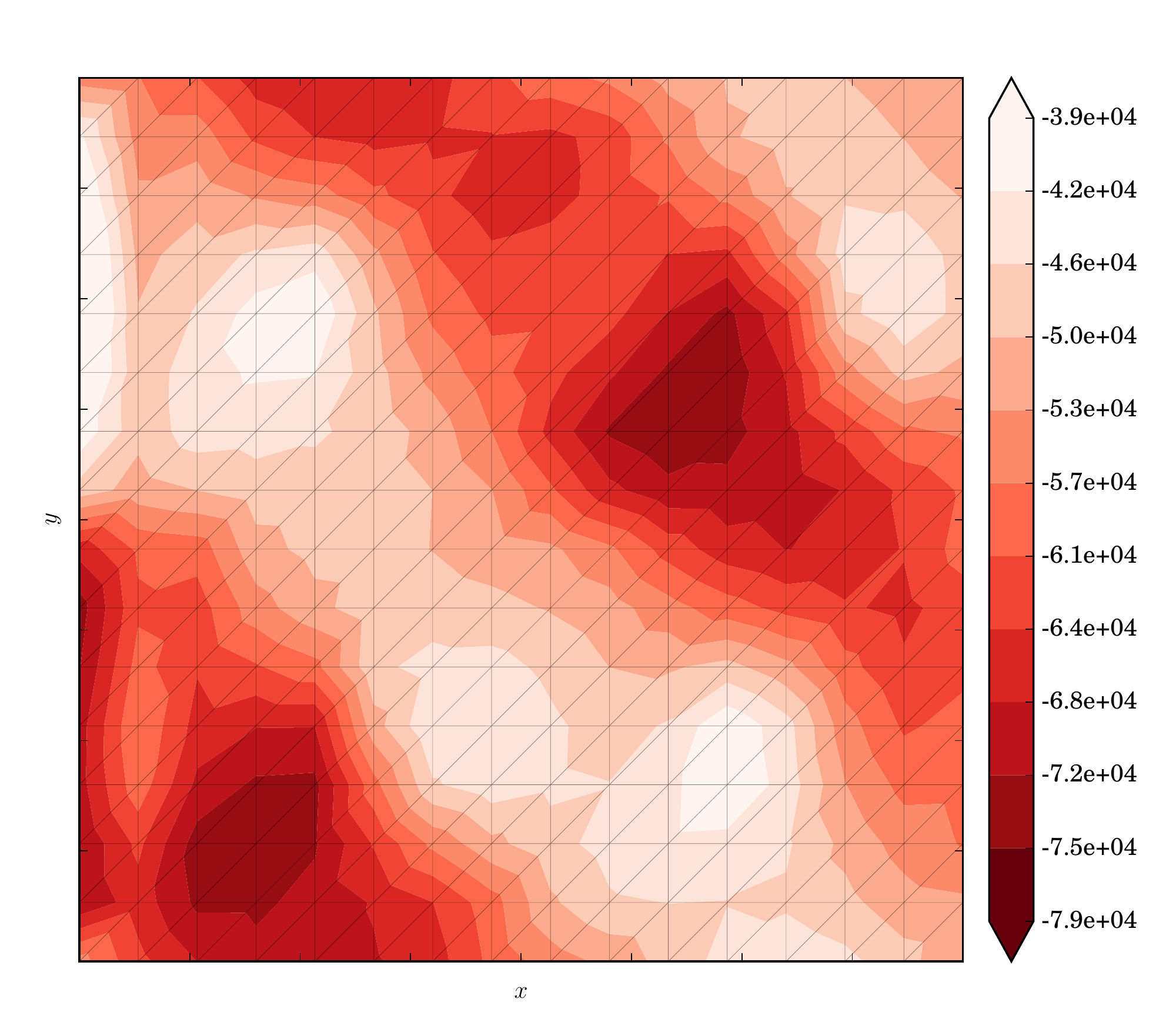}
  \caption{$M_{iz}$}
  \label{fs_M_iz}
  \end{subfigure}

  \begin{subfigure}[b]{0.3\linewidth}
    \includegraphics[width=\linewidth]{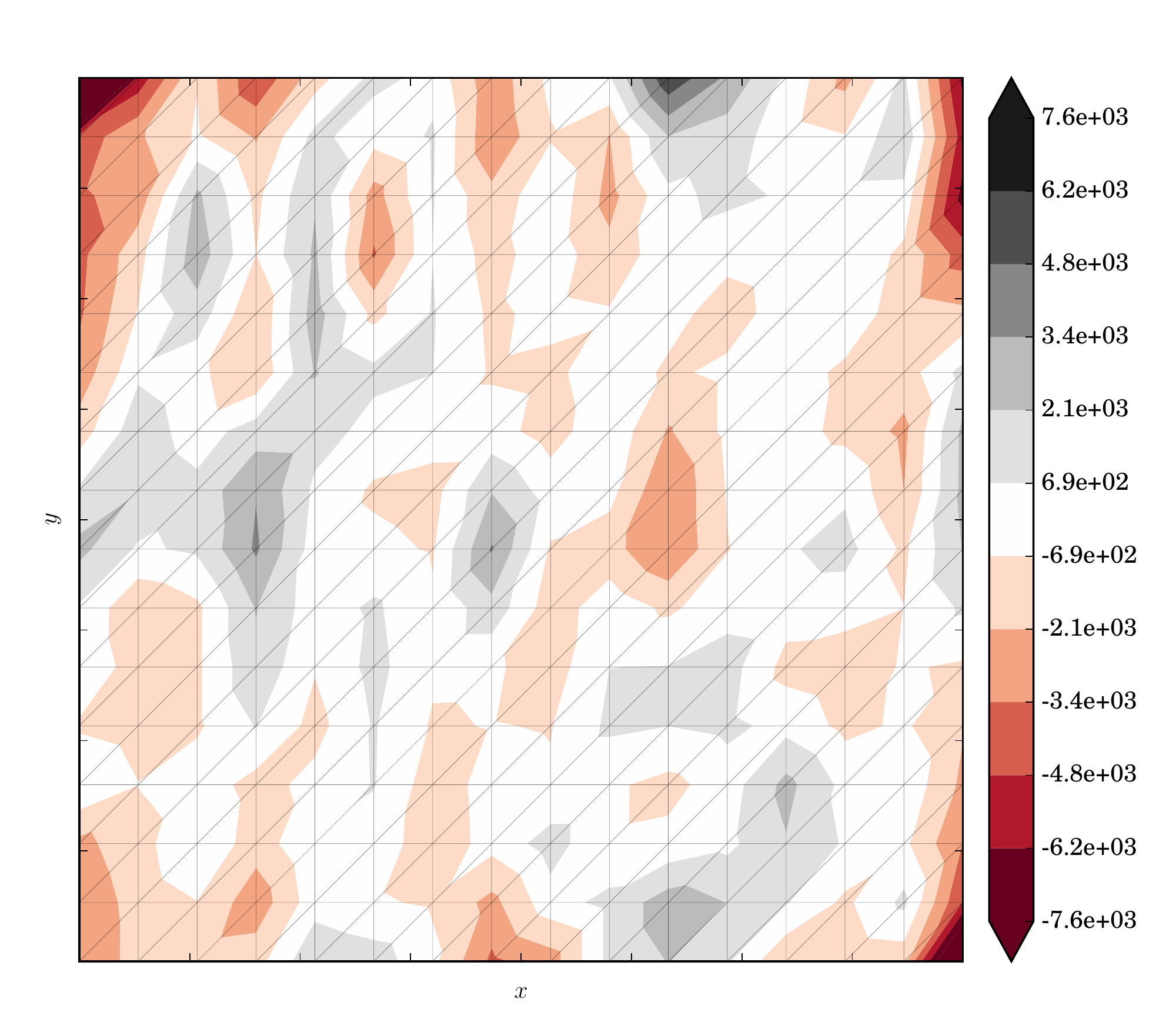}
  \caption{$M_{ji}$}
  \label{fs_M_ji}
  \end{subfigure}
  \begin{subfigure}[b]{0.3\linewidth}
    \includegraphics[width=\linewidth]{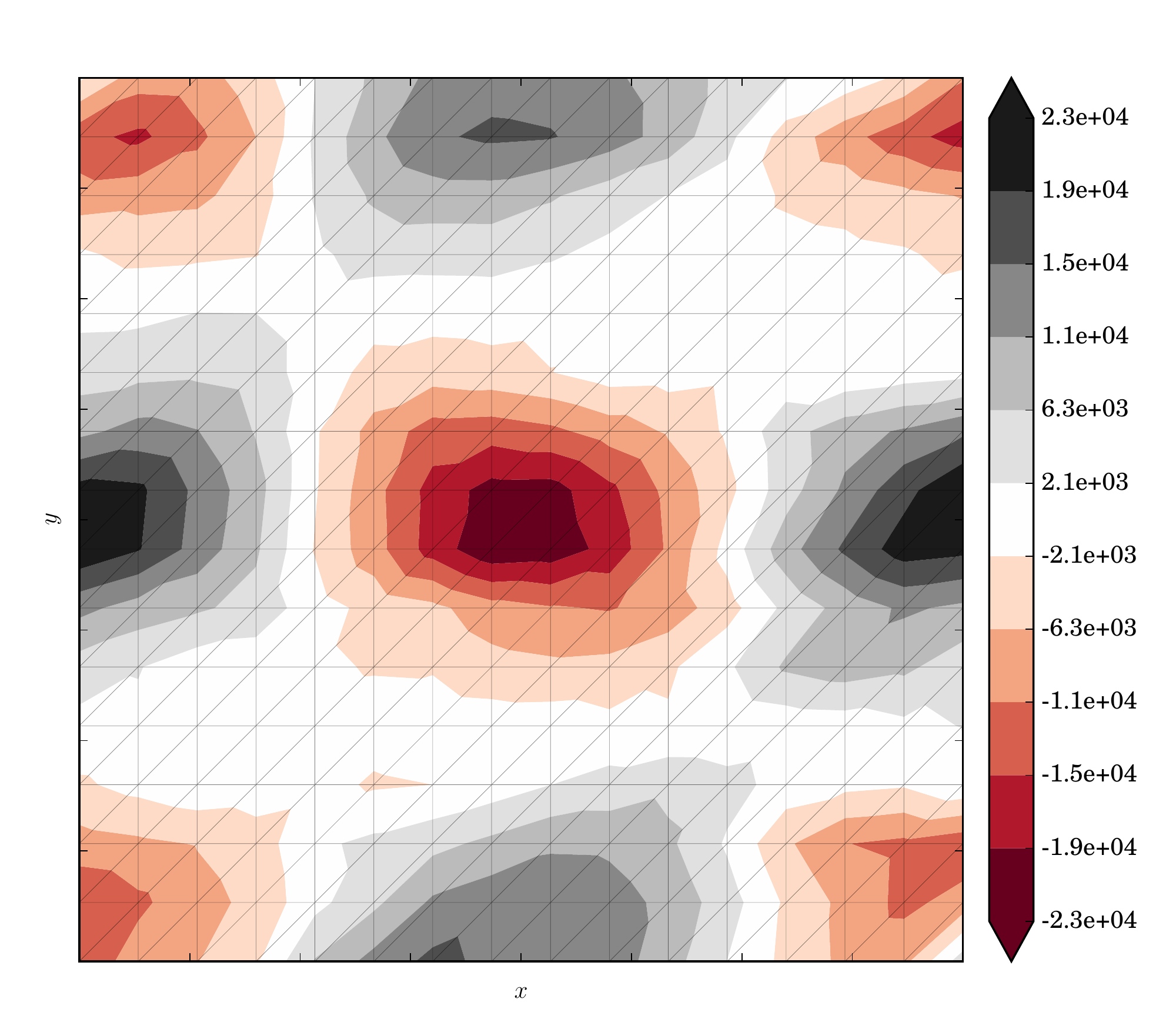}
  \caption{$M_{jj}$}
  \label{fs_M_jj}
  \end{subfigure}
  \begin{subfigure}[b]{0.3\linewidth}
    \includegraphics[width=\linewidth]{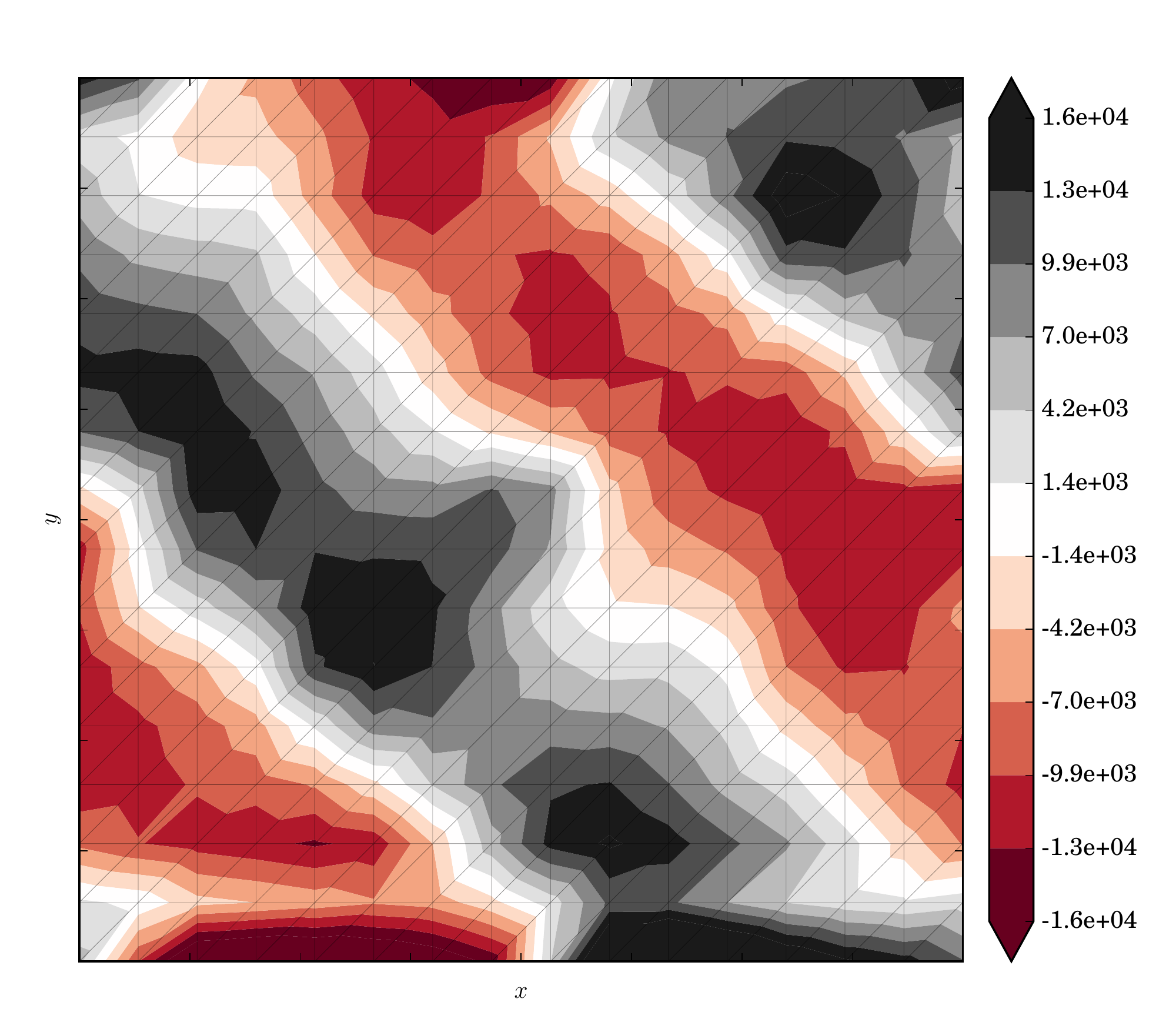}
  \caption{$M_{jz}$}
  \label{fs_M_jz}
  \end{subfigure}

  \begin{subfigure}[b]{0.3\linewidth}
    \includegraphics[width=\linewidth]{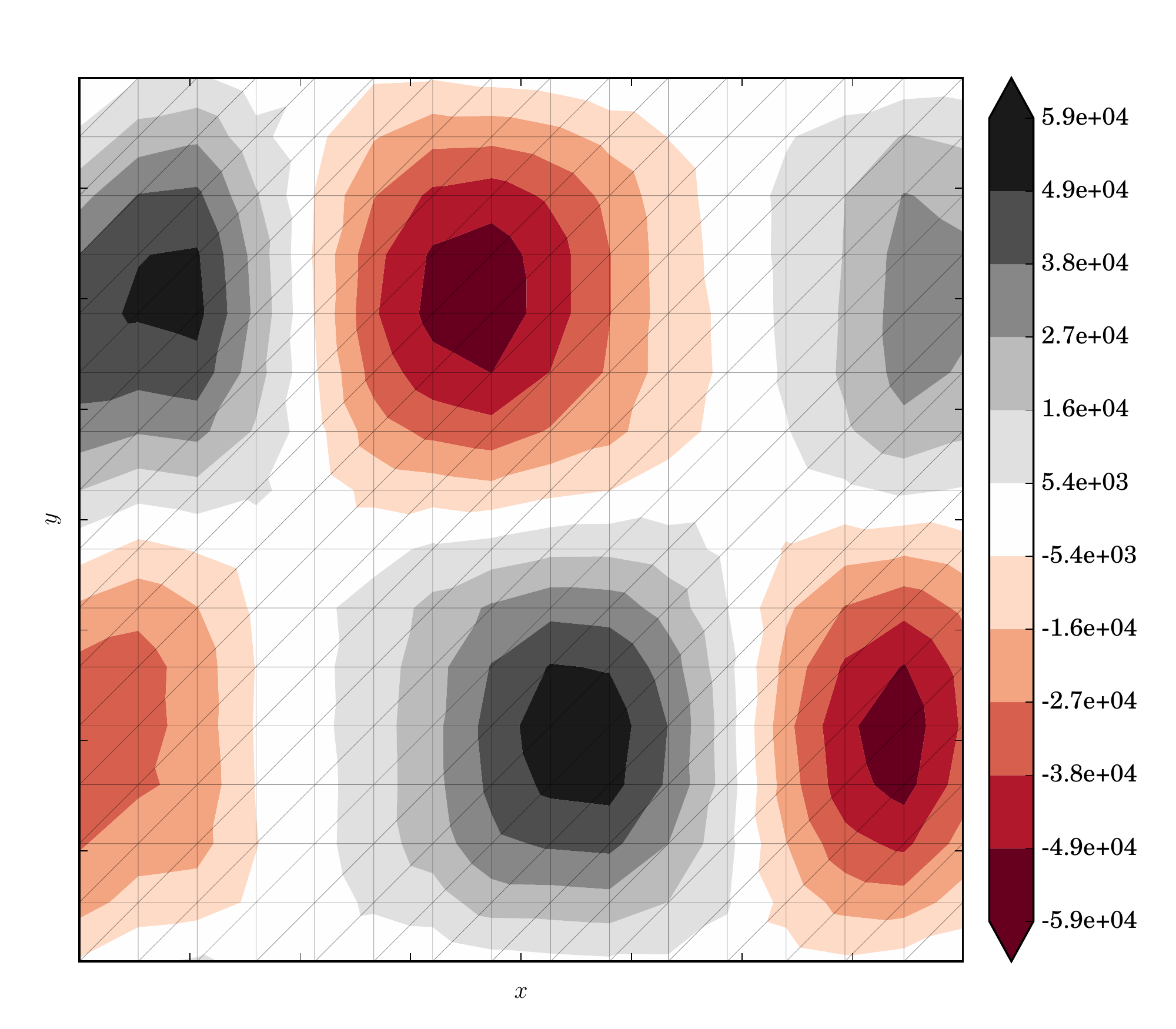}
  \caption{$M_{zi}$}
  \label{fs_M_zi}
  \end{subfigure}
  \begin{subfigure}[b]{0.3\linewidth}
    \includegraphics[width=\linewidth]{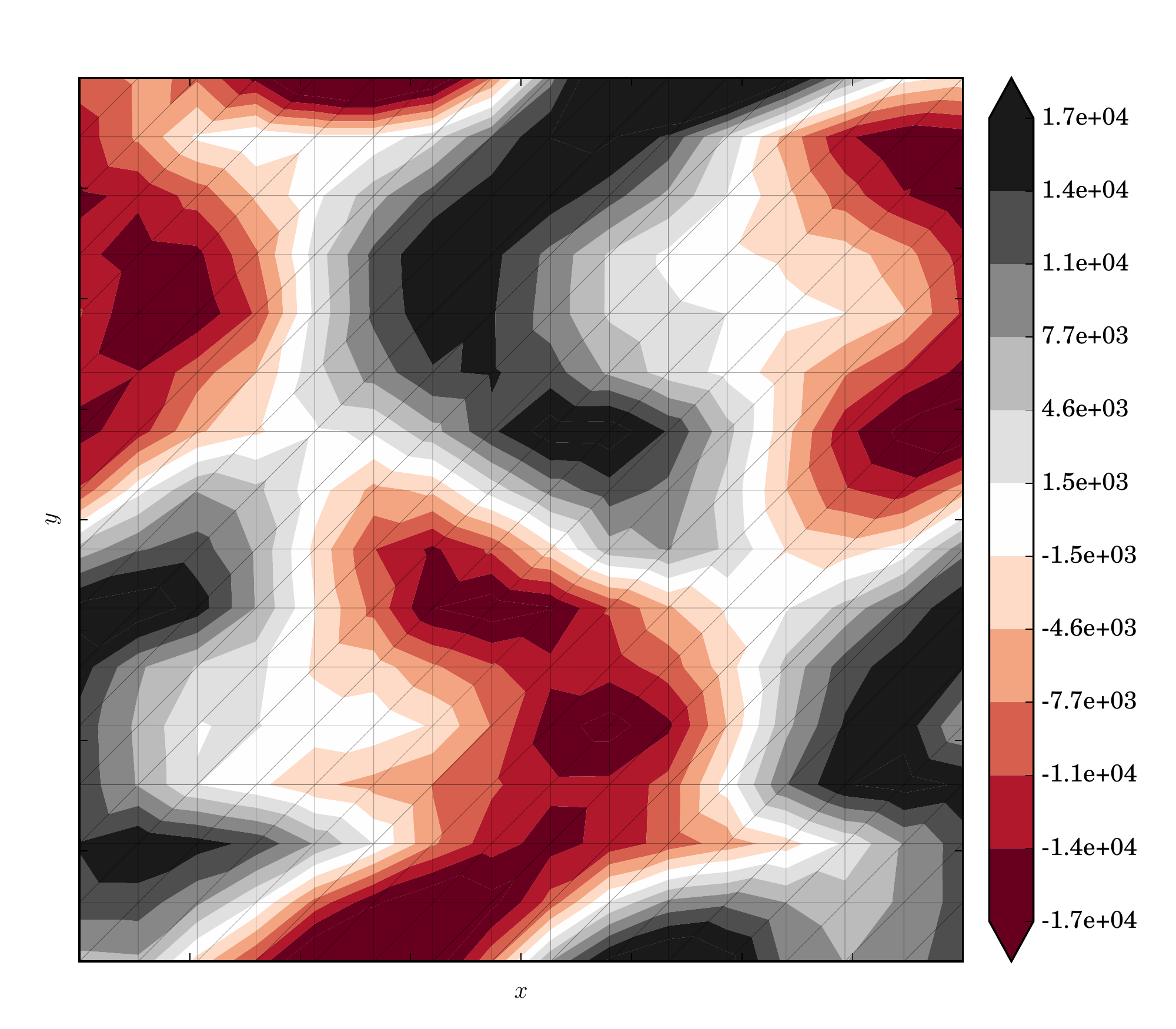}
  \caption{$M_{zj}$}
  \label{fs_M_zj}
  \end{subfigure}
  \begin{subfigure}[b]{0.3\linewidth}
    \includegraphics[width=\linewidth]{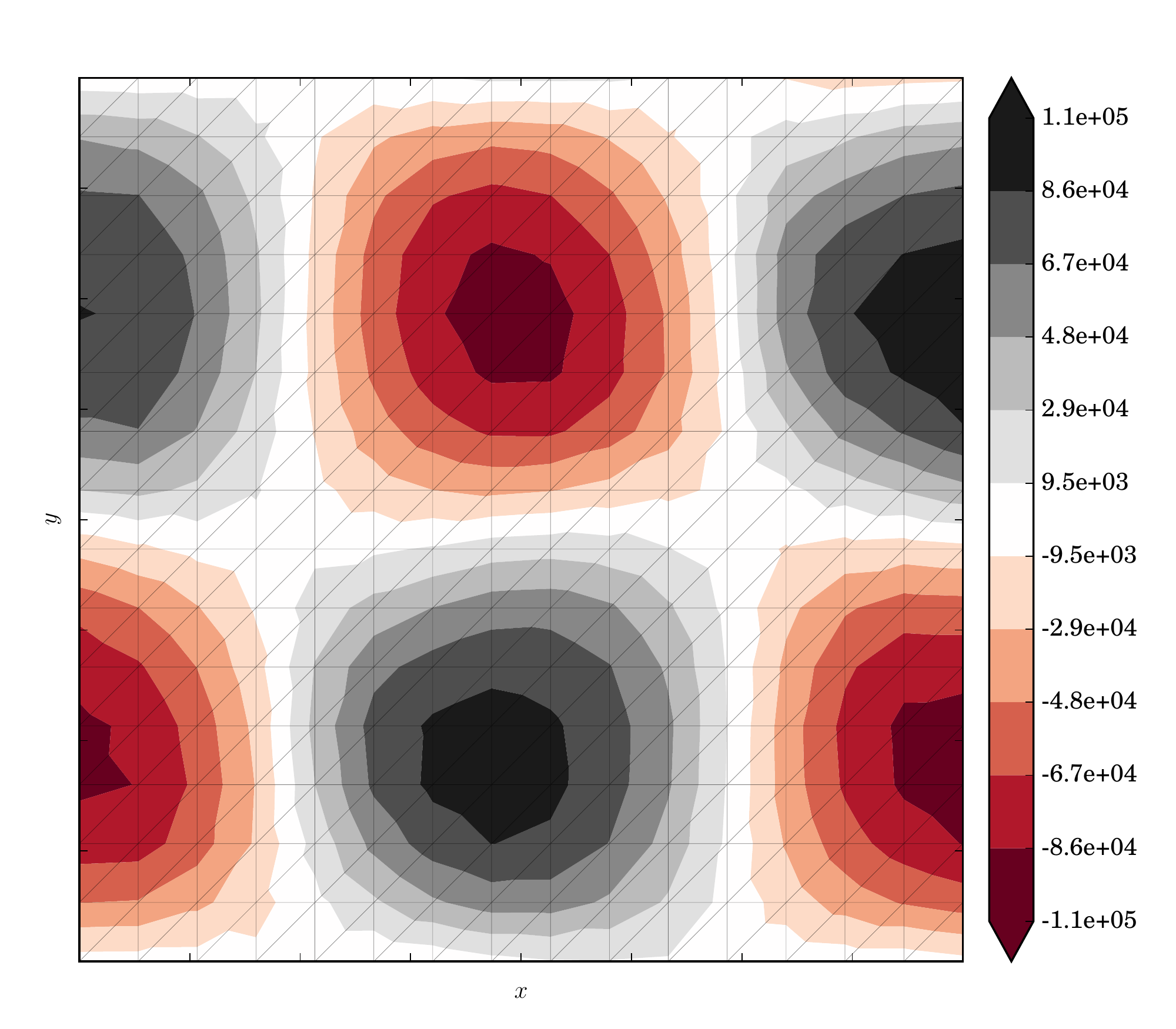}
  \caption{$M_{zz}$}
  \label{fs_M_zz}
  \end{subfigure}
 
  \caption[ISMIP-HOM full-Stokes membrane stress balance]{Full-Stokes membrane stress balance $M_{kk}$.}

  \label{fs_membrane_stress_balance}

\end{figure*}

\begin{figure*}
  
  \centering 

  \begin{subfigure}[b]{0.3\linewidth}
    \includegraphics[width=\linewidth]{images/stress_balance/BP/U_mag.pdf}
  \caption{$\mathbf{u}_S$}
  \label{bp_msb_U}
  \end{subfigure}
  \begin{subfigure}[b]{0.3\linewidth}
    \includegraphics[width=\linewidth]{images/stress_balance/BP/p.pdf}
  \caption{$p |_B$}
  \label{bp_msb_p}
  \end{subfigure}

  \begin{subfigure}[b]{0.3\linewidth}
    \includegraphics[width=\linewidth]{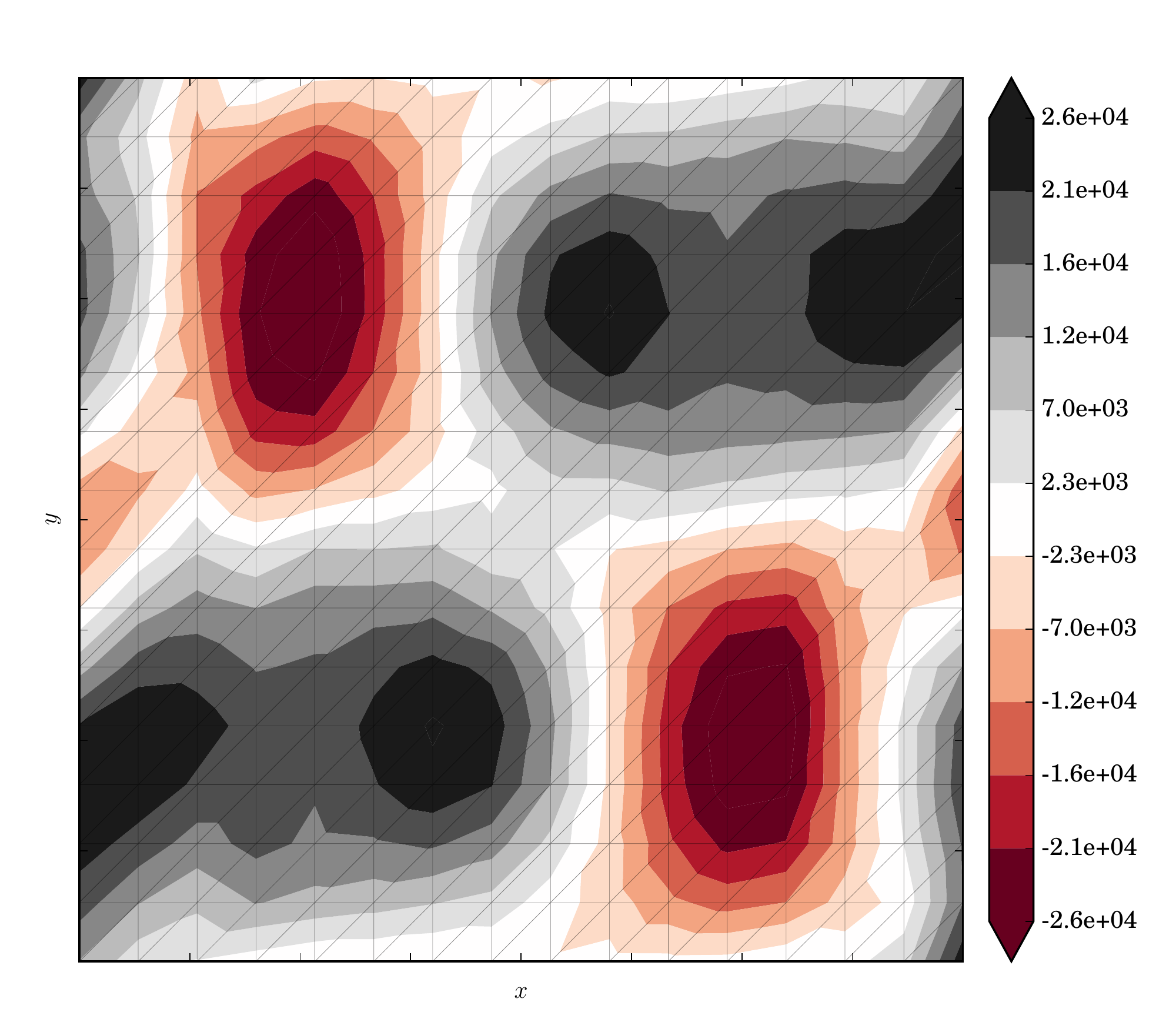}
  \caption{$M_{ii}$}
  \label{bp_M_ii}
  \end{subfigure}
  \begin{subfigure}[b]{0.3\linewidth}
    \includegraphics[width=\linewidth]{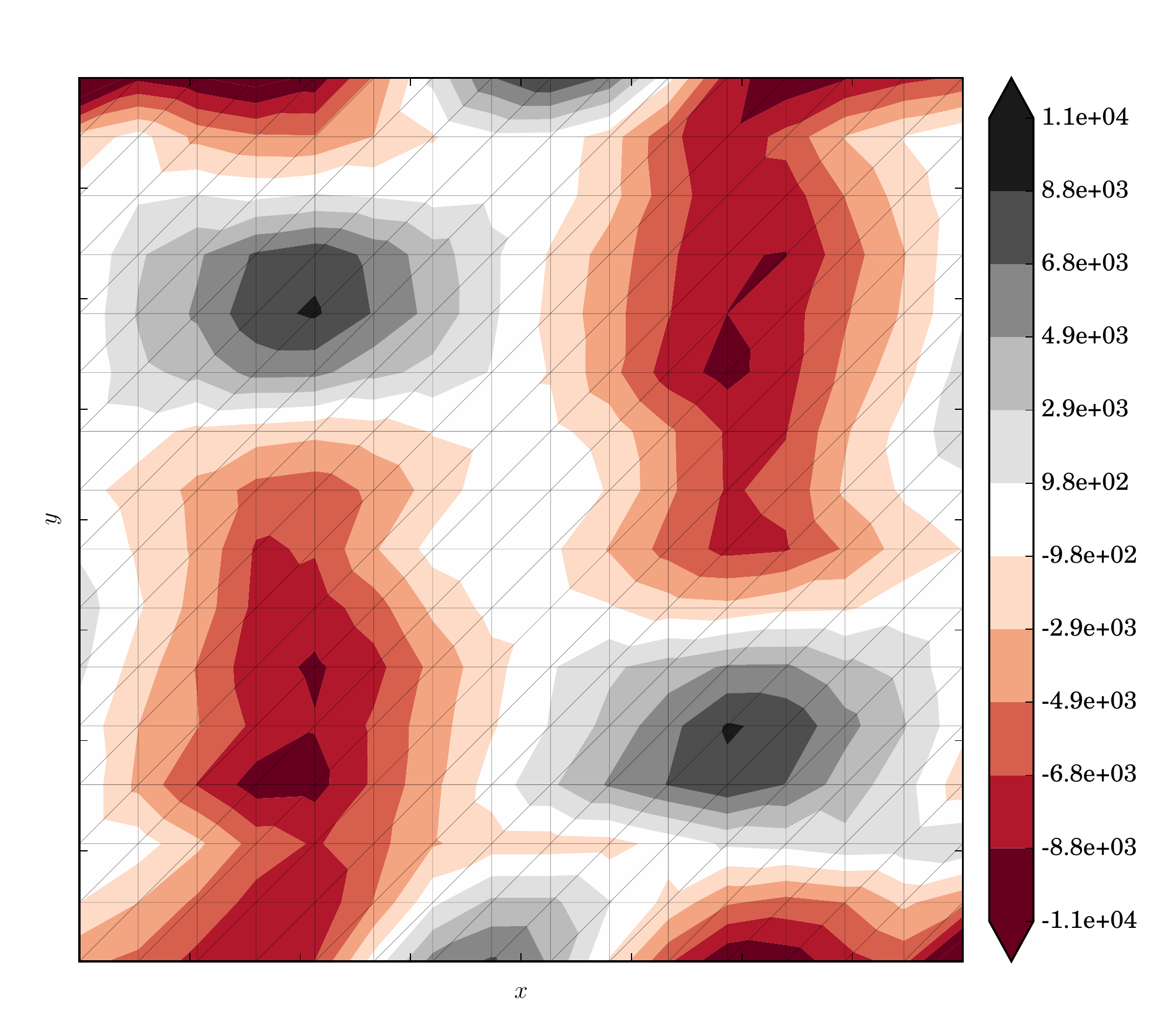}
  \caption{$M_{ij}$}
  \label{bp_M_ij}
  \end{subfigure}
  \begin{subfigure}[b]{0.3\linewidth}
    \includegraphics[width=\linewidth]{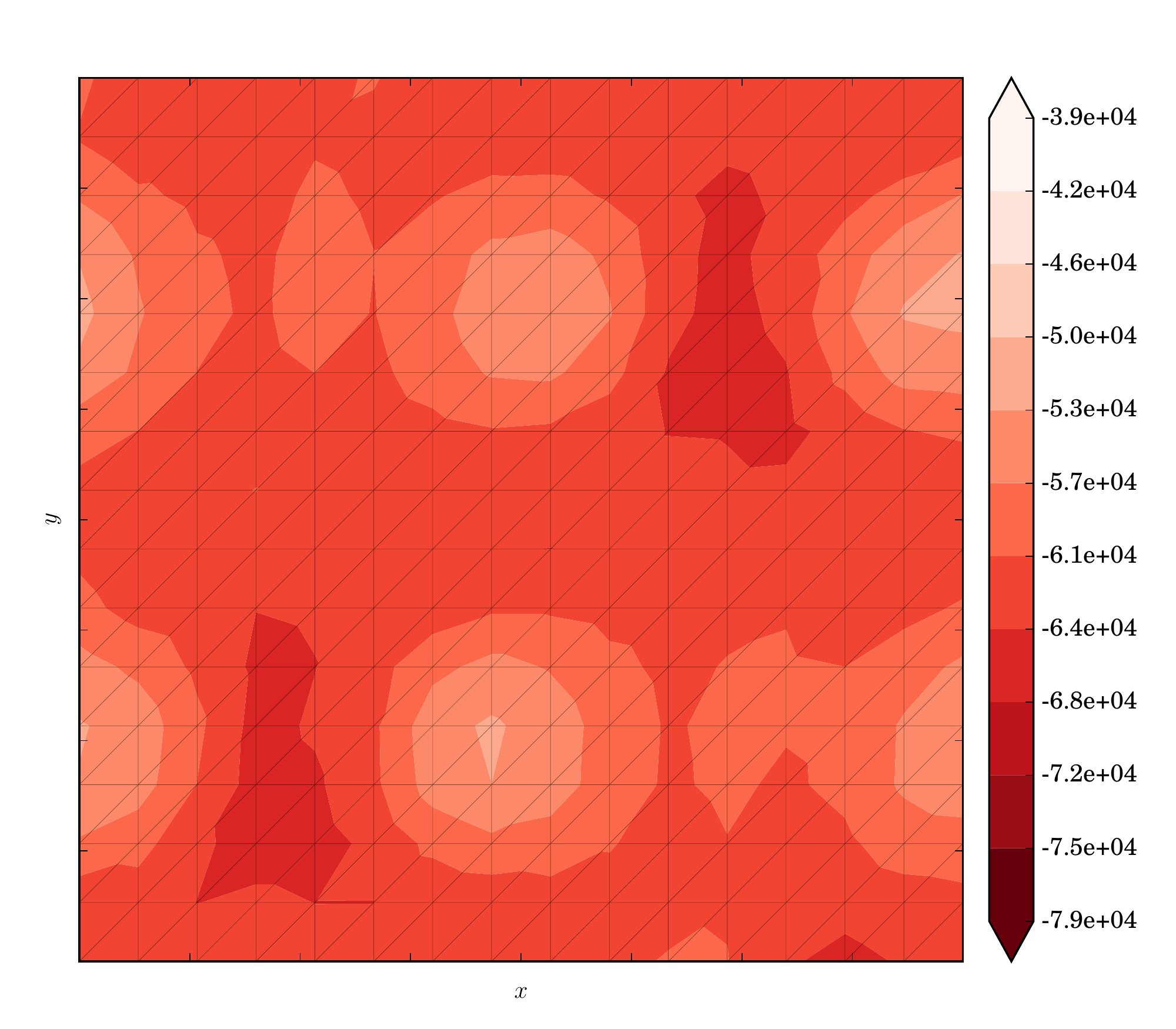}
  \caption{$M_{iz}$}
  \label{bp_M_iz}
  \end{subfigure}

  \begin{subfigure}[b]{0.3\linewidth}
    \includegraphics[width=\linewidth]{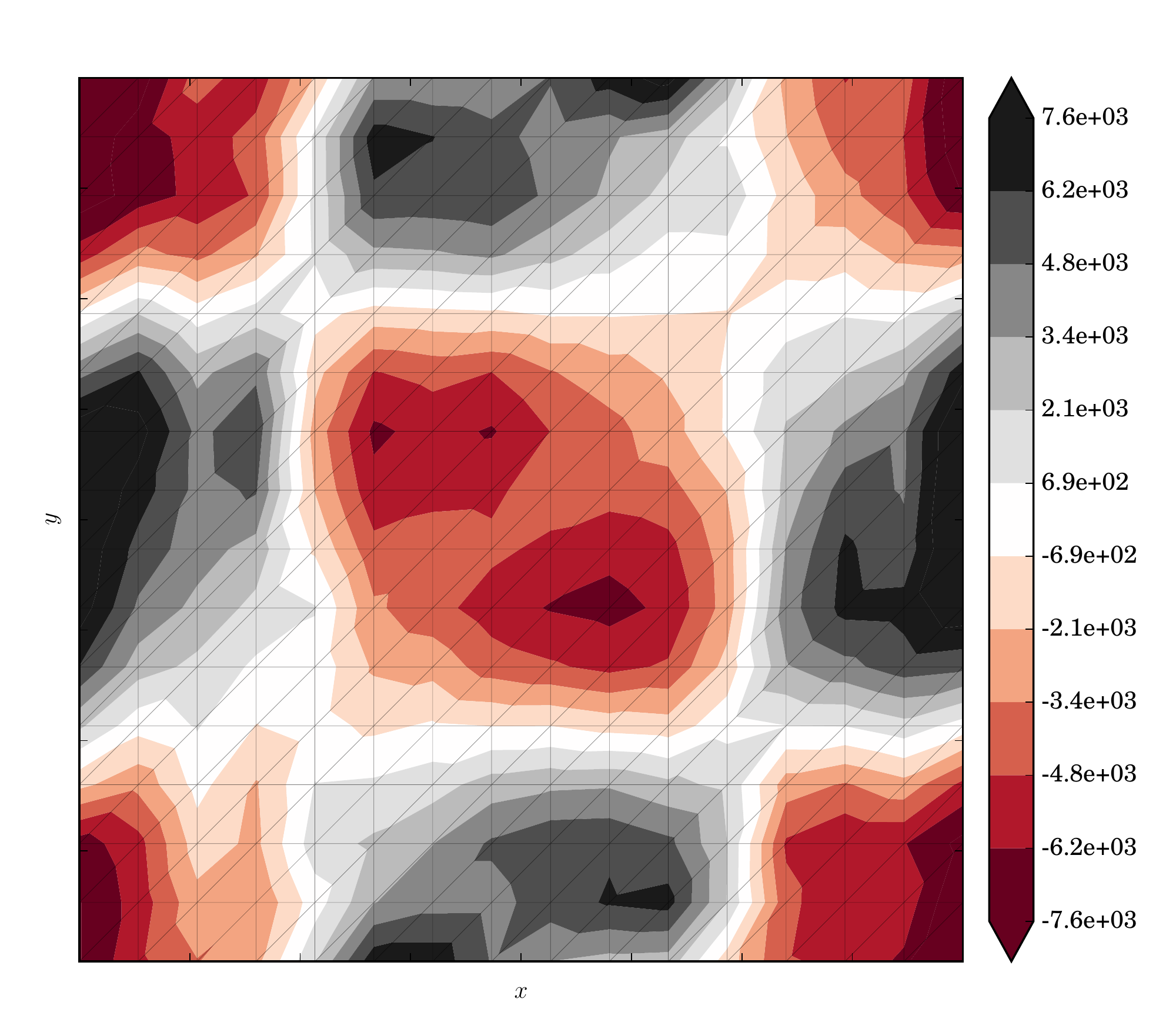}
  \caption{$M_{ji}$}
  \label{bp_M_ji}
  \end{subfigure}
  \begin{subfigure}[b]{0.3\linewidth}
    \includegraphics[width=\linewidth]{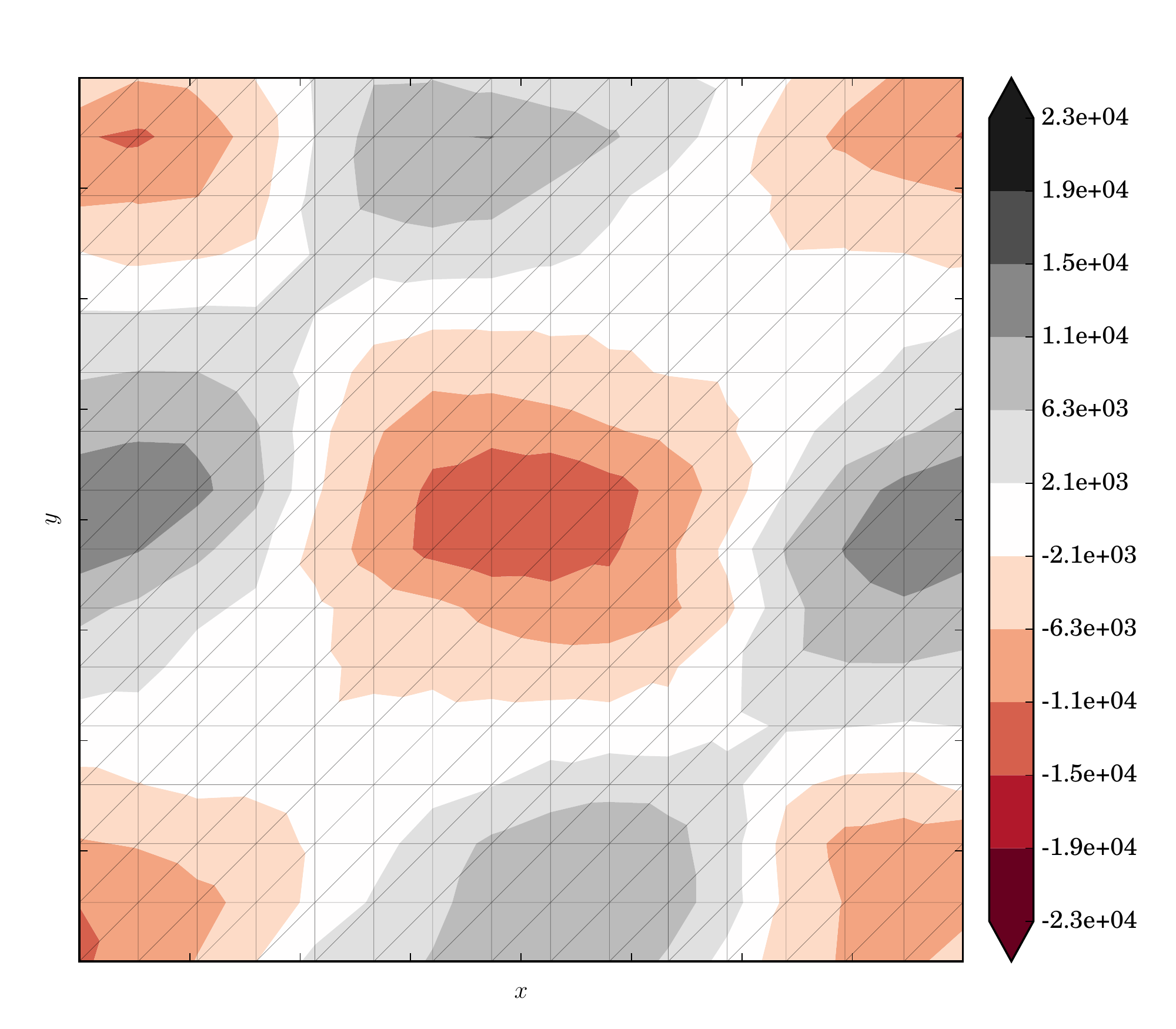}
  \caption{$M_{jj}$}
  \label{bp_M_jj}
  \end{subfigure}
  \begin{subfigure}[b]{0.3\linewidth}
    \includegraphics[width=\linewidth]{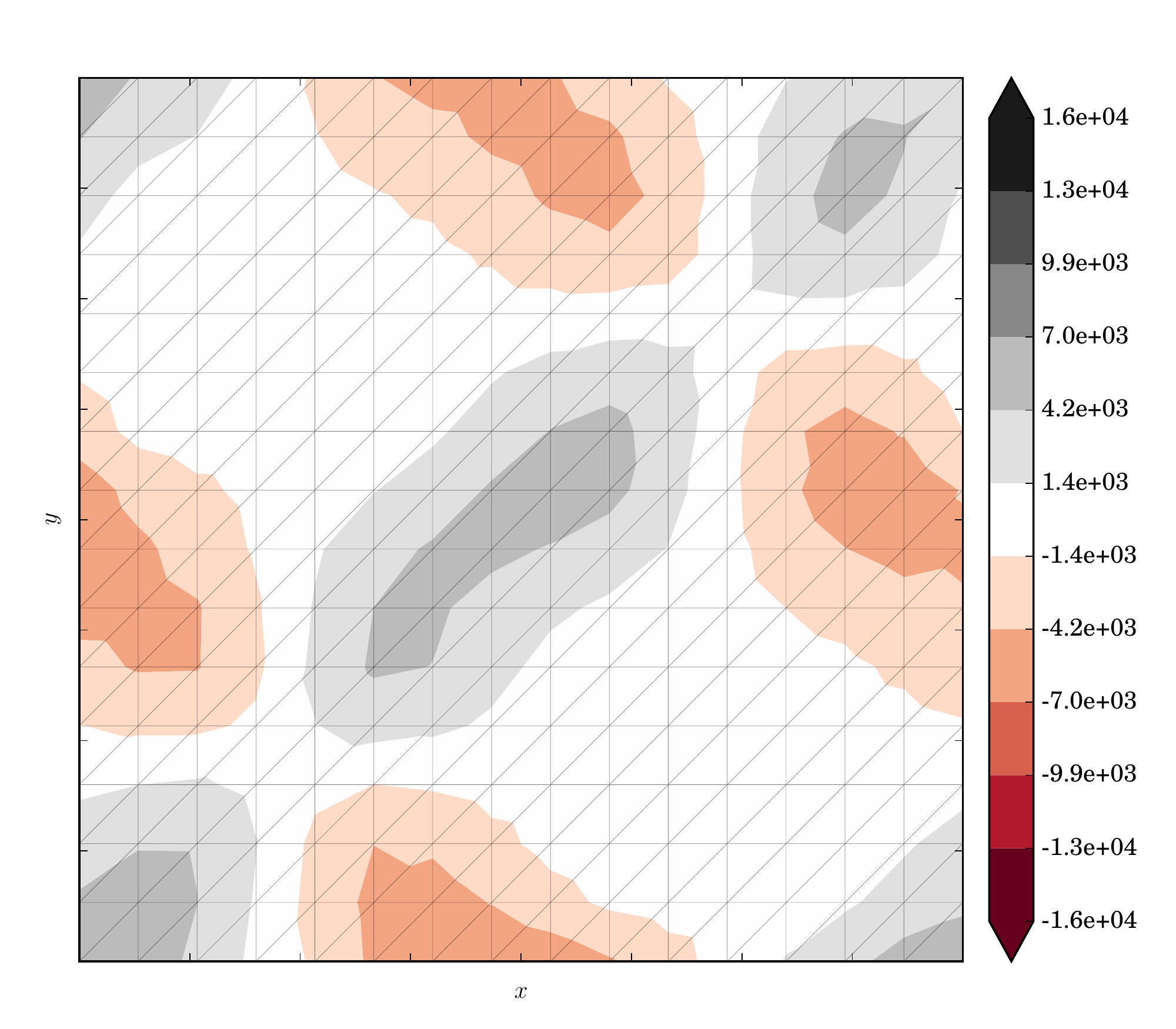}
  \caption{$M_{jz}$}
  \label{bp_M_jz}
  \end{subfigure}

  \begin{subfigure}[b]{0.3\linewidth}
    \includegraphics[width=\linewidth]{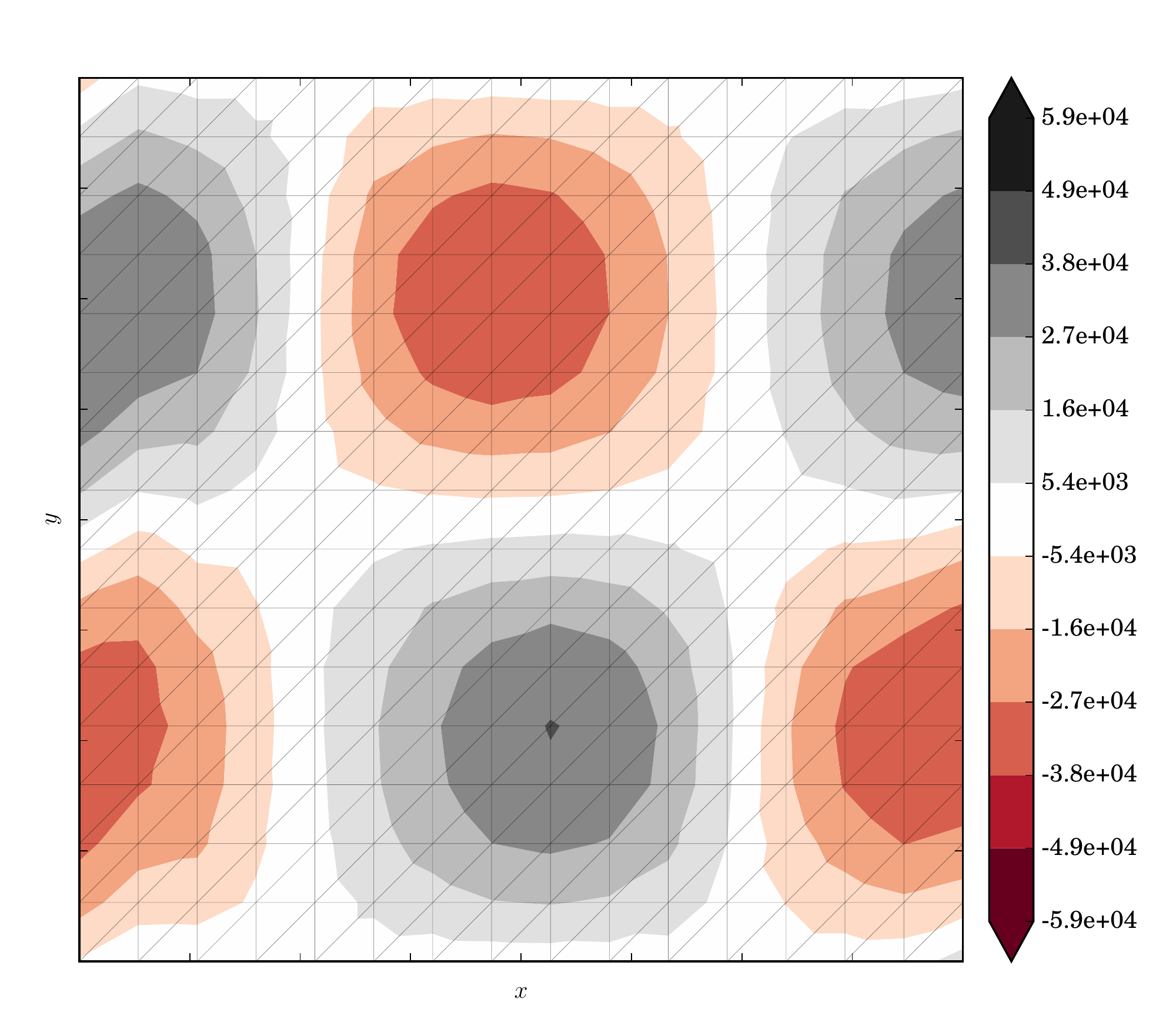}
  \caption{$M_{zi}$}
  \label{bp_M_zi}
  \end{subfigure}
  \begin{subfigure}[b]{0.3\linewidth}
    \includegraphics[width=\linewidth]{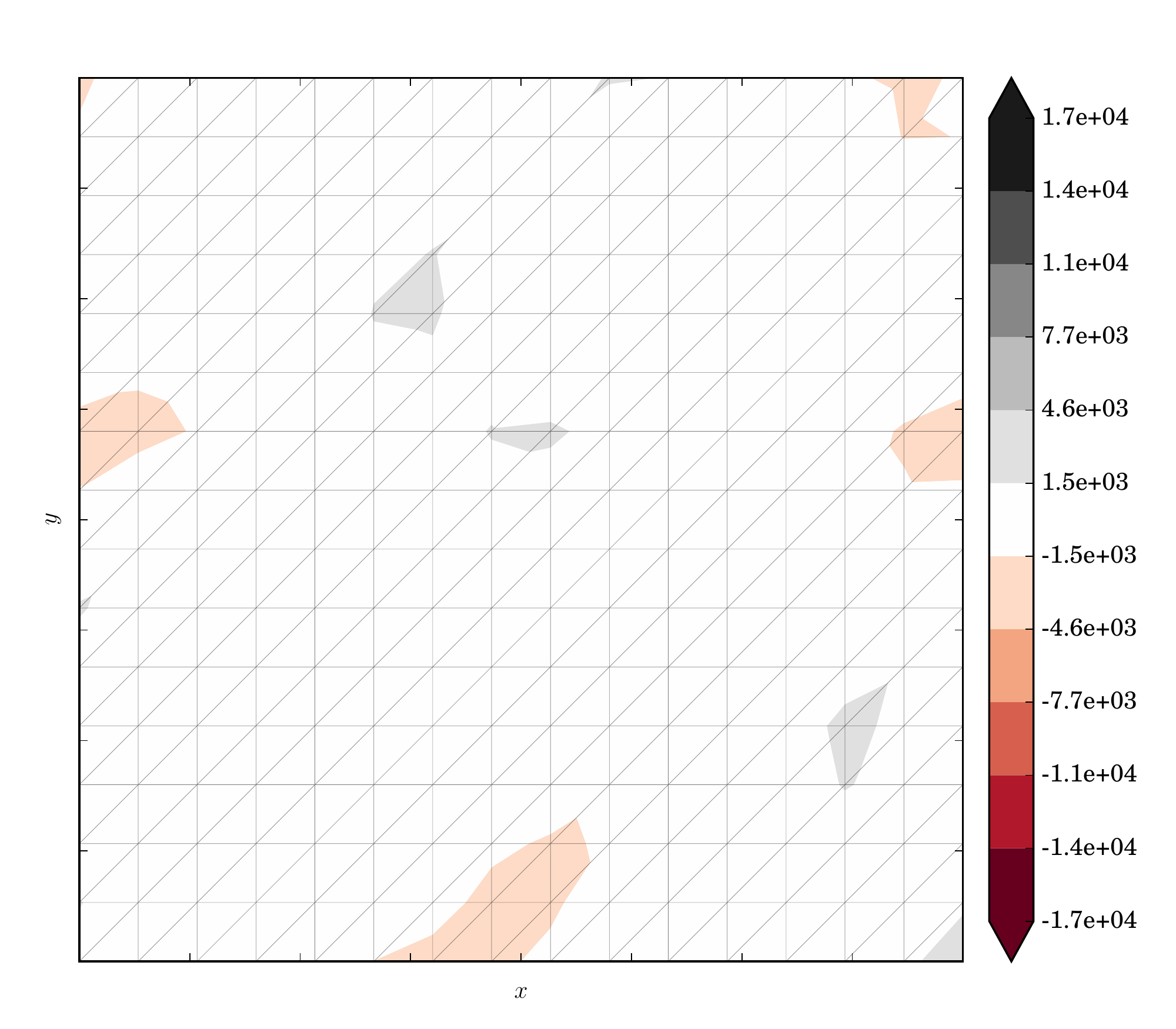}
  \caption{$M_{zj}$}
  \label{bp_M_zj}
  \end{subfigure}
  \begin{subfigure}[b]{0.3\linewidth}
    \includegraphics[width=\linewidth]{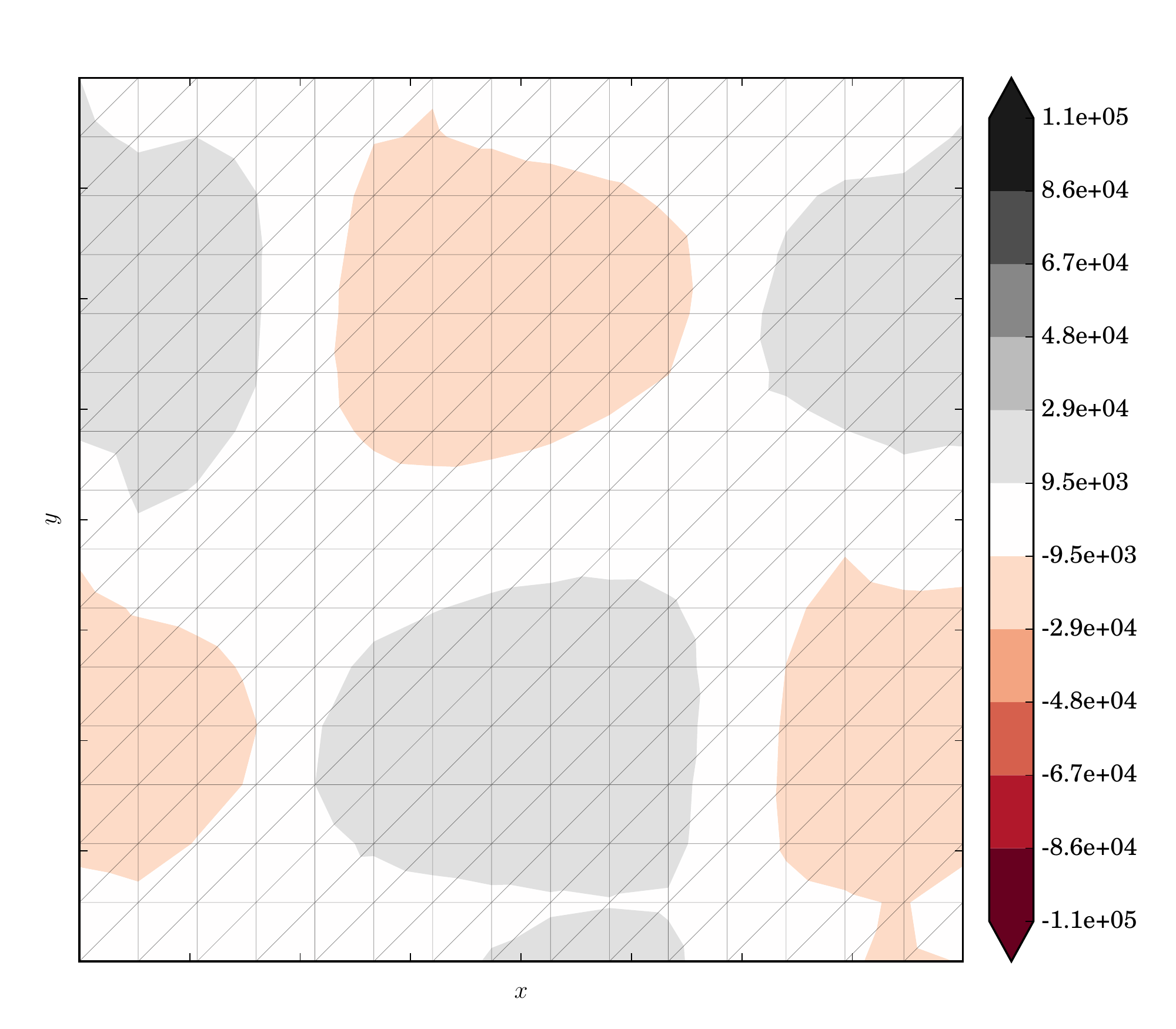}
  \caption{$M_{zz}$}
  \label{bp_M_zz}
  \end{subfigure}
 
  \caption[ISMIP-HOM first-order membrane stress balance]{First-order membrane stress balance $M_{kk}$.}

  \label{bp_membrane_stress_balance}

\end{figure*}

The distribution and magnitude of the membrane-stress-balance results associated with the first-order model (Figure \ref{bp_membrane_stress_balance}) are both remarkably different from the full-Stokes results (Figure \ref{fs_membrane_stress_balance}).  This model appears to over-estimate the contribution of along-flow normal stress $M_{ii}$ (Figure \ref{bp_M_ii}); lateral-shear $M_{ij}$ and $M_{ji}$ (Figures \ref{bp_M_ij} and \ref{bp_M_ji}); and vertical-shear $M_{iz}$ (Figure \ref{bp_M_iz}), while under-estimating the remaining terms.

\begin{table}
\centering
\caption[Stress-balance ISMIP-HOM variables]{ISMIP-HOM stress-balance variables.}
\label{ismip_hom_stress_balance_values}
\begin{tabular}{llll}
\hline
\textbf{Variable} & \textbf{Value} & \textbf{Units} & \textbf{Description} \\
\hline
$\dot{\varepsilon}_0$ & $10\sups{-15}$ & a\sups{-1}   & strain regularization \\
$\beta$   & $1000$          & kg m\sups{-2}a\sups{-1} & basal friction coef. \\
$A$       & $10\sups{-16}$  & Pa\sups{-3}a\sups{-1}   & flow-rate factor \\
$\ell$    & $8$             & km & width of domain \\
$F_b$     & $0$             & m a\sups{-1}            & basal water discharge \\
%$\dot{a}$ & $0$             & m a\sups{-1}            & surface accumulation \\
$a$       & $0.5$           & $\circ$                 & surface gradient mag. \\
$\bar{B}$ & $1000$          & m & average basal depth \\
$b$       & $500$           & m & basal height amp.\\
$k_x$     & $15$            & -- & number of $x$ divisions \\
$k_y$     & $15$            & -- & number of $y$ divisions \\
$k_z$     & $5$             & -- & number of $z$ divisions \\
$N_e$     & $6750$          & -- & number of cells \\
$N_n$     & $1536$          & -- & number of vertices \\
\hline
\end{tabular}
\end{table}

\pythonexternal[label=cslvr_stress_balance_script, caption={CSLVR script that performs the stress-balance calculation for the ISMIP-HOM problem of \S \ref{ssn_stress_balance_ismip_hom_test_simulations}.}]{scripts/stress_balance/ISMIP_HOM_A_FS.py}

%===============================================================================

%===============================================================================

%===============================================================================

%===============================================================================

\begin{figure*}
  
  \centering 
  
  \begin{subfigure}[b]{0.3\linewidth}
    \includegraphics[width=\linewidth]{images/stress_balance/RS/U_mag.pdf}
  \caption{$\mathbf{u}_S$}
  \label{rs_msb_U}
  \end{subfigure}
  \begin{subfigure}[b]{0.3\linewidth}
    \includegraphics[width=\linewidth]{images/stress_balance/RS/p.pdf}
  \caption{$p |_B$}
  \label{rs_msb_p}
  \end{subfigure}

  \begin{subfigure}[b]{0.3\linewidth}
    \includegraphics[width=\linewidth]{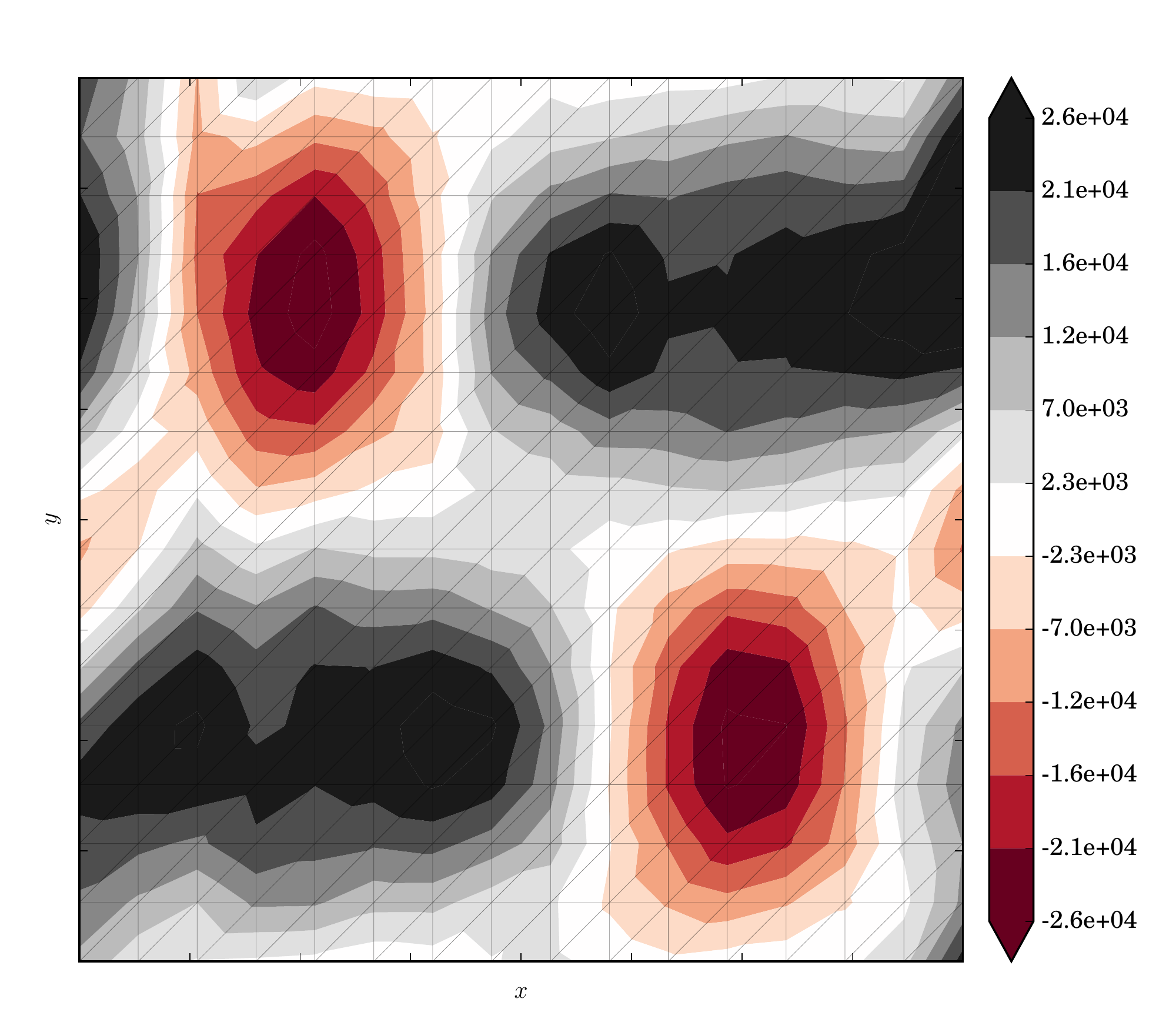}
  \caption{$M_{ii}$}
  \label{rs_M_ii}
  \end{subfigure}
  \begin{subfigure}[b]{0.3\linewidth}
    \includegraphics[width=\linewidth]{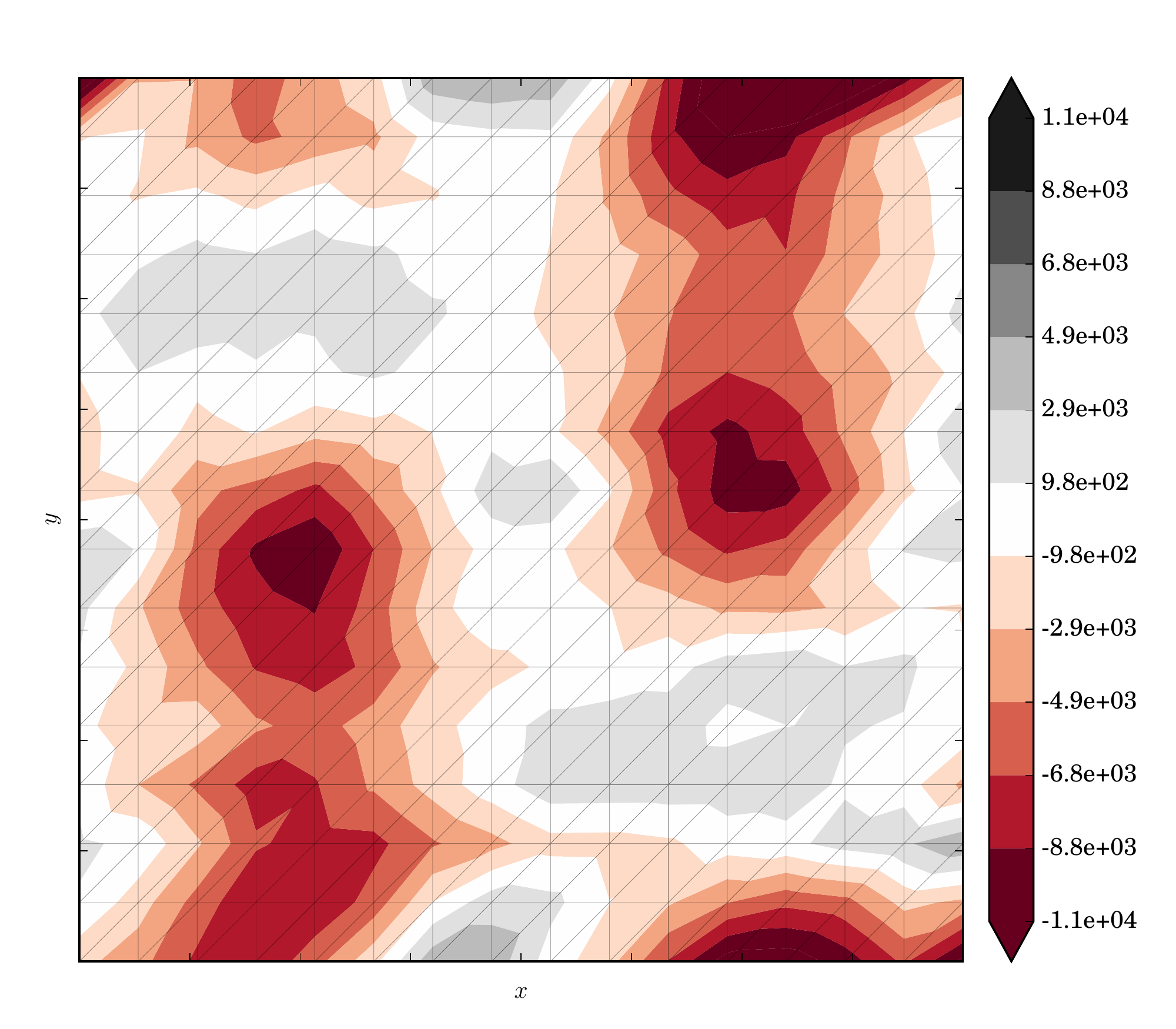}
  \caption{$M_{ij}$}
  \label{rs_M_ij}
  \end{subfigure}
  \begin{subfigure}[b]{0.3\linewidth}
    \includegraphics[width=\linewidth]{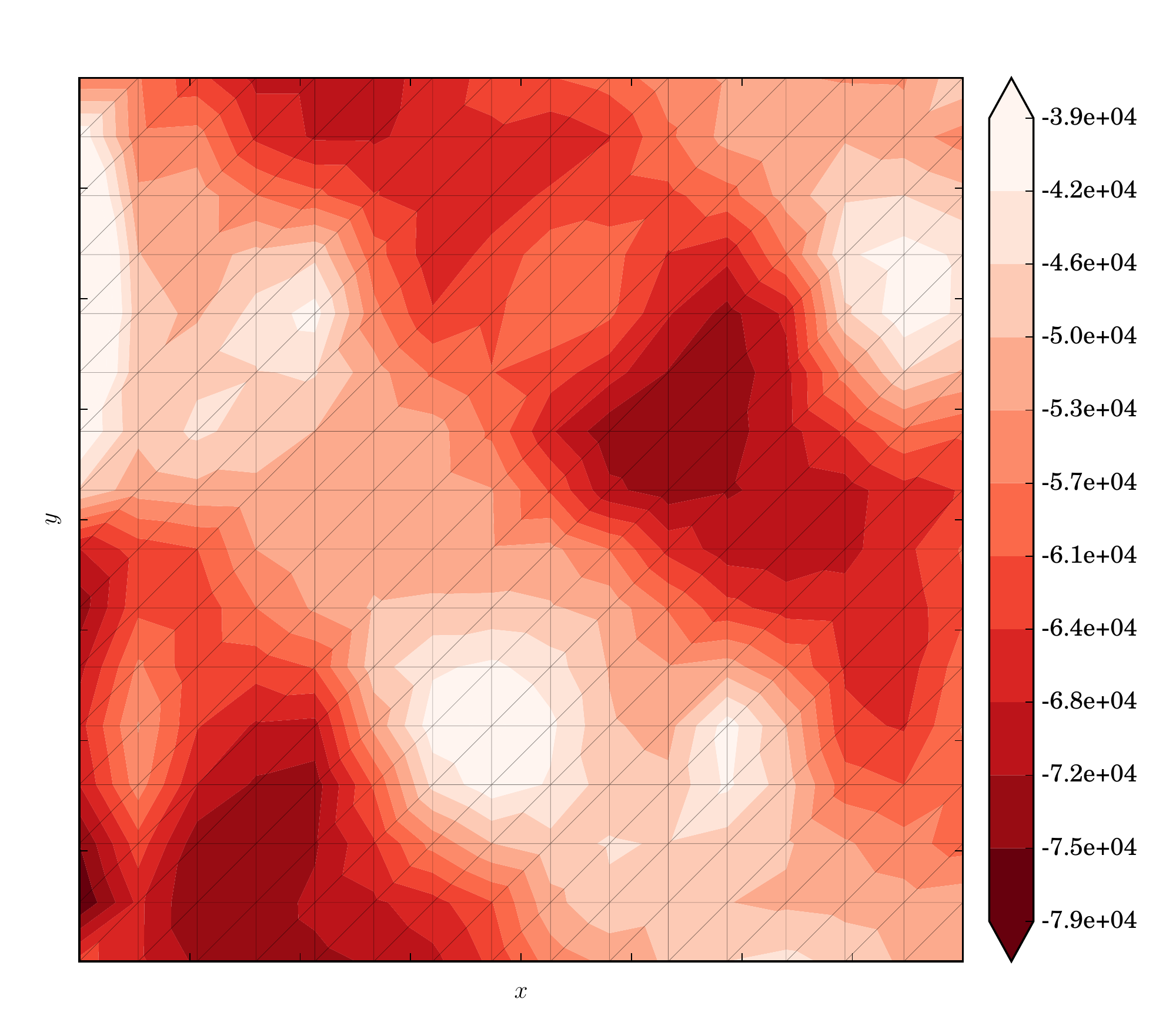}
  \caption{$M_{iz}$}
  \label{rs_M_iz}
  \end{subfigure}

  \begin{subfigure}[b]{0.3\linewidth}
    \includegraphics[width=\linewidth]{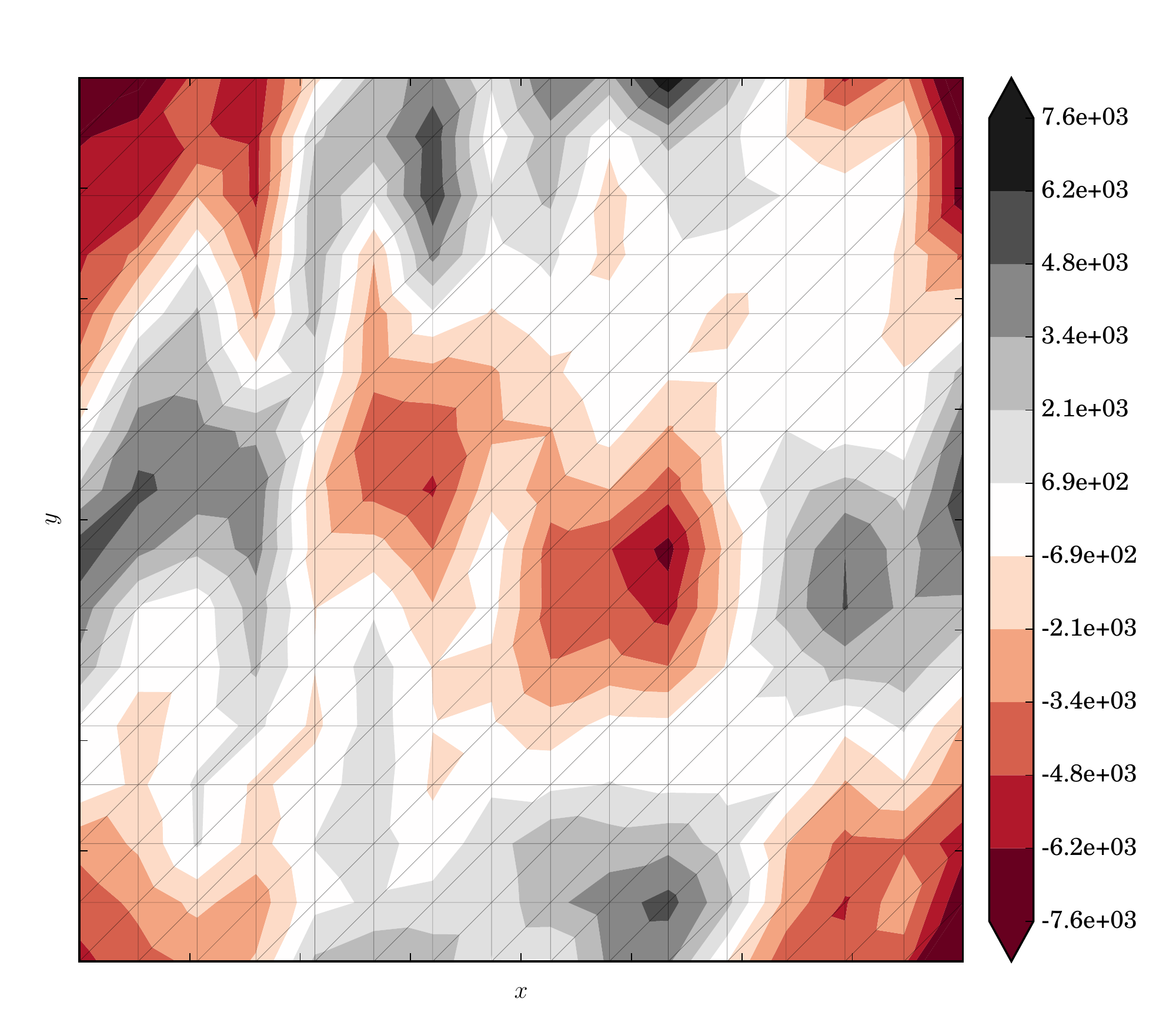}
  \caption{$M_{ji}$}
  \label{rs_M_ji}
  \end{subfigure}
  \begin{subfigure}[b]{0.3\linewidth}
    \includegraphics[width=\linewidth]{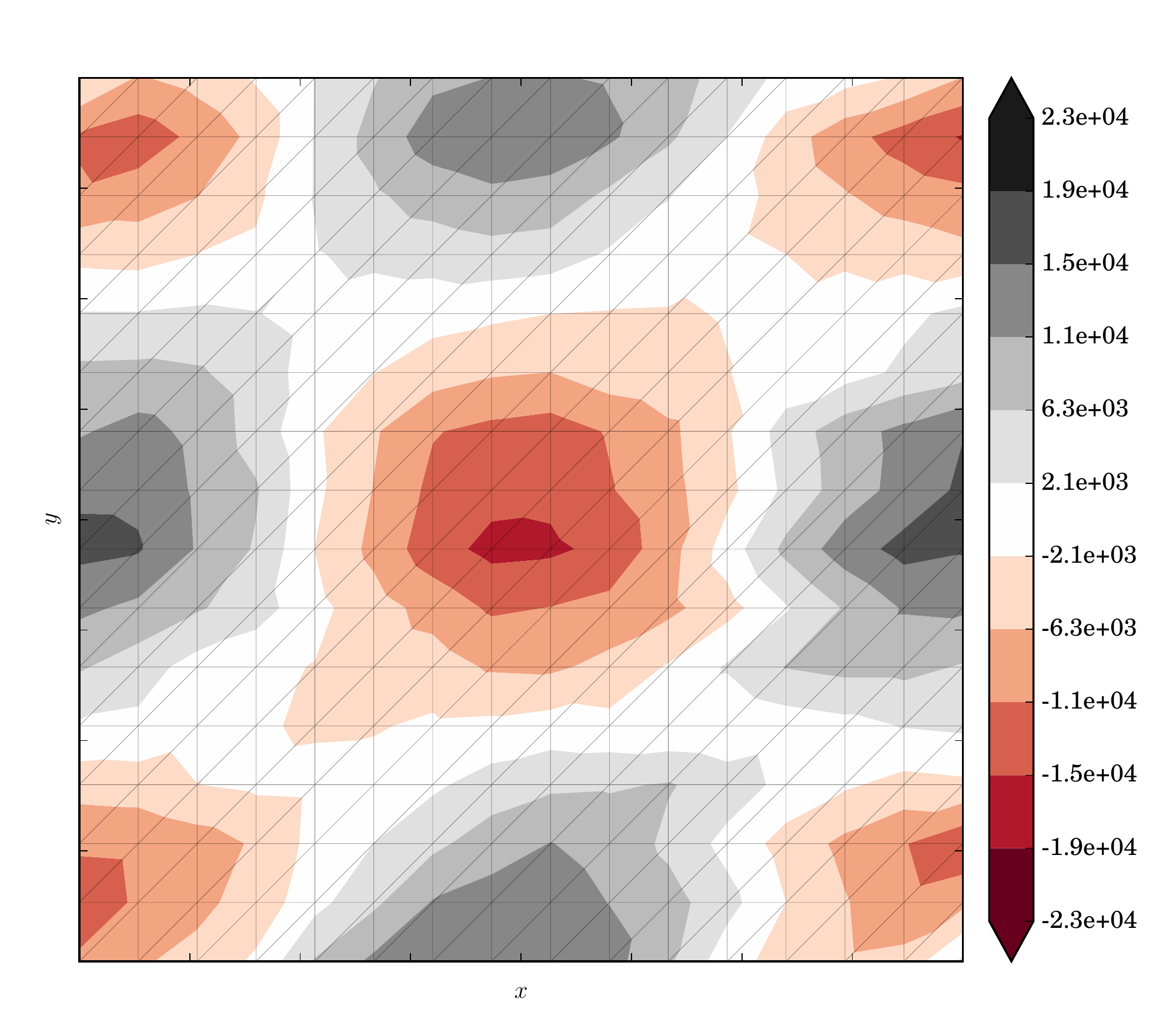}
  \caption{$M_{jj}$}
  \label{rs_M_jj}
  \end{subfigure}
  \begin{subfigure}[b]{0.3\linewidth}
    \includegraphics[width=\linewidth]{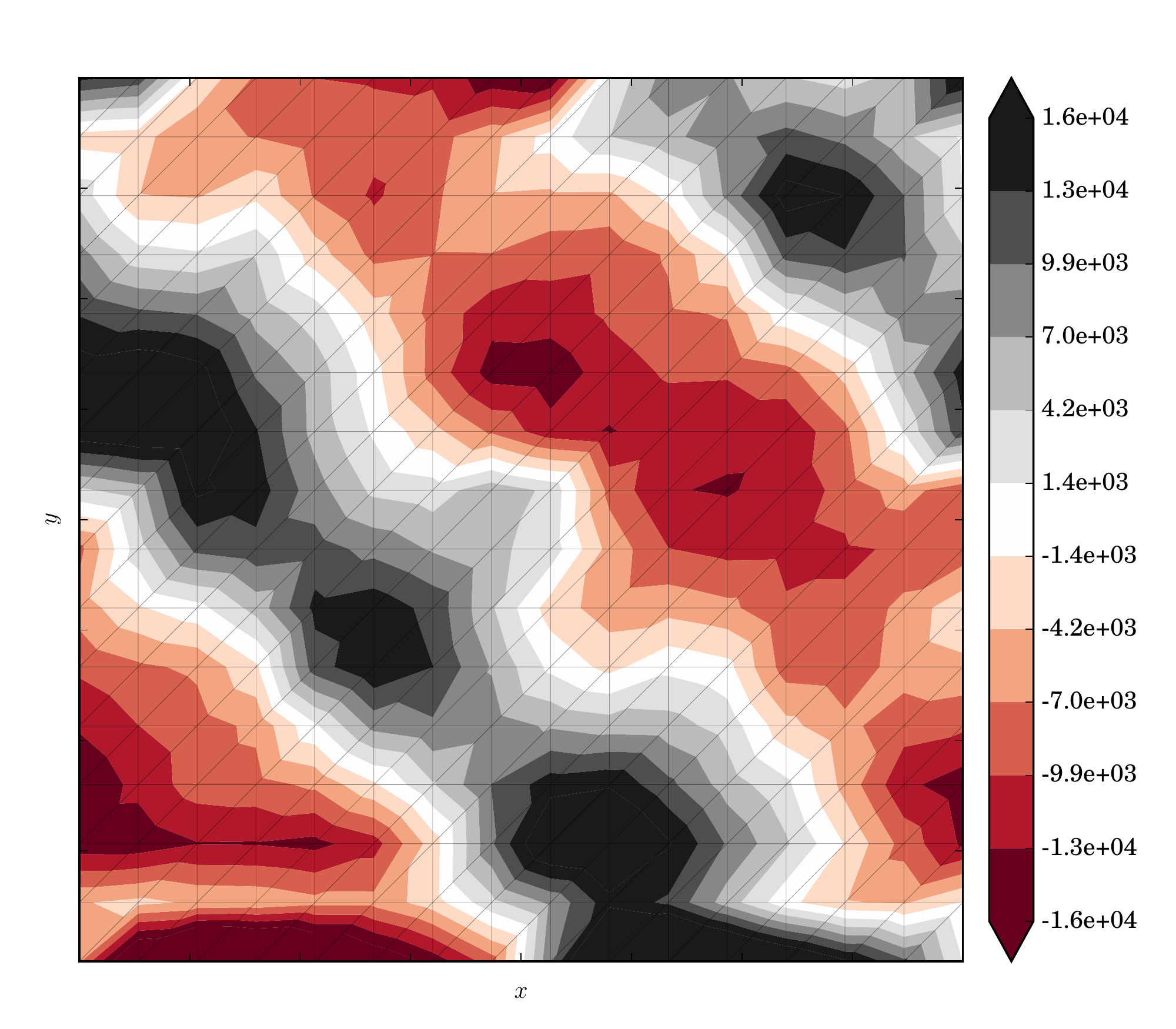}
  \caption{$M_{jz}$}
  \label{rs_M_jz}
  \end{subfigure}

  \begin{subfigure}[b]{0.3\linewidth}
    \includegraphics[width=\linewidth]{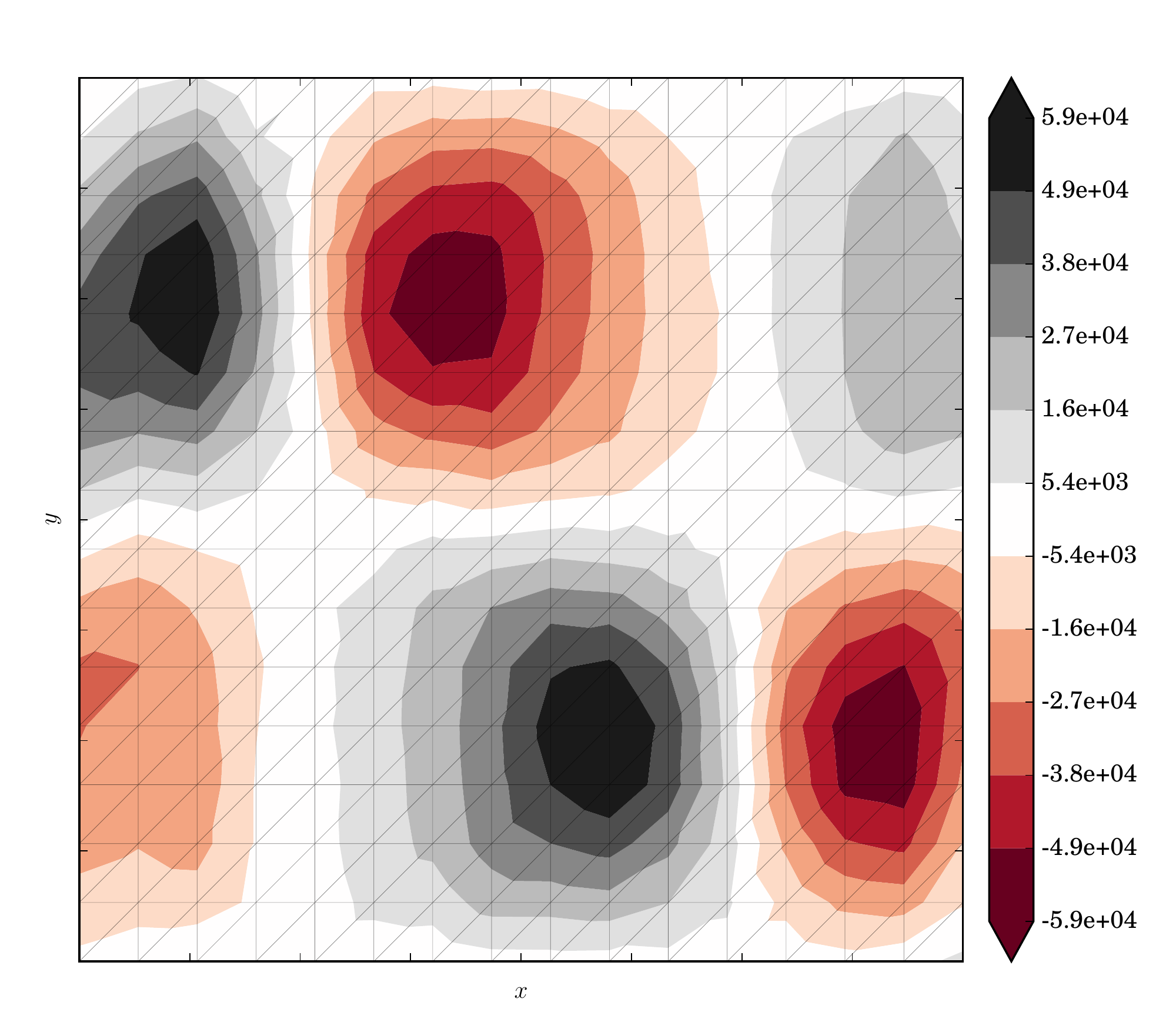}
  \caption{$M_{zi}$}
  \label{rs_M_zi}
  \end{subfigure}
  \begin{subfigure}[b]{0.3\linewidth}
    \includegraphics[width=\linewidth]{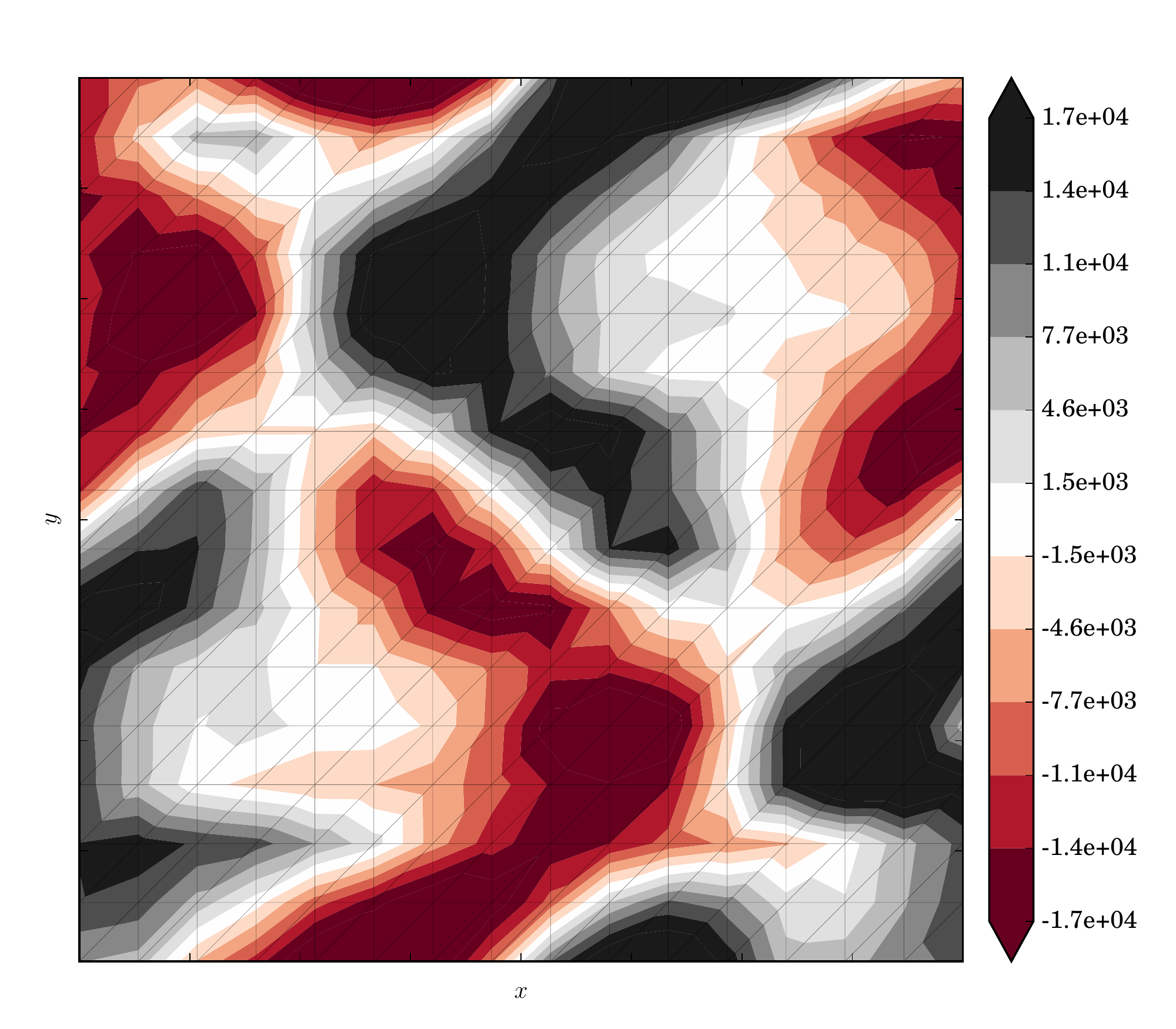}
  \caption{$M_{zj}$}
  \label{rs_M_zj}
  \end{subfigure}
  \begin{subfigure}[b]{0.3\linewidth}
    \includegraphics[width=\linewidth]{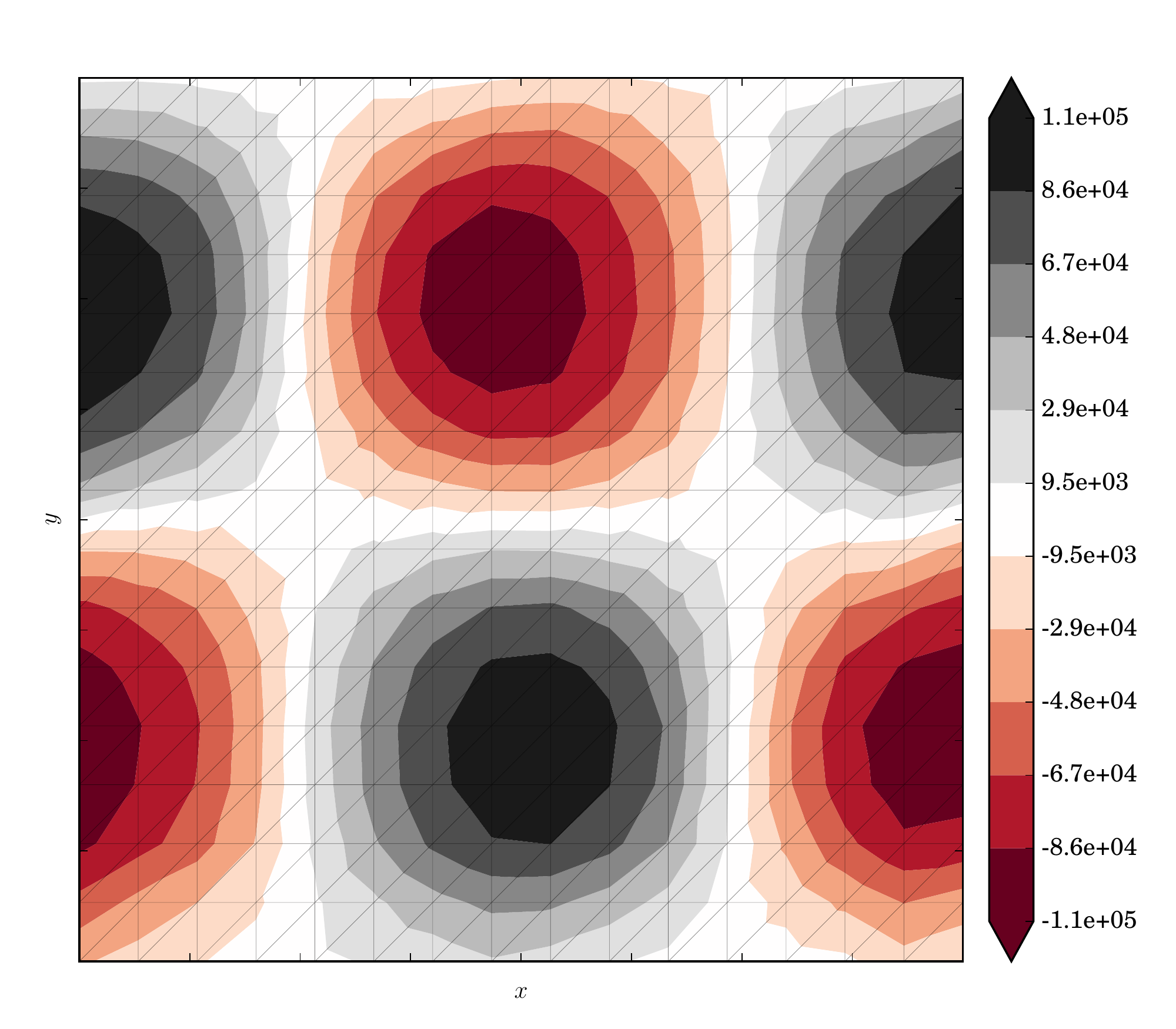}
  \caption{$M_{zz}$}
  \label{rs_M_zz}
  \end{subfigure}
 
  \caption[ISMIP-HOM reformulated-Stokes membrane stress balance]{Reformulated-Stokes membrane stress balance $M_{kk}$.}

  \label{rs_membrane_stress_balance}

\end{figure*}

%===============================================================================

%===============================================================================
%===============================================================================

\chapter{Ice age} \label{ssn_ice_age}

\index{Ice age}
The total change in age with time is always equal to unity; for every step forward in time, the age of ice will age by an equivalent step.  In order to quantify this change in the Eulerian coordinate system, the dependence of age on both time \emph{and} space requires the evaluation of the material derivative of age $a_{\text{ge}}$ with the \index{Chain rule} \index{Material derivative} \emph{chain rule} : 
\begin{align}
  \label{age_equation}
  \frac{da_{\text{ge}}}{dt} &= 1 \notag \\
  \frac{\partial a_{\text{ge}}}{\partial t} \frac{\partial t}{\partial t} + \frac{\partial a_{\text{ge}}}{\partial x} \frac{\partial x}{\partial t} + \frac{\partial a_{\text{ge}}}{\partial y} \frac{\partial y}{\partial t} + \frac{\partial a_{\text{ge}}}{\partial z} \frac{\partial z}{\partial t} &= 1 \notag \\
  \frac{\partial a_{\text{ge}}}{\partial t} + u \frac{\partial a_{\text{ge}}}{\partial x} + v \frac{\partial a_{\text{ge}}}{\partial y} + w \frac{\partial a_{\text{ge}}}{\partial z} &= 1 \notag \\
  \frac{\partial a_{\text{ge}}}{\partial t} + \mathbf{u} \cdot \nabla a_{\text{ge}} &= 1,
\end{align}
where velocity $\mathbf{u} = [u\ v\ w]\T$ is the solution to one of the momentum-balance models described in Chapter \ref{ssn_momentum_and_mass_balance}.

In areas where the accumulation/ablation rate $\dot{a}$ is positive, the ice-age on the surface will be new at the current time step.  Hence a homogeneous, essential boundary condition is present there.  Due to the fact that age equation (\ref{age_equation}) is hyperbolic, we cannot specify any other boundary conditions on the other `outflow' surfaces \citep{hughes_VII}.  Therefore, the only boundary condition is the essential condition
\begin{align}
  \label{age_boundary_condition}
  a_{\text{ge}} &= 0 &&\text{ on } \Gamma_{S} \big|_{\dot{a} > 0} &&\leftarrow \text{new snow}.
\end{align}
Note that in steady-state this equation becomes
\begin{align}
  \label{ss_age_equation}
  \mathbf{u} \cdot \nabla a_{\text{ge}} &= 1.
\end{align}

\section{Variational form}

This problem is purely advective, and as such the weak solution to this problem using standard Galerkin methods is numerically unstable (read to \S \ref{ssn_stabilized_methods}).  To solve this issue, streamline upwind/Petrov-Galerkin (SUPG) stabilization \citep{brooks} is applied, which has the effect of adding artificial diffusion to the variational form in areas of high velocity.

First, note that the intrinsic-time parameter for this problem is identical to (\ref{tau_supg}) with $\xi = 1$ due to the fact that no diffusion is present.  Making the appropriate substitutions in (\ref{tau_supg}) results in the intrinsic time parameter \index{Intrinsic-time parameter!Age equation}
\begin{align}
  \label{tau_age}
  \tau_{\text{age}} = \frac{h}{2 \Vert \mathbf{u} \Vert},
\end{align}
where $h$ is the element size.

Therefore, using general stabilized form (\ref{generalized_form}) with the operator $\Lu v = \mathbf{u} \cdot \nabla v$, SUPG operator (\ref{bubble_supg_operator}), and intrinsic-time parameter (\ref{tau_age}), the stabilized variational form corresponding to steady-state age equation (\ref{ss_age_equation}) consists of finding $a_{\text{ge}} \in \trialspace$ (se trial space (\ref{trial_space})) such that
\begin{align}
  \label{age_variational_form_intermediate}
  \int_{\Omega} \mathbf{u} \cdot \nabla a_{\text{ge}}\ \phi\ d\Omega + \int_{\Omega} \tau_{\text{age}} \mathbf{u} \cdot \nabla \phi\ d\Omega &= \int_{\Omega} \phi\ d\Omega
\end{align}
for all $\phi \in \testspace$ (see test space (\ref{test_space})), subject to essential boundary condition (\ref{ss_age_equation}).

The implementation of this problem by CSLVR is shown in Code Listing \ref{cslvr_age}.

\pythonexternal[label=cslvr_age, caption={CSLVR source code for the \texttt{Age} class.}, firstline=1, lastline=148]{cslvr_src/age.py}

%===============================================================================
%===============================================================================

\chapter{Application: Jakobshavn} \label{ssn_application_jakobshavn}

\begin{figure}
  \centering
    \includegraphics[width=\linewidth]{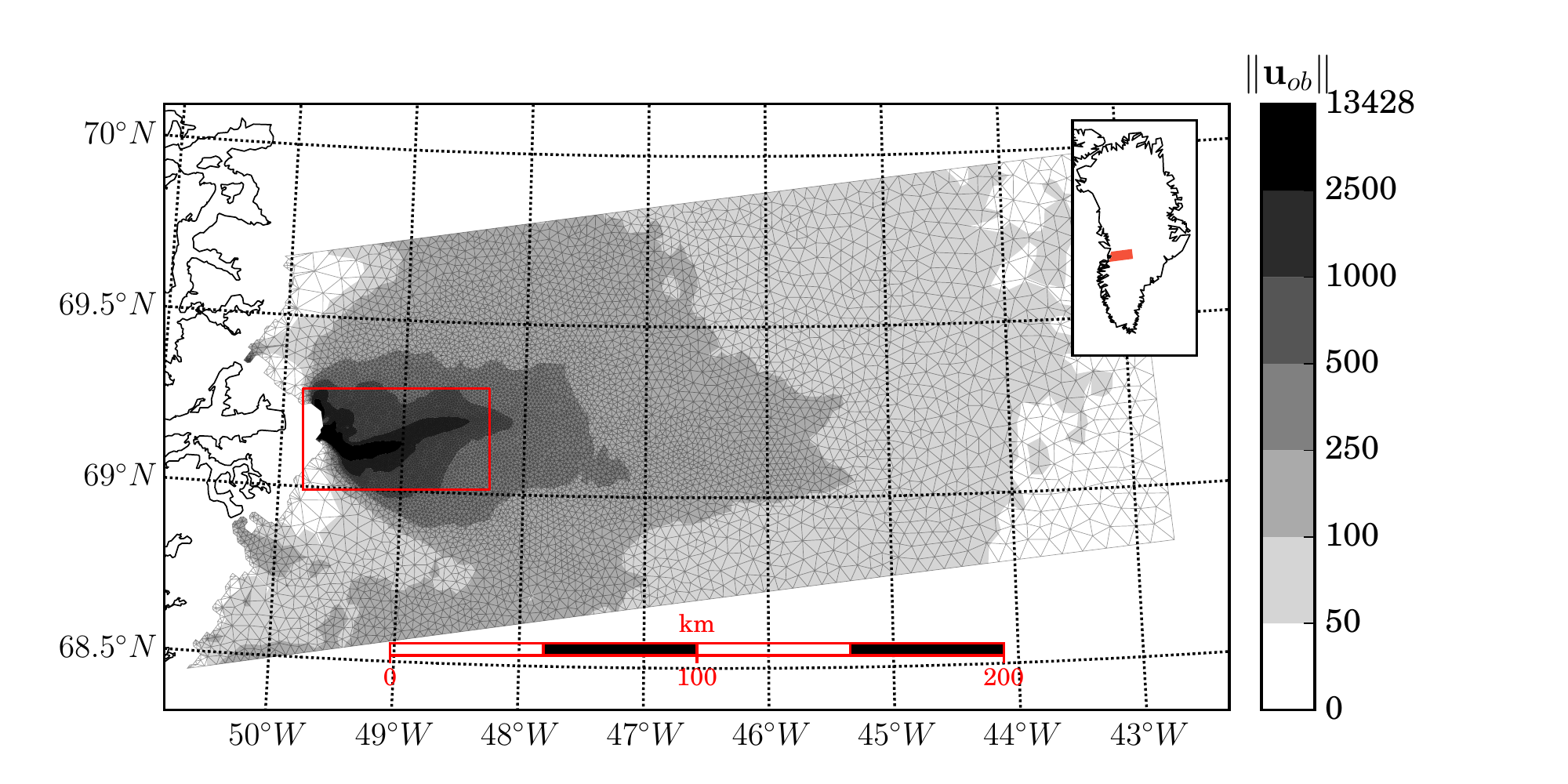}
  \caption[Jakobshavn Glacier mesh]{Surface velocity magnitude observations in m a\sups{-1} from \citet{rignot_greenland} over the Jakobshavn region highlighted in {\color[RGB]{245, 81, 58}orange} in the inlaid map.}
  \label{jakobshavn_region}
\end{figure}

\begin{figure*}
  \centering
    \includegraphics[width=0.8\linewidth]{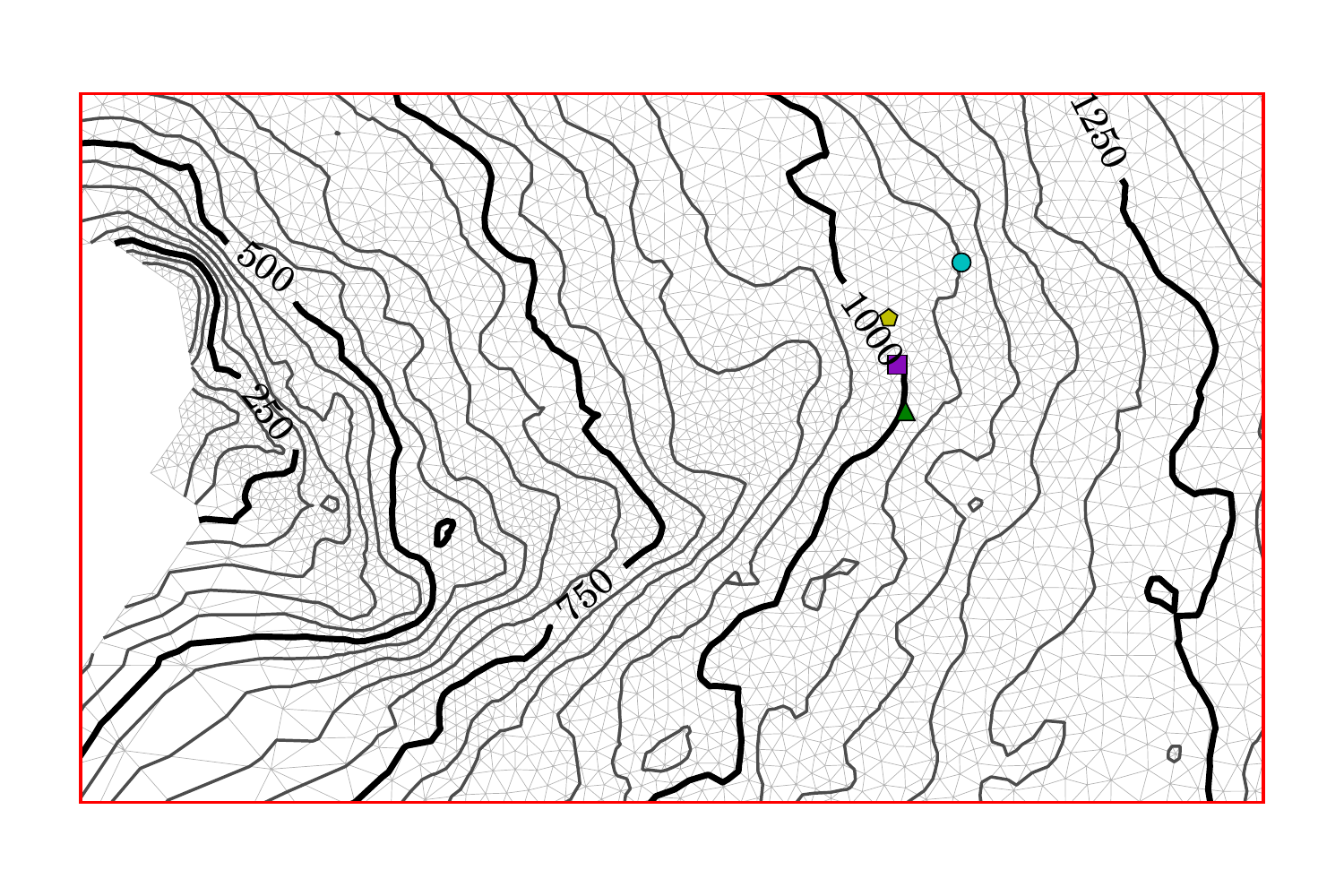}
    \includegraphics[width=0.8\linewidth]{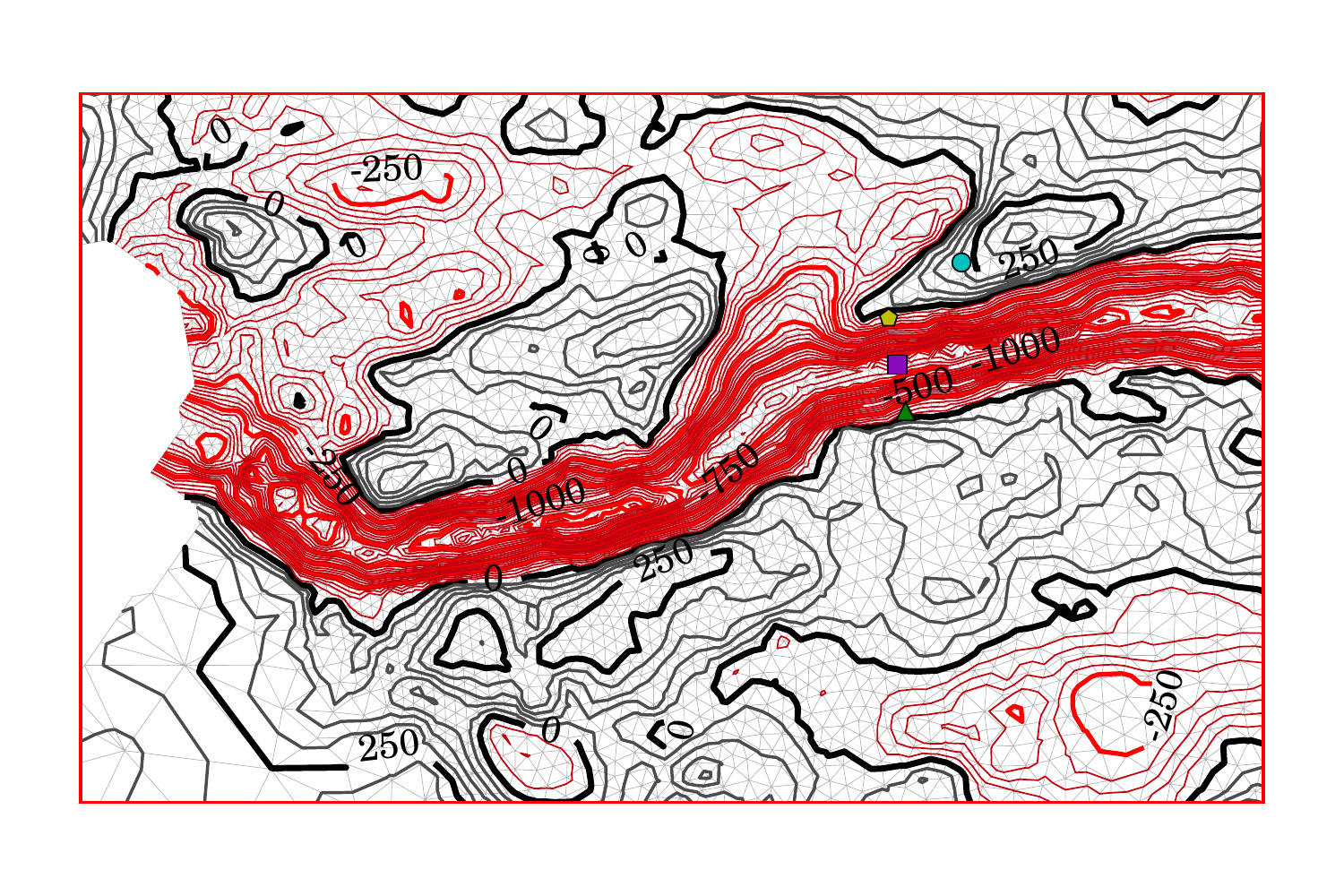}
  \caption[Jakobshavn Glacier topography]{Topography of surface $S$ in m (top) and basal topography in m (bottom) provided by \citet{bamber} for the rectangular region outlined in {\color{red}red} in Figure \ref{jakobshavn_region}.  Topography contours are spaced 50 m, with negative heights colored {\color{red}red}.  The colored points roughly correspond to the locations examined by \citet{luthi}.}
  \label{jakobshavn_topography}
\end{figure*}

\index{Non-linear differential equations!3D}

The region surrounding Jakobshavn Glacier was chosen to test the applicability of the methods of Chapters \ref{ssn_momentum_and_mass_balance}, \ref{ssn_internal_energy_balance}, and \ref{ssn_inclusion_of_velocity_data} to real-world data.  To simplify computations while simultaneously producing high-resolution results, the domain was restricted to a rectangular region with Eastern-most boundary located close to the divide, with an along-flow width of approximately 120 km.  A triangular mesh was first created with a minimum cell diameter of 500 m in the fastest-flowing regions and maximum cell diameter of 15 km in the slowest-flowing regions.  This mesh was then extruded vertically into ten layers of tetrahedra and deformed to match the geometry supplied by \citet{bamber} (see Code Listing \ref{cslvr_jakobshavn_gen_mesh} and Figures \ref{jakobshavn_region} and \ref{jakobshavn_topography}).

In order to retain the well-posedness of thermo-mechanical system (\ref{energy_euler_lagrange}), (\ref{gamma_a_ebc} -- \ref{energy_flux}), (\ref{cons_momentum} - \ref{cons_mass}), and (\ref{surface_stress} -- \ref{impenetrability}), extra boundary conditions must be applied along the interior regions of ice having been cut by the specification of the mesh boundary, denoted $\Gamma_D$ (Figure \ref{ice_profile_domain}).  These extra conditions are
\begin{align}
  \label{gamma_d_stress}
  \sigma \cdot \mathbf{n} &= - f_c \mathbf{n} &&\text{ on } \Gamma_D &&\leftarrow \text{ full-Stokes cry'st'c str.} \\
  \label{gamma_d_bp_stress}
  \sigma_{\text{BP}} \cdot \mathbf{n} &= \mathbf{0} &&\text{ on } \Gamma_D &&\leftarrow \text{ first-order cry'st'c str.} \\
  \label{gamma_d_energy}
  \theta &= \theta_S &&\text{ on } \Gamma_D &&\leftarrow \text{ Dirichlet},
\end{align}
with cryostatic stress $f_c$ defined by (\ref{exterior_pressure}).  As depicted by the $f_c \mathbf{n}$ vectors in Figure \ref{ice_profile_domain}, full-Stokes stress condition (\ref{gamma_d_stress}) and first-order stress condition (\ref{gamma_d_bp_stress}) represent the contribution of stress presented to the ice column by the surrounding ice removed from the domain.

While it is unlikely that the energy is unchanging from the surface throughout the interior of the ice along the surface $\Gamma_D$ as stated by `Dirichlet' condition (\ref{gamma_d_energy}), the region of interest lies $\approx$ 75 km from this boundary, with very low velocity magnitude (Figure \ref{jakobshavn_region}).  Thus it is expected that the effect of this inconsistency is small.

To initialize the basal traction, we used the SIA approximate field $\beta_{\text{SIA}}$ described in \S \ref{ssn_dual_optimization}.  In order to initialize flow-rate factor (\ref{rate_factor}), the initial energy values $\theta^i$ throughout the interior $\Omega$ were initialized using quadratic energy (\ref{energy_quad}) with initial water content $W^i(z) = 0$ and surface temperatures $T^i(x,y,z) = T_S(x,y), \forall z$ provided by \citet{fausto}.  Also, pressure melting temperature (\ref{temperature_melting}) requires that pressure $p$ be initialized; here we applied cryostatic pressure $f_c$ in (\ref{exterior_pressure}) such that $p^i(z) = f_c(z) = \rho g (S - z)$.  Finally, the basal-water discharge $F_b$ across the entire basal domain was initialized to zero.

The $L^2$ and logarithmic cost functional coefficients in (\ref{momentum_objective}), $\gamma_1$ and $\gamma_2$, respectively, were determined as described in \S \ref{ssn_ismip_hom_inverse_sims}; by completing Algorithm \ref{tmc_da} several times and adjusting their relative values such that at the end of the process their associated functionals were of approximately the same order (Figure \ref{convergence}).

An appropriate value for the regularization parameters $\gamma_3$ and $\gamma_4$ in momentum objective (\ref{momentum_objective}) was determined from an L-curve process (\S \ref{ssn_l_curve}).  First, it was noted that when performing this procedure for only one regularization term, i.e., setting one of either $\gamma_3$ or $\gamma_4$ to zero, this process resulted in choosing $\gamma_3 = \gamma_4 = 10$.  We then took $\gamma_4 = 10$ and increased $\gamma_3$ until the Tikhonov regularization functional began to affect the regularization of optimal traction $\beta^*$, resulting in choosing $\gamma_3 = 0.1$.

Finally, the geothermal heat flux was set to the average Greenland value of $\bar{q}_{geo} = 4.2 \times 10^{-2}$ W m\sups{-2} \citep{patterson} so as to minimize its effect on basal melt rate (\ref{basal_melt_rate}) and hence also basal water discharge $F_b$ as expressed by basal energy flux (\ref{energy_flux}).  For a complete listing of initial variables and coefficients used, see Table \ref{initial_values}.  The CSLVR scripts used to generate the mesh, data, and results are shown in Code Listings \ref{cslvr_jakobshavn_gen_mesh}, \ref{cslvr_jakobshavn_gen_data}, and \ref{cslvr_jakobshavn_da}, respectively.

\begin{table}
\centering
\caption[Jakobshavn simulation variables]{Initial values used for Jakobshavn sim's.}
\label{initial_values}
\begin{tabular}{llll}
\hline
\textbf{Variable} & \textbf{Value} & \textbf{Units}\footnote{All units including the dimension of time have been converted from per second to per annum in order to account for the very slow speed of ice.} & \textbf{Description} \\
\hline
$D$ & $0$ & m & ocean height \\
$q_{geo}$ & $4.2 \times 10\sups{-2}$ & W m\sups{-2} & geothermal heat flux \\
$\dot{\varepsilon}_0$ & $10\sups{-15}$ & a\sups{-1} & strain regularization \\
$\gamma_1$ & $5 \times 10^3$ & kg m\sups{-2}a\sups{-1} & $L^2$ cost coefficient \\
$\gamma_2$ & $10^{-2}$ & J a\sups{-1} & log. cost coefficient \\
$\gamma_3$ & $10^{-1}$ & m\sups{6}kg\sups{-1}a\sups{-1} & Tikhonov reg. coef. \\
$\gamma_4$ & $10$ & m\sups{6}kg\sups{-1}a\sups{-1} & TV reg. coeff. \\
$\theta^i$ & $\theta(T_S, 0)$ & J kg\sups{-1} & initial energy \\
$\beta^i$ & $\beta_{\text{SIA}}$ & kg m\sups{-2}a\sups{-1} & initial friction coef. \\
$F_b^i$ & $0$ & m a\sups{-1} & ini.~basal water flux \\
$p^i$   & $f_c$ & Pa & initial pressure \\
$k_0$   & $10^3$ & -- & energy flux coefficient \\
\end{tabular}
\end{table}

%===============================================================================
%===============================================================================

\section{Results}

\begin{figure*}
  \centering
    \includegraphics[width=\linewidth]{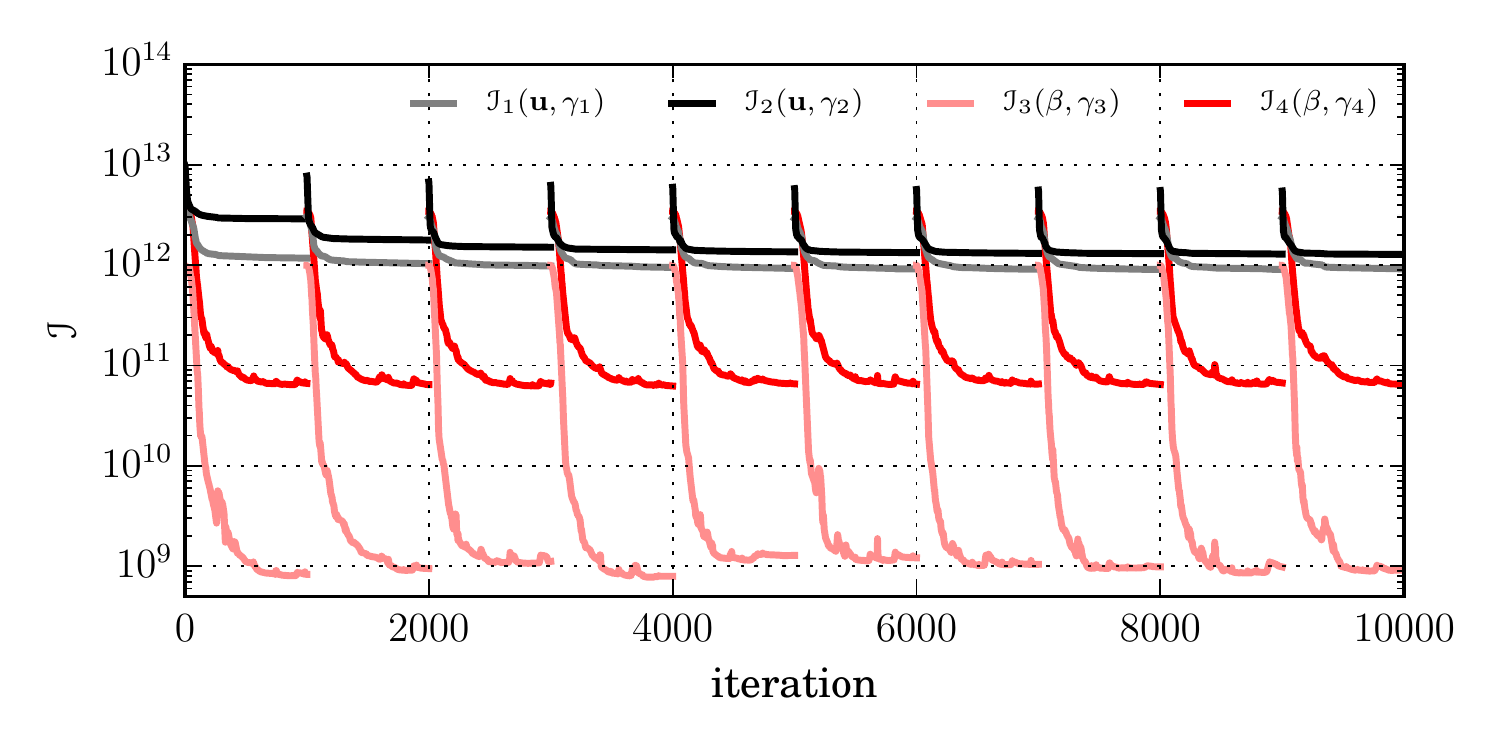}
  \caption[Jakobshavn simulation convergence plot]{Convergence plots of momentum objective functional (\ref{momentum_objective}) for Algorithm \ref{tmc_da} using maximum basal water content $W_c = 0.01$.  The individual components include logarithmic cost functional (black), $L^2$ cost functional ({\color[RGB]{128,128,128}dark gray}), Tikhonov regularization functional ({\color[RGB]{255,142,142} light red}), and total variation regularization functional ({\color[RGB]{255,0,0}red}). The peaks located every 1000 iterations indicate iterations of Algorithm \ref{tmc_da}.}
  \label{convergence}
\end{figure*}

The first set of results were generated using a maximum basal water content of $W_c = 0.01$, corresponding to the maximum observed in this area \citep{luthi}.  To ensure convergence, Algorithm \ref{tmc_da} was run for 10 iterations using 1000 iterations of momentum barrier problem (\ref{momentum_barrier}) (Figure \ref{convergence}).

\begin{figure*}
  \centering
    \includegraphics[width=0.48\linewidth]{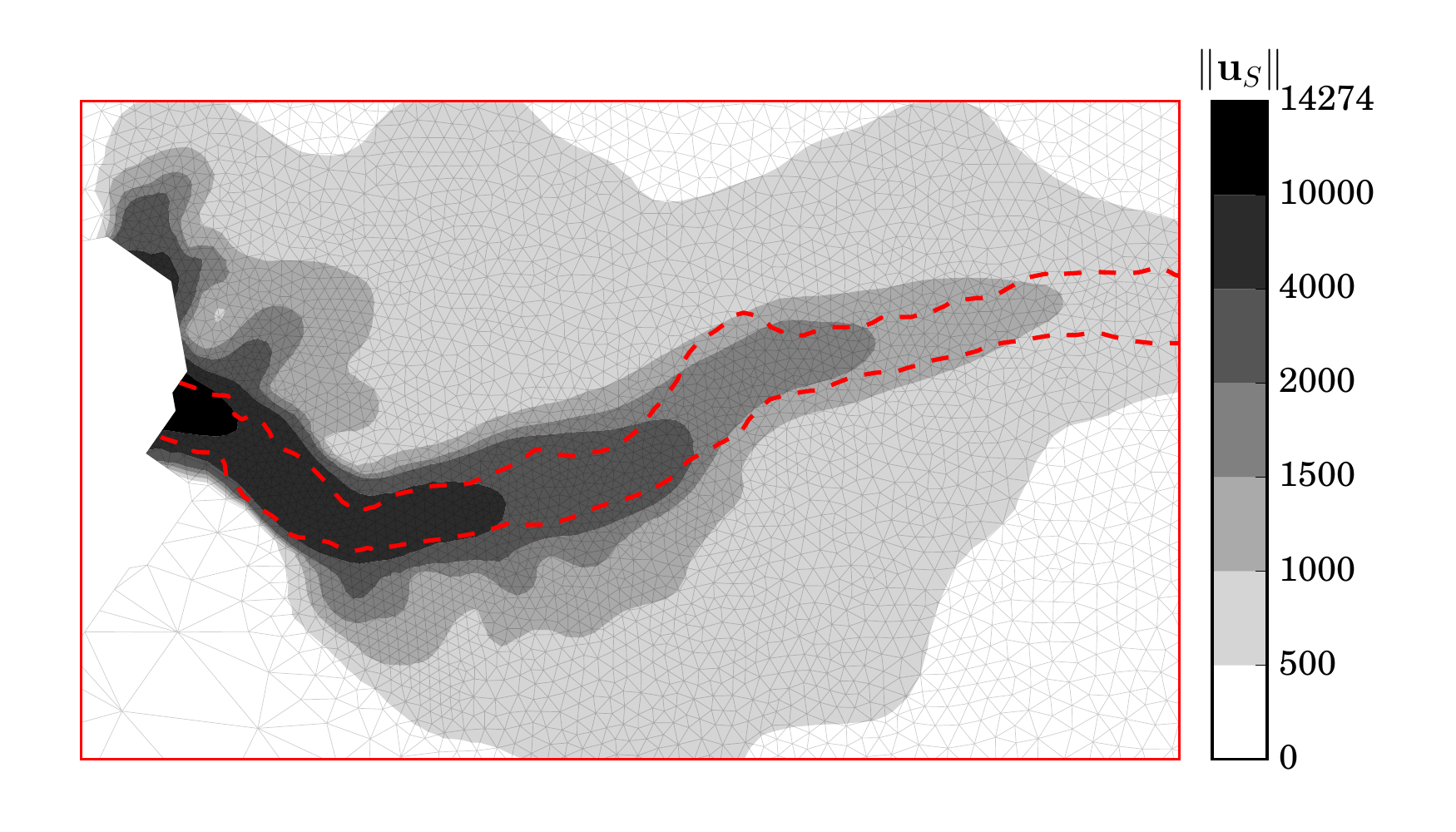}
    \quad
    \includegraphics[width=0.48\linewidth]{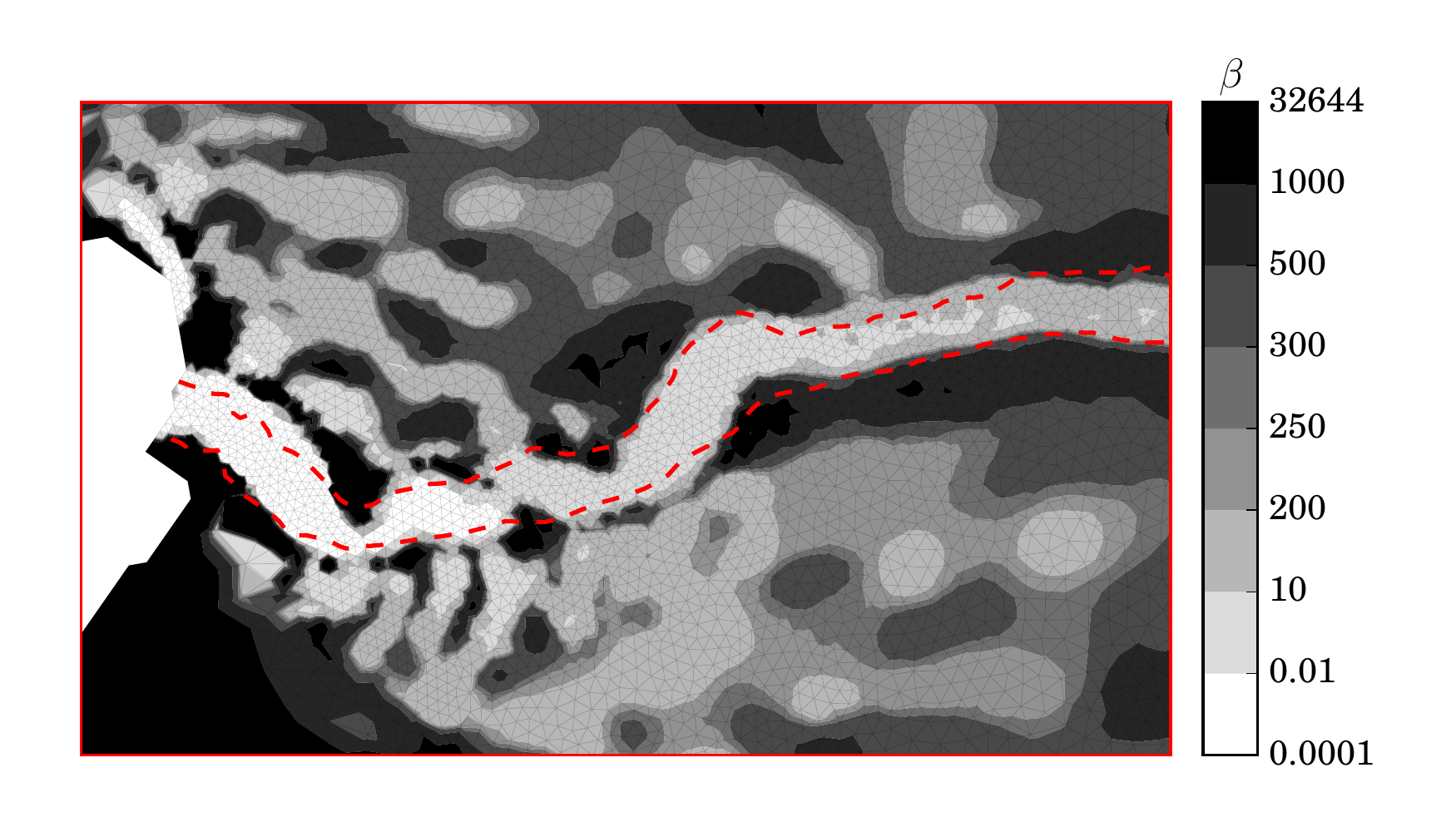}

    \includegraphics[width=0.48\linewidth]{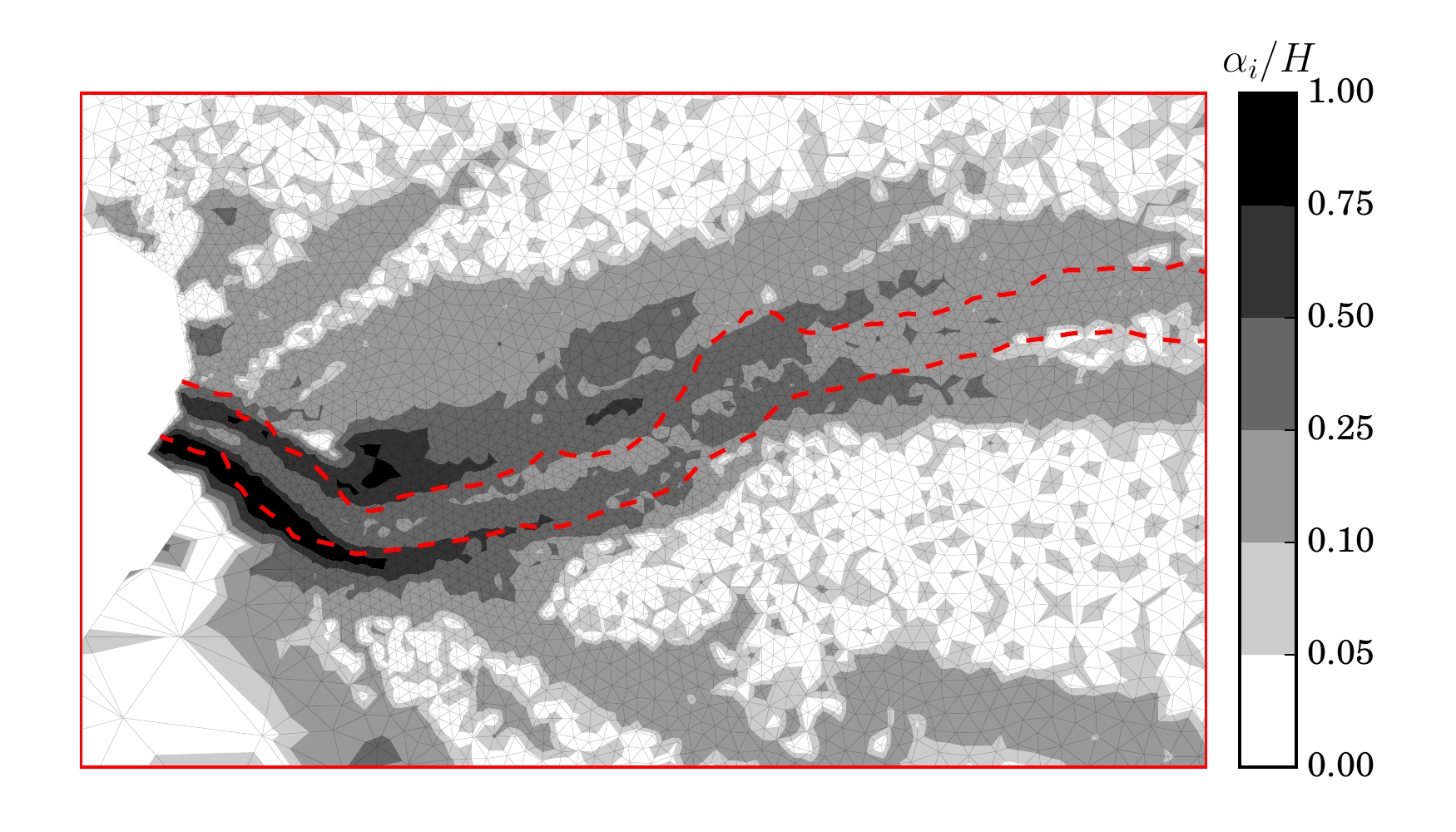}
    \quad
    \includegraphics[width=0.48\linewidth]{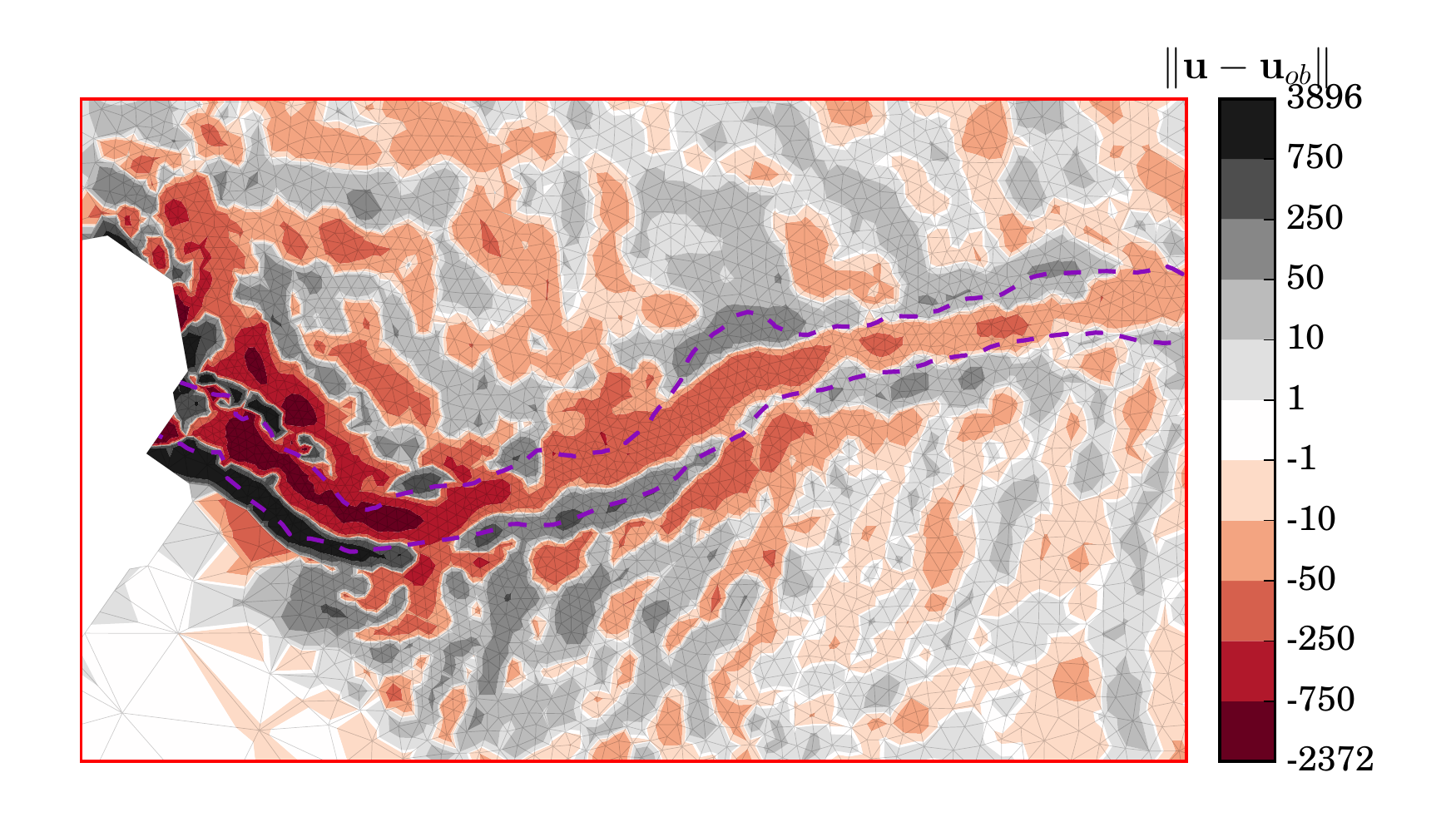}

    \includegraphics[width=0.48\linewidth]{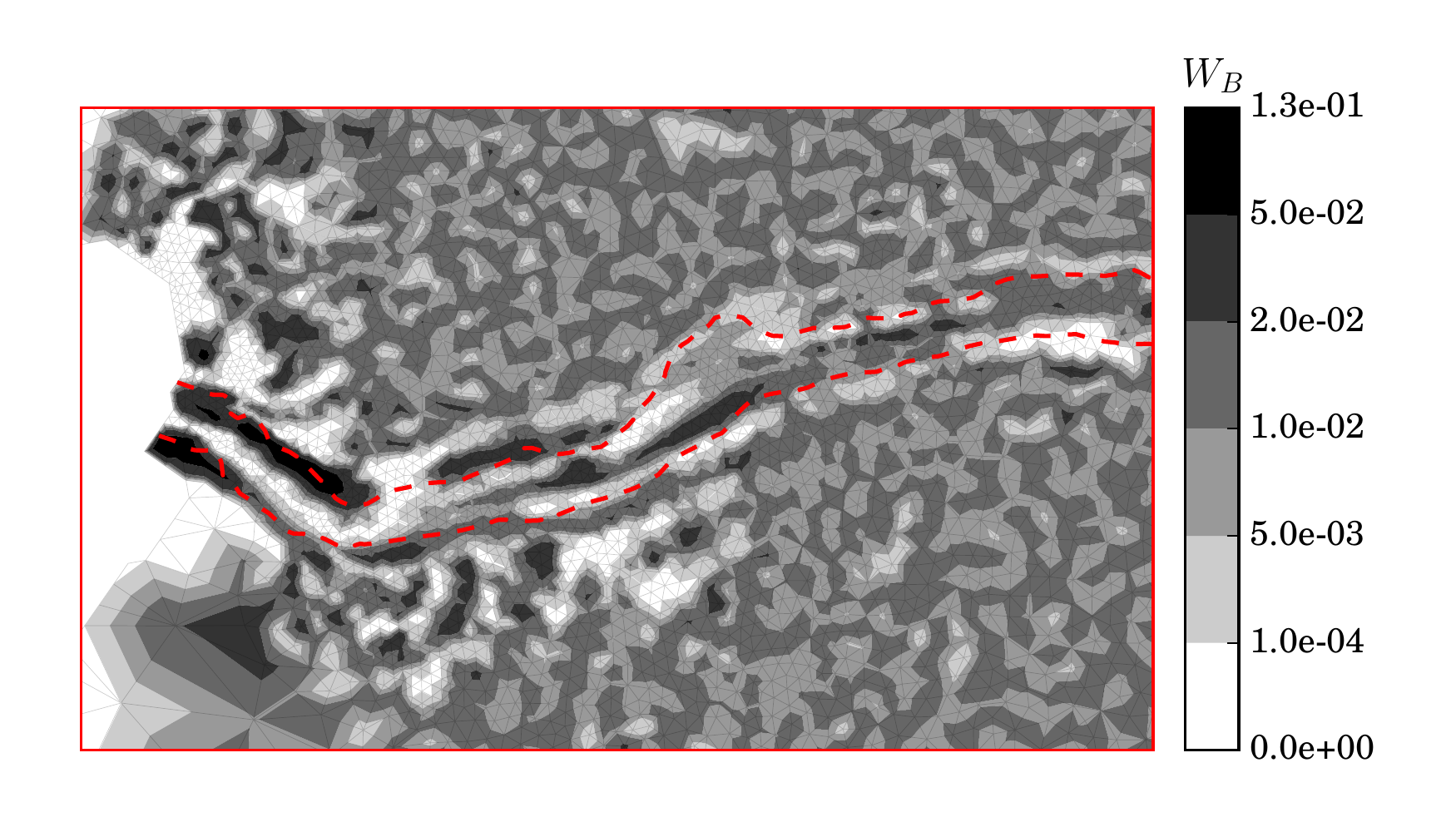}
    \quad
    \includegraphics[width=0.48\linewidth]{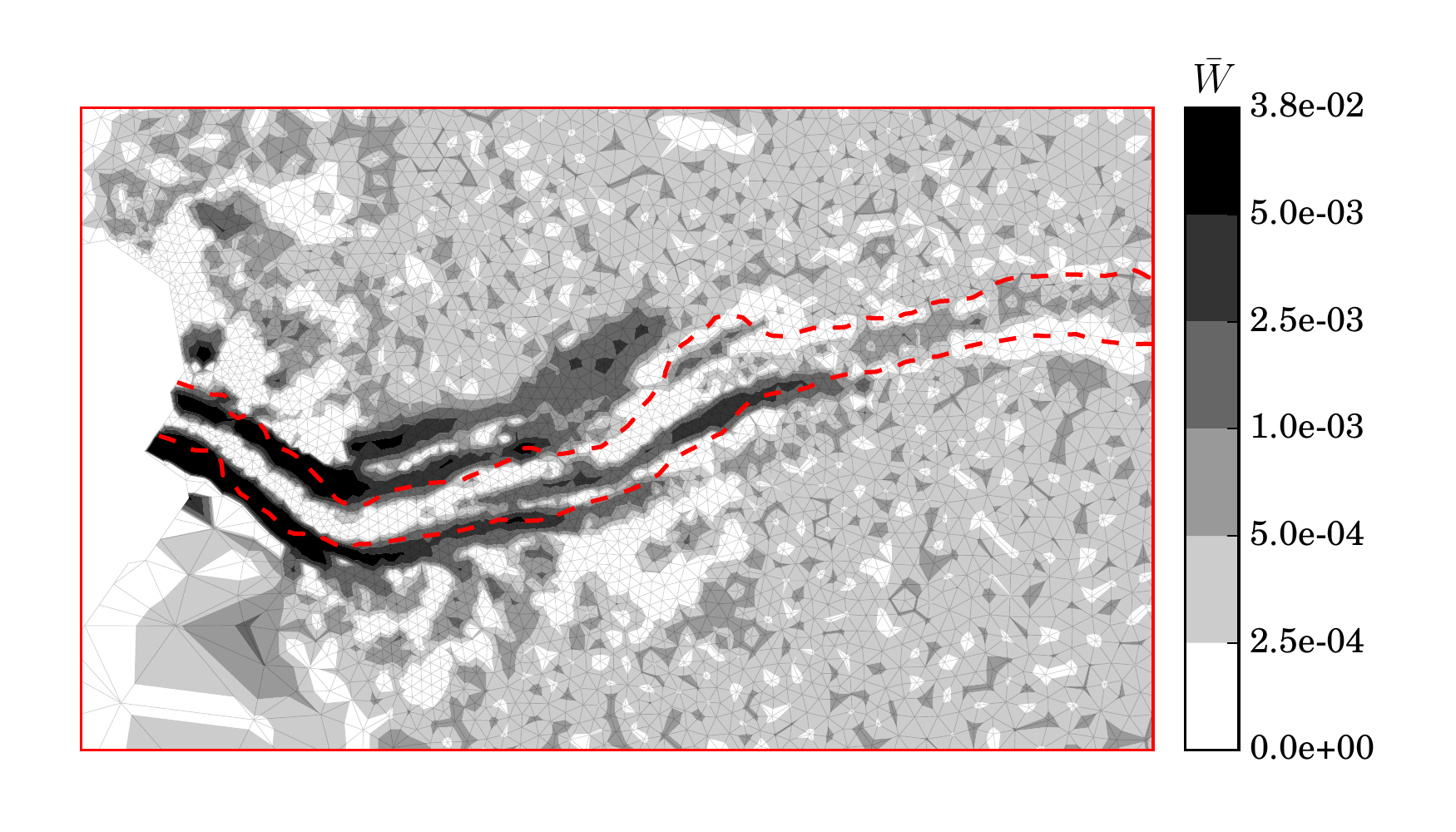}

    \includegraphics[width=0.48\linewidth]{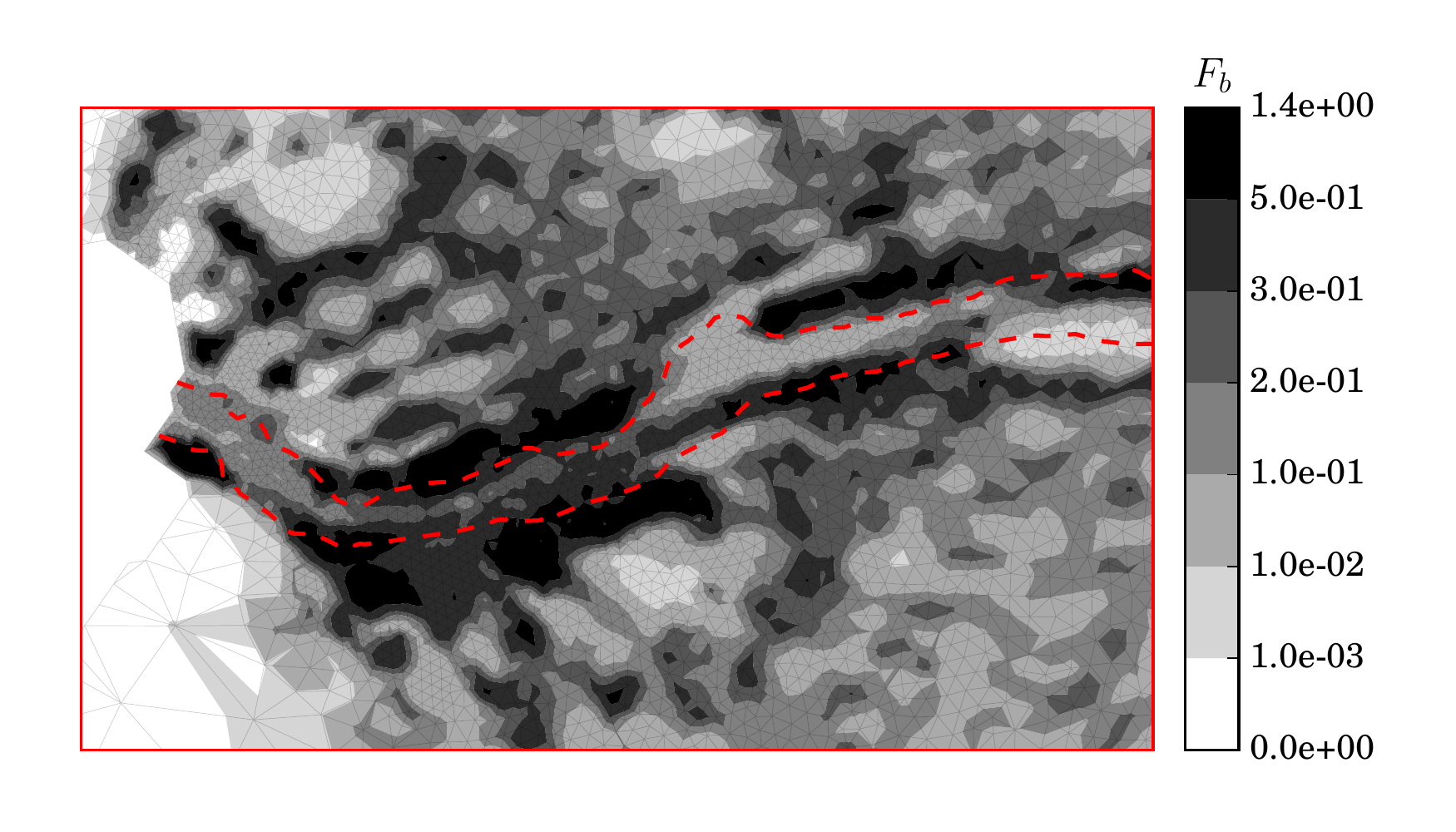}
    \quad
    \includegraphics[width=0.48\linewidth]{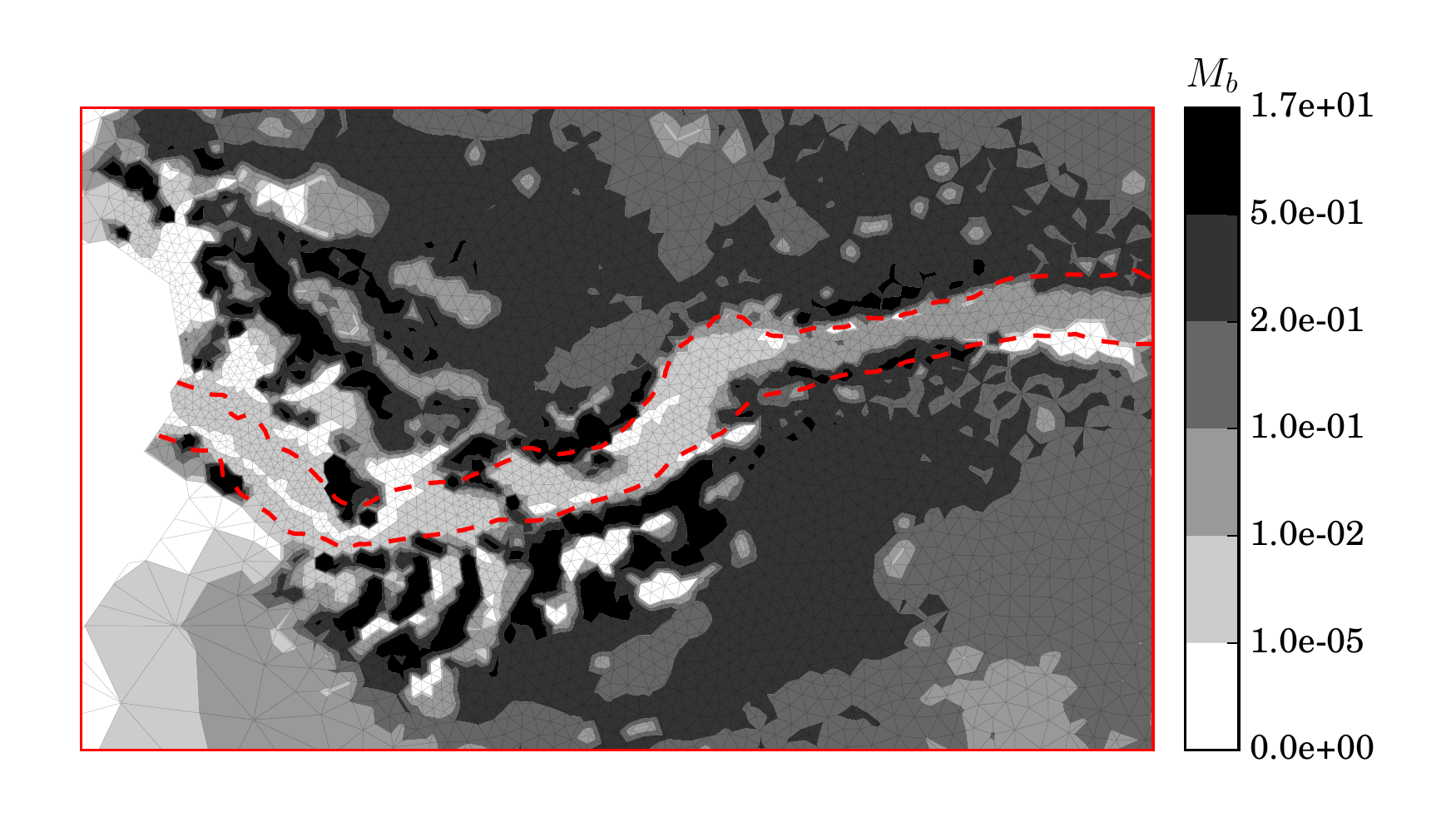}
  \caption[Jakobshavn simulation results]{Results of TMC-inversion procedure \ref{tmc_da} with maximum basal water content $W_c = 0.01$ for region of Western Greenland's Jakobshavn Glacier.  The optimized surface velocity magnitude $\Vert \mathbf{u}_S \Vert$ (top left), optimized basal traction $\beta$ (top right), ratio of ice column that is temperate (top middle left), velocity magnitude misfit (top middle right), basal water content $W$ (bottom middle left), vertically averaged water content (bottom middle right), optimized basal water discharge $F_b$ (bottom left) and basal melt rate $M_b$ (bottom right).  The dashed lines indicate the $-500$ m depth contour of basal topography $B$ depicted in Figure \ref{jakobshavn_topography}.}
  \label{jakobshavn_results}
\end{figure*}

For each iteration of this simulation, the surface velocity mismatch between modeled results and observations supplied by \citet{rignot_greenland} remained at least one order of magnitude lower than the surface speed, with greatest error near the terminus (Figure \ref{jakobshavn_results}).  As evident by Figure \ref{convergence}, both regularization functionals (\ref{tikhonov_regularization}, \ref{tv_regularization}) and the $L^2$ cost functional (\ref{l2_cost}) of momentum objective (\ref{momentum_objective}) reach approximately the same respective minimums at the end of each iteration, while the logarithmic cost functional (\ref{logarithmic_cost}) -- which is most affected by velocity mismatches in slower regions of flow distant from the terminus -- continues to decrease.  This may be due to the fact that for each iteration, both basal-melting rate $M_b$ and basal traction $\beta$ remain quite low near the region of fast flow nearest the terminus, and thus $L^2$ cost (\ref{l2_cost}) is little affected by changes in internal energy.

It was observed that for each iterative change in basal traction $\beta$ -- and thus also basal melt-rate (\ref{basal_melt_rate}) and internal friction (\ref{strain_heat}) -- the optimization procedure for basal-water discharge $F_b$ would reach a value required to flush out water generated by both of these sources.  As a result of this, at each iteration the basal water content remained close to $W_c$ and the vertically averaged water content $\bar{W}$ remained in all but a few areas below 5\% (Figure \ref{jakobshavn_results}).

\begin{figure*}
  \centering
    \includegraphics[width=0.48\linewidth]{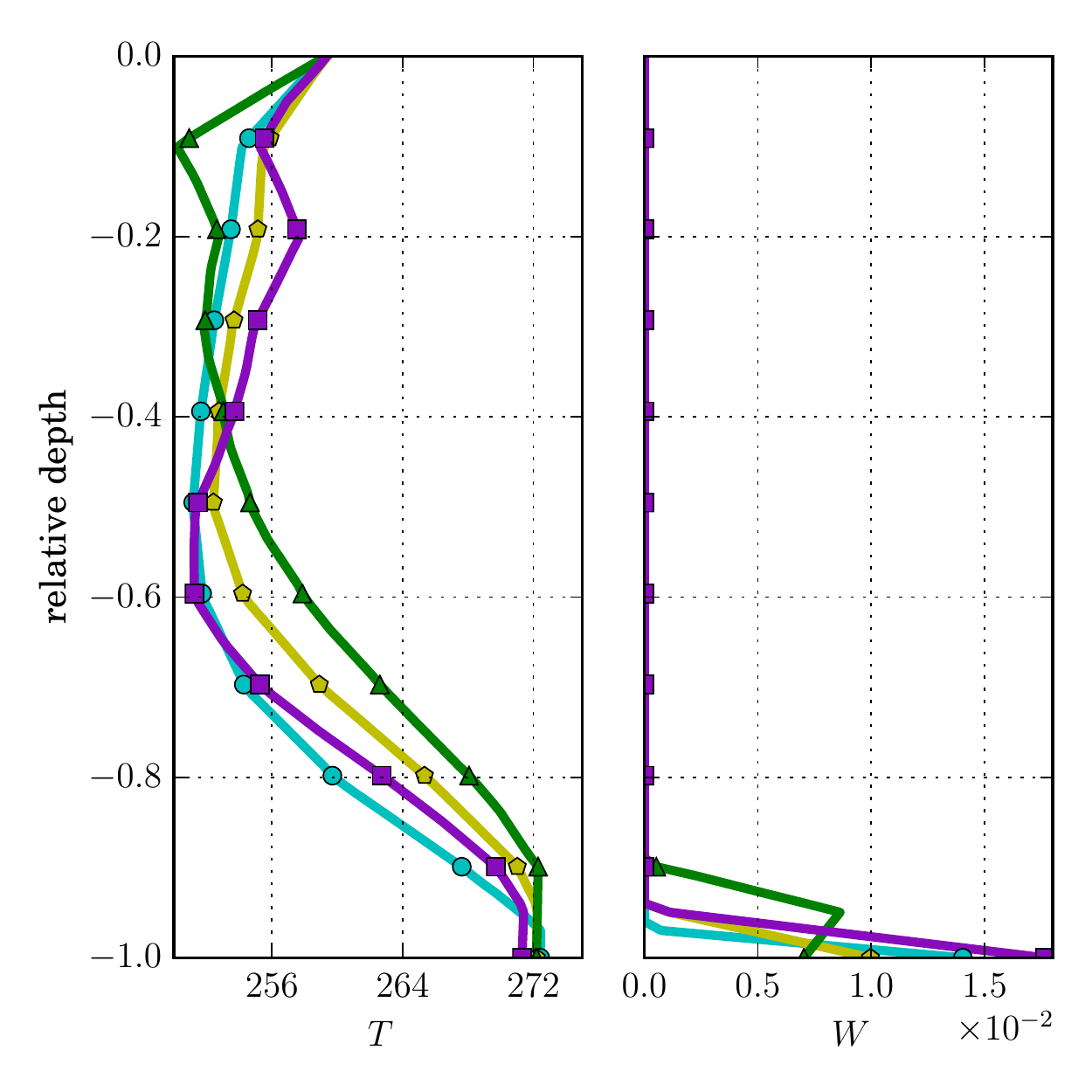}
    \quad
    \includegraphics[width=0.48\linewidth]{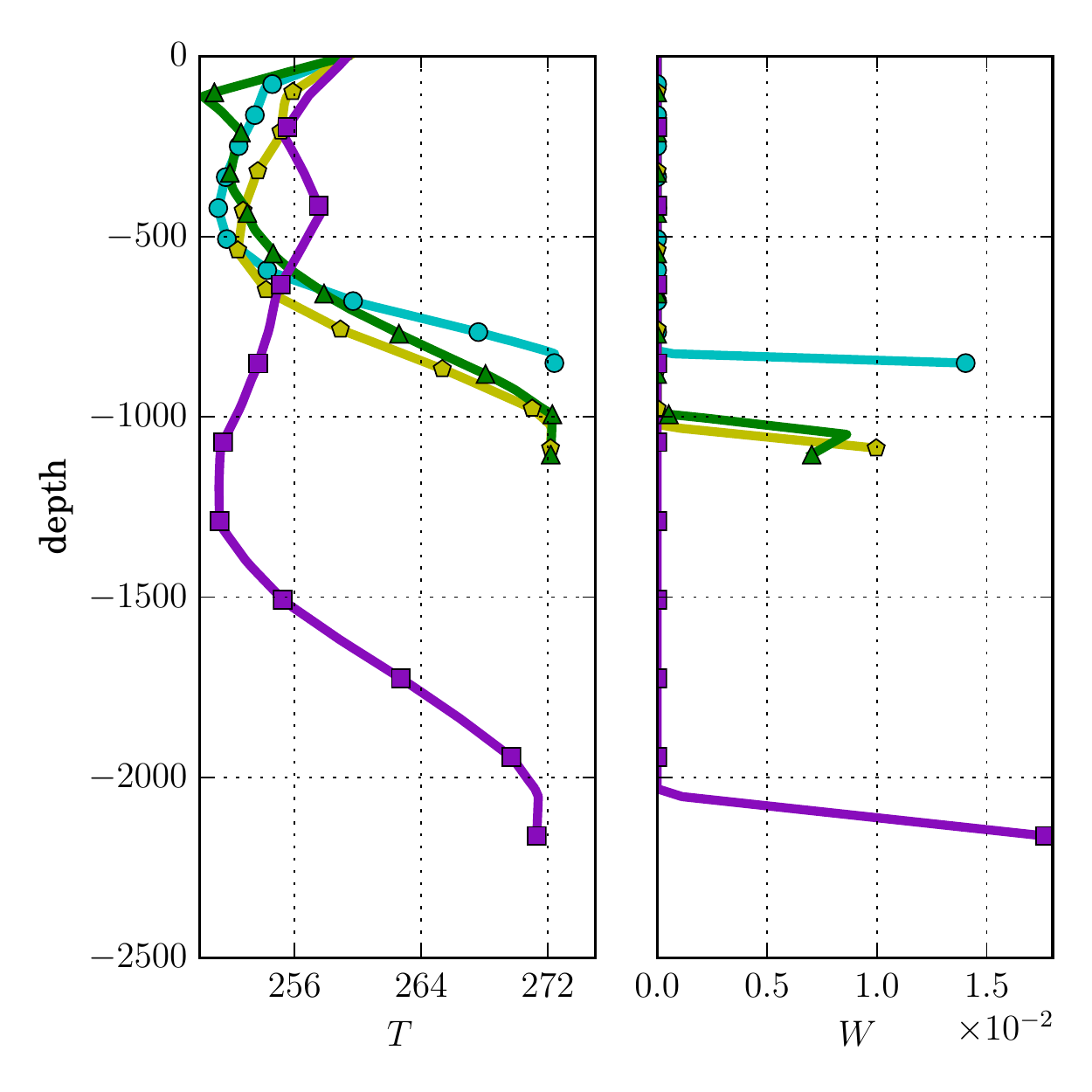}
  \caption[Jakobshavn energy profiles]{Temperature $T$ in degrees K and unit-less water content $W$ profiles with relative-depth-vertical coordinates (left) and actual depth coordinates (right).  The colored points correspond to the latitude/longitude coordinates in Figure \ref{jakobshavn_topography}, and the points indicate vertex locations of the mesh from which the data were interpolated.}
  \label{jakobshavn_profile}
\end{figure*}

The column percentage of temperate ice resulting from this simulation is on average about twice that reported by \citet{luthi}, despite the imposition of an approximately 10$^{\circ}$ K lower surface temperature.  This is likely due to both the improper specification of the CTS constraints, as explained by \citet{blatter_ArXiv}, and the low vertical resolution of the mesh (Figure \ref{jakobshavn_profile}).  The sharp decline in temperature near the surface, followed by the sharp increase at the next vertex of the mesh, suggests that the thermal gradient near the surface is creating numerical oscillations.  In such a case, increasing the vertical resolution of the mesh will reduce the instability and result in a higher-magnitude temperature gradient near the surface, hence reducing the temperature to a level closer to expectations.

Remarkably, nearly the entire column of ice is temperate along the shear margins nearest the terminus (Figure \ref{jakobshavn_results}).  As a result, basal traction $\beta$ must increase in order to compensate for the approximately three-fold enhancement of flow-rate factor (\ref{rate_factor}) and relatively low surface speed there.  Additionally, the basal water discharge optimization process does not appear to compensate for the relatively high basal water content of these regions; it may be necessary to append energy objective (\ref{energy_objective}) with an additional term favoring areas with largest misfit, similar to momentum objective (\ref{momentum_objective}).  Finally, we note that an entirely different ice-flow model applies to the shear margins due to the extreme levels of strain-heat and damaged ice in this region.

\begin{figure*}
  \centering
    \includegraphics[width=\linewidth]{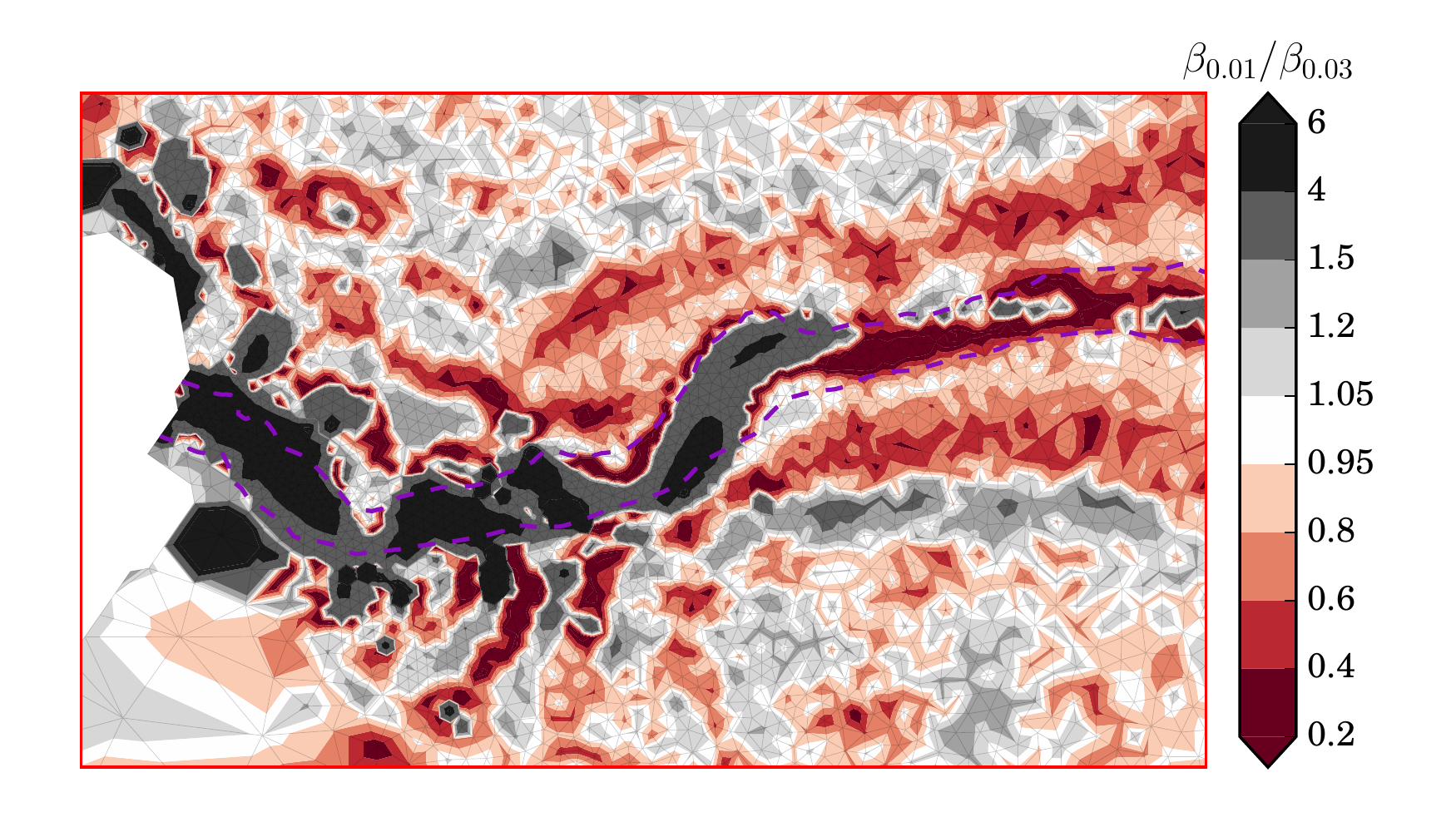}
  \caption[Effect of water content on basal traction]{Ratio of basal traction attained by Algorithm \ref{tmc_da} between using maximum basal water content $W_c = 0.01$ and $W_c = 0.03$.  These results were obtained using fewer momentum optimization iterations, resulting in traction fields which are more irregular than that depicted in Figure \ref{jakobshavn_results}.}
  \label{deltas}
\end{figure*}

For an additional test, the data-assimilation process described by Algorithm \ref{tmc_da} was performed once more for ten iterations with two simulations; one using a maximum basal water content $W_c =0.03$ and one using $W_c = 0.01$, in the hope of comparison.  The ratio of basal traction fields derived using maximum water content $W_c = 0.01$ and $W_c = 0.03$ is shown in Figure \ref{deltas}.  While the traction appears to be up to six times greater along the main trench, it is important to note that the friction is quite small in this region (Figure \ref{jakobshavn_results}) and thus has little effect on the velocity field.  However, the traction along the flanks of the trench approximately 20 km inland are up to half has large when the maximum basal water content is reduced from 3\% to 1\%.  This may be explained by both the reduction in height of the CTS and the reduction in water content to a point below that required by the empirically-constrained water content relation $W_f$ in rate factor (\ref{rate_factor}).

\section{Conclusion}

The methods presented in \S \ref{ssn_water_content_optimization} and \S \ref{ssn_dual_optimization} offer a new approach to approximate the energy and momentum distributions of an ice-sheet or glacier.  This method, having been derived from established balance equations, provides the ability to match energy distributions to observations of internal-water content.  This method results in a consistent estimate of velocity, energy, basal-water discharge, and basal traction.  Applying this procedure over the region of Jakobshavn indicated that altering the maximum-allowed basal-water content has a significant effect on basal traction values derived by data-assimilation methods.

\pythonexternal[label=cslvr_jakobshavn_gen_mesh, caption={CSLVR script used to generate the mesh for the Jakobshavn simulation.}]{scripts/jakobshavn/gen_jakobshavn_mesh_small_block.py}

\pythonexternal[label=cslvr_jakobshavn_gen_data, caption={CSLVR script used to generate the data used by Code Listing \ref{cslvr_jakobshavn_da}.}]{scripts/jakobshavn/gen_jakobshavn_vars_small.py}

\pythonexternal[label=cslvr_jakobshavn_da, caption={CSLVR script used to simultaneously assimilate data $\mathbf{u}_{ob}$ and $W_c$.}]{scripts/jakobshavn/data_assimilation.py}

\begin{figure*}
  
  \centering
  
  \begin{subfigure}[b]{0.33\linewidth}
    \includegraphics[width=\linewidth]{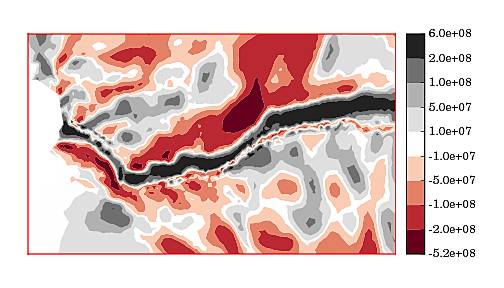}
  \caption{$N_{ii}$}
  \label{N_ii}
  \end{subfigure}
  \begin{subfigure}[b]{0.33\linewidth}
    \includegraphics[width=\linewidth]{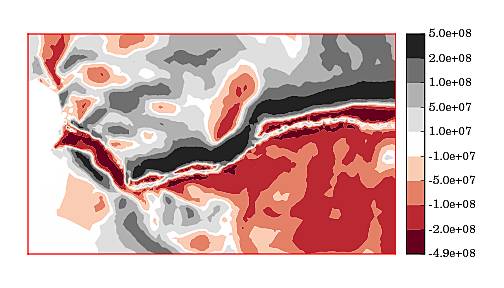}
  \caption{$N_{ij}$}
  \label{N_ij}
  \end{subfigure}
  \begin{subfigure}[b]{0.33\linewidth}
    \includegraphics[width=\linewidth]{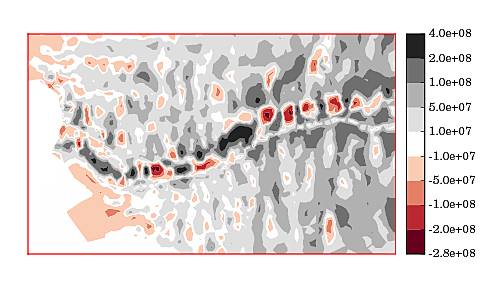}
  \caption{$N_{iz}$}
  \label{N_iz}
  \end{subfigure}

  \begin{subfigure}[b]{0.33\linewidth}
    \includegraphics[width=\linewidth]{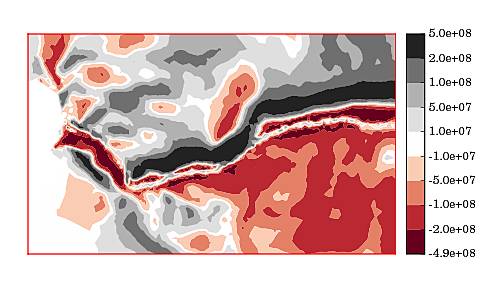}
  \caption{$N_{ji}$}
  \label{N_ji}
  \end{subfigure}
  \begin{subfigure}[b]{0.33\linewidth}
    \includegraphics[width=\linewidth]{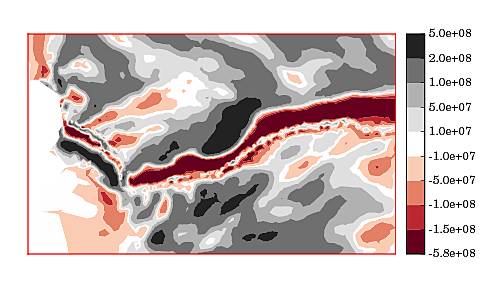}
  \caption{$N_{jj}$}
  \label{N_jj}
  \end{subfigure}
  \begin{subfigure}[b]{0.33\linewidth}
    \includegraphics[width=\linewidth]{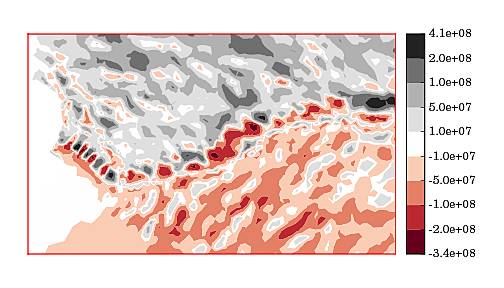}
  \caption{$N_{jz}$}
  \label{N_jz}
  \end{subfigure}

  \begin{subfigure}[b]{0.33\linewidth}
    \includegraphics[width=\linewidth]{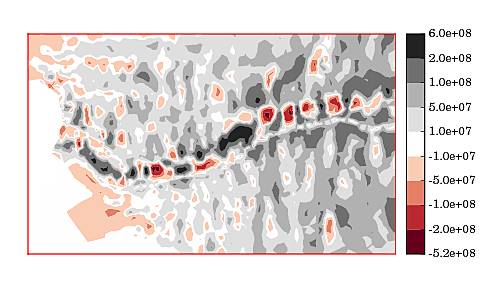}
  \caption{$N_{zi}$}
  \label{N_zi}
  \end{subfigure}
  \begin{subfigure}[b]{0.33\linewidth}
    \includegraphics[width=\linewidth]{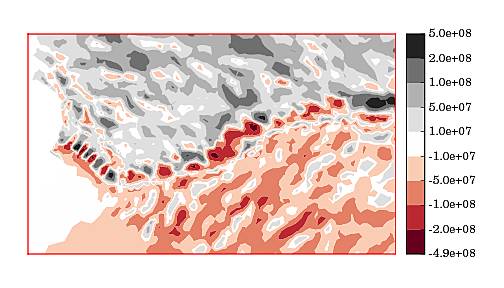}
  \caption{$N_{zj}$}
  \label{N_zj}
  \end{subfigure}
  \begin{subfigure}[b]{0.33\linewidth}
    \includegraphics[width=\linewidth]{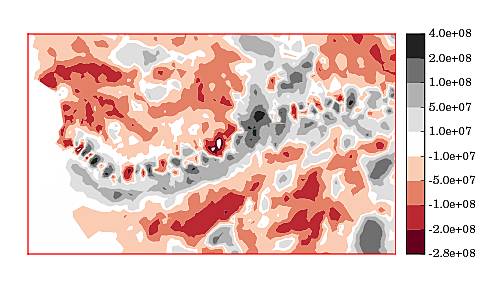}
  \caption{$N_{zz}$}
  \label{N_zz}
  \end{subfigure}

  \caption[Jakobshavn membrane stress]{Membrane stress $N_{kk}$.}
  \label{jakobshavn_membrane_stress}

\end{figure*}

\begin{figure*}
  
  \centering
  
  \begin{subfigure}[b]{0.33\linewidth}
    \includegraphics[width=\linewidth]{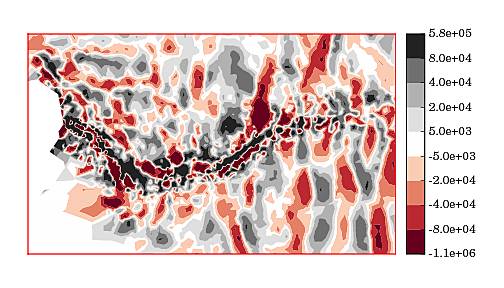}
  \caption{$M_{ii}$}
  \label{M_ii}
  \end{subfigure}
  \begin{subfigure}[b]{0.33\linewidth}
    \includegraphics[width=\linewidth]{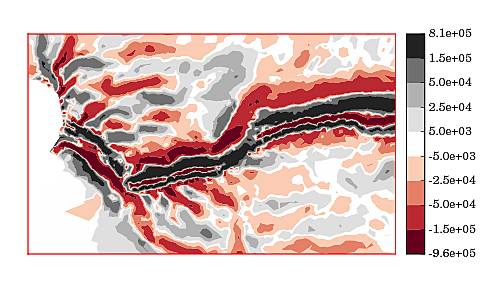}
  \caption{$M_{ij}$}
  \label{M_ij}
  \end{subfigure}
  \begin{subfigure}[b]{0.33\linewidth}
    \includegraphics[width=\linewidth]{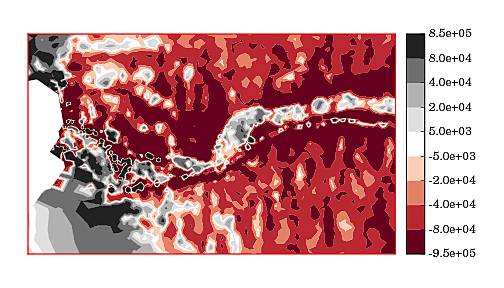}
  \caption{$M_{iz}$}
  \label{M_iz}
  \end{subfigure}

  \begin{subfigure}[b]{0.33\linewidth}
    \includegraphics[width=\linewidth]{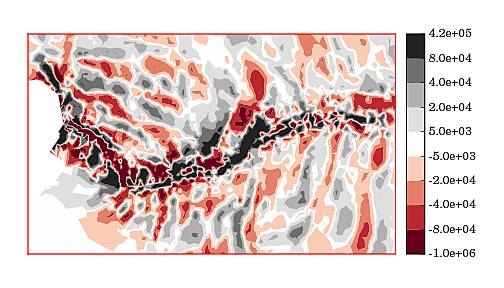}
  \caption{$M_{ji}$}
  \label{M_ji}
  \end{subfigure}
  \begin{subfigure}[b]{0.33\linewidth}
    \includegraphics[width=\linewidth]{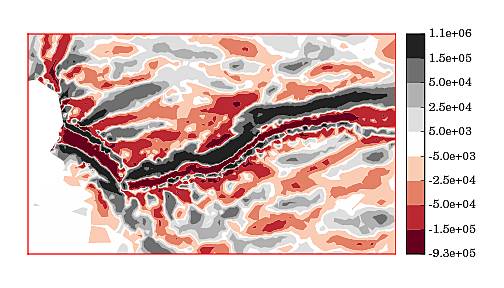}
  \caption{$M_{jj}$}
  \label{M_jj}
  \end{subfigure}
  \begin{subfigure}[b]{0.33\linewidth}
    \includegraphics[width=\linewidth]{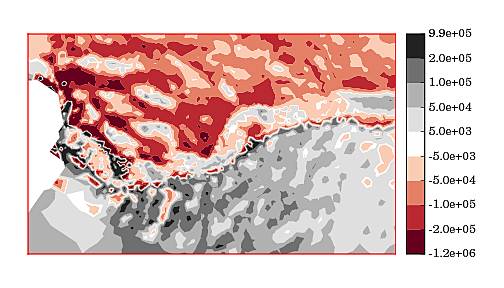}
  \caption{$M_{jz}$}
  \label{M_jz}
  \end{subfigure}

  \begin{subfigure}[b]{0.33\linewidth}
    \includegraphics[width=\linewidth]{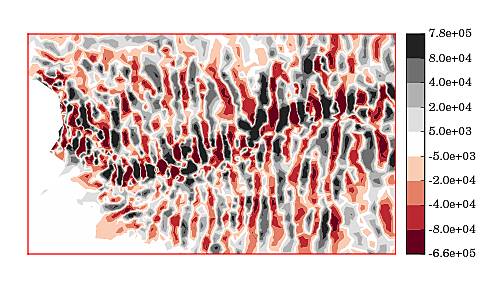}
  \caption{$M_{zi}$}
  \label{M_zi}
  \end{subfigure}
  \begin{subfigure}[b]{0.33\linewidth}
    \includegraphics[width=\linewidth]{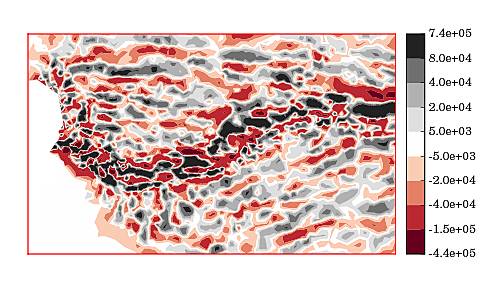}
  \caption{$M_{zj}$}
  \label{M_zj}
  \end{subfigure}
  \begin{subfigure}[b]{0.33\linewidth}
    \includegraphics[width=\linewidth]{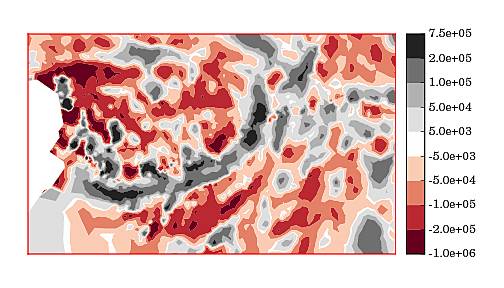}
  \caption{$M_{zz}$}
  \label{M_zz}
  \end{subfigure}

  \caption[Jakobshavn membrane stress balance]{Membrane stress balance $M_{kk}$.}
  \label{jakobshavn_membrane_stress_balance}

\end{figure*}

\pythonexternal[label=cslvr_jakobshavn_sb, caption={CSLVR script used calculate the stress-balance for the Jakobshavn simulation results as depicted in Figures \ref{jakobshavn_membrane_stress} and \ref{jakobshavn_membrane_stress_balance}.}]{scripts/jakobshavn/stress_balance.py}

\pythonexternal[label=cslvr_jakobshavn_region, caption={CSLVR script used generate Figures \ref{jakobshavn_region} and \ref{jakobshavn_topography}.}]{scripts/jakobshavn/plot_region.py}

%===============================================================================
%===============================================================================

\appendix
\renewcommand\chaptername{Appendix}

\setcounter{equation}{0}
\renewcommand\theequation{A\arabic{equation}}
\chapter{Jump condition at the basal surface}

In order to clearly illustrate the derivation of latent energy flux (\ref{latent_flux}), the process described in \S~9.3 of \citet{greve} is rewritten using the differing notation and energy definition (\ref{temperate_energy}) used here.

First, the general jump condition of a singular surface $\sigma$ located within the material volume $\Omega$ is
\begin{align}
  \label{general_jump_condition}
  \forall \mathbf{x} \in \sigma : && \llbracket \psi \rrbracket (\mathbf{x},t) = \psi^+(\mathbf{x},t) - \psi^-(\mathbf{x},t),
\end{align}
where $\mathbf{x} = [x\ y\ z]^\intercal$ is the position vector of the surface, $t$ is time, and
\begin{align*}
  \forall \mathbf{x} \in \sigma : && \psi^- (\mathbf{x},t) = \lim_{\mathbf{y} \rightarrow \mathbf{x}, \mathbf{y} \in \Omega^-} \psi(\mathbf{y},t) \\
  && \psi^+ (\mathbf{x},t) = \lim_{\mathbf{y} \rightarrow \mathbf{x}, \mathbf{y} \in \Omega^+} \psi(\mathbf{y},t).
\end{align*}
By convention, the positive side of the ice base is identified with the lithosphere and negative side with the ice.

Next, the \emph{mass jump condition} of a substance $s$ with density $\rho_s$ on singular surfaces is defied as 
\begin{align}
  \label{mass_jump_relation}
  \llbracket \rho_s (\mathbf{u}_s - \mathbf{w}) \cdot \mathbf{n} \rrbracket = P,
\end{align}
where $\mathbf{u}_s$ is the substance velocity, $\mathbf{w}$ is the velocity of the singular surface, $\mathbf{n}$ is the outward-pointing normal vector, and $P$ is the rate of production of the substance on the singular surface.  For the mixture here, we have some component of ice and water, denoted with subscripts $i$ and $w$, respectively, with \emph{barycentric velocity}
\begin{align}
  \label{barycentre}
  \mathbf{u} = \frac{1}{\rho} \left( \tilde{\rho}_i \mathbf{u}_i + \tilde{\rho}_w \mathbf{u}_w \right).
\end{align}
with partial densities $\tilde{\rho}_i$ and $\tilde{\rho}_w$ defined as the mass of ice and water per unit volume of the mixture.  Thus, the water content of the mixture is defined as
\begin{align}
  \label{mass_fraction_water_content}
  W = \frac{\tilde{\rho}_w}{\rho}, \hspace{10mm} (1-W) = \frac{\tilde{\rho}_i}{\rho}.
\end{align}
In addition, a non-advective water mass flux $\mathbf{j}$ describes the water motion relative to the motion of barycentre (\ref{barycentre}),
\begin{align}
  \label{diffusive_latent_flux}
  \mathbf{j} = \tilde{\rho}_w (\mathbf{u}_w - \mathbf{u}) = \rho W (\mathbf{u}_w - \mathbf{u}).
\end{align}

Next, the \emph{mass balance for the component water} is defined as
\begin{align*}
  \frac{\partial \tilde{\rho}_w}{\partial t} + \nabla \cdot \left( \tilde{\rho}_w \mathbf{u}_w \right) = M,
\end{align*}
where $M$ is the rate of water mass produced per unit mixture volume.  Applying water content definition (\ref{mass_fraction_water_content}) and non-advective water mass flux (\ref{diffusive_latent_flux}), this is equivalent to
\begin{align}
  \label{water_mass_balance}
  \rho \dot{W} = - \nabla \cdot \mathbf{j} + M,
\end{align}
where the Newton-overdot notation ($\dot{\ }$) denotes time differentiation.  The constitutive relations used to close the system are
\begin{align}
  \label{energy_time_deriavative}
  \dot{\theta} &= a \dot{T}_m + b T_m \dot{T}_m + L_f \dot{W} \\
  \label{water_fick_law}
  \mathbf{j} &= - \tilde{\nu} \nabla W \\
  \label{temperate_sensible_energy}
  \mathbf{q}_s &= - k(T_m) \nabla T_m,
\end{align}
where $\tilde{\nu} = \nu / L_f$ is the `water diffusivity' as presented in \citet{greve97}, and energy definition (\ref{temperate_energy}) was used to derive energy time derivative (\ref{energy_time_deriavative}).  The second relation is Fick's diffusion law for the motion of water, and the last term is the sensible energy flux using Fourier's law of heat conduction.  Using this notation, the total heat flux is
\begin{align*}
  \mathbf{q} = \mathbf{q}_s + \mathbf{q}_l = \mathbf{q}_s + L_f \mathbf{j},
\end{align*}
and using the stress and strain constitutive relation expressed through shear viscosity (\ref{viscosity}) in strain-heat definition (\ref{strain_heat}), the \emph{mixture energy balance} is thus \citep{greve}
\begin{align}
  \label{mixture_energy_balance}
  \rho \dot{\theta} = - \nabla \cdot \left( \mathbf{q}_s + L_f \mathbf{j} \right) + Q.
\end{align}
Introducing constitutive relations (\ref{energy_time_deriavative} -- \ref{temperate_sensible_energy}, \ref{strain_heat}) into water mass balance (\ref{water_mass_balance}) and mixture energy balance (\ref{mixture_energy_balance}) yield respectively
\begin{align}
  \label{water_mass_balance_revised}
  \rho \dot{W} = \tilde{\nu} \nabla \cdot \nabla W + M
\end{align}
and
\begin{align}
  \label{revised_mixture_energy_balance}
  \rho \left( a \dot{T}_m + b T_m \dot{T}_m + L_f \dot{W} \right) = \nabla \cdot \left( k \nabla T_m \right) + L_f \tilde{\nu} \nabla \cdot \nabla W + Q.
\end{align}
Solving for the water content time derivative term, 
\begin{align*}
  L_f \rho \dot{W} = \nabla \cdot \left( k \nabla T_m \right) + L_f \tilde{\nu} \nabla \cdot \nabla W + Q - \rho a \dot{T}_m - \rho b T_m \dot{T}_m,
\end{align*}
and inserting (\ref{water_mass_balance_revised}), the expression for the water production rate is therefore
\begin{align*}
  L_f M = \nabla \cdot \left( k \nabla T_m \right) + Q - \rho a \dot{T}_m - \rho b T_m \dot{T}_m.
\end{align*}

Next, the \emph{mass jump relation for the component water} is defined using (\ref{mass_jump_relation}),
\begin{align*}
  \llbracket \tilde{\rho}_w (\mathbf{u}_w - \mathbf{w}) \cdot \mathbf{n} \rrbracket = \rho M_b,
\end{align*}
where the water-production term $P = \rho M_b$ in (\ref{mass_jump_relation}) has been defined using basal melting rate (\ref{basal_melt_rate}).  Using general jump condition (\ref{general_jump_condition}), we have
\begin{align*}
  \tilde{\rho}_w^- F_b - \tilde{\rho}_w^+ (\mathbf{u}_w^+ - \mathbf{w}) \cdot \mathbf{n} = \rho M_b,
\end{align*}
with \emph{water mass flux into the base} $\tilde{\rho}_w^- F_b$ defined with basal water discharge
\begin{align*}
  F_b = (\mathbf{u}_w^- - \mathbf{w}) \cdot \mathbf{n}.
\end{align*}
Using water content definition (\ref{mass_fraction_water_content}) and assuming the water content on the lithosphere side is composed entirely of water,
\begin{align}
  \label{component_water_jump}
  \rho W (\mathbf{u}_w - \mathbf{w}) \cdot \mathbf{n} = \rho_w F_b - \rho M_b,
\end{align}

Next, the \emph{mass jump relation for the component ice} is similarly defined as
\begin{align*}
  \llbracket \tilde{\rho}_i (\mathbf{u}_i - \mathbf{w}) \cdot \mathbf{n} \rrbracket = - \rho M_b.
\end{align*}
Because the lithosphere is impermeable to ice, as evident by impenetrability condition (\ref{impenetrability}), this simplifies to
\begin{align}
  \label{component_ice_jump}
  \tilde{\rho}_i (\mathbf{u}_i - \mathbf{w}) \cdot \mathbf{n} = \rho (1-W)(\mathbf{u}_i - \mathbf{w}) = \rho M_b.
\end{align}

Next, using barycentric velocity (\ref{barycentre}) and water content (\ref{mass_fraction_water_content}), it follows that
\begin{align*}
  \mathbf{u} - \mathbf{w} = W(\mathbf{u}_w - \mathbf{w}) + (1-W)(\mathbf{u}_i - \mathbf{w}),
\end{align*}
which upon scalar multiplication by $\mathbf{n}$ and use of water jump (\ref{component_water_jump}) and ice jump (\ref{component_ice_jump}), we have
\begin{align*}
  (\mathbf{u} - \mathbf{w}) \cdot \mathbf{n} &= \frac{\rho_w}{\rho} F_b. 
\end{align*}
Using non-advective water mass flux (\ref{diffusive_latent_flux}), component water jump (\ref{component_water_jump}), and component ice jump (\ref{component_ice_jump}), we have the flux of water normal to the basal boundary
\begin{align*}
  \mathbf{j} \cdot \mathbf{n} &= \rho W (\mathbf{u}_w - \mathbf{u}) \cdot \mathbf{n} \notag \\
  &= \rho W (\mathbf{u}_w - \mathbf{w}) \cdot \mathbf{n} - \rho W (\mathbf{u} - \mathbf{w}) \cdot \mathbf{n} \notag \\
  &= \rho_w F_b - \rho M_b - W \rho_w F_b \notag \\
  &= (1 - W) \rho_w F_b - \rho M_b \approx \rho_w F_b - \rho M_b.
\end{align*}

Finally, from water-flux constitutive relation (\ref{water_fick_law}) we have 
\begin{align*}
  \left( \nu \nabla W \right) \cdot \mathbf{n} &= \left( L_f \tilde{\nu} \nabla W \right) \cdot \mathbf{n} = - L_f \mathbf{j} \cdot \mathbf{n}  \notag \\
  &= \rho L_f M_b - (1-W) \rho_w L_f F_b \notag \\
  &\approx \rho L_f M_b - \rho_w L_f F_b, 
\end{align*}
Hence water flux boundary condition (\ref{latent_flux}) has been derived. $\qed$

%===============================================================================
%===============================================================================

\setcounter{equation}{0}
\renewcommand\theequation{B\arabic{equation}}
\chapter{Leibniz formula} \label{leibniz_formula}

\index{Leibniz's Rule}
Leibniz formula, referred to as \emph{Leibniz's rule for differentiating an integral with respect to a parameter that appears in the integrand and in the limits of integration}, states that
\begin{align*}
  \frac{d}{dx} \int_{a(x)}^{b(x)} F(x,y) dy = &+ \int_{a(x)}^{b(x)} F_x(x,y) dy \\
  &+ F(x,b(x))b'(x) - F(x,a(x))a'(x),
\end{align*}
where $F$ and $F_x$ are both continuous over the domain $[a,b]$.

\vspace{10mm}

\noindent \textbf{Proof:}

Let
$$I(x,a,b) = \frac{d}{dx} \int_{a(x)}^{b(x)} F(x,y) dy.$$
Then
$$\frac{dI}{dx} = \frac{\partial I}{\partial x} \frac{\partial x}{\partial x} + \frac{\partial I}{\partial a} \frac{\partial a}{\partial x} + \frac{\partial I}{\partial b} \frac{\partial b}{\partial x},$$
and
\begin{align*}
  \frac{\partial I}{\partial x} &= \frac{\partial}{\partial x} \int_{a}^{b} F(x,y) dy = \int_{a(x)}^{b(x)} F_x(x,y) dy \\
  \frac{\partial x}{\partial x} &= 1 \\
  \frac{\partial I}{\partial a} &= \frac{\partial}{\partial a} \int_{a(x)}^{b(x)} F(x,y) dy = \frac{\partial}{\partial a} \Big[ f(x,b(x)) - f(x,a(x)) \Big] = - F(x,a(x)) \\
  \frac{\partial a}{\partial x} &= a'(x) \\
  \frac{\partial I}{\partial b} &= \frac{\partial}{\partial b} \int_{a(x)}^{b(x)} F(x,y) dy = \frac{\partial}{\partial b} \Big[ f(x,b(x)) - f(x,a(x)) \Big] = F(x,b(x)) \\
  \frac{\partial b}{\partial x} &= b'(x),
\end{align*}
and thus
$$\frac{dI}{dx} = \int_{a(x)}^{b(x)} F_x(x,y) dy + F(x,b(x))b'(x) - F(x,a(x))a'(x) \qed$$

\backmatter

\addcontentsline{toc}{chapter}{References}
\printbibliography

\addcontentsline{toc}{chapter}{Index}
\printindex

\end{document}